\documentclass[preprint]{elsart}
\usepackage{graphicx}
\usepackage{hyperref}
\usepackage{natbib}
\usepackage{amssymb}
\usepackage{amsmath}
\usepackage{booktabs}
\usepackage[usenames, dvipsnames]{xcolor}

\newcommand{\Ang}{\mbox{\AA}}

\newcommand{\avg}[1]{\ensuremath{\langle #1 \rangle}}
\newcommand{\bma}{\begin{math}}
\newcommand{\ema}{\end{math}}
\newcommand{\beq}{\begin{equation}}
\newcommand{\eeq}{\end{equation}}
\newcommand{\beqa}{\begin{eqnarray}}
\newcommand{\eeqa}{\end{eqnarray}}
\newcommand{\bc}{\begin{center}}
\newcommand{\ec}{\end{center}}
\newcommand{\bit}{\begin{itemize}}
\newcommand{\eit}{\end{itemize}}

\font\BFd=cmmib10
\font\BFt=cmmib10
\font\BFs=cmmib10 scaled 700
\font\BFss=cmmib10 scaled 500

\def\bbox#1{%
\relax\ifmmode
\mathchoice
{{\hbox{\BFd #1}}}
{{\hbox{\BFt #1}}}
{{\hbox{\BFs #1}}}
{{\hbox{\BFss #1}}}
\else \mbox{#1} \fi }

\def\k{{\bbox{k}}}

\def\x{{\bbox{x}}}

\begin{document}

\begin{frontmatter}
\title{Line-Intensity Mapping}
\author[ad1,ad2]{Tzu-Ching Chang\thanksref{corr1}},
\author[ad3]{Adam Lidz\thanksref{corr2}}
\thanks[corr1]{E-mail: {\tt
    tzu@caltech.edu}.}
\address[ad1]{Jet Propulsion Laboratory, California Institute of
  Technology \\ 4800 Oak Groove Drive,
  \cty Pasadena, CA 91109, \cny USA}
\address[ad2]{California Institute of Technology, 1200 E. California Boulevard, Pasadena, CA 91125, USA}
\thanks[corr2]{E-mail: {\tt
    alidz@sas.upenn.edu}.}
\address[ad3]{Department of Physics and Astronomy, University of
  Pennsylvania \\ 209 South 33rd Street, \cty Philadelphia, PA 19104,
  \cny USA}

\begin{abstract}
Line-Intensity Mapping (LIM) has emerged as a powerful technique for studying large-scale structure and the high-redshift universe, enabling three-dimensional maps of
line emission across vast cosmological volumes. 
In this review, we summarize the LIM framework, its key scientific goals, and its future prospects. We describe
the landscape of emission line tracers, theoretical modeling approaches, anticipated signals, and data-analysis methodologies.  We also discuss
experimental challenges, particularly those posed by astrophysical foregrounds, and review possible mitigation strategies. Further, we highlight a range of cross-correlation science cases, linking LIM with other cosmological surveys. Finally, we summarize current and upcoming experiments and early results, including recent first detections, while outlining the outlook for future discoveries. Specifically, LIM may offer new insights into galaxy formation and evolution and cosmology, while revealing the Epoch of Reionization, Cosmic Dawn, and possibly the Cosmic Dark Ages. LIM enables cosmological measurements that complement other probes and provide unique access to the high-redshift universe, potentially shedding light on dark matter, dark energy, and cosmic inflation.   
\end{abstract}

\begin{keyword}
Cosmology: theory - intergalactic medium - interstellar medium - line:
formation \sep large-scale structure \sep technique: intensity mapping 

\PACS 98.80.-v \sep 96.30.Ks \sep 96.40.Jw
\end{keyword}
\end{frontmatter}

\tableofcontents

\section{Introduction}
\label{S:intro}

This review aims to provide a broad overview of the emerging field of line-intensity mapping (LIM). We will define this technique and describe its primary science goals; identify promising target lines; summarize the relevant line emission physics; discuss line-intensity mapping signal models; present simulation efforts, analysis
techniques, and statistical methodologies; comment on the potential for cross-correlation measurements; describe science forecasts and existing constraints; discuss systematic challenges and the approaches for mitigating them; and summarize the current experimental landscape.

\subsection{What is Line-Intensity Mapping?}

The aim of LIM surveys is to measure the fluctuations in
the intensity of line radiation as a function of angle on the sky and
observed frequency. The line radiation stretches in wavelength as it propagates through the expanding universe and so there is a direct translation between observed frequency and redshift for each line. Under an assumed cosmological model, the redshift and angle on the sky can be converted into radial and transverse distances, and so the joint spectral and angular information probes the line-intensity fluctuations in three spatial dimensions.
In general, the three-dimensional line-intensity fluctuations trace the underlying large-scale structure of the universe; that is, on large spatial scales the matter is distributed across a ``cosmic web'' of sheets, filaments, and nodes, with empty voids in between. Furthermore, the line-intensity fluctuation measurements may be performed across a broad range of observed frequencies to trace the evolution of large-scale structure across redshift and hence cosmic time. 

This approach may be contrasted with that of traditional galaxy surveys, which 
identify individual bright sources of emission above a flux limit within some
narrow observing aperture. In LIM, rather than detecting individual sources, one measures the spatial intensity fluctuations from {\em all} of the photons received as opposed to using only those above some flux threshold. Typically, LIM surveys measure the collective emission from many individually unresolved sources. One motivation for this is that it is economical: in order to measure large-scale intensity variations, high angular resolution is unnecessary, and so this avoids the need for expensive large-aperture telescopes. LIM also probes the collective impact of low luminosity sources which may be too faint to detect in targeted surveys,  economically traces large cosmic volumes, and is well-suited for studying diffuse and extended sources of emission. 

As we will discuss, LIM is a broadly applicable technique which encompasses a variety of spectral line targets towards diverse scientific goals. Ultimately, the power of this method will arise partly from combining line-intensity measurements across a full suite of different lines spanning common volumes of the universe over as much of cosmic history as possible. This will allow one to trace atomic, molecular, and ionized gas, probing
the ensemble-averaged properties of the interstellar, circumgalactic, and intergalactic media across cosmic time.

This line-intensity mapping technique may be viewed as an extension of two-dimensional intensity mapping measurements. The prime example of two-dimensional intensity mapping is the case of cosmic microwave background (CMB) anisotropy
measurements. The primary CMB anisotropies refer to angular variations in the CMB temperature (or equivalently intensity) observed across the sky. These offer a snapshot view of the surface of last scattering at a redshift of $z \sim 1,100$ when the universe was around $380,000$ years old. The precise CMB anisotropy measurements and the simple linear physics involved in their modeling yield a wealth of information regarding the composition, structure, geometry, and history of our universe.
In addition to the CMB, there are other successful two-dimensional measurements 
of intensity variations, which trace the {\em projected} fluctuations in the distribution of light \cite{Hill:2018trh}. In these cases, the light is typically generated by continuum processes rather than line emission\footnote{Although, in some cases, line-emission makes a significant contribution to the broad-band emission as well.}, and structure along the line-of-sight across a range of redshifts (e.g. \cite{Cooray16}) is probed. Examples here include fluctuation measurements in the near-infrared
background \cite{Zemcov14,Kashlinsky15,Helgason16}, observations of far-infrared fluctuations \cite{Puget96, Hauser01, Dole06, PlanckXXX, Viero13}, and measurements of variations in the optical \cite{Mitchell-Wynne:2015rha}, X-ray \cite{Dijkstra04}, and gamma-ray backgrounds \cite{Fermi-LAT:2012pez}.

The CMB anisotropy and other two-dimensional intensity mapping measurements, along with expertise acquired from traditional galaxy redshift surveys, provide a rich heritage that LIM efforts hope to utilize and extend upon. As we will see, although the analysis, modeling, and instrumental expertise acquired in these earlier efforts are invaluable, LIM has its own unique set of challenges. This review is focused on the opportunities and challenges of LIM. This work complements the recent review article in reference~\cite{Bernal:2022jap}, which overlaps with this article but is written from a different perspective. Both reviews update an earlier 2017 community-wide summary in reference~\cite{Kovetz:2017agg}. For shorter white paper overviews, we refer the reader to references~\cite{Kovetz:2019uss} and \cite{Chang:2019xgc}.

\subsection{Why Line-Intensity Mapping?}

In considering  ``Why Line-Intensity Mapping?'', we first note that this is a new and emerging field, whose diverse applications are still being explored and developed.
Indeed, the general idea has stimulated broad interest
with LIM surveys currently underway, in the planning stages, or being considered using: the HI 21 cm line \cite{HERA:2022wmy,CHIME:2022dwe}, the Lyman-$\alpha$ (Ly-$\alpha$) line \cite{Dore2014,Pullen:2013dir,Silva13}, rotational transitions
from CO molecules \cite{COMAPI2022,Keating:2020wlx}, the [CII] 158 $\mu$m transition and other atomic fine-structure lines \cite{Sun:2020mco,Bethermin2022,FYSTI2022,Cataldo21,TIM2020,Padmanabhan:2021tjr}, the HeII $1640 \, \Ang$ line \cite{Visbal:2015sca,Parsons:2021qyw}, and rotational and vibrational transitions from molecular hydrogen \cite{Gong13}, among others.
Furthermore, the planned surveys aim to probe a wide range of different time periods, across almost the entire 13.8 billion year history of our universe, and are motivated by
a diverse set of scientific topics, including studies of:
the Epoch of Reionization (EoR); the first luminous sources; the large-scale structure in the distribution of galaxies near
the epoch of peak cosmic star formation; the bulk properties of the interstellar medium -- averaged over large ensembles of galaxies -- across cosmic time; the
primordial power spectrum of density fluctuations; dark matter properties; and the accelerating expansion of the universe.  
Although the scientific applications are wide-ranging, all of the various LIM surveys share a common methodology. Furthermore, recurring themes
emerge as one considers the advantages of the LIM approach in tackling this broad range of open astrophysical and cosmological questions.  
It is therefore helpful and timely to review the subject as a unified whole.

\begin{figure}
\begin{center}
\includegraphics[width=\textwidth]{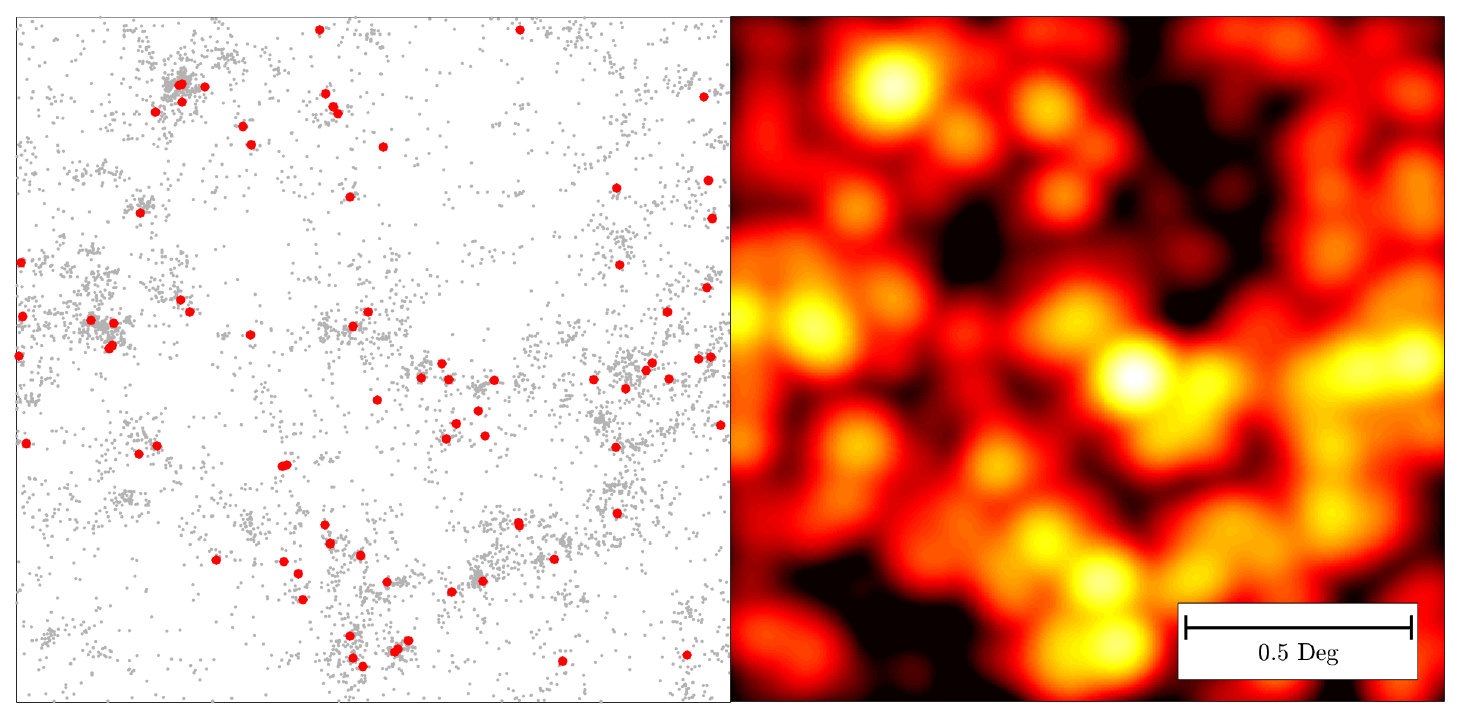}
\caption{Illustration of the line-intensity mapping approach. {\em Left panel}: The gray points show a model for the CO(1-0)
emission from simulated galaxies near $z \sim 3$ across a 2.5 deg$^2$ field. The red points show the sources that would be sufficiently bright to observe with a traditional galaxy survey (using an extremely long integration with the VLA). {\em Right panel}: For contrast, this panel shows the (smoothed, yet noise-free) CO(1-0) brightness temperature fluctuations from the same region, as one would measure in a LIM survey. The line-intensity fluctuations trace the total clustered emission, including the collective impact of galaxies that are too faint to detect individually in even an ambitious traditional galaxy survey. 
From Patrick Breysse, \cite{Kovetz:2017agg}.}
\label{fig:illust}
\end{center}
\end{figure}

Stated generally, LIM is advantageous for studying diffuse radiation backgrounds and/or large-scale fluctuations in the line emission from discrete sources.  As noted above, unlike traditional surveys that target only sources above a flux limit, LIM measures the aggregate emission from all sources, resolved and unresolved. This methodology underpins the diverse scientific goals we outline below.

Figure~\ref{fig:illust} provides an illustration, contrasting the case of a traditional galaxy survey targeting galaxies emitting in the CO(1-0) line with what a LIM survey in the same line might capture.
As an example, note that none of the CO emitters in the lower left-hand corner of the left panel of the figure is bright enough to make it into the traditional galaxy survey catalog, but their collective emission nevertheless produces a bright spot in the CO line-intensity map (right panel).
By probing spectral lines and by covering a wide range in observed frequency, LIM 
observations proceed tomographically to map out emission fluctuations as a function of redshift.
LIM may thereby provide an economical way of probing large cosmological volumes: unlike in the case of
traditional surveys, high angular resolution is not required, and so small telescope apertures generally suffice. On the other hand, large collecting areas are often
needed to achieve sensitivity to the faint emission fluctuations of interest. 

\subsubsection{Science Motivations}

In order to illustrate the broad promise of LIM surveys more concretely, it is helpful to briefly describe several possible applications of the approach and the associated scientific goals. In this list, we include both concrete science targets as well as more general themes of interest for LIM. The topics below hence vary widely in their level of specificity, but we believe they nevertheless help to illustrate the promise and general aims of LIM. 
We will
further elaborate on these examples throughout this review.

\begin{itemize}

\item {\bf The Largest Data Set on the Sky.} A primary goal of observational cosmology is to map out as much of the observable universe as possible. One motivation
for this is that -- in the regime where detector and discreteness noise (shot-noise) are negligible -- the statistical precision at which the primordial matter 
power spectrum may be measured depends on the square-root of the number of Fourier modes enclosed within a survey volume (see \S \ref{sec:pk_var}). Since the number of modes surveyed itself
scales with volume, this motivates covering as much of the sky over the broadest redshift range possible. This may then provide insights regarding the
inflationary epoch and the nature of dark matter, among other topics. A further, more general, motivation is that surveying the universe at (thus far) unexplored times may reveal
surprises: e.g., unanticipated astrophysical sources of radiation, unexpected contributions to the energy budget of the universe, and/or clues as to what lies beneath the now ``standard'', Lambda Cold Dark Matter (LCDM) cosmological model. 

\begin{figure}
\begin{center}
\includegraphics[width=\textwidth]{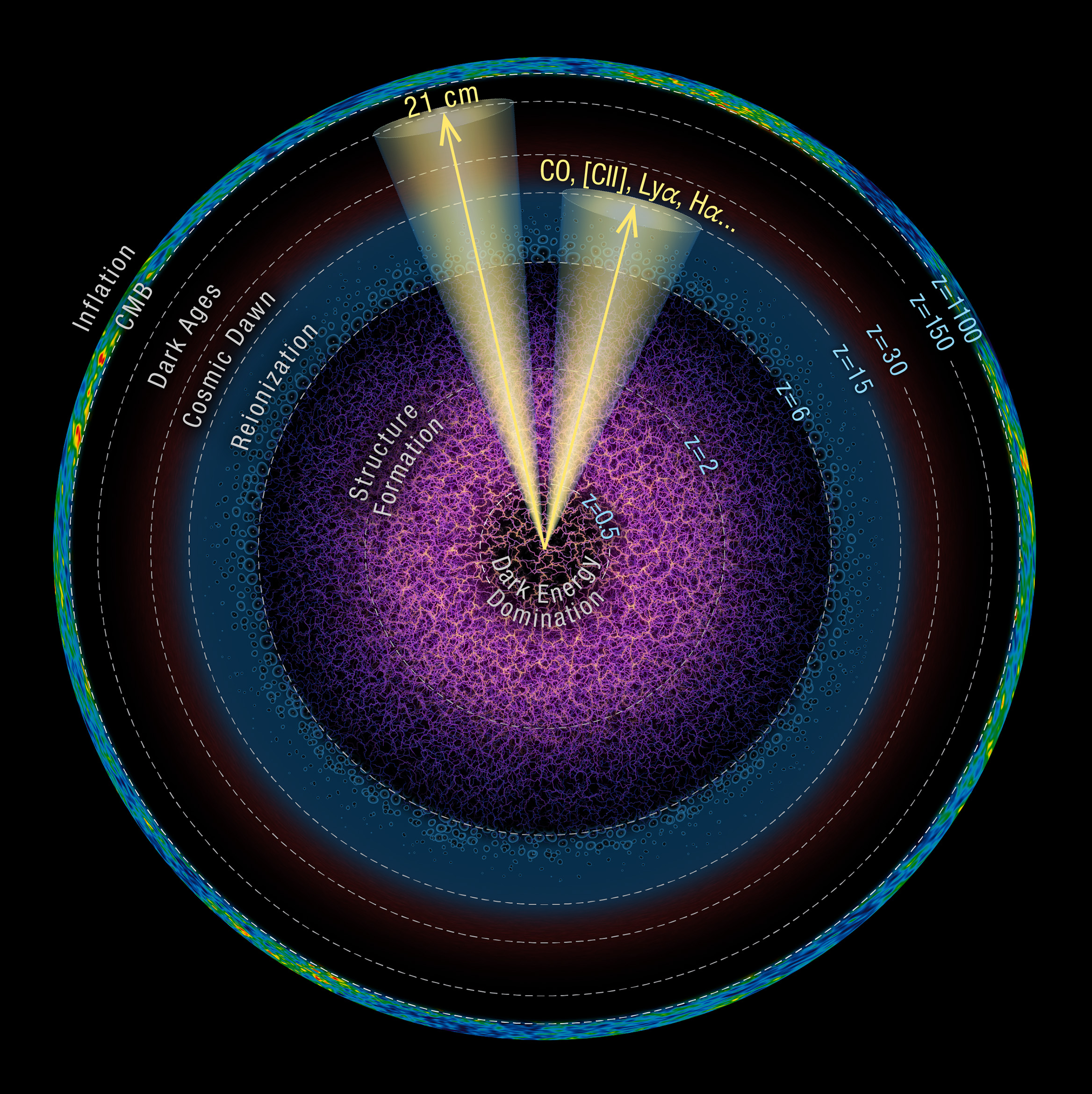}
\label{fig:modes_illust}
\caption{Illustration of the potential observational reach of LIM. The goal of LIM is to map most of the observable universe with multiple spectral line tracers. The circles approximately mark key eras in our cosmic history, and are
scaled according to the comoving distance out to the corresponding redshifts. 
The primary CMB anisotropies probe a thin spherical shell on the surface of last scattering at $z \sim 1,1000$. Traditional galaxy and quasar surveys generally span only a small fraction of the volume of the observable universe out to roughly $z \sim 2-3$.  
The cone marked ``21 cm'' illustrates the wide redshift range accessible (in principle) to LIM surveys using the 21 cm line, spanning from the present day into the EoR, Cosmic Dawn, and ultimately the Cosmic Dark Ages. 21 cm LIM can potentially probe a far greater fraction of the observable universe than otherwise possible.
LIM surveys in other lines can span large portions of the same volume, cross-checking the 21 cm observations and providing valuable additional information. 
The precise redshift range accessible here depends on the lines employed, and how rapidly chemical enrichment occurs, but it should be possible to probe well into the reionization era.
Adapted from C. Chiang (unpublished), and reference~\cite{Tegmark:2008au}.}
\end{center}
\end{figure}
 
LIM studies with the redshifted 21 cm line may ultimately be unparalleled as a probe of raw cosmological volume. In principle, this line can be used to probe almost the entire history of the universe below
$z \lesssim 200$ when the CMB temperature decouples from the gas kinetic temperature \cite{Loeb:2003ya,Pen:2003yv}. At extremely high redshift, the 21 cm line probes neutral hydrogen
in the bulk of the universe before galaxies form. Then, as the first luminous sources turn on, one has an early pre-reionization period in which the Wouthuysen-Field effect and
inhomogeneous X-ray heating determine the structure of the redshifted 21 cm signal \cite{Madau97,Furlanetto:2006pg} (see \S \ref{S:21cm_overview}). Next, the line probes diffuse neutral gas during the EoR, and finally the 21 cm line
can be used to study the post-reionization era using neutral hydrogen left over in individually unresolved galaxies and damped Ly-$\alpha$ systems \cite{Chang:2007xk,Wyithe:2007rq}. In principle, more than $\sim 10^{16}$ Fourier
modes are accessible to 21 cm LIM \cite{Loeb:2003ya}. For contrast, the number of modes probed using Planck observations of the (primary) CMB temperature fluctuations is $\sim$ a few $\times 10^6$ \cite{Tegmark:2008au}. The greater number of modes accessible to 21 cm owes to it being a three-dimensional tracer, while the primary CMB anisotropies probe only a two-dimensional surface, and the CMB fluctuations are washed-out on small scales by Silk damping.  Figure~\ref{fig:modes_illust} provides a schematic illustration of the fraction of the volume of the observable universe accessible to future 21 cm LIM observations and using some other spectral lines. 
Many practical challenges will likely prevent even highly futuristic 21 cm surveys from accessing all of the Fourier modes that may, in principle, be studied, but this nevertheless provides
strong motivation for future efforts.  

Although no other spectral line has the same
potential as the 21 cm line for probing large cosmological volumes, it should be possible to perform related LIM measurements using additional lines across overlapping parts of the
Hubble volume. This will provide a valuable cross-check on the 21 cm line observations and complementary information, as we will discuss.\\

\item {\bf Baryon Acoustic Oscillations.}  Baryon acoustic oscillations (BAOs) are features imprinted in the clustering of galaxies and other large-scale structure tracers, 
sourced by acoustic waves which
propagated in the early universe before cosmological recombination \cite{Eisenstein:1998tu}. The physical scales of the BAOs are set by the sound horizon, the distance acoustic waves travel between the big bang and recombination. The sound horizon provides a standard ruler (i.e., an object of known physical size): the angles spanned by the BAOs, and their redshift extent, then yield clean information regarding the expansion history of the universe.
Detecting BAOs across different redshifts may help in understanding
the accelerating expansion of the universe and dark energy.
The BAO features have traditionally been measured by identifying individual galaxies, and measuring their two-point
correlation function or power spectrum. However, it is unnecessary to actually resolve the galaxies for such measurements: the BAOs should leave their imprint on the large-scale power spectrum of emission fluctuations from individually unresolved galaxies. This realization has motivated efforts to measure the BAO features using the redshifted 21 cm line in the post-reionization
era \cite{Chang:2007xk,Wyithe:2007rq,CHIME2022}, while other emission lines may also provide valuable tracers \cite{Bernal:2019gfq}. These surveys offer an attractive means for mapping large volumes across a range of cosmic times, especially at relatively high redshifts, which are difficult to probe by other means.\\

\item{\bf Velocity-Induced Acoustic Oscillations.} The supersonic relative velocity between dark matter and baryons after recombination can suppress the abundance of the first stars and galaxies \cite{Tseliakhovich10}. By modulating the abundance of the first luminous sources on the coherence scale of these relative velocities, this effect should lead to distinctive $k$-dependent wiggles in the 21 cm power spectrum during Cosmic Dawn \cite{Fialkov13,Munoz:2019fkt}. 
These features, dubbed ``velocity-induced acoustic oscillations'' (VAOs), will be prominent provided the star formation efficiency in small mass halos is sufficiently large, and may supply a standard ruler at $z \sim 15-20$ \citep{Munoz:2019fkt}.\\

\item{\bf Primordial Non-Gaussianity.} In the simplest models of cosmic inflation, the initial fluctuations responsible for structure formation follow a Gaussian distribution. It is important to test this prediction as sharply as possible: departures from Gaussianity offer a potential handle on the physics of inflation, or alternative scenarios, and precious information regarding the earliest phases of our cosmic history. An interesting signature of primordial non-Gaussianity is that it leads to a distinctive scale-dependent clustering signal for dark matter halos and the galaxies/line-emitters that reside within them \cite{Dalal:2007cu}. In the case of so-called ``local-type non-Gaussianity'', the variance of the density fluctuations is enhanced in regions of excess gravitational potential on large scales. In this case, the power spectrum of the dark matter halo distribution has a distinctive $k$-dependence with $P(k) \propto 1/(k^2 T(k))$ where $T(k)$ is the transfer function \cite{Dalal:2007cu}. Other forms of non-Gaussianity, beyond the local-type, also imprint distinctive scale-dependent halo clustering signatures. LIM is a potentially powerful technique to search for these signatures since it promises to capture extremely large cosmological volumes \cite{MoradinezhadDizgah:2018zrs,MoradinezhadDizgah:2018lac}. It will nevertheless be challenging to robustly measure the large-scale Fourier modes of interest where primordial non-Gaussianity leaves the most pronounced effects.\\

\item{\bf Dark Matter Decays.} In a number of plausible models, dark matter particles may decay by emitting photons. A prominent example is the case of axionic dark matter. The axion is a well-motivated particle, originally introduced to solve the strong CP problem, yet the axion is also a good dark matter candidate\footnote{The strong CP problem is that the strong force apparently respects charge-parity (CP) symmetry, although one would generally expect a CP-violating term in the Lagrangian of quantum chromo-dynamics (QCD). This term would give the neutron an electric dipole moment, in strong conflict with observations unless a parameter of the theory, $\bar{\theta}$, is chosen to be unnaturally small. A possible solution to this problem is that $\bar{\theta}$ is a field that dynamically relaxes to zero \cite{PQ77}. This leads to a new particle, the axion, which might also be the dark matter \cite{Wilczek78,Weinberg78}.}. There also might be additional axion-like particles which are irrelevant for the strong CP problem (and do not couple to QCD), but may nevertheless act as dark matter and have observable astrophysical signatures. 
 Axions couple weakly to electromagnetism, and an axion may decay by emitting two photons. If the axion has a mass $m_\chi$, then each photon produced in a decay has a frequency $\nu = m_\chi c^2/2 h$, where $h$ is Planck's constant. This scenario produces interesting and potentially observable signatures in LIM experiments \cite{Creque-Sarbinowski:2018ebl,Bernal:2020lkd,Gong:2015hke}. In particular, the decay photons will source additional fluctuations in line-intensity maps that are not associated with known atomic or molecular emission lines. In addition, these fluctuations would directly track the dark matter distribution and, in general, be less biased tracers than atomic or molecular emission, since the latter lines will be sourced primarily by galaxies residing in massive, highly clustered dark matter halos. 

The intensity fluctuations sourced by dark matter decays may show up in LIM surveys as unexpected interloper contaminants (i.e. as line emission in addition to the usual atomic/molecular transitions targeted by the survey). Hence the search for dark matter decays will be enabled by the currently planned 
LIM surveys, 
designed for other purposes, and this provides an exciting secondary science case. Notably, reference ~\cite{Bernal:2020lkd} forecasts that future LIM observations from e.g. SPHEREx, HETDEX, and AtLAST can place competitive or leading constraints on the QCD axion and axion-like particles in some regions of parameter space (see \S \ref{S:dm_lim} and Fig.~\ref{fig:axion_lim}). Another well-motivated example of a dark matter candidate that may be constrained with this method is sterile neutrino dark matter.\\

\item{\bf Neutrino Mass and Light Relic Abundance.} Neutrino oscillation experiments have provided strong evidence for neutrino mass, in departure from the standard model of particle physics in which -- without extensions -- neutrinos are strictly massless. 
Specifically, neutrino oscillation experiments have measured the difference between the squares of each of two pairs of mass eigenstates, $m_1^2 - m_2^2$, $|m_2^2 - m_3^2|$, but the absolute mass scales remain uncertain \cite{Formaggio:2021nfz}\footnote{Note that only the absolute value of the second pair of mass-squared differences is known.}. 
A key goal in cosmology is hence to determine the sum of the masses of the different neutrino species by exploiting their impact on the growth of large-scale structure and the expansion history of the universe. Although massive neutrinos are relativistic at early times, their temperature drops as the universe expands; they eventually become non-relativistic and comprise a small fraction of the overall matter density of the universe. A key effect here is that neutrinos have significant thermal velocities and so do not cluster on scales smaller than their {\em free-streaming lengths}. The free-streaming scale is the length traveled in the age of the universe by a neutrino moving at its thermal velocity (which varies with cosmic time). This leads to a characteristic suppression in the matter power spectrum relative to a case where all of the matter density is comprised of cold dark matter. LIM may play a valuable role here by surveying large volumes of the universe over a wide redshift range \cite{MoradinezhadDizgah:2021upg}. LIM may also help constrain the abundance of light relics which contribute to the energy density in radiation in the early universe (see \S \ref{S:nu_lim}).\\

\item {\bf The Redshifted 21 cm line and Neutral Hydrogen During the EoR.} Historically, the reionization-era 21 cm signal was the first application of LIM to be considered in
detail (e.g. \cite{Madau97}).
Many ongoing experiments have made remarkable progress in the last several years, and are currently poised to make a first detection of neutral hydrogen in the IGM during reionization, hopefully in the relatively near future (e.g. \cite{HERA:2022wmy,Mertens2020,Trott2020}). In the midst of the EoR, the IGM is expected to resemble a two-phase medium:
highly ionized ``bubbles'' of hydrogen gas form around clustered UV-luminous sources, while regions of mostly neutral hydrogen are intermixed.
The ionized regions imprint large-scale
fluctuations in the redshifted 21 cm signal, which will vary with redshift as the ionized regions grow and merge, with the signal dropping off as reionization completes and essentially the entire
volume of the IGM becomes filled with ionized gas. After reionization, significant quantities of neutral hydrogen remain only in galaxies and damped Ly-$\alpha$ systems. In the earliest phases of the reionization history, the 21 cm signal is also sensitive to the thermal state of the IGM and the intensity of the radiation field near the Ly-$\alpha$ resonance (see \S \ref{S:21cm_overview}).
Early redshifted 21 cm
experiments are expected to make statistical detections by binning together many individually noisy Fourier modes and measuring, for example, the power spectrum of 21 cm fluctuations.
As we will discuss, strong foreground emission makes these measurements -- and those of many other LIM experiments -- challenging. \\

\item {\bf Cross-correlation with Redshifted 21 cm Data Cubes.}  Given the challenge of separating the redshifted 21 cm signal from foregrounds, it may be difficult to establish that a putative
redshifted 21 cm signal truly originates from the high redshift universe, and is not instead the result of residual foregrounds. One powerful way of verifying the high redshift nature of a possible redshifted 21 cm signal
would be to demonstrate that it correlates (or anti-correlates) with another tracer of high redshift structure: for example, a galaxy survey \cite{Furlanetto:2006pg,Lidz:2008ry}. In the case of the reionization-era 21 cm signal, the type of galaxy survey required for this cross-correlation measurement
is unfortunately quite challenging, especially if a traditional galaxy survey is employed. First, the 21 cm surveys cover hundreds of square degrees on the sky or more. By contrast, traditional reionization-era galaxy surveys mostly span only small fields-of-view. For example, deep fields from the Hubble Space Telescope (HST), the Atacama Large Millimeter Array (ALMA), and the James Webb Space Telescope (JWST) generally cover only a few arcminutes on a side.  These fields span angles smaller than even the angular resolution of the 21 cm observations, and are poorly suited for 21 cm cross-correlation measurements, although see \cite{Visbal:2022qxo} for further discussion. 
Furthermore, wide-field {\em photometric} surveys are also not well-matched for measuring cross-power spectra with redshifted 21 cm surveys. The problem here is that photometric surveys measure mostly transverse modes on the sky. These modes, however,
will be lost to foreground cleaning in the redshifted 21 cm experiments\footnote{Note that {\em some} foreground cleaning must be applied to measure the 21 cm-galaxy cross-correlation. This is because, for example, synchrotron emission
from the high redshift galaxies surveyed constitutes some small portion of the redshifted 21 cm foreground. The foreground emission is also a {\em noise source} for the cross-correlation measurement unless cleaned, and so residual foregrounds may shrink the expected signal-to-noise ratio of cross-correlation estimates.  
In other words, cross-correlations do not remove the need for foreground cleaning, they just erase or strongly reduce the average bias from residual foregrounds and may thus help in verifying initial auto-power spectrum detections.}.
It may nevertheless be possible to consider suitable higher-order cross-correlation statistics between the redshifted 21 cm and photometric surveys \cite{Zhu:2015zlh,Sun:2024vhy}, beyond the usual cross-power spectrum, but this requires a more complex analysis and additional investigations.

Another possible solution is to perform a LIM survey in {\em some other line}: more specifically, one desires a bright emission line or lines that may be used to trace-out the large-scale structure in the
(unresolved) reionization-era galaxies \cite{Lidz11,Gong11}. These surveys may be better-matched to the 21 cm observations and probe large angular scales with relatively high frequency resolution, while fine angular resolution is not required as such scales are unresolved in 21 cm anyway. The resulting
cross-correlation measurement could then be used to verify an initial 21 cm detection. Furthermore, as we will discuss, such a cross-correlation measurement would facilitate the interpretation of the redshifted 21 cm signal during reionization (\S \ref{S:xcorr}).
The cross-correlation directly targets the interplay between the ionizing sources and the surrounding IGM, which is fundamental to the process of reionization itself. Candidate emission lines for cross-correlation studies
include Ly-$\alpha$, H-$\alpha$, the [CII] fine-structure line, and rotational transitions from CO molecules. If multiple such lines can be measured across common volumes, then the cross-correlations between each of the the different pairs of lines will also be valuable \cite{Sun2022}.\\

\item {\bf The Ly-$\alpha$ Line.}  The Ly-$\alpha$ line is an interesting target for LIM surveys across a wide range of redshifts. In principle, the Ly-$\alpha$ LIM signal may receive
contributions sourced by both recombination cascades occurring within interstellar HII regions, which form around young, massive stars, as well as from recombinations in the more diffuse gas in the IGM. In addition, continuum photons redshifting into Lyman-series lines will be absorbed by neutral hydrogen and a radiation background will be produced as these atoms re-emit Ly-$\alpha$ photons. We discuss these processes further, and additional ways of producing Ly-$\alpha$ photons, in \S \ref{S:lya_rt}.
During the EoR, the Ly-$\alpha$ fluctuations will be modulated by the presence of ionized bubbles and provide information regarding the spatial fluctuations in the ionization field (e.g. \cite{Visbal:2018dsi}).  Since Ly-$\alpha$ is a resonant line, even the photons that are produced within HII regions in the interstellar medium (ISM) of a galaxy scatter many times in the ISM and the circumgalactic
medium (CGM) of the host galaxy. These photons only gradually diffuse spatially and in frequency space (\S \ref{S:lya_rt}). Consequently, many anticipated sources of Ly-$\alpha$ photons may lead to extended, low surface brightness
emission that is missed by traditional narrow band selection techniques. Hence Ly-$\alpha$ LIM may help provide a complete census of Ly-$\alpha$ photons and in studying the extended low surface brightness halos
around Ly-$\alpha$ emitting galaxies and quasars. These studies may lead to insights regarding the ISM/CGM/IGM and the EoR.\\

\item{\bf Faint Galaxy Populations.}  Since LIM surveys are sensitive to the total clustered emission in the line of interest, they can potentially be used to probe the collective impact of faint galaxies that are too dim to detect individually. One interesting application here may be to map-out faint galaxy populations during the EoR. 
Indeed, current observations find that the galaxy luminosity function (at rest-frame UV wavelengths) rises very steeply towards the faint end, especially at $z \gtrsim 6$ \cite{Bouwens22}. It is then challenging to detect the bulk of the ionizing sources, even using the JWST (e.g. \cite{Sun16}). It may be possible
to nevertheless constrain their bulk impact using LIM, provided, of course, that the low UV-luminosity galaxies emit appreciably in the lines probed by the LIM survey. For instance, if the low luminosity galaxies are extremely metal-poor then hydrogen or helium lines will be required to trace them.\\

\item {\bf Tracers of the Star Formation Rate Density. } 
The cosmic star formation history -- characterized by the average star formation rate per unit comoving volume (the star formation rate density, SFRD) as a function of redshift -- is also fundamental to our understanding of galaxy formation and evolution \cite{Madau:2014bja}.
The luminosities of many of the lines of interest for LIM are well-correlated with the star formation rates of the emitting galaxy populations (see \S \ref{S:landscape}).
The strength of the LIM signal in each line versus redshift can then be used to infer the redshift evolution of the cosmic mean SFRD (see \S \ref{S:SFRD}). 
In the case of more traditional SFRD estimates, the inferences rely on extrapolations towards low-luminosities from the brighter systems that are directly detected. The LIM technique is sensitive to the total clustered line emission and can help cross-check the traditional estimates. In addition, rest-frame UV-based determinations of the SFRD miss dust-obscured galaxy populations and/or require dust attenuation corrections; the contributions from dust-obscured populations can be accounted for using, for example, [CII] LIM (see \S \ref{S:SFRD}).\\

\item{\bf Probing the Interstellar Medium Across Cosmic Time.} A complete description of galaxy formation and evolution across cosmic time requires understanding not only the stellar populations of galaxies, but
also the cosmic history of their multi-phase interstellar and circumgalactic media. Of particular importance is the cold molecular gas out of which stars form. LIM using rotational
transitions from CO molecules can probe the bulk cold gas properties of galaxies over cosmic time (see \S \ref{S:signals}). This may potentially be combined with [CII] LIM surveys and measurements in other atomic fine-structure lines, along with HI LIM data, which each trace warmer atomic phases of the ISM. In conjunction, these surveys will help determine the bulk properties
of the gas in galaxies as a function of redshift. Along with tracing the gas content of galaxies, recent work has shown that LIM may also be used to map out the dust content of galaxies by measuring the fluctuating emission background from polycyclic aromatic hydrocarbons (PAHs) \cite{Cheng25}.
\\

\item{\bf The Chemical Enrichment History of the Universe.} The heavy elements in the universe, beyond helium in the periodic table -- referred to 
as ``metals'' by astronomers -- are synthesized via: nuclear fusion reactions in the cores of stars, in supernova explosions, planetary nebulae, and in neutron star mergers\footnote{In addition to hydrogen and helium, trace amounts of lithium are produced in the big bang.}. Outflows and stellar explosions spread these elements into the surrounding gas, and the metal content builds up over cosmic time as subsequent generations of stars live and die. Measurements of the metal content of galaxies in currently available samples (mostly at $z \sim 0-3$, but with rapidly expanding redshift coverage from JWST observations) reveal a correlation between stellar mass and gas-phase metallicity: galaxies with smaller stellar mass tend to have lower metallicities than those with higher stellar mass (see e.g. \cite{Ma:2015ota} and references therein). This is thought, in part, to reflect outflows which drive gas and metals out of low-mass galaxies with shallow potential wells, while larger galaxies mostly retain their metals. LIM measurements using metal-line tracers may help to explore the overall ensemble-averaged metal content of galaxies across cosmic time, and potentially place constraints on the mass-metallicity relation at low stellar masses, where it is challenging to detect galaxies individually. Further, it may be possible to develop specialized analyses to constrain the metal content of the circumgalactic and intergalactic media using LIM.\\

\item{\bf Galaxy Properties as a Function of Large-Scale Environment.} It is difficult for targeted observations of individual galaxies to study trends in galaxy evolution with large-scale environment, because this requires spanning a large volume at high sensitivity. In contrast, LIM observations capture the bulk emission properties across a full range of environments, from under-dense voids to rare high-density peaks. Specialized analyses, perhaps combining LIM data cubes with optical galaxy surveys, may be used to study how line emission varies with environment. Further work is required to flesh-out the detailed science case and measurement prospects here.\\

\item{\bf Sources of Redshift Information.} LIM surveys provide tracers of large-scale structure with full redshift information. These data cubes may be cross-correlated with photometric catalogs and other two-dimensional fields (i.e., projected maps lacking redshift information), to extract redshift distributions for the photometric sources of interest. A technical issue for such efforts is that low $k_\parallel$ modes will be lost in LIM data cubes owing to foreground contamination, yet these are the only modes available in a photometric survey. It should nevertheless be possible to extract redshift information using suitable higher-point statistics, beyond the usual two-point cross-correlations, which survive after filtering low $k_\parallel$ modes from the LIM data \cite{Pen12}, although further investigation is required here. \\

\item{\bf LIM Lensing.} As photons propagate through the universe, they are deflected by gradients in the gravitational potential. This gravitational lensing effect imprints distinctive statistical signatures in CMB temperature and polarization anisotropy measurements, and these imprints have been measured at high statistical significance \cite{Planck:2015mym,ACT:2023kun}. Alternatively, the lensing effect can be measured through the spatial correlations it induces between the shapes of distant galaxies, known as ``cosmic shear'' (e.g.\cite{DES:2021gua}). The CMB and cosmic shear measurements are valuable since they can be used to construct maps of the projected gravitational potential out to the redshifts of the sources (i.e. out to the surface of last scattering at $z \sim 1,100$ in the case of the CMB or out to the redshifts of the lensed galaxies surveyed in the case of cosmic shear analyses). Since these maps trace the projected matter distribution directly, they avoid complications and uncertainties related to galaxy biasing and the galaxy-halo connection.
Furthermore, the projected mass maps yield a wealth of cosmological information regarding: the matter power spectrum, the growth of large-scale structure, and the composition and expansion history of the universe. 
Line-intensity maps provide additional lensing source ``screens'' across a broad range of precisely known redshifts: combined with CMB lensing and cosmic shear, this may allow tomographic extractions of the projected mass distribution across different redshift intervals \cite{Pen:2003yv,Zahn:2005ap,Foreman:2018gnv}. For example, reference~\cite{Maniyar:2021arp} shows that a combination of CMB lensing maps and two line-intensity lensing maps at $z=5$ and $z=6$ can be used to extract the projected mass distribution from $z=5-1,100$ alone, i.e. one can ``null'' out the lower redshift $z < 5$ CMB lensing contributions. The resulting 
CMB $\times$ LIM nulling measurements offer potentially interesting probes of the matter power spectrum at high redshift, with prospects for detecting the BAO features at $z \geq 5$ \cite{Fronenberg:2023juh}, and for constraining cosmological parameters, including neutrino mass \cite{Fronenberg:2023qtw}.\\
 
\item{\bf Low Surface Brightness Galaxies.}  Recent studies using the Dragonfly Telephoto Array have uncovered populations of ultra-diffuse (low surface brightness) galaxies, and also enabled measurements of the outskirts of galaxies \cite{Abraham14,2015ApJ...798L..45V}, both missed by traditional galaxy surveys. The collective emission from low surface brightness galaxies will be captured in LIM data cubes, allowing their global impact to be assessed. 
Specialized stacking analyses using LIM surveys may also enable studies of line emission from the outskirts of galaxies.\\

\item{\bf Line Emission Signals as CMB Foregrounds.} A key goal for future CMB surveys is to measure spectral distortions, departures of the CMB spectrum from a perfect blackbody. LIM signals constitute an important foreground contaminant for these efforts (see \S \ref{S:spectral_disto}, \cite{Mashian:2016bry,Serra:2016jzs,Chung:2023ncd}.) 
Another example of a LIM foreground for CMB observations is that extragalactic CO emission fluctuations may contaminate efforts to measure the kinetic Sunyaev-Zel'dovich (kSZ) effect \cite{Maniyar:2023cuj}. In this context, these authors show that although the total CO emission fluctuation power in current models is small relative to the kSZ signal, the CO-Cosmic Infrared Background (CIB) cross-correlation is important and must be accounted for.\\

\item{\bf Removing $\tau$ as a CMB Nuisance Parameter.} The optical depth to electron scattering, $\tau$, quantifies the integrated probability that CMB photons scatter off of free electrons produced during and after reionization. This leads to two main signatures in the CMB. First, electron scattering sources a large-scale bump in the angular power spectra of the CMB polarization anisotropies at $\ell \lesssim 10-20$, whose amplitude depends on $\tau$ (e.g. \cite{Zaldarriaga:1996ke}). Second, $\tau$ leads to a damping of the primary CMB anisotropies at $\ell \gtrsim 20$, which is largely degenerate with the amplitude of primordial density fluctuations, $A_s$. Due to this degeneracy, $\tau$ is a critical nuisance parameter in cosmological inferences from the CMB. 
An appealing possibility, however, is to use reionization-era 21 cm LIM measurements to map out the reionization history and thereby infer $\tau$, potentially removing this quantity as a CMB nuisance parameter \cite{Liu:2015txa}. This would help, for example, in determining the sum of the neutrino masses through a combination of CMB lensing and primary CMB anisotropy measurements, which otherwise suffer from the degeneracy between $A_s$ and $\tau$ \cite{Liu:2015txa}. Thus, the redshifted 21 cm measurements may benefit our understanding of cosmological parameters, while also revealing the reionization history of the universe. A potential concern, in addition to the challenges of the 21 cm measurements themselves, is that inferring $\tau$ requires knowledge of the {\em average} reionization history, while near-term 21 cm experiments will likely measure only the 21 cm power spectrum as a function of redshift. There may be important modeling uncertainties in relating the observed fluctuation power spectra to the mean reionization history.\\ 

\item{\bf Bonus Science Enabled by LIM Surveys.} Finally, it is worth emphasizing that LIM surveys enable interesting ancillary science. A prime example here is detecting fast radio bursts (FRBs) with 21 cm LIM surveys. In particular, after the CHIME radio telescope was equipped with a back-end for detecting FRBs across its frequency coverage of $\sim 400-800$ MHz, it has become an excellent FRB detector. Indeed, the vast majority of known FRBs have been discovered by CHIME with outstanding prospects for further advances \cite{CHIMEFRB:2021srp}. This, in turn, enables interesting studies of the unknown progenitors/central engines of FRBs, their radiation-generating mechanisms, and cosmological applications including studying missing baryons, the CGM, and magnetic fields. FRBs also provide unprecedented probes of dark matter properties via their gravitational lensing, among other exciting science prospects \cite{Petroff:2021wug}.

Further ancillary science enabled by LIM surveys include improved maps of the emission from our Milky Way galaxy across a range of frequencies. For example, CO line-intensity maps at $\sim 30$ GHz from the COMAP survey may help to understand the origin of anomalous microwave emission, which may arise from spinning dust grains in 
our galaxy \cite{COMAPI2022}.\\

\end{itemize}

\subsection{When Is Line-Intensity Mapping Advantageous?}
\label{sec:when_lim}

At this point it is worth discussing under which circumstances LIM is preferable to more traditional surveys using targeted measurements of individual discrete objects such as galaxies. Here we give only some general and qualitative remarks and address these issues in more detail in \S \ref{sec:lim_v_traditional}. First, note that some LIM surveys will also allow detections of individual bright objects. One may also be able to carry-out LIM analyses using traditional galaxy survey data. As current examples of the latter, \cite{Croft:2015nna,Croft:2018rwv} used SDSS spectroscopic fibers centered on LRGs and ``sky fibers'' (which are not centered on particular target objects) to look for diffuse Ly-$\alpha$ emission at $z \sim 2-3$ in cross-correlation with quasar and Ly-$\alpha$ forest measurements (see \S \ref{sec:lya_current}).
Hence, in some cases, both LIM and discrete object analyses may be carried out on the same data set and profitable. 

In order to detect individual discrete objects in a spectroscopic survey, a source must be bright enough to detect above both the detector noise, and the confusion noise from fainter emitters, in a voxel. Here a voxel refers to a three-dimensional pixel or volume element in a LIM survey. Each voxel is generally specified by a comoving volume, whose transverse dimensions are set by the angular resolution of the survey and whose line-of-sight extent depends on spectral resolution and the spectral channel bandwidth. 
The confusion noise here may result from either Poisson fluctuations or the clustering of faint sources below the survey detection limits. Residual foreground contamination (leftover after any filtering and other mitigation steps are applied) may also contribute to the effective noise per pixel and challenge source detection efforts. As will be discussed in more detail in \S \ref{sec:lim_v_traditional}, traditional surveys generally require: relatively small voxels, low detector and confusion noise in each voxel, and a sufficient abundance of bright emitters (e.g., \cite{Uzgil:2014pga}). In contrast, LIM is the preferred method in the regime of large voxels (e.g. in the case of small aperture telescopes with limited angular resolution). 
Further, even in the regime where the detector noise per voxel is large, noisy estimates may nevertheless be combined to detect, for example, the power spectrum of line-intensity fluctuations in particular $k$-bins (see \S \ref{sec:pk_var}). However, traditional surveys may fail to detect sources in this regime. Note also that in a LIM analysis the confusion noise is essentially the signal of interest, as opposed to an obstacle. 

As alluded to earlier in the previous section, the LIM technique may help in scenarios where there are abundant low-luminosity sources of emission, which lie beneath the detection thresholds of targeted surveys. Moreover, it should be the method of choice for studying diffuse and/or spatially-extended emission. On the other hand, in some cases the emission in a line of interest may be dominated by luminous and spatially-compact sources: in this case even the aggregate population-averaged statistics may be reliably traced by bright objects which can be detected individually in the survey.  
In more detail, the optimal analysis technique depends on the scientific aims and the qualitative considerations mentioned here can be sharpened for particular goals (\S \ref{sec:lim_v_traditional}, \cite{Cheng:2018hox,Schaan:2021hhy}).

In this context, it is also interesting to note a recent generalization of the conventional LIM analysis technique (discussed further in \S \ref{S:3dlightcone}).  Specifically, reference \cite{Cheng:2022ani} considers fitting the angular auto and cross-power spectra between maps at different frequencies, $\nu$ and $\nu^\prime$, $C_{\ell, \nu, \nu^\prime}$, to joint models for the source spectral energy distributions (SEDs), luminosity densities, spatial clustering, and their redshift evolution. 
The SED properties identified here describe the characteristics of collective emission fluctuations and may generally differ from those of individual emitters. 
In this approach, the model simultaneously fits for information regarding both the line and continuum emission from the ensemble of sources. In this way, it moves towards the ``no photon left behind'' ideal of LIM studies by considering information both from spectral lines and the underlying continuum emission components. Hence, novel analyses of upcoming LIM data sets may allow one to extract information beyond the emission fluctuations in the primary lines targeted by the survey.

\section{The Landscape of Lines}
\label{S:landscape}

A wide range of different spectral lines have been considered for LIM applications. Here we examine some of the most prominent examples under current consideration, discuss the key physics of each line, and mention some primary science targets for specific cases.

\begin{figure}
\begin{center}
\includegraphics[width=\textwidth]{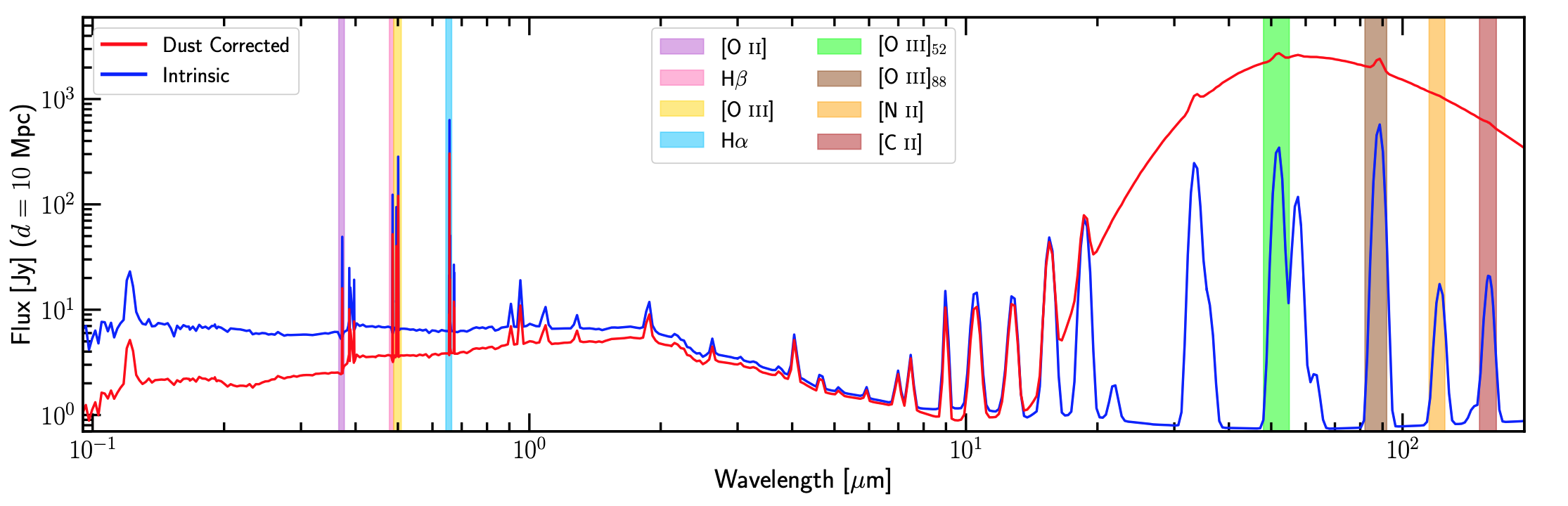}
\label{fig:thesan_sed}
\caption{A model SED from the THESAN simulation suite, illustrating some of the possible line targets for LIM surveys. The blue curve shows the intrinsic emission line spectrum from the simulated galaxy, while the red curve is reprocessed by dust extinction/emission. Rest-frame optical transitions from [OII] ($3726, 3729 \, \Ang$), [OIII] ($4959, 5007 \, \Ang$), H-$\alpha$ ($6563 \, \Ang$), and H-$\beta$ ($4861 \, \Ang$) are indicated, along with fine-structure emission lines from [OIII] ($52 \mu$m, $88 \mu$m), [NII] ($122 \mu$m), and [CII] ($158 \mu$m). Other promising line-emission targets for LIM surveys include CO rotational transitions, Ly-$\alpha$ (included here, but unmarked on the left hand side of the SED), HeII recombination lines, and others. In this section, we discuss many of these emission lines and their scientific utility.  
From \cite{Kannan:2021ucy}.}
\end{center}
\end{figure}

In order to place this discussion in context, Figure \ref{fig:thesan_sed} shows an example simulated galaxy SED (from the THESAN simulation suite, \citep{Kannan:2021ucy}). Specific emission lines are marked, spanning from the rest-frame ultraviolet out to sub-mm wavelengths. Measurements toward individual galaxies have long used this rich set of spectral lines, among others, to help understand the ISM properties and stellar populations of the observed galaxies. Specifically, spectroscopic measurements of the lines simulated in Figure~\ref{fig:thesan_sed} provide
valuable information regarding ISM gas-phase metallicities, densities, temperatures, and ionization states, the incident radiation spectra, and how these quantities vary across diverse galaxy populations. 

An aim of LIM is to detect some of these same emission lines, averaged over ensembles of line-emitting galaxies and measured as a function of redshift. This should access some of the same information that has been extracted towards individual galaxies via modeling their emission lines, yet in a population-averaged sense. Although early LIM surveys may focus on one or a few particular lines, the ultimate power of this method may come from observing many separate lines across common and vast regions of the universe. In the case of LIM observations spanning many different lines, each arising primarily from the ISM of the emitting galaxies, the information content will resemble that of an ensemble-averaged SED.

The precise physical interpretation of the coarse-grained information obtained in a LIM observation remains an unsolved problem for the field. The challenge is in part due to the huge dynamic range of spatial scales involved in the problem. In addition, the average quantities returned in LIM often involve non-linear functions of ISM properties, such as the gas-phase metallicity. The average of a non-linear function generally departs from the function at the average value, and so the precise meanings of the coarse-grained averages returned by LIM surveys remain complex and unclear. In this section, our scope is hence limited to discussing some of the emission line
targets and their broad-brush astrophysical dependencies.

\subsection{The Multi-Phase ISM}
\label{S:multi_ism}

The high redshift ISM, as well as that in our own Milky Way and nearby galaxies, is multi-phase: interstellar gas spans a broad range in temperature, density, and ionization state. LIM promises to capture a diverse set of emission lines and probe ISM gas across a variety of phases, averaged over ensembles of emitting regions and galaxies. The redshift evolution may also be measured, providing insight into how the ISM properties develop over cosmic time. 
We start with a brief overview of some of the key ISM phases and the primary LIM emission line targets proposed to capture
these phases. As we will discuss, in some cases the emission lines arise from multiple different parts of the ISM, as well as from the CGM and IGM.  
We focus on the lines discussed in the current literature, and so our description of the ISM is incomplete (see, e.g. \cite{Draine11,Osterbrock06} for additional information) and only intended to provide a broad-brush overview. In the future, we expect additional line targets to emerge and capture still further properties of the interstellar, circumgalactic, and intergalactic gas across cosmic time. After providing a brief description of the most relevant ISM phases for current LIM studies, we turn to take a closer look at relevant aspects of the emission line physics for various cases of interest. 

\begin{figure}
    \centering
    \includegraphics[width=\textwidth]{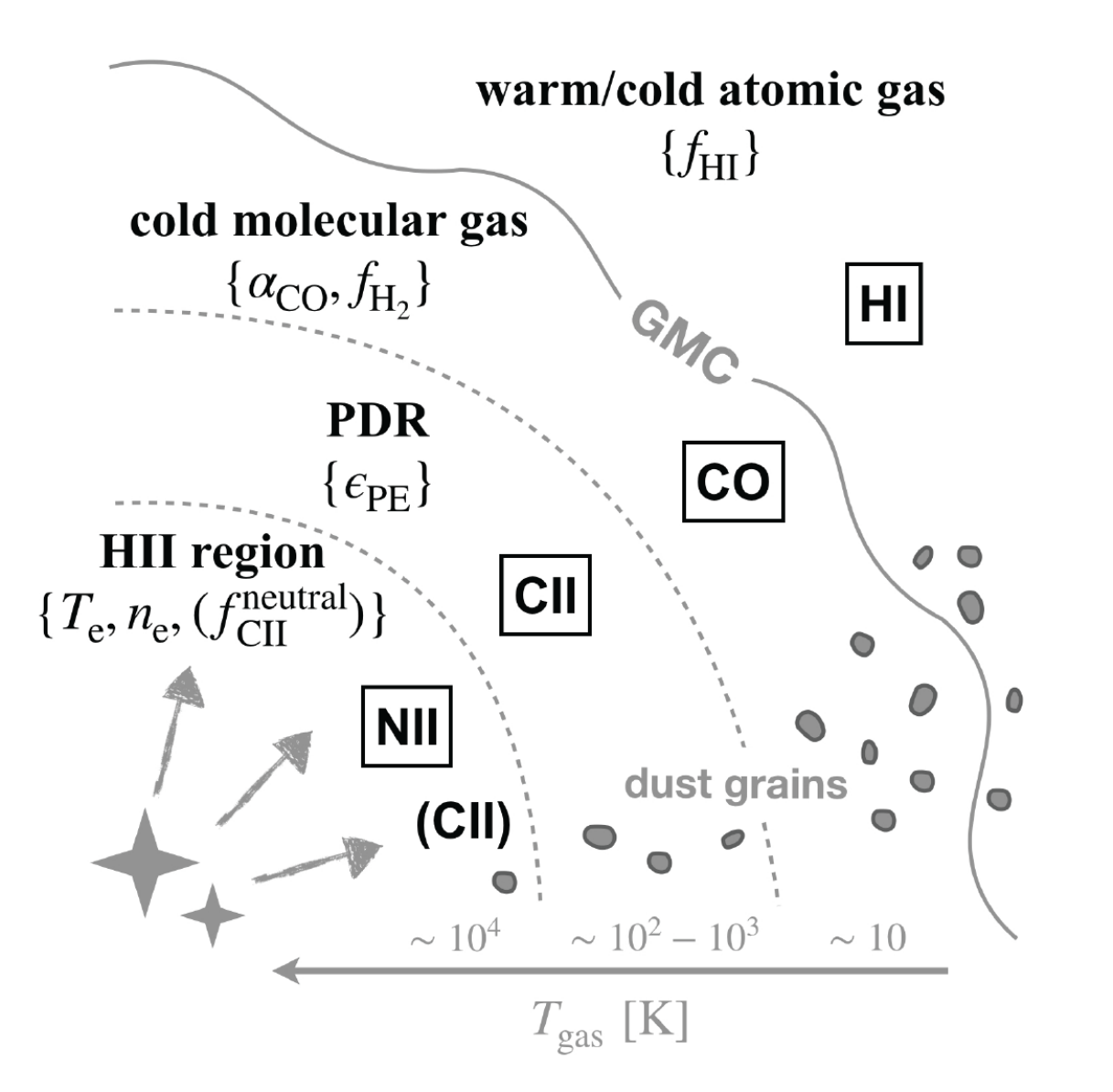}
    \caption{Illustration of some of the key ISM phases and associated emission lines. In {\em the HII region} phase, newly formed O and B stars -- i.e. massive stars with high surface temperature which emit copious numbers of hydrogen ionizing photons --  photoionize surrounding neutral hydrogen gas. Here transitions between energy states in various ions are collisionally excited, leading to long wavelength fine-structure transitions as well as optical and ultraviolet emission lines. Examples include emission from CII, NII, OII, and OIII ions, and many additional lines. Further, recombination cascades from residual neutral hydrogen in HII regions lead to Lyman-series and Balmer-series lines (among others) at rest-frame optical and ultraviolet wavelengths.  In {\em the PDR} phase hydrogen is largely neutral, but ultraviolet photons singly-ionize carbon atoms and dissociate molecular gas. Further out, in {\em the cold molecular gas} phase, CO molecules are shielded from dissociating UV radiation by dust grains. Finally, outside of the dense cloud (the ``GMC'' in the diagram), lies -- along with other gas and dust -- {\em warm/cold atomic gas}, which mainly consists of neutral hydrogen and is best traced by the 21 cm transition. From \citep{Sun19}.}
    \label{fig:my_label}
\end{figure}

{\bf Cold Neutral Medium}: This gas consists mostly of neutral atomic hydrogen. In the Milky Way, the typical temperature of the cold neutral gas is $\sim 100$ K and the hydrogenic number density is around $n_{\rm H} \sim 30$ cm$^{-3}$ \citep{Draine11}. A key tracer of this gas is the 21 cm line. 

{\bf HII regions}: Massive, hot, and young O and B stars photoionize regions of hydrogen gas around them. These HII regions are mainly embedded in the dense clouds of gas in which stars form. The temperature of the gas in an HII region is typically around $10^4$ K, set by the balance between photoionization heating and cooling from line, bound-free, and free-free emission. The HII regions in the Milky Way span a wide range in density, including more diffuse gas with $n_{\rm H} \lesssim 1$ cm$^{-3}$ as well as clouds with density as high as $n_{\rm H} \sim 10^4$ cm$^{-3}$ \citep{Draine11}. A variety of optical/ultraviolet and longer wavelength emission lines arise from HII regions, including Ly-$\alpha$, H-$\alpha$, [OII] and [OIII] lines, [NII], and [CII] lines, among others.  

{\bf Photo-dissociation regions (PDRs)}: In part, the photo-dissociation regions in a galaxy lie in between the HII regions around young stars and denser gas that consists largely of molecules. Although hydrogen is predominantly neutral in this phase, elements with lower ionization potentials are significantly ionized within the PDRs. For instance, the first ionization potential of carbon is 11.26 eV and so photons between 11.26 eV and the hydrogen ionization potential, 13.6 eV, may penetrate into the PDR and singly ionize carbon (while the higher energy photons are instead mostly absorbed within the HII regions). Hence the [CII] emission line arises partly from PDRs, although other ISM phases may emit [CII] as well.  

{\bf Molecular Clouds}: In the molecular phases of the ISM the gas densities and column densities of gas and/or dust are large enough that the gas is shielded against photo-dissociating UV radiation. Among the best emission line tracers of molecular gas are lines emitted when CO molecules transition between different rotational states. Although molecular hydrogen is generally the most abundant ISM molecule, H$_2$ lacks a permanent electric dipole moment and so radiates weakly, unlike CO and other tracer molecules. In the Milky Way, molecular gas spans a wide range in temperature ($T \sim 10-50$ K) and gas density ($n_{\rm H} \sim 100-10^6$ cm$^{-3}$) \cite{Draine11}. A number of different physical processes are thought to play a role in setting the temperature and excitation state of CO-emitting molecular gas, including photoelectric heating off of dust grains, cosmic ray ionization, heating from the CMB, AGN feedback, shock heating from supernova explosions, and cooling from line emission, among others (e.g. \cite{Obreschkow09}). 

{\bf Additional ISM Phases}: Warm neutral gas in the ISM of other galaxies may also be traced by the 21 cm line, while hot diffuse phases heated by supernovae winds in distant galaxies may be difficult to trace with LIM surveys. 

\subsection{Emission Line Tracers} We can now consider different types of line emission and their utility for LIM measurements. We briefly review relevant aspects of atomic physics and the properties of some key categories of emission lines. 

\subsubsection{Fine-Structure Emission Lines} 
\label{S:fine_structure}
Fine-structure emission lines provide an important coolant for interstellar gas; the small energy splitting between atomic fine-structure energy levels allows these transitions to be excited in cool phases of the ISM and the resulting emission is accessible at relatively long sub-mm wavelengths. Physically, fine-structure emission lines arise from interactions between the spin and orbital angular momentum of electrons in atoms and ions. In the rest frame of an orbiting electron, the electron ``sees'' a magnetic field from the circulating nuclear charge which, in turn, interacts with the electron's spin magnetic moment. 

An order-of-magnitude estimate for the energy of the spin-orbit interaction is:
\begin{equation}
\Delta E_{\rm so} \sim \alpha^4 m_e c^2 \sim 10^{-3} {\rm eV},
\label{eq:so_oom}
\end{equation}
where $\alpha=e^2/(\hbar c)$ is the fine-structure constant and $m_e c^2$ is the rest energy of an electron. This estimate follows from considering the magnetic field produced by a circulating proton situated one Bohr radius away from an electron, and the energy of interaction with the electron's spin magnetic moment. The precise spin-orbit interaction energy depends on the electric charge of the nucleus, although the inner electrons will partly shield the outer electrons from the nuclear charge. 
Numerically, the spin-orbit interaction is on the order of $\sim 10^{-3}-10^{-2}$ eV and so transitions between different fine-structure states lie at millimeter (mm) or sub-mm wavelengths. The temperatures required to collisionally excite such transitions are roughly in the $\sim 10-100$ K range and trace, in part, relatively cool regions where collisions are insufficiently energetic to excite rest-frame optical or ultraviolet lines.  

In order to understand the role of this interaction in more detail, it is useful to recall some pertinent aspects of the quantum theory of multi-electron atoms (see, e.g. \citep{Osterbrock06,Draine11}). Although inexact, a first approximation is to consider each orbiting electron as moving in an average effective potential determined by the other electrons and the nuclear charge. This effective potential is approximately spherically symmetric and so each electron may be described 
in terms of their individual, single-electron states characterized by a principal quantum number, $n$, and an orbital angular momentum, $l$. Each quantum number here is specified by an integer, with $l$ giving the orbital angular momentum in units of $\hbar$ and lying between $0 \leq l \leq n-1$. Furthermore, the spin of each electron is $\hbar/2$ and the projection of the orbital and spin angular momenta onto the z-axis are described by $m_l$ and $m_s$, respectively. The value of $m_l$ is restricted to lie between $-l, -l + 1, ...0, 1, ...l$ and so there are $2 l +1$ different possible $m_l$ states, while $m_s$ can be $\pm 1$ (corresponding to z-components of the spin angular momentum with $s_z = \pm \hbar/2$.) Note that for a hydrogenic atom, states with different $l$ but the same $n$ have nearly identical energies; the same is not generally true for multi-electron atoms as the effective potential deviates from a simple $1/r$ form.
The Pauli exclusion principle forbids two electrons from occupying the same quantum mechanical state, and so an ``orbital'' characterized by $(n,l)$ can be occupied by at most $2 (2 l+1)$ electrons. The angular momentum quantum number is usually denoted by a letter, with $l=(0,1,2,3,..)=(s,p,d,f,...)$\footnote{The usual mnemonic is ``(s)mart (p)hysicists (d)on't (f)orget''. After this, the angular momenta are described alphabetically, e.g. $l=4$ is denoted by ``g''.}. 

The ground state of an atom or an ion can then be described in terms of electrons occupying orbitals, or shells/subshells of increasing energy. For example, singly-ionized carbon (CII) consists of five electrons: in the ground state, two electrons reside in the $(n,l)=(1,0)$ state; two electrons in the $(n,l)=(2,0)$ subshell; and one electron in the $(n,l)=(2,1)$ subshell. This state is denoted by $1s^2 2s^2 2p^1$ where e.g. $2p^1$ denotes an $(n,l)=(2,1)$ state and the superscript indicates that one electron resides in the subshell. Closed shells, where the electron occupancy is complete, are spherically symmetric and have total orbital and spin angular momenta of zero. The complete shells have little interaction with external electrons and can generally be neglected. The electrons outside of closed shells are referred to as valence electrons. 

In the Russell-Saunders coupling approximation, the states can be further described by {\em terms} specified by their total orbital angular momentum, ${\bf L} = \sum_i {\bf l_i}$ and total spin angular momentum, ${\bf S} = \sum_i {\bf s_i}$ where $l_i$ and $s_i$ are the orbital and spin angular momenta of individual electrons. Further, the states can be labeled by the components of these angular momenta in a particular direction, e.g. by $L_z$ and $S_z$. The Russell-Saunders coupling approximation is a good description when the Coulomb interaction between individual electrons is large compared to spin-orbit interactions. In general, this is a good approximation for light atoms/ions. In this case, ${\bf L}, {\bf S}, L_z$, and $S_z$ are all approximate constants of motion and hence the eigenvalues of these operators are good quantum numbers\footnote{Although note that $L_z$ and $S_z$ are not strictly conserved in the presence of spin-orbit interactions.}.
In addition, the total, orbital plus spin, angular momentum is denoted by ${\bf J} = {\bf L} + {\bf S}$. Note that $J$ is an exact constant of motion and hence a good quantum number. The angular momenta obey the usual quantum mechanical rules for the addition of individual angular momenta. The states with different values of $J$ are split into levels with slightly different energies (``fine-structure splitting'') owing to the spin-orbit interaction referred to earlier. Terms, characterized by $S, L, J$, are denoted as: $^{2S+1} L_J$, with angular momenta in units of $\hbar$ and $L=(S,P,D,F,...)$ for $L=(0,1,2,3,...)$. The allowed states here must respect the Pauli exclusion principle. 

The energy ordering of the fine-structure states may be determined through Hund's rules. Specifically, Hund's rules describe the energy ordering of different terms, specified by $L, S, J$, for states with the same electronic configurations. 
The general idea behind these rules is that the orbiting electrons are kept as far apart as possible in the lower-energy, more tightly bound states, since this minimizes their repulsive Coulomb interactions. For example, Hund's first rule tells us that the lowest-energy state has the maximum possible spin: when each electron spin is aligned in the same direction, the electrons must avoid each other spatially by the Pauli principle. 

Hund's second rule says that in the ground state, after maximizing the spin, the orbital angular momentum should also be as large as possible, subject to the Pauli exclusion principle. This arises because electrons orbiting in the same sense -- maximizing the total orbital angular momentum -- tend to spend more time apart from each other, while electrons traveling in opposite directions pass near to each other more often.  Finally, for electrons whose outermost sub-shells are less than half filled, states with smaller total $J$ are lower in energy. On the other hand, higher $J$ states have lower energy in cases where the outermost sub-shell is more than half full. This rule results from the spin-orbit interaction. 

\begin{figure}
    \centering
    \includegraphics[width=\textwidth]{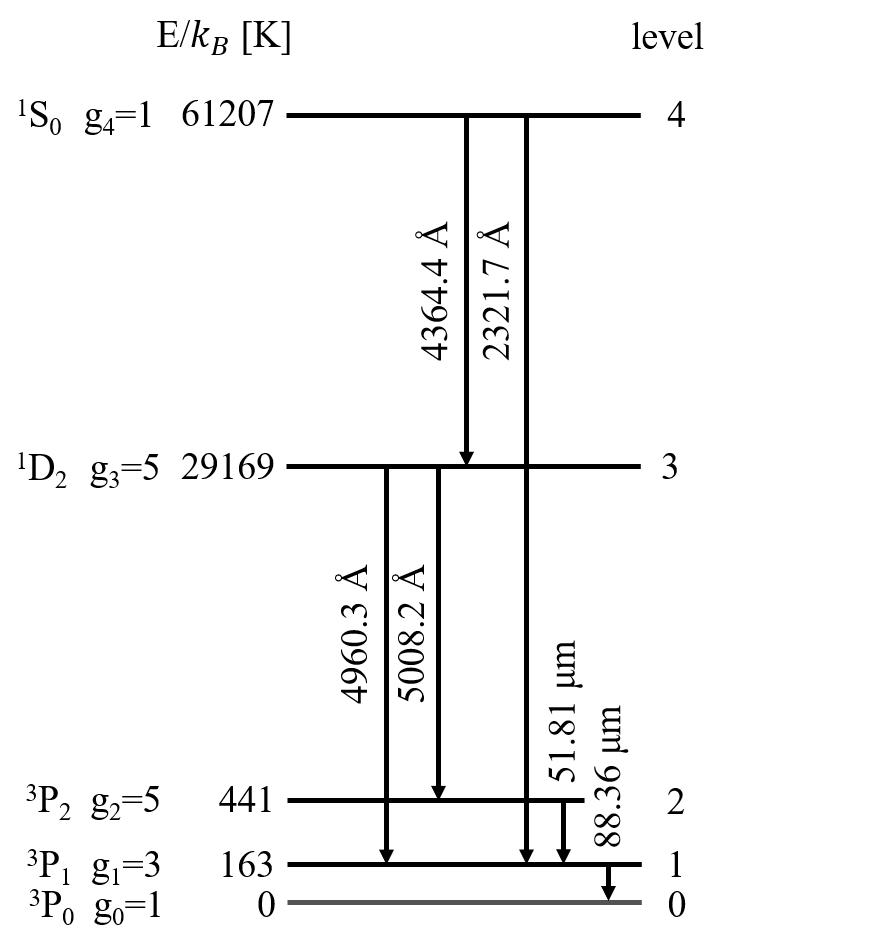}
    \caption{Energy levels of OIII.  The right-most column specifies the  spin, orbital, and total angular momentum of each state. The next column gives the degeneracy of states, with $g=2J+1$ distinct states in each level. The third column is the energy relative to the ground state in temperature units, while the arrows indicate transitions between different states and the associated emission wavelengths. The three lowest levels are split by the spin-orbit fine-structure interaction; transitions between these levels lead to relatively long wavelength emission and their small energy splitting makes the results insensitive to the gas temperature. From \cite{Yang2020}, adapted from \cite{Draine11}.}
    \label{fig:oiii_levels}
\end{figure}

For example, consider the illustrative case of OIII, doubly-ionized oxygen. This ion has six electrons, with ground state configuration $1s^2 2s^2 2p^2$. Note that the ground states of CI, NII, and other six electron ions have identical ground state configurations, but we focus on OIII here. 
The $2p^2$ valence electrons can combine to form a total spin of $S=1, 0$ with degeneracy $2 S + 1 = 3, 1$, respectively. Hund's first rule tells us that the state with $S=1$ lies lower in energy.  The total orbital angular momentum of the two valence electrons can be $L=(0,1,2)$. The Pauli principle allows $(L,S) = (1,1); (2,0); (0,0)$, since $S=1$ states have symmetric spin wave-functions and must have spatially anti-symmetric wave functions (and so $L=1$). For $S=0$, the spin wave function is anti-symmetric, and therefore either of the spatially symmetric states with $L=0$ and $L=2$ are permitted. The lowest energy level is a $^3 P_0$ state; Hund's third rule tells us that the total angular momentum of the ground state is $J=0$ since the relevant $2p$ sub-shell is less than half filled. The rules of angular momentum addition allow states with $J=1$ and $J=2$ as well, which are slightly higher in energy than the $J=0$ ground state. That is, the fine-structure interaction splits the $^3 P$ states in energy as: $^3 P_0$, $^3 P_1$, and $^3 P_2$, in order of increasing energy. The next higher-energy state is a $^1 D_2$ state, which is lower in energy than the $^1 S_0$ level by Hund's second rule. 

Figure \ref{fig:oiii_levels} provides a summary of the key OIII energy levels. The three lowest lying states, split by the fine-structure interaction, differ in energy only at the $\sim 100$ K level (i.e. $\sim 10^{-2}$ eV in energy units), which is small relative to the $\sim 10^4$ K temperature of the HII region gas. The transitions between the different levels here are forbidden by electric dipole selection rules, which require changes in the electronic configurations (i.e., one of the individual orbital angular momentum quantum numbers, $l$, must change by 
$\Delta l = \pm 1$ in an electric dipole transition). The two transitions between the three lower levels in Figure \ref{fig:oiii_levels} are, however, allowed by magnetic dipole selection rules and proceed accordingly, albeit at a slow rate relative to electric dipole transitions (e.g. \cite{Osterbrock06})\footnote{Transitions that are forbidden by electric dipole selection rules, but are allowed by -- for example -- slower magnetic dipole or electric quadrupole selection rules are denoted with brackets. For instance, the $4364\,  \Ang$ ($^1 S_0 \rightarrow \, ^1 D_2$) [OIII] transition (Figure~\ref{fig:oiii_levels}) is an electric quadrupole transition, while the [OIII] $88\, \mu$m ($^3 P_1 \rightarrow \, ^3 P_0$) jump is a magnetic dipole transition. At the densities obtained in Earth-bound laboratories, the spontaneous decay rates of ``forbidden'' transitions are slow relative to collisional de-excitations. However, in the rarefied conditions of the ISM, spontaneous decays may outpace collisional de-excitations and ``forbidden'' becomes a misnomer.}.
The 88 $\mu$m line from jumps between the $^3 P_1$ and $^3 P_0$ states has recently been detected towards a handful of Lyman-break selected galaxies out to $z \sim 6-9$ using ALMA (e.g. \cite{Harikane20}). These detections have recently been extended all the way to $z=14.2$, with a $\sim 6-\sigma$ [OIII] 88 $\mu$m line measurement towards a JWST photometric candidate galaxy \cite{Schouws24}.
Hence, [OIII] lines have provided spectroscopic confirmation of some of the highest redshift galaxies detected to date and insights into the properties of the ISM in these early galaxies \cite{Jones20,Yang2020}. In addition, this line is a potentially attractive target for future LIM studies, as are some of the rest-frame optical OIII transitions \cite{Moriwaki18,Padmanabhan:2021tjr,Gong17,Dore2014}. 

As another example of importance for LIM studies, we consider the case of the CII ion. As mentioned earlier, CII has a single $2p$ valence electron, and so $S=1/2$. The spin-orbit coupling splits the state into a higher energy level with $L=1, J=3/2$ and a lower level with $L=1, J=1/2$. The ordering follows from Hund's third rule: CII's outermost sub-shell is less than half filled, and so the state with higher $J$ has higher energy. Hence, the [CII] 158 $\mu$m emission line results from the radiative decay from the excited $^1 P_{3/2}$ to the lower $^1 P_{1/2}$ state. The energy difference between these two states in temperature units is only $91$ K and so, again, the upper level may be collisionally-excited in cool regions of the ISM. In the case of CII, collision partners include electrons, hydrogen atoms, and hydrogen molecules. 

\subsubsection{Rest-Frame Optical and Ultraviolet Lines from HII Regions} \label{S:rest_opt_uv_hii}

In addition to sourcing (in part) the fine-structure lines at longer wavelengths, HII regions also produce rest-frame optical and ultraviolet transitions. Several such transitions are  promising targets for LIM experiments. 
It is useful to split these lines into two loose categories: {\em collisionally-excited transitions} and {\em recombination lines}. The distinction here is not precise, as we will see, since the same emission line may be produced either through a collisional excitation, or via a recombination, yet one mode or the other is generally dominant for present purposes.   
The collisionally-excited line emission is sourced when thermal electrons in an HII region collide with surrounding ions. These collisions can raise the ion to an excited level with an energy of order $\sim {\rm eV}$ above the ground state in the case of an optical line.  The collisionally-excited line radiation is, in turn, produced when the excited ions decay into lower energy states. For example, the $^1 D_2 \rightarrow ^3 P_1$ and $^1 D_2 \rightarrow ^3 P_2$ transitions in OIII produce emission lines at $4960 \, \Ang$ and $5008 \, \Ang$, respectively (see Figure~\ref{fig:oiii_levels}). Other important collisionally-excited lines from HII regions include [OII] and [NII] transitions, among many more.

The luminosity ratio between two lines of the same species can be used as either a temperature or a density diagnostic, depending on the energy separation of the relevant states. For example, if the energy difference between the states involved is small compared to the temperature of the gas, then the line ratio provides a density diagnostic. In this small energy separation limit, the level populations are insensitive to temperature but will depend on the gas density: below the critical density (see \S \ref{S:lev_pops}) spontaneous decays will outpace collisional de-excitations, while much above the critical density (and at high temperature) the level populations will reflect the degeneracies of the different energy states. At intermediate densities, the line ratios will vary with the density, smoothly interpolating between the two limiting cases. The luminosity ratios here provide valuable information regarding the gas density. In the opposite case, where the energy separation is comparable to the temperature of the gas, the level populations and resulting emission line ratios provide a temperature diagnostic. For example, in [OIII] the luminosity ratio between the $4364 \, \Ang$ transition and the $4960 \, \Ang$/$5008 \, \Ang$ transitions provides a temperature indicator. {For further details, see for example reference \cite{Yang2020}, their Eq. 33 and the associated discussion.}

In the context of LIM, we should keep in mind that we measure only the emission fluctuations averaged across ensembles of galaxies, each typically containing many different HII regions. These may span a range of densities, metallicities, temperatures, and other properties, and the luminosity in a line of interest may be a non-linear function of some of these parameters. While the line luminosities and HII region properties may correlate with dark matter halo mass, there may also be substantial scatter at fixed host mass.
It is then unclear how to interpret the ratio of, e.g., the power spectra of the line-intensity fluctuations in two lines. In the case of temperature diagnostics, for example, one would naturally expect the ratio of these power spectra to contain information regarding the typical HII region temperatures,
but further investigation is required to make a more precise statement here.  

The recombination lines are produced when an ion in an HII region combines with a previously free electron; if the electron is captured into an excited state, it then undergoes a rapid radiative cascade, emitting lines as it decays through a set of lower energy levels. Some of the key recombination lines for LIM include the Ly-$\alpha$ line ($n=2 \rightarrow 1$), the Balmer H$-\alpha$/H-$\beta$ lines ($n=3 \rightarrow 2$/$n=4 \rightarrow 2$), and the HeII $1640 \, \Ang$ line (i.e. the Balmer-$\alpha$ line for singly-ionized helium). For example, LIM in each of the HeII $1640 \, \Ang$ and H$-\alpha$ lines provides a spectral hardness diagnostic and a potential probe of Pop-III stars (see \S \ref{s:balmer}, \cite{Visbal:2015sca,Parsons:2021qyw,Oh:2000sg}). 

In order to calculate, for instance, the Ly-$\alpha$ emission produced in recombining gas, one first computes the probability that an electron is captured into an excited state with quantum numbers $(n,l)$ \cite{Osterbrock06}. Next, one considers all of the possible decays through intermediate states obeying the quantum mechanical selection rules, which ultimately lead to a $2p \rightarrow 1s$ transition and a Ly-$\alpha$ photon. Finally, the Ly-$\alpha$ emission rate follows from summing over all possible initial capture states, $(n,l)$, weighted by the probability of a recombination into this state, and the probability of producing a Ly-$\alpha$ photon from the initial state $(n,l)$. In case-B recombinations (see \S \ref{S:ion_vol}), the results of these calculations can be easily summarized (see e.g. \cite{Dijkstra17} for more details). At a typical temperature of $T=10^4 K$, $68\%$ of recombinations lead to a Ly-$\alpha$ photon while $45\%$ of recombinations lead to an H-$\alpha$ photon \cite{Osterbrock06}. Consequently, the expected luminosity in the Ly-$\alpha$ line is simply $L_\alpha = 0.68 h \nu_\alpha \dot{N}_{\rm rec}$, where $\dot{N}_{\rm rec}$ is the rate of radiative recombinations. Assuming photoionization equilibrium in the HII region, the rate of recombinations balances the rate of interior photoionizations, $\dot{N}_{\rm HI}$. Hence:
\begin{equation}
    L_\alpha = 0.68\, h \nu_\alpha (1 - f_{\rm esc}) \dot{N}_{\rm HI}
    \label{eq:lalpha}
\end{equation} 
In other words, 68\% of the (non-escaping) ionizing radiation (in terms of photon number) is re-processed into emission in the Ly-$\alpha$ line alone. As long recognized, the Ly-$\alpha$ line hence provides a powerful probe of high redshift galaxy formation \cite{Partridge67}. 
Here $f_{\rm esc}$ is the escape fraction, i.e. $1-f_{\rm esc}$ of the ionizing photons are absorbed within the galaxy. Strictly speaking, even the ionizing photons that are absorbed outside of the galaxy (in the CGM or IGM) will lead -- in photoionization equilibrium -- to recombinations and Ly-$\alpha$ photons that might be detected in a LIM experiment. These Ly-$\alpha$ photons are interesting, but should make sub-dominant contributions to the average specific intensity since the escape fraction is expected to be small on average, $f_{\rm esc} << 1$ \cite{Inoue:2006um,Flury2022,Giovinazzo26}.

In addition, some Ly-$\alpha$ photons may be sourced by other processes. For example, atoms in dense and warm yet (partly) neutral gas may be collisionally excited and produce Ly-$\alpha$ in the subsequent radiative decay \cite{Haiman2000}. This requires warm $T \sim 10^4-10^5$ K gas so that collisions can excite atoms from the ground state to the first excited level while the gas should not be too hot -- otherwise, it will be too highly ionized to emit appreciably in Ly-$\alpha$. 
A further mechanism is that Ly-$\alpha$ may be produced ``fluorescently'' \cite{Gould1996} as gas is ionized by a nearby {\em external} source, or via a photon from the ultraviolet background radiation. This gas then emits Ly-$\alpha$ photons as it recombines. Note that fluorescent re-emission and CGM/IGM Ly-$\alpha$ radiation are specific examples of recombination cascade emission. Fluorescence differs from the more conventional recombination radiation of Eq.~\ref{eq:lalpha} only in that the photo-ionizing source in fluorescence is external, while in CGM/IGM emission the recombining gas is outside of the galaxy itself and so differs in location (and
may also be triggered by an external source). 
Finally, Ly-$\alpha$ photons may be produced as ultraviolet continuum photons redshift into a Lyman-series resonance and are absorbed, with Ly-$\alpha$ produced in the subsequent radiative cascade \cite{Silva13,Pullen:2013dir,Chen04}. 
In this sub-section, we focus on the contribution from recombinations within galaxies. Although Ly-$\alpha$ sourced by redshifting continuum photons can dominate the average specific intensity at some post-reionization redshifts, the recombinations within galaxies are always expected to make the largest contribution to the spatial fluctuations in the Ly-$\alpha$ specific intensity on accessible scales \cite{Pullen:2013dir}. 

It is also useful to relate the rate of ionizing photon production to the star formation rate in an HII region. Since the ionizing photons are produced by rare, short-lived, massive stars, this relationship depends on the population synthesis model 
-- especially the age of the stellar populations -- and the stellar initial mass function. It also varies with stellar metallicity, which impacts the opacity in the stellar atmospheres and the stellar surface temperatures. A fitting formula from \cite{Schaerer03} to population synthesis models in the case of a constant star formation rate and age larger than 6 Myr gives \cite{Yang2020}:
\begin{equation}\label{eq:q_sfr}
\begin{split}
\log_{10}\!\left(
    \frac{\dot{N}_{\mathrm{HI}}~[\mathrm{s}^{-1}]}{\mathrm{SFR}~[M_\odot\,\mathrm{yr}^{-1}]}
\right)
&= -0.0029 \left[ \log_{10}\!\left( \frac{Z}{Z_\odot} \right) + 7.3 \right]^{2.5} \\
&\quad +\, 53.81 - \log_{10}(2.55).
\end{split}
\end{equation}
This assumes a Salpeter IMF \cite{Salpeter55} with masses between $0.1 M_\odot$ and $100 M_\odot$\footnote{The factor of 2.55 adjusts the results of \cite{Schaerer03} to account for stars with mass between $0.1-1 M_\odot$ \cite{Raiter10}.}. 
Although we will proceed with this formula in what follows, it is worth commenting on some of its possible limitations.
Specifically, recent population synthesis modeling work has emphasized the importance of binarity and stellar rotation, which are neglected in this fitting formula, yet may enhance the ionizing photon production rate \cite{Stanway20,Topping15} and impact its evolution with time after stellar birth. According to Eq.~\ref{eq:q_sfr}, at $Z=0.2 \, Z_\odot$, for example, the rate of ionizing photon production is $\dot{N}_{\rm HI} = 1.2 \times 10^{53} \, {\rm s}^{-1}$ for an SFR of $1 M_\odot/{\rm yr}$. 
This corresponds to $3,900$ ionizing photons produced for each stellar nucleon, assuming a mean mass per stellar nucleon of $\mu  m_p = 1.22 \, m_p$ (with $m_p$ denoting the proton mass), appropriate for gas of primordial composition with a helium mass fraction of $Y_{\rm He} = 0.24$.

Plugging $\dot{N}_{\rm HI}$ into Eq.~\ref{eq:lalpha} gives:
\begin{equation}
    L_\alpha = 9.1 \times 10^{41}\, \frac{{\rm erg}}{{\rm s}} \left[\frac{1-f_{\rm esc}}{1}\right] \left[\frac{{\rm SFR}}{{\rm 1 M_\odot/yr}}\right],
    \label{eq:power_lya}
\end{equation}
again assuming a stellar metallicity of $Z=0.2 \, Z_\odot$ and a gas temperature of $10^4$ K. This neglects any dust extinction. Although Eq.~\ref{eq:power_lya} is instructive, it describes only the overall power output in Ly-$\alpha$, while -- as mentioned earlier -- radiative transfer effects will have prominent, and interesting, effects on the Ly-$\alpha$ LIM signal (see \S \ref{S:lya_rt}.) 
The luminosity in H-$\alpha$ will follow a formula similar to Eq.~\ref{eq:power_lya}, but H-$\alpha$ will be about 8.2 times smaller in luminosity than Ly-$\alpha$ (assuming case-B recombination at $T=10^4$ K).

\subsection{Level Populations}\label{S:lev_pops} In order to calculate the line emission from gas in various spectral lines, we need to determine the fraction of atoms, ions, or molecules in different energy states. In an effort to understand some of the ingredients involved, it is instructive to work out the simplified yet illustrative case of the two-level atom\footnote{In what follows, we refer to the system as an ``atom'' for brevity, but the description of course applies more generally to ions, while related considerations also apply to molecular states or other two-level systems.}. Let us denote the number density of atoms in the upper energy level by $n_1$ and the number density in the lower energy state as $n_0$. For starters, we consider the simplest case where: 1) the gas is optically thin to the radiation emitted in decays from the excited state to the lower energy level, so that we can ignore the possibility that emitted radiation is absorbed elsewhere in an emitting cloud, and 2) any external radiation field has negligible intensity. That is, we neglect the impact of absorption and stimulated emission of radiation on the level populations. 

In this case, the relevant processes that can take an atom from the excited state to the lower energy level are: a) collisional de-excitations between the atom and another particle with number density $n_{\mathrm{c}}$ ({\em without photon emission}; the collision partner gains energy as the atom loses it) and b) spontaneous decays from state $1 \rightarrow 0$ with a photon carrying off the energy difference between the two states. The rate of spontaneous decays is described by the Einstein A-coefficient, $A_{10}$, with units of ${\rm s}^{-1}$. Ignoring radiative excitations, as mentioned above, collisional excitations are solely responsible for moving atoms from the ground state to the excited level. In a collisional excitation, the collision partner loses the energy it transfers to the atom. 

Provided these are the only processes that change the abundance of atoms in the two states, one can write:
\begin{equation}
    \frac{dn_0}{dt} = n_1 A_{10} + n_{\mathrm{c}} n_1 C_{10} - n_{\mathrm{c}} n_0 C_{01}.
    \label{eq:levels}
\end{equation}
The first term on the right-hand side of the equation describes the increase in the population of atoms in the ground state from spontaneous decays, the second is the rate of collisional de-excitations, while the third term characterizes the depopulation of the ground state from collisional excitations to the higher energy level. Here we have assumed that a single dominant collision partner, with number density $n_{\mathrm{c}}$, is responsible for the excitations/de-excitations, and $C_{10}$, $C_{01}$ give de-excitation/excitation rates in units of ${\rm cm}^3 \, {\rm s}^{-1}$. 
In some applications, multiple different collision partners may play a role, and one must generalize the above equation to sum over multiple species, $i$, with different number densities, $n_{\mathrm{c},i}$ and collisional rates, $C_{01,i}$, $C_{10,i}$. Here, we focus on the simpler case where collisions with a single partner are dominant.

In many applications, the timescales over which the cloud properties vary are long compared to the atomic physics timescales governing collisions and spontaneous decays. In this case, an equilibrium will be established in which $\frac{dn_0}{dt}=0$. The equilibrium solution for the level populations is then:
\begin{equation}
    \frac{n_1} {n_0} = \frac{n_{\mathrm{c}} C_{01}}{A_{10} + n_{\mathrm{c}} C_{10}}
    \label{eq:equilib}
\end{equation}
Further, it is useful to define the so-called {\em critical density} at which the rate of spontaneous decays matches that of collisional de-excitations:
\begin{equation}
    n_{\rm crit} = A_{10}/C_{10}.
    \label{eq:ncrit}
\end{equation}
In a system with more than two levels, the equation for the critical density must be suitably generalized. 
In the limit that the the density of collision partners is much smaller than the critical density, $n_{\mathrm{c}} << n_{\rm crit}$, Eq.~\ref{eq:equilib} gives $n_1/n_0 \sim n_{\mathrm{c}} C_{01}/A_{10}$. In this case, if spontaneous decays also proceed rapidly compared to collisional excitations, most of the atoms will be in the ground state. In the opposite limit, $n_{\mathrm{c}} >> n_{\rm crit}$ spontaneous decays occur at a negligible rate relative to collisional de-excitations and $n_1/n_0 \sim C_{01}/C_{10}$. 

Provided that that the collision partners follow a Maxwell-Boltzmann distribution at a temperature $T$, the principle of detailed balance gives:
\begin{equation}
    \frac{C_{01}}{C_{10}} = \frac{g_1}{g_0} e^{-\frac{E_{0 1}}{k_{\mathrm{B}} T}}.
    \label{eq:detailed_rates}
\end{equation}
Here, $g_0$ and $g_1$ count the number of distinct states with energy $E_0$ and $E_1$, respectively, while $E_{01}= E_1 - E_0$ is the difference in energy between the first-excited state and the ground state. The rate of collisional excitations will be heavily suppressed 
relative to de-excitations by the exponential if $E_{01} >> k_{\mathrm{B}} T$. On the other hand, if $k_{\mathrm{B}} T >> E_{01}$ then the energy cost of an excitation is small compared to that available from the thermal energy of the surrounding gas reservoir and the relative rates of excitations/de-excitations are governed by the ratio of degeneracy factors, $g_1/g_0$.

In cases where the collision partners are thermal electrons, i.e. $n_{\mathrm{c}}=n_{\mathrm{e}}$, the electrons (by definition) follow a Maxwell-Boltzmann distribution at temperature $T$. Then the collisional excitation coefficient may be written as \citep{Draine11}:
\begin{equation}
    C_{01} = 8.629 \times 10^{-8}\, {\rm cm}^3 {\rm s}^{-1} \left[\frac{T}{10^4 K}\right]^{-1/2} \frac{\Omega(0,1)}{g_0} {\rm exp}\left(-\frac{E_{10}}{k_{\mathrm{B}} T}\right).
\end{equation}
The collision strength, $\Omega(0,1)$, is determined through quantum mechanical calculations which account for the fact that the incident electron influences the wave function of the bound electrons. It is often of order unity and a weak function of temperature \citep{Draine11}. 

The luminosity of the line emission in the transition from $1 \rightarrow 0$ may be written as:
\begin{equation}
    L_{10} = \int dV n_1 h \nu_{10} A_{10},
    \label{eq:lum_10}
\end{equation}
where the integral is over the volume of emitting gas. In general, the number density of atoms in the excited state, $n_1$, may vary across the emitting region. Again, it is useful to contrast cases where the gas density is small and large relative to the critical density. In the low-density case,
\begin{equation}
    L_{10} = \int dV n_0 n_{\mathrm{c}} C_{01} h \nu_{10}; \, n_{\mathrm{c}} << n_{\rm crit},
    \label{eq:lum_lowden}
\end{equation}
while in the high-density limit,
\begin{equation}
    L_{10} = \int dV n_0 \frac{g_1}{g_0} e^{-\frac{E_{01}}{k_{\mathrm{B}} T}} h \nu_{10} A_{10}; \, n_{\mathrm{c}} >> n_{\rm crit}.
    \label{eq:lum_Hden}
\end{equation}
Note that in the low density limit the luminosity scales with $n_0 n_{\mathrm{c}}$, i.e. as gas density squared, while at high density it only scales
with $n_0$, linear in the gas density and independent of the density of the collision partners ($n_{\mathrm{c}}$). At low densities, every collisional excitation is balanced by a radiative decay but at higher densities some of the collisional excitations are followed by collisional de-excitations. At high densities the level populations are governed by the Boltzmann distribution. Notice also that in the low density limit the luminosity is independent of the decay rate, $A_{10}$, but depends on the collisional excitation rate, $C_{01}$. On the other hand, in the high density limit the luminosity is independent of the collisional rate coefficient, but does depend on $A_{10}$.

\subsection{Ionized Volume}\label{S:ion_vol} For emission lines that trace, in whole or in part, the HII regions in a galaxy, it is useful to estimate the volume of ionized hydrogen across the galaxy. Although in reality this emission arises from a collection of discrete HII regions, for determining the total ionized volume across the galaxy we can approximately consider the entire ionizing luminosity as concentrated in a single effective source at the center of a gigantic HII region. The volume of this region can be determined assuming photoionization equilibrium, i.e., a steady state has been reached where the rate of photoionizations within the HII region balance the rate of interior recombinations. This steady-state balance should hold as long as the average time between recombinations is short compared to the timescale over which the rate of ionizing photon production varies, as will usually be an excellent approximation under the prevailing ISM conditions. 

In this calculation, we adopt what is known as the ``case-B recombination rate''. To understand the circumstances under which this rate applies, note that recombinations to the ground state produce photons with energies greater than the $13.6$ eV ionization threshold which can, in turn, ionize hydrogen atoms themselves. The case-B approximation assumes that these ionizing photons are absorbed nearby and so have no net effect on the overall ionization state of the gas; only recombinations to higher energy levels are successful in counteracting photoionizations\footnote{The opposite case, in which photons produced in recombinations to the ground state escape without being absorbed in the HII region, is called ``case-A recombination''. In case-A scenarios, recombinations to the ground state have a net impact on the ionization state and so the case-A recombination rates are a bit higher than the case-B rates.}. Likewise in the case-B approximation, all Lyman-series photons produced in the recombination cascade are absorbed by nearby neutral hydrogen atoms (and these atoms are essentially all in the ground state).

Under these assumptions, the ionized volume across a galaxy follows the classic Str\"{o}mgren sphere form \cite{Stromgren39}, which may be written as:
\begin{equation}
(1 - f_{\rm esc}) \dot{N}_{\rm HI} = C \alpha_{\rm B} \langle n_{\rm e} \rangle \langle n_{\rm H} \rangle V_{\rm HII}.
\label{eq:vol_stromgren}
\end{equation}
In this equation, $\dot{N}_{\rm HI}$ is the total rate of hydrogen ionizing photons produced across the entire galaxy; $f_{\rm esc}$ is the escape fraction of ionizing photons, which accounts for the fact that some ionizing photons escape the galaxy;  $\alpha_{\rm B}$ is the case-B recombination coefficient which depends on the temperature of the gas in the HII regions; and $C$ is a clumping factor defined as $C=\langle n_{\rm e} n_{\rm H} \rangle/\langle n_{\rm e} \rangle \langle n_{\rm H} \rangle$. Here $n_{\rm e}$ is the number density of free electrons in the HII region, while the number density of protons is well-approximated by the total density of hydrogen, $n_{\rm H}$, within the highly-ionized HII region.  The averages here can be thought of as volume-averages over the ensemble of HII regions across a galaxy. The clumping factor arises because recombination is a two-body process, and so the recombination rate scales as density-squared, and in general $\langle n^2 \rangle \neq \langle n \rangle^2$. 
Note that Eq.~\ref{eq:vol_stromgren} neglects any absorption of hydrogen ionizing photons by helium, dust, or metals within an HII region. 

It is instructive to further calculate the radius of this effective HII region as $R_{\rm HII} = \left(3 V_{\rm HII}/4\pi\right)^{1/3}$.
Inserting some typical numbers gives:
\begin{align}\label{eq:rad_stromgren}
R_{\rm HII} = 0.14\,{\rm kpc} 
&\left[\frac{1 - f_{\rm esc}}{0.9}\right]^{1/3}
\left[\frac{\dot{N}_{\rm HI}}{10^{53}\,{\rm s}^{-1}}\right]^{1/3}
\left[\frac{\alpha_{\rm B}}{2.59 \times 10^{-13}\,{\rm cm}^3\,{\rm s}^{-1}}\right]^{-1/3} \nonumber \\
&\times
\left[\frac{C \langle n_{\rm e} \rangle \langle n_{\rm H} \rangle}{10^4\,{\rm cm}^{-6}}\right]^{-1/3}.
\end{align}
The rate of ionizing photon production, $\dot{N}_{\rm HI}$, may be related to the star formation rate in the galaxy. This depends on the stellar population synthesis model, the stellar metallicity,  stellar IMF, binarity, and stellar rotation. In the case of a continuous SFR model, with an age $\gtrsim 10$ Myr or so, a rough estimate -- accurate only at the factor of $\sim$ two level -- is that an SFR of $1 M_\odot/{\rm yr}$ produces ionizing photons at about a rate of $\dot{N}_{\rm HI} \sim 10^{53}$ s$^{-1}$ (see Eq.~\ref{eq:q_sfr} for a more detailed relation). Thus, the characteristic $\dot{N}_{\rm HI}$ in Eq.~\ref{eq:rad_stromgren} corresponds to about $1 M_\odot/{\rm yr}$ of star formation; this rate scales linearly with the SFR. The case-B recombination rate number above is for gas at $T=10^4$ K, and the electron/hydrogen number densities are set to plausible HII region values.

This equation can be coupled with the level population analysis of the previous section to estimate the luminosity of various emission lines from HII regions. Generically, in the case of the low-density limit (i.e. $n_{\mathrm{c}} << n_{\rm crit}$) we can combine Eq.~\ref{eq:lum_10} with Eq.~\ref{eq:vol_stromgren} to write:
\begin{equation}
L_{10} \approx (1 - f_{\rm esc}) \dot{N}_{\rm HI} h \nu_{10} C_{01} \frac{n_{\mathrm{c}} n_0}{C \alpha_{\rm B} n_{\mathrm{e}} n_{\mathrm{p}}}.
    \label{eq:lum_Hii_simple}
\end{equation}
Here we have supposed typical values -- i.e., we ignore any radial dependence -- for the number densities of ions in the lowest energy state, $n_0$, and of collision partners, $n_{\mathrm{c}}$. A more accurate expression would involve an additional volume-average. Likewise, $n_{\mathrm{e}}$ and $n_{\mathrm{p}}$ are volume-averaged quantities while the recombination coefficient, $\alpha_{\rm B}$, here adopts a typical temperature for the HII region. 

This expression can be further simplified in some cases of interest. For instance, in HII regions, the collision partners will generally be thermal electrons and so $n_{\mathrm{c}} \sim n_{\mathrm{e}}$. Furthermore, suppose that most of the atoms of a given species are in a particular ionization state, with $n_{\mathrm{X}}$ denoting the number density of the element of interest, and that these ions are primarily in the ground state such that $n_0 \sim n_{\mathrm{X}}$. 
In this case, we can write (e.g. \cite{Sun19,Yang2020}):
\begin{equation}\label{eq:l10_simple}
L_{10} \approx (1 - f_{\rm esc}) h \nu_{10} \frac{\dot{N}_{\rm HI}C_{01}}{C \alpha_B}  \left(\frac{n_{\mathrm{X}}}{n_{\mathrm{p}}}\right)_\odot \frac{Z}{Z_\odot}.
\end{equation}
This equation is generally applicable for metal lines.
As written, this expression supposes solar abundance ratios: in principle, this supposition can be tested if lines from multiple elements can be measured. Under the assumptions adopted here, the luminosity scales linearly with the gas phase metallicity, although note that the photoionization rate, $\dot{N}_{\rm HI}$, will depend somewhat on the stellar metallicity (Eq.~\ref{eq:q_sfr}), while the temperature (and hence $\alpha_{\rm B}$) will also vary a bit with gas-phase metallicity.  

This equation is handy for rough estimates, but should be modified if: i) more than two energy levels are important, ii) the gas density is comparable to or larger than the critical density, iii) the HII region gas is not mainly in the ionization state of interest, iv) upper energy states have appreciable populations, iv) the gas is not optically thin, v) the HII region properties are significantly inhomogeneous, or vi) external radiation fields play an important role in determining the level populations. The equation also ignores absorption of hydrogen ionizing photons by other elements, and any dust extinction. That is, Eq.~\ref{eq:l10_simple} contains a number of simplifying assumptions and so must be applied with care. 

Nevertheless, Eq.~\ref{eq:l10_simple} illustrates important features of the collisionally excited emission lines from HII regions, and a number of extensions to this expression are straightforward to implement \cite{Yang2020}. In general, this approach starts to connect the line luminosity to the star formation rate of a galaxy and the properties of its ISM. This is complementary to often-employed empirical methods which connect line luminosity and star formation rate based on current observations. In the case of LIM, the empirical approach generally requires significant extrapolations towards high redshift and to small star formation rates. Physical models for the line luminosity can help check such extrapolations, while also making explicit connections to the properties of the interstellar media of the emitting galaxies. As we will see in \S \ref{S:modeling}, however, understanding the correlation between line luminosity and star formation rate is just a first step in predicting LIM signals and this relation will only be indirectly probed by the upcoming measurements. 

\subsection{Specifics of Important Example Lines}

We now turn to examine, more closely, specific emission lines of interest for LIM studies. Here we will discuss these cases, starting from long (rest-frame) wavelength and moving to shorter wavelengths. This discussion will partly draw on the concepts introduced in the previous subsections of this section, while some new topics will be introduced as relevant. 
{At the beginning of each of the following sub-sections, we provide a brief italicized header to specify which previous sub-sections from this chapter are being applied and to help introduce new concepts that are covered.}

\subsubsection{The 21 cm Line} 
\label{S:21cm_overview}

{\textit{This section introduces the hyperfine transition and its astrophysical context; the level population framework of \S\ref{S:lev_pops} provides relevant background.}}

The 21 cm line arises from the hyperfine interaction between the proton and electron spin in the ground state of a hydrogen atom. That is, the electron spin magnetic moment interacts with the magnetic field produced by the spin magnetic moment of the proton. 
The total spin angular momentum (the orbital angular momentum is zero in the ground state of the hydrogen atom) of the electron-proton system is denoted by $F$ while the individual proton and electron spin contributions are given by $I$ and $J$, respectively. The triplet state with $F=1$ has slightly higher energy than the singlet state with $F=0$. The order of magnitude of the hyperfine interaction energy is $\sim 10^{-6}$ eV, down by roughly a factor of $\sim m_e/m_p$ from the energy of the fine-structure interaction in Equation~\ref{eq:so_oom}. A more detailed calculation (e.g. including the proton $g$-factor) gives an energy splitting between the hyperfine triplet and singlet states of $5.9 \times 10^{-6}$ eV, corresponding to a wavelength of $\lambda_{21} = 21$ cm and a frequency of $\nu_{21} = 1420$ MHz\footnote{For full expressions and derivations see e.g. reference \cite{Bethe:1957ncq}.}.

\begin{figure}
    \centering
    \includegraphics[width=\textwidth]{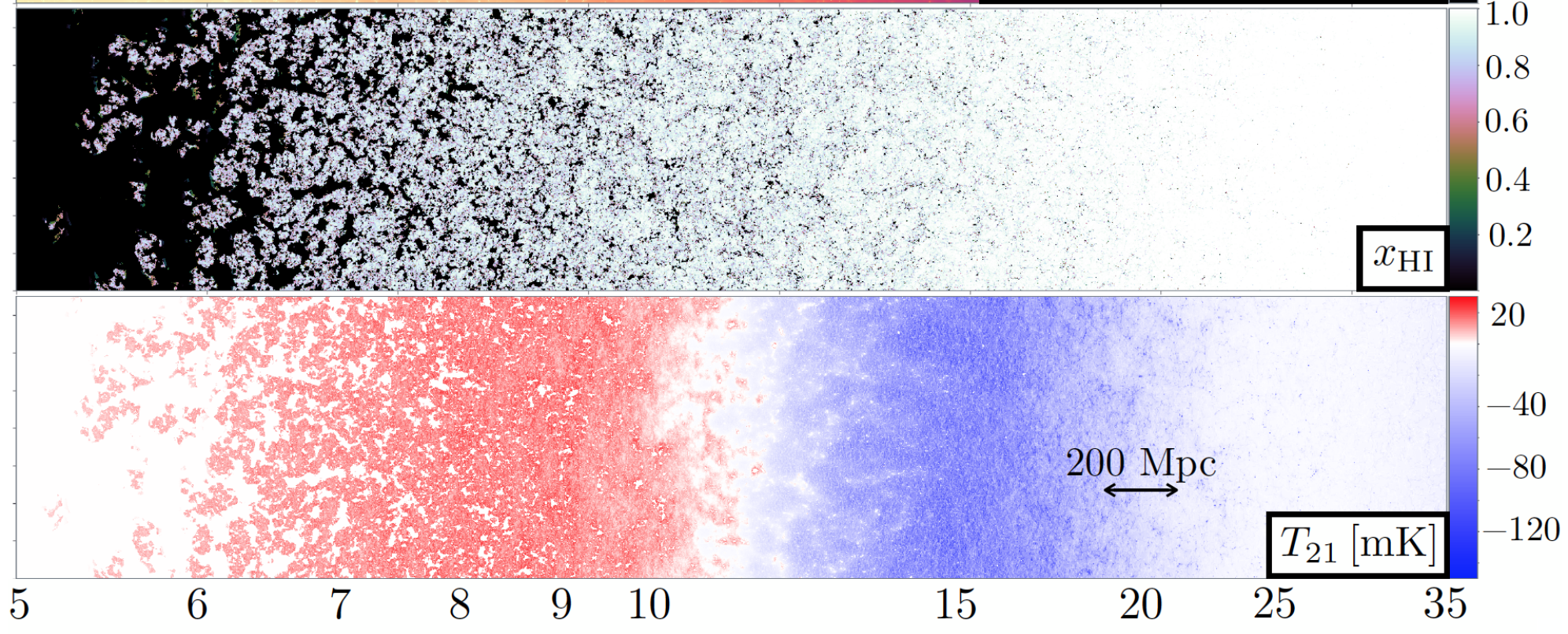}
    \caption{Models of redshifted 21 cm fluctuations during the Epoch of Reionization and the Cosmic Dawn. The panels show a lightcone through a 21cmFAST \citep{Mesinger11} simulation model. The thickness of each slice (i.e., the dimension into and out-of the page) is 1.5 co-moving Mpc, while the height is 750 co-moving Mpc. The top panel shows the neutral hydrogen fraction, $x_{\rm HI}$, across the lightcone, while the bottom panel is the 21 cm brightness temperature (relative to the CMB). The 21 cm signal appears in absorption against the CMB at $z \sim 10-20$ in this model (blue regions in the bottom panel), and then emission (red regions), until the universe is progressively ionized (dark regions in top panel and white regions in the bottom panel). From \cite{Munoz:2021psm}.}
    \label{fig:21cm_evolution}
\end{figure}

As already alluded to, the 21 cm transition is a key tracer of neutral hydrogen and holds tremendous promise for LIM studies. It can be used to study diffuse neutral gas in the IGM before and during reionization, and to trace neutral phases of the ISM/CGM in and around galaxies across cosmic time. Although we refer the reader to excellent reviews on the 21 cm signal from the EoR and the Cosmic Dawn in the current literature for details \cite{Furlanetto:2006jb,Pritchard12}, we mention a few main points here since this is a key science target for LIM surveys. This case highlights the utility of LIM for studying diffuse gas originating in the IGM and, potentially, in the pre-galactic medium before galaxies form. 

Some of the main features of 21 cm LIM may be understood by examining the equation for the 21 cm brightness temperature contrast with respect to the CMB (e.g. \cite{Furlanetto:2006jb}):
\begin{equation}
    \delta T_{\rm b} = 28\, {\rm mK} \, x_{\rm HI} (1 + \delta_\rho) \left[\frac{T_{\rm s} - T_{\rm cmb}}{T_{\mathrm{s}}}\right] \left[\frac{1+z}{10}\right]^{1/2}.
    \label{eq:t21}
\end{equation}
This equation gives the brightness temperature of a neutral hydrogen cloud at observed frequency $\nu_{21}/(1+z)$, where $\nu_{21}= 1420$ MHz is the rest-frame frequency of the 21 cm line. It is the solution of the radiative transfer equation\footnote{This result assumes the optically thin limit, which should be an excellent approximation across most of the volume of the universe.} (e.g \citep{Rybicki86}), with the specific intensity of a neutral hydrogen cloud specified in temperature units, as is common in radio astronomy (see Eq.~\ref{eq:tb}). 
The brightness temperature here is expressed relative to the CMB temperature, as the CMB is an unavoidable observing background for these measurements. The 21 cm brightness temperature depends on the neutral hydrogen fraction, $x_{\rm HI}$, the gas density contrast, $\delta_\rho$, the CMB temperature at the redshift of the cloud (i.e. $T_{\rm cmb} = 2.73 \, {\rm K} (1+z)$), and on the so-called spin temperature, $T_{\rm s}$\footnote{This equation also assumes that the CMB is the only non-negligible source of radio background photons in the frame of the cloud. For alternative scenarios, see \cite{Feng:2018rje,Ewall-Wice:2018bzf,Ewall-Wice:2019may}.}.
The larger the neutral fraction and/or gas density, the larger the 21 cm brightness temperature contrast (at fixed $z, T_{\rm s}$). 
For simplicity, Eq.~\ref{eq:t21} neglects the effects of redshift space distortions from peculiar velocities: these should be included for fully accurate results \citep{Mao2012}, but do not change the main qualitative features considered here. 

The spin temperature depends on the relative population of atoms in the excited triplet state (with abundance $n_1$) and in the lower singlet level (abundance $n_0$) via:
\begin{equation}
    \frac{n_1}{n_0} = 3 \, {\rm exp}\left(-\frac{T_\star}{T_{\rm s}}\right).
    \label{eq:tspin}
\end{equation}
Here $T_\star$ is the separation between the singlet and triplet states in temperature units, $T_\star = h \nu_{21}/k_{\rm B} = 68 \, {\rm m K}$, and the factor of 3 reflects the relative statistical weights of the triplet/singlet states. 

From Eq.~\ref{eq:t21} it follows that a neutral hydrogen cloud is: observable in 21 cm emission with $\delta T_{\rm b} > 0$ if its spin temperature is larger than the CMB temperature (at the redshift of the cloud), detectable in absorption ($\delta T_{\rm b} < 0$) if the cloud is cooler than the CMB, or unobservable if the cloud's spin temperature matches the CMB temperature ($\delta T_b = 0$). The level populations, and hence the spin temperature (Eq.~\ref{eq:tspin}), are determined by three main processes (e.g. \citep{Furlanetto:2006jb}). The first process is the absorption of, and stimulated emission by, CMB photons. These tend to drive the spin temperature of the 21 cm transition to the CMB temperature. In the absence of other processes, the hyperfine states equilibrate with the CMB and the 21 cm brightness temperature contrast tends to zero. Second, collisions between neutral hydrogen atoms and either other hydrogen atoms, electrons, or protons, couple the 21 cm spin temperature to the gas kinetic temperature. However, near the cosmic mean density, these collisions are only efficient at extremely high redshift, roughly $z \gtrsim 30$ \citep{Pritchard12}. 

The third important process in setting the spin temperature is called the Wouthuysen-Field effect \cite{Wouthuysen52,Field58}. In the Wouthuysen-Field effect, a hydrogen atom in one of the hyperfine states absorbs a UV photon as it redshifts into a Lyman-series resonance. 
The excited atom subsequently decays back down to the ground state, but may swap hyperfine states after absorbing and re-emitting Lyman-series photons. That is, this process mixes the hyperfine states. It turns out that the Wouthuysen-Field effect also tends to couple the 21 cm spin temperature to the gas kinetic temperature. In order for the Wouthuysen-Field coupling to be efficient, a background of UV photons is required. Specifically,
on the order of one Lyman-$\alpha$ photon per logarithmic frequency interval (in the frame of the gas) for every ten hydrogen atoms are needed \cite{Chen04}. This should be achieved much before reionization completes \cite{Chen04,Madau97}, which requires $\sim$ a few ionizing photons per hydrogen atom (after accounting for recombinations). 

The redshift evolution of the 21 cm brightness temperature fluctuations (Eq.~\ref{eq:t21}) reflects a combination of these three processes, which determine the spin temperature, along with reionization (via $x_{\rm HI}$), the growth of structure (through $\delta_\rho$), and sources of heating/cooling which impact the gas kinetic temperature. The current expectation is that X-ray photons will heat the gas up to temperatures much above the CMB temperature significantly before the universe is fully reionized, but after the Wouthuysen-Field coupling is achieved \cite{Pritchard12}. The long mean free paths of the X-ray photons allow them to partially ionize and photo-heat gas in the largely neutral medium before this gas is fully reionized by UV photons. The X-ray emission only needs to warm the gas above $T_{\rm cmb} \sim 30$ K, and this will cause the (previously coupled with $T_{\rm s} \sim T_{\rm gas}$) neutral hydrogen to evolve from being observable in absorption to an emission signal. 
There are a number of plausible sources for early X-rays, including AGN, supernova remnants, and high-mass X-ray binaries. The latter sources, powered by the accretion of matter onto a compact object from the wind of a massive companion star, may be dominant. 

The bottom panel of Figure~\ref{fig:21cm_evolution} shows a simulated model for the evolution of the 21 cm brightness temperature field, and its spatial variations. 
First, there is an early phase -- mostly above the redshifts shown in the figure -- where the spin temperature is collisionally coupled to the gas temperature and thermally decoupled from the CMB. In this regime (spanning roughly $30 \lesssim z \lesssim 150$), known as the Cosmic Dark Ages, the 21 cm signal is potentially observable in absorption \cite{Loeb:2003ya}. This is an appealing target for fundamental cosmology since astrophysical sources have yet to form, and the 21 cm fluctuations should be mostly gentle and hence well-described by linear perturbation theory, yet it will be extremely challenging to observe in practice\footnote{In part because the brightness temperature of the synchrotron emission foreground from our own galaxy scales with frequency as $\propto \nu^{-2.6}$. The large foreground contamination and inevitable background-limited detector noise at low frequencies make observing strongly redshifted 21 cm fluctuations from the Cosmic Dark Ages difficult.}.
At the right-most end of the panel ($z \sim 35$) the 21 cm transition is largely unobservable (with most simulation voxels near $\delta T_{\rm b} \sim 0$). This occurs because the gas density has declined enough throughout most of the universe for collisional-coupling to be inefficient, while the UV background has not yet built up (at least in the particular simulation model shown in the figure). Next, the Wouthuysen-Field coupling kicks-in and leads to an absorption signal throughout most of the simulation volume at $z \sim 12-20$ in the model of Figure~\ref{fig:21cm_evolution}. Here, the contrast with the CMB reaches a minimum value of $\delta T_{\rm b} \sim -120$ mK near $z \sim 15$.  Subsequently, around $z \lesssim 12$ (in the scenario shown) X-ray heating boosts the kinetic temperature of the gas above the CMB temperature, and the 21 cm signal becomes observable in emission. 
Finally, hydrogen is progressively reionized by UV photons and the 21 cm signal gradually fades as the neutral hydrogen vanishes. After reionization, neutral hydrogen will remain in galaxies and their circumgalactic media, although this gas is not captured in the model of Figure~\ref{fig:21cm_evolution}.

{In the post-reionization regime, neutral hydrogen resides in dense, self-shielded systems where collisional coupling and Ly-$\alpha$ pumping ensure $T_{\rm s} \sim T_{\rm k} \gg T_{\rm CMB}$. Hence, the average brightness temperature contrast in Eq.~\ref{eq:t21} takes on a simple form. 
Specifically, from the expression above one can write:
\begin{equation}
\delta T_{\rm b}= T_0 \frac{\Omega_{\mathrm{HI}}(z)}{10^{-3}}\left[\frac{\Omega_{\mathrm{m}} +  \Omega_{\Lambda}(1+z)^{-3}} {0.29}\right]^{-1/2}\left(\frac{1+z}{2.5}\right)^{1/2},
\label{eq:t21b}
\end{equation}
where $\Omega_{\mathrm{HI}}$ is the average neutral hydrogen mass-density in units of the critical density, and $T_0=0.39$ mK \cite{Chang:2007xk, Switzer2015}\footnote{Note that there is a typo in \cite{Chang:2007xk}.}. The characteristic $\Omega_{\mathrm{HI}} \sim 10^{-3}$ number gives the order of magnitude for the HI mass density indicated by measurements of the damped Ly-$\alpha$ absorber column densities towards background quasars, which suggest relatively weak
$z$ dependence \cite{Crighton:2015pza}.
}

In addition to the overall average behavior alluded to above, the spatial fluctuations in the 21 cm signal contain a wealth of information. For example, the spatial variations in the 21 cm brightness temperature during the reionization era encode interesting information regarding the clustering of the ionized sources, and the spatial distribution of dense sinks of ionizing photons. We refer the reader to 21 cm-focused reviews \citep{Furlanetto:2006jb,Pritchard12} for further discussion.

In broad summary, the 21 cm signal is expected to reveal several landmarks in the formation of cosmic structure and the first luminous sources. First, the Cosmic Dark Ages reflect mainly linear physics and are hence a potential treasure trove of information regarding fundamental cosmology. Second, 
the Wouthuysen-Field coupling epoch is set by the time when the first luminous sources produce on the order of one Ly-$\alpha$ photon, per log frequency, for every ten hydrogen atoms \cite{Chen04}. Third, the X-ray heating epoch produces an emission signal after on the order of one X-ray photon for every $10^4-10^5$ hydrogen atoms are emitted. Fourth, the reionization epoch completes when around a $\sim$ few hydrogen-ionizing photons are emitted per hydrogen atom (accounting for recombinations). Finally, the 21 cm signal in the post-reionization era arises from neutral gas within galaxies, their circumgalactic media, and any other sufficiently dense regions of the universe that retain neutral hydrogen.

\subsubsection{CO Rotational Transitions} 
\label{S:co_lum_model}

{\textit{The rotational transition physics is introduced here; the level population and critical density framework of \S\ref{S:lev_pops} is relevant background, and is extended to an optically thick case here.}}

A primary tracer of molecular gas, and a promising target for LIM studies, is provided by the transitions between different rotational states of the CO molecule \cite{Righi:2008br,Visbal10,Gong11,Lidz11}. The rotational energy levels may be calculated by treating the CO molecule as a quantized, rigid rotator. Denoting the reduced mass of the CO molecules by $\mu$ and the spatial separation between the two atoms by $R$, the rotational energy levels are given by (e.g. \cite{Draine11}):
\begin{equation}
E_J = \frac{\hbar^2}{2 \mu R^2} J(J+1),
\label{eq:E_J}
\end{equation}
where $J$ is the orbital angular momentum quantum number with $J=0,1,2,...$ Transitions between two adjacent levels $J \rightarrow J-1$ emit photons with a frequency of:
\begin{equation}
    \nu_{J,J-1} = J \times 115 \, {\rm GHz}.
\end{equation}
It is also helpful to note the energy difference between rotational transitions in temperature units, $h \nu_{J, J-1}/k_{\mathrm{B}} = J \times 5.5\, {\rm K}$. The transitions are described by the shorthand $\mathrm{CO}(J\text{--}J\!-\!1)$ so that CO(1-0), for instance, indicates a jump between the $J=1$ and $J=0$ states. 

In the general spirit of this section, we can obtain a rough estimate of the luminosity in the CO(1-0) transition following \cite{Draine11,Sun19}. 
Specifically, we can determine an approximate relationship between the total mass of molecular hydrogen in a galaxy and its CO(1-0) luminosity. This involves a number of simplifying assumptions, clarified below. We only sketch the calculation, and refer the reader to \cite{Draine11} (their Chapter 19) for details. Here we would like to account for the fact that molecular clouds are often, at least in the relatively nearby universe, optically thick in the CO(1-0) line. In this case, radiation emitted from within a molecular cloud will often be absorbed or induce stimulated emission elsewhere in the cloud. 
In principle, this complicates calculations significantly since the level populations/emission at one location of the cloud depend on the properties/emission at other positions. Here we assume that the level populations can be approximated as spatially uniform and that the abundances of CO molecules in the lowest two rotational levels are characterized by a (spatially uniform) excitation temperature, $T_{\rm ex}$:
\begin{equation}
    \frac{n_1}{n_0} = \frac{g_1}{g_0} {\rm exp}\left(-\frac{h \nu_{10}}{k_{\mathrm{B}} T_{\rm ex}}\right).
\end{equation}
In the escape probability approximation we consider the probability, averaged over angle and frequency, that an emitted photon escapes from within a uniform, spherical cloud. This 
can be related to the line-center optical depth from the cloud center to the cloud edge. 
The escape probability, $\beta$, may be approximated as \cite{Draine11}: $\beta \approx 1/(1 + 0.5 \tau_0)$. The line-center optical depth across the cloud, of radius R, is:
\begin{equation}\label{eq:tau_zero}
    \tau_0 = n_0 \frac{g_1}{g_0} \frac{\lambda^2_{10} A_{10}}{8 \pi} \frac{\lambda_{10}}{\sqrt{\pi} b} R\, \Bigg[1 - {\rm exp}\left(-\frac{h \nu_{10}}{k_{\mathrm{B}} T_{\rm ex}}\right)\Bigg].
\end{equation}
Here $n_0$ is the number density of CO molecules in the lowest ($J=0$) level, the series of factors before the radius ($R$) give the line-center absorption cross section ($b$ is the $b$-parameter, related to the frequency width of the line emission), and the second term in the square brackets accounts for stimulated emission. 

As in \cite{Draine11}, we further assume that the line-broadening is dominated by turbulent motion, and that the CO(1-0) emitting molecular cloud is in virial equilibrium with turbulent kinetic energy balancing gravitational potential energy. In this case, the b-parameter is given by $b = (2 G M/5 R)^{1/2}$. This enters into the equation for the line-center optical depth (Eq.~\ref{eq:tau_zero}) and partly determines the escape probability, $\beta$.  
In the optically thick limit ($\tau_0 >> 1$, $\beta \sim 2/\tau_0$), the CO(1-0) luminosity is
\begin{equation}
    L_{10}= \frac{4 \pi R^3}{3} h \nu_{10} A_{10} n_1 \frac{2}{\tau_0}. 
    \label{eq:lco_thick}
\end{equation}
This equation has a simple interpretation. 
The optical depth is just the path length (here $R$) divided by the mean-free path, $\lambda$ (not to be confused with the transition wavelength, $\lambda_{10})$, and so in the optically thick limit the emerging radiation comes from an outer shell of width $\sim 2 \lambda/3$. That is, the number of molecules that emit escaping CO(1-0) radiation, per second, is $4 \pi R^2 (2 \lambda/3) n_1 A_{10}$ and the energy emitted in each transition is $h \nu_{10}$ (thus yielding the result in Eq.~\ref{eq:lco_thick}).

Strictly speaking, this is the emission from a single molecular cloud. However, if the clouds are non-overlapping both spatially and in frequency, then the total CO emission is just a sum over that from individual clouds. In this case, one can relate the CO(1-0) luminosity to the total mass of molecular hydrogen across the entire galaxy. Specifically, after inserting Eq.~\ref{eq:tau_zero} into Eq.~\ref{eq:lco_thick}, the CO(1-0) luminosity per unit mass in molecular hydrogen is given by \cite{Draine11}:
\begin{equation}
\frac{L_{10}}{M_{\rm H_2}} = 32 \pi^2 \frac{h c}{\lambda^4_{10}} \left(\frac{2 G}{15}\right)^{1/2} \frac{1}{\sqrt{\rho}} \frac{M}{M_{\rm H_2}} \frac{1}{{\rm exp}\left(\frac{h \nu_{10}}{k_{\mathrm{B}} T_{\rm ex}}\right) -1}.
\label{eq:lco_mh2}
\end{equation}
Here, $\rho$ is the average total gas density in the molecular cloud, $M$ is the total mass of the molecular cloud, and $M_{\rm H_2}$ is its mass in molecular hydrogen. Inserting numbers, we find: 
\begin{equation}\label{eq:lco_mh2_numbers}
    \frac{L_{10}}{M_{\rm H_2}} = 1.4 \times 10^{-5} \frac{L_\odot}{M_\odot} \left[\frac{n_{\rm H_2}}{10^3\, {\rm cm}^{-3}}\right]^{-1/2} \left[\frac{2 (1 + f_{\rm He})}{2.72}\right]^{1/2}  \frac{1}{\rm{exp}\left(\frac{h \nu_{10}}{k_{\mathrm{B}} T_{\rm ex}}\right) -1}. 
\end{equation}
Here we have adopted a typical molecular hydrogen number density, $n_{\rm H_2}$ for a molecular cloud (at least in our Milky Way). The factor $2 (1 + f_{\rm He})$
accounts for helium \citep{Sun19}: the total mass in the cloud is $M = (1 + f_{\rm He}) M_{\rm H_2}$, while the  gas density is related to the number density of molecular hydrogen as $\rho = (1 + f_{\rm He}) 2 m_{\rm H} n_{\rm H_2}$, with $m_{\rm H}$ denoting the mass of a hydrogen atom (and $2 m_{\rm H}$ is the mass per hydrogen molecule). 

As a brief aside, we can connect this equation to additional quantities that are commonly considered in the current literature on CO emission and molecular gas. Specifically, rather than the CO(1-0) luminosity (denoted by $L_{10}$ here), it is common to consider the velocity-integrated brightness temperature multiplied by the source area\footnote{The velocity here is defined in the source rest frame.}. We will denote this quantity, referred to as the brightness-temperature luminosity, by $L^\prime_{10}$ and express it in the conventional units of ${\rm K\, km\, s^{-1}\, pc^2}$. The brightness-temperature luminosity and the normal luminosity (i.e. the total radiated power in units of $L_\odot$ or erg/s) are related by $L^\prime_{10} = \lambda^3_{10} L_{10}/(8 \pi k_{\mathrm{B}}) $ \cite{Obreschkow09}. In the literature, a variant of Eq.~\ref{eq:lco_mh2} is often considered: the ratio, $\alpha_{\rm CO}$, of the molecular hydrogen mass to the brightness-temperature luminosity. In the model of Eq.~\ref{eq:lco_mh2} this is given by:
\begin{align}\label{eq:alpha_co}
\alpha_{\rm CO} =  \frac{M_{\rm H_2}}{L^\prime_{10}} = 3.6\, \frac{M_\odot}{\rm{K\, km\, s^{-1}\, pc^2}}  & \left[\frac{n_{\rm H_2}}{10^3\, {\rm cm}^{-3}}\right]^{1/2} \left[\frac{2.72}{2 (1 + f_{\rm He})}\right]^{1/2} \nonumber \\
\times & \left[\rm{exp}\left(\frac{h \nu_{10}}{k_{\mathrm{B}} T_{\rm ex}}\right) - 1 \right].
\end{align}
For a slightly larger molecular density of $n_{H_2} = 2 \times 10^{3}$ cm$^{-3}$ and $T_{\rm ex} = 10$ K \cite{Sun19}, this gives $\alpha_{\rm CO} = 3.8$ in the above units.

Finally, a related quantity -- denoted $X_{\rm CO}$ -- is also frequently used. This is the $\mathrm{H}_2$ column density divided by the velocity-integrated brightness temperature. The quantity $X_{\rm CO}$ may be expressed in units of ${\rm cm^{-2}\, \left(K \, km\, s^{-1}\right)^{-1}}$ and is found to be \cite{Obreschkow09}:
\begin{equation}\label{eq:xco}
    X_{\rm CO} = \frac{\alpha_{\mathrm{CO}}}{2 m_\mathrm{H}} = 6.2 \times 10^{19}
    \frac{{\rm cm^{-2}}}{{\rm K \, km\, s^{-1}}}
    \frac{\alpha_{\rm CO}}{M_\odot/{\rm K\, km\, s^{-1}\, pc^2}}.
\end{equation}
For $\alpha_{\rm CO} = 3.8$, this gives $X_{\rm CO} = 2.4 \times 10^{20}$, with each quantity in the above units.
Essentially, these relations (Eqs. \ref{eq:alpha_co} and \ref{eq:xco}) are expressions for the mass-to-light ratio connecting molecular hydrogen mass and alternative CO(1-0) emission observables.

{Empirically, $\alpha_{\rm CO}$ varies significantly across different galaxy populations and environments. This variation is an important source of uncertainty in using CO as a molecular gas tracer. For reference, the commonly adopted Milky Way value is $\alpha_{\rm CO} = 4.3 \, M_\odot \, ({\rm K \, km \, s^{-1} \, pc^2})^{-1}$ 
\cite{Bolatto13}, broadly consistent with the model estimate above. However, in starburst galaxies and ultra-luminous infrared galaxies (ULIRGs), where the molecular gas tends to be warmer, more turbulent, and less self-gravitating than in normal spirals, empirical values are around $\alpha_{\rm CO} \sim 0.8 \, M_\odot \, ({\rm K \, km \, s^{-1} \, pc^2})^{-1}$ \cite{Bolatto13}, about a factor of $\sim 5$ smaller than in the Milky Way. On the other hand, low metallicity dwarf galaxies typically lack the large dust columns required to shield CO from dissociating UV radiation. In such galaxies, empirical estimates of $\alpha_{\rm CO}$ can be well above the Milky Way value \cite{Bolatto13}. 
A challenge in using CO LIM signals to determine a census of molecular gas density will be to account for the diversity of plausible conversion factors, and their redshift evolution. 
}

Although instructive, we should also keep in mind the limitations of Eq.~\ref{eq:lco_mh2} and its related forms. First, these equations assume spatially uniform, optically thick, molecular clouds and ignore spatial/frequency overlap of different clouds across each galaxy. Next, in order to actually predict the CO(1-0) emission from a high redshift galaxy with Eq.~\ref{eq:lco_mh2} additional modeling/assumptions are required regarding: i) the reservoir of molecular hydrogen in a galaxy, ii) the number density of molecular hydrogen molecules in typical clouds, and iii) the excitation temperature, $T_{\rm ex}$. 

There are additional issues that merit careful attention, especially for LIM applications at high redshift \cite{Munoz:2013tv,Mashian15,Lidz11,Obreschkow09}. First, the optically thick assumption may be inappropriate at high redshift, especially in galaxies with low dust abundances. This is because a sufficient dust column is required to shield CO molecules from dissociating UV radiation. Second, the CMB provides an observing background and there may be little contrast between cool molecular clouds and the CMB at high redshift. A related issue is that the lowest level rotational states may be de-populated if the gas is relatively warm at high redshift, potentially favoring transitions between higher rotational levels.

We refer the reader to \cite{Munoz:2013tv} and \cite{Mashian15} for works that pursue more detailed CO emission models. Briefly, these two studies consider global models for the structure of high redshift galaxies, and exploit correlations between star formation rate surface densities and ISM gas densities. In addition, they employ a variant of the escape probability approximation mentioned above to model radiative transfer effects and to self-consistently predict the CO level populations and the resulting emission. Ultimately, these studies make predictions for CO luminosities as a function of host halo mass, including the emission in the higher-J transitions. 

Further work is necessary to test these models and the cruder case of Eq.~\ref{eq:lco_mh2}. Nevertheless, note that Eq.~\ref{eq:lco_mh2} implies -- assuming molecular clouds with a typical molecular hydrogen number density and excitation temperature -- that CO LIM can provide a census of molecular hydrogen gas versus redshift (e.g. \citep{Sun19}, see \S \ref{s:molecular}). 
That is, up to the luminosity-weighted bias factor, the clustering term (see Eqs.~\ref{eq:avg_inu},\ref{eq:twop},\ref{eq:blum}) in CO LIM yields the CO luminosity density. This can be converted into the average mass density in molecular hydrogen using Eq.~\ref{eq:lco_mh2} to relate CO luminosity and the mass in molecular hydrogen. Explicitly, if the average CO(1-0) brightness temperature is $\avg{T_{10}}$ the molecular hydrogen mass density, $\rho_{\rm H_2}$ follows as:
\begin{equation}\label{eq:rhoh2_from_tb}
\rho_{\rm H_2} = \alpha_{\rm CO}\frac{H(z)}{(1+z)^2} \avg{T_{10}}.
\end{equation}
{We refer the reader to reference \cite{Breysse:2021ecm} for further discussion regarding modeling the connection between CO LIM signals and the molecular hydrogen mass density.}

\subsubsection{The [CII] Fine-Structure Line}
\label{sec:cii_line}

{\textit{The fine-structure origin of this line is discussed in \S\ref{S:fine_structure}; its multi-phase ISM origin draws on \S\ref{S:multi_ism}; the level population framework of \S\ref{S:lev_pops} is also relevant.}}

A prime target for LIM studies (with a number of surveys underway or in the planning stages, see \S \ref{S:projects}) is the [CII] emission line at a rest-frame wavelength of $158\, \mu$m. As discussed earlier, this fine-structure line arises from transitions between the upper $^1 P_{3/2}$ level and the lower $^1 P_{1/2}$ state. These levels are separated by $91$ K in temperature units, and the ionization potential for singly-ionized carbon (i.e. to go from CI to CII) is $11.26$ eV. In principle, the [CII] $158 \, \mu$m line can hence arise from a broad range of ISM phases including PDRs, HII regions, portions of the cold neutral medium, and others. Empirically, at least in low redshift galaxies, the [CII] line is responsible for as much as $0.1-1\%$ of the total far-infrared luminosity of a galaxy \cite{Stacey91} and so is a bright line. In addition, in spite of the fact that [CII] emission may arise from diverse environments, current data shows that the luminosity in this line is fairly well correlated with a galaxy's star formation rate \cite{DeLooze:2014dta}. Recent ALMA data have started to detect this line back into the EoR, and test whether these correlations apply and evolve at early cosmic times \cite{Capak15,Harikane18,LeFevre20}.

It is challenging to construct first-principles models of the [CII] emission from galaxies, in part because of the diverse set of phases that may give rise to this line emission. In spite of these difficulties, recent work has made progress in combining semi-analytic galaxy formation models with some of the important line-emission physics. These works predict correlations between [CII] luminosity and star formation rates, [CII] luminosity functions, and other observable properties of [CII] emitters \cite{Lagache18,Ferrara19,Popping19,Liang:2023sxx}. For example, \cite{Ferrara19} models the [CII]-emitting region in a galaxy as a plane-parallel slab, with a total gas column density, a uniform number density of gas particles, and a constant metallicity. This gas is illuminated by a source of radiation, and the model calculates how much of the slab will be in each of three zones. The first zone consists of an ionized hydrogen phase, in which most of the carbon is doubly-ionized (CIII), yet with traces of CII. Next, in the second zone hydrogen is mostly neutral but carbon is still singly-ionized by photons between $11.26$ eV and $13.6$ eV.
Finally, in the third zone the gas is mostly molecular, and the carbon mostly neutral, as the dust and molecular hydrogen columns between this zone and the source are large enough to absorb most of the carbon-ionizing photons. 

In addition to modeling the ionization structure of the galaxy, it is important to consider the absorption and stimulated emission of CMB photons as these processes may play an important role in determining the relevant CII level populations \cite{Lagache18}, especially at high redshift. Further, one must account for the CMB as an observing background which can reduce the contrast with the [CII] emitting regions. In some parts of parameter space, one should also account for the optical depth of the galaxy to the [CII] radiation it emits \cite{Lagache18}. The results from PDRs/regions of neutral hydrogen (i.e. \cite{Ferrara19}'s Zone II) are also somewhat sensitive to the precise temperature of the relatively cool gas in this zone. Hence, even modeling fairly coarse-grained properties of the [CII] emission from high redshift galaxies is somewhat involved. The details of these models are beyond our scope here, and so we refer the reader to the above publications for further information. 

Instead, we briefly discuss the simpler approach described in \cite{Sun19}, although this simplicity comes with caveats. In this global model, the cooling rate in the ISM is assumed to be entirely dominated by [CII] emission. This is further assumed to have reached an equilibrium balance with the heating rate; it is, in turn, supposed that the heating is dominated by photo-electric heating off of dust grains. That is, the heat input is determined by the kinetic energy of electrons ejected from dust grains which absorb UV photons\footnote{Note that the electrons lose energy as they escape the grains, and so only a fraction of the incoming photon energy is carried into the gas by the outgoing photo-electron. The heating then depends on the electric charge of the dust grains, which determines, in part, the energy lost by the photo-electrons in escaping the Coulomb potential of the charged grains (e.g. \cite{Rubin09}.)}.   

Specifically, the photo-electric heating rate is $\epsilon_{\rm PE} L_{\rm abs}$ where $\epsilon_{\rm PE}$ is a (small) photo-electric heating efficiency factor, and $L_{\rm abs}$ is the power absorbed from UV photons. Given the assumption that the heating is balanced by cooling, which is dominated by emission in the [CII] line, $L_{\mathrm{CII}} \sim \epsilon_{\rm PE} L_{\rm abs}$. Next, it is supposed that the power absorbed from UV light is ultimately re-radiated by dust grains, so that $L_{\rm abs} \sim L_{\rm IR}$ and hence
$L_{\mathrm{CII}} \sim \epsilon_{\rm PE} L_{\rm IR}$. Finally, the IR luminosity can also be related to the star formation rate -- which powers the UV luminosity that is absorbed and re-radiated. The conversion between IR luminosity and SFR, denoted $\mathcal{K}_{\rm IR}$ -- with ${\rm SFR} = \mathcal{K}_{\rm IR} L_{\rm IR}$ -- depends on the stellar IMF, how much of the star formation is dust-obscured, the star formation history, and the metallicity, among other factors (see \cite{Madau:2014bja} for a discussion). In the case of a Salpeter IMF, solar metallicity, and a constant star formation rate of age 300 Myr, \cite{Madau:2014bja} give $\mathcal{K}_{\rm IR} = 1.73 \times 10^{-10} M_\odot/{\rm yr}/L_\odot$. Assuming a photo-electric heating efficiency of $\epsilon_{\rm PE} = 5 \times 10^{-3}$ \cite{Sun19}, this yields a rough but useful estimate of the $L_{\mathrm{CII}}-{\rm SFR}$ relation:
\begin{equation}
    L_{\rm CII} \approx 2.9 \times 10^7 L_\odot \left[\frac{{\rm SFR}}{M_\odot/{\rm yr}}\right] \left[\frac{\epsilon_{\rm PE}}{5 \times 10^{-3}}\right] \left[\frac{\mathcal{K}_{\rm IR}}{1.7 \times 10^{-10} M_\odot/{\rm yr}/L_\odot}\right].
    \label{eq:lcii_sfr}
\end{equation}
 
 Although useful, we should briefly emphasize the limitations of Eq.~\ref{eq:lcii_sfr}, as discussed further in \cite{Sun19} and references therein. First, this relation takes [CII] as the sole coolant for the interstellar gas, although other lines may be important or even dominant in some galaxies. Second, the photo-electric heating efficiency depends to some extent on environment and galaxy properties. Third, the [CII] emission may saturate at high temperatures, densities, and for strong radiation fields. Finally, the mapping between IR luminosity and star formation rate ($\mathcal{K}_{\rm IR}$) depends on the metallicity and other properties of each galaxies' stellar populations \cite{Madau:2014bja}.

\subsubsection{[OIII] Fine-Structure Lines}

{\textit{The fine-structure energy levels (Figure\,\ref{fig:oiii_levels}) and Russell-Saunders coupling are introduced in \S\ref{S:fine_structure}; the luminosity estimate here is an application of Eq.\,\ref{eq:l10_simple} from \S\ref{S:ion_vol}.}}

As briefly mentioned earlier, [OIII] fine-structure lines at a rest-frame wavelength of 88 $\mu$m have been detected from a number of $z \sim 6-9$ Lyman-break galaxies and this has provided important spectroscopic confirmation of these photometrically-selected galaxies (e.g. \cite{Harikane20} and references therein). In addition, current samples suggest that the luminosity of these high redshift galaxies -- which emitted their light less than $\sim$ 1 Gyr after the big bang -- is comparable to that of galaxies with similar star formation rates in the more nearby universe. Consequently, [OIII] 88 $\mu$m emission and the nearby 52 $\mu$m transition (see Figure \ref{fig:oiii_levels}) may be useful for LIM studies \cite{Padmanabhan:2021tjr,Moriwaki18}.

A simple estimate of the luminosity in the [OIII] 88 $\mu$m line comes from a small extension of Eq.~\ref{eq:l10_simple}, and so provides an interesting example application of this formula \cite{Yang2020}. Specifically, this estimate assumes the low density limit so that collisional de-excitations are negligible and it further assumes that each galaxy is optically thin to its [OIII] 88 $\mu$m emission. Additionally, it assumes that all of the oxygen within the effective HII region is doubly-ionized. Reference \cite{Yang2020} discusses extensions and refinements to this treatment, and cross-checks their results against the \textsc{CLOUDY} code. The estimate below slightly modifies Eq.~\ref{eq:l10_simple} to account for collisional excitations from the ground-state (0th level in Figure \ref{fig:oiii_levels}, the $^3 P_0$ state) to the 2nd excited state (i.e. the $^3 P_2$ state in Figure \ref{fig:oiii_levels}) as well as the first excited state ($^3 P_1$). 
Quantitatively, assuming $T=10^4$ K and
the low-density limit, the relationship between a galaxy's [OIII] 88 $\mu$m luminosity and its ionizing luminosity is \cite{Yang2020}:
\begin{equation}
    L_{88} = 1.8 \times 10^7 \, L_\odot \, \left[\frac{1-f_{\rm esc}}{1}\right] \left[\frac{Z/Z_\odot}{0.2}\right] \left[\frac{\dot{N}_{\rm HI}}{1.2 \times 10^{53} {\rm s}^{-1}}\right],
    \label{eq:loiii_yang}
\end{equation}
where the fiducial value of $\dot{N}_{\rm HI}$ above assumes
a stellar metallicity of $Z=0.2 \, Z_\odot$ and a star formation rate of ${\rm SFR} = 1 M_\odot/{\rm yr}$ (Eq.~\ref{eq:q_sfr}). 
In this limit, the [OIII] 88 $\mu$m luminosity scales linearly with the gas-phase metallicity and is a linear function of the SFR, since $\dot{N}_{\rm HI} \propto {\rm SFR}$ (Eq.~\ref{eq:q_sfr}). There is a weaker dependence on the stellar metallicity, which influences the relationship between the star formation rate and the rate of ionizing photon production. Note that the oxygen-to-iron ratio may be super-solar at high redshifts, as expected for enrichment dominated by core-collapse supernovae; effectively, this corresponds to a case where the gas-phase metallicity exceeds the stellar metallicty \cite{Steidel2016}.

\subsubsection{The Lyman-$\alpha$ Line}\label{S:lya_rt}

{\textit{Ly-$\alpha$ production via recombination cascades is discussed in \S\ref{S:rest_opt_uv_hii}; this section focuses on the radiative transfer effects not covered there, which are specific to this line.}}

Although we discussed the generation of Ly-$\alpha$ radiation in recombination cascades within HII regions in \S \ref{S:rest_opt_uv_hii}, this line provides an interesting case for LIM and further discussion of the important radiative transfer effects that shape this emission are warranted. Here we will focus on a short summary of a few key points and refer the reader to the excellent lecture notes of reference~\cite{Dijkstra17} for further details. 
As discussed briefly in \S \ref{S:rest_opt_uv_hii}, Ly-$\alpha$ photons may also be produced by other mechanisms including \cite{Silva13,Pullen:2013dir}: collisional excitations, fluorescent re-emission, recombinations in the CGM/IGM, and in the radiative cascade following the absorption of continuum photons which redshift into a Lyman-series resonance. 

Let us start by considering the path of a Ly-$\alpha$ photon that is produced in an HII region, following a recombination, within the ISM of a galaxy. Although the hydrogen within the HII region is largely ionized, there is some residual neutral hydrogen gas, essentially all of which is in the ground state energy level. The cross-section for absorption in the Ly-$\alpha$ line is extremely large, and so the emitted Ly-$\alpha$ photon has a high probability of being absorbed by a nearby hydrogen atom, exciting this atom from the ground state $1 s$ level into a $2 p$ state.
The excited atom will spontaneously decay back to the ground state, on a timescale of order $\sim A^{-1}_{\rm Ly-\alpha} \sim 10^{-9}$ s, emitting an additional Ly-$\alpha$ photon in a different direction. That is, this is essentially a scattering process: each Ly-$\alpha$ photon will effectively be emitted, absorbed, and re-emitted -- typically many times over -- until it either escapes or is destroyed. 
In general, a hydrogen atom that scatters a Ly-$\alpha$ photon will have some random and/or bulk velocity relative to the gas that emitted the photon. This will imprint a Doppler shift: the frequency of the photon received at the atom will differ from the emitted frequency owing to the relative velocity between the emitting gas and the scattering atom. Likewise, the scattered (re-emitted) photon will receive a second Doppler shift owing to the motion of the scattering atom. In the frame of the scattering atom, the outgoing frequency is nearly identical to the incoming frequency since atomic recoil is generally negligible \cite{Dijkstra17}. In contrast, in the frame of the surrounding gas and that of any outside observer, there are Doppler shifts which depend on the velocity vector of the atom and the incoming/outgoing directions of the photon.  
Consequently, as the Ly-$\alpha$ photons scatter around in the HII region (and sometimes beyond), they undergo a random walk in both frequency and spatial position: these photons gradually migrate away from their starting location and initial frequency (see, e.g. \citep{Dijkstra17,Laursen2010}). 

The scattering of Ly-$\alpha$ photons within an HII region depends on the column density of neutral hydrogen gas across the region. This may vary significantly from sightline-to-sightline and across different HII regions within a galaxy. In addition, the distribution of column densities are likely impacted by supernova explosions -- which can blow low density channels through the ISM -- and perhaps other feedback processes. The distribution of dust -- which can destroy Ly-$\alpha$ photons -- in the galaxy and the temperature of the gas are also important. Furthermore, the kinematics of the ISM gas plays a crucial role in shaping the emergent spectrum. 

The effects of kinematics can be nicely illustrated through three toy examples \cite{Santos2004,Dijkstra2007,Verhamme2006}. The first is an expanding cloud of material. In this case, photons that start out blueward of line center are more likely to be absorbed, because photons from the central source appear redshifted in the frame of the outflowing gas and the blue-side photons are hence shifted into resonance. 
On the other hand, photons that start redward of line center escape more readily than in a static case because they are redshifted further from resonance.
Consequently, an expanding cloud will tend to emit a single red peak, which is enhanced by the outflow. Conversely, the second toy scenario is a contracting cloud -- in the frame of the inflowing gas, the photons appear blue-shifted and this tends to deplete the red-side of the line. That is, now the red-side photons get shifted closer to line center while photons that start on the blue-side are shifted away from resonance and preferentially escape (although the blue-side photons may subsequently be redshifted into the Ly-$\alpha$ line in the frame of lower redshift gas).

The third case we consider here is that of the popular ``shell-model'' description which partly captures the effect of winds from star-forming galaxies \cite{Verhamme2006,Dijkstra17}. In this model, an expanding shell of gas moves at an outflow speed, $v_{\rm out}$. The neutral hydrogen is assumed to
be concentrated entirely in this thin shell, with an emitting HII region at the center. In this scenario, most of the detected Ly-$\alpha$ photons are received after they scatter back towards the observer off of the far side of the expanding shell. 
Photons that start from the central HII region at the Ly-$\alpha$ frequency enter the shell with a redshift -- the central region is receding from the vantage point of a hydrogen atom in the shell -- and then receive an additional redshift upon back-scattering to the observer. That is, in net, a photon back-scattering off of a receding shell drops in frequency by $\Delta \nu/\nu = - 2\,  v_{\rm out}/c$. 

Finally, Ly-$\alpha$ photons may be absorbed by dust grains and destroyed before making it out of the galaxy or even their original HII region.
Hence, the escape and spectral shape of the Ly-$\alpha$ photons emerging from a galaxy's ISM depends on the distribution of neutral hydrogen and dust grains in the galaxy \cite{Neufeld91,Hansen2006}, and its kinematics. All of these may possess complex and clumpy sub-structures on small spatial scales, and are connected with some of the feedback processes which regulate galaxy formation. 

Finally, the emerging Ly-$\alpha$ photons may also scatter off of neutral hydrogen gas in the CGM and IGM. The CGM scattering offers potentially valuable information about the extended gas reservoirs around galaxies, and the role of inflows and outflows in shaping galaxy formation. The IGM scattering is especially interesting before reionization completes, because the patchy ionization structure expected during the EoR will imprint spatial variations in the specific intensity of the observed Ly-$\alpha$ radiation \cite{Visbal:2018dsi,Ambrose:2025mcg,Almualla:2025pix}. As mentioned earlier, Ly-$\alpha$ may also be produced by other processes in addition to those originating in recombination cascades within a galaxy's HII region. Hence, the Ly-$\alpha$ LIM signal is rich, yet complex.

\begin{figure}
\begin{center}
\includegraphics[width=\textwidth]{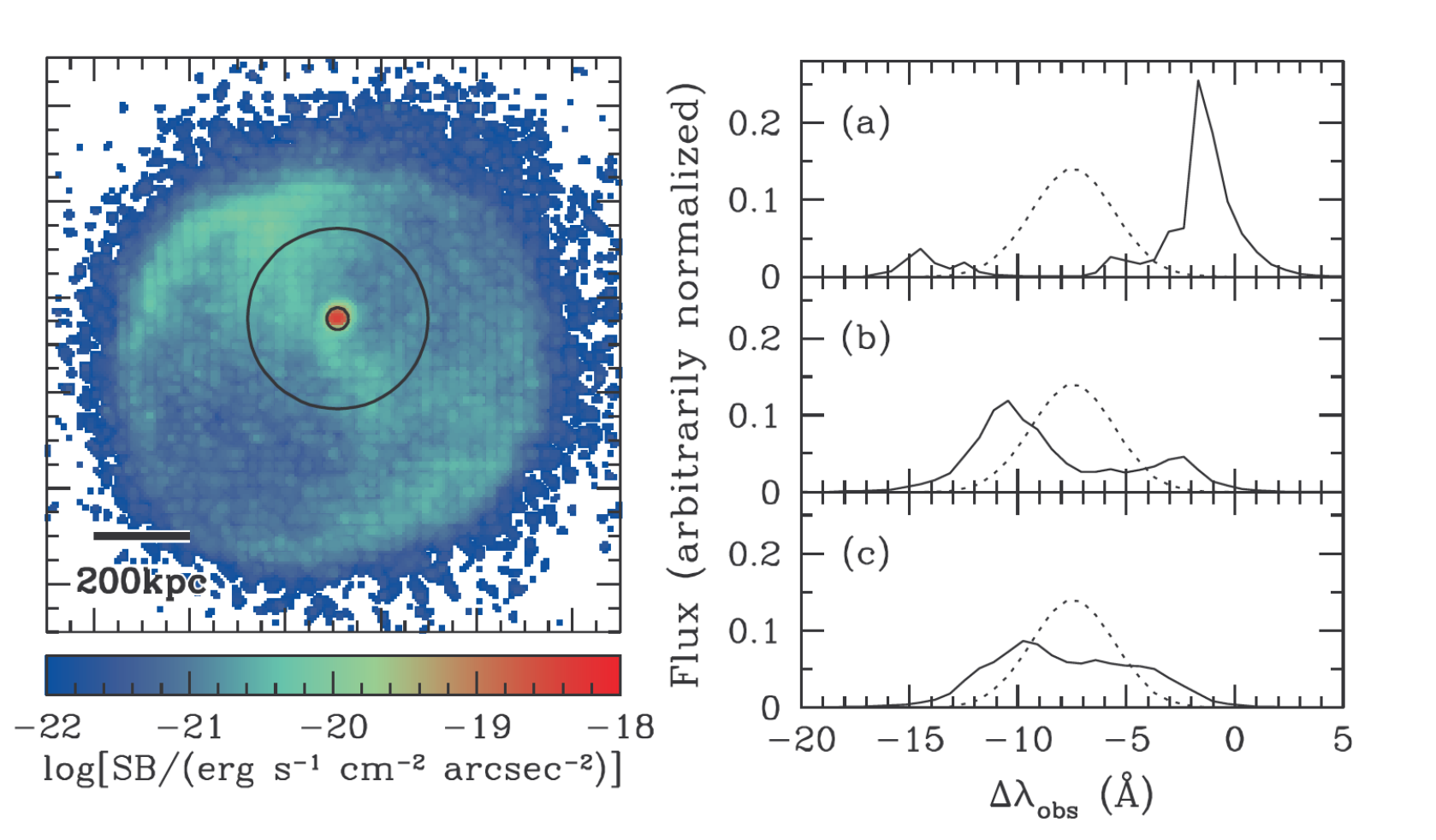}
\caption{Extended Ly-$\alpha$ emission around a simulated galaxy at $z=5.7$. {\em Left panel}: The surface brightness profile of Ly-$\alpha$ emission, with the inner red circle spanning a region comparable to the virial radius of the galaxy. {\em Right panel:} Example frequency profiles of the Ly-$\alpha$ emission. The solid line in the top row shows a line profile from emitting gas within the inner red circle. The solid line in the second row shows gas in between the red circle and the other (black) circle shown. The bottom row gives an example line profile from outside of the black circle. The intrinsic emission from the galaxy is modeled as a point source (i.e., effectively unresolved by the simulation) at the center of the halo with a Gaussian line profile, and an intrinsic linewidth set by the halo virial velocity. In each row, the dotted lines show this Gaussian model for the intrinsic line profile. 
From \cite{Zheng10}.}
\label{fig:lya_zheng}
\end{center}
\end{figure}

In any case, part of what makes Ly-$\alpha$ an appealing target for LIM surveys is that the scattering process broadens the Ly-$\alpha$ emission both spatially and in frequency. As a result, Ly-$\alpha$ emitting galaxies may be surrounded by low surface brightness ``halos'' of Ly-$\alpha$ emission that are effectively discarded in traditional narrow-band Ly-$\alpha$ surveys.  That is, in traditional surveys for Ly-$\alpha$ emitting galaxies (LAEs), candidate LAEs are selected on the basis of their flux in a narrow aperture, while the low surface brightness halos are ignored. These extended halos are shaped by the Ly-$\alpha$ scattering process and offer insights into the surrounding distribution of neutral hydrogen gas, and its kinematics, that are hard to access by other means. In fact, specialized stacking measurements targeting the low surface brightness emission have detected the average diffuse Ly-$\alpha$ emission halos around $z \sim 2.5$ Lyman-break galaxies for over a decade now \cite{Steidel2011}. 

Figure~\ref{fig:lya_zheng} shows an illustrative example: here the authors model the radiative transfer of Ly-$\alpha$ photons around a simulated galaxy at $z=5.7$ (corresponding to a wavelength window of high atmospheric transmission and the center of a commonly employed narrow-band filter). Spatially, the Ly-$\alpha$ scattering leads to an extended low surface brightness halo of emission around the galaxy ({\em left panel}). The scattering also plays an important role in shaping the emission line profile ({\em right panel}). See \cite{Zheng10} for further discussion regarding the simulated Ly-$\alpha$ surface brightness and frequency profile. 

A few slightly more detailed remarks regarding Ly-$\alpha$ scattering may be instructive. One illuminating (excuse the pun) example is to consider
the frequency at which a Ly-$\alpha$ photon will typically escape an extremely optically thick medium \cite{Adams72,Dijkstra17}. First, however, we should recall a few salient facts and numbers regarding the Ly-$\alpha$ absorption cross section. In this context, it is useful to consider the frequency offset from line center in units of the Doppler broadening width. That is,
\begin{equation}
    x = \frac{\nu - \nu_\alpha}{\Delta \nu_D},
    \label{eq:xfreq}
\end{equation}
where $\nu_\alpha$ is the central frequency of the Ly-$\alpha$ line at $\nu_\alpha = 2.466 \times 10^{15}$ Hz, $\lambda_\alpha = 1215.67  \, \Ang$ and $\Delta \nu_D$ is the Doppler width, $\Delta \nu_D/\nu_\alpha = b/c$. Here $b= \sqrt{2 k_{\mathrm{B}} T/m_{\mathrm{H}}}$ is the $b$-parameter characterizing thermal broadening, with $m_{\mathrm{H}}$ being the mass of a hydrogen atom. The line profile, $\phi(x)$, describes how the absorption is spread out over frequency. In our convention, $\int dx \phi(x) = 1$, i.e. $\phi(x)$ is normalized such that the integral of the line profile over all frequencies is unity. 
The absorption cross-section at frequency offset $x$ may then be written as $\sigma_\alpha(x) = \sigma_\alpha(0)  \phi(x)/\phi(0)$,
where $\sigma_\alpha(0)$, $\phi(0)$ respectively describe the line-center cross-section and line profile function. 
In our application, $\phi(x)$ has a Doppler core set by thermal motions and damping wings owing to the (short) lifetime for an atom in a $2 p$ state to spontaneously decay to the $1 s$ level. The spontaneous decay rate implies a fuzziness to the energy of the upper level by virtue of the energy-time Heisenberg uncertainty relation. 
The resulting line profile may be well-described by a Voigt profile. For present purposes, this profile may be approximated by \cite{Dijkstra17}:
\begin{align}
    \sigma_\alpha(x) \approx & \, \sigma_\alpha(0) \, {\rm exp}(-x^2) ; \;  {\rm core} \nonumber \\
    & \sigma_\alpha(0) \, \frac{a_v}{\sqrt{\pi} x^2}; \; {\rm wing}
    \label{eq:voigt_approx}
\end{align}
Here $a_v$ is the dimensionless Voigt parameter, $a_v = A_\alpha/(4 \pi \Delta \nu_D) = A_\alpha \lambda_\alpha/(4 \pi b)$, where $A_\alpha$ is the Einstein-A coefficient for the Ly-$\alpha$ transition. Numerically, $a_v  = 4.7 \times 10^{-4} (T/10^4 {\rm K})^{-1/2}$. According to Eq.~\ref{eq:voigt_approx}, the line profile follows a Gaussian form at small $x$ (in the ``core'' of the line), with the width of the Gaussian set by thermal broadening. At large $|x|$, in the ``wings'' of the line, the profile varies as $1/x^2$ owing to the natural line width. 

It is also useful to note the numerical value of the line-center cross-section. In terms of the $b$-parameter and fundamental constants \cite{Rybicki86},
\begin{equation}
    \sigma_\alpha(0) = \sqrt{\pi} \frac{e^2}{m_e c^2}\frac{c}{b} f_\alpha \lambda_\alpha,
\end{equation}
with $f_\alpha$ being the oscillator strength for Ly-$\alpha$ absorption, $f_\alpha=0.4162$. Inserting numbers,
\begin{equation}
    \sigma_\alpha(0) = 5.9 \times 10^{-14}\, {\rm cm}^2 \left(\frac{T}{10^4 {\rm K}}\right)^{-1/2}.
    \label{eq:line_center_sig_alpha}
\end{equation}
That is, the cross-section for Ly-$\alpha$ absorption at line center is famously large, about 11 orders of magnitude larger than the Thomson scattering cross-section. 

With this prelude, we can now return to the question of the escape of Ly-$\alpha$ photons from an optically thick medium \cite{Adams72}. We consider the simple case of a static, homogeneous medium of unspecified geometry and negligible dust content, as we are only after general trends here. In addition, we suppose that all of the Ly-$\alpha$ photons start at the center of the region and at line center. 
If the optical depth at line center is very large, e.g. $\tau_0 = N_{\rm HI} \sigma_\alpha(0) = 10^7 \left(N_{\rm HI}/1.7 \times 10^{20} {\rm cm}^{2}\right) \left(T/10^4 K\right)^{-1/2}$, then
any photons that escape the medium will be those that -- by chance -- manage to scatter out into the wings of the line. 
The photons that are instead close to line center have a tiny mean free path for scattering and remain in the medium. The photons that make it into the wing in an optically thick region still generally undergo additional scatterings and perform a random walk in both frequency and spatial position. In the case of scattering in the wing of the line, it can be shown that the rms frequency shift in a single scattering event is $x = 1$ Doppler width \cite{Osterbrock62,Dijkstra17}. There is also a slight tendency for these photons to return toward the core of the line, such that the mean shift per scattering is $-1/|x|$ \cite{Osterbrock62,Dijkstra17}. One consequence of this is that a photon that begins at frequency $x$ will typically return to the core of the line after $N_s \sim x^2$ scatterings, if it does not escape the medium in the process. As a photon scatters, it diffuses spatially in a random walk and propagates an rms distance of $d \sim \sqrt{N_s} \lambda(x)$ away from the center of the region, where $\lambda(x)$ is the mean free path of a photon at frequency $x$. In the line wings, the cross-section is a relatively slowly varying function of frequency, while the mean free path becomes tiny close to the core of the line. We can hence approximate
$\lambda(x) \sim \lambda(0) \sqrt{\pi} x^2/a_v$, with $\lambda(0) = 1/[n_{\rm HI} \sigma_\alpha(0)]$, being the mean free path at line center. 

The typical frequency of escaping photons will be those with $d \sim R$, where $R$ is the distance from the center of the medium to its edge. Photons at smaller frequency offsets (from line center) will generally be scattered back into the core, and require another ``excursion'' into the wings to potentially escape. Photons at larger frequency will be rare, since diffusing photons will have already migrated far enough to escape, while it is unlikely for a chance scattering to displace a photon from the core out to such high frequency. Hence, at very high line-center optical depth the typical frequency of escaping photons is set by the condition that $d \sim R$ and is of order \cite{Adams72,Dijkstra17}:
\begin{equation}
    x_\star \approx \pm \left(\frac{N_{\rm HI} \sigma_\alpha(0) a_v}{\sqrt{\pi}}\right)^{1/3}.
    \label{eq:x_emerge}
\end{equation}
Numerically, this gives $x_\star = 13.9$ for $N_{\rm HI} = 1.7 \times 10^{20} {\rm cm}^2$, and $T=10^4$ K -- the result scales as $x_\star \propto N_{\rm HI}^{1/3} T^{-1/3}$. In the case of a static medium, the spectrum consists of two distinct peaks which emerge symmetrically around line center (i.e., the two peaks correspond, respectively, to each of the $\pm$ solutions of Eq.~\ref{eq:x_emerge}). The emergent spectrum close to line center ($x \sim 0$) is entirely suppressed by the enormous optical depth to such photons. 
Note that the width of the line in velocity units is $2\, x_\star b \sim 360$ km/s. This highlights the important role of scattering in shaping the line profile, which in the optically thick illustration here, is vastly broader than the thermal width of the line. As an aside, Eq.~\ref{eq:x_emerge} 
also suggests that studies of Ly-$\alpha$ line profiles may help in understanding the escape fraction of ionizing photons, $f_{\rm esc}$: Ly-$\alpha$ photons traversing across smaller neutral hydrogen columns will scatter less and produce peaks closer to line center (i.e., at lower $x_\star$ in Eq.~\ref{eq:x_emerge}). This may then indicate a more porous and ionized ISM, and one in which hydrogen ionizing photons can also escape more easily. Further, in the case of sufficiently low-density/ionized channels, Ly-$\alpha$ photons can escape the galaxy close to the systemic redshift/line center. See reference \cite{Naidu:2021ryj} and references therein for more discussion regarding the connection between Ly-$\alpha$ line profiles and $f_{\rm esc}$.

Note that Eq.~\ref{eq:x_emerge} only captures the case of a static medium: as alluded to earlier, the kinematics of the scattering medium may strongly impact the emergent spectrum. As discussed previously, in outflowing gas we expect a suppressed blue peak and an enhanced red peak. In the case of an inflow, the blue peak is enhanced while the red peak is diminished. 

As with many LIM signals, it is challenging to model all of the potentially important aspects of the Ly-$\alpha$ line-intensity fluctuations from first principles. In the case of Ly-$\alpha$, the photons may be produced across a range of environments, while the radiative transfer effects reprocess this emission and depend on the properties of the ISM, CGM, and IGM on diverse spatial scales. In general, current models predict that the dominant contribution to the Ly-$\alpha$ intensity power spectrum is from photons produced after recombinations within the ISM, followed by those sourced by CGM/IGM recombinations, while other sources of photons are mainly sub-dominant (see \cite{Silva13,Pullen:2013dir} for more details here). Nevertheless, specialized analyses and modeling efforts may help to separate various effects in the Ly-$\alpha$ intensity fluctuation power spectrum. For instance, the fluctuations in the polarization of the Ly-$\alpha$ emission signal may be observable in the future: this can help in breaking parameter degeneracies that are present when analyzing only 
the total Ly-$\alpha$ intensity \cite{Mas-Ribas:2020wkz,Dijkstra2008,Rybicki99}. 

An especially interesting target for Ly-$\alpha$ LIM is the reionization-era signal. As mentioned earlier, the ionization state of the IGM during the EoR is expected to contain an evolving mixture of highly-ionized bubbles and intervening, largely neutral gas. This inhomogeneous ionization structure should modulate the reionization-era Ly-$\alpha$ LIM signal and imprint potentially observable effects on the Ly-$\alpha$ emission power spectrum \cite{Visbal:2018dsi}. Some of the Ly-$\alpha$ photons will scatter off of residual neutral hydrogen within ionized bubbles in the IGM (as well as in the ISM/CGM), but -- during the EoR -- some photons will scatter off of diffuse highly neutral gas in the IGM, which lies outside of the ionized regions. The photons that make it into the neutral gas will generally need to redshift well into the wing of the line to escape. Reference \cite{Visbal:2018dsi} finds that Ly-$\alpha$ scattering in the neutral IGM during reionization induces a scale-dependent damping of the intensity fluctuation power spectrum. This may provide a valuable handle on the spatial distribution of neutral gas during reionization, and its evolution with redshift.

\subsubsection{Balmer Lines}
\label{s:balmer}

{\textit{Balmer line emission follows directly from the recombination cascade framework of \S\ref{S:rest_opt_uv_hii} and Eq.\,\ref{eq:lalpha}; this section mainly applies those results.}}

Calculating hydrogen Balmer line emission is essentially a direct application of the formulas previously discussed in \S \ref{S:rest_opt_uv_hii}. Specifically, the luminosity in one of these transitions essentially follows Eq.~\ref{eq:lalpha} -- modified only to account for the fraction of recombinations producing the line of interest and its energy -- $L_{\rm Balmer} = \epsilon_{\rm Balmer} h \nu_{\rm Balmer} \left(1-f_{\rm esc}\right) \dot{N}_{\rm HI}$. Here it is supposed that the HII regions are in photo-ionization equilibrium, that $1-f_{\rm esc}$ of the ionizing photons are absorbed within the galaxy, and that a fraction $\epsilon_{\rm Balmer}$ of the recombinations lead to the Balmer line of interest, with energy $h \nu_{\rm Balmer}$. The $\epsilon_{\rm Balmer}$ factor depends on the temperature of the gas and the particular transition. 

These lines likely provide the most direct tracers of ionizing photon production and the overall star formation rate density. Although the link between star formation rate and ionizing photon production rate (see e.g. Eq.~\ref{eq:q_sfr}) depends on the stellar IMF, metallicity, and star formation history, these hydrogen lines are independent of the gas-phase metallicity, and avoid the complex multi-phase origin of [CII] emission. The Balmer line luminosity from a galaxy will, however, depend somewhat on the escape fraction of ionizing photons and on whether case-A or case-B recombination coefficients are a better description in the galaxy of interest. 
Furthermore, the Balmer line luminosities should be corrected for dust extinction. Towards an individual galaxy, the luminosity ratio between different Balmer lines can help determine the amount of dust extinction, since in addition to dust these ratios depend only on the physics of the radiative recombination cascade with a weak temperature dependence. In the context of LIM, dust corrections may need to be incorporated into the models of the ensemble-averaged fluctuation signals. 
Another interesting direction is to compare and cross-correlate the LIM signals in the Balmer lines with the Ly-$\alpha$ fluctuations \cite{Heneka:2016kss,Heneka21}. Since the Balmer-series lines travel from emission to the observer without scattering off of intervening neutral hydrogen, unlike Ly-$\alpha$, these analyses should help in isolating the impact of radiative transfer effects on the Ly-$\alpha$ fluctuation signal. 

Another promising aim is to target the recombination lines produced when doubly-ionized helium recombines (HeIII $\rightarrow$ HeII) \cite{Visbal:2015sca,Parsons:2021qyw,Oh:2000sg}. The strongest such line is the HeII Balmer-$\alpha$ line (i.e., the transition between the $n=3 \rightarrow 2$ states in singly-ionized helium) \cite{Oh:2000sg}. The ratio of the luminosity in the HeII Balmer-$\alpha$ line (at $1640 \, \Ang$) to that in the hydrogen Balmer-$\alpha$ line is \cite{Oh:2000sg}:
\begin{equation}
    \frac{L_{\rm HeII-\alpha}}{L_{\rm H-\alpha}} = \frac{j_{\rm HeII-\alpha}}{j_{\rm H-\alpha}} \frac{\alpha_{\rm B, HII} \, n_{\rm HII}}{\alpha_{\rm B, HeIII} \, n_{\rm HeIII}} \frac{\dot{N}_{\rm HeII}}{\dot{N}_{\rm HI}} = 4.7 \times \frac{\dot{N}_{\rm HeII}}{\dot{N}_{\rm HI}}.
    \label{eq:lheii}
\end{equation}
In the first equality above, the first term is the ratio of the emission coefficients in the HeII and hydrogen Balmer lines, the second term involves the recombination coefficients for ionized hydrogen and doubly-ionized helium, the third term is the ratio of ionized hydrogen number density to that of doubly-ionized helium, and the last factor depends on the spectral shape of the ionizing radiation. Specifically, $\dot{N}_{\rm HeII}/\dot{N}_{\rm HI}$ gives the ratio between the rate of production of photons capable of doubly-ionizing helium to that of hydrogen ionizing photons. The above relation assumes that a negligible fraction of the HeII and HI ionizing photons escape the galaxy, while the second equality gives a number relating the luminosity ratio and the spectral shape \cite{Oh:2000sg}. This value assumes that the gas temperature is $10^4$ K but the results are very weakly sensitive to temperature. It also supposes that the HeIII-region (HII region) gas is mostly doubly-ionized helium (ionized hydrogen), with a primordial helium-to-hydrogen abundance ratio. 

A primary motivation for HeII LIM is that these emission fluctuations may trace Pop-III star formation \cite{Oh:2000sg,Visbal:2015sca,Parsons:2021qyw}. Pop-III stars are the (so far hypothetical) metal-free first stars: these stars are expected to mostly be massive with high surface temperatures \cite{Bromm2009}, although their detailed IMF remains highly uncertain. In the likely massive/high temperature case, Pop-III stars have hard ionizing spectra, capable of doubly-ionizing helium, unlike most Pop-II stars. Although Pop-III stars are expected to be too faint to detect individually in the foreseeable future, their collective emission may potentially be studied using the HeII LIM technique, particularly when performed in conjunction with H-$\alpha$ LIM \cite{Visbal:2015sca,Parsons:2021qyw}. The HeII LIM signal may, however, also receive contributions from Pop-II Wolf-Rayet stars and quasars or mini-quasars. These may be separable from the Pop-III contribution using additional metal line tracers, and the broader-line width for quasar ``contaminants'', while X-ray emission may also help in quantifying any quasar contributions to the HeII fluctuations \cite{Visbal:2015sca,Parsons:2021qyw}. 

\subsubsection{Other Assorted Cases}
\label{sec:other_lines}

In addition to the examples discussed explicitly above, some other promising targets include rest-frame optical collisional excitation lines from OII and OIII ions. These will be measured by SPHEREx (see \S \ref{S:projects}), along with hydrogen Balmer lines. It may be especially interesting to combine SPHEREx LIM studies of OII/OIII/H-$\alpha$/H-$\beta$ with recent and ongoing JWST observations, which have detected these same lines towards individual reionization-era galaxies \cite{Tacchella22,Curti22,Schaerer22}. 

Among the other cases considered in the current literature are LIM surveys using (quadrupolar) rotational and vibrational lines from molecular hydrogen \cite{Gong13}, ${\rm H}_2$, and rotational transitions from hydrogen deuteride molecules \cite{Breysse:2021utr}, HD. Molecular hydrogen is thought to be the sole coolant responsible for the formation of the first stars, enabling primordial gas, condensing in the center of early dark matter halos (with total mass of order $M \sim 10^6 M_\odot$, collapsing around $z \sim 20-30$), to fragment and form Pop-III stars \cite{Bromm2009}. Hence, the LIM fluctuations from molecular hydrogen lines may allow one to probe this time period and the first stars themselves. This could potentially be combined with the HeII recombination lines discussed previously. However, the molecular hydrogen signal is faint and challenging to detect \cite{Breysse:2021utr}.
In addition, dissociating UV radiation \cite{Haiman:1996rc} may make for only a brief era where molecular hydrogen emission is strong. Hydrogen deuteride molecules may have provided an important coolant for the second generation of stars (termed ``Pop-III.2'') that are expected to form from gas which is ionized by some of the very first (``Pop-III.1'') stars. Although a tantalizing possibility, HD rotational line emission from the Cosmic Dawn will also be challenging to detect, while HD-mapping at lower redshifts is more within reach of current instrumental capabilities \cite{Breysse:2021utr}. 

Another interesting case is the Fe ${\rm K}\alpha$ line at a rest-frame energy of 6.4 keV \cite{Hutsi2012}. This line is produced around active galactic nuclei (AGN) and is thought to arise when iron (Fe) in an accretion disk is illuminated by X-rays emitted from a surrounding corona of hot gas \cite{Fabian2000}. Some of the X-rays are absorbed, ejecting one of the tightly bound inner two electrons in the lowest $n=1$ energy level (known as a K-shell electron). An $n=2$ electron can then transition into the vacancy in the $n=1$ shell, while emitting a 6.4 keV photon. LIM with the Fe ${\rm K} \alpha$ line would provide a means to study the
clustering properties of AGN and the total luminosity density in the 6.4 keV line, as a function of redshift. As a tracer of AGN activity across cosmic time, it would nicely complement the other lines discussed above which -- in most cases considered -- are expected to trace the emission from star-forming galaxies.

\subsection{Empirical Line-luminosity Results}
\label{s:empirical_line}

The previous sub-sections have emphasized important features of line emission models. Here we briefly compile some current results from the complementary approach of using empirically-derived correlations between line luminosity and star formation rate. This section focuses on the sub-mm lines, where there have been a number of recent studies. 

\subsubsection{Empirical Fits: CO Transitions} 

In the CO literature, it is common to consider the correlation between the velocity-integrated brightness temperature (\S \ref{S:co_lum_model}) in various CO transitions and the total far-infrared luminosity of a galaxy. The far-infrared luminosity can, in turn, be related to the star formation rate \cite{Kennicutt98}. These results may be combined to connect the CO luminosity to the star formation rate \cite{Carilli11,Lidz11,Pullen13,Li:2015gqa}. One must keep in mind, however, that the properties of the molecular gas, dust, star formation, and radiation backgrounds in the galaxy samples used to establish these empirical correlations may differ from those probed in a LIM survey. This is especially the case at low luminosity and high redshift, which lie beyond the range of current calibration samples. In what follows we assume that the relationship between far-infrared luminosity and SFR follows $L_{\rm IR} = 10^{10} L_\odot\, ({\rm SFR}/{1 M_\odot/{\rm yr}})$ \cite{Li:2015gqa}. We combine this with the best-fit correlation between IR luminosity and the velocity-integrated CO brightness temperature luminosity from \cite{Carilli:2013qm}, fit to a range of galaxy types to obtain:
\begin{equation}
    L_{\rm CO(1-0)} = 1.8 \times 10^4 L_\odot \left[\frac{{\rm SFR}}{1 M_\odot/{\rm yr}}\right]^{0.73}.
\label{eq:lco_10_empirical}
\end{equation}

This result is for the case of the CO(1-0) transition, although in many cases only higher-J transitions are observed directly. The measurements from higher-J lines are scaled to predict the CO(1-0) luminosity, assuming typical excitation conditions, yet allowing the rescaling factor to vary with galaxy type.  See \cite{Carilli:2013qm} for more details: their Table 2 gives the assumed conversion factors relating the velocity integrated brightness temperature in the CO(1-0) line and higher order transitions. The same table can hence be used to estimate the luminosities in the higher-order transitions from the CO(1-0) case above. Reference \cite{Carilli:2013qm} also considers separate fits for different galaxy types. 

{Another useful related reference is \cite{Keating20}, who adopt a different compilation of CO observations. The precise calibration sample employed can shift the predicted CO luminosity at a given SFR significantly. A careful accounting of this uncertainty is required when interpreting CO LIM measurements.}

\subsubsection{Empirical Fits: The [CII] $158 \, \mu$m Line} 

The local relation between [CII] luminosity and SFR is studied in \cite{DeLooze:2014dta}. Across their entire sample of 530 galaxies, spanning different populations including: metal-poor dwarf galaxies, starbursts, galaxies with AGN, ultra-luminous infrared galaxies (ULIRGs), and $z > 0.5$ sources, their best-fit relation may be written as:
\begin{equation}
L_{\rm CII, local} = 8.3 \times 10^6 L_\odot\, \left[\frac{\rm SFR}{1 M_\odot/{\rm yr}}\right]^{0.99}.
    \label{eq:delooze_cii_sfr}
\end{equation}
The $1-\sigma$ dispersion around this relation is 0.42 dex. 
These authors also give separate fits for different galaxy populations and we refer the reader to that work for the details here.

The ALPINE survey has recently measured the [CII] luminosity and estimated star formation rates towards $118$ galaxies at $z=4-6$. They have combined these with other measurements at $z \sim 6-8$ in the literature to explore how this relation evolves with redshift \cite{Schaerer20}. Their total sample consists of $153$ galaxies. Interestingly, after accounting for dust-obscured star formation, they find that the $L_{\rm CII} - {\rm SFR}$ relation in these galaxies is broadly consistent with the local relation from \cite{DeLooze:2014dta}. Quantitatively, depending a bit on the treatment of [CII] upper limits, \cite{Schaerer20} find a best-fit relation of:
\begin{equation}
    L_{\rm CII, high-z} = 1.1 \times 10^7 L_\odot\, \left[\frac{\rm SFR}{1 M_\odot/{\rm yr}}\right]^{1.0}.
    \label{eq:alpine_cii_sfr}
\end{equation}
These authors show how their fit results depend on the SFR estimate employed, on the dust correction procedure, and on the treatment of non-detections. We refer the reader to the original work for details. 

\subsubsection{Empirical Fits: The [OIII] $88 \mu$m Line}

The work of \cite{DeLooze:2014dta} also measures the local relation between the [OIII] $88 \mu$m luminosity and star formation rate in their galaxy sample. Across the same diverse galaxy sample discussed in the previous subsection, but in this case including just 83 detections, \cite{DeLooze:2014dta} find:
\begin{equation}
    L_{\rm 88, local} = 4.8 \times 10^6 L_\odot\, \left[\frac{\rm SFR}{1 M_\odot/{\rm yr}}\right]^{0.89}.
    \label{eq:delooze_oiii}
\end{equation}
These results show a 1-$\sigma$ dispersion of about 0.66 dex, although the scatter among the sub-classes of galaxies considered in that work is substantially smaller than this. 

A recent analysis of $\sim 10$ $z \sim 6-9$ galaxies from \cite{Harikane20} finds that these galaxies are comparable to, or slightly more luminous in the [OIII] $88 \mu$m than local low-metallicity dwarf galaxies, and much more luminous than local starburst galaxies. The best fit to current measurements from these authors is:
\begin{equation}
    L_{\rm 88, high-z} = 2.5 \times 10^7\, L_\odot\, \left[\frac{\rm SFR}{1 M_\odot/{\rm yr}}\right]^{0.97}.
    \label{eq:harikane_oiii}
\end{equation}
Note the broad similarity of the empirical relation above with the simple model of Eq.~\ref{eq:loiii_yang}. See \cite{Yang2020} for further discussion.

{
\subsubsection{Halo Mass-Luminosity Relations}
\label{S:lm_scaling}}

{
An alternative empirical approach is to connect line luminosity directly to host dark matter halo mass (e.g. \cite{Padmanabhan:2017ate,Padmanabhan18}), generally by using abundance matching to relate line luminosity functions and halo mass functions. This is convenient for LIM models because the halo mass function and the spatial clustering of dark matter halos are well-understood theoretically, while the LIM statistics can be expressed directly in terms of the luminosity-halo mass relation, $L(M)$ (see \S \ref{S:modeling}). 
A commonly adopted functional form for $L(M)$ is a double power law in halo mass. In this case, the luminosity increases as $L \propto M^\alpha$ before turning over at high masses and scaling as $L \propto M^\beta$.  Physically, the turnover at high masses reflects feedback from AGN and/or other quenching processes which suppress star formation and line emission in massive halos. Reference \cite{Padmanabhan:2017ate} presents fits of this form for the CO(1-0) line, while \cite{Padmanabhan18} gives results for [CII]. 
These fits are calibrated against compilations of available luminosity and SFR data, and represent one approach to interpolating between the data and predicting average luminosities at unprobed halo masses. As with the L(SFR) fits, one must be cautious about extrapolating results to high redshift and low luminosities. 
}

\section{Modeling LIM Signals}
\label{S:modeling}

The previous section provided an overview of the physics of an assortment of emission line targets and described models, as well as current empirical estimates, for the luminosity in these lines. More directly relevant for LIM studies are quantities that describe the {\em ensemble-averaged} emission from many individually unresolved galaxies and the spatial fluctuations in this emission. Here our focus is on emission lines from galaxies, as opposed to cases such as the 21 cm line where the emission -- at least before reionization completes -- comes primarily from diffuse gas in the IGM. 

A central goal of LIM studies is to measure the power spectrum of the spatial fluctuations in the line emission. Hence, our first main objective is to derive a formula for this power spectrum. We will subsequently consider additional statistical measures (\S~\ref{S:statistics}). In our first treatment here, we make several simplifying assumptions that we will later relax:

\begin{itemize}
    \item We will ignore the impact of redshift space distortions from peculiar velocities.\\
    \item We consider only spatial scales much larger than the virial radius of the galaxy-hosting dark matter halos.\\
    \item We start with a simplified case in which every dark matter halo contains either one or zero galaxies. This leads to a simplified treatment of shot-noise, which we will generalize subsequently.\\
    \item We describe the large-scale fluctuations in a linear-biasing approximation.\\
    \item We treat the frequency profile of each emission line as a delta function in frequency.\\
\end{itemize}

The first quantity of interest for LIM is the average specific intensity in the emission line of interest,
$\avg{I_\nu}$. This quantity has units of $\left[I_\nu\right] =   \rm{ergs}\ \rm{cm}^{-2}\ \rm{s}^{-1}\ $\\ $\rm{Hz}^{-1}\ \rm{str}^{-1}$, often expressed
in terms of $\rm{Jy}/\rm{str}$ where $1\, \rm{Jy} = 10^{-23}\, \rm{ergs}\ \rm{cm}^{-2}\ \rm{s}^{-1}\ $ \\ $ \rm{Hz}^{-1}$. The specific intensity
hence describes the energy per unit time per unit area per unit frequency interval per unit solid angle. Put differently, one can consider the energy passing through an area $dA$ from light rays propagating nearly in the direction normal, along $\bf{\hat{n}}$, to the area element. If the light rays span a solid angle $d \Omega$ in direction around $\bf{\hat{n}}$, and a frequency range $d \nu$ around $\nu$, then the energy passing through $dA$ in time $dt$ is
$dE = I_\nu \, dt dA d\nu d\Omega$. This defines the specific intensity \cite{Rybicki86}. 

For simplicity, we first approximate the profile
of each emission line as a delta function in frequency:
\begin{equation}
L_\nu = L\, \delta_D\left(\nu - \frac{\nu_r}{1+z}\right),
\label{eq:lum_profile}
\end{equation}
where $\nu$ is the frequency in the observer's frame, $\nu_r$ is the rest-frame wavelength of the transition, and $z$ is the redshift of the emitting source.
Let us further suppose that the average abundance of galaxies per comoving volume with luminosity between $L$ and $L+dL$ is described by a luminosity
function, $\phi(L) dL$. 
Provided the emission is not scattered or absorbed on its way to us, the average specific intensity received in the line is described by:
\begin{equation}
\avg{I_\nu} = \int dz^\prime \frac{d \chi}{dz^\prime} \frac{(1+z^\prime)^2}{\nu_r} \delta_D(z - z^\prime) \int dL \phi(L) \frac{L}{4 \pi D_{\mathrm{L}}^2(z^\prime)} D_{\mathrm{A,co}}^2(z^\prime).
\label{eq:avi_integral}
\end{equation}
Here $\chi(z)$ is the comoving distance to redshift $z$, $D_{\mathrm{L}}$ is the luminosity distance, and $D_{\mathrm{A,co}}$ is the comoving angular diameter distance to redshift
$z$ (equivalent to $\chi(z)$ in a flat universe). 

The interpretation of this equation is straightforward. First, the integral over $z^\prime$ sums over all sources whose 
emission is observed at frequency $\nu$. 
Next, $\frac{L}{4\pi D_{\mathrm{L}}^2(z^\prime)}$ is the flux observed from a source
at redshift $z^\prime$. The integral over luminosity sums over all sources with luminosity function $\phi(L)$, while $D_{\mathrm{A,co}}^2(z^\prime) d\chi/dz^\prime$ is
the comoving volume per unit solid angle per unit redshift. The factor of $(1+z^\prime)^2$ arises in moving from the specific luminosity in the rest frame of the source to that in the observed frame, and after converting the Dirac delta function in frequency to one in redshift. 
It is also useful to note that $d\chi/dz = c/H(z)$, 
where $H(z)$ is the Hubble parameter at redshift $z$. We can then carry out the integral over $z^\prime$ for the delta function line profile of Eq.~\ref{eq:lum_profile}, noting that $D_{\mathrm{L}}=(1+z) D_{\mathrm{A,co}}$. This yields:
\begin{equation}
\avg{I_\nu} =  \frac{1}{4\pi} \frac{c}{H(z) \nu_r} \int dL L \phi(L)
 = \frac{1}{4\pi}\frac{c\, \epsilon_L}{H(z) \nu_r}.
\label{eq:avg_inu}
\end{equation}

Here $\epsilon_L$ denotes the comoving emissivity of the line emission, i.e. the average luminosity per unit volume in the line at the redshift of interest. The average specific intensity hence depends on the total emissivity, including sources with individually small luminosities which are hard to detect in traditional galaxy surveys.
Note that for some applications, the specific intensity is conventionally expressed in terms of an equivalent Rayleigh-Jeans brightness temperature, $T_{\mathrm{B}}$:
\begin{equation}
\avg{T_{\mathrm{B}}} = \frac{c^2 \avg{I_\nu}}{2 k_{\mathrm{B}} \nu^2},
\label{eq:tb}
\end{equation}
where $\nu$ is frequency in the observed frame, and $k_{\mathrm{B}}$ is Boltzmann's constant.

Although the average specific intensity characterizes the strength of the line emission, this quantity is not directly observable with a LIM survey. The main reason for this is that in addition
to the line emission of interest, each resolution element in a LIM survey will typically be swamped by foreground emission, especially spectrally-smooth continuum emission (and possibly
interloper line emission).
In order to separate the foreground continuum emission, one uses its spectral smoothness; essentially, the line emission signal has a relatively small amplitude but a great deal of spectral structure
and can be distinguished from large-amplitude foregrounds which vary slowly with frequency (see \S \ref{S:challenges}).
This allows one to measure spatial fluctuations in the emission signal (although long wavelength modes along the line of sight
are sacrificed), but not the average emission.

In order to characterize the -- more readily observable -- spatial fluctuations, we turn to
the two-point correlation function, and its Fourier transform, the power spectrum. Here we start in configuration space and generalize Eq.~\ref{eq:avg_inu} to calculate the spatial correlation between the specific intensities at two separate points:
\begin{align}
& \avg{I_\nu(\x_1) I_\nu(\x_2)}  - \avg{I_\nu}^2  = \left[\frac{1}{4\pi} \frac{c}{H(z) \nu_r}\right]^2 \times \nonumber \\
&  \int dL dL^\prime \, L  L^\prime  \bigg[\phi(L) \phi(L^\prime) b(L) b(L^\prime) \xi_{\delta,\delta}(|\x_1-\x_2|)  
+ \, \phi(L)\delta_D(L - L^\prime) \delta_D(\x_1 - \x_2)\bigg].
\label{eq:twop}
\end{align}
Here $b(L)$ denotes the linear bias of a line-emitting source at luminosity $L$, while $\xi_{\delta,\delta}(|\x_1 - \x_2|)$ is the linear density correlation function.
This expression involves two terms, the first of which arises when $\x_1$ and $\x_2$ are two separate points, and the second of which takes $\x_1=\x_2$. The first term here assumes pure linear
biasing and reflects the clustering of the host halos of galaxies with luminosity $L$. The second term is a Poisson contribution arising from the discrete nature of the emitting sources.
Eq.~\ref{eq:twop} can be rewritten after factoring out the spatially-averaged specific intensity as:
\begin{equation} 
\avg{I_\nu(\x_1) I_\nu(\x_2)} - \avg{I_\nu}^2 = \avg{I_\nu}^2 \left[\avg{b_L}^2 \xi_{\delta,\delta}(|\x_1 - \x_2|) + \frac{\avg{L^2}}{\avg{L}^2} \delta_D(\x_1-\x_2) \right].
\label{eq:twop_fac}
\end{equation}
Here, $\avg{b_L}$ denotes the average luminosity-weighted bias of the emitting sources:
\begin{equation}
\avg{b_L} = \frac{\int dL\, b(L) L \phi(L)}{\int dL\, L \phi(L)},
\label{eq:blum}
\end{equation}
while
\begin{equation}
\frac{\avg{L^2}}{\avg{L}^2} = \frac{\int dL\, L^2 \phi(L)}{\left[\int dL\, L \phi(L)\right]^2}
\label{eq:poss}
\end{equation}
has units of volume and depends on the second-moment of the luminosity function.

Equivalently, in Fourier space, the power spectrum of emission fluctuations is:
\begin{equation}
P_I(k) = \avg{I_\nu}^2 \left[\avg{b_L}^2 P_{\delta,\delta}(k) + \frac{\avg{L^2}}{\avg{L}^2}\right].
\label{eq:pofk}
\end{equation}
Assuming linear biasing, the first term is proportional to the linear matter power spectrum and arises because the line-emitting host halos are biased tracers of the underlying density field.
One of the main goals of LIM surveys is to measure $P_I(k)$ and Eq.~\ref{eq:pofk} provides a compact description of the main science that may be extracted from
these observations. The auto spectrum of the line intensity fluctuations depends on: the power spectrum of density fluctuations, the average specific intensity of the emission line of interest, on the luminosity-weighted bias of the
underlying host halos, and on Poisson fluctuations in the abundance of discrete line-emitting sources. These quantities are in turn determined by the first and second moments of the luminosity function, on the galaxy-halo connection (i.e. on how galaxies of luminosity $L$ populate halos of mass $M$), and on the matter power
spectrum at the redshift of the emitting gas. Note that some references do not explicitly factor $\avg{I_\nu}^2$ out of the Poisson term (e.g. \cite{Bernal:2022jap}), and Eq.~\ref{eq:pofk} therefore appears in various different, yet equivalent, forms across the literature.

\begin{figure}
\begin{center}
\includegraphics[width=\textwidth]{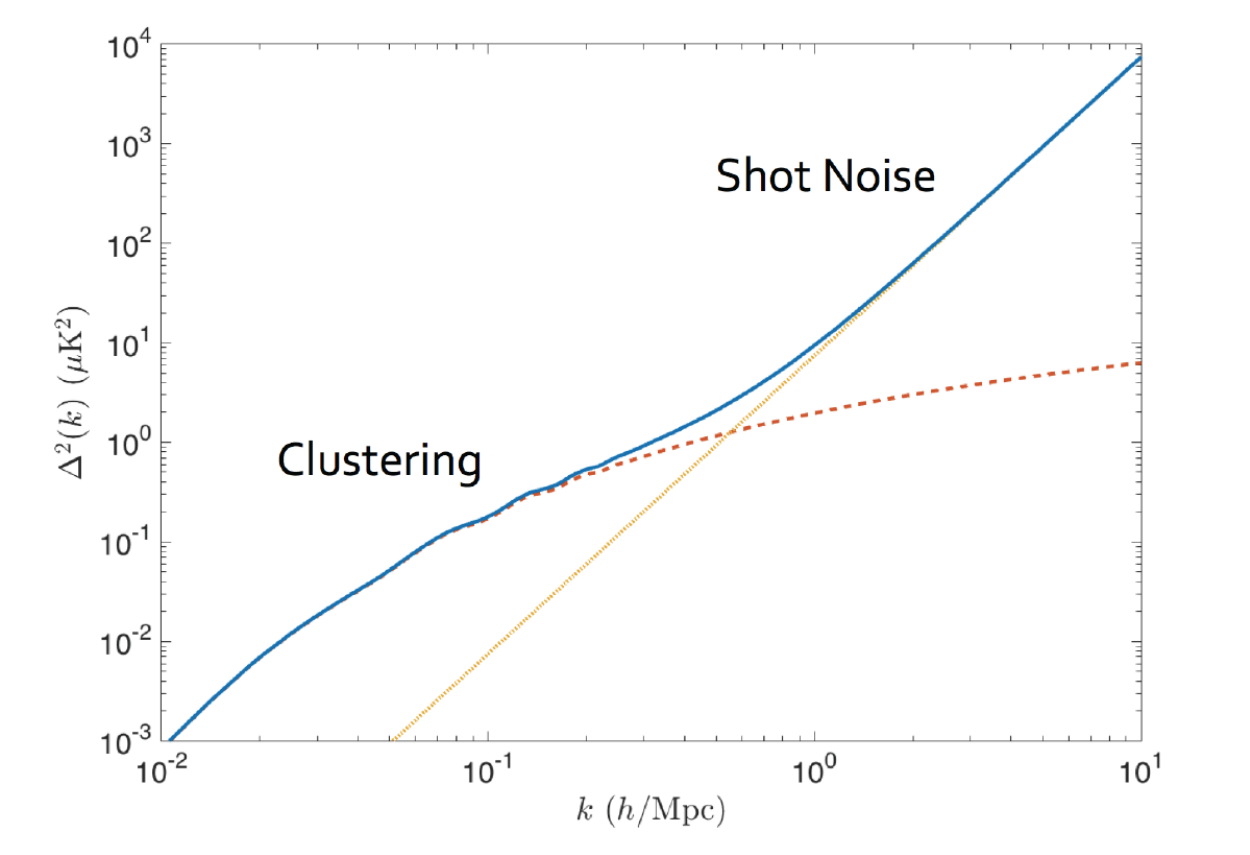}
\caption{An illustrative example model of the LIM power spectrum. The blue curve shows the total power spectrum of the CO(1-0) brightness temperature field at $z=3$. The red dashed-line shows the clustering term of Eq.~\ref{eq:pofk}, while the orange dotted-line gives the shot-noise contribution. From \cite{Kovetz:2017agg}.}
\label{fig:lim_clustering_shot}
\end{center}
\end{figure}

The power spectrum is often expressed in terms of $\Delta^2_I(k) = k^3 P_I(k)/(2 \pi^2)$. The quantity $\Delta^2_I(k)$ describes the contribution to the variance of the field $I_\nu(\x)$ per logarithmic interval in $k$, i.e. $\Delta^2_I(k) = d \sigma^2_I/d \ln k$. Figure~\ref{fig:lim_clustering_shot} shows an example of the line intensity mapping power spectrum for the case of CO(1-0) line fluctuations. In this case, the fluctuations in the brightness temperature field (Eq.~\ref{eq:tb}) are shown. As expected from Eq.~\ref{eq:pofk}, the fluctuation power $\Delta^2(k)$ scales with the average brightness temperature squared and is given here in $\mu {\rm K}^2$ units. On large scales (low $k$) the power spectrum is dominated by the clustering term, i.e. the first term in Eq.~\ref{eq:pofk}, while the shot-noise fluctuations take over at higher $k$. In the particular case of the CO(1-0) model of Figure~\ref{fig:lim_clustering_shot}, the two contributions become comparable at wavenumbers a little below $k \sim 1$ h Mpc$^{-1}$. 
In other cases of interest, the cross-over between the two terms may move to larger or smaller scales depending on: the redshift (with the matter power spectrum dropping towards higher redshifts), the host dark matter halo masses of the emitters (through their impact on $\avg{b_{\rm L}}$), and the abundance of the line-emitting galaxies (via the Poisson term). 

In order to construct a model along the lines of that in Figure~\ref{fig:lim_clustering_shot}, following Eq.~\ref{eq:pofk}, we need to specify the luminosity function of the line emitters and the source bias as a function of luminosity. The latter generally involves relating the luminosity
of each galaxy to its host halo mass, since the clustering of dark matter halos as a function of their mass and redshift is well characterized from numerical simulations and analytic modeling (e.g. \citep{Sheth02,Tinker:2008ff}.)

\subsection{Line Luminosity Function Modeling}

A number of different approaches for modeling line luminosity functions have been considered in the literature. One method starts by assuming a correlation between a galaxy's line luminosity and its star formation rate (e.g. \S \ref{s:empirical_line}):
\begin{equation}
L = L_0 \left[\frac{\rm{SFR}}{1\, M_\odot/\rm{yr}}\right]^\beta,
\label{eq:lum_sfr}
\end{equation}
where $L$ is the total luminosity in the line, $L_0$ is the luminosity of a galaxy that forms stars at a rate of $\rm{SFR}=1\, M_\odot/\rm{yr}$, and $\beta$
is a power-law index. Often, Eq.~\ref{eq:lum_sfr} has been fit from observations rather than predicted from first principles. As discussed in \S \ref{S:landscape}, an important issue in forecasting the sensitivity
of upcoming LIM surveys is that Eq.~\ref{eq:lum_sfr} is often calibrated using nearby, relatively bright galaxy samples. This relation may not apply for the galaxy populations that
produce most of the emission fluctuations in many of the planned LIM observations. Nevertheless, adopting the locally calibrated relation provides a useful baseline. Alternatively, this correlation could be predicted or studied using (partly) first-principles simulated models\footnote{The ``partly'' qualifier is included here because current simulations require sub-grid models that are themselves empirically calibrated.}. 

The $L-\rm{SFR}$ relation can be combined with measurements of the abundance of star-forming galaxies as a function of their star formation rate (termed ``star formation rate functions'') at various redshifts. This has been fairly well-characterized from observations of the UV luminosity functions at $z \sim 0-8$ (e.g. \citep{Bouwens15,Bouwens22}, with constraints reaching beyond $z \gtrsim 10$ from recent JWST observations \cite{Bouwens:2022gqg,Gardner:2006ky}.   
Thus far, the star formation rate function is well fit by a Schechter function \cite{Schechter76} form:
\begin{equation}
\Phi(\rm{SFR}) d\rm{SFR} = \Phi_\star \left(\frac{\rm{SFR}}{\rm{SFR}_\star}\right)^\alpha \rm{exp}\left[-\frac{\rm{SFR}}{\rm{SFR}_\star}\right] \frac{d\rm{SFR}}{\rm{SFR}_\star}.
\label{eq:sfr_schech}
\end{equation}
Here $\Phi_\star$ is a characteristic comoving abundance, while $\rm{SFR_\star}$ is the ``break'' star formation rate: at higher star formation rates, the abundance is exponentially suppressed, while the abundance increases as a power-law towards lower star formation rates. 

If all galaxies above some minimum star formation rate ($\rm{SFR_{min}}$) are luminous in the line of interest, and follow the relationship
of Eq.~\ref{eq:lum_sfr} without scatter, the comoving luminosity density follows from integrating over the SFR function (Eq.~\ref{eq:sfr_schech}) as:
\begin{equation}
\epsilon_L = \Phi_\star L_0 \left(\frac{\rm{SFR}_\star}{1 M_\odot/\rm{yr}}\right)^\beta \Gamma\left[\alpha+\beta+1, \frac{\rm{SFR_{min}}}{\rm{SFR}_\star}\right].
\label{eq:eps_co}
\end{equation}
Here $\Gamma$ denotes an Incomplete Gamma Function. The average specific intensity in the line may then be calculated according to Eq.~\ref{eq:avg_inu}. Note that the emissivity scales with $L_0 \Phi_\star$, and so this expression has the required luminosity density units. 
As expected, a larger value of $\beta$ weights galaxies with large star formation rates more strongly, while galaxies with low star formation rates become more important in the case of a weaker (e.g. sub-linear) scaling of line luminosity with star formation rate. Likewise, a steeper faint-end slope (i.e., a more negative value of $\alpha$) weights the small star formation rate galaxies more heavily in the Incomplete Gamma Function (until $\rm{SFR}_{\rm min}$).

\begin{figure}
\begin{center}
\includegraphics[width=\textwidth]{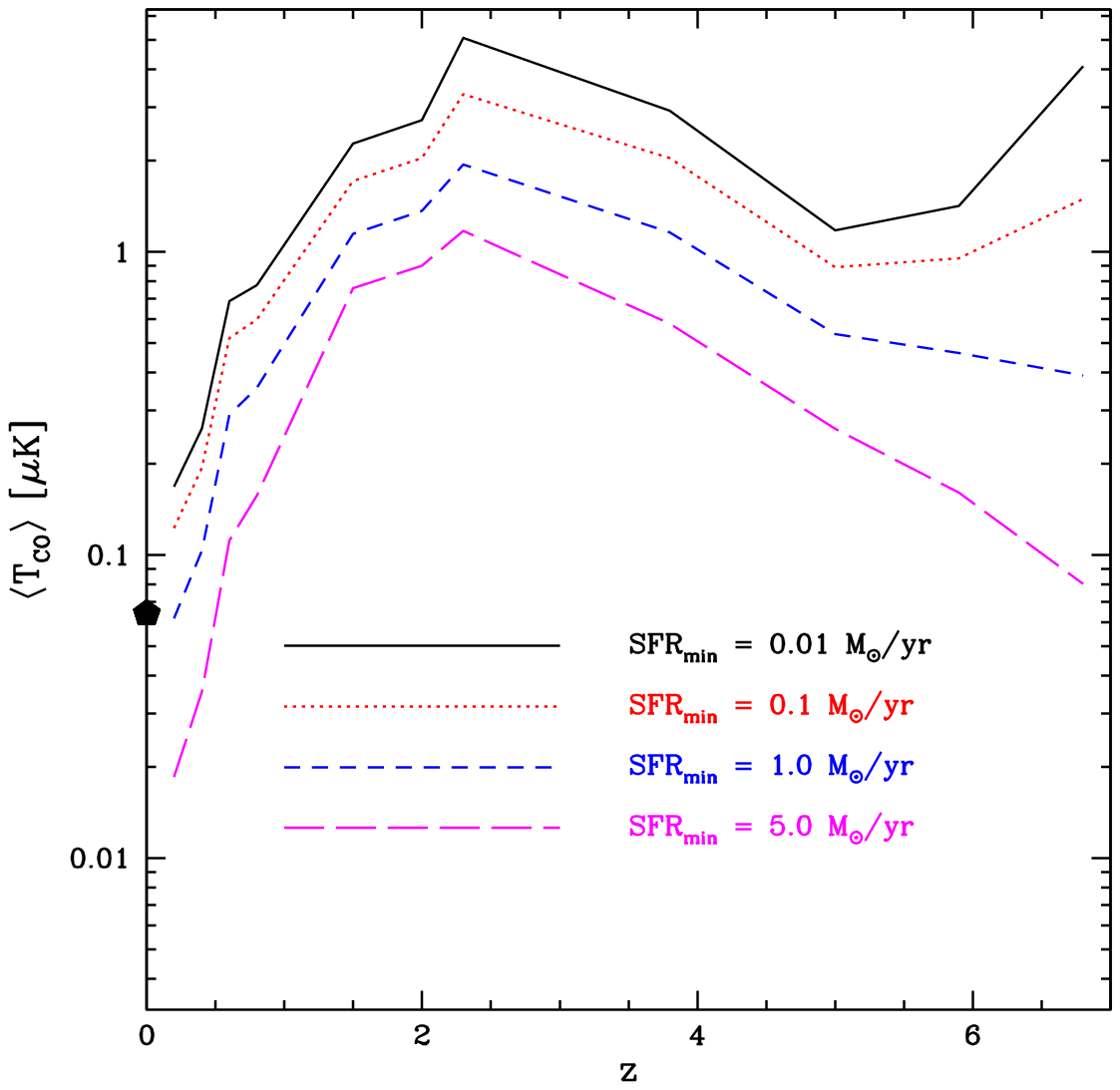}
\caption{Empirically-calibrated models for the average brightness temperature of CO(1-0) emission as a function of redshift. These models assumes a local calibration between CO(1-0) line-luminosity and SFR \citep{Lidz11}, the star formation rate functions from \cite{Smit12}, and Eqs.~\ref{eq:avg_inu},\ref{eq:tb},\ref{eq:eps_co}. The different curves show examples where galaxies with low star formation rates, below the threshold value ${\rm SFR_{min}}$ do not contribute to the CO(1-0) emission. The pentagon shows the average brightness temperature inferred from an observational estimate of the CO luminosity function near $z=0$. The model curves are noisy because they are calibrated from empirical measurements of the star formation rate functions in a few discrete redshift bins. From \cite{Pullen13}.}
\label{fig:tco_examp}
\end{center}
\end{figure}

Figure~\ref{fig:tco_examp} shows an example of the average CO(1-0) brightness temperature in this type of model. Here, the models each assume the following correlation between CO(1-0) luminosity and SFR (see \cite{Lidz11} for details) derived from fairly local galaxy samples:
\begin{equation}
    L_{\rm CO(1-0)} = 3.2 \times 10^4 L_\odot \left[\frac{{\rm SFR}}{M_\odot/{\rm yr}}\right]^{3/5}.
    \label{eq:lco_sfr}
\end{equation}
Note that this is slightly different than the more recent empirical result in Eq.~\ref{eq:lco_10_empirical}.
As mentioned earlier, this model provides an interesting baseline assumption. However, it is unlikely to be applicable in galaxies at moderately high redshifts or at low SFRs, where CO emission may be suppressed owing to a combination of decreasing dust abundance (which shields CO from dissociating UV radiation), declining metallicity, and the increasing CMB temperature (which reduces the contrast between the CO emitting galaxies and the CMB) \cite{Obreschkow09,Lidz11}. If galaxies with low star formation rates are deficient in CO, this may have a strong impact on the average CO(1-0) brightness temperature, denoted by $\avg{T_{\mathrm{CO}}}$ in Figure~\ref{fig:tco_examp}. This is especially the case given the sub-linear scaling of CO(1-0) luminosity with SFR in Eq.~\ref{eq:lco_sfr} and at high redshift, where the faint-end slope of the star formation rate function (i.e., the parameter $\alpha$ in Eq.~\ref{eq:sfr_schech}) is steep. These features are illustrated in the redshift evolution of the average brightness temperature in this model, as shown in Figure~\ref{fig:tco_examp}.

\begin{figure}
\begin{center}
\includegraphics[width=\textwidth]{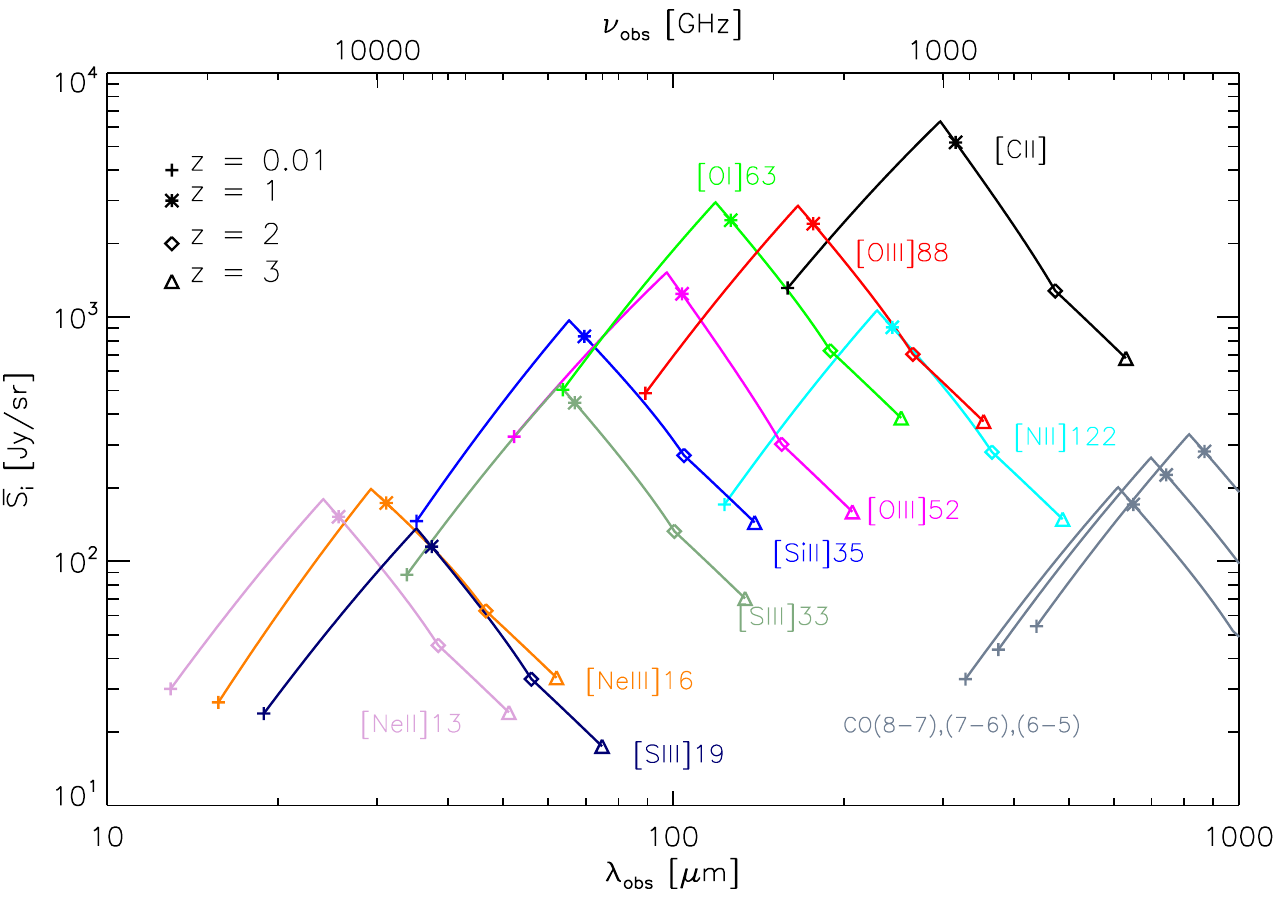}
\caption{Empirically-calibrated models for the average specific intensity in various fine-structure lines and higher-order CO transitions as a function of observed wavelength and frequency. The results of these calculations are shown for $z=0-3$. From \cite{Uzgil:2014pga}.}
\label{fig:avgi_uzgil}
\end{center}
\end{figure}

A slightly different, but closely related approach, is illustrated for the case of multiple different fine-structure emission lines, as well as some higher-J CO transitions, in Figure~\ref{fig:avgi_uzgil}. In this case, instead of using the UV-derived star formation rate functions as described above, the authors adopt infrared luminosity functions \cite{Bethermin11} and local, empirical correlations between infrared and line luminosity \cite{Spinoglio12}. In this model, the average specific intensity in each line shows a similar rise and fall with increasing observed wavelength/redshift. This mainly reflects cosmic evolution in the star formation rate density; tracing-out the history of cosmic star formation using various LIM targets is an important science goal, as further motivated in \S \ref{S:intro}, \S \ref{S:SFRD}. 

Figure~\ref{fig:avgi_uzgil} also partly illustrates the issue of line-confusion in these surveys, as a given observed wavelength or frequency typically receives contributions from multiple different emission lines from gas at a range of redshifts. In much of the redshift range shown, however, it is interesting to note that the average [CII] intensity dominates that of the other lines, at least for this particular model. 

We can also compute the Poisson term, $P_{\rm shot} = \avg{I_\nu}^2 \avg{L^2}/\avg{L}^2$ in a similar way. The result of this calculation is:
\begin{equation}
P_{\rm shot} = \frac{\avg{I_\nu}^2}{\Phi_\star}\frac{\Gamma\left(\alpha + 2 \beta + 1, \frac{\rm SFR_{min}}{\rm{SFR}_\star}\right)}{\left[\Gamma\left(\alpha + \beta + 1, \frac{\rm SFR_{min}}{\rm{SFR}_\star}\right)\right]^2}.
\label{eq:pkshot}
\end{equation}
The Poisson term scales inversely with the number density, $P_{\rm shot} \propto 1/\Phi_\star$, as expected for a shot-noise contribution. The Incomplete Gamma Function terms here give more weight to the luminous galaxies than in the case of the emissivity (Eq.~\ref{eq:eps_co}), since the Poisson contribution is determined by the second moment of the luminosity function.

\subsection{Bias Factor Models}

In order to calculate the luminosity-weighted bias (Eq.~\ref{eq:blum}), we need to consider the relationship between line luminosity and host halo mass, or between SFR and halo mass.
That is, the clustering of the line emission depends on the galaxy-halo connection. 
This is in contrast
to the average specific intensity which may be computed directly from the observed star formation rate functions and the line luminosity-SFR correlation, without explicitly connecting to the underlying host dark matter halo masses. In the case that exactly one or zero galaxies reside in each host halo, the shot-noise term can also be modeled without exploring precisely how galaxies inhabit their host halos. In \S \ref{S:refinements} we discuss how the shot-noise calculation must be refined for more general halo occupation scenarios. 
In any case, in order to model the clustering term it is necessary to consider the line emitter-host halo connection. This is also useful for extrapolating predictions of the average intensity beyond the redshift and SFR range captured by current SFR function measurements.

Here we discuss an empirically-motivated abundance matching approach, which starts from the observed star formation rate functions (Eq.~\ref{eq:sfr_schech}), as used in the previous section
to model the average specific intensity. Although a number of abundance matching models appear in the literature, here we follow the approach of \cite{Mashian16}. This work updates the observed star formation rate functions from
\cite{Smit12}, which are in turn based on dust-corrected UV luminosity function measurements. These are combined with model dark matter halo mass function models from \cite{Sheth02} to determine 
the ${\rm SFR}-M_{\rm halo}$ relation at $z=4-8$.

In the variant considered by \cite{Mashian16}, a monotonic relation between SFR and halo mass is assumed; this is calibrated by matching abundances according to:
\begin{equation}
\int_{{\rm SFR}}^{\infty} d{\rm SFR}^\prime \Phi({\rm SFR}^\prime) = f_{\rm duty} \int_{M}^{\infty} dM^\prime n(M^\prime).
\label{eq:abund_match}
\end{equation}
Here we allow for the possibility of bursty star formation with a duty cycle, $f_{\rm duty}$, less than unity. In this case, only a fraction ($f_{\rm duty}$) of the dark matter halos above a mass $M$ actively host star forming galaxies at a given redshift\footnote{Note that \cite{Mashian16} consider only $f_{\rm duty}=1$, but here we explore models with smaller duty cycles as well. It is impossible to distinguish models with varying duty cycles on the basis of UV luminosity functions alone, and so sharp constraints on scenarios with smaller duty cycles await more precise clustering measurements.}.  
In what follows, we assume that galaxies form only in host halos above $M_{\rm min} = 10^8 M_\odot$. This minimum mass is comparable to the atomic cooling mass (e.g. \citep{Barkana:2000fd}), above which primordial
gas can cool by line emission from atomic transitions and ultimately form stars.  We further assume that the duty cycle in halos above the minimum mass is independent of host halo mass.

Note that the duty cycle
is only an approximate model for the effects of bursty star formation, while other sources of line-emission stochasticity may also be relevant for LIM models including time variable dust obscuration, excitation conditions, optical depth effects, and viewing angle dependencies, among others. 
Also note, however, that lognormal scatter in the SFR-M relation (Eq.~\ref{eq:sfr_scatter})
may not capture strongly quenched galaxies as effectively as an explicit non-unity duty cycle \cite{Yang22}. We refer the interested reader to additional recent studies which discuss improved models for bursty star formation, and explore the impact on LIM signals \cite{Liu:2024fti,Munoz26,Kovetz26}.

\begin{figure}
\begin{center}
\includegraphics[width=\textwidth]{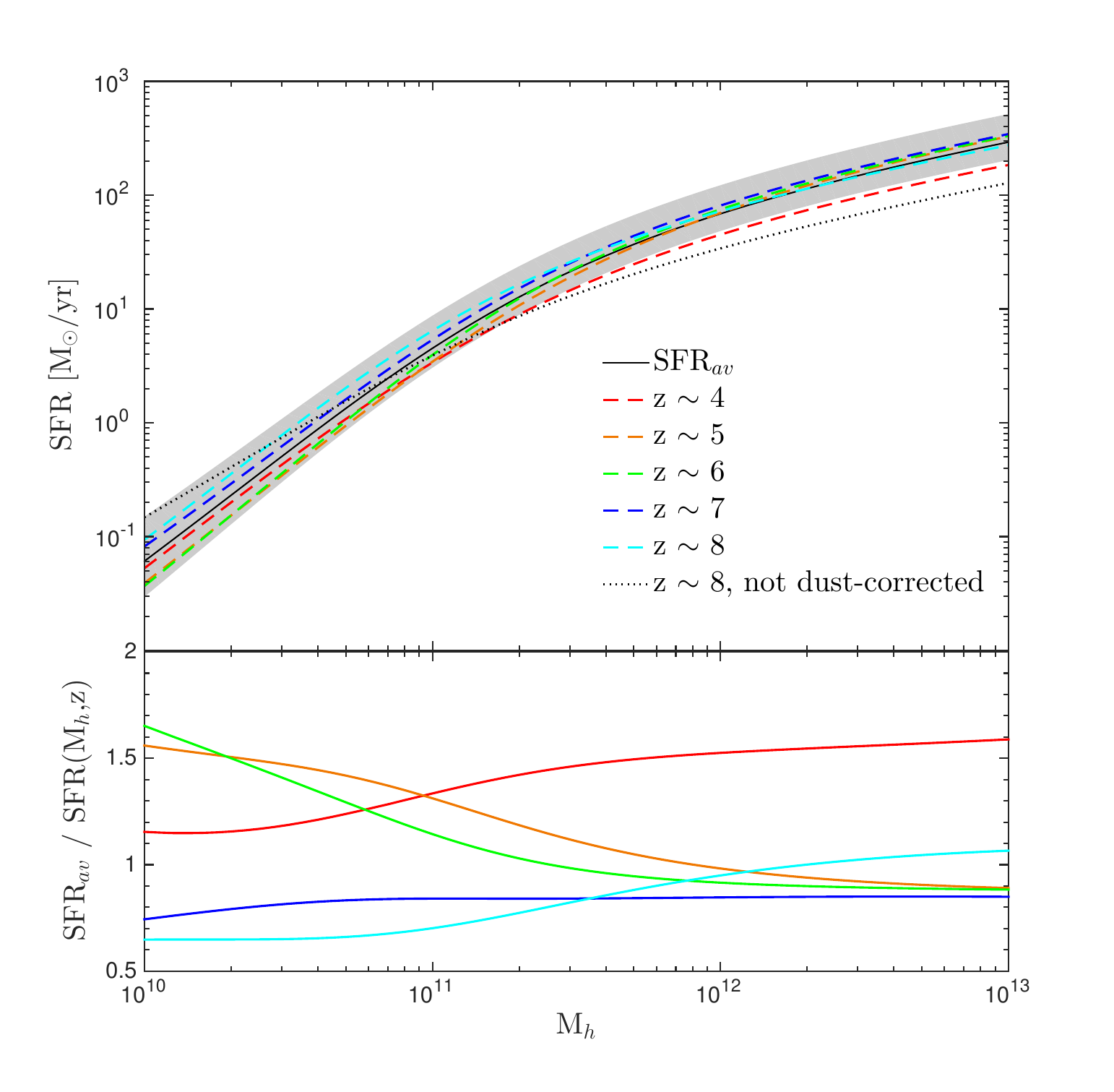}
\caption{Average SFR-$M_{\rm halo}$ relationship at $z \sim 4-8$, calibrated from abundance matching. The different colored lines show the results obtained using
the observed luminosity functions at various redshifts. The black dotted line shows the $z=8$ result in the absence of any dust correction. The gray shaded band gives an estimate of the $1-\sigma$
uncertainty on the average SFR-$M_{\rm halo}$ relation.  The bottom panel shows the ratio of the SFR at each individual redshift to the average SFR over the entire redshift range, each considered as a function of halo mass.  From \cite{Mashian16}.}
\label{fig:sfr_mhalo_average}
\end{center}
\end{figure}

Reference~\cite{Mashian16} uses Eq.~\ref{eq:abund_match} (assuming $f_{\rm duty} = 1$) to calibrate the average SFR-$M_{\rm halo}$ relationship from $z=4-8$ (see Figure \ref{fig:sfr_mhalo_average}). To within current observational uncertainties,
they find little evolution over this redshift range. It is therefore a useful starting point for extrapolating predictions to higher redshifts, smaller star formation rates, and fainter
luminosities than currently observable.  Following \cite{Mashian16}, we further suppose that there is a lognormal scatter around the median SFR-$M_{\rm halo}$ (SFR-M) relation:
\begin{equation}
P({\rm SFR}|M) = \frac{1}{\sqrt{2 \pi \sigma^2}} {\rm exp} \left[-\frac{{\rm ln}^2\left({\rm SFR}/{\rm SFR}_{\rm median}(M)\right)}{2 \sigma^2}\right] \frac{d{\rm SFR}}{{\rm SFR}},
\label{eq:sfr_scatter}
\end{equation}
The dispersion is taken to be $\sigma = \ln(10) \times 0.5\, {\rm dex}$. The SFR function at any redshift may be deduced from (\cite{Mashian16}):
\begin{equation}
\Phi({\rm SFR},z) = \int dM n(M) P({\rm SFR}|M)
\label{eq:phi_sfr_infer}
\end{equation}
This expression is useful for LIM models, provided the line luminosity is correlated with the SFR (e.g. Eq.~\ref{eq:lum_sfr}, which may be easily generalized to incorporate any additional scatter in the luminosity-SFR relationship). 

At the moment, our main interest is in modeling the luminosity-weighted bias of Eq.~\ref{eq:blum}. Combining Eqs.~\ref{eq:lum_sfr}, \ref{eq:sfr_scatter}, \ref{eq:phi_sfr_infer}, this is given by:
\begin{equation}
\avg{b} = \frac{\int_{\rm M_{min}}^\infty dM n(M) b(M) \left[\frac{{\rm SFR}_{\rm median}(M)}{1 M_\odot/\rm{yr}}\right]^\beta}{\int_{\rm M_{min}}^\infty dM n(M) \left[\frac{{\rm SFR}_{\rm median}(M)}{1 M_\odot/\rm{yr}}\right]^\beta}.
\label{eq:blum_sfr_model}
\end{equation}
Note that the scatter is independent of halo mass in this model (Eq.~\ref{eq:sfr_scatter}) and so cancels out in the average luminosity-weighted bias calculation 
of Eq.~\ref{eq:blum_sfr_model}. In the context of the model considered here, the luminosity-weighted bias is sensitive to how the median SFR, ${\rm SFR}_{\rm median}$, scales with host halo mass, $M$, and to
the power law index $\beta$, which determines the relative importance of lower SFR and higher SFR galaxies. Another important quantity is the duty cycle, $f_{\rm duty}$ (Eq.~\ref{eq:abund_match}): fixing the star formation rate function, galaxies with a given average SFR are populated in halos of decreasing mass as the duty cycle drops.
In other words, one can match the observed abundance of star-forming galaxies either if star formation has a short duty cycle but occurs in abundant, low-mass halos, or if it has a longer duty cycle yet takes place mostly in rarer, high-mass halos. 

\begin{figure}
\begin{center}
\includegraphics[width=\textwidth]{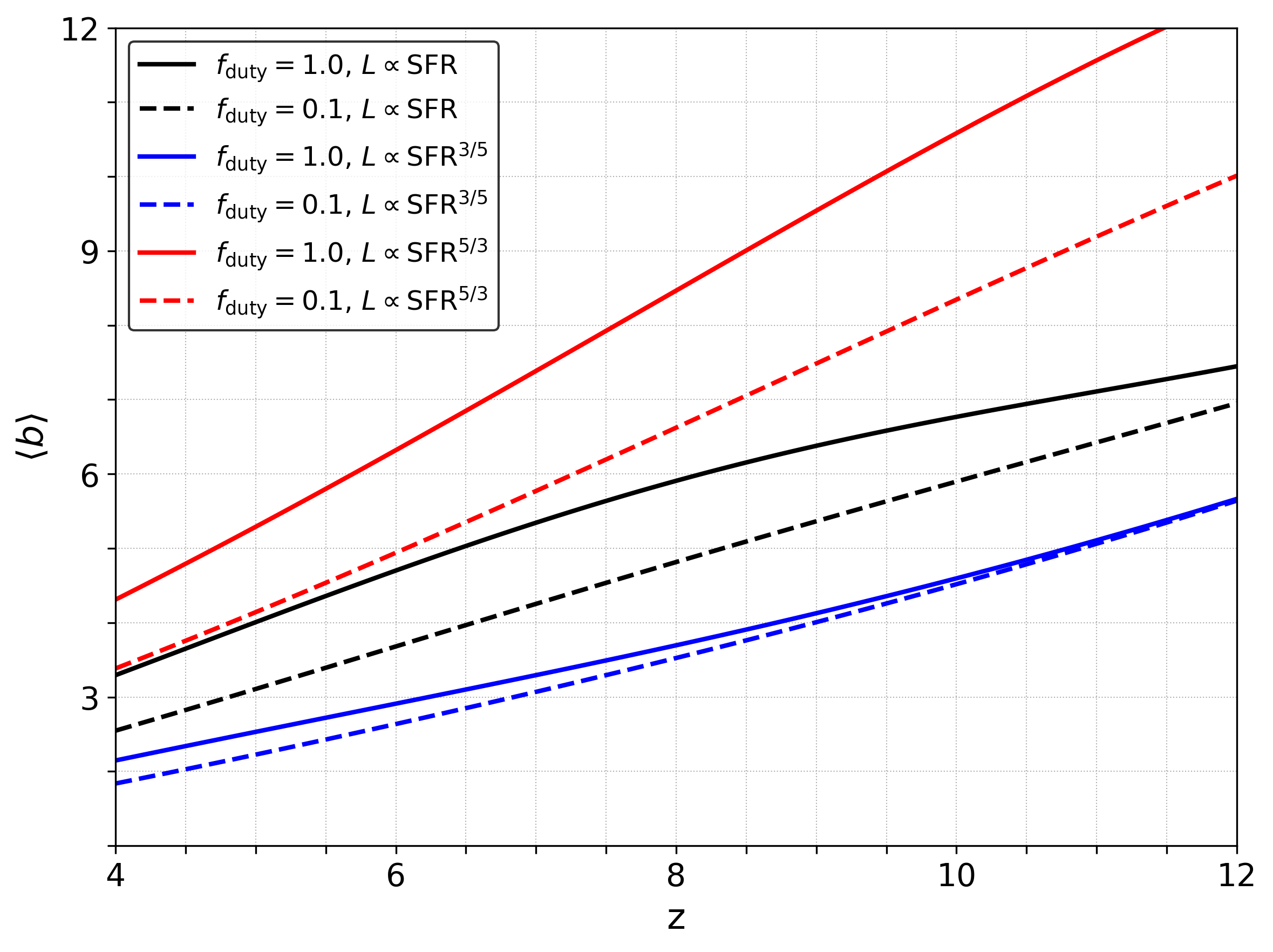}
\caption{Models for the redshift evolution of the luminosity-weighted bias. The solid lines assume a duty cycle of unity, while the dashed lines adopt $f_{\rm duty}=0.1$. The red, black, and blue
lines for each duty cycle consider cases where the line luminosity follows $L \propto {\rm SFR}^{5/3}$, $L \propto {\rm SFR}$, and $L \propto {\rm SFR}^{3/5}$, respectively. Note that the models here presume a 
correlation between line luminosity and SFR, Eq.~\ref{eq:lum_sfr}, but are otherwise independent of the emission line specifics. 
In each case, a minimum host halo mass
of $M_{\rm min} = 10^8 M_\odot$ is assumed.}
\label{fig:bias_model_v_z}
\end{center}
\end{figure}

Some example models for the redshift evolution of the luminosity-weighted bias are shown in Fig.~\ref{fig:bias_model_v_z}.  Here we consider $f_{\rm duty} = 0.1$ and $1$ and lines with luminosity
$L \propto {\rm SFR}^{5/3}$, $L \propto {\rm SFR}$, and $L \propto {\rm SFR}^{3/5}$. These cases leave a wide range of possibilities for the redshift evolution of the luminosity-weighted bias factors. As discussed above, decreasing the duty cycle lowers the characteristic host dark matter halo masses, and the resulting bias factors. Likewise, the bias factors are significantly smaller
if the line luminosity scales
sub-linearly with SFR (as is observed, for example, in samples of low redshift galaxies emitting in low-$J$ CO transitions, see \S \ref{s:empirical_line} and Eq.~\ref{eq:lco_10_empirical}) than if luminosity scales linearly or super-linearly with SFR. This occurs
because the product $n(M) b(M) \left[\frac{{\rm SFR}_{\rm av}(M)}{1 M_\odot/{\rm yr}}\right]^\beta$ determines how halos of different mass and bias factor contribute to the weighted
average of Eq.~\ref{eq:blum_sfr_model}, and the sub-linear scaling gives more weight to the lower mass, less-clustered host halos. 

\subsection{Conditional Luminosity Function}
\label{S:clf}

Another approach is to make use of the conditional luminosity function (CLF) \cite{Yang2008}, which specifies the average number of emitters with line luminosity between $L$ and $L + dL$ residing in a halo of mass $M$. This is closely related to the above model but the CLF description directly links the line luminosity and halo mass without necessarily connecting to the star formation rate. Note that in general the CLF will differ depending on the emission line being studied, and so one might add an index, $L_i$, to specify the luminosity in the $i$th line. We will usually suppress such indices for brevity of notation.  
Here the CLF is denoted $\phi_{\mathrm{c}}(L|M)$, and connects the galaxy luminosity function $\phi(L)$ to the halos mass function, $n(M)$ via:
\begin{equation}
\phi(L) = \int dM\, n(M) \phi_{\mathrm{c}}(L|M).
    \label{eq:clf}
\end{equation}
Note that, given this definition, $\phi_{\mathrm{c}}(L|M)$ has dimensions of $1/L$ while
$\phi(L)$ has dimensions of $1/({\rm volume} \times L)$. 
Of course, the CLF itself is generally non-trivial to predict and model. One approach is to follow the strategy of the previous section and empirically link the conditional luminosity function to the star formation rate functions. In this case, the two methods are essentially identical. 

The average number of emitters residing in halos of mass $M$, and the average luminosity of these emitters may be determined by taking moments of the CLF. Specifically, 
\begin{equation}
    \avg{N(m)} = \int_{L_1}^{L_2} dL\, \phi_{\mathrm{c}}(L|M),
    \label{eq:clf_number}
\end{equation}
gives the average number of galaxies with luminosity between $L_1$ and $L_2$ in a halo. For LIM the relevant thing is generally the total number of galaxies across all luminosities, and so $L_1=0$, $L_2 = \infty$. The average luminosity of these sources is:
\begin{equation}
    \avg{L(m)} = \int_0^\infty dL\, L \phi_{\mathrm{c}}(L|m).
    \label{eq:clf_lum}
\end{equation}
Likewise, the average luminosity density follows directly from integrating Eq.~\ref{eq:clf_lum} over the halo mass function:
\begin{equation}
    \epsilon_L = \int dM\, n(M) \avg{L(m)}.
    \label{eq:clf_epsilon}
\end{equation}

The form of the CLF generally depends on whether the galaxies are {\em centrals}, i.e. residing in the center of their host dark matter halos, or {\em satellites}, which live at the center of sub-halos and orbit inside a larger halo. The total CLF can then be written as the sum of contributions from centrals and satellites,
$\phi_{\rm tot}(L|M) = \phi_{\rm cen}(L|M) + \phi_{\rm sat}(L|M)$.
In the literature, it is common to assume a lognormal probability distribution for the central galaxy CLF and a modified Schechter function for the satellite contribution. These forms are found to provide a good fit to the galaxy luminosity functions from Sloan Digital Sky Survey (SDSS) data at low redshift \cite{Yang2008}. Explicitly,
\begin{equation}
    \phi_{\rm cen}(L|M) dL = \frac{1}{\sqrt{2 \pi \sigma^2}} {\rm exp} \left[-\frac{\left[{\rm ln}\, L - {\rm ln}\, L_{\mathrm{c}}\right]^2}{2 \sigma^2}\right] \frac{dL}{L}
    \label{eq:clf_centrals}
\end{equation}
gives the CLF for central galaxies. The satellite CLF follows \cite{Yang2008}:
\begin{equation}
    \phi_{\rm sat}(L|M) dL = \phi_{\rm s} \left(\frac{L}{L_{\rm s}}\right)^{\alpha_{\rm s}} {\rm exp}\left[-\left(\frac{L}{L_{\rm s}}\right)^2\,\right] \frac{dL}{L_{\rm s}},
\end{equation}
which resembles a Schechter function \cite{Schechter76}, but the satellite abundance falls off more steeply with luminosity here.
Note that $\phi_{\rm s}$ in the above equation is a dimensionless normalization factor, since in our convention $\phi_{\rm sat}(L|M) dL$ is dimensionless. 
Although we have suppressed the mass dependence for brevity of notation, $L_{\mathrm{c}}, L_{\mathrm{s}}, \phi_{\mathrm{s}}, \alpha_{\mathrm{s}}$ and $\sigma$ may depend on halo mass. The full CLF model hence involves a fairly large number of free parameters. 

In the context of LIM, a multi-variate generalization of the CLF is useful \cite{Schaan:2021gzb,Schaan:2021hhy}. The multi-variate CLF may be written as \\ 
$\phi_{\mathrm{c}}(L_1, L_2, ..., L_n | M) dL_1 dL_2 ... dL_n$, specifying the number of galaxies with luminosities near $L_1$, $L_2, ..., L_n$ residing in a halo of mass $M$. We refer the reader to \cite{Schaan:2021gzb} for a discussion regarding multi-line CLF models. 

\subsection{Cross-Power Spectra}
\label{S:px_multiline}

In addition to the auto-power spectrum of line-intensity emission fluctuations, we will frequently be interested in the cross-power spectrum between the emission fluctuations in two separate lines. 
In analogy with Eq.~\ref{eq:pofk}, the cross-power spectrum between two lines is:
\begin{equation}
P_{1,2}(k) = \avg{I_{1}}\avg{I_{2}} \left[\avg{b_1}\avg{b_2} P_{\delta,\delta}(k) + \frac{\avg{L_1 L_2}}{\epsilon_1 \epsilon_2}\right].
\label{eq:px}
\end{equation}
This equation makes the same set of simplifying assumptions as in the case of the auto-power spectrum, which we will discuss further and refine shortly (\S \ref{S:refinements}). Note that the shot-noise term that enters the cross-power spectrum depends on whether the same set of galaxies are luminous in each line, or whether the line emission is mainly produced by distinct sources. 
The quantity $\avg{L_1 L_2}/(\epsilon_1 \epsilon_2)$ has {\em units of volume} under this notation.  
Under the modeling assumptions of this section, each dark matter halo hosts exactly one or zero luminous sources. That is, $\avg{L_1 L_2}$ is a short-hand for the quantity:
\begin{equation}
    \avg{L_1 L_2} = \int dM n(M) \int dL_1 dL_2 L_1 L_2 \,\phi_{\mathrm{c}}(L_1,L_2|M),
    \label{eq:lone_ltwo}
\end{equation}
As usual $\epsilon$ denotes the luminosity-density in a line with:
\begin{equation}
    \epsilon_1 = \int dM n(M) \int dL_1 L_1 \, \phi_{\mathrm{c}}(L_1|M),
    \label{eq:lone}
\end{equation}
and
\begin{equation}
    \epsilon_2 = \int dM n(M) \int dL_2 L_2 \, \phi_{\mathrm{c}}(L_2|M).
    \label{eq:ltwo}
\end{equation}

We caution that the notation $\avg{L_1 L_2}$, while suggestive of a luminosity correlation, actually denotes a volume-averaged moment of the joint luminosity-distribution with dimensions of $[\text{lum}^2/\text{vol}]$. The same applies to the auto-spectrum shot-noise term involving $\avg{L^2}$ (Eq.~\ref{eq:poss}), and we retain this notation for consistency.
Thus, the Poisson term has units of specific intensity-squared times volume, as expected.
In the case that the emitters in each of the two lines are entirely hosted by distinct sets of dark matter halos, $\avg{L_1 L_2}=0$, and there is no shot-noise contribution to the cross-power spectrum. In the other limiting case, the emitters in each line occupy exactly the same set of halos, and the shot-noise contribution to the cross-power may be strong. 

It is also convenient to define the correlation coefficient between the fluctuations in the two lines:
\begin{equation}
    r_{1,2}(k) = \frac{P_{1,2}(k)}{\sqrt{P_{1,1}(k) P_{2,2}(k)}},
\end{equation}
which ranges between $r_{1,2}(k)=-1$ for the case of strongly anti-correlated lines and $r_{1,2}(k)=1$ for highly correlated lines. In the shot-noised dominated limit $r_{1,2} \rightarrow \avg{L_1 L_2}/\sqrt{\epsilon^2_1 \epsilon^2_2}$. Note that in the case that two emission lines trace the same gas with $L_1 \propto L_2$, $r_{1,2} \rightarrow 1$ in the shot-noise dominated regime. 

It is further interesting to consider the cross-power spectrum between a line-intensity map and a traditional galaxy survey. This case may also be generalized to consider additional tracers of large-scale structure, beyond that of a traditional galaxy survey, such as quasar abundance fluctuations \citep{Breysse19}, Ly-$\alpha$ forest absorption \citep{Qezlou:2023fle,Croft:2018rwv}, and cosmic shear/lensing mass maps \citep{Chung:2022lpr}. Here, let us consider the cross-correlation between a line-intensity map, with specific intensity $I(\x_1)$ (here we leave out the $\nu$ subscript for brevity) and a galaxy-fluctuation field $\delta_{\mathrm{g}}(\x_2)$, where $\delta_{\mathrm{g}}(\x) = \left(n_{\mathrm{g}}(\x) - \avg{n_{\mathrm{g}}}\right)/\avg{n_{\mathrm{g}}}$ is the fractional fluctuation in the galaxy abundance around the cosmic mean. The cross-power spectrum, under the usual set of assumptions/approximations in this section, is:
\begin{equation}
P_{I,g}(k) = \avg{I} \left[\avg{b_L}\avg{b_{\mathrm{g}}} P_{\delta,\delta}(k) + \frac{\avg{L|g}}{\avg{L}}\right].
\label{eq:px_gal}
\end{equation}
In this equation, $\avg{I}$ is the average specific intensity of the line emission and $\avg{b_L}$ is the luminosity-weighted bias, as usual, while $\avg{b_{\mathrm{g}}}$ is the number-weighted bias factor of the galaxies in the traditional galaxy survey. 
The Poisson term depends on the line luminosity density of the emitters that are also contained within the traditional galaxy survey. That is, the notation $\avg{L|g}$ indicates that what is relevant here is the luminosity of the line-emission {\em conditioned} on the emitter being a member of the traditional survey. The relationship here, again, ignores contributions from satellite galaxies.  
This term may be calculated as
\begin{equation}
\frac{\avg{L|g}}{\avg{L}} = \frac{\int dL L  \, \phi_{\mathrm{c,g}}(L|g)}{\int dM n(M) \int dL L \, \phi_{\mathrm{c}}(L|M)},
    \label{eq:px_gal_poisson}
\end{equation}
where $\phi_{\mathrm{c,g}}(L|g)$ is analogous to the usual CLF, except here the conditioning is on the voxel containing a galaxy in the traditional survey, rather than on halo mass. 
As in the cross-power spectrum between two lines, the shot-noise contribution to the cross-spectrum of Eq.~\ref{eq:px_gal} will be small if the line-emitter and galaxy samples are largely disparate.

The cross shot-noise piece here can also be written entirely in terms of the probability that galaxies/line emitters occupy halos of mass $M$ as \citep{Schaan:2021gzb}:
\begin{equation}
\frac{\avg{L|g}}{\avg{L}} = \frac{\int dM n(M) \int dL L \, \phi_{\mathrm{c}}(L|g,M)}{\left[\int dM n(m) \int dL L \, \phi_{\mathrm{c}}(L|M) \right]\left[\int dM n(M) \avg{N_{\rm gal}(M)}\right]}.
\label{eq:px_gal_poisson_halo}
\end{equation}
In this description, $\phi_{\mathrm{c}}(L|g,M)$ gives the probability that a line emitter has a luminosity $L$, conditioned on it residing in a halo of mass $M$ and that the same halo contains a galaxy in the traditional survey. Here $\avg{N_{\rm gal}(M)}$ is the average number of galaxies in a halo of mass $M$, which may be computed using an analogue of Eq.~\ref{eq:clf_number}. Hence the quantities in the denominator of Eq.~\ref{eq:px_gal_poisson_halo} are, respectively: the line emissivity in the LIM survey (Eq.~\ref{eq:clf_epsilon}), and the abundance of galaxies per unit comoving volume in the traditional galaxy survey. 

It may also be interesting to extend the cross-correlation measurements considered here by splitting the traditional survey into sub-samples of galaxies with different properties. For instance, one could cross-correlate line-intensity maps and galaxy samples with different estimated star formation rates, stellar masses, metallicities, or other properties. On small scales, where cross shot-noise dominates, this would help quantify how line emission correlates with galaxy properties. It may also be interesting to use the traditional survey to characterize large-scale environmental properties, and explore how line emission varies with environment.

\subsection{Summary: Equivalent Forms of the Power Spectrum Integrals}
\label{S:pspec_forms_summary}

The power spectrum and related quantities derived above, including the luminosity density $\epsilon_L$, mean intensity $\avg{I}$, the luminosity-weighted bias $\avg{b_L}$, the shot-noise $P_{\rm shot}$, and the cross-shot term $\avg{L_1 L_2}/(\epsilon_1 \epsilon_2)$ appear in various different forms in the literature. Specifically, these can be expressed in terms of integrals over several different distributions. Although these are mathematically equivalent, given appropriate relations between their ingredients, each has distinct practical advantages. Tables~\ref{tab:ps_forms1} and \ref{tab:ps_forms2} collect these forms explicitly.

The $\phi(L)$ form ties power spectrum models directly to
line luminosity functions, although evaluating the luminosity-weighted bias factor requires
a relation between luminosity and large-scale clustering. 
The SFR function $\Phi(\rm SFR)$ form is useful when
the $L$-SFR relation is well-calibrated and SFR functions are available (e.g.\
from UV or IR observations), and it avoids the need to model the galaxy-halo
connection explicitly when computing the mean intensity and shot noise. The $L(M)$
formulation integrates directly over the halo mass function $n(M)$ with a
deterministic (or mean) luminosity-halo mass relation. In some cases, a non-unity duty
cycle may be incorporated, as discussed previously.

The CLF formulation $\phi_c(L|M)$ is the most general of the four: it subsumes the $L(M)$ model as a special case in which each halo contains exactly one galaxy with a
delta-function luminosity, $\phi_c(L|M) = \delta_D(L - L(M))$, while also
accommodating scatter, satellite populations, and allowing the natural multi-variate generalizations required for cross-correlations, as discussed in the previous sub-sections.

\begin{table*}
\centering
\small
\caption{Equivalent forms for the luminosity density
$\epsilon_L$ and luminosity-weighted bias $\langle b_L\rangle$.
The $L(M)$ row assumes a deterministic mean relation; scatter is incorporated
by replacing $L(M)$ with $\int dL\,L\,\phi_{\rm c}(L|M)$, as in the bottom row.}
\label{tab:ps_forms1}

\renewcommand{\arraystretch}{1.9}
\setlength{\tabcolsep}{6pt}

\begin{tabular}{lcc}
\addlinespace[6pt]
\toprule
Formulation
&
$\epsilon_L$
&
$\langle b_L\rangle$
\\
\midrule
\addlinespace[4pt]

$\phi(L)$
&
$\int dL\,L\,\phi(L)$
&
$\dfrac{\int dL\,b(L)L\phi(L)}
{\epsilon_L}$
\\[6pt]

$\Phi({\rm SFR})$
&
$\int d{\rm SFR}\,L\,\Phi$
&
$\dfrac{\int d{\rm SFR}\,bL\Phi}
{\epsilon_L}$
\\[6pt]

$L(M)\, n(M)$
&
$\int dM\,n\,L$
&
$\dfrac{\int dM\,n\,b\,L}
{\epsilon_L}$
\\[6pt]

$\phi_{\rm c}(L|M)$
&
$\int dM\,n\int dL\,L\,\phi_{\rm c}$
&
$\dfrac{\int dM\,n\,b
\int dL\,L\,\phi_{\rm c}}
{\epsilon_L}$
\\
\addlinespace[4pt]
\bottomrule
\end{tabular}
\end{table*}

\begin{table*}
\centering
\small
\caption{Equivalent forms for the LIM shot-noise amplitude and
cross-shot noise integrals.}
\label{tab:ps_forms2}

\renewcommand{\arraystretch}{1.9}
\setlength{\tabcolsep}{6pt}

\begin{tabular}{lcc}
\addlinespace[6pt]
\toprule
Formulation
&
$P_{\rm shot}/\langle I\rangle^2$
&
$\langle L_1L_2\rangle/
(\epsilon_1\epsilon_2)$
\\
\midrule
\addlinespace[4pt]

$\phi(L)$
&
$\dfrac{\int dL\,L^2\phi(L)}
{\epsilon_L^2}$
&
$\dfrac{\int dL_1\,dL_2\,
L_1L_2\,\phi(L_1,L_2)}
{\epsilon_1\epsilon_2}$
\\[8pt]

$\Phi({\rm SFR})$
&
$\dfrac{\int d{\rm SFR}\,
L^2\Phi}
{\epsilon_L^2}$
&
$\dfrac{\int d{\rm SFR}\,
\langle L_1L_2 | {\rm SFR} \rangle \Phi}
{\epsilon_1\epsilon_2}$
\\[8pt]

$L(M)\,n(M)$
&
$\dfrac{\int dM\,n\,L^2}
{\epsilon_L^2}$
&
$\dfrac{\int dM\,n\,L_1L_2}
{\epsilon_1\epsilon_2}$
\\[8pt]

$\phi_{\rm c}(L|M)$
&
$\dfrac{\int dM\,n
\int dL\,L^2\phi_{\rm c}}
{\epsilon_L^2}$
&
$\dfrac{\int dM\,n
\int dL_1\,dL_2\,
L_1L_2\,\phi_{\rm c}(L_1, L_2|M)}
{\epsilon_1\epsilon_2}$
\\
\addlinespace[4pt]
\bottomrule
\end{tabular}
\end{table*}

\subsection{Refinements}
\label{S:refinements}

As discussed at the beginning of this section, the power spectrum models discussed thus far have made a number of simplifying assumptions. Here we further discuss -- and partly relax -- some of these simplifications. 

\subsubsection{Redshift Space Distortions: Large Scales}

One key issue is that we have thus far neglected the impact of redshift space distortions \cite{Sargent77,Kaiser87}. In this context, redshift space distortions arise from the fact that the line-emitting galaxies traced in LIM surveys may move towards nearby mass concentrations, i.e. the line-emitters may have peculiar velocities, in addition to their Hubble recession velocities owing to the expansion of the universe. These
peculiar velocities impact the mapping between the comoving distance along the line-of-sight to an emitting galaxy and its observed redshift. On the other hand, peculiar velocities do not influence the relationship between transverse distance and angle on the sky. Consequently, peculiar velocities break the statistical isotropy of the observed LIM fluctuations and the power spectrum actually depends on both the magnitude of the wavevector, $|{\bf k}|=k$, and the line-of-sight component of this vector, $k_\parallel = \mu k$. Although this complicates the modeling involved, it also implies that further information may potentially be extracted from LIM data cubes if the angular ($\mu$) dependence is well-measured.

On large spatial scales, the dominant effect is that matter, and line-emitting galaxies, tend to fall into large-scale overdensities. Therefore, based on their observed redshifts, such galaxies appear closer together along the line-of-sight than in actuality. That is, the observed galaxy density in ``redshift space'' is enhanced along the line-of-sight and hence so is the power spectrum of line-intensity fluctuations,  relative to the underlying ``real space'' matter power spectrum (which neglects peculiar velocities.) On large scales, the analysis of \cite{Kaiser87} implies a modified form for the power spectrum of LIM fluctuations (Eq.~\ref{eq:pofk}), with a dependence on both the magnitude of a wavevector and its direction.  
Similarly, the cross-power spectra of Eqs.~\ref{eq:px} and \ref{eq:px_gal} also become anisotropic. For the case of the auto-power spectrum, in the large-scale limit where the two-halo term dominates \cite{Kaiser87}:
\begin{equation}
    P_I(k, \mu) = \avg{I_\nu}^2 \avg{b_L}^2 \left(1 + \beta \mu^2\right)^2 P_{\delta,\delta}(k),
    \label{eq:pk_rspace_twoh}
\end{equation}
where $\beta=f_\Omega/\avg{b_L}$. Here we work in the distant observer, plane-parallel approximation where the line-of-sight can be approximated as lying along a Cartesian axis in, for instance, the $\bf{\hat{z}}$-direction. The line-of-sight component of the wave-vector is $k_\parallel = \mu k$, where $\mu$ is the cosine of the angle between the direction of $\bf{k}$ and $\bf{\hat{z}}$. 
The quantity $f_\Omega$ denotes the logarithmic derivative of the linear growth factor of density fluctuations with respect to scale factor, $f_\Omega = d\,{\rm ln D}/d\,{\rm ln} a$, and is well approximated by $f_\Omega \approx \left[\Omega_m(z)\right]^{0.55}$ in a LCDM cosmological model \cite{Linder:2005in}. 
As expected, the fluctuations in  Eq.~\ref{eq:pk_rspace_twoh} are enhanced for Fourier modes with large line-of-sight components (i.e., higher $\mu$) relative to transverse modes (at smaller $\mu$). See \cite{Scoccimarro:2004tg} for a discussion regarding the limitations of the Kaiser \cite{Kaiser87} formula, and issues related to commonly employed models on small-scales (the redshift space power spectrum on small-scales is discussed here below in \S \ref{S:rspace_fog}). 

One interesting feature of Eq.~\ref{eq:pk_rspace_twoh} is that the $\mu$-dependent coefficient, $\beta$, varies with the luminosity-weighted bias, $\avg{b_L}$, but is independent of the average specific intensity, $\avg{I_\nu}$. Thus, if the angular dependence of the line-intensity power spectrum may be measured accurately and precisely enough, one can break the degeneracy between $\avg{b_L}$ and $\avg{I_\nu}$ (assuming LCDM and that $f_\Omega$ is well determined by other cosmological probes) (e.g. \cite{Lidz11}). 

Similarly, the redshift space cross-power spectrum between two line-intensity maps on large scales becomes:
\begin{equation}
    P_{1,2}(k,\mu) = \avg{I_1} \avg{I_2} \avg{b_1} \avg{b_2}  \left(1 + \beta_1 \mu^2\right) \left(1 + \beta_2 \mu^2\right)
    P_{\delta, \delta}(k),
    \label{eq:px_rspace_twoh}
\end{equation}
where $\beta_1 = f_\Omega/\avg{b_1}$ and $\beta_2 = f_\Omega/\avg{b_2}$. Likewise, the two-halo contribution to the redshift space cross-power spectrum between a line-intensity map and a galaxy survey (Eq.~\ref{eq:px_gal}) can be described in an analogous manner to Eq.~\ref{eq:px_rspace_twoh}. 

\subsubsection{Redshift Space Distortions: Small Scales}
\label{S:rspace_fog}

On small scales, the effect of redshift space distortions depends on how the galaxies are distributed within their host dark matter halos. It is useful to consider two simple cases that should bracket the range of possibilities here. First, in the dark matter halos that host galaxies, there could be multiple line-emitting galaxies on average. These galaxies will be distributed throughout the halo, perhaps tracking the NFW profile \cite{Navarro97}, and in random motion around the halo center. In this case, the typical peculiar velocities of the satellite galaxies exceed the differential Hubble recession velocity across the halo: galaxies on the near-side of the halo appear displaced and on the opposite side of the halo center, while the reverse occurs for galaxies on the far-side. In other words, the galaxies appear elongated in the line-of-sight direction in redshift space and there is a corresponding power spectrum suppression at high $k_\parallel$ \cite{Hamilton:1997zq}. This ``finger-of-god'' effect may be modeled as a damping of the form $\delta_s(\k) \rightarrow e^{-k^2 \mu^2 \sigma^2/2} \delta_r(\k)$, where the dispersion $\sigma$ reflects the random motion of the line emitters within their host halos, while $\delta_s(\k)$ and $\delta_r(\k)$ are the redshift space and real space fluctuation fields, respectively. This dispersion will increase with halo mass, as can be incorporated into models of the LIM power spectrum (see \cite{Schaan:2021gzb}, \S \ref{sec:master_halo_model} for the details here). Note that in some models the halo mass dependence of $\sigma$ is neglected for simplicity. In any case, the exact damping scale is uncertain as it depends on the unknown mass profile of each host halo, among other properties. 

In the second example scenario, each line-emitting galaxy occupies the center of its host dark matter halo. In this case, which may be a good approximation at high redshift -- where most galaxies are centrals as opposed to satellite galaxies -- the usual finger-of-god effect should be absent. However, it is nevertheless important to account for the finite frequency width of the spectral lines under consideration \cite{COMAP:2021rny}. In general, this will lead to a damping of high $k_\parallel$ modes: this is qualitatively similar to a finger-of-god effect, yet differs in its physical origin. For example, in the case that the line emission comes from gas in a rotationally-supported disk, the width of each emission line will be broadened by the Doppler effect (unless the disk is seen face-on). The line-width then depends on the circular velocity of the emitting gas and the inclination angle of the disk relative to the line-of-sight. In this case, the damping scale should also vary with halo mass since galaxies in deeper potential wells will have larger circular velocities and more rotational broadening. In some cases, the emitting gas may instead arise mainly from gas clouds in random motion, rather than in rotationally-supported disks. Then, the line-broadening depends on the velocity dispersion of the emitting clouds, which should also increase with halo mass.  

Finally, note that the damping scale may, in general, reflect both contributions from multiple emitting galaxies within each host halo as well as from rotational broadening/random motions within each galaxy. That is, although we considered two limiting cases -- the traditional finger-of-god effect sourced by multiple galaxies in random motion around their host halo centers and the line broadening from central galaxies -- both of these effects may in fact contribute to the damping. There are hence some uncertainties in the modeling of the damping term. In any event, measuring the damping effect would provide some broad-brush information regarding the typical line-widths and random motions. Accessing this information will require high spectral resolution to capture the modes impacted, with $k_\parallel \sigma \gtrsim 1$. We refer the reader to \cite{Schaan:2021gzb,COMAP:2021rny} and \S \ref{sec:master_halo_model} for further details regarding finger-of-god/finite line-width models.

\subsubsection{One-Halo Term}
\label{S:one-halo}

In the halo model description \cite{Cooray02}, the two-point function of the emission fluctuations at large separations arises from spatial correlations between halos (the ``two-halo term''). At small separations, the two-point function instead arises from correlations within halos (the ``one-halo term''). That is, on small scales the one-halo term dominates and the LIM power spectrum is sensitive to how the line-emitting gas populates the host dark matter halos.
As in the discussion of the previous section, two simple cases are: one where the line emission comes primarily from satellite galaxies, and scenarios where central galaxies dominate the emission. In the latter case, the spatial distribution of the line emission is highly concentrated towards the halo centers (i.e., each galaxy is well-approximated by a point source) and the one-halo term is nearly a pure shot-noise contribution (and generally accounted for in a separate ``shot-noise term'').  In the opposite, satellite-dominated limit, the small-scale fluctuations will depend on how the satellites and their line emission are distributed throughout each halo. A reasonable and commonly adopted assumption is that this emission traces the overall matter distribution, which is typically assumed to follow the NFW profile \cite{Navarro97}. It is further supposed that the satellite galaxy population is drawn from a Poisson distribution, with the average spatial distribution matching the NFW profile (e.g. \citep{Schaan:2021gzb}). In this case, there is an additional shot-noise contribution from Poisson fluctuations in the abundance of satellite galaxies populating halos of a given mass \cite{Schaan:2021gzb}, and their resulting line luminosities. 

\subsection{Master Formulae}
\label{sec:master_halo_model}

After this overview, it is useful to summarize our discussion with some master formulae that capture the main effects \cite{Schaan:2021gzb}. 
Our expressions differ in notation from those in \cite{Schaan:2021gzb}, but there are additional small differences. We will comment briefly on some of these differences (generally in footnotes to the text). 
Here we will give expressions for the LIM auto-power spectra and the cross-power spectrum between two LIM data cubes. The auto-power may be written as:
\begin{equation}
    P_I(k,\mu) = P_{\mathrm{2h}}(k,\mu) + P_{\mathrm{1h}}(k,\mu) + P_{\rm shot}(k,\mu),
    \label{eq:pauto_grand}
\end{equation}
i.e., as the sum of a two-halo term, a one-halo term, and a shot-noise contribution. 

\begin{figure}
\begin{center}
\includegraphics[width=\textwidth]{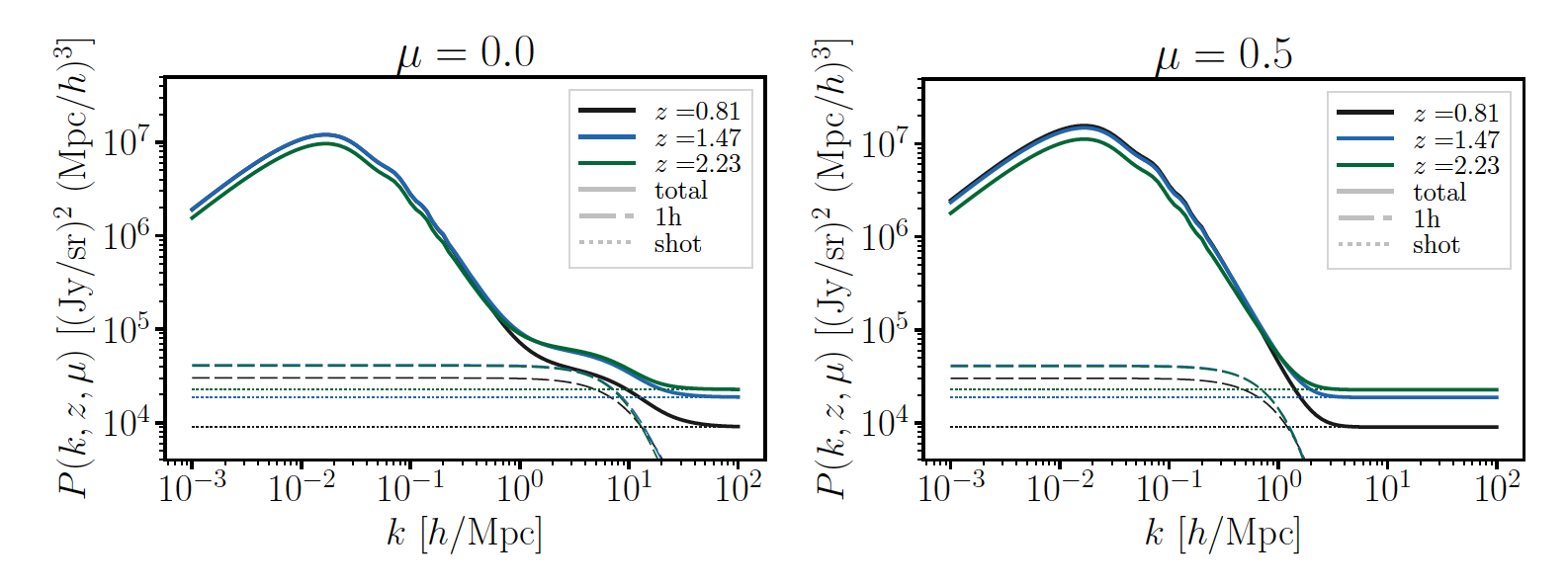}
\caption{Example halo model calculation of the LIM power spectrum. The solid line shows the total power spectrum of specific intensity fluctuations in the H-$\alpha$ line at various redshifts (see legend). In each model, the emitting galaxies are assumed to follow the NFW profile and so effectively only the satellite contributions are included. The line-broadening is assumed to be negligible. The dashed and dotted lines show the one-halo and shot-noise contributions, respectively. The {\em left panel} shows the case of transverse Fourier modes, while the {\em right panel} takes modes with $\mu=0.5$ (i.e. modes inclined by $60^\circ$ to the line of sight). On large scales the two-halo term dominates, on intermediate scales the one-halo term is largest, while shot-noise is important at still higher $k$. The finger-of-god effect suppresses the one-halo term for the inclined modes in the right panel. 
From \cite{Schaan:2021gzb}. }
\label{fig:halo_model_example}
\end{center}
\end{figure}

The two-halo term is:
\begin{equation}
    P_{\mathrm{2h}}(k,\mu) = \avg{I}^2 \avg{b(k,\mu)}^2 \left[1 + \frac{F(k,\mu)}{\avg{b(k,\mu)}} \mu^2\right]^2 P_{\rm lin}(k).
    \label{eq:ptwoh_grand}
\end{equation}
The structure of this equation is similar to our previous model including the Kaiser effect (Eq.~\ref{eq:pk_rspace_twoh}), but here the luminosity-weighted bias has some scale dependence while the parameter $\beta$ has been replaced by a generalized 
scale-dependent term, $F(k,\mu)/\avg{b(k,\mu)}$. The $k$ and $\mu$ dependence in the luminosity-weighted bias factor owes to: the spatial distribution of satellite galaxies, line-broadening, and the finger-of-god effect. The finger-of-god effect only impacts the redshift space clustering of satellite galaxies, since central galaxies are assumed fixed at the center of their dark matter host halos. We suppose that line-broadening is sub-dominant to the finger-of-god effects for the smaller satellite galaxies, but that it is important for the central galaxies. The (dimensionless) Fourier transform of the spatial distribution of the satellite galaxies in a host halo of mass $M$ is denoted by $u(k|M)$. This distribution is assumed to follow an NFW profile, i.e. it is supposed that the satellite galaxies track the underlying mass distribution. The
line-of-sight smoothing of the satellite galaxies is described by $\sigma_{\mathrm{s}}$. This quantity is related to the line-of-sight velocity dispersion of the satellite galaxies via $\sigma_{\mathrm{s}} = (1+z) \sigma_{\mathrm{1 d}}/H(z)$, which maps between the velocity units of $\sigma_{\mathrm{1 d}}$, the line-of-sight dispersion, and comoving distance units.
The line-broadening is assumed to follow a Gaussian smoothing with a standard deviation of $\sigma_{\rm lb}$ in comoving distance units. Each of these smoothing factors is mass-dependent \cite{Schaan:2021gzb,COMAP:2021rny}, but we suppress this in our notation for brevity. For example, under the approximation that the satellite galaxies move in the potential of a singular isothermal sphere, $\sigma^2_{\mathrm{1 d}} = G M/(2 r_{\rm vir})$ \cite{White2001}, where $r_{\rm vir}$ is the virial radius of the host halo of mass $M$. 

Note that 
Eq.~\ref{eq:ptwoh_grand} still assumes a linear-biasing/linear-order matter density fluctuation model and so the two-halo power above is proportional to the linear theory matter power spectrum, $P_{\rm lin}(k)$.
In this case the bias factor becomes:
\begin{equation}
    \avg{b(k,\mu)} = \frac{1}{\epsilon} \int dM n(M) b(M) \left[\avg{L_{\mathrm{c}}(M)} {\rm e}^{-k^2 \mu^2 \sigma^2_{\mathrm{lb}}/2} + \avg{L_{\mathrm{s}}(M)} {\rm e}^{-k^2 \mu^2 \sigma^2_{\mathrm{s}}/2} u(k|M)\right]
    \label{eq:blum_grand},
\end{equation}
where $\epsilon$ is the luminosity density in the line, $\avg{L_{\mathrm{c}}(M)}$ is the average luminosity of central galaxies residing in host halos of mass $M$, and $\avg{L_{\mathrm{s}}(M)}$ is the
same for the satellite galaxies. These may be computed from the CLFs (Eqs.~\ref{eq:clf}-\ref{eq:clf_centrals}). The quantity $F(k,\mu)$ in Eq.~\ref{eq:ptwoh_grand} is referred to as the effective growth factor and is given by:
\begin{equation}
    F(k,\mu) = \frac{f_\Omega}{\avg{\rho_M}} \int dM M n(M) \rm{e}^{-k^2 \mu^2 \sigma^2_{\mathrm{s}}/2} u(k|M),
    \label{eq:fkmu}
\end{equation}
where $\avg{\rho_M}$ is the comoving matter density. Note that the effective growth factor is {\em mass-weighted rather than luminosity-weighted} \cite{Schaan:2021gzb}, as the anisotropic term arises from peculiar motions which are sourced by mass-density inhomogeneities. Here the halo profile and finger-of-god contributions describe the mass distribution in redshift space, and so have a different origin than the related quantities in $\avg{b(k,\mu)}$ which arise from line-emitting satellite galaxies. Nevertheless, their functional forms will be identical in the case that the satellites trace the mass distribution within halos. In the large-scale limit, $u(k|M) \rightarrow 1$, $\sigma_{\mathrm{s}} \rightarrow 1$, $\sigma_{\mathrm{lb}} \rightarrow 1$, and the two-halo term will approach the pure Kaiser formula limit of Eq.~\ref{eq:pk_rspace_twoh}. At higher wavenumbers the above formula accounts for the damping from the halo profiles, finger-of-god, and line-broadening effects. 

Next, as discussed in \S~\ref{S:one-halo}, the one-halo term reflects the distribution of satellite galaxies within the halo, along with the finger-of-god and line-broadening effects. This may be modeled as\footnote{Reference \cite{Schaan:2021gzb} -- responsible for firming up much of the formalism discussed here -- accidentally omits the $u(k|M)$ factor in the second line of Eq.~\ref{eq:pk_oneh_grand}, describing the central-satellite cross term (their Eq. A.20), (E. Schaan, private communication). Throughout, that work also neglects the
line-broadening effects included here.}:
\begin{align}
    P_{\mathrm{1h}}(k,\mu) &= \frac{\avg{I}^2}{\epsilon^2} \int dM n(M) \bigg[\avg{L_{\mathrm{s}}(M)}^2 {\rm e}^{-k^2 \mu^2 \sigma^2_{\mathrm{s}}} |u(k|M)|^2 \nonumber \\ 
    & + 
    2 \avg{L_{\mathrm{c}}(M)} \avg{L_{\mathrm{s}}(M)} {\rm e}^{-k^2 \mu^2 (\sigma^2_{\mathrm{s}} + \sigma^2_{\mathrm{lb}})/2} u(k|M)\bigg].
    \label{eq:pk_oneh_grand}
\end{align}
In order to understand this expression, it is helpful to consider its Fourier counterpart, the configuration-space correlation function between the specific intensity fluctuations at two points (with each point lying within the same dark matter halo in the case of the one-halo term). The first term reflects the contributions from pairs of satellite galaxies, while the second term accounts for the correlated emission between central and satellite galaxies. 
Since the central galaxies are modeled as point sources at the center of each halo, central-central correlations arise (in configuration space) only from a single point and this contributes to the shot-noise rather than the one-halo term. The contributions to $P_{\mathrm{1h}}(k,\mu)$ from satellite pairs are weighted by the average satellite luminosity squared and the satellite/finger-of-god profile squared, as well as the halo mass function, and summed over all halo masses. The central-satellite pair term depends on the average luminosity of centrals and satellites, the line-broadening and finger-of-god effects, and on a single factor of the satellite profile. In this model, the one-halo term vanishes in the limit that the line emission is entirely dominated by central galaxies.

Finally, the shot-noise contribution is slightly modified from our previous expression (Eq.~\ref{eq:pkshot}). Here the shot-noise term may be written as:
\begin{equation}
    P_{\rm shot}(k,\mu) = \frac{\avg{I}^2}{\epsilon^2} \int dM n(M) \left[\avg{L^2_{\mathrm{c}}(M)} \rm{e}^{-k^2 \mu^2 \sigma^2_{\mathrm{lb}}}
    + \avg{L^2_{\mathrm{s}}(M)}\right].
    \label{eq:pkshot_grand}
\end{equation}
In this equation $\avg{L^2_{\mathrm{c}}(M)}$ is the average squared-luminosity of the central galaxies, while $\avg{L^2_{\mathrm{s}}(M)}$ is the same for satellites. The line-broadening smooths out the contribution from centrals, but this effect is taken to be negligible for the satellites. Note that line-broadening impacts the shot-noise, but finger-of-god effects do not \cite{COMAP:2021rny}. The quantities 
$\avg{L^2_{\mathrm{c}}(M)}$ and $\avg{L^2_{\mathrm{s}}(M)}$ may be computed via the CLF of the centrals and satellites:
\begin{equation}
    \avg{L^2_{\mathrm{c}}(M)} = \int dL L^2 \phi_{\rm cen}(L|M),
\end{equation}
and
\begin{equation}
    \avg{L^2_{\mathrm{s}}(M)} = \int dL L^2 \phi_{\rm sat}(L|M).
\end{equation}

An example of the LIM power spectrum in the halo model is shown in Figure~\ref{fig:halo_model_example} (from \cite{Schaan:2021gzb}). Here we show the power spectrum of H-$\alpha$ intensity fluctuations around $z \sim 1-2$ for transverse ($\mu=0$) Fourier modes and wavenumbers with a component along the line of sight ($\mu=0.5$). These models consider a slightly simplified case where all emitting galaxies track the NFW distribution and so the cases shown do not make the satellite/central distinction in Eqs.~\ref{eq:pauto_grand}-\ref{eq:pkshot_grand}. The results illustrate the expected dominance of the two-halo term on large scales, the one-halo contribution on intermediate scales, and the shot-noise at high $k$. 

Analogous expressions hold for the cross-power spectrum between two LIM data cubes \cite{Schaan:2021gzb}. Specifically, the two-halo term for the cross-power between lines 1 and 2 (under the same set of assumptions as in Eq.~\ref{eq:ptwoh_grand}) is:
\begin{align}
    P^{\mathrm{2h}}_{1,2}(k,\mu) & = \avg{I_1} \avg{b_1(k,\mu)} \avg{I_2} \avg{b_2(k,\mu)} \times \nonumber \\
    & \left[1 + \frac{F(k,\mu)}{\avg{b_1(k,\mu)}} \mu^2\right]\left[1 + \frac{F(k,\mu)}{\avg{b_2(k,\mu)}} \mu^2\right] P_{\rm lin}(k).
    \label{eq:px_twoh_grand}
\end{align}
This is a straightforward generalization of Eq.~\ref{eq:ptwoh_grand}, with the luminosity-weighted bias in each line following formulae similar to Eq.~\ref{eq:blum_grand}.
The one-halo contribution to the cross-power is\footnote{Reference \cite{Schaan:2021gzb} assumes the latter two central-satellite terms are equal on average.}:
\begin{align}
    P^{\mathrm{1h}}_{1,2}(k,\mu) &= \frac{\avg{I_1} \avg{I_2}}{\epsilon_1 \epsilon_2} \int dM n(M) \bigg[\avg{L^{\mathrm{s}}_1(M)} \avg{L^{\mathrm{s}}_2(M)} {\rm e}^{-k^2 \mu^2 \sigma^2_{\mathrm{s}}} |u(k|M)|^2 \nonumber \\ 
    & + 
    \avg{L^{\mathrm{c}}_1(M)} \avg{L^{\mathrm{s}}_2(M)} {\rm e}^{-k^2 \mu^2 (\sigma^2_{\mathrm{s}} + \sigma^2_{\mathrm{lb}})/2} u(k|M) \nonumber \\
    & + 
    \avg{L^{\mathrm{s}}_1(M)} \avg{L^{\mathrm{c}}_2(M)} {\rm e}^{-k^2 \mu^2 (\sigma^2_{\mathrm{s}} + \sigma^2_{\mathrm{lb}})/2} u(k|M)
    \bigg].
    \label{eq:px_oneh_grand}
\end{align}
The first term accounts for contributions from when the emitters in each line are in satellites of the same halo. 
The cross terms in the above expression involve pairs where the galaxy emitting in line 1 is a central and that in line 2 is a satellite, and vice-versa. This equation assumes that the line-broadening and finger-of-god smoothing scales in each line have the same functional dependence on host halo mass. 

Finally, the shot-noise contribution to the cross-power is given by:
\begin{align}
    P^{\rm shot}_{1,2} &= \frac{\avg{I_1} \avg{I_2}}{\epsilon_1 \epsilon_2} \int dM n(M) \times \nonumber \\ 
    & \int dL_1 dL_2 L_1 L_2 \left[\phi_{\rm cen}(L_1, L_2|M) {\rm e}^{-k^2 \mu^2 \sigma^2_{\mathrm{lb}}}
    + \phi_{\rm sat}(L_1, L_2|M)\right]. 
\end{align}
Here $\phi_{\rm cen}(L_1, L_2|M)$ and $\phi_{\rm sat}(L_1, L_2|M)$ are the bi-variate extensions of the CLF \cite{Schaan:2021gzb} for central and satellite galaxies, respectively. 
All of the above expressions have fairly natural extensions to the case of cross-correlations between LIM and traditional galaxy surveys (see \cite{Schaan:2021gzb} for details).

\subsubsection{Beyond Linear Biasing and Other Issues}

In the previous several sub-sections we considered several refinements to the simplest LIM power spectrum model (e.g. Eq.~\ref{eq:pofk}): including the Kaiser effect from peculiar velocities on large scales \cite{Kaiser87}, finger-of-god/finite line-width effects on smaller scales, and the small-scale impact of the distribution of line-emitting gas within each host halo (i.e. the ``one-halo term''). There are nevertheless several remaining shortcomings to the model, which we briefly discuss here. Recent work has started to address these issues, drawing on methods developed in other sub-fields of large-scale structure (e.g. the effective field theory of large-scale structure, see \cite{MoradinezhadDizgah:2021dei,Ivanov:2022mrd} and references therein).  

One main defect is that the power spectrum model discussed thus far neglects both non-linearities and non-locality in the halo biasing relation and only partly treats non-linearities in the underlying matter density fluctuations. A further issue is that of ``halo exclusion'': the basic halo model calculations should be adjusted to enforce that each mass element belongs to only one dark matter halo, i.e. that halos are non-overlapping \cite{Baldauf:2013hka}. 
A particular challenge for the halo model is capturing the
transition regime, on intermediate length scales, in between where the two-halo and one-halo term dominate. This is the regime where non-linearities, halo exclusion, and other effects are challenging to model yet important. 

Another potential issue is that for LIM, we care specifically about the distribution of the line-emitting gas and this may depend on environmental variables beyond the usual host halo mass. See references \cite{Croton:2006ys,Wechsler:2018pic,Yuan:2022rsc} for discussions of related environmental effects in the context of traditional galaxy surveys. 
For example, in the case of, say, CO emission, 
the sources may preferentially reside in field environments while galaxies in groups or clusters may not retain the necessary reservoir of cold molecular gas. That is, even across halos of similar mass, CO line emission may vary with large-scale environment. This feature might be important in modeling the overall auto-power spectrum of CO emission, for instance. It might also be an interesting signal: one could imagine studying the environmental dependence of line-emission by measuring, for example, the PDF of line-intensity conditioned on the large-scale galaxy density (or some other large-scale structure environmental proxy).

\section{Line-intensity Mapping vs  Spectroscopic Galaxy Surveys}
\label{sec:lim_v_traditional}

As discussed briefly in the Introduction, a natural question is when is a line-intensity mapping analysis preferable to a traditional galaxy survey? Now that we have discussed LIM models, we can address this question in detail. Specifically, we start by summarizing the work of \cite{Cheng:2018hox}, which quantitatively determines the optimal strategy
for extracting information regarding the underlying galaxy density fluctuations, $\delta_{\mathrm{g}}$. 

In both LIM analyses and traditional galaxy redshift surveys, one aim is to
determine the statistical properties of $\delta_{\mathrm{g}}$ from the observed luminosities in a number of survey voxels. The data cubes in LIM and traditional surveys are each generated by applying a mapping to the voxel luminosities, denoted by
$\hat{O}(L)$, where $\hat{O}$ stands for ``observable''. The resulting fields
are proxies for the underlying galaxy density fluctuations, $\delta_{\mathrm{g}}$. For example, in a traditional galaxy survey a voxel is considered to host a detected galaxy when its luminosity exceeds some threshold, $L_{\mathrm{th}}$. The threshold luminosity might be set to five times the rms voxel noise, for instance, corresponding to a $5-\sigma$ detection. The resulting field is then ``digitized'' to consist of 1's (detections) and 0's (non-detections). Hence, the traditional survey amounts to adopting a step function at $L_{\mathrm{th}}$ for $\hat{O}(L)$. On the other hand, in line-intensity mapping one analyzes specific intensity fluctuations, and these are proportional to the voxel emissivities, which are linearly proportional to the total luminosity in a voxel, and hence $\hat{O}(L) = L$ in this case.

More generally, \cite{Cheng:2018hox} determine the ``optimal observable'' mapping
which maximizes the Fisher information regarding $\delta_{\mathrm{g}}$. They first consider a toy scenario where all emitting sources have an identical luminosity, $l$, although a given voxel may nevertheless contain multiple emitters. 
The optimal observable here depends on the joint probability distribution function (PDF) that a voxel contains a galaxy density, $\delta_{\mathrm{g}}$, and a luminosity between
$L$ and $L + dL$. This is denoted by $P(L,\delta_{\mathrm{g}})$ and the optimal observable
may be constructed using a $P(D)$ analysis formalism \cite{Lee:2008fm}. See also the closely related formalism discussed in \S \ref{sec:pdf}.
First, let $P_k(L,\delta_{\mathrm{g}})$ be the probability that a voxel has a luminosity between $\left[L, L+dL\right]$ given that it contains $k$ sources.
In the case that the $k$ sources are spatially uncorrelated and follow a Poisson distribution, $P(L,\delta_{\mathrm{g}})$ is given by:
  \begin{equation}\label{E:PL}
  P(L,\delta_{\mathrm{g}})=\sum_{k=0}^{\infty} \mathcal{P}(k,N(\delta_{\mathrm{g}}))P_k(L,\delta_{\mathrm{g}}),
  \end{equation}
  where
  \begin{equation}
  \mathcal{P}(k,N(\delta_{\mathrm{g}}))=\frac{e^{-N(\delta_{\mathrm{g}})}N^k(\delta_{\mathrm{g}})}{k!}
  \end{equation}
  is a Poisson distribution, and $N(\delta_{\mathrm{g}})$ is the average number of sources in a voxel.  Both $N(\delta_{\mathrm{g}})$ and $P_k(L,\delta_{\mathrm{g}})$ can be derived from a given luminosity function model, $\Phi(\ell,\delta_{\mathrm{g}})$, and voxel 
  volume, $V_{\mathrm{vox}}$: $N(\delta_{\mathrm{g}}) = V_{\mathrm{vox}}\int\Phi(\ell,\delta_{\mathrm{g}})d\ell$, while $\ell$ denotes the luminosity of a single source.

  The effect of measurement noise is easily included. In the case that the noise follows
  a Gaussian distribution, with a constant ($L$-independent) rms noise, $\sigma_L$, the noisy
  distribution $P(L,\delta_{\mathrm{g}},\sigma_L)$ is given by a convolution with the intrinsic distribution:
  \begin{equation}\label{E:PLG}
  \begin{split}
  P&(L,\delta_{\mathrm{g}},\sigma_L)=P(L,\delta_{\mathrm{g}})\ast G(\sigma_L)\\
  &\equiv \int dL'P(L',\delta_{\mathrm{g}})\frac{1}{\sqrt{2\pi}\sigma_L}e^{-(L-L')^2/2\sigma_L^2}.
  \end{split}
  \end{equation}
Equivalently, one can perform a convolution with each of the $P_k$ terms in Eq.~\ref{E:PL}. 

This setup can then be used to find the optimal observable for determining $\delta_{\mathrm{g}}$ and its statistics, given information regarding the luminosity function and measurement noise. To develop intuition, reference~\cite{Cheng:2018hox} first considers a toy scenario in which each galaxy emits with the same luminosity.
In this case, the authors demonstrate that the optimal analysis is largely shaped by the average effective number of sources per voxel, $N(\delta_{\mathrm{g}})$. For instance, when $N(\delta_{\mathrm{g}}) \ll 1$, most voxels are devoid of sources and the optimal analysis is the traditional, individual source detection approach. On the other hand, when $N(\delta_{\mathrm{g}}) \gg 1$, the line-intensity mapping strategy extracts the maximal amount of information regarding
$\delta_{\mathrm{g}}$.

The authors further consider the more realistic situation where the galaxies follow a luminosity function, rather than
each emitting galaxy having an identical luminosity. Specifically, 
a modified Schechter function form is assumed for the galaxy luminosity function \cite{Cheng:2018hox}:
\begin{equation}
    \Phi(\ell)=\Phi_* \ell^{\alpha} e^{-\ell} e^{-\ell_{\mathrm{min}}/\ell},
\end{equation}
where $\ell \equiv l/l_*$, i.e., the luminosity $l$ is expressed in units of $l_*$, a characteristic luminosity. The two exponential cut-offs, at the bright and faint-ends (with $\ell_{\mathrm{min}}$ giving the low-luminosity cut-off), ensure convergence. 

The first three moments of the luminosity function define the following quantities of interest:
\begin{eqnarray}
N &=& V_{\mathrm{vox}} \int d\ell \Phi(\ell) \\
\langle \hat{L} \rangle &=& V_{\mathrm{vox}} \int d\ell\Phi(\ell)\ell  \\ 
\sigma_{\mathrm{SN}}^2 &=& V_{\mathrm{vox}} \int d\ell \Phi(\ell)\ell^2,  
\end{eqnarray}
where N is the mean number of sources in a voxel of volume $V_{\mathrm{vox}}$, $\langle \hat{L} \rangle$ is the mean luminosity and $\sigma_{\mathrm{SN}}^2$ is the shot-noise due to the finite number of sources contributing to the intensity signal. Note that the effective number of sources $N_{\mathrm{eff}}$ can then be expressed as
\begin{equation}
  N_{\mathrm{eff}} = \frac{\langle \hat{L} \rangle^2}{\sigma_{\mathrm{SN}}^2}, 
\end{equation}
which is approximately the number of sources per $\log(\ell)$ at $l_*$. 
It is also useful to consider the shot-noise contributions from sources less luminous than luminosity $\ell$,
\begin{equation}
\sigma_{\mathrm{SN}}^2(\ell) = V_{\mathrm{vox}} \int_0^{\ell} d\ell^\prime \Phi(\ell^\prime)\ell^{\prime 2}
\label{eq:sn_cumulative}
\end{equation}
We can then define a luminosity scale, $L_{\mathrm{SN}}$, where the shot-noise matches the source luminosity, $\sigma_{\mathrm{SN}}(L_{\mathrm{SN}})=L_{\mathrm{SN}}$.
Specifically, when $\ell < L_{\mathrm{SN}}$, $\sigma_{\mathrm{SN}}(\ell) > \ell$, which implies that shot-noise sourced confusion noise from faint sources dominates. On the other hand, when $\ell > L_{\mathrm{SN}}$, $\sigma_{\mathrm{SN}}(\ell) < \ell$, confusion noise from faint sources becomes negligible. %

More generally, \cite{Cheng:2018hox} show that the relative amplitudes of three quantities: the characteristic source luminosity, $l_*$, the (Gaussian) instrumental noise, $\sigma_L$, and the shot-noise ``crossover'' luminosity, $L_{\mathrm{SN}}$, serve to characterize the optimal analysis strategy. The authors then find that LIM
is optimal when $L_{\mathrm{SN}} > l_*$, given the prominent confusion noise in this case. Conversely, when the faint source populations contribute negligibly, $L_{\mathrm{SN}} < l_*$, the optimal observable depends on the instrumental sensitivity. In particular, if $L_{\mathrm{SN}} < \sigma_L < l_*$, the traditional mode of individual source detections (i.e., in which $\hat{O}(L)$ is a step-function) is optimal. On the other hand, at large instrumental noise, $L_{\mathrm{SN}} < l_* < \sigma_L$, LIM is again preferable. Finally, when the instrumental noise is extremely low, $\sigma_L <  L_{\mathrm{SN}} < l_*$, there are two solutions: the traditional approach is optimal in the regime $L_{\mathrm{SN}} < l_*$, while LIM is better between $\sigma_L <  L_{\mathrm{SN}}$.

Naturally, some data sets may profit from both LIM and traditional analyses. 
Although the above analysis adopts idealized descriptions of the source populations and the observational surveys, they nevertheless provide a useful framework for comparing traditional galaxy redshift surveys and LIM analyses. Reference~\cite{Schaan:2021hhy} further compares traditional galaxy redshift surveys and LIM. In contrast to the above treatment from \cite{Cheng:2018hox}, reference~\cite{Schaan:2021hhy} considers the LIM/traditional survey signal-to-noise-ratios for different Fourier modes (see also \cite{Uzgil:2014pga}), while the above formalism focuses on individual voxels. See \cite{Schaan:2021hhy} for further discussion, details, and comparisons between
the Fourier mode-based and voxel-based approaches.

\section{Simulated Signals}
\label{S:signals}

We now turn to consider current approaches for simulating LIM signals. These
efforts, while challenging, are crucial for a range of investigations including:
capturing the full range of physical processes involved in line emission, accounting for quasi-linear and non-linear clustering, producing more realistic forecasts of signal strengths, providing tests of data analysis methodologies and for developing software pipelines, and ultimately for interpreting the upcoming measurements and performing parameter inference. For example, simulations are needed to reliably account for the multi-phase ISM structure which shapes the emission in a number of the lines of interest for LIM surveys (\S \ref{S:landscape}). They are also important for treating radiative transfer effects which, for example, strongly influence
the propagation of Ly-$\alpha$ photons (\S \ref{S:lya_rt}) and are crucial for modeling the reionization process itself. As far as their role in guiding LIM data analyses, simulated skies can help: identify potential measurement biases, in computing transfer functions to account for signal loss in the analysis, and more generally in considering the coupling between systematic effects and the underlying signals. Ultimately, as discussed further in what follows, the interpretation of forthcoming LIM measurements and the associated parameter inference are also challenging problems, necessitating a hierarchy of different modeling efforts.

The first challenge is one of dynamic range: in many cases the line emission is
shaped by $\sim$ parsec-scale structure in the ISM of individual galaxies, while the
aggregate emission fluctuations may be measured on cosmological length scales as large as a $\sim$ Gpc. It is beyond the reach of current simulations to fully capture the relevant dynamic range here, and so a diverse set of investigations and hybrid techniques -- effectively stitching together calculations focusing on different length scales -- are required to attack this problem. One approach
combines numerical hydrodynamic simulations of galaxy formation with radiative
transfer, photo-ionization, and PDR models (e.g. \cite{Olsen17,Moriwaki18,Leung20,Olsen21}). In simulations capturing cosmological length scales, this requires sub-grid models. That is, the largest scale hydrodynamic cosmological simulations of galaxy formation, which now produce extensive samples of simulated galaxies with diverse properties (\cite{Vogelsberger:2014dza,Schaye:2014tpa,Springel18,Vogelsberger:2019ynw}),
lack
the resolution required to capture the multi-phase ISM in the simulated galaxies.
Other efforts focus instead on performing self-consistent models for metal and dust production, while partly resolving HII regions, PDRs, and molecular gas in galaxy-scale simulations \cite{Katz19,Pallottini:2019uil,Kannan:2021ucy,Kannan22,Nakazato23,Yang23}. 
The cost of these calculations, however, allows modeling only a small handful of galaxies. Alternatively, semi-analytical models (e.g. \cite{Lagache18,Popping19,Yang2021}) 
include physically motivated descriptions of the ISM across millions of galaxies, but are computationally expensive and sometimes capture only limited cosmological volumes. 

A connected challenge is that the predictions from the most detailed simulations become tied to the set of ``sub-grid parameters'' adopted to approximately account for unresolved, yet important, physical processes including, for instance, those related to supernova and AGN feedback. 
These simulations, while most faithfully capturing the full range of scales and physical processes at play, are too expensive to investigate wide variations around the sub-grid recipes assumed.
Furthermore, the essential ingredients shaping the LIM signals in such simulations are not always transparent. Ultimately, flexible and efficient simulation frameworks are required to model multiple LIM signals simultaneously, while allowing model testing and parameter inference.

\subsection{Semi-Empirical approaches}

Currently, most LIM simulations can be categorized as ``semi-empirical''. Typically, these assume empirically-motivated scaling relations (\S \ref{s:empirical_line}) to link
the mass of simulated dark matter halos to the total luminosities of the galaxies residing within a halo, for each emission line of interest.
These efforts sometimes assume direct correlations between line luminosity and halo mass (e.g. \cite{Yue:2015sua,Chung2020}), or
exploit halo occupation distribution models for the galaxy populations (e.g., \cite{Yue2019,Yang2022}). The semi-empirical approach yields rapid predictions for
LIM signals across cosmological volumes and accounts for quasi-linear and non-linear clustering effects. However, the empirical calibrations adopted are
generally extrapolated from lower redshift observations and/or towards low luminosities. Further, they do not fully connect with the galaxy and ISM properties that shape the line emission signals. Finally, this approach often adopts mean relations and neglects scatter in the galaxy properties, spectral energy distributions, and (for example) the CO rotational line intensity ladder.

With this prelude, we turn to review a few representative simulation efforts in the current literature. The semi-empirical cases often involve three key ingredients:
\begin{itemize}
    \item First, a simulation is used to generate realizations of the cosmological
    density and velocity fields and/or the distribution of dark matter halos.\\

    \item Second, a prescription is used to populate dark matter halos with 
    star-forming galaxies. This often involves assuming empirically-calibrated
    parametric functions for the star formation rate as a function of halo mass and redshift, $\rm{SFR}(M,z)$. For example, the UniverseMachine \cite{Behroozi2019}
    provides star formation histories at $0 < z < 10$ based on parameterized correlations between galaxy and halo assembly histories, which are in turn calibrated to a broad suite of current observations.\\ 
    
    \item Third are the previously mentioned empirical relationships between
    line luminosity and SFR (see also \S \ref{s:empirical_line}). In some cases, this involves assuming a galaxy SED. A useful tool here is the Flexible Stellar Population Synthesis (FSPS) model \cite{ConroyFSPS2009}, which calculates stellar population synthesis models, the effects of dust attenuation, and the resulting SED.\\
     
\end{itemize}

In the first step, different approaches have been used in the literature. First, are full cosmological hydrodynamic or N-body (i.e. gravity only) simulations. These give accurate models for cosmological density, velocity, and halo fields
down to the resolution limit of the simulations. Owing to their computational expense, however, this approach is usually confined to a handful of simulation runs. A complementary technique -- generally known as the ``semi-numerical'' simulation method -- uses first or second-order Lagrangian perturbation theory to evolve simulated realizations of the initial density/velocity fluctuations into the quasi-linear regime. Further, the excursion set formalism \cite{Bond91} is used to identify dark matter halos -- or, alternatively, to model the overall fraction of matter in a simulation cell which has collapsed into halos above some minimum threshold mass -- across the simulation volume. The advantages of the semi-numeric approach are its speed and flexibility, although it is less accurate than a full simulation, especially on small spatial scales.

Examples of the semi-numeric approach in the literature include
21cmFAST \cite{Mesinger11} and SimFast21 \cite{Santos10}. These use 2LPT or the Zel'dovich approximation to evolve the density and velocity fields from suitable cosmological initial conditions, and the
excursion set model of reionization \cite{Furlanetto:2004nh} to model the ionization field and the reionization history. 
Additional semi-numerical studies employing variants of the excursion set model of reionization include \cite{Zahn07,Choudhury2009,Alvarez12}.
The DexM code \cite{Mesinger2007}, included as an option within 21cmFAST, produces realizations of the dark matter halo distribution based on the excursion set formalism. These codes have been used in numerous LIM studies, including \cite{Silva13} who use SimFast21 to model reionization-era signals in the 21 cm and Ly-$\alpha$ lines and \cite{Heneka:2016kss} which uses DexM and 21cmFAST for the same purposes.

Another example of this approach is the work of \cite{Dumitru2019}, which models the [CII] and 21 cm signals during the EoR. This study combines an excursion set treatment of reionization with a hydrodynamic cosmological simulation from the Sherwood simulation suite\footnote{\url{https://www.nottingham.ac.uk/astronomy/sherwood/}} \cite{Bolton2017}. 
Here, the authors also make use of the study of \cite{Lagache18}, which uses a combination of semi-analytic galaxy formation models \cite{Cousin2015, Cousin2016, Cousin2019}  and observations to determine the relationship between [CII] luminosity and SFR at high redshift. The \cite{Lagache18} model assumes that [CII] emission at high redshift arises predominantly from PDRs, and is characterized by three key parameters describing: the mean hydrogen density, gas metallicity, and the intensity of the interstellar radiation field.
The [CII] line emission is then computed using the photoionization code \textsc{Cloudy} \cite{Ferland2017}. The model is able to reproduce the mean L$_{\mathrm{[CII]}}$–SFR relation observed from 50 star-forming galaxies at z $\ge$ 4, although there is a large scatter (0.51 to 0.62 dex from z = 4 to z = 7.6) owing to variations in the PDR parameters across simulated galaxies in the semi-analytic models. This result is then applied to the dark matter halo catalog from the Sherwood simulation, after assuming a correlation between SFR and halo mass, to model the [CII] emission fluctuations and their cross-power spectra with the 21 cm fields. 

The \textsc{WebSky} simulations, based on the peak patch algorithm, have been used to make efficient full sky maps of CMB secondary anisotropies and the CIB \cite{Alvarez16,Stein20,Lee24}.
A recent paper adds [CII] emission models to \textsc{WebSky}, which are publicly available \cite{Carlson2025} \footnote{\url{https://mocks.cita.utoronto.ca/index.php/WebSky_CII}}. The \textsc{WebSky} [CII] luminosity-halo mass relation follows the work of \cite{Horlaville2024}.

Another approach for modeling sub-mm LIM signals is The Simulated Infrared Dusty Extragalactic Sky (SIDES) simulation\footnote{\url{https://gitlab.lam.fr/mbethermin/sides-public-release}}. These simulations were originally designed to model the far-infrared continuum emission from dusty galaxies \cite{Bethermin2017}, but have since been updated to include sub-mm LIM signals including CO, [CII], and [CI] lines \cite{Bethermin2022}. The authors use an abundance-matching technique to populate simulated halos and sub-halos with galaxies, including models
for star-forming and passive populations. The star-forming galaxies are further distributed along the star-forming main sequence, based on observational constraints from \cite{Schreiber15}, and assigned star formation rates. 
The authors include empirically-calibrated models for the far-infrared dust continuum emission and the aforementioned line luminosities (as a function of SFR). For example, the authors implement the locally observed L$_{\mathrm{[CII]}}$-SFR relation from \cite{DeLooze:2014dta} with 0.2 dex of scatter (see also \S \ref{s:empirical_line}). The authors argue that this relation is, somewhat surprisingly, consistent with measurements up to $z \sim 6$ from the 
ALMA ALPINE large program \cite{LeFevre20,Bethermin20,Faisst20}. 
For CO, the authors note that correlations between CO(1-0) luminosity and bolometric infrared luminosity are redshift invariant among main-sequence galaxies \cite{Sargent14}. They further assume templates for the higher-$J$ CO transitions, which are linked to the intensity of the UV radiation field and
the authors' dust continuum model. This leads to diverse predictions for the line
luminosity ratios in various CO transitions, which evolve with redshift. The authors further present new empirical calibrations for modeling [CI] emission.

Reference \cite{Yang2022} presents another empirically-calibrated simulation framework for sub-mm
LIM, including [CII], CO, and [CI] lines, based on semi-analytic galaxy formation models (SAMs) (\cite{Somerville99,Somerville15}). 
Here, the authors
assume a double power-law form for the average line luminosity-halo mass relationships, along with redshift-dependent scatter drawn from a lognormal distribution. These relations are calibrated from the results of \cite{Popping19}, which applies a sub-grid model for molecular clouds on top of SAMs, and then uses the DESPOTIC code \cite{Krumholz14} to model the resulting line luminosities. 
The SAM tracks galaxy properties for millions of simulated galaxies, and includes
physically-motivated models for partitioning the ISM into its multiple phases,
for the size distribution of molecular clouds, and their density profiles. These calculations reproduce observed correlations between luminosity and SFR for [CII], multiple CO transitions, and [CI] lines at $0 \le z \le 6$. However, the authors find departures from the empirical relations at luminosities smaller than directly observed, highlighting the importance of physically-motivated modeling and the danger of extrapolating empirical trends beyond their reach. In the case of [CII], the hybrid SAM approach yields results that are broadly consistent with simulations employing sub-grid line emission models on top of a cosmological hydrodynamic simulation \cite{Leung20}, although the SAM calculations are orders of magnitude faster.

\begin{figure}
\begin{center}
\includegraphics[width=\textwidth]{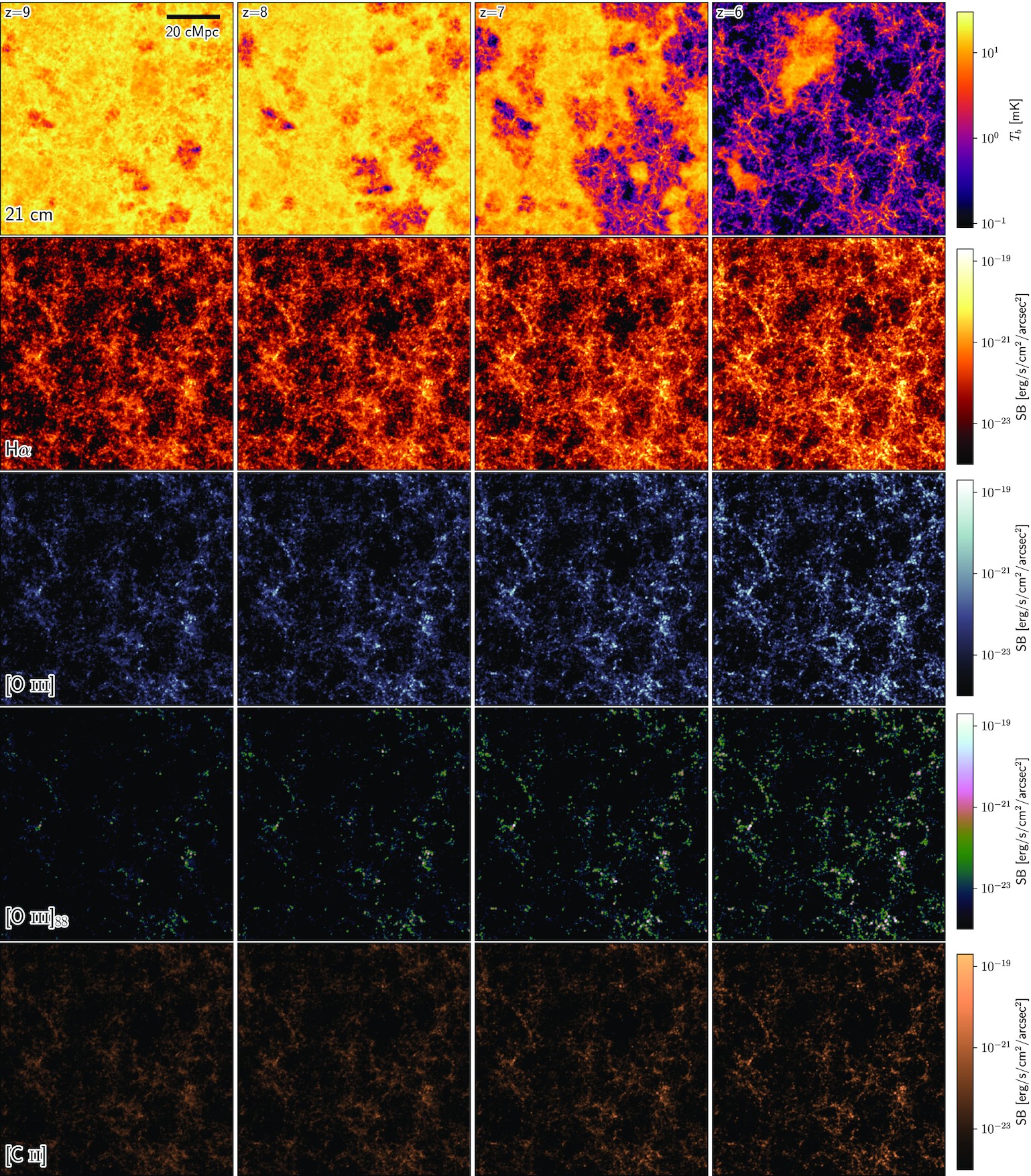}
\caption{Slices through multi-line LIM signal calculations from the THESAN-1 simulation.  The columns show snapshots at $z= 9, 8, 7$ and $6$, from left to right. The top row shows the 21 cm signal, while subsequent rows show: H-$\alpha$, a rest-frame optical [OIII] line, the 88 $\mu$m [OIII] line, and [CII] emission. The 21 cm signal is shaped by the growth of ionized regions, which imprint large-scale variations, and lead to a gradually diminishing signal. The other lines mainly trace the overdensities and filamentary structure across the simulation box, with
the H-$\alpha$ emission including a more extended and diffuse component sourced by radiative recombinations within the IGM. 
From \cite{Kannan:2021ucy}.
\label{fig:thesan}
}
\end{center}
\end{figure}

Two recent simulation efforts present reionization-era LIM signal models across multiple emission lines, including the sub-mm lines discussed above, while capturing additional lines as well. Each of these works incorporates sub-grid models for ISM-scale and line emission physics, while spanning cosmological volumes, yet the two studies adopt rather different methodologies. First, the THESAN simulations
\cite{Kannan2022a} are radiation-magneto-hydrodynamic simulations, spanning 95.5 comoving Mpc on a side, which self-consistently model hydrogen reionization and galaxy formation, while also including recipes for black hole growth and accretion. Sub-grid models are included to help follow line emission across multiple lines of interest \cite{Kannan:2021ucy}. This approach provides detailed, high resolution models but the computational expense limits these runs to a few plausible scenarios. The other effort is LIMFAST \cite{Mas-Ribas2022,Sun2022},
which is a semi-numerical toolkit based on 21cmFAST, yet extended to cover a multitude of LIM signals from the EoR. It provides flexible and efficient large-scale LIM models: its computational speed allows one to explore many different reionization history and galaxy formation scenarios.

Specifically, the THESAN simulations are performed with \textsc{Arepo-rt} \cite{Kannan:2018frm}, a radiation hydrodynamic extension to the moving mesh code \textsc{Arepo} \cite{Springel10,Weinberger:2019tbd}. 
The authors employ the IllustrisTNG (\cite{Springel18,Pillepich18})
galaxy formation model to describe sub-resolution processes such as star formation, black hole accretion, and feedback from supernovae and AGN. 
Both simulated stars and AGN act as sources of radiation, although on average AGNs make only small contributions (producing $\le 1\%$ of the ionizing photon budget) to reionization in the simulation.
The THESAN simulations are then post-processed to predict the luminosities
for multiple emission lines. This post-processing step uses stellar population
synthesis models from FSPS \cite{ConroyFSPS2009}, and nebular emission line
calculations from \cite{Byler17}, which are in turn based on \textsc{Cloudy} \cite{Ferland2017}.
The sub-grid models here assume that a negligible fraction of Lyman continuum photons escape from HII regions, and neglect any dust absorption of such photons.
It is assumed that only stellar particles of age less than 10 Myr remain surrounded by their birth clouds and produce nebular emission. The gas phase metallicities of sub-grid HII regions are taken to match that of nearby stellar particles, while the ionization parameters and gas densities are fixed to plausible values \cite{Byler18}. 
The resulting calculations produce LIM signal models
for H-$\alpha$, [OII], rest-frame optical [OIII] lines, sub-mm [OIII] lines, [NII],
and [CII] lines. The impact of dust attenuation on the line emission signals is accounted for using the Monte Carlo radiative transfer code \textsc{SKIRT} \cite{Camps20}. 
Figure~\ref{fig:thesan} provides an illustration of the simulated signals, and shows the rich potential of multi-line LIM observations during the EoR. These calculations help motivate further LIM survey efforts.

\begin{figure}
\begin{center}
\includegraphics[width=\textwidth]{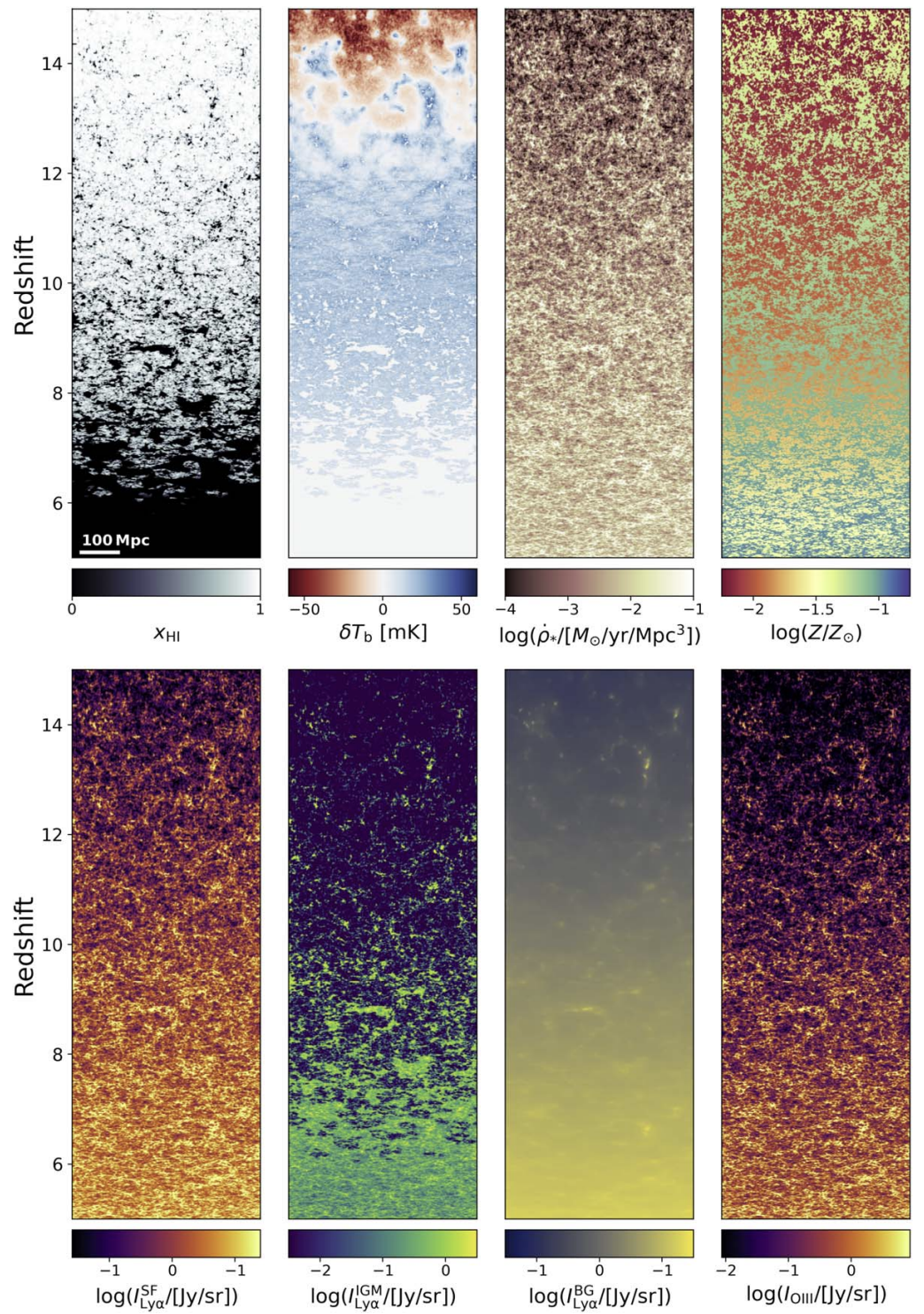}
\caption{
Light-cones extracted from LIMFAST simulations, spanning $5 \le z \le 15$, showing
the simulated ionization, chemical enrichment, and star formation histories, along with
models of the emission in various lines. Specifically, the rows show (from top to bottom) the evolution of:
the neutral hydrogen fraction, the 21 cm brightness temperature, the star formation rate density, the metallicity in collapsed structures, the specific intensity of Ly-$\alpha$ emission sourced by star formation, the ionized IGM, and the scattering background, respectively, and the specific intensity of [OII] $5007 \Ang$ line emission.
From \cite{Mas-Ribas2022}.
\label{fig:LIMFAST}}
\end{center}
\end{figure}

LIMFAST builds on the publicly available 21cmFAST code \cite{Mesinger11, Park2019}. As mentioned earlier, this is a semi-numerical tool tailored mainly for modeling the large-scale 21 cm emission during reionization and this code is commonly used in the 21 cm LIM community. LIMFAST thus shares the efficient, semi-numerical configuration of 21cmFAST to approximately track the formation of large-scale structure, the partitioning of mass into dark matter halos, and the reionization process, as we discussed previously. 
LIMFAST employs a more sophisticated treatment of galaxy properties based
on the quasi-equilibrium galaxy formation model for high redshift, star-forming galaxies from \cite{Furlanetto2017,Furlanetto2021}. This captures a range of physical properties that are important for LIM studies, including the gas mass, stellar mass, metallicity, and SFR of each model galaxy. The line-intensity fields
for multiple lines of interest are then calculated based on \textsc{Cloudy} line
luminosity simulations. Illustrative examples from LIMFAST calculations are shown in Figure~\ref{fig:LIMFAST}. These models follow a broad range of LIM signals, while rapidly capturing some of the important physics involved, and their efficiency allows one to explore many different galaxy formation scenarios. 

Finally, reference~\cite{Yang24} presents another approach for simulating multiple lines of interest. Specifically, these authors introduce a novel technique for modeling a number of [OII], [OIII], and Balmer series lines, including lines that should be detectable with SPHEREx (see \S \ref{S:projects}). Their approach combines three levels of modeling. First, a semi-analytic treatment is used to calculate the equilibrium ionization structure, level populations, and the emission in multiple lines around isolated and uniform HII regions (\cite{Yang23}, extending on the work of \cite{Yang2020}). Next, FIRE zoom-in simulations of individual galaxies are employed \cite{Hopkins:2017ycn}. These simulations partly resolve ISM scales, and include detailed models for star formation and feedback. It is assumed that each young stellar particle in the FIRE simulations is surrounded by an HII region, whose properties are determined from neighboring gas particles in the simulation. The semi-analytic HII region line emission calculations mentioned previously are carried-out for each such HII region. This model hence accounts for variations in the gas and stellar population properties across each simulated galaxy. A machine-learning approach is then used to determine the multi-line conditional probability distribution function for the line-luminosity to stellar mass ratio in each line of interest, given the simulated stellar particle age, metallicity, and stellar mass. Essentially, the FIRE simulations are used to produce a sub-grid line emission model which can be applied on top of the stellar particles in a larger volume, yet lower resolution cosmological hydrodynamic simulation. In \cite{Yang24}, the authors apply this model on top of the Illustris-TNG simulations \cite{Springel18,Pillepich18}. This modeling framework hence partly tackles the challenging dynamic range problem for LIM surveys by combining calculations of individual HII regions, FIRE-simulated galaxies with partly-resolved interstellar media, and ensembles of simulated galaxies across cosmological volumes from the IllustrisTNG simulation suite.

\subsection{Semi-analytical/empirical approaches}

The semi-numerical methods discussed previously start from realizations of the matter density fluctuations, generated either using the Zel'dovich approximation, 2LPT, or N-body simulations. There are additional modeling frameworks in the literature, which we term ``semi-analytical/empirical'' methods. In this case, the matter distribution is often modeled as a lognormal random field or with linear perturbation theory, while star formation, radiation fields, and line luminosities are included based on empirical calibrations and models for the halo mass function and linear biasing. In the lognormal approximation, many statistics can be calculated analytically without the need for explicit three-dimensional realizations of the matter distribution, although realizations are also straightforward to produce in the lognormal case. Since the observables can be quickly predicted in these semi-empirical approaches, they are well-suited for Bayesian inference, emulator training, and the exploration of vast astrophysical and cosmological parameter spaces.

Prominent examples include ARES, Zeus21, and oLIMpus. ARES is a flexible semi-analytic framework for modeling early galaxy populations, radiation backgrounds, and reionization histories that has been widely used to predict high-redshift 21 cm and galaxy observables \cite{ARES2012,ARES2014}. ARES has recently been extended to model multi-wavelength LIM observations across cosmic time, with specific application to multi-line SPHEREx datasets \cite{Bock:2025ijl}. In the same spirit, Zeus21 \cite{Zeus21} provides a fully analytic description of the 21 cm signal during Cosmic Dawn based on lognormal density modeling. It allows rapid calculations while retaining good agreement with semi-numerical calculations. This approach has been extended in the oLIMpus code \cite{oLIMpus},
which produces mock [O II], [O III], H$\alpha$, H$\beta$, [C II], CO, and 21 cm line signals, including auto-power spectra, cross-power spectra, lightcones, and mock survey realizations within seconds. These high-speed frameworks
are well-suited for MCMC analyses, survey optimization, and joint inference using multiple correlated tracers of reionization and galaxy evolution.

Several complementary semi-analytical/empirical tools further enrich the current modeling landscape. LIMpy \cite{LIMpy} emphasizes flexible prescriptions linking halo mass, star formation rate, and multi-line luminosities, allowing users to generate intensity maps for [C II], [O III], and the full CO rotational ladder using either analytic halo models or external halo catalogs, making it particularly useful for assessing astrophysical uncertainties and interfacing with cosmological simulations. SkyLine \cite{SkyLine} provides an efficient end-to-end framework for constructing mock LIM observations and survey forecasts with an emphasis on observational realism. SIMPLE \cite{SIMPLE} offers a modular semi-empirical approach for generating consistent predictions for multiple emission lines and their covariance across a broad range of galaxy formation scenarios. The optimal modeling approach therefore depends on the specific application, although multiple frameworks offer complementarity and valuable cross-checks.

\section{Statistics}
\label{S:statistics}

Having described a range of simulation and analytic modeling techniques, we turn to consider additional statistical properties of the LIM signals. 
In \S \ref{S:modeling} we described a halo-model-based approach for calculating: the LIM power spectrum, the cross-power spectra between two lines, and the cross-power between a LIM survey and a traditional galaxy survey. Here our goal is to discuss statistics beyond the power spectrum, including: the variance of the auto and cross-power spectra (\S \ref{sec:pk_var}, \S \ref{sec:px_var}),
novel statistical analyses using multiple emission lines (\S \ref{sec:multi_tracer}, \S \ref{sec:pk_from_px}), and additional higher-order statistics including 
the one-point probability distribution of the specific intensity (\S \ref{sec:pdf})\footnote{In general, the one-point probability distribution depends on the full set of moments of the specific intensity field, where $\avg{I^n}$ describes the nth moment. It thus contains information beyond the mean and variance. This is the sense in which it is a ``higher-order'' statistic.}, and the bispectrum (\S \ref{sec:bispectra}). 

\subsection{Power Spectrum Error Bars}
\label{sec:pk_var}

An important step in forecasting the capabilities of future LIM surveys is to determine the expected error bars on measurements of the power spectrum of the line-intensity fluctuations. Here, we outline the calculations involved in this step. Related considerations apply to forecasting higher-order statistics, but we focus on the spherically-averaged power spectrum since this is generally the simplest case and is a key metric for all upcoming LIM surveys. 

The error bar on a power spectrum measurement includes both sample variance and detector noise contributions. The sample variance term, also known as cosmic variance in this context, arises because the survey region might contain an excess of overdensities or underdensities compared to a typical region.  That is, the survey volume may not be representative of the underlying power spectrum that would be measured after averaging over a large ensemble of such regions. Likewise, the power spectrum estimated from a single Fourier mode within the survey may not reflect the average power spectrum at that wavenumber. 
The sample variance contribution to the power spectrum error bar is reduced by increasing the survey volume, since this provides more independent measurements of each Fourier mode that is sampled. The detector noise contribution to the power spectrum variance depends on the intensity of the sky background at the observed frequencies of interest, thermal noise within the telescope itself, and on other properties of the instrument. It may also be reduced by sampling additional Fourier modes. In the case of interferometric observations, the power spectrum of the detector noise (aka the ``noise power spectrum'') will depend on the number of telescope pairs at various separations: that is, on the baseline distribution of the interferometric array \cite{Liu:2019awk}. 

Let us start with the sample-variance dominated limit, in which detector noise contributions are negligible. We further consider the angle-averaged power spectrum and Fourier modes that are small in wavelength relative to the survey dimensions. In this case, we can neglect the mode-coupling that results from the finite survey size\footnote{In a finite survey, the estimated power spectrum is a convolution between the true power spectrum and the Fourier transform of the survey window function (squared). That is, the measured power spectrum is a running average of the true power spectrum, where the width of the wavenumber-averaging narrows with increasing survey dimensions.}. For further simplicity, we also ignore the smoothing effects from the finite angular and spectral resolution of the instrument here: implicitly, we consider well-resolved Fourier modes. (See below and Eq.~\ref{eq:pkvar_weight} for a discussion regarding the impact of angular and spectral resolution.)
We then consider estimates of the spherically-averaged power spectrum in bins of width $\Delta k$, centered on various modes with wavenumber $k_m$, where $m$ indexes the bin. The power spectrum estimate, for the line-intensity fluctuation field, at $k_m$ may be written as:
\begin{equation}
    \hat{P_I}(k_m) = \frac{\mathcal{A}}{N_{k_m}} \sum_{i=1}^{N_{k_m}} \delta_I^*(k_i) \delta_I(k_i),
    \label{eq:pk_est}
\end{equation}
where $N_{k_m}$ is the number of Fourier modes that fit into the wavenumber bin of width $\Delta k$, and $\mathcal{A}$ is a normalization constant that depends on the Fourier convention employed. 
This estimate averages over all of the Fourier modes -- with wavevectors in different orientations -- in the Fourier-space shell, and for simplicity we assume here that the noise does not vary across this bin in wavenumber\footnote{It is straightforward to generalize to the case that the noise does vary across each bin. In this case, the best estimate weights each mode in the sum by the inverse-variance of the power spectrum estimate at that $k_i$. See below and e.g. \cite{Liu:2019awk}.}.
Note that we are using the notation $\delta_I(\k)$ to describe the specific intensity variations, but we are not, in fact, normalizing here 
by $\avg{I}$ as might be implied by the $\delta_I$ notation. We consider the un-normalized fluctuations here because LIM surveys do not generally measure $\avg{I}$ itself -- at least not directly -- as discussed earlier.  

Since the line-intensity fluctuation, $\delta_I(\x)$, is a real-valued field, the Fourier counterparts $\delta_I(\k_i)$ obey the condition $\delta_I^*(\k_i) = \delta_I(-\k_i)$. In this case, only the modes in the upper-half hemisphere contain independent information and the sum in Eq.~\ref{eq:pk_est} may be restricted to such modes. Then the number of Fourier modes contained in the survey is approximately:
\begin{equation}
    N_{k_m} = \frac{1}{2}\, 4 \pi k_m^2 \Delta k \frac{V_{\rm survey}}{(2 \pi)^3}. 
    \label{eq:nmode}
\end{equation}
Here $4 \pi k_m^2 \Delta k$ is the volume of a spherical bin in Fourier space, while $(2 \pi)^3/V_{\rm survey}$ reflects the spacing of the Fourier modes that fit within the volume of the survey. The factor of $1/2$ out front arises because the sum is restricted to the upper-half hemisphere. 

The variance of the power spectrum estimate then follows from
\begin{equation}
    {\rm var}\left[\hat{P}_I(k_m)\right] = \avg{\hat{P_I}^2} - \avg{\hat{P}_I}^2 = \frac{{\mathcal{A}}^2}{N^2_{k_m}}
    \sum_{i=1}^{N_{k_m}}\sum_{j=1}^{N_{k_m}} \left[\avg{\delta^*_i \delta_i \delta^*_j \delta_j} - \avg{\delta^*_i \delta_i}\avg{\delta^*_j \delta_j}\right],
    \label{eq:pkvar}
\end{equation}
where we use the shorthand notation $\delta_i = \delta_I(k_i)$. Expanding the correlator involving four factors of $\delta$ in Eq.~\ref{eq:pkvar}, we have:
\begin{equation}
    \avg{\delta^*_i \delta_i \delta^*_j \delta_j} = \avg{\delta^*_i \delta_i \delta^*_j \delta_j}_c + \avg{\delta^*_i \delta_i} \avg{\delta^*_j \delta_j} +
    \avg{\delta^*_i \delta_j} \avg{\delta^*_j \delta_i} +
    \avg{\delta^*_i \delta^*_j} \avg{\delta_i \delta_j},
    \label{eq:pkvar_expand}
\end{equation}
where the first term is a connected four-point function. The second term cancels with the latter correlator in Eq.~\ref{eq:pkvar}, while $\mathcal{A} \avg{\delta^*_i \delta_j} = \mathcal{A} \avg{\delta^*_j \delta_i} = \delta^K_{i,j} \avg{\hat{P_I}}$, with $\delta^K_{i,j}$ denoting a Kroenecker delta symbol. The final term in Eq.~\ref{eq:pkvar_expand} vanishes since $\delta(\k_i)$ and $\delta(\k_j)$ are uncorrelated unless $\k_i=-\k_j$, which is excluded given our restriction to the upper-half hemisphere. Finally, in what follows we neglect the connected four-point contribution to Eq.~\ref{eq:pkvar_expand}, under the Gaussian statistics approximation. 

In this case, we arrive at a simple expression for the power spectrum variance:
\begin{equation}
    {\rm var}\left[\hat{P}_I(k_m)\right] = \frac{1}{N_{k_m}} \avg{P_I(k_m)}^2.
    \label{eq:pkvar_sample}
\end{equation}
Here $\avg{P_I(k_m)}$ is the expected average signal auto-power, without noise bias.
In this sample-variance dominated limit, the fractional error bar on a power spectrum estimate is simply $\sigma_P/\avg{P} = 1/\sqrt{N_{k_m}}$ and hence scales with the inverse square-root of the volume surveyed (Eq.~\ref{eq:nmode}). It is straightforward to generalize this to the more realistic case, which includes detector noise. Specifically, denoting the power spectrum of the detector noise by $\mathcal{N}(k_m)$, the result is:
\begin{equation}
    {\rm var}\left[\hat{P}_I(k_m)\right] = \frac{1}{N_{k_m}} \left[\avg{P_I(k_m)} + \mathcal{N}(k_m)\right]^2.
    \label{eq:pkvar_full}
\end{equation}
In practice, this result should be modified to account for the finite angular and spectral resolution of the surveys. We summarize the main modifications here (see e.g. reference \cite{Bernal:2019jdo} for further details). Briefly, let $W(k,\mu)$ describe the signal attenuation from the angular and spectral resolution of the survey with the observed
power spectrum being reduced by a factor of the window squared, $P_{\rm obs}(k,\mu) = W^2(k,\mu) \, P(k,\mu)$. For the window-deconvolved intensity fluctuation power, the best estimate follows from inverse-variance weighting. This may be written as:
\begin{equation}
\frac{1}{\mathrm{Var}\left[\hat{P}_I(k_m)\right]} = 
\int_{\mu_{\rm min}}^{\mu_{\rm max}} \! d\mu\, 
\frac{k_m^2\, \Delta k\, V_{\mathrm{survey}}}{4\pi^2} \cdot 
\frac{1}{\left[ \left\langle P_I(k_m, \mu) \right\rangle + 
\frac{\mathcal{N}(k_m, \mu)}{W^2(k_m, \mu)} \right]^2}.
\label{eq:pkvar_weight}
\end{equation}
This expression considers the inverse-variance weighted and spherically-averaged power spectrum in a narrow wavenumber bin of width $\Delta k$ around $k_m$.
The integral ranges over all wavevector orientations, $\mu$, spanned by the survey within the upper-half hempisphere. The quantity in the denominator of the right-hand side is an effective noise per Fourier mode and down-weights modes that are poorly sampled given the finite angular and spectral resolution of the survey. {That is, Fourier modes that are heavily attenuated with $W^2(k_m, \mu) \ll 1$ have high effective noise
variance and contribute negligible amounts of information to an estimate of the spherically-averaged power spectrum.}

In the case of single-dish LIM experiments, the noise power spectrum may be described by: 
\begin{equation}
    \mathcal{N}(k_m) = \sigma^2_N V_{\rm pix},
    \label{eq:noise_pk}
\end{equation}
where $\sigma^2_N$ is the noise variance per voxel in specific intensity units (squared; typically Jy$^2$/sr$^2$) and $V_{\rm pix}$ is the co-moving volume per survey voxel. The noise variance can be written as $\sigma^2_N = \rm{NEI}^2/t_{\rm pix}$ where $\rm{NEI}$ is the noise-equivalent intensity and $t_{\rm pix}$ is the observing time per voxel. As mentioned earlier, the noise power spectrum for an interferometric measurement is more complicated and depends on the baseline distribution of the radio array (see e.g. \citep{Liu:2019awk} for details).  

Eq.~\ref{eq:pkvar_full} has important implications for planning LIM surveys. Specifically, given a fixed total observing time, $t_{\rm obs}$, it is interesting to determine whether it is better to fix this entire time on a single region of the sky, or to instead observe $N_{\rm sub}$ sub-regions for a shorter time, $t_{\rm sub}$ per region, with $t_{\rm obs} = N_{\rm sub} t_{\rm sub}$. That is, is it better to go deep on a single patch or shallow across a wider area? The noise power spectrum, $\mathcal{N}(k_m)$, generally averages down as one observes a given patch on the sky for longer. However, the error bar on the power spectrum will no longer improve significantly once the noise power $\mathcal{N}(k_m)$ falls much below the expected signal power spectrum, $\avg{P_I(k_m)}$ (see Eq.~\ref{eq:pkvar_full}). At this point the measurement becomes sample-variance limited and to reduce the error bar further one should observe a different patch on the sky. This will increase the number of modes sampled, $N_{k_m}$, and reduce the error bar. 

More quantitatively, we can minimize the variance in Eq.~\ref{eq:pkvar_full} to determine the optimal survey strategy. Here we imagine observing a number of contiguous sub-regions, although a generalization is to consider various sparse sampling strategies where the sub-regions are distributed across a larger volume, but with gaps in the coverage (e.g. \citep{Chiang:2013ksa}). The noise power spectrum will generally scale as $\mathcal{N}(k_m) \propto 1/t_{\rm sub}$, while provided the Fourier modes of interest are small in wavelength compared with the size of each sub-region, the number of modes will vary with $N_{\rm sub} \propto 1/t_{\rm sub}$. Taking the derivative of Eq.~\ref{eq:pkvar_full} with respect to $t_{\rm sub}$, one can see that the optimal strategy sets the observing time per sub-region, $t_{\rm sub}$, such that $\mathcal{N}(k_m) = \avg{P_I(k_m)}$ \cite{Knox95,Tegmark:1997vs}, i.e. so that the noise power matches the expected signal power for the Fourier mode of interest.  This confirms the intuition of the previous paragraph, and the above criterion provides a valuable benchmark for optimizing LIM survey design. Of course this is only one of many considerations that may impact the experimental design, but it is nevertheless helpful for understanding the expected statistical power of different survey configurations. 

\subsection{Cross-Power Spectrum Error Bars}
\label{sec:px_var}

A very similar calculation to the auto-power spectrum analysis above gives the variance of the cross-power spectrum estimated between two line-intensity maps. Omitting the details, the result is \cite{Lidz11}:
\begin{equation}
{\rm var}\left[\hat{P}_x(k_m)\right] = \frac{1}{2 N_{k_m}}
\left\{\avg{\hat{P}_x(k_m)}^2 + \left[\avg{\hat{P}_A(k_m)} + \mathcal{N}_A(k_m)\right] \left[\avg{\hat{P}_B(k_m)} + \mathcal{N}_B(k_m)\right]\right\}.
\label{eq:px_var}
\end{equation}
Here $P_x(k_m)$ is the cross-power spectrum between lines $A$ and $B$, while $P_A(k_m)$ and $P_B(k_m)$ are their respective auto-power spectra, and $\mathcal{N}_A(k_m)$, $\mathcal{N}_B(k_m)$ are the noise power spectra in each line. The above formula assumes that the noise in lines $A$ and $B$ are uncorrelated. The factor of $1/2$ enters because we count only modes in the upper-half hemisphere. The mode-count, $N_{k_m}$, here should include only the common, overlapping portion of the survey volumes in each line. 

Analogous expressions also apply to the cross-power spectrum between a line-intensity map and a galaxy survey (e.g. \citep{LidzTaylor2016}). In that case, the noise power spectrum for galaxy survey arises from shot-noise in the galaxy distribution, with $\mathcal{N} = 1/\bar{n}_g$, where $\bar{n}_g$ is the average abundance of galaxies per unit co-moving volume. {Note that the optimal survey strategy for measuring 
the cross-power spectrum between a line-intensity map and an external survey may be quite different than that for
the LIM auto-power spectrum, as will be discussed further in the next sub-section.}

\subsection{Further One-Point and Higher-Order Statistics}

The line emission fluctuation fields are non-Gaussian and so the upcoming data sets will contain information beyond the usual power spectra of line-intensity fluctuations. This is especially the case for the reionization-era 21 cm signal, where the ionized regions imprint large non-Gaussian variations in the 21 cm data cube. In that context, a range of statistics have been considered including the one-point probability distribution function \cite{Mellema:2006pd}, the 21 cm bispectrum \cite{Bharadwaj05}, trispectrum \cite{Mondal:2015oga}, matched-filters for bubble finding \cite{Datta2008,Malloy2013}, and others. Here we will discuss the one-point probability distribution function of non-21 cm data cubes and the 21 cm bispectrum/cross-bispectrum. 

\subsubsection{Probability Distribution Function}
\label{sec:pdf}

The one-point probability distribution function (PDF) of the specific intensity field across a LIM data cube contains information regarding the full luminosity function of the line emission \cite{Breysse16}. This is in contrast to the line-intensity fluctuation power spectrum, which depends only on the first two moments of the luminosity function (\S \ref{S:modeling}). In this context, the PDF is referred to as the voxel intensity distribution (VID) \cite{Breysse16,Breysse:2016szq}, as it describes the probability distribution of specific intensities within different volume elements (voxels) across the survey. In some cases, the brightness temperature is considered instead of the specific intensity. The VID measurements may also be performed as a function of smoothing scale. 

The VID can be measured directly from simulated line-intensity data cubes, or modeled in the following manner \cite{Breysse16,Breysse:2016szq,Bernal:2022jap}. 
First, consider the specific intensity in a voxel containing exactly one line-emitting source:
\begin{equation}
    I_1 = \frac{c}{4 \pi H(z) \nu_r} \frac{L}{V_{\rm vox}}.
    \label{eq:i_single}
\end{equation}
Here, and in the following parts of this subsection, we neglect the frequency label on $I$ for brevity, yet it is a specific intensity (i.e., intensity per unit frequency). However, the luminosity, $L$, is the total line luminosity, rather than a specific luminosity.  Eq.~\ref{eq:i_single} follows from reasoning similar to that which led to Eq.~\ref{eq:avg_inu}. Here, and throughout, we neglect the possibility that a single source contributes to more than one voxel, which should be a very good approximation given the coarse voxels expected for most LIM surveys. 

The differential probability distribution of the specific intensity, conditioned on a voxel containing a single source is:
\begin{equation}
    P_I(I|1) =\frac{1}{\avg{n}} \frac{4 \pi H(z) \nu_r V_{\rm vox}}{c}\phi\left(L = \frac{4 \pi H(z) \nu_r V_{\rm vox}}{c} I\right).
    \label{eq:dpdi_single}
\end{equation}
Here, $\avg{n} = \int dL \phi(L)$ is the number density of emitters, and $\phi(L)$ is the luminosity function. The probability per unit luminosity of a single source residing in a voxel is $\phi(L)/\avg{n}$, while Eq.~\ref{eq:dpdi_single} maps between luminosity and specific intensity and hence the expression for $P_I(I|1)$ includes the necessary Jacobian factor. 

Naturally, a given voxel may have more than one emitter, or contain zero sources. If the overall average abundance of sources contained within a voxel, and emitting in a given line, is $\avg{N}$, then the abundance of sources in a voxel at spatial position $\x$ is:
\begin{equation}
    N(\x) = \avg{N} \left[1 + \delta_{\mathrm{e}}(\x)\right].
\end{equation}
Here $\delta_{\mathrm{e}}(\x)$ describes the fluctuations in the emitter abundance, smoothed on the scale of the voxel. (For brevity, we omit the smoothing scale in our notation here). In each voxel, one should additionally take a Poisson sample, drawn from a distribution with $N(\x)$ galaxies on average. We denote this by $P_{\mathrm{p}}(M|N)$, specifying the Poisson probability that a voxel contains M sources, when we expect $N$ on average. 

Next, we need to find the probability distribution for the specific intensity in a voxel with $M$ sources inside. This is determined by taking $M$ convolutions between the single-source intensity PDFs \cite{Breysse16,Breysse:2016szq,Bernal:2022jap}:
\begin{equation}
P_I(I|M) = P_I(I|1) \ast P_I(I|1) \ast .... \ast P_I(I|1),
\label{eq:dpdi_msource}
\end{equation}
where the ``...'' indicates additional convolutions, with $M$ separate convolutions in total. 
Finally, we arrive at an expression for the VID:
\begin{equation}
    P_I(I) = \int d \delta_{\mathrm{e}} P_{\mathrm{e}}(\delta_{\mathrm{e}}) \sum_{M=0}^{\infty} P_I(I|M) P_{\mathrm{p}}(M|N).
    \label{eq:vid_final}
\end{equation}
Here we average over the smoothed emitter-density PDF, $P_{\mathrm{e}}(\delta_{\mathrm{e}})$, which is sometimes approximated by a lognormal distribution \cite{Breysse16}. Alternatively, in the case of large voxels, one can assume a linear-biasing description \cite{Breysse:2022alx,Thiele:2018jdl}: this has the virtue of connecting the VID with halo bias models.  
As mentioned earlier, and can now be seen explicitly from Eqs.~\ref{eq:dpdi_single}-\ref{eq:vid_final}, the VID depends on the full luminosity function and captures more information than available in the power spectrum alone. As one example, it may help in breaking the degeneracy between $\avg{b}$ and $\avg{I}$ in auto-power spectrum measurements (\S \ref{S:modeling}, Eq.~\ref{eq:pofk}) \cite{Breysse:2022alx}. 

\begin{figure}
\begin{center}
\includegraphics[width=\textwidth]{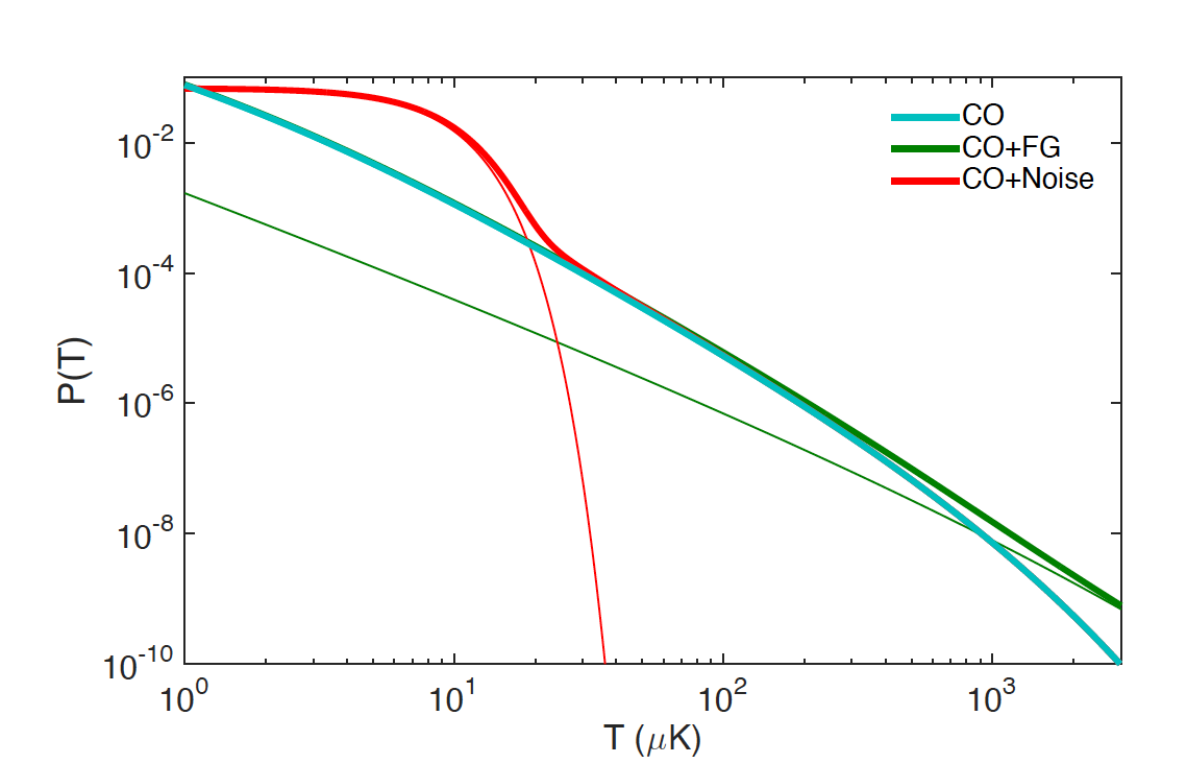}
\caption{Model for the one-point PDF (aka the ``VID'') of the CO(1-0) LIM brightness temperature signal at $z = 2.6$. The thin lines show the contaminant PDFs from instrumental noise (red) and HCN line-interlopers (green). The red solid line gives the convolution between the signal PDF and the instrumental noise, while the dark green curve is the signal convolved with the HCN foreground. The impact of continuum foreground contamination is neglected in this illustration. The signal PDF is non-Gaussian and future VID measurements may allow one to extract information beyond the first two moments of the luminosity function. 
From \cite{Breysse16}.
\label{fig:vid_illust}}
\end{center}
\end{figure}

In practice, however, one does not directly access the {\em signal} VID; instead, the LIM data cubes will be corrupted with instrumental noise and foreground contamination. In general, the foreground contamination will include both continuum foregrounds and line interlopers (see \S \ref{S:challenges}). Ignoring foreground contamination for the moment, the total intensity in a voxel has signal and noise contributions with $\tilde{I}(\x) = I(\x) +\, n(\x)$\footnote{Note that here we are using $n(\x)$ to denote the instrumental noise, not to be confused with the number density of line-emitters. In this section, the number of line-emitters in a voxel is given by $N(\x)$ while the instrumental noise in the same location is $n(\x)$.}. In this case, the observed PDF, $P_{\tilde{I}} = P_I \ast P_n$, is a convolution between the intrinsic signal PDF and that of the instrumental noise. In practice, exploiting the VID hence requires a good understanding of the instrumental noise PDF. 
Furthermore, the LIM data cube must be filtered to avoid continuum foregrounds and masks may be required to mitigate interloper contaminants. The effect of foreground filtering on the VID model of Eqs.~\ref{eq:i_single}-\ref{eq:vid_final} is not easy to include analytically, but this is straightforward to study using simulated LIM data cubes. Figure~\ref{fig:vid_illust} provides an illustration of the VID model in the case of the CO(1-0) brightness temperature distribution, and also shows the impact of instrumental noise and line-interloper foregrounds. {For clarity, we note that updates to the HCN foreground model in reference \cite{Chung:2017uot} make this particular foreground less important than implied by the figure, but the general concern about foreground contamination to the VID remains.}

One approach for a VID-style analysis that should avoid many of the foreground contamination issues is developed in \cite{Breysse:2019cdw}. These authors consider combining a line-intensity map with a traditional galaxy survey using the conditional VID, $P_{\tilde{I}}(\tilde{I}|N_{\mathrm{g}})$. This is the PDF of voxel intensities conditioned on the same voxel containing $N_{\mathrm{g}}$ galaxies in the traditional galaxy survey. This approach is broadly analogous to taking the cross-power spectrum between a LIM data cube and a galaxy survey to avoid foreground bias in the LIM data. 
The intensity, $\tilde{I}$, here includes foreground contaminants and instrumental noise. The PDF, $P_{\tilde{I}}(\tilde{I}|N_{\mathrm{g}})$, can hence be factorized into the convolution of an intrinsic signal piece and the PDF of the contaminants. Unlike the signal itself, the contaminant piece does not depend on $N_{\mathrm{g}}$, provided the LIM foregrounds and noise are themselves uncorrelated with the traditional galaxy survey. In this case, it is convenient to consider the conditional VID in Fourier space, exploiting the fact that the Fourier transform of a convolution is a product of Fourier transforms. 
Conveniently, the foreground contamination divides out on average in the ratio between the Fourier transforms of conditional VIDs, for voxels with two different galaxy counts (\cite{Breysse:2019cdw}; see their Eq. 5 for details). This statistic is therefore unbiased on average by foreground contamination, although residual foregrounds will boost the error bars on its measurement. 

A novel generalization of this method has recently been developed in References~\cite{Breysse:2022fdi,Chung:2022zeu}. In those works the authors consider the joint PDF between any two random fields. They show that a suitably normalized Fourier transform of the joint PDF may be estimated without any average bias from foregrounds and noise, provided these are uncorrelated between the two fields. This statistic hence has some of the same virtues as the cross-power between two fields, while also accessing additional non-Gaussian information (which is not contained in the cross-power spectrum itself).

\subsubsection{Bispectra and Cross-Bispectra}
\label{sec:bispectra}

A number of papers have considered the 21 cm bispectrum during the EoR and Cosmic Dawn (e.g. \cite{Bharadwaj05,Watkinson:2017zbs,Majumdar:2017tdm,Giri:2018dln}), while \cite{Beane:2018pmx} model the 21 cm-[CII]-[CII] cross-bispectrum. In the future, it will also be fruitful to consider the bispectra of other LIM data sets as well. 
The 21 cm signal during reionization is an interesting case given the strong non-Gaussianities expected in the reionization-era signal. The main challenges with the 21 cm bispectrum are that: even the noise and foreground-free signal can be difficult to interpret, while properly calculating the bispectrum error covariance matrix and convolving theory models with survey window functions are computationally expensive. 

The above works have measured the 21 cm bispectrum from reionization simulations, and started to address the first challenge regarding its physical interpretation. An interesting case is that of the squeezed-triangle bispectrum, $B(\k_1, \k_2, \k_3)$, in which one wavevector is much smaller in magnitude than the other two, i.e. $k_1 << k_2, k_3$ (e.g. \cite{Giri:2018dln}). This statistic quantifies the extent to which the small-scale 21 cm power spectrum is enhanced or diminished in a large-scale region of above-average 21 cm brightness temperature. That is, it describes a coupling between long- and short-wavelength modes: while this would vanish in the case of a Gaussian random field, the 21 cm fluctuation fields are non-Gaussian, and this statistic should yield information regarding the reionization process. Reference~\cite{Giri:2018dln} models the squeezed-triangle bispectrum, considering the response of the small-scale 21 cm power spectrum to a local, long-wavelength enhancement or depression in the 21 cm fluctuations. The authors show that this mode-coupling signal evolves in an interesting and distinctive way as reionization proceeds, partly reflecting the tendency for large-scale overdense regions to reionize before typical regions. This tendency arises because the sources of ionizing photons are expected to be more highly biased tracers of the density fluctuations than the sinks of ionizing photons: quantifying this trend using the squeezed-triangle bispectrum may help in determining the properties of the sources/sinks and their evolution during reionization. We refer the reader to the above reference for details and to the other aforementioned works for a discussion of the 21 cm bispectrum in additional triangle configurations.

Reference~\cite{Beane:2018pmx} discusses the 21 cm-[CII]-[CII] cross-bispectrum. Recall that as a consequence of homogeneity/translation invariance, the bispectrum/cross-bispectrum is non-vanishing only for triplets of wavenumbers $\k_1, \k_2, \k_3$ that sum to zero, and so form a closed triangle. Ignoring anisotropies from redshift space distortions, the bispectrum may be specified in terms of
the length of two of the wavevectors and the angle between them, i.e. on $k_1, k_2, \theta_{12}$. In the large-scale structure literature, it has long been recognized that the galaxy bispectrum contains potentially valuable information regarding galaxy biasing, especially through the dependence of the bispectrum on the shape of the wavevector triangles (e.g. \cite{Fry94,Scoccimarro2001}). Likewise, bispectra of the line-intensity fluctuations provide a possible handle on the biasing relation between the line emission fluctuations and the underlying variations in the density field \cite{Beane:2018pmx}. The 21 cm bias factor is expected to evolve strongly as reionization proceeds, and so measurements at different frequencies/redshifts may help to extract information regarding the timing of reionization. The important advantage of a cross-bispectrum -- such as that between a single 21 cm field and two [CII] fields covering the same redshifts and portions of the sky -- is that this statistic is less prone to bias from foreground contamination and other systematics (provided they are not shared between the two surveys).

\subsection{Stacking, Two-Point Cross-Correlation Functions, and Cross-Power Spectra}
\label{sec:stacking}

Stacking is a widely used technique for detecting faint, spatially localized signals in diffuse maps: small patches from a map of, for example, CMB temperature, 21 cm, or X-ray surface-brightness, among others, are cut-out and averaged at the locations of galaxies in a catalog. This can boost the signal-to-noise ratio when individual sources are too faint to detect in the diffuse map. This approach has been employed successfully in the LIM context by  CHIME \cite{CHIME2022} and COMAP \cite{COMAP:2025mov}. The output of these stacking analyses is an average emission profile as a function of the distance from the center of each galaxy. In general, this profile includes emission from the catalog galaxy itself and from neighboring sources. This output is, in fact, mathematically equivalent to the two-point cross-correlation function between the two data sets, and also to their cross-power spectrum in Fourier space. Although formally equivalent, there can be practical advantages for a particular choice among these three analyses, depending on the application. The information content may be more transparent in one case, while systematics can be easier to handle in one of these equivalent statistical measures.  

\subsubsection{Equivalence}

We start by demonstrating the equivalence of the three statistics. 
Let $T({\bf x})$ be a diffuse map (or data cube) defined over a survey volume $V$, and consider a galaxy catalog containing $N_g$ objects at positions ${{\bf x}_i}$. 
We defined mean-subtracted fluctuation fields, $\delta T$ for the diffuse radiation and $\delta_g$ for the galaxy distribution: 
\begin{equation}
    \delta_g({\bf{x}}) = \frac{n_g({\bf{x}})}{\bar{n}_g} -1, \,  
    \delta T({\bf{x}}) = T({\bf{x}}) - \langle T \rangle, 
\end{equation}
where $n_g({\bf{x}}) = \sum_i \delta_D({\bf{x}}-{\bf{x}}_i)$ is the galaxy number density field and $\bar{n}_g = N_g / V$ is the mean abundance in the galaxy catalog; $\langle \delta_g \rangle = \langle \delta_T \rangle = 0$ where $\langle \rangle$ denotes a volume average.

The stacking measurement averages the fluctuation field at each galaxy position, as a function of the offset ${\bf r}$ from the galaxy center:
\begin{equation}
    S({\bf{r}})=\frac{1}{N_g}\sum_i^{N_g} \delta T({\bf{x}}_i+{\bf{r}}).
\end{equation}
On the other hand, the galaxy–diffuse radiation cross-correlation function is defined as the spatial average of the product of the two fluctuation fields, separated by $r$:
\begin{equation}
\xi({\bf{r}}) = \frac{1}{V} \int_V \delta_g({\bf{x}}) \delta T({\bf{x}}+{\bf{r}}) d^3x.
\label{eq:crossg}
\end{equation}
We can rewrite $\delta_g$ as
\begin{equation}
\delta_g({\bf{x}}) = \frac{1}{\bar{n}_g}\sum_i^{N_g} \delta_D({\bf{x}}-{\bf{x}}_i)-1.
\end{equation}
Substituting this expression back into Eq.~\ref{eq:crossg}, we have
\begin{equation}
    \xi({\bf{r}}) = \frac{1}{V} \int_V \left[\frac{1}{\bar{n}_g}\sum_i^{N_g} \delta_D({\bf{x}}-{\bf{x}}_i)-1\right] \delta T({\bf{x}}+{\bf{r}}) d^3x.
\end{equation}
Evaluating the spatial integrals and using $N_g = \bar{n}_g V$, we then have
\begin{equation}
    \xi({\bf{r}}) = \frac{1}{N_g} \sum_i^{N_g} \delta T({\bf{x}}_i+{\bf{r}}) - \langle \delta T \rangle.
\end{equation}
The first term is exactly the stacking profile, and since $\langle \delta T \rangle =0$ by construction, we have 
\begin{equation}
    {S}({\bf{r}}) = \xi_{\delta_g\delta T}({\bf{r}}).
\end{equation}
Finally, by the Wiener–Khinchin theorem, the correlation function and the cross-power spectrum P({\bf{k}}) are a Fourier transform pair, and we therefore have the equivalence chain:
\begin{equation}
    {S}({\bf{r}}) = \xi_{\delta_g\delta T}({\bf{r}}) = \int \frac{d^3k}{(2\pi)^3} P_{\delta_g,  \delta T}({\bf{k}}) e^{i\bf{k} \cdot {\bf{r}}}.
\end{equation}
At zero separation (zero lag), i.e. at the position of the galaxy, this reduces to
\begin{equation}
    {S}(0) = \xi_{\delta_g\delta T}(0) = \int \frac{d^3k}{(2\pi)^3} P_{\delta_g, \delta T}(\bf{k}). 
\end{equation}
The stacking amplitude measured directly at the galaxy positions is simply the zero-lag cross-correlation amplitude, which is equivalent to the total cross-power spectrum integrated over all Fourier modes.
In practical applications, these expressions require straightforward generalizations to account for the finite spectral and angular resolution of the LIM measurements and window functions to account for masking and the geometries of each survey.

\subsubsection{Comparison}

Since the three statistics are equivalent, evaluating one of them is in principle sufficient. However, measuring more than one allows internal cross-checks, as certain systematic errors may be easier to isolate in one of the three representations.

Real-space statistics evaluated with an aperture much smaller than the survey footprint are, by construction, local and largely insensitive to the global shape of the survey mask,  irregular boundaries, holes, and varying depth. By contrast, these effects need to be explicitly corrected for with matched random catalogs for the Fourier-space statistics. In Fourier space, the observed fluctuations involve a convolution between the underlying fluctuation field and the Fourier transform of a survey window function, which couples different $k$-modes together.
This therefore requires either a deconvolution step, forward-modeling the convolution with the survey window, or a full mode-decoupling correction. For highly irregular or discontinuous footprints, stacking (or the two-point cross-correlation function) can be more forgiving. 

On the other hand, while real-space statistics are simple and intuitive, in the linear regime different Fourier modes evolve independently whereas the two-point correlation function at a given separation $r$ receives contributions from a broad range of scales, including non-linear high-$k$ modes. Instrumental angular and spectral resolution are also naturally incorporated as multiplicative windows in $k$. 
Finally, 
the one-halo and two-halo contributions (see \S \ref{S:modeling}) typically dominate at different $k$, making them easier to distinguish in Fourier space than configuration space.

The variance of the cross-power spectrum is also more transparent to compute and interpret than the corresponding estimates for the two-point cross-correlation function. The uncertainties in the correlation function show strong bin-to-bin correlations. In this case, simple diagonal error bar estimates are untrustworthy and the full covariance matrix needs to be determined, either empirically or from simulations. On the other hand, for the cross-power spectrum in the Gaussian limit, different $k$-bins are independent, at least for modes with wavelengths that are small compared to the length scales of the survey. Here, the power spectrum variance follows from simple mode counting (Eq.~\ref{eq:px_var}) with $\rm{var}\left[P_{\rm x}(k) \right] \propto P_{\rm x}^2(k)/N_{\rm m}(k)$. Of course, this limit is idealized since non-linear structure growth and irregular survey geometries generate off-diagonal terms in the cross-power covariance matrix. Therefore, cross-power spectrum analyses still often require jackknife or mock covariance estimates in practice, although the bin-to-bin coupling is less severe than for real-space estimators.

Systematic effects also manifest themselves differently in these statistics. In stacking and cross-correlation analyses, it is often easy to guard against residual backgrounds, calibration offsets, and certain types of foreground contamination: one can simply subtract off LIM measurements at random sky positions from those around the targeted
galaxies. On the other hand, foreground cleaning or avoidance may generally be easier to treat in Fourier space. For example, continuum foregrounds are spectrally smooth and mostly corrupt low $k_\parallel$ modes (see \S \ref{S:challenges}).
In the cross-power spectrum, slowly varying calibration drifts may also be confined to low $k$, while beam or instrumental smoothing leads to a controllable suppression at high $k$, and so specific $k$-bins can frequently be excised or corrected without discarding further data. In real-space, such systematics may often be spread across a broad range of scales and hard to separate cleanly. As mentioned previously, the survey geometry is easier
to account for in stacking or cross-correlation analyses than in cross-power spectrum estimates where the survey window must be deconvolved for direct comparisons with theoretical models.

Reference \cite{Dunne2025} notes the equivalence of stacking to the zero-lag two-point cross-correlation function, as discussed above. For their particular estimator, the authors find that the stacking results are insensitive to larger-scale Fourier modes, nothing that such modes can be probed with low-$k$ cross-power spectrum measurements. Alternatively, the stacking estimator can be generalized to non-zero lags and weights can be applied to help optimize the analysis for large-scale modes.

Interestingly, reference \cite{Kramer2026} compares stacking, the cross-power spectrum, and the conditional VID (CVID, see \S \ref{sec:pdf}) using simulated EXCLAIM [CII] maps. The paper forecasts mean detection significances of $3.9\sigma$ (stacking), $3.7\sigma$ (CVID), and $4.2\sigma$ (cross-power spectrum) in its fiducial noise scenario, but finds the ordering reverses in a lower-noise scenario ($35.4\sigma, 76.6\sigma$, and $38.8\sigma$, respectively). In their fiducial noise-dominated case, all three estimators give comparable forecasts, with only a small edge to the cross-power spectrum. In the low-noise limit, the CVID roughly doubles the signal-to-noise ratio compared to the other statistics. This points to the ability of the CVID to extract non-Gaussian information, which is lost in the two-point statistical analyses, from the LIM data in the low-noise limit.

\subsection{The Multi-Tracer Technique}
\label{sec:multi_tracer}

In the context of the LIM $\times$ galaxy cross-power spectrum, it is important to note that the ratio between $\avg{b_1}\avg{I_1}$ and the average galaxy bias, $\avg{b_{\mathrm{g}}}$, may be measured without sample variance (at least provided each field traces the matter density field with negligible stochasticity, i.e., $r \approx 1$) \cite{Seljak09,McDonald:2008sh,Bernstein11,Switzer:2017kkz,Liu:2020izx,Oxholm:2021zxp}. The ability to measure bias ratios between different probes without sample/cosmic variance is termed the ``multi-tracer'' method, and promising applications include extracting the scale-dependent biasing signature induced by primordial non-Gaussianity \cite{Dalal:2007cu,Seljak09} and information about the growth of structure via redshift space distortions \cite{McDonald:2008sh}. In the LIM context, the multi-tracer method is also relevant for determining $\avg{b_1} \avg{I_1}$ itself \cite{Switzer:2017kkz}. 

In what follows, we assume both negligible stochasticity ($r=1$) and purely linear evolution, as should be good approximations on large scales.  In this case, consider the same Fourier mode in the line-intensity fluctuation field, $I_1(\k) = \avg{b_1} \avg{I_1} \delta_\rho(\k)$, and the galaxy distribution, $\delta_{\mathrm{g}}(\k) = \avg{b_{\mathrm{g}}} \delta_\rho(\k)$: these fluctuations each trace the same density variations, $\delta_\rho(\k)$. Therefore, the ratio of $I_1(\k)$ and $\delta_{\mathrm{g}}(\k)$ is independent of $\delta_\rho(\k)$ and may hence be estimated without sample variance. Similarly, considering the same Fourier mode, the ratio of the line-galaxy cross-power spectrum and the galaxy auto-spectrum is
$P_x(\k)/P_{\mathrm{g}}(\k) = \avg{b_1}\avg{I_1}/\avg{b_{\mathrm{g}}}$, which also does not suffer from sample variance. 

To make this quantitative, one approach is to use the Fisher matrix formalism (e.g. \cite{Tegmark97}). Specifically, the Fisher matrix here quantifies the information in the data covariance matrix -- here in Fourier space -- regarding the model parameters. It can be used to determine the best possible parameter errors that can be determined given the specifications of upcoming surveys. The Fisher information is given by the expectation value of the curvature matrix, $F_{ij} = - \avg{\partial^2 {\rm ln \, L}/\partial p_i \partial p_j}$, where ${\rm L}$ denotes the likelihood function, evaluated around the (assumed) true parameter values \cite{Tegmark97}. This measures how rapidly the likelihood function falls off in the vicinity of the maximum likelihood parameter values. The inverse of the Fisher matrix then gives the expected error covariance matrix. 

Here we start from the covariance matrix for measurements of a single Fourier mode in each of the LIM survey and the galaxy survey:
\begin{equation}
\Sigma =  \begin{pmatrix}
P_I + \mathcal{N}_I & P_x\\
P_x & P_{\mathrm{g}} + \mathcal{N}_{\mathrm{g}}
\end{pmatrix}
\label{eq:cov_ig}
\end{equation}
This matrix includes the auto-power spectra of both the signal and noise for each of 
the LIM and galaxy surveys (along the diagonal), as well as the cross-power spectrum between the two data sets (the off-diagonal $P_x$ term). For simplicity, we neglect cross-shot noise, but refer the reader to \cite{Oxholm:2021zxp,Liu:2020izx} for a treatment of this and further generalizations to the illustrative case considered here. 

In this case, assuming a Gaussian likelihood function, the Fisher matrix may be written as \cite{Tegmark97}:
\begin{equation}
    F_{i j} = \frac{1}{2} \rm{Tr} \left[ \frac{\partial \Sigma}{\partial p_i} \Sigma^{-1} \frac{\partial \Sigma}{\partial p_j} \Sigma^{-1} \right].
    \label{eq:fisher}
\end{equation}
The error bar on the $i$th parameter, marginalized over the other parameters in the problem, is given by $\sigma^2_i = (F^{-1})_{i i}$, i.e., by the $i, i$ element of the inverse of the Fisher information matrix. 
For example, consider a case where we are interested in two main parameters, $A \equiv \avg{b}{I}$ and $P_{\rm lin}(k)$\footnote{A further generalization would be to vary the galaxy bias, $b_{\mathrm{g}}$, and marginalize over that as well. Here we consider the simpler case that the galaxy bias is already well-known or well-measured through other means.}.

Applying Eq.~\ref{eq:fisher}, the expected error on $A$, marginalized over $P_{\rm lin}(k)$, is \cite{Oxholm:2021zxp}:
\begin{equation}
    \frac{\sigma^2_A}{A^2} =  W_I + W_{\mathrm{g}} + W_I W_{\mathrm{g}}
    \label{eq:var_a_multi}
\end{equation}
We mostly follow here the notation of \cite{Oxholm:2021zxp}, in which $W_I = \mathcal{N}_I/(A^2 P_{\rm lin})$ and $W_{\mathrm{g}} = \mathcal{N}_{\mathrm{g}}/(b_{\mathrm{g}}^2 P_{\rm lin})$. That is, these quantities give the noise-to-signal ratio for the power spectra in each of the LIM and galaxy survey, respectively (for the Fourier mode of interest). This result can easily be generalized to determine the error bar on $A$ estimated from a bin with $N_{k_m}$ independent modes (Eq.~\ref{eq:nmode}): this follows simply from dividing Eq.~\ref{eq:var_a_multi} by $N_{k_m}$. Equivalently, the Fisher matrix describing the information contained in a set of independent measurements is given by the sum of their individual Fisher matrices. 

The first key feature of Eq.~\ref{eq:var_a_multi}, anticipated earlier, is that sample variance is absent. That is, the error on $A \equiv \avg{b}\avg{I}$ improves without bound as the noise in the LIM survey and the shot-noise in the galaxy survey decrease.  This is in contrast to the error bar on an auto-power spectrum measurement which eventually saturates with decreasing noise owing to sample variance (Eq.~\ref{eq:pkvar_full}). 
This can then impact the optimal survey strategy for LIM observations. As a prime example, imagine an upcoming LIM survey that targets a portion of an existing traditional galaxy survey. In this case, the strategy that minimizes the error bar on $A$ for the LIM survey is to observe a sub-region of the galaxy survey until $W_I$ drops to $W_{\mathrm{g}}$ \cite{Oxholm:2021zxp}. Beyond this limit, the error bar on $A$ is limited by shot-noise in the galaxy distribution, and the LIM telescope should move to cover additional portions of the galaxy survey. This will help by increasing the number of modes sampled. The scenario discussed here is analogous to the case of optimizing an auto-power spectrum measurement (Eq.~\ref{eq:pkvar_full}). In the case of the auto-power, the optimal approach is to match the noise power to the signal power, mitigating sample variance and detector noise. In the multi-tracer case analyzed here, 
the best strategy matches the line-intensity noise term ($W_I$) to the shot-noise term in the galaxy distribution ($W_{\mathrm{g}}$).  

The discussion above can be generalized to consider the cross-power spectra between different LIM data cubes, the impact of any correlated shot-noise on the cross-power spectrum, and other science targets such as primordial non-Gaussianity (e.g. \cite{Liu:2020izx}).  The multi-tracer method may, in some cases, motivate different choices for the survey and instrumental design than would otherwise be considered.

\subsection{Auto-Power from Cross-Power}
\label{sec:pk_from_px}

Here we consider another possible advantage of tracing the same cosmological volume using multiple different emission lines. Among other benefits and applications, if $N$ separate lines may be measured in a shared volume, then this will enable measurements of $N(N-1)/2$ unique cross-power spectra. As we emphasized previously, the cross-power spectra are expected to be less sensitive to residual foreground contamination and other systematic concerns, apart from common or spatially-correlated foregrounds. Specifically, independent contaminants will generally increase the variance of cross-power spectrum estimates without introducing an average bias.

As discussed in the previous section, on sufficiently large scales we expect the line-intensity fluctuations to trace the underlying density fluctuations, with negligible stochasticity, and for a linear-biasing description to be adequate. That is, the fluctuations in the line emission are well-described by $\delta_I(\k) = \avg{I} \avg{b} \delta_{\rm lin} (\k)$ at low $k$. As usual, $\avg{I}$ is the average specific intensity, $\avg{b}$ is the luminosity-weighted bias factor, and $\delta_{\rm lin}(\k)$ denotes the density fluctuation field in linear perturbation theory. We have neglected redshift space distortions here for simplicity, and assumed that the cross-correlation coefficient $r$ between the line-emission fluctuations and the linear density variations is unity. Furthermore, we have assumed negligible cross shot-noise on the scales of interest. If these assumptions apply to two lines, $i$ and $j$, the cross-power spectrum, $P_{i,j}(k)$, may be described simply by:
\begin{equation}
P_{i,j}(k) = \avg{I_i}\avg{I_j} \avg{b_i}\avg{b_j} P_{\rm lin}(k).
    \label{eq:px_simple}
\end{equation}

The focus of this section is on an interesting application of this formula to the case where three (or more) lines are measured in a common volume. We denote the three lines by $1, 2, 3$, respectively. Then three separate cross-power spectra can be measured between the different pairs of fields: $P_{1, 2}(\k)$, $P_{1, 3}(\k)$, $P_{2, 3}(\k)$\footnote{As a brief aside, note that redshift evolution in each signal breaks a symmetry in the two-point cross-correlation function. In particular, the configuration-space cross-correlation function is no longer symmetric under the exchange of the line-of-sight positions in the two fields. This leads to an imaginary part in the cross-power spectrum $P_{i,j}(\k)$ which has further interesting applications in the context of LIM : see \cite{Zhou:2020hqh,Sato-Polito:2020qpc,Bonvin:2013ogt}.}. 
Provided the conditions of Eq.~\ref{eq:px_simple} are applicable, one can estimate the auto-power spectrum of any of the lines from this set of three cross-power spectra \cite{Beane:2018dzk}. For example, the auto-power spectrum of field $1$, $P_{1,1}(\k)$, may be obtained from:
\begin{equation}
    P_{1,1}(\k) = \frac{P_{1,3}(\k) P_{1,2}(\k)}{P_{2,3}(\k)} =
    \avg{I_1}^2 \avg{b_1}^2 P_{\rm lin}(\k).
    \label{eq:pauto_from_px}
\end{equation}
Suitable permutations of this relation may, of course, be used to derive the auto-power spectra of lines $2$ and $3$, while one can also imagine extensions to the case of more than three fields.

One special case of interest may be to derive the auto-power spectrum of the redshifted 21 cm fluctuations during, for example, the EoR. This case is notable given the strong foreground contamination expected for such measurements, which may bias 21 cm auto-power spectrum measurements if foreground avoidance/cleaning algorithms are imperfect. If two additional lines, tracing the large-scale distribution of the galaxies during the EoR (for example), may be measured as well, then one can potentially measure the 21 cm auto-power spectrum {\em without foreground bias} using Eq.~\ref{eq:pauto_from_px}. One main limitation of this approach is that Eq.~\ref{eq:pauto_from_px} only applies on large spatial scales, while surveys with sufficiently high spatial/spectral resolution measure higher $k$-modes more precisely (given the larger number of modes, $N_{\mathrm{modes}}(k)$ -- see Eq.~\ref{eq:nmode} -- that fit within high $k$-bins). Nevertheless, the three-field approach may be more robust than a conventional auto-power spectrum estimate at low $k$ and its range of applicability may be tested with simulated models (see \citep{Beane:2018dzk,Schaan:2021gzb,McBride:2023exl} for details).

Another potential application of the three-field approach is to help avoid line-interloper contamination bias (\S \ref{S:challenges}).  One approach here would be to infer the auto-power in the target line of interest from Eq.~\ref{eq:pauto_from_px} \cite{Beane:2018dzk}: this should be immune to interloper bias provided each line is widely separated from the target line in redshift. In addition, the interlopers in fields $1, 2, 3$ must themselves be separated in redshift and hence uncorrelated or weakly correlated.  In contrast to other cross-power spectrum approaches for avoiding interloper bias (e.g. \citep{LidzTaylor2016}), the three-field technique would allow one to infer the large-scale auto-power in the line of interest. That is, one can move past measuring only the overall product between the matter power spectrum and the product of the average intensities, luminosity-weighted bias factors in the two lines (which is what may be inferred from the line-line cross-power spectrum alone, e.g. Eq.~\ref{eq:px_simple}).  Furthermore, in principle, one might also be able to apply the three-field estimate at the redshifts {\em of the interlopers themselves}. To allow this, the survey would need to capture three separate emission lines at the redshift of the interloper in question (yet with differing observed frequencies). Then the auto-power spectrum for an interloper might be inferred using three cross-power spectra, as in Eq.~\ref{eq:pauto_from_px}. This appears to be a relatively clean approach for disentangling the contribution of known interlopers from the target line at the power spectrum level. In practice, however, this requires surveys which span a relatively broad range in frequency such that multiple lines can be measured at the redshifts of each important interloper. 

Alternatively, two of the fields might be traditional large-scale structure catalogs, as opposed to LIM data cubes. For example, suppose field $1$ is the line emission from a known interloper at redshift $z_1$. Then one can use a traditional galaxy survey and a quasar catalog at the same redshift, $z_1$, to infer the large-scale auto-power spectrum of the interloper emission at redshift $z_1$:
\begin{equation}
    P_{1,1}(\k) = \frac{P_{\mathrm{1,g}}(\k) P_{\mathrm{1,q}}(\k)}{P_{\mathrm{g,q}}(\k)} =
    \avg{I_1}^2 \avg{b_1}^2 P_{\rm lin}(\k),
    \label{eq:pint_auto_from_cross}
\end{equation}
where $P_{\mathrm{1,g}}$ and $P_{\mathrm{1,q}}$ are, respectively, the interloper line-galaxy and interloper line-quasar cross-correlations, while $P_{\mathrm{g,q}}$ is the galaxy-quasar cross-power spectrum.

\section{Cross-Correlation Potentials}
\label{S:xcorr}

One motivation for LIM surveys is that the resulting data cubes will enable a broad range
of cross-correlation studies. The cross-correlation measurements are less prone to systematic effects, such as foreground
contamination, than auto-correlation analyses. In addition, cross-correlations may help in extracting physical quantities of interest, as we will discuss.
Here we split our discussion into a post-reionization section and a subsequent section that focuses on the EoR. 

\subsection{Post-reionization Correlations}

It is notable that the first LIM detections have been achieved through cross-correlation measurements (see \S \ref{S:measurements}). This is the case because
cross-correlations partly circumvent the formidable challenges from foreground contamination that must be overcome for auto-spectrum measurements; it is therefore
easier to demonstrate the robustness of a cross-spectrum detection. The current cross-spectrum analyses are made possible by traditional galaxy and quasar
survey data sets, which are currently plentiful at $z \lesssim 3$ and may be fruitfully combined with LIM surveys.

Another potential set of applications involves using the redshift information from LIM surveys to extract the redshift distribution of photometrically-selected galaxy and quasar samples, and
the redshift dependence of diffuse radiation backgrounds, such as the thermal Sunyaev-Zel'dovich (tSZ) effect background or the CIB. While promising, this type of analysis needs to account for the loss of long-wavelength modes in the line-of-sight direction in LIM surveys owing to foreground contamination. Note that photometric galaxy surveys, and other two-dimensional data sets, {\em only} measure long-wavelength variations in the line-of-sight direction, i.e., precisely the modes one loses to foreground contamination in the LIM surveys. This limitation may be overcome, however, by turning to higher-point statistics \cite{Zhu:2015zlh}.

A further interesting direction is that existing cross-correlation analyses can help validate foreground cleaning algorithms and auto-spectrum measurements. 
As an explicit example, in the post-reionization universe the large-scale 21 cm
auto-power spectrum is proportional to the product of the mass-weighted neutral hydrogen bias with the cosmological mass density in neutral hydrogen and the power spectrum of density fluctuations, i.e., $P_{21,21}(k) \propto \left(\avg{b_{\mathrm{HI}}} \Omega_{\mathrm{HI}}\right)^2
P_{\delta,\delta}(k)$.  On the other hand, the cross-power spectrum with a traditional spectroscopic galaxy survey follows $P_{\mathrm{21,g}}(k)  \propto r \avg{b_{\mathrm{HI}}} \Omega_{\mathrm{HI}} \avg{b_{\mathrm{g}}} P_{\delta,\delta}(k)$. On large scales, the stochasticity parameter, $r \rightarrow 1$, while the galaxy bias may be inferred from the galaxy auto-spectrum. That is, both measurements should allow one to infer $\avg{b_{\mathrm{HI}}} \Omega_{\mathrm{HI}}$; this provides
a powerful consistency test. In fact, as discussed further in \S \ref{S:measurements} this type of consistency test 
was carried out for the first time in 21 cm by CHIME near $z \sim 1$ \cite{CHIME:2025cee}.
This provides a powerful demonstration of the efficacy of foreground cleaning algorithms. Note that the foreground-to-signal-strength ratio is broadly comparable here to that for the reionization-era 21 cm signal, and so this success also provides encouragement for reionization-era 21 cm efforts as well.    
That is, although the foreground emission from our own galaxy is brighter at the frequencies of interest for EoR-era 21 cm studies, the signal should also be larger during most of the EoR, and so the foreground-to-signal ratio is expected to be similar.

\subsection{Reionization-Era Cross-Correlations}

Although challenging to detect, the cross-correlation between a reionization-era redshifted 21 cm survey and an additional LIM data cube should prove valuable. First, as discussed in the Introduction, the cross-correlation can help confirm the high redshift origin of each signal. Second, the cross-correlation directly probes the interplay between the ionizing sources (as traced by the LIM data cube in the non-21 cm emission line) and neutral hydrogen in the intergalactic medium (as delineated by the redshifted 21 cm line). More specifically, the cross-correlation encodes information about the typical sizes of the ionized bubbles during reionization, which is otherwise challenging to extract from upcoming data sets \cite{Lidz:2008ry,Lidz11}.

\begin{figure}
\begin{center}
\includegraphics[width=\textwidth]{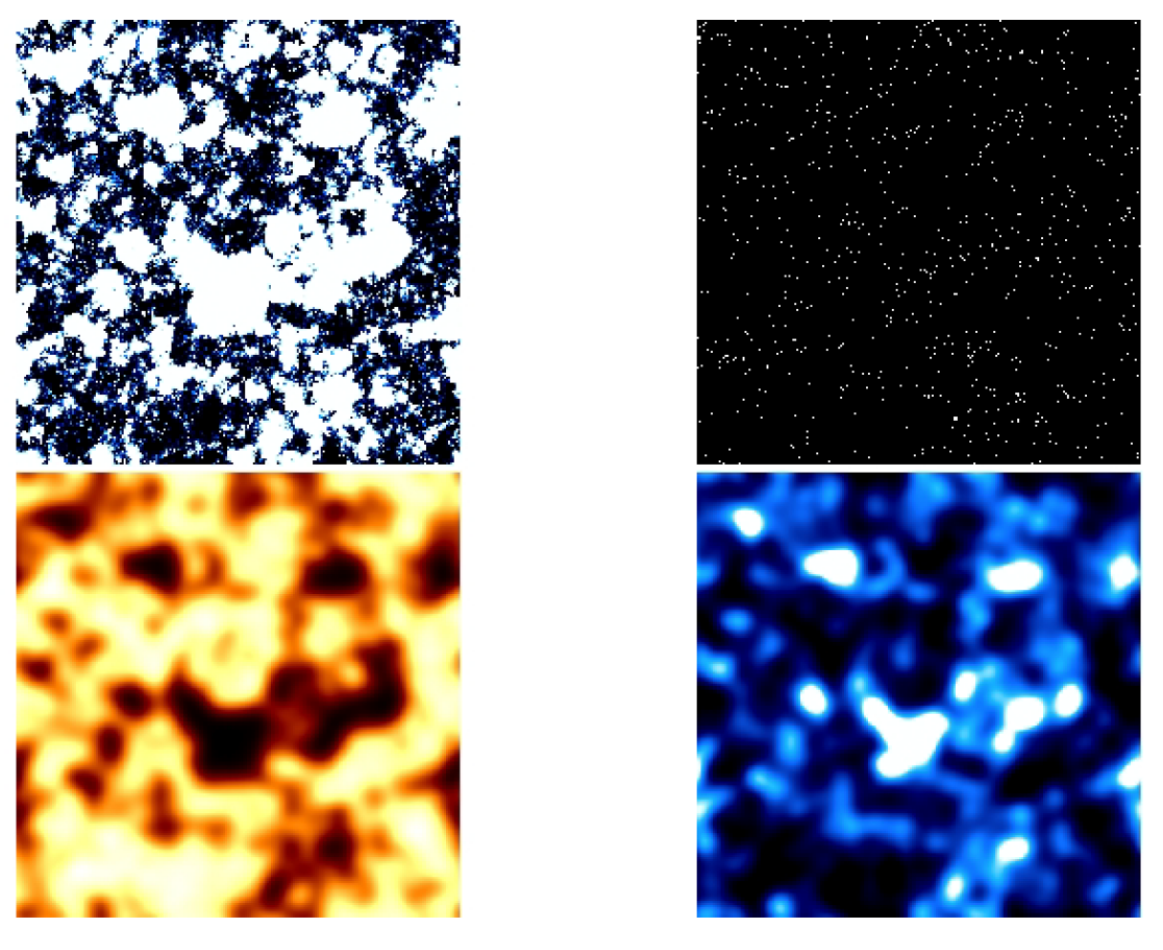}
\caption{Simulated CO and redshifted 21 cm maps. Top left: The ionization field from a numerical simulation of reionization. Top right: The galaxies in the same simulation.
Bottom left: The simulated smoothed redshifted 21 cm field. Bottom right: The smoothed CO map from the simulation. The comoving length of each side of the simulation box is $130$ Mpc/$h$. From \cite{Kovetz:2017agg,Lidz11}.}
\label{fig:co_21}
\end{center}
\end{figure}

Figure~\ref{fig:co_21} provides a qualitative impression of the expected 21 cm LIM signal during reionization at $z \sim 7$. This figure
shows slices through a simulated data cube with the ionization field and galaxy distribution, along with mock 21 cm and CO emission fields.  The smoothed CO emission traces the galaxy distribution, with peaks in the galaxy field matching up with bright regions in the CO emission field. Likewise, the large ionized regions in the ionization slice trace large-scale overdensities around which galaxy formation and reionization occur earlier than in typical parts of the universe. These regions hence correspond to ``holes'' (i.e., dim regions) in the 21 cm field since they lack neutral hydrogen. Therefore, the 21 cm and CO emission fields are anti-correlated on large scales, with the bright regions in the CO data cube aligning with dark regions in the 21 cm survey and vice versa. On small scales, however, the 21 cm and CO emission fields are uncorrelated. This occurs because all galaxies in the model ionize gas around them and the gas within the ionized regions is highly ionized irrespective of the precise galaxy overdensity.

\begin{figure}
\begin{center}
\includegraphics[width=\textwidth]{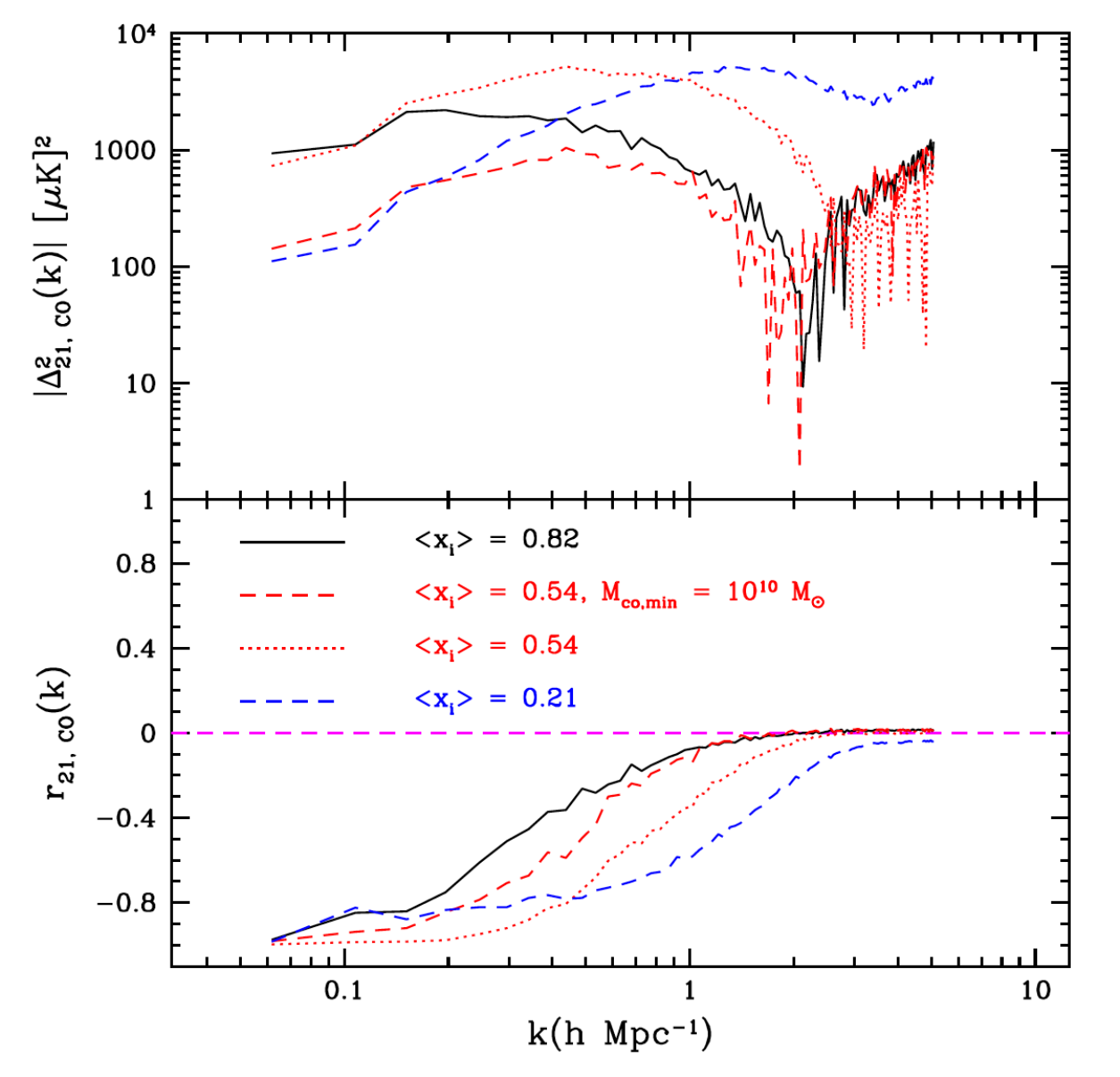}
\caption{Cross-correlation between CO and redshifted 21 cm fluctuations. The top panel shows the absolute value of the CO-21 cm cross-power spectrum. The red dashed line assumes that CO emitters reside in halos above $M_{\rm co, min} = 10^{10} M_\odot$, while the other curves include CO emitters in halos down to $M_{\rm co, min} = 10^8 M_\odot$ near the atomic cooling limit where $T_{\rm vir} = 10^4$ K. The bottom panels shows the cross-correlation coefficient between the two fields. On large scales the two fields are strongly anti-correlated with $r(k) \sim -1$, while the two fields are close to uncorrelated, $r(k) \sim 0$ on smaller scales. The transition scale from anti-correlated to uncorrelated moves to larger scales (lower $k$) as reionization proceeds, and the bubbles grow.  From \cite{Lidz11}.}
\label{fig:xcorr_co_21cm}
\end{center}
\end{figure}

These trends are quantified in Figure \ref{fig:xcorr_co_21cm}. The bottom panel shows the cross-correlation coefficient between the 21 cm and CO fields as a function of wavenumber, for a few different stages in the reionization process. That is, it shows $r(k) = P_x(k)/\left[P_{21}(k) P_{\mathrm{CO}}(k)\right]^{1/2}$, where $P_x(k)$ is the cross-power spectrum and $P_{21}(k)$ and $P_{\mathrm{CO}}(k)$ are, respectively,  the 21 cm and CO auto-power spectra. As reionization proceeds, a larger and larger fraction of the volume of the IGM enters the ionized phase and the ionized regions grow in size. The correlation coefficient between the 21 cm and CO fields turns over at smaller wavenumbers as the ionized regions grow. That is, if the 21 cm-CO cross-power spectrum can be measured, this provides a direct tracer of bubble growth during reionization. Similar general behavior is expected if emission lines other than CO are used (see e.g. \cite{Sun2022}). Here, CO(2-1) just provides one illustrative case, although other lines may be brighter at the redshifts of interest and prove superior for EoR cross-correlation measurements.

Note that the turnover scale depends on the size of the ionized regions around the particular dark matter halos that host CO emitters \cite{Lidz:2008ry,Lidz11,Sun2022}.  The sizes of the ionized regions generally depend on the collective impact of many individual galaxies, with larger ionized bubbles forming around large-scale overdensities where galaxy formation is accelerated. Statistically, larger ionized bubbles will thus tend to form around more biased emitters, and so the turnover scale in the cross-spectrum will depend somewhat on how biased the CO emitters are and hence on how they populate dark matter halos. For instance, the red dashed line in Figure~\ref{fig:xcorr_co_21cm} and red dotted lines compare the cross-power/cross-correlation coefficients in two different models for the CO emitters: at the same stage of reionization (near the mid-point when $\sim 50\%$ of the volume is ionized), the correlation coefficient turns over at lower $k$ when the CO emitters live in larger host halos. 

This also implies that the cross-power spectrum between 21 cm and another line-intensity map will depend partly on which particular emission line one correlates the 21 cm data with. For example, consider an emission line that is sourced primarily from galaxies in rather massive dark matter halos: these sources will be surrounded by larger bubbles than in other lines where the emitters reside in more typical host halos. Consequently, the cross-correlation coefficient (with the 21 cm field) will turn over on larger scales when using the line that traces the more massive host halos \cite{Lidz:2008ry,Sun2022,Kannan:2021ucy}, although in both cases the turn-over scale will still evolve strongly as reionization proceeds and the bubbles grow. 
Other interesting cases include the cross-correlations between 21 cm and narrow-band surveys for Ly-$\alpha$ emitters \cite{Lidz:2008ry,Hutter:2023rja,Hutter:2025phq}
or Ly-$\alpha$ line-intensity maps. These measurements will be affected by the scattering of Ly-$\alpha$ photons from neutral hydrogen atoms in the IGM. The combination of Ly-$\alpha$, H-$\alpha$ (which will not be scattered by intergalactic neutral hydrogen atoms, essentially all of which will lie in the ground state), 21 cm line-intensity maps, and their mutual cross-correlations, should provide a powerful probe of bubble sizes and the impact of Ly-$\alpha$ scattering \cite{Heneka21,Heneka:2016kss}. 

It is important to note that residual foregrounds may still prohibit cross-correlation detections. While only shared foregrounds between two lines lead to an average bias, any residual foreground contamination will still contribute to the {\em variance} of a cross-power spectrum estimate. Although this contrasts with the case of the auto-power spectrum, where residual foregrounds lead to a bias, strong foreground suppression may nevertheless be required to achieve reasonable signal-to-noise ratio detections of cross-power spectra. In general, for reionization-era LIM surveys to be useful counterparts to 21 cm surveys, rather wide-field surveys (spanning $\sim 100$s of square degrees on the sky), and high spectral resolution are required \citep{Fronenberg:2024olu}. Reference~\citep{Fronenberg:2024olu} develops end-to-end simulations for 21 cm $\times$ [CII] emission cross-correlations. As in the case of full auto-power spectra measurement pipelines, end-to-end cross-power simulations can help to more robustly forecast the signal-to-noise ratio of measurements and in optimizing future joint survey designs. Among other benefits, these simulations help to include residual foregrounds, instrumental effects, the impact of survey window functions, and to account for the error covariance between cross-power estimates in different $k$-bins.

Although cross-correlation measurements are not a ``magic bullet'' -- they do not fully evade the challenges of LIM observations -- we nevertheless expect them to play an important role in the field moving forward. They will help validate auto-power spectrum measurements and ultimately will help in extracting additional science.

\section{Science Cases and Forecasts}
\label{S:science}

Having described LIM science goals, modeling approaches, and statistical tools, we turn to consider how well upcoming surveys can achieve some of their key science objectives. We also discuss the overall science goals here in further detail.  
We caution here that LIM is a relatively young field, and so our understanding of: systematic effects, optimal survey design, approaches for scaling-up and boosting the sensitivity of current instruments, and signal strength are all evolving. Consequently, we expect measurement and parameter constraint forecasts to also be refined over the next several years as the field matures. In addition, different researchers adopt disparate assumptions for predicting LIM signals and varied levels of optimism and sophistication into their planned analyses, so it is challenging to compare forecasts or even to place the various science prospects on a common footing. It is nevertheless instructive to consider some of the current forecasts. We caution that in some cases these involve sizable extrapolations from what is currently feasible.

\subsection{Cosmology}

Although LCDM is arguably a good fit to all current cosmological data sets, many of the key ingredients of this model remain completely mysterious, including the nature of dark energy, the properties of dark matter, and the physics of the inflationary epoch or alternatives. Indeed, LCDM may turn out to be only an effective description of a deeper underlying theory with additional, likely unforeseen, ingredients. 
LIM may help in testing LCDM and perhaps in revealing cracks in its foundations, and this may offer guidance towards a deeper understanding of the laws of nature.  

Current LIM research at the interface with fundamental cosmology has focused on: using BAO features in the LIM power spectrum to constrain the expansion history and dark energy/modified gravity (\S \ref{S:bao_lim}),  primordial non-Gaussianity and inflation (\S \ref{S:png_lim}), dark matter properties (\S \ref{S:dm_lim}), and as a probe of neutrino mass and light relic abundances (\S \ref{S:nu_lim}). Line intensity data cubes may also provide source screens with known redshifts for gravitational lensing studies, which may be used to determine the projected mass distribution across a range of redshifts (\S \ref{S:lim_lensing}). 
On a slightly different theme, LIM signals act as foregrounds which must be cleaned in efforts to measure spectral distortions in the CMB and other effects (\S \ref{S:spectral_disto}). 
Here, we briefly discuss the motivation and prospects for these applications of LIM. 

\subsubsection{Expansion History of the Universe: Dark Energy and Modified Gravity}
\label{S:bao_lim}

\begin{figure}
    \begin{center}
    \includegraphics[width=\textwidth]{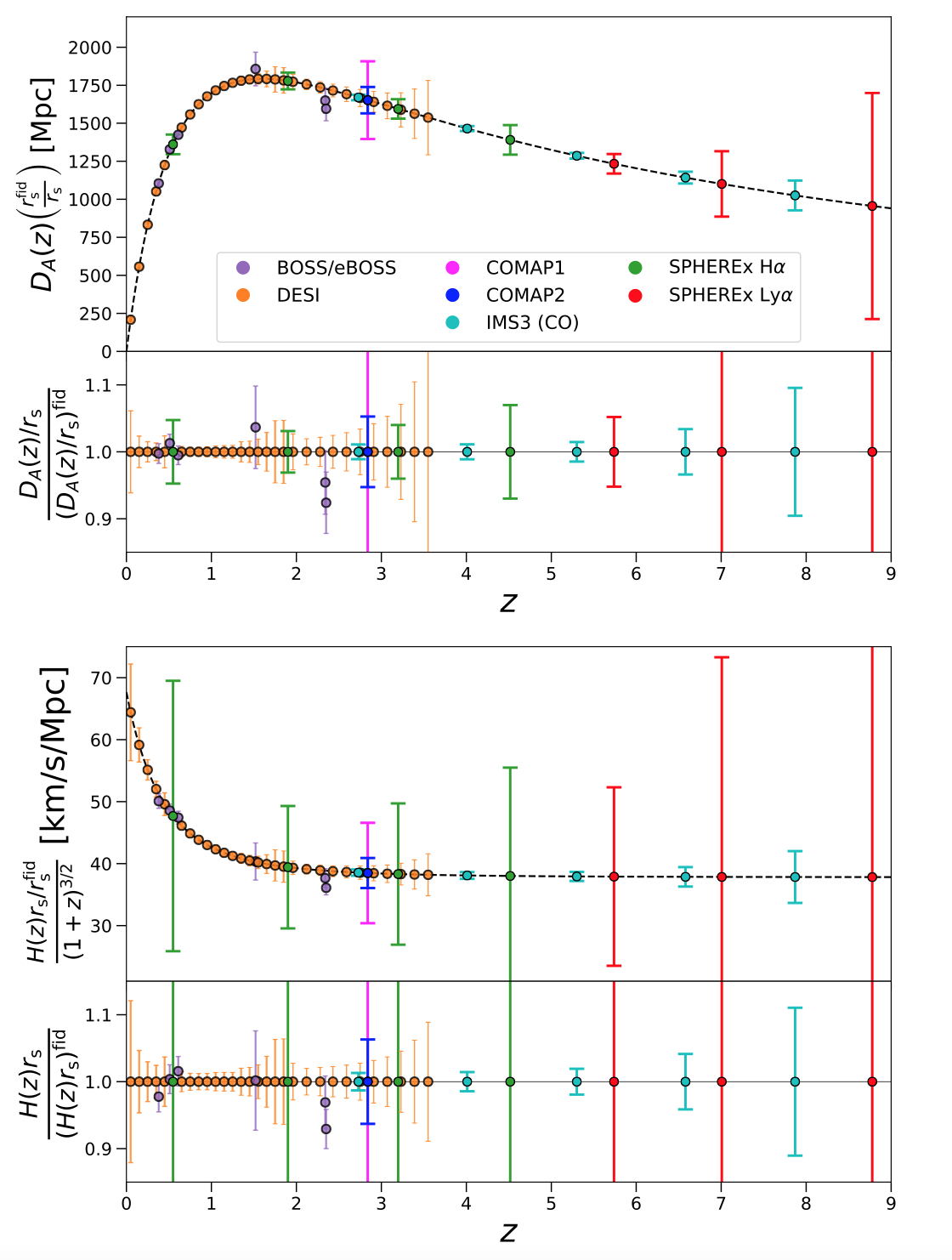}
    \caption{Constraint forecasts on the angular diameter distance ({\em top}) and the Hubble parameter ({\em bottom}), as a function of redshift, from upcoming LIM
    measurements. In each case the forecasted constraints come from fitting the BAO features in the line emission power spectrum. The dark purple points and error bars show current measurements using optically-selected galaxy samples from the BOSS/eBOSS surveys. The orange points and error bars show forecasts for DESI. The other points show expectations for future LIM surveys: COMAP1 (magenta), COMAP2 (blue), a futuristic third-generation CO survey (cyan), and SPHEREx using H-$\alpha$ (green) and Ly-$\alpha$ (red) lines. From \cite{Bernal:2019gfq}.}
    \label{fig:bao_lim_forecasts}
    \end{center}
\end{figure}

As mentioned in the Introduction, the BAO features from acoustic waves in the early universe provide standard rulers which may be measured economically using LIM surveys \cite{Chang:2007xk,Wyithe:2007rq}. For example, the CHIME project aims to use post-reionization 21 cm LIM to measure the BAO features from $z \sim 0.8-2.5$. In an idealized, foreground and systematics-free limiting case, CHIME is forecast to measure a spherically-averaged distance measure, $D_{\mathrm{V}}$, to percent-level precision in each of $\sim 15$ redshift bins across this redshift range \cite{CHIME:2022dwe}. This will enable measurements of the expansion history at and around the redshift where dark energy apparently starts to dominate the overall energy budget of the universe. These measurements will probe the universe at higher redshift than possible with current optically-selected galaxy surveys, and overlap with quasar and Ly-$\alpha$ forest measurements \cite{CHIME:2022dwe}.

It may also be possible to probe the BAO features using additional emission lines \cite{Bernal:2019gfq}, which will allow expansion history measurements all the way out to $z \sim 9$. One particular motivation for this, as emphasized in \cite{Bernal:2019gfq}, is the apparent discrepancy between the value of the Hubble parameter, $H_0$, determined from local distance-ladder measurements and that inferred from CMB anisotropies. This is referred to in the literature as the ``Hubble tension'' (see, e.g., \cite{Kamionkowski:2022pkx} for a recent review): it may point to unaccounted-for systematic errors in the local and/or CMB measurements or it could indicate a breakdown of the LCDM model. Furthermore, recent DESI results may provide a hint for dynamical dark energy \cite{DESI:2025zgx}, although see reference~\cite{Efstathiou:2025tie} for a different point of view. 
In any case, LIM BAO measurements may help to fill-in constraints on the expansion history at intermediate redshifts -- the epoch between $z \sim 3 - 1,100$ is essentially terra incognita -- and this might provide clues to help resolve the Hubble discrepancy and/or further test evolving dark energy scenarios. 

Reference \cite{Bernal:2019gfq} forecasts the constraints on the Hubble parameter and the angular diameter distance versus redshift for upcoming and futuristic CO LIM surveys, as well as the prospects for H-$\alpha$ and Ly-$\alpha$ line-intensity maps from SPHEREx. The ratio between the line emission signal and continuum foreground contamination is expected to be more favorable in these lines than in 21 cm (see \S \ref{S:challenges}), and so this may be an important advantage of surveys in CO, H-$\alpha$, and Ly-$\alpha$. The line-intensity power spectrum, as a function of line-of-sight and transverse wavenumber, $P_I(k_\parallel,k_\perp)$, depends on the sound horizon scale, $r_{\mathrm{s}}$. That is, on the distance acoustic waves travel between the big bang and recombination. The line-of-sight fluctuation power also depends on the mapping between the redshift extent of the sound horizon and its comoving length scale, and hence on $H(z)$. Finally, the relationship between the observable angular extent of the sound horizon and its comoving size depends on the angular diameter distance, $D_{\mathrm{A}}(z)$. Hence the forecasts of \cite{Bernal:2019gfq} in Fig.~\ref{fig:bao_lim_forecasts} consider $D_{\mathrm{A}}(z)/r_{\mathrm{s}}$ and $H(z) r_{\mathrm{s}}$ relative to some fiducial cosmological model. These results marginalize over other uncertain quantities for each line, including suitable products of the specific intensity in each line, their luminosity-weighted bias factors, the amplitude of density fluctuations (e.g. parameterized by $\sigma_8$), and $f_\Omega$ (see Eq.~\ref{eq:pk_rspace_twoh}), as well as the level of finger-of-god smoothing and the shot-noise term. These forecasts show that LIM may significantly extend the reach of the expansion history measurements. In addition, observations of the 21 cm signal during Cosmic Dawn may offer related constraints out to $z \sim 15-20$ via the VAO features mentioned in the Introduction (see \cite{Munoz:2019fkt} for details). 
LIM will hence further help in testing models of dark energy and modified gravity, each of which may make distinctive predictions for the expansion history and related distance measures. These issues are further elucidated in the literature in \cite{Karkare:2018sar}, which forecasts LIM constraints on early dark energy, and \cite{Scott:2022fev}, which considers future LIM bounds on modified gravity models, {and \cite{MoradinezhadDizgah:2023src} which explores both early dark energy and modified gravity scenarios.}

\subsubsection{Primordial Non-Gaussianity and Inflation}
\label{S:png_lim}

In the simplest models of inflation, the perturbations responsible for cosmological structure follow a Gaussian distribution. However, more complex models of inflation, such as those involving multiple scalar fields, may give rise to departures from Gaussianity, i.e. to primordial non-Gaussianity (PNG), as might alternatives to inflation. Hence, PNG may provide a further handle on the physics governing inflation or alternatives. 
At leading order, PNG may be characterized by the primordial bispectrum of perturbations, parameterized by an amplitude, $f_{\rm NL}$, and a shape\footnote{More precisely, in the case of local type non-Gaussianity, $f_{\rm NL}$ is defined so that the Newtonian potential is given by $\Phi({\bf x}) = \phi_{\rm G}({\bf x}) + f_{\rm NL} \left[\phi_{\rm G}^2({\bf x}) - \avg{\phi_{\rm G}^2({\bf x})}\right]$. Here, $\phi_{\rm G}({\bf x})$ is a Gaussian random field. Note that the fractional level of non-Gaussianity is $\sim 10^{-5} \, |f_{\rm NL}|$.}. For example, in so-called ``local-type'' models of PNG, the primordial bispectrum is enhanced for squeezed triangle configurations where one wavenumber is much smaller than the other two.  On the other hand, for ``equilateral'' type PNG, the bispectrum is peaked for equilateral triangle configurations. Physically, multi-field inflation may give rise to local-type PNG, while single-field inflation with higher derivative interactions generates equilateral-type PNG (e.g. \cite{Green:2022bre} and references therein.) 

An important signature of PNG is that it gives rise to a distinctive scale-dependent biasing of dark matter halos. This may be detectable, or constrained, through measuring the power spectrum of biased tracers of the matter distribution \cite{Dalal:2007cu}.
In the case of local-type PNG (with positive $f_{\rm NL}$), the small-scale variance of the density field is enhanced in large-scale regions with above average gravitational potential, due to the coupling between short and long wavelength modes. This excess variance boosts the abundance of dark matter halos in local regions with large gravitational potential: owing to the mode-coupling from PNG, the halo abundance is modulated by the fluctuations in the gravitational potential. 
Let us denote the primordial gravitational potential fluctuation, the matter density field, and the matter transfer function in Fourier space by $\Phi(k)$, $\delta_\rho(k)$, and $T(k)$, respectively. Using Poisson's equation, $k^2 T(k) \Phi(k) \propto \delta_\rho(k)$, and the extra scale-dependent biasing varies with wavenumber as
$\propto 1/(k^2 T(k))$ \cite{Dalal:2007cu}. That is, the primordial three-point function generates an interesting and distinctive signature in the two-point function of biased tracers of the matter distribution. 

In the case of local-type non-Gaussianity, the current constraints from Planck 2018 CMB data are: $f_{\rm NL} = -0.9 \pm 5.1$ \cite{Planck:2019kim}. This bound partly motivates the target for future large-scale structure surveys, which seek to measure $f_{\rm NL}$ with a statistical precision of $\sigma(f_{\rm NL}) \sim \mathcal{O}(1)$. This is in principle achievable with future large-scale structure surveys, although reaching this target demands exquisite control over systematic effects. 

Alternatively, this scale-dependent biasing signature may be observable using LIM measurements of the power spectrum of line emission fluctuations \cite{MoradinezhadDizgah:2018lac}. LIM offers the potential to economically survey large volumes of the universe with full three-dimensional information.
In the case of the scale-dependent biasing signature, reference \cite{MoradinezhadDizgah:2018lac} finds that futuristic single-tracer surveys for CO or [CII] may in principle reach the ambitious target of $\sigma(f_{\rm NL}) \sim 1$ in the case of local-type non-Gaussianity. The multi-tracer method \cite{Seljak09} should also help (see \S \ref{sec:multi_tracer}), as the relative scale-dependent bias of two tracers may be measured without cosmic variance, while the line-intensity bispectrum should add further constraining power as well. One of the main challenges for detecting the scale-dependent biasing signature using LIM surveys is that these data sets will lose low-$k$ Fourier modes owing to continuum foreground avoidance/removal \cite{Lidz:2013tra,MoradinezhadDizgah:2018lac}, yet the large-scale modes are the ones most impacted by PNG. 
The degradation from foreground contamination may, however, vary considerably depending on the emission line considered, especially with the continuum foreground to line emission ratio (see \S \ref{S:challenges}). The prospects here will be clarified as first-generation LIM surveys return results, as this will help to better understand the role of foreground contamination, and strategies for mitigating it.

\subsubsection{Dark Matter Properties}
\label{S:dm_lim}

The 21 cm line has long been recognized as a potentially valuable probe of dark matter properties. First, it may be employed to extract information regarding the thermal and ionization history of the universe from the Cosmic Dark Ages through to the Cosmic Dawn and the EoR \cite{Furlanetto:2006jb}. These, in turn, may reveal or constrain modifications to the thermal and ionization states of the universe from decaying or annihilating dark matter (e.g. \cite{Furlanetto:2006wp}). Second, the 21 cm line studies may show signatures of dark matter models in which the power spectrum of initial density fluctuations is suppressed/enhanced on small scales (which are hard to probe by other means), or strongly constrain such scenarios (e.g \cite{Sitwell:2013fpa,Lidz:2018fqo,Munoz:2019hjh,Jones:2021mrs}). For example, in warm dark matter (WDM) scenarios free-streaming will suppress the small-scale power spectrum. Alternatively, in the case of fuzzy dark matter (FDM) \cite{Hu:2000ke,Hui:2016ltb} the dark matter particles are ultralight bosons (with masses approximately in the range of
$m_{\mathrm{FDM}} \sim 10^{-22} - 10^{-18}$ eV leading to astrophysical signatures). In FDM, the de Broglie wavelengths of the dark matter particles are large enough to be astrophysically relevant. Among other consequences of these enormous de Broglie wavelengths, the FDM particles will resist confinement in small-mass dark matter halos. In WDM, FDM, and related alternative models, the resulting suppression of small-mass halos leads to a delay in galaxy formation and the timing of Cosmic Dawn and the EoR \cite{Barkana:2001gr}. These effects will also modify the spatial variations in the Cosmic Dawn and reionization processes \cite{Jones:2021mrs}. The timing and fluctuations in the Cosmic Dawn/EoR signals should be measured 
by upcoming 21 cm surveys, and so these observations may provide invaluable information regarding dark matter properties.

\begin{figure}
    \begin{center}
    \includegraphics[width=\textwidth]{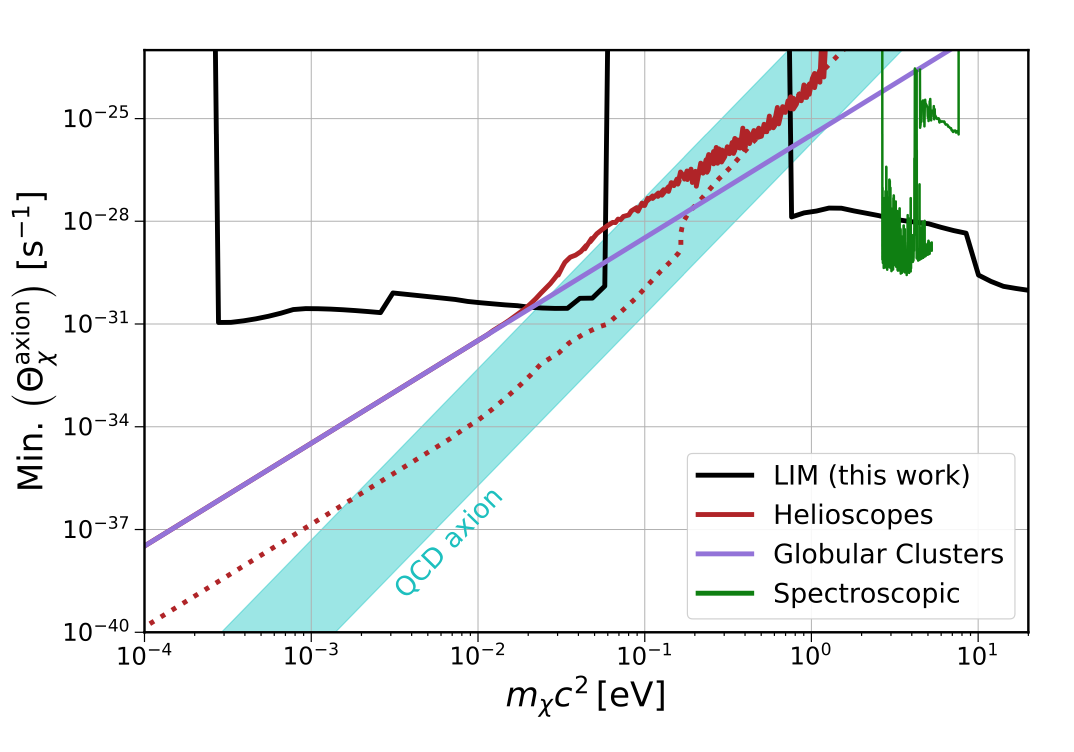}
    \caption{Forecasted constraints on axion models with LIM surveys. Future measurements of the line-intensity power spectrum and VID may be used to constrain the possibility that dark matter particles decay into two photons, (or into a photon and another daughter particle). The y-axis gives the (suitably scaled, see \cite{Bernal:2020lkd} for details) decay time of the dark matter particles. The cyan shaded region shows the expected parameter space for QCD axions, including motivated extensions to the canonical QCD axion parameter space. LIM surveys potentially offer a stronger handle on axions with mass $m_\chi c^2 \sim 1-20$ eV than other planned techniques.  From \cite{Bernal:2020lkd}.}
    \label{fig:axion_lim}
    \end{center}
\end{figure}

LIM surveys using additional lines may also enable further tests of dark matter properties. For example, reference \cite{Libanore:2022ntl} studies the CO LIM VID (see \S \ref{sec:pdf}) and shows that this statistic is indirectly sensitive to the high-$k$ linear matter power spectrum at $k \gtrsim 0.5$ Mpc$^{-1}$, which is relatively hard to probe by other means. These are scales that may show departures from CDM if the dark matter is instead WDM, FDM, or another interesting alternative candidate. In this context, the CO emission VID is sensitive to the high-$k$ matter power spectrum because the CO emitters may lie in relatively small-mass dark matter halos, while the small-mass halo abundance and clustering bias depend on the shape and amplitude of the linear matter power spectrum at relatively high wavenumber. 

Another promising direction is that if dark matter particles decay, then a LIM survey may be able to directly detect the electromagnetic radiation emitted \cite{Creque-Sarbinowski:2018ebl,Bernal:2020lkd,Gong:2015hke}. 
This is complementary to the effects mentioned previously where the 21 cm line is used to probe the impact of dark matter on the surrounding IGM. For example, consider the case that a dark matter particle with rest-mass energy $m_\chi c^2$ decays into two photons, each with a rest-frame frequency of $\nu = m_\chi c^2/(2 h)$. The main idea proposed in \cite{Bernal:2020lkd} is that this radiation would appear as unexpected interloper line contamination in future LIM surveys targeting conventional emission lines. In this way, dark matter decay lines might be detected or constrained in LIM surveys designed for other purposes. 
Suppose that the LIM survey targets an atomic or molecular emission line at a rest-frame frequency of $\nu_r$ and a redshift $z_t$: this line will be observed at $\nu_{\rm obs} = \nu_r/(1+z_t)$. The same observed frequency might also include emission from dark matter decaying at redshift $z_{\rm dm}$, such that $\nu_{\rm obs} = \nu/(1+z_{\rm dm})$, where $\nu=m_\chi c^2/(2 h)$ is the rest-frame frequency of the decay emission. The dark matter decay emission may potentially be isolated using techniques suggested in the literature for separating interloper contamination (see \S \ref{S:challenges}).
For example, assuming the redshift of the target line to map between frequencies/angles and comoving distances, the dark matter decay contribution to the power spectrum will appear anisotropic. The power spectrum anisotropy can then be used to identify interlopers. In this respect, the dark matter decay emission behaves like an interloper line, yet at an unexpected rest-frame frequency, generally distinct from the frequencies of standard atomic and molecular emission lines. The dark matter decay photons are also expected to make a distinctive contribution to the VID \cite{Bernal:2020lkd}.

Using this approach, the forecasted sensitivity from reference \cite{Bernal:2020lkd} to general axion-like particles for various LIM surveys is shown in Figure~\ref{fig:axion_lim}. The forecasts here consider a broad range of current and planned LIM surveys, spanning a number of different lines including 21 cm, CO, [CII], Ly-$\alpha$, and H-$\alpha$. Taken together, the wide frequency coverage of these surveys should allow
axion decay signature searches across a broad parameter space in axion mass and decay lifetime (into two photons). At the upper frequency end, \cite{Bernal:2020lkd} confine their forecasts to $m_\chi c^2 \lesssim 20$ eV, since more massive particles will produce decay photons that will interact with neutral hydrogen atoms (e.g. 10.2 eV photons will be absorbed in the Ly-$\alpha$ resonance). An especially notable finding, illustrated in Figure~\ref{fig:axion_lim}, is that SPHEREx and HETDEX should provide state-of-the-art sensitivity to axion decays in the $m_\chi c^2 \sim 1-20$ eV mass range. This overlaps partially with some motivated extensions of the QCD axion parameter space (cyan band in the figure), raising the possibility that LIM observations could detect or constrain the QCD axion\footnote{For context, it helpful to consider the case that QCD axions make up all of the dark matter and are produced by the misalignment mechanism, without fine-tuning. In this case, provided Peccei-Quinn symmetry breaking occurs before inflation, the expected axion masses are $m_\chi c^2 \sim 10^{-6}-10^{-5}$ eV \cite{Marsh:2015xka}, although lower masses are possible if the misalignment angle is fine-tuned.  
This range lies outside of the LIM detection region in Figure~\ref{fig:axion_lim}. In the case of Peccei-Quinn symmetry breaking after inflation and again supposing that QCD axions make up the dark matter, the expected mass is $m_\chi c^2 \sim 10^{-6}-10^{-4}$ eV, which still lands beyond the LIM forecast in Figure~\ref{fig:axion_lim}. There are also, however, other possible production mechanisms -- i.e., alternatives to the misalignment mechanism --  for QCD axions \cite{Marsh:2015xka}.} or constrain the properties of this hypothetical particle.

Related to this, recent studies have reported a possible excess in the Cosmic Optical Background (COB) intensity relative to expectations from deep galaxy counts~\cite{Lauer:2022fgc,Symons:2022lke}. These analyses used data from the \emph{New Horizons} spacecraft, which is located nearly 60~AU from the Sun. At this distance, zodiacal light foregrounds from sunlight scattered by interplanetary dust are significantly reduced, allowing a more direct measurement of the COB at wavelengths around $\lambda \sim 0.6\,\mu\mathrm{m}$. The \emph{New Horizons} measurements reported by~\cite{Lauer:2022fgc,Symons:2022lke} suggest that the COB intensity is roughly a factor of two higher than expected based on deep \emph{HST} galaxy counts. This discrepancy motivated the proposal that axion-like particle decays might contribute to the excess COB photons~\cite{Bernal:2022wsu}. That study identified a viable region of parameter space in which dark matter decays could explain the apparent COB excess and emphasized that future LIM surveys, such as SPHEREx and HETDEX, could provide decisive tests of this scenario.  However, a more recent analysis by~\cite{Postman:2024erl} revisits the contribution from diffuse Galactic light (DGL)---starlight in the Milky Way scattered by interstellar dust---finding that it was previously underestimated in~\cite{Lauer:2022fgc}. 
After updating the DGL model,~\cite{Postman:2024erl} suggest that the revised COB intensity is consistent with the level inferred from deep galaxy counts, although they note that a small anomalous component may still lie within the measurement uncertainties. These results appear to differ, however, with the conclusions of~\cite{Symons:2022lke}. Future LIM observations will be valuable in further constraining the origin of the COB and assessing the presence of any residual excess.

\subsubsection{Neutrino Mass and Light Relic Abundance}
\label{S:nu_lim}

Some key goals of upcoming large-scale structure and next-generation CMB surveys include determining the sum of the neutrino masses, and measuring the energy density in relativistic particles that were present in the early universe (``radiation''). The latter quantity is generally parameterized by $N_{\rm eff}$, the effective number of neutrino species in the early universe, although any particles moving relativistically contribute to the energy density in radiation and $N_{\rm eff}$, regardless of whether they are neutrinos. Indeed, many extensions to the standard model of particle physics include additional light relic particles which may be detectable indirectly through cosmological observations. 
In detail, $N_{\rm eff}$ has several interesting effects with observable consequences in cosmology \cite{Dvorkin:2022jyg}. First, the contribution of light relics to the energy density increases the expansion rate of the universe and this is a significant effect during radiation-dominated phases in the early universe. Holding other parameters fixed, this changes the redshift of matter-radiation equality (and hence the horizon size at matter-radiation equality), the sound horizon, and the Silk damping (photon diffusion) scale.
Second, perturbations in the light-relic abundance may influence the growth of fluctuations in the photon, baryon, and dark matter density distributions via gravitational interactions. Specifically, before recombination, perturbations in the photon-baryon fluid travel at the sound speed in the plasma, $c_{\rm s} \sim c/\sqrt{3}$, while free-streaming light relics travel at nearly the speed of light. That is, the acoustic waves lag behind the light relic perturbations.
The gravitational influence of the supersonic light relic perturbations then leads to a phase shift in the position of the BAOs \cite{Bashinsky:2003tk,Baumann:2019keh}, which may be measurable in the LIM power spectrum \cite{MoradinezhadDizgah:2021upg}. The analogous phase shift of the CMB acoustic peaks from $N_{\rm eff}$ has been measured at high significance \cite{Follin:2015hya}, but three-dimensional large-scale structure probes -- including LIM -- may help tighten constraints \cite{Baumann:2017gkg}. The phase shift predictions may be modified if the light relics possess self-interactions or other interactions: in such cases, the perturbations in the light-relic density will not travel faster than the sound speed \cite{Dvorkin:2022jyg}. This provides a test of whether the light relics are truly free-streaming.

Neutrino masses impact the matter power spectrum because neutrinos contribute (a sub-dominant) part of the mass density once they become non-relativistic, while their random thermal velocities wipe out small-scale fluctuations in their density distribution. More specifically, there is a neutrino-mass-dependent free-streaming wavenumber $k_{\rm fs}$: at scales of $k << k_{\rm fs}$ neutrinos cluster just like CDM, while on small-scales $k >> k_{\rm fs}$ the neutrinos do not cluster. The larger the sum of the neutrino masses, the more they contribute to the matter density and this reduces the amount of small-scale clustering. There are
actually two effects here: on small-scales only the CDM (and not the neutrinos) clusters, and secondly, the growth of small-scale fluctuations in the CDM is itself slowed owing to the presence of neutrinos. Together, a handy rough estimate is that the matter power spectrum at $k >> k_{\rm fs}$ is suppressed by roughly a factor of $\delta P/P \sim (1 - 8 f_\nu)$ -- relative to the case of zero neutrino mass -- where $f_\nu = \Omega_\nu/\Omega_{\mathrm{m}}$ is the fraction of the mass density in neutrinos \cite{Hu:1997mj}. 

LIM may offer a handle on the sum of the neutrino masses through future measurements of the power spectrum of line emission fluctuations. Again, LIM may offer an economical approach for accessing large volumes of the universe across broad ranges in redshift. These surveys may potentially measure many Fourier modes and return precise power spectrum measurements, which may help in determining the sum of the neutrino masses \cite{MoradinezhadDizgah:2021upg}. The main challenges for using LIM to probe neutrino mass relate to the uncertain bias factors and average specific intensities of the line emission, and their redshift evolution. In addition, accurate and precise models for the scale-dependent biasing of the line emission are required. 
See reference \cite{MoradinezhadDizgah:2021upg} for a discussion of the quantitative prospects and challenges here. 

\subsubsection{LIM Lensing}
\label{S:lim_lensing}

As mentioned in the Introduction, LIM maps provide source screens of known distance which can be used for lensing analyses. A full discussion of LIM lensing is beyond the scope of this review, but we mention a few pertinent issues and cite some key reference papers here.
There are a number of different ways in which LIM data may be used directly for lensing analyses or as an aid to other lensing efforts. One approach is to adapt and extend the quadratic estimator formalism used in CMB lensing analyses to the case of three-dimensional LIM data sets \cite{Hu:2001kj,Pen:2003yv,Zahn:2005ap,Foreman:2018gnv}. Other possible LIM lensing analyses include: measuring the correlated shapes between individually detected/resolved line-emitting regions as a function of their separation \cite{Pen:2003yv}; and using the lensing magnification-induced correlations between foreground LIM data and background galaxy samples \cite{Witzemann20};  while traditional cosmic shear maps estimated from galaxy surveys may also be correlated with LIM data \cite{Chung:2022lpr}. 

Let us briefly discuss the quadratic estimator formalism \cite{Dodelson17}. The CMB photons, as well as the photons measured in LIM surveys, undergo deflections as they propagate through the large-scale structure on their way to the observer. This results in a remapping between the
observed (lensed) and original (unlensed) angular coordinates of the radiation intensity field. The remapping here is related to the projected gravitational potential. Consider then the correlation between the intensity fluctuations observed in two different directions, ${\bf \hat{n}_1}$ and ${\bf \hat{n}_2}$. The lensing effect implies that these correlations vary with the actual location on the sky, depending on the projected gravitational potential traversed, as well as with the angular separation between the two directions.
In Fourier space, this breaking of translation invariance leads to distinctive correlations between different Fourier modes. 
The standard quadratic estimator formalism exploits these distinctive correlations to reconstruct the projected gravitational potential fluctuations. LIM data can be handled analogously to the CMB lensing case \cite{Pen:2003yv,Zahn:2005ap,Foreman:2018gnv}, while providing
many more source screens with well-known distances to help tomographically reconstruct the gravitational potential fluctuations \cite{Fronenberg:2023juh,Fronenberg:2023qtw}. A challenge for the LIM lensing case, however, is that non-linear gravitational evolution itself induces Fourier mode correlations that must be separated from the lensing effects \cite{Foreman:2018gnv}. (This is in contrast to the CMB case where the source field is a Gaussian random field.) See reference~\cite{Schaan:2018yeh} for investigations of related non-Gaussian effects in the context of the CIB. 
In any case, future mass maps from LIM lensing studies promise to be invaluable for understanding the cosmic expansion history and the growth of large-scale structure, and for detecting or bounding neutrino masses. 

Another lensing-related application of LIM surveys is to CMB-delensing \cite{Sigurdson:2005cp,Karkare:2019qla}. 
Among the current goals for CMB cosmology is to measure, or at least bound, the presence of primordial curl-type $B$-modes in the polarization of the CMB \cite{Kamionkowski:1996ks,Seljak:1996gy}. The $B$-mode polarization signal may be generated by gravitational waves produced during an inflationary epoch: a successful $B$-mode polarization measurement would help confirm the inflationary paradigm and pin-down the energy scale of inflation. One challenge for such measurements is that gravitational lensing deflections convert polarization $E$-modes into $B$-modes. It is possible, however, to reconstruct the intervening lensing potential and use this to remove the lensed $B$-modes in a procedure known as ``delensing'' \cite{Smith:2010gu}. LIM surveys may help here by tracing the matter fluctuations across a broad range of redshifts, and allowing cross-correlation measurements with the reconstructed lensing potential (from the CMB itself or from LIM data).  The advantage of LIM here is that it may allow fluctuation measurements out to higher redshift than feasible with other large-scale structure tracers, and help span the broad CMB lensing kernel, thereby improving the fidelity of CMB delensing efforts \cite{Karkare:2019qla}. It may be important, however, to devise analysis approaches here that circumvent the loss of long wavelength line-of-sight modes in the LIM data owing to foreground cleaning \cite{Moodley:2023lmu}. On the other hand, recent work shows that mode-coupling induced by redshift evolution along the past light cone in the LIM data should allow
direct LIM $\times$ lensing cross-correlations even after filtering foreground corrupted low $k_\parallel$ modes from the LIM measurements \cite{Shen25}. 
Analogous light-cone evolutionary effects may also allow direct cross-correlations between foreground-filtered LIM data and other two-dimensional tracers as well, e.g., photometric galaxy surveys, kSZ, tSZ, and others, in addition to CMB lensing.

\subsubsection{Line Emission Signals as CMB Foregrounds} 
\label{S:spectral_disto}

One key goal for future CMB surveys -- currently in the planning/proposal stage -- is to measure spectral distortions, i.e. departures of the sky-averaged CMB radiation from a blackbody spectral shape \cite{Chluba:2019nxa,Kogut:2019vqh}\footnote{Some spectral distortions, such as the thermal Sunyaev-Zel'dovich effect \cite{Sunyaev1970} from hot electrons in clusters, groups, galaxies, and the circumgalactic/intergalactic medium will vary spatially, while distortions generated at early times, e.g. before recombination, are expected to have negligible spatial fluctuations.}. Although measurements from COBE/FIRAS placed stringent upper limits on spectral distortions \cite{Fixsen96}, expected and plausible processes will inevitably drive matter and radiation out of perfect thermal equilibrium and produce small deviations in the CMB from a blackbody spectrum. One prominent example, expected within the standard model of cosmology, is the damping of acoustic waves in the photon-baryon plasma at redshifts between $5 \times 10^4 < z < 2 \times 10^6$. At these times, photon-generating processes are inefficient, and the energy injection from the damped acoustic waves leads to a Bose-Einstein distribution for the radiation field, characterized by a chemical potential of $\mu \sim 2 \times 10^{-8}$ \cite{Chluba:2019nxa}\footnote{Note that at $z \geq 2 \times 10^6$ the photons thermalize efficiently and any initial spectral distortion is removed.}. The precise level of this distortion turns out to probe the integrated primordial power spectrum on scales of 
$k \sim 10^2-10^4 \, {\rm Mpc}^{-1}$ \cite{Hu:1994bz,Chluba:2019nxa}. This hence offers a valuable handle on small-scale fluctuations (in the linear regime), which are inaccessible with other techniques. At lower redshifts, $z \lesssim 5 \times 10^4$,
Compton scattering becomes inefficient at maintaining equilibrium between electrons (which efficiently share their energy with surrounding baryonic matter through collisions) and photons. In this case, energy injection into the electron distribution produces a so-called $y$-distortion. This distortion is characterized by the Compton $y$-parameter which quantifies the fractional energy boost to CMB photons after scattering off of hot free electrons along the line-of-sight. 
The expected level of this distortion is $y \sim 2 \times 10^{-6}$ \cite{Hill:2015tqa}, mainly sourced by hot gas in clusters and groups at relatively low redshifts, only about an order of magnitude below current bounds. In addition to these expected signals and others, more exotic energy injection processes, such as from decaying or annihilating dark matter, may also generate spectral distortions.

These prospects have motivated a number of proposals for future high spectral resolution CMB surveys, which can improve on the statistical precision of spectral distortion measurements by about a factor of $10^3$ \cite{Chluba:2019nxa,Kogut:2019vqh}. Foreground contamination provides a challenge, however, for extracting the subtle spectral distortion signatures expected in the standard model of cosmology and beyond. In particular, one will need to separate both spectrally-smooth foregrounds from synchrotron radiation, free-free emission, and emission from dust grains, and also the contributions from line emission in various transitions from gas at a wide range of redshifts. The line emission presents a particularly daunting challenge because its specific intensity will not be described by a simple power law in frequency \cite{Mashian:2016bry,Serra:2016jzs,Chung:2023ncd}. LIM surveys may help pin down models for the extragalactic contributions to the line emission foregrounds here and thereby assist efforts to measure small CMB spectral distortions. Specifically, a good understanding of the sky-averaged brightness temperature from multiple CO transitions, as well as [CII], [NII], and other lines -- with each line contributing to a range of observed frequencies through gas emitting at various redshifts -- will be required to robustly extract the primordial $\mu$-distortion signal. 
In turn, the spectral distortion surveys themselves may also be employed for LIM investigations, especially through exploiting cross-correlations with other surveys \cite{Chluba:2019nxa,Switzer:2017kkz}.

As mentioned in the Introduction, LIM fluctuations may also be an important contaminant for some efforts to measure CMB secondary anisotropies. The most important example here may be the case of the kSZ effect \cite{Sunyaev1970}, which can be extracted from multi-frequency measurements of the CMB angular power spectrum. The kSZ effect arises as CMB photons scatter off of free electrons participating in bulk flows: the CMB photons receive a redshift or blueshift depending on the sign of the line-of-sight peculiar velocity field. This leads to secondary CMB anisotropies, with a ``patchy reionization'' contribution from spatial fluctuations in the ionization fraction during reionization \cite{Gruzinov:1998un} and a post-reionization contribution sourced by density/velocity inhomogeneities. A challenge for detecting this signal at the power spectrum level is to separate it from other sources of anisotropy, including radio sources, the CIB from dusty star-forming galaxies, the tSZ effect, and others. The kSZ effect preserves a blackbody spectrum while these other contributions differ spectrally, yet their angular power spectra are much larger in amplitude than the kSZ signal and so need to be carefully separated. Line-intensity emission fluctuations, which are themselves correlated with the CIB and tSZ signals, become an important contaminant for efforts to extract the kSZ power spectrum. Specifically, a recent study finds that the CIB-CO cross-power spectrum is comparable in amplitude to the kSZ auto-power spectrum and is hence an important contaminant that must be accounted for \cite{Maniyar:2023cuj}. 
Those authors further discuss how CO fluctuations may also contaminate various cross-correlations between CMB maps and other tracers of large-scale structure.

\subsection{Astrophysics}

Although LIM was first conceived as an efficient approach for probing large-scale structure without resolving individual galaxies, LIM can nevertheless offer unique insights into the aggregate properties of the emitting galaxies themselves. Here we discuss the prospects for providing a census
of the intensity of spectral line emission as a function of redshift (\S \ref{S:census}), for constraining luminosity functions (\S \ref{S:LF}), for
inferring the cosmic star formation rate density (\S \ref{S:SFRD}), for measuring the history of cosmic reionization and the growth of ionized bubbles (\S \ref{S:reion}),
and for tracking the abundance of molecular gas across cosmic time (\S \ref{s:molecular}). 

\subsubsection{Census of Light}
\label{S:census}

As LIM is sensitive to all sources of emission in the atomic or molecular transitions of interest, it can be used to measure the mean luminosity density (that is, the emissivity, see Eq.~\ref{eq:avg_inu}) as a function of redshift. This, in turn, informs our understanding of the chemical enrichment history of the universe, the census of baryons across multiple gas phases, and the overall cosmic energy inventory, in much the same spirit as the classic ``Cosmic Energy Inventory'' of \cite{Fukugita2004}. Ultimately, fully exploiting this approach will require measurements from a diverse set of lines that trace both atomic and molecular gas, across a broad variety of ionization states and gas densities. Ideally, one will measure the fluctuations over a wide range of spatial scales and observing frequencies, providing a complete census of emission line activity across cosmic history with line emission measurements complementing observations of the radiation backgrounds produced by continuum processes \cite{Hill:2018trh}.

In this context, it is important to first note that the average luminosity density is difficult to measure via traditional galaxy surveys owing to their limited survey depth and area. 
LIM surveys, however, can extract the luminosity density via measurements
of the power spectrum of line-intensity fluctuations. Specifically, the two-halo term is proportional to the average specific intensity squared (see e.g. Eq.~\ref{eq:ptwoh_grand} and/or the simplified version of Eq.~\ref{eq:pofk}).
The mean specific intensity is, in turn, related to the average luminosity density (denoted there by $\epsilon_L$) via Eq.~\ref{eq:avg_inu}. The degeneracy between the clustering bias and the average intensity can be broken by measuring the angular dependence of the line-intensity fluctuation power spectrum (see \S \ref{S:refinements} and Eq.~\ref{eq:pk_rspace_twoh}, Eq.~\ref{eq:ptwoh_grand}). Alternatively, the line luminosity function can be extracted using an integral constraint from LIM power spectrum measurements across a broad range of wavenumbers, as discussed in \S \ref{S:LF} below. Although degeneracies will remain between the clustering bias and the luminosity density, accessing the LIM signal in both the clustering and shot-noise dominated regimes helps. The shot-noise signal depends on the second moment of the luminosity function, while the luminosity density is determined by the first moment and is traced in the clustering regime (see Eq.~\ref{eq:blum} - \ref{eq:poss}, Eq.~\ref{eq:pk_est}). The one-point PDF provides another potential handle (\S \ref{sec:pdf}). 

\begin{figure}
    \begin{center}
    \includegraphics[width=\textwidth]{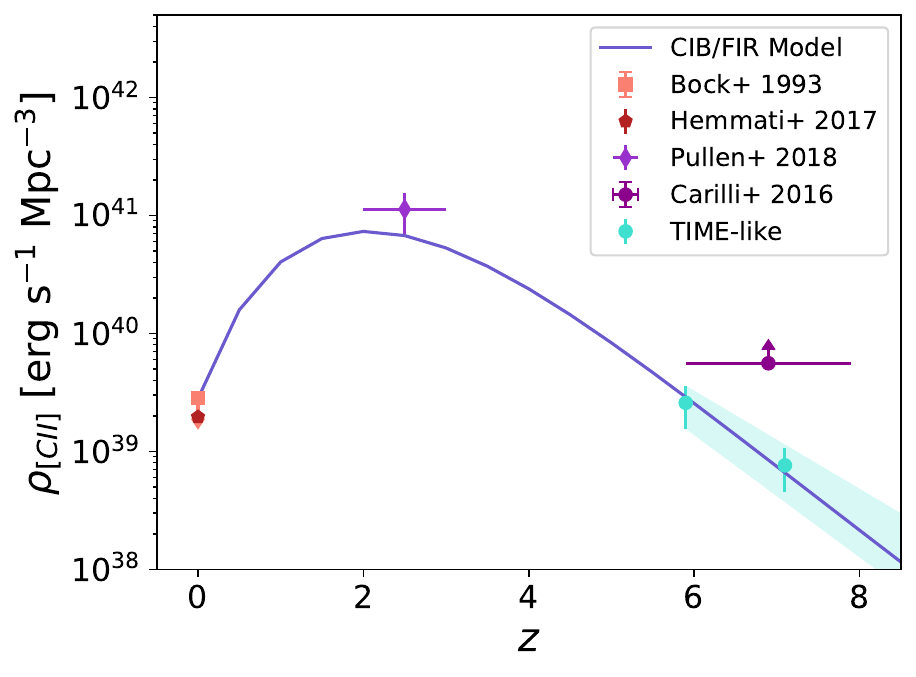}
    \caption{The [CII] luminosity density evolution inferred from several
    observations in the current literature, compared to forecasts for a LIM survey. The CIB/FIR theory curve is based on the model of \cite{Sun19},
    while the data points are discussed in the text. The cyan band shows
    forecasts for a TIME-like [CII] LIM survey. From \cite{Sun19}. 
    }
    \label{fig:luminosity_density}
    \end{center}
\end{figure}

As an example, Figure~\ref{fig:luminosity_density} shows a model for the redshift evolution of the [CII] luminosity density compared to current observations in the literature, as well as forecasts for future LIM measurements. The evolution spans from the local universe, through cosmic noon, and into the EoR. The local measurements include \cite{Hemmati2017} based on Herschel observations of more than 500 nearby galaxies, and an upper limit from a sounding rocket measurement \cite{Bock1993} towards the Lockman Hole. The $z=2.6$ data point from \cite{Pullen:2017ogs,Yang:2019eoj} is estimated using Planck high-frequency data sets cross-correlated with SDSS quasars and CMASS galaxies 
(see \S \ref{S:measurements}).
The $z \sim 7$ data point comes from \cite{Carilli2016} and is based on a blind ALMA survey near 242 GHz for [CII] emitters. In this case, the limited survey depth may result in incompleteness, and there may also be contamination from CO rotational transitions at lower redshifts. Finally, sample variance in the limited sky region covered by the blind ALMA survey may be another concern. 
This estimate does not, therefore, provide a strict lower limit on the [CII] emission signal. The cyan shaded band illustrates the reach of a nominal TIME-like experiment, showcasing the potential constraining power of LIM \cite{Sun19}. Ultimately, related measurements can be performed for a range of different line tracers.

\subsubsection{Spectral Line Luminosity Functions}
\label{S:LF}

As discussed above and in \S \ref{S:modeling} the line-intensity power spectrum is determined largely by the clustering and shot-noise contributions (e.g. Eq.~\ref{eq:pofk}). The mean specific intensity is proportional to the line luminosity density, and hence the first moment of the spectral line luminosity function (Eq.~\ref{eq:avg_inu}). The shot-noise contribution is set by the second moment of the line luminosity function (Eq.~\ref{eq:poss}). The clustering and shot-noise components hence provide integral constraints on the luminosity function, with the shot-noise piece receiving most of its weight from the bright end of the luminosity function. As discussed in \S \ref{sec:pdf}, the PDF or VID may offer a further handle on the line luminosity functions.

For example, reference~\cite{Sun:2020mco} provides [CII] line luminosity function forecasts for future LIM power spectrum measurements. 
Specifically, those authors parameterize the correlation between [CII] and UV luminosity, including some scatter around the
mean trend, and consider the impact of various star formation models on the resulting UV and [CII] luminosity functions. They then give forecasts for
future TIME measurements, and extensions, near $z \sim 6$ (see their Figure 8 for the detailed results here). Although the LIM constraints are indirect, they
allow luminosity function estimates at high redshifts where the requisite survey depth and sky coverage are hard to achieve. Further, the LIM measurements help determine the role of faint-galaxy populations, as we have emphasized throughout this review. 

\subsubsection{Cosmic Star Formation History}
\label{S:SFRD}

As mentioned in the Introduction, LIM surveys may help deduce the cosmic star formation history, commonly characterized by the redshift evolution of the star formation rate density (SFRD) (i.e., the average star formation rate per unit comoving volume). As discussed in \S \ref{S:landscape}, a range of different emission line tracers may provide handles on the SFRD. The SFRD tracers
include Balmer lines, which are produced during recombination cascades in HII regions, offering a relatively direct tracer of ionizing photon production. The ionizing photon production rate can, in turn, be linked to star formation rate (see e.g., Eq.~\ref{eq:q_sfr}), subject to some uncertainties discussed in \S \ref{S:rest_opt_uv_hii}. An important issue here is to account for dust attenuation, which is accomplished most cleanly through measurements of multiple Balmer lines at a given redshift.
[CII] emission provides another promising SFRD tracer. This line can be used to detect emission from dusty galaxies, which are obscured in the rest-frame ultraviolet and optical. The challenge for using [CII] as a SFRD tracer is that it can generally arise from multiple ISM phases, and the connection with star formation may hence be complex (\S \ref{sec:cii_line}). 

In the context of the model discussed in the previous section, the SFRD can nevertheless be derived from the [CII] luminosity density. Specifically, the assumed correlation between [CII] and UV continuum luminosity allows one to translate between the [CII] and UV luminosity densities, which can be further connected to the SFRD. 
For example, these connections may be used to fit parameterized models that describe the star formation rate as a function of halo mass and redshift,
${\rm SFR}(M,z)$. The SFRD then follows as:
\begin{equation}
    \dot{\rho_*}(z) = \int_{M_{\rm min}}^{\infty} dM \, n(M) {\rm SFR}(M,z), 
\end{equation}
where $M_{\rm min}$ is the smallest mass halo hosting star formation, which is often set to $M_{\rm min} = 10^8 M_{\odot}$, corresponding roughly to the atomic cooling threshold mass \cite{Barkana:2000fd}. 
A key goal of current and forthcoming [CII] experiments is to determine the SFRD
at high redshift \cite{Sun:2020mco,CONCERTO2020,TIM2020}. 

\begin{figure}
    \begin{center}
    \includegraphics[width=\textwidth]{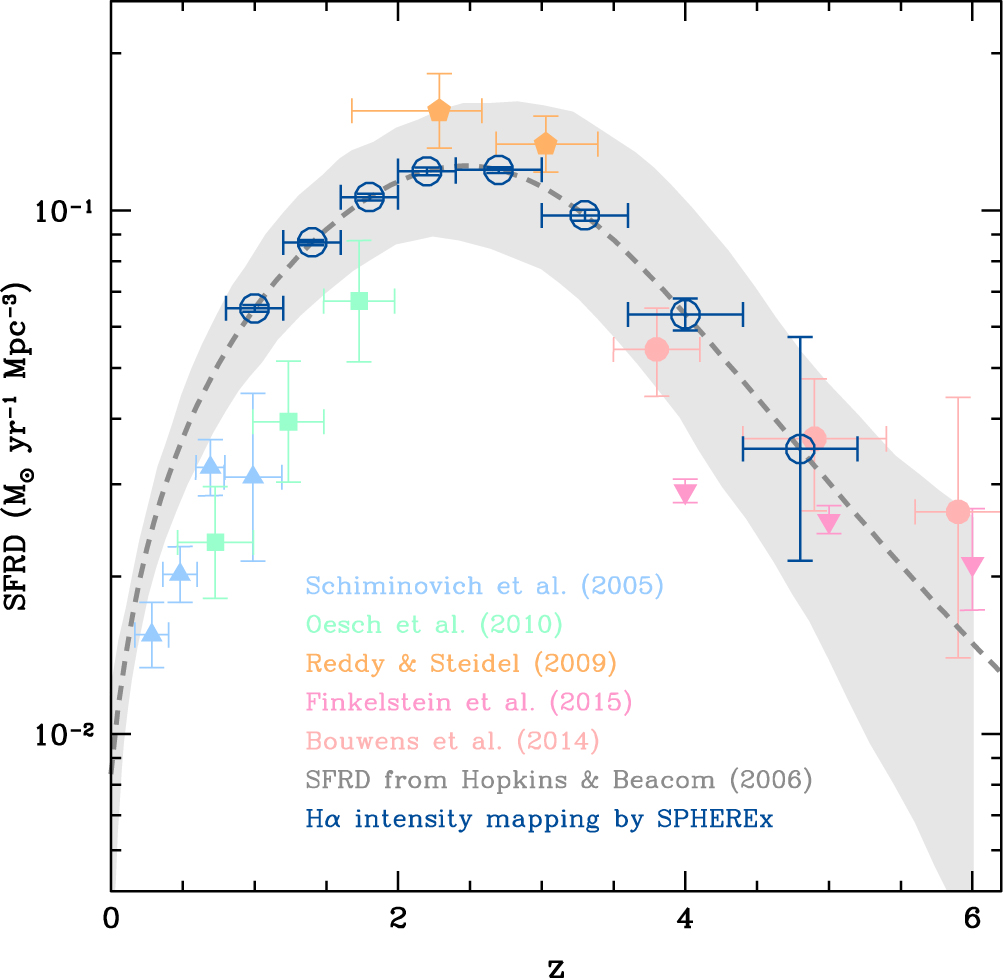}
    \caption{
Forecasted constraints on the cosmic star formation history from upcoming SPHEREx
H-$\alpha$ LIM measurements. The dashed lines and and shaded band give measurements and uncertainties from the compilation of \cite{Hopkins:2006bw}, while the other points and error bars are complied from other sources in the literature, as indicated in the legend. The blue points and error bars show the SPHEREx forecasts for H-$\alpha$ LIM, which promise fairly stringent limits from a complementary approach. From \cite{Gong17}.
    }
    \label{fig:sfrd_spherex}
    \end{center}
\end{figure}

Similarly, reference \cite{Gong17} forecasts the SFRD constraints which may be obtained using H-$\alpha$ LIM measurements from the SPHEREx mission. These authors assume a linear relationship between H-$\alpha$ line luminosity and SFR (see also \S \ref{S:rest_opt_uv_hii}), and then link the H-$\alpha$ luminosity density to the SFRD. 
The SFRD forecast for SPHEREx H-$\alpha$ LIM measurements is shown in Figure~\ref{fig:sfrd_spherex}. The forecasts suggest that tight constraints
on the SFRD will be possible from $0.5 \lesssim z \lesssim 4$ and that looser
bounds will be placed out to $z \sim 5$.

\subsubsection{Reionization History and Bubble Growth}
\label{S:reion}

As discussed in the Introduction and previous sections, LIM may potentially play a key role in understanding cosmic reionization. One aspect of this is that LIM can trace the distribution of the star-forming galaxies that reionize the universe across large spatial scales, and complement JWST measurements, which are confined to small fields-of-view. The process of reionization involves the growth and percolation of ionized bubbles, which can reach tens of comoving Mpc in length scale, and so measurements spanning large regions of the sky are required to obtain representative samples. 
As emphasized earlier, LIM surveys in multiple lines can trace different phases of the ISM, CGM, and IGM to help reveal the physical processes driving reionization.

As outlined in \S \ref{S:xcorr}, joint LIM surveys in H-$\alpha$, Ly-$\alpha$ and the 21 cm line will provide a comprehensive view of the spatial structure and evolution of the reionization process. These can be further combined with additional spectral lines, including [OII], [OIII], [CII], and CO transitions, to trace out the properties of the ionizing sources, the nature of their interstellar media, and the chemical enrichment history of the universe during reionization. Furthermore, cross-correlations with 21 cm can help extract information regarding bubble sizes (see \S \ref{S:xcorr}), \cite{Lidz11,Gong11,Heneka:2016kss,Dumitru2019,Heneka21,Sun:2020mco,Sun2022}.

\begin{figure}
    \begin{center}
    \includegraphics[width=\textwidth]{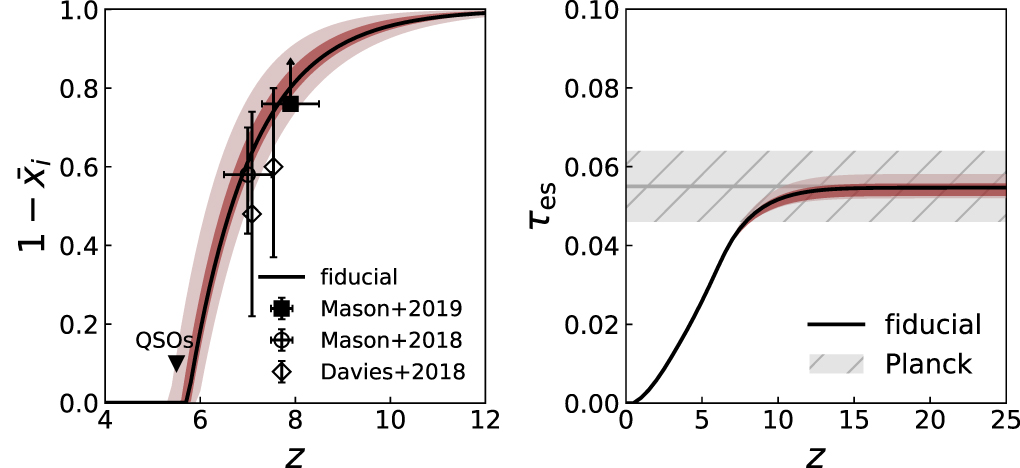}
    \caption{
    Forecasted constraints on the reionization history and the Thomson-scattering optical depth from [CII] LIM surveys. The [CII] LIM measurements will extract information regarding the redshift evolution of the ionizing sources, which will help determine the reionization history and the optical depth to CMB photons. These inferences require models to connect the [CII] measurements with the ionizing sources and the reionization history. The forecasts for TIME are shown in light red, while results for an extension to TIME, TIME-EXT, are shown in dark red. The points with error bars in the left hand panel show some current reionization history measurements in the literature, while the grey band in the right hand panel shows CMB optical depth measurements from Planck. 
    From \cite{Sun:2020mco}.}
    \label{fig:TIME_reionization}
    \end{center}
\end{figure}

In terms of the reionization history itself, measurements of the redshift evolution of the luminosity density in lines tracing star formation can be used
to inform models for the ionizing sources and help deduce the average ionization history. For example, reference \cite{Sun:2020mco} forecasts the reionization history constraints possible from future TIME measurements of the [CII] luminosity density. In a commonly adopted model, the filling factor of the ionized regions, $Q_{\mathrm{HII}}$, evolves as \cite{Shapiro87,Madau99,Barkana:2000fd,Sun:2020mco}:
\begin{equation}
    \frac{dQ_{\mathrm{HII}}}{dz} = \zeta \frac{df_{\mathrm{coll}}}{dz} + \frac{C(z) \alpha_B(T_e)}{H(z)}(1+z)^2 \bar{n}_H^0 Q_{\mathrm{HII}},
\label{eq:dqdz}
\end{equation}
where $\bar{n}_H^0$ is the mean comoving number density of hydrogen, $C(z)=\avg{n_e^2}/\avg{n_e}^2$ is the IGM clumping factor, and $\alpha_B(T_e)$ is the case-B recombination coefficient. The first term supposes that the rate of ionizing photon production is proportional to the time derivative of the halo collapse fraction (for halos above the atomic cooling mass threshold), while the second term accounts for recombinations. The rate of ionizing photon production can be
related to the SFRD, $\dot{\rho_*}(z)$, which follows from the [CII] luminosity density as discussed in the previous section \cite{Sun:2020mco}: 
\begin{equation}
    \zeta \frac{df_{\mathrm{coll}}}{dz} = \frac{A_{\mathrm{He}}f_{\mathrm{esc}}f_{\gamma}\Omega_{\mathrm{m}}}{\bar{\rho}\Omega_{\mathrm{b}}} \times \dot{\rho_*}(z) \times \frac{dt}{dz}.
\end{equation}
Here $\zeta$ describes the ionizing efficiency which itself is a product of the star formation efficiency $f_*$, the escape fraction of ionizing photons $f_{\mathrm{esc}}$, the average number of ionizing photons produced per stellar baryon $f_\gamma = 4000$, and a correction factor $A_{\mathrm{He}} = 4/(4-3Y_{\mathrm{He}}) = 1.22$ for the presence of helium, i.e., $\zeta = A_{\mathrm{He}} f_* f_{\mathrm{esc}} f_\gamma$. 
Therefore, one can use the SFRD inferences to deduce the reionization history after making plausible assumptions regarding: the recombination rate in the IGM (set by
the clumping factor in Eq.~\ref{eq:dqdz} and the gas temperature), $f_{\mathrm{esc}}$, and $f_\gamma$. In general, one also needs to extrapolate the SFRD beyond the redshift range of the LIM
surveys. These calculations also account for a likely minor correction from free electrons outside of HII regions, produced by X-ray photoionizations (not discussed here explicitly; see \cite{Sun:2020mco} for details). 

The resulting forecasts for TIME [CII] measurements of the average neutral fraction, denoted here by $1-\bar{x_i}$, are shown (as a function of redshift) in the left panel of Figure~\ref{fig:TIME_reionization}. Some representative current measurements from the literature are shown for contrast. The improved knowledge of the SFRD from future [CII] LIM measurements can help narrow the plausible reionization history constraints considerably. Likewise, the Thomson-scattering optical depth can be estimated from the inferred reionization history (right panel), and this can potentially sharpen current/future CMB-based constraints. Although these inferences are model dependent, as they require linking the SFRD to the [CII] luminosity density, and knowledge of $C$, $f_{\mathrm{esc}}$, $f_\gamma$, the future wealth of LIM survey data should offer opportunities to cross-check some of these assumptions.

\subsection{Cosmic Dawn and Pop-III stars}

LIM may also enable probes of the first generation of stars, known as Population-III (Pop-III) stars, which may be key drivers of the Cosmic Dawn era (e.g, \cite{Barkana:2000fd,Bromm2009}). These stars formed from metal-free gas and are thought to be massive, with typical stellar masses of up to several hundred solar masses or more \cite{Abel2002,Bromm2002}, although the IMF remains highly uncertain.

Reference \cite{Visbal:2015sca} suggests using He II 1640 $\rm{\AA}$ LIM as a diagnostic for Pop-III dominated galaxies, since this line helps trace the hardness of the stellar ionizing spectrum (see \S \ref{s:balmer}). Pop-III stars are expected to have high masses and surface temperatures, and should therefore emit copious numbers of photons capable of doubly-ionizing helium. This will, in turn, lead to He II 1640 $\rm{\AA}$ emission from the resulting recombination cascades. This contrasts with Pop-II stars, which are mostly incapable of doubly-ionized helium due to their softer ionizing spectra. AGN contributions can potentially be distinguished from their broad lines, or X-ray measurements \cite{Visbal:2015sca}, and may be less important at early Pop-III dominated eras \cite{Ogata26}, depending on the history of supermassive black hole growth. 

\begin{figure}
\begin{center}
\includegraphics[width=\textwidth]{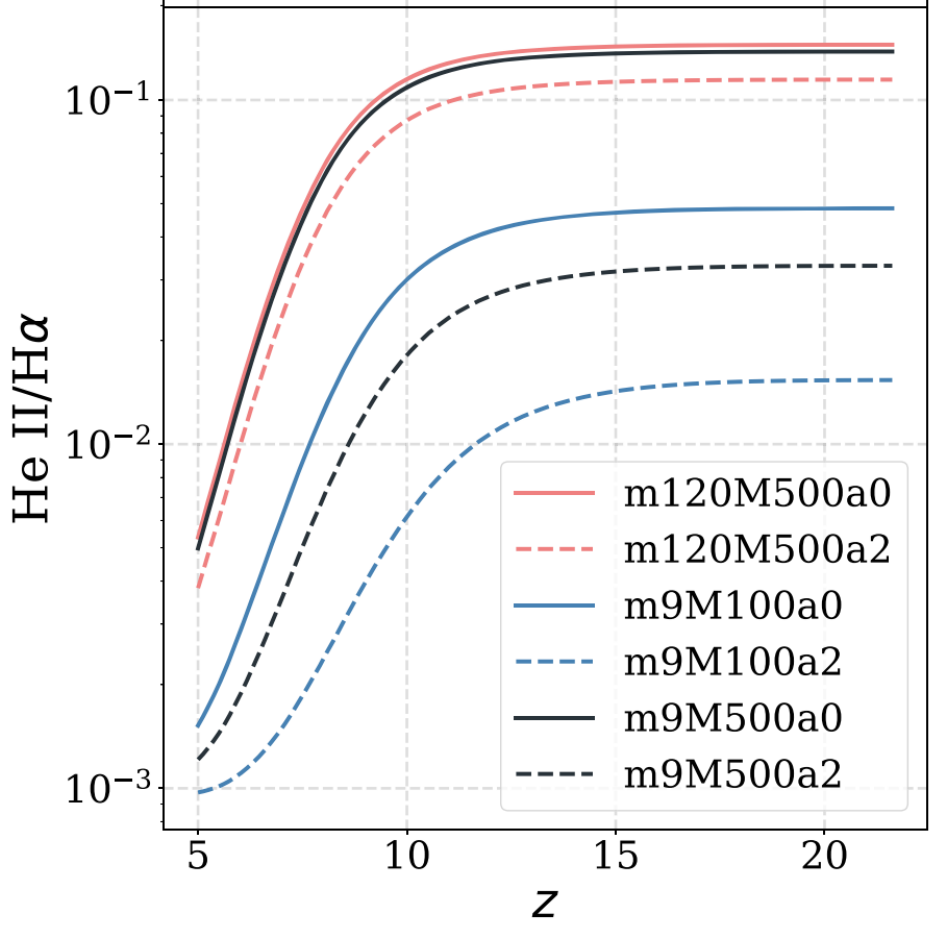}
\caption{Line-intensity ratio of He II/H-$\alpha$ inferred from LIM measurements,
as a function of redshift for six Pop-III IMF models (different curves). The decline in these ratios towards low redshift, at $z \lesssim 12$ in the models shown here, indicate the demise of Pop-III star formation and the emergence of Pop-II stars. The normalization of the line ratios and the redshift evolution are sensitive to the IMF model (see the original work for details). From \cite{Parsons:2021qyw}.
\label{fig:heIIha}}
\end{center}
\end{figure}

Reference \cite{Parsons:2021qyw} further demonstrates the potential for LIM to constrain the Pop-III IMF via the
average HeII 1640 $\rm{\AA}$/H-$\alpha$ line-intensity ratio, inferred from their respective LIM power spectrum amplitudes.  The Pop-III IMF controls the hardness of the ionizing spectrum, and the resulting line-intensity ratio.
The authors use LIMFAST \cite{Mas-Ribas2022} (discussed in \S \ref{S:signals}) to estimate the line emission from Pop-II and Pop-III dominated galaxies at $5 \leq z \leq 20$. Figure~\ref{fig:heIIha} shows the simulated line ratios, as a function of redshift, for different Pop-III IMFs. They find three important regimes in the line ratio:
(1) ratios of He II/H$\alpha \gtrsim 0.1$ indicate top-heavy Pop-III IMFs with typical stellar masses of several hundred solar masses, which occur at $z \gtrsim 12$ when Pop-III stars dominate star formation. (2) Ratios of He II/H$\alpha \lesssim 0.01$ indicate low levels of Pop-III star formation and signal the end of the Pop-III dominated star formation era. (3) Finally, ratio values of $0.01 \lesssim$ He II/H$\alpha \lesssim 0.1$ are complex to interpret because of several competing effects. In this regime, the line ratio may be shaped by different upper-mass bounds or slopes for the Pop-III IMF, by Pop-II star formation contributions, by variations in the escape fraction of ionizing photons, or by the star formation duty cycle, among other effects. Nevertheless, the He II/H-$\alpha$ ratio inferred from LIM measurements is a promising tracer of the elusive Pop-III IMF.

Other possible tracers of Pop-III stars and Cosmic Dawn include H2 and HD lines (see \S \ref{sec:other_lines}). These measurements can potentially be combined with 21 cm observations from the same era.

\subsubsection{Molecular Gas Abundance}
\label{s:molecular}

\begin{figure}
\begin{center}
\includegraphics[width=\textwidth]{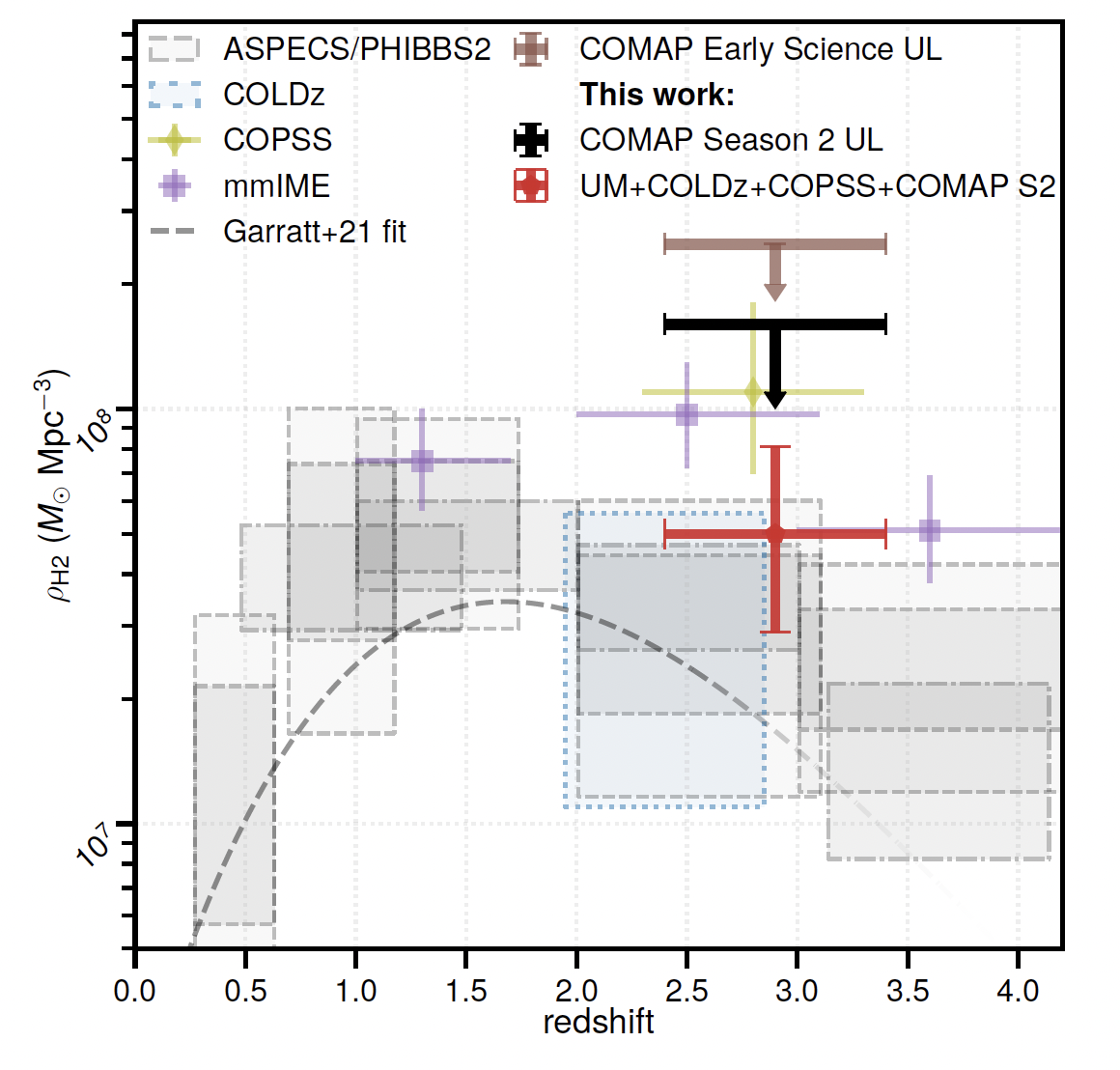}
\caption{{Current constraints on molecular gas density as a function of redshift. The CO LIM-derived constraints on the density of molecular hydrogen,
 $\rho_{\mathrm{H_2}}$, from COMAP Season 2 data are shown by the thick black downward pointing arrow, the COMAP early science bound is given in brown, along with those from COPSS \cite{Keating16} (yellow) and mmIME \cite{Keating:2020wlx} (purple). The red point and error bars show the $\rho_{\mathrm{H_2}}$ inferred from COMAP Season 2 data, after including informative priors based on \textsc{UniverseMachine} models of the galaxy-halo connection \cite{Behroozi2019} and COLDz data \cite{Riechers19}. All error bars and bounds are $1-\sigma$ confidence levels.  
 These LIM bounds are compared with CO-based measurements from ASPECS \cite{Decarli2020}, PHIBBS2 \cite{Lenkic2020}, and COLDz \cite{Riechers19}. Finally, the dashed line is from the analysis of \cite{Garratt21}, which is based on stacked 850 $\mu$m measurements and an assumed correlation with CO(1-0) luminosity. 
 All results assume $\alpha_{\mathrm{CO}} = 3.6 \, M_\odot/{\rm{K\, km\, s^{-1}\, pc^2}}$ except COPSS, which -- as a fiducial choice -- uses a Milky Way-like conversion of $\alpha_{\mathrm{CO}} = 4.3 \, M_\odot/{\rm{K\, km\, s^{-1}\, pc^2}}$, and the \cite{Garratt21} results which adopt $\alpha_{\mathrm{CO}} = 6.5 \, M_\odot/{\rm{K\, km\, s^{-1}\, pc^2}}$.
 From \cite{Chung:2024iob}.}}
\label{fig:rho_H2}
\end{center}
\end{figure}

As discussed in \S \ref{S:co_lum_model}, the average CO(1-0) brightness temperature measured in a LIM survey may be used to estimate the cosmic average abundance of molecular gas. In particular, Eq.~\ref{eq:rhoh2_from_tb} gives
the mass density in molecular hydrogen from the average CO(1-0) brightness temperature, under the assumption of a typical conversion factor, $\alpha_{\mathrm{CO}}$.
Several current and ongoing LIM surveys are poised to determine the molecular gas density near cosmic noon around $z \sim 2$, including the COMAP experiment \cite{COMAPI2022}, and surveys primarily targeting [CII] fluctuations at higher redshifts, for which CO transitions are interesting interloper contaminants. Current bounds and measurements from LIM surveys and targeted observations are shown in Figure~\ref{fig:rho_H2} and discussed further in \S \ref{S:measurements}. Ongoing COMAP measurements and future [CII] experiments, including CONCERTO \cite{CONCERTO2020} and TIME \cite{Sun:2020mco}, will improve the estimates here. Additional work is required to understand and calibrate the $\rm{CO-H_2}$ conversion factor, $\alpha_{\mathrm{CO}}$, its correlations with galaxy properties, and redshift dependence. 
In the future, CO LIM measurements can be combined with measurements of the abundance of neutral atomic gas (extracted from 21 cm LIM surveys), observations of the abundance of ionized gas, and estimates of the stellar mass density and SFRD versus redshift. This will help in understanding the evolutionary connections between the buildup/depletion of gas reservoirs and the cosmic star formation history.

\section{Experimental Challenges}
\label{S:challenges}

Ambitiously, LIM experiments aim to make use of all detected photons, even while operating in typically foreground- and noise-dominated regimes. This necessitates both exquisite control over instrumental artifacts, and precise foreground mitigation schemes. Furthermore, these challenges are coupled together, as instrumental features may be imprinted on top of bright astrophysical foregrounds. Indeed, separating the LIM signal from the astrophysical foregrounds, as observed with a realistic telescope, is the primary challenge for LIM. 
Circumventing these difficulties requires careful experimental design along with novel data analysis methods. Although some of the details vary across experiments, we provide a general overview of the main astrophysical foreground culprits and describe several promising mitigation strategies from the current literature.

In addition, as most current LIM experiments operate in the noise dominated regime, with the exception of SPHEREx on small scales, we discuss below the nature of the detectors used and how they shape the noise properties and mitigation strategies.

\subsection{Instrumental Noise}

\subsubsection{Detectors}
\label{sec:detectors}

Before discussing noise, we note that detectors are separated into two broad regimes depending on wavelength:  
coherent and incoherent detectors. For LIM applications, 21 cm and CO experiments such as CHIME, HIRAX and COMAP are in the coherent detector regime, while [CII], H$-\alpha$, and Ly-$\alpha$ LIM experiments such as TIME, SPHEREx, and HETDEX use incoherent detectors.  These represent two fundamentally different strategies for turning electromagnetic radiation into a measurable signal. The difference is largely dictated by how the photon energy $h\nu$ compares to the thermal energy $k_B T$ of the measurement system.

At radio, microwave, and millimeter wavelengths, $h\nu << k_BT$, the photon occupation number is high, and the field behaves like a classical electromagnetic wave, characterized by amplitude and phase.
Coherent (heterodyne) detectors exploit this by mixing the incoming field with a local oscillator in a nonlinear element, such as a Schottky diode \cite{CondonRansom}, an SIS (superconductor-insulator-superconductor) junction \cite{Tucker}, or a HEMT (High Electron Mobility Transistor) low-noise amplifier \cite{HEMP}, and down-converting the signal to a lower frequency while keeping the phase information intact. This makes interferometry and heterodyne spectroscopy possible: because phase information is preserved, signals from spatially-separated telescopes can be combined interferometrically,  and the spectral content across a receiver bandpass can be resolved coherently.  
The cost is a fundamental noise penalty due to the standard quantum limit (SQL; \S\ref{sec:sensitivity}), which scales with $h\nu$ and is increasingly severe towards higher frequencies.

At infrared, optical, ultraviolet, X-ray, and gamma-ray energies, $h\nu > k_B T$ and it becomes difficult or impossible to track phase information, and detectors instead measure deposited energy or power directly. Incoherent (direct-detection) detectors take this approach. In CMB or LIM applications such as TIME, TES (transition-edge sensor) bolometers \cite{Richards} measure the temperature increase from absorbed power. Kinetic inductance detectors (KIDs), which are superconducting microresonators where an absorbed photon breaks Cooper pairs and shifts the resonant frequency of an LC circuit, offer comparable sensitivity with simpler frequency-domain multiplexing. KIDs have been adopted in instruments such as CONCERTO, TIM, EXCLAIM, and the CCAT EoR-Spec (see \S \ref{S:projects}). Photoconductors and photodiodes, used in SPHEREx for example, exploit the photoelectric or photoconductive effect. These detectors do not preserve phase information, so a wave-like interferometric combination is not possible in the usual sense, but they scale more readily to large multiplexed arrays. At the highest energies in the X-ray and gamma-ray wavebands, the sensitivity is set by Poisson photon-counting statistics rather than an amplifier noise temperature.

At millimeter wavelengths in between these two regimes, it is interesting to compare SIS and TES detectors. SIS junctions are coherent, near-quantum-limited heterodyne mixers. They enable interferometry (e.g., ALMA\footnote{\url{https://www.almaobservatory.org/en/home/}}) and high-resolution spectroscopy, but their bandwidth is capped by the superconducting gap frequency (roughly 1 THz for niobium) and they don't scale well to large pixel counts. On the other hand, TES bolometers are incoherent, and they scale well to large SQUID-multiplexed arrays for wide-field continuum imaging and CMB surveys, but sacrifice phase and spectral resolution (unless coupled with a spectrometer). They are limited by phonon/photon noise plus the 1/f bath-temperature drift discussed below in \S \ref{sec:1/f}.

\subsubsection{Sensitivity}
\label{sec:sensitivity}

Fundamentally, the two detector paradigms are bounded by different noise floors. The core fact is the standard quantum limit (SQL) for coherent amplification. Because a coherent receiver must preserve both amplitude and phase, it is measuring two non-commuting quadratures of the field simultaneously. Quantum mechanics (via the amplifier noise theorem \cite{HausMullen}) requires that any phase-preserving linear amplifier adds at least half a photon of noise per mode, which translates into a minimum achievable system noise temperature of order $T_q \simeq h\nu/k_B$. This sets a fundamental floor that scales linearly with frequency. At radio and microwave frequencies, $h\nu/k_B$ is a small fraction of a Kelvin, so the SQL penalty is negligible compared to other noise sources, and state-of-the-art SIS mixers and cryogenic HEMT amplifiers routinely approach within a factor of a few of quantum-limited performance. At sub-mm and far-IR frequencies, however, $h\nu/k_B$ reaches tens of Kelvin and beyond, and the SQL floor itself becomes comparable to or exceeds the temperatures achievable with practical cryogenics. At this point, coherent detection simply cannot compete with direct detection on a per-pixel noise basis.

Incoherent (direct) detectors avoid the SQL noise penalty because they do not preserve phase information. Instead, their fundamental noise floor is set by classical (statistical) fluctuation limits such as photon shot noise,  arising from the Poisson statistics of photon arrival times. In addition, for thermal detectors, important noise contributions include phonon noise, arising from thermodynamic fluctuations in energy exchange with the heat bath, and noise from  statistical fluctuations in the arrival rate of background photons. The latter is the limiting noise term for space-based bolometers in a low-background environment. Consequently, bolometer arrays become more sensitive than heterodyne receivers in the sub-mm and far-IR.

In summary, coherent receivers are characterized by a system noise temperature $T_{\rm sys}$, often decomposed into a receiver temperature, sky/atmosphere contributions, and spillover. Their sensitivity follows the radiometer equation, $\Delta T \propto T_{\rm sys}/\sqrt{\Delta \nu t}$, where noise $\Delta T$ averages down with bandwidth $\Delta \nu$ and integration time $t$. Incoherent bolometric detectors are instead characterized by noise-equivalent power (NEP) in $\rm{W}/\sqrt{\rm{Hz}}$, typically broken into photon-noise NEP, phonon-noise NEP, and read-out/Johnson-noise NEP contributions added in quadrature, with the point-source analog being NEFD (noise-equivalent flux density). Photon-counting detectors at UV through gamma-ray energies use yet another vocabulary to describe noise: quantum efficiency, dark-count rate, and energy resolution.

An important effect beyond the fundamental noise floor is the array scalability, which affects overall survey sensitivity. In the previous example, note that coherent receivers are expensive and complex to multiplex and so are usually limited to modest pixel counts
since each pixel requires its own mixer, local-oscillator distribution, and intermediate frequency (IF) chain, even though an SIS detector can in principle match a bolometer's noise temperature.
Bolometers and photoconductors are read-out through SQUID \cite{SQUID} or CMOS \cite{CMOS} multiplexers, respectively, and scale to thousands of pixels relatively cheaply. KIDs push this scalability further, as each superconducting microresonator can be fabricated with a distinct resonant frequency and thousands of pixels can be read-out over a single microwave feedline using frequency-domain multiplexing \cite{Day2003}, avoiding much of the per-pixel wiring complexity that SQUID- or CMOS-based read-outs require. 
Survey mapping speed scales with both the number of independent detectors and the per-detector sensitivity, and so incoherent arrays often achieve greater total survey sensitivity even where per-pixel noise floors are comparable.  
This is a major reason why wide-field continuum surveys such as LIM and CMB experiments at milimeter/sub-mm wavelengths almost universally use bolometer or KID arrays rather than heterodyne receiver arrays, reserving coherent detection for applications that specifically require phase information such as interferometry and high-resolution spectroscopy.

\subsubsection{1/f Noise}
\label{sec:1/f}

The power spectral density of $1/f$ noise (flicker noise) rises roughly as $1/f$ toward low temporal frequencies ($f$), in contrast to white noise, which is a flat function of frequency. Flicker noise is a rather universal phenomenon; it shows up in electronics, thermal sensors, and even biological and financial time series, essentially anywhere that something drifts slowly relative to a faster measurement. In the LIM context, $1/f$ noise arises in any real detector chain from long-timescale drifts in gain, responsivity, or bias, which appear as excess power at low frequencies. The key figure of merit is the knee frequency, $f_{\rm knee}$, the frequency at which the $1/f$ contribution equals the white-noise floor: drift dominates at $f < f_{\rm knee}$, while in the $f > f_{\rm knee}$ regime, the detector behaves like an ideal white-noise system. Both the physical origin and the numerical values of $f_{\rm knee}$ are determined by the detector architecture and operating wavelength. For coherent detectors, examples include amplifier gain instability and local-oscillator phase noise for heterodyne receivers in the radio or millimeter wavebands. For incoherent detectors, thermal-bath drift, cold read-out electronics, and atmospheric fluctuations are important for bolometers and photoconductors. Other examples include trap-based generation-recombination noise in IR photoconductive arrays and read-out flicker noise in CCDs \cite{CCD} and CMOS imagers. At X-ray/gamma-ray energies, the overall sensitivity is dominated by Poisson counting statistics rather than a classical continuous 1/f spectrum, and the front-end pre-amplifier's dielectric 1/f term sets a floor on energy resolution.

In any scanned or time-ordered measurement, this long-timescale drift translates into spatial or physical scales, above which the measurement is contaminated by this excess noise.
The appropriate mitigation strategy depends on the two detector paradigms discussed in \S \ref{sec:detectors}:

\begin{itemize}
    \item In coherent systems, gain fluctuations in an amplification chain rescale the measured signal multiplicatively. Standard mitigation strategies therefore exploit the ability to compare two coherent signals directly, for example by Dicke switching \cite{Dicke} between the sky and a reference load, or by cross-correlating independent amplifier chains to suppress uncorrelated gain fluctuations.\\
    
    \item For incoherent detectors, $1/f$ noise instead arises from thermal-bath drift, cold read-out electronics (e.g., SQUID multiplexers), and atmospheric loading for ground-based instruments. Mitigation relies on common-mode subtraction across a focal-plane array or chopping/nodding the beam between source and background. In photon-counting X-ray and gamma-ray detectors, since the $1/f$ floor is a fixed property of the front-end pre-amplifier, mitigation shifts from noise cancellation to calibration: periodically tracking and correcting gain drift from detector aging, radiation damage, etc.\\
\end{itemize}
    
    Across all detector regimes, the common thread is the same: identify what is shared or predictable about the drift, and either difference it away through coherent switching or correlation, incoherent common-mode subtraction or chopping, or calibrate it out in the photon-counting regime. 
    
Furthermore, survey strategies and data analysis techniques can also help determine time drift.
Here, one treats the slowly varying instrumental baseline as an extra unknown to fit out, using redundant crossings of the same sky pixel from different scan angles as a constraint, commonly referred to as cross-linking \cite{Crawford2007}. Destriping exploits the fact that any pixel hit from multiple scan directions must show the same sky signal despite different drifts at each visit, and the density and angular diversity of these crossings set a hard limit on performance. Too few crossings, or crossings restricted to similar angles, leave the largest-scale modes in the map degenerate with an uncorrectable slow drift. Related approaches include maximum-likelihood map-making that models the noise covariance explicitly, and polynomial or PCA/common-mode template fitting (see \S \ref{S:PCA}), where a slowly varying polynomial or a shared low-frequency template, built from many detectors or repeated scans of the same field, is fit and subtracted per scan.

In the destriping algorithms used by Planck and most ground-based CMB surveys to remove the drift (e.g., \cite{destripping}), each scan segment is assigned a constant offset parameter, and a joint least-squares fit solves simultaneously for the sky map and all of the offsets. Software implementations of this approach include MADAM \cite{MADAM2005,MADAM2010}, used operationally in the Planck LFI mapmaking pipeline; SANEPIC \cite{SANEPIC} which uses a maximum likelihood approach to solve for correlated noise jointly across many detectors; and TOAST \cite{TOAST}, a general, actively maintained map-making framework with a built-in destriper used in LiteBIRD\footnote{\url{https://www.isas.jaxa.jp/en/missions/spacecraft/future/litebird.html}} and Simons Observatory\footnote{\url{https://simonsobservatory.org}} pipelines. 
A recent LIM-related application is the use of a MADAM-like analysis to generate all-sky microwave and diffuse radio maps using cruise data from the Juno mission \cite{Anderson2025}, illustrating the portability of CMB map-making techniques to LIM and LIM-adjacent efforts.

\subsection{Astrophysical Foreground Contamination}

The key astrophysical foregrounds generally fall into two categories: continuum emission foregrounds and line interlopers.
Typically, the continuum foregrounds are much larger in amplitude than the LIM signal, yet spectrally smooth. The line interloper foregrounds are from additional spectral lines, whose emission happens to land at the same observed frequencies as the
primary line targeted by the survey. 
This line confusion problem usually comes from sources at lower redshifts than the target line, although in principle it can involve emission at higher redshifts (``extraloper'' may be a better term in this case) too. In contrast to the continuum foregrounds, the interloper foregrounds contain a great deal of spectral structure and so different techniques are  required to separate the interlopers from the signal.
Although the distinction between continuum and line interloper foregrounds is conceptually useful, it may be less practically relevant in the optical and near-infrared where crowded emission lines can appear as a continuum-like foreground when observed at low spectral resolution. Nevertheless, we will discuss these two categories separately since this distinction remains useful for most current LIM programs.

\subsubsection{Continuum Foreground-to-Signal Ratios}

The foreground-to-signal ratio is a useful metric as it determines the fidelity required of foreground mitigation
algorithms. Here, we compare estimates for the strengths of key continuum foregrounds and LIM signals across much of the electromagnetic spectrum. We calculate both the mean intensity and the spatial fluctuations in the intensity for the foregrounds and LIM signals, considering radio, far-IR, and near-IR frequencies from 10 MHz to beyond 1 PHz. To characterize the strength of the fluctuations, we consider the angular power spectrum of each foreground 
at a multipole moment of $\ell=1000$, while for the 3D LIM signals we assume a spectral resolution of $R=100$ in the near- and far-IR. In the radio waveband, we consider the 21 cm fluctuations at $k=0.1$ Mpc$^{-1}$. 
Note that some of these choices are a bit arbitrary and we aim only for rough estimates here. 
These estimates nevertheless provide useful intuition into the challenges of foreground mitigation efforts, and how these difficulties vary across wavebands.

\paragraph{Radio} 

Continuum foregrounds have been studied extensively in the radio in the context of 21 cm LIM, where Galactic and extragalactic synchrotron and free–free emission overwhelm the expected 21 cm signal by several orders of magnitude, depending on the observed frequency/redshift (e.g., \cite{Furlanetto:2006jb, Morales2006, Bowman2009, Liu2012, Parsons2012, Switzer2015}). 
Reference \cite{Liu:2019awk} provides an excellent review of foreground 
mitigation techniques, focusing on interferometric measurements in the radio. While some of the details vary
across wavebands and with the specifics of the experiment and instrumental design, many of the basic principles are broadly applicable.
Since the 21 cm foregrounds have been studied most extensively, we start with this case. 

\begin{figure}
    \begin{center}
    \includegraphics[width=\textwidth]{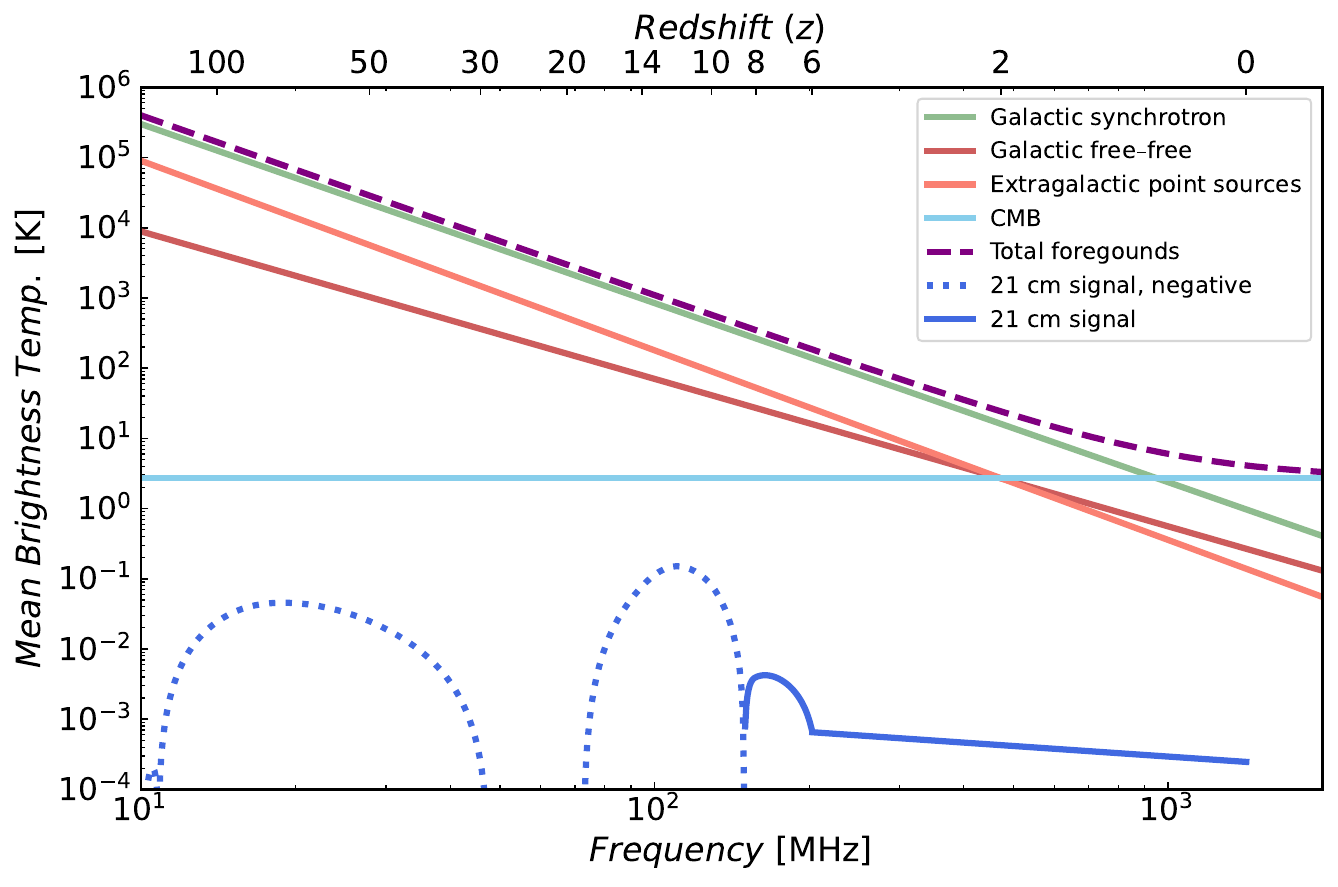}
    \caption{The mean brightness temperature of the 21 cm LIM signal and foregrounds as a function of frequency. We indicate the corresponding redshift of the 21 cm signal along the top x-axis. The foregrounds include contributions from Galactic synchrotron (green) and free-free (red) emission, extragalactic point sources (salmon), and the CMB (light blue). The 21 cm signals (blue) are mostly negative against the CMB at high redshifts (blue dotted curve) and near zero at around 50 MHz in this model. The solid blue parts of the 21 cm signal curve shows frequencies where the 21 cm signal is observable in emission against the CMB.}
    \label{fig:21 cm_mean_fg}
    \end{center}
\end{figure}

\textit{Mean Intensity} Figure \ref{fig:21 cm_mean_fg} shows the mean brightness temperature of Galactic synchrotron ($T_{\mathrm{sync}}$) and Galactic free-free ($T_{\mathrm{ff}}$) emission, extragalactic point source synchrotron radiation ($T_{\mathrm{ps}}$), and the CMB ($T_{\mathrm{CMB}}$). Each component is modeled as a simple power-law function: 
\begin{equation}
T_i = A_i \left(\frac{\nu}{\nu_0}\right)^{-\beta_i} [\mathrm{K}],
\end{equation}
with a reference frequency $\nu_0 = 150\,$MHz. We adopt the values of ($A_i$, $\beta_i$) = (300, 2.55) for $T_{\mathrm{sync}}$ \cite{Shaver1999, Oliveira-Costa2008}, (30, 2.10) for $T_{\mathrm{ff}}$ \cite{Jester2009}, and (60, 2.70) for $T_{\mathrm{ps}}$ \cite{Oliveira-Costa2008}, whereas $T_{\mathrm{CMB}} = 2.725$K \cite{Fixsen2009}. The total radio foreground is then 
\begin{equation}
T_{\mathrm{tot}} = T_{\mathrm{sync}} + T_{\mathrm{ff}} + T_{\mathrm{ps}} + T_{\mathrm{CMB}},
\end{equation}
dominated by Galactic synchrotron radiation $T_{\mathrm{sync}}$.

The high-redshift ($z>6$) global (mean)  21 cm signal, during the EoR and Cosmic Dawn, is taken from Figure 5 of \cite{Mesinger2016}. We assume their bright galaxy model. The post-reionization era 21-cm signal follows the analytic formula, Eq. \ref{eq:t21b}, (except denoted here by $T_{\rm 21 cm}$ rather than $\delta T_{\rm b}$), discussed in \S \ref{S:21cm_overview}. 
The figure illustrates that the mean brightness temperature of the total continuum foreground is many orders of magnitude larger than the average 21 cm signal, with the precise ratio depending on frequency.

\begin{figure}
    \begin{center}
    \includegraphics[width=\textwidth]{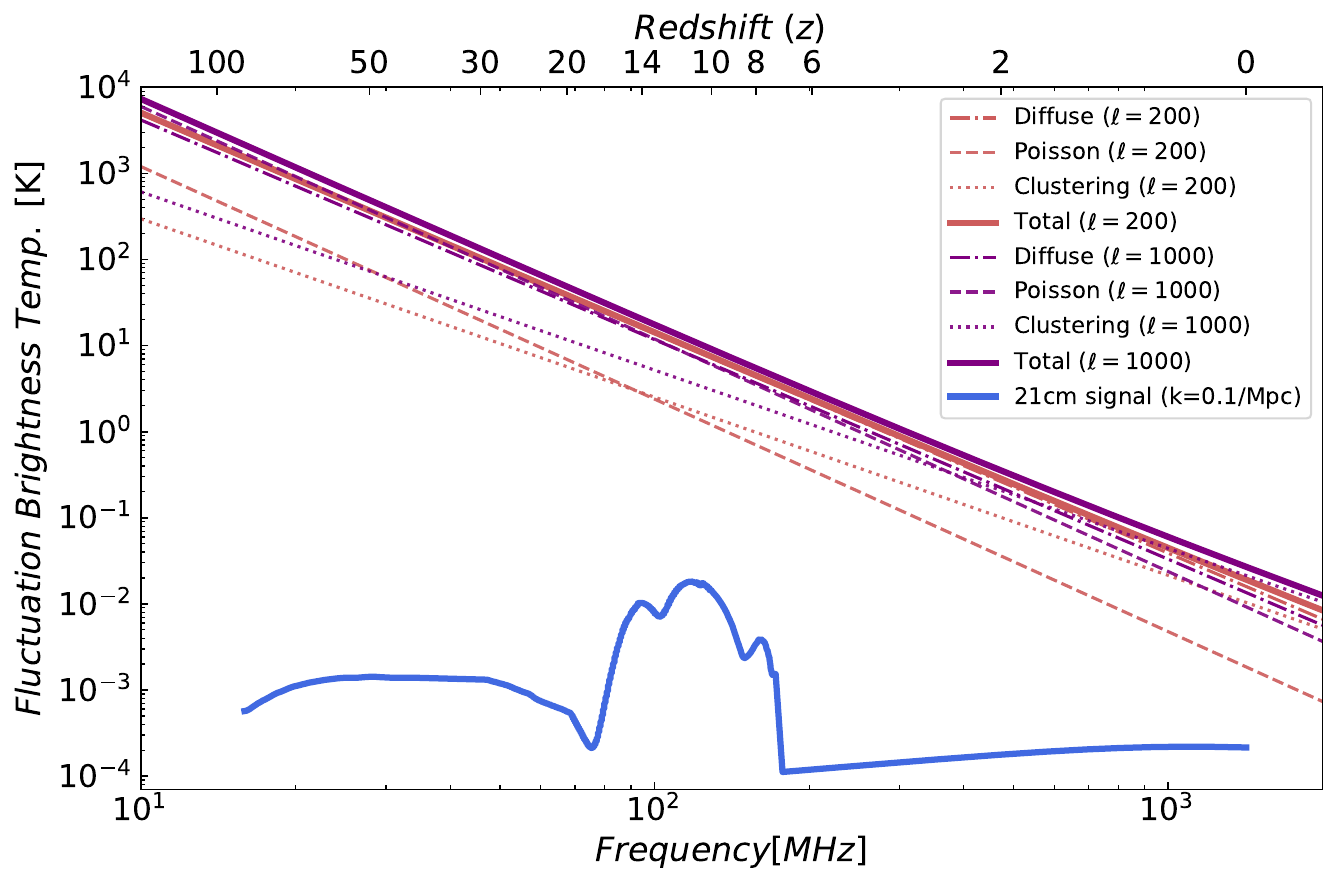}
    \caption{The fluctuation brightness temperature of the 21 cm signal and its foregrounds. The top x-axis marks the corresponding redshifts of the 21 cm signal. We plot the extragalactic (synchrotron) foregrounds at angular scales of $\ell=200$ (salmon) and $\ell=1000$ (purple), and the 21 cm signal (blue) at a 3D Fourier wavenumber of $k=0.1 \, \mathrm{Mpc}^{-1}$. The various curves, as described in the legend, further indicate the contributions from diffuse emission, shot-noise (``Poisson''), and extragalactic source clustering.}
    \label{fig:21 cm_fluct_fg}
    \end{center}
\end{figure}

\textit{Fluctuation Intensity} Figure \ref{fig:21 cm_fluct_fg} shows the rms fluctuations in the Galactic and extragalactic foregrounds and the 21 cm signal as a function of observing frequency. The Galactic foreground estimates are based on the measurements of reference \cite{Bernardi2009} at 150 MHz in the Fan Region at low Galactic latitude. For diffuse Galactic emission, measured on $>10$ arcmin scales, we adopt an angular power spectrum of the following form:
\begin{equation}
C_\ell^{\mathrm{Diff}}(\nu) = C_{G}(\nu_G) \left(\frac{\ell}{\ell_G}\right)^{-\alpha_G}\left(\frac{\nu}{\nu_G}\right)^{-2\beta_G},
\end{equation}
where $\nu_G=150$ MHz, $\ell_G=400$, $\alpha_G=2.2$, $\beta_G=2.55$, and $C_{\ell_G}^{\mathrm{Diff}}(\nu_G) = C_{G}(\nu_G) = 8.2 \times 10^{-4}$ K$^2$ \cite{Bernardi2009}. This serves as an upper bound on the diffuse Galactic synchrotron emission at high latitude. 
Reference \cite{Bernardi2009} determines the power-law $\ell$-dependence
based on estimates at $\ell=300-900$, but we extrapolate slightly beyond this to $\ell=1000$ in our figure. 

For extragalactic point sources, we describe their Poisson shot-noise angular power spectrum $C_\ell^{\mathrm{shot}}$ as a constant in $\ell$ with a frequency dependence of $\nu^{-2\beta_{s}}$, i.e., 
\begin{equation}
    C_\ell^{\mathrm{shot}} = C_{s}(\nu_s) \left(\frac{\nu}{\nu_s}\right)^{-2\beta_{s}}, 
\end{equation}
where $C_s(\nu_s)=10^{-4}$K$^2$, $\nu_s=150$ MHz, and $\beta_{s}=2.7$ \cite{Bernardi2009}. 

For the extragalactic point source clustering power spectrum, we follow \cite{Santos2005} and assume
\begin{equation}
C_\ell^{\mathrm{Clus}}(\nu) = C_c(\nu_c) \left(\frac{\ell}{1000}\right)^{-\alpha_{c}}\left(\frac{\nu}{\nu_c}\right)^{-2\beta_c}. 
\end{equation}
Upon imposing a relatively deep flux cut at $S=0.1$ mJy, the parameters are $C_c(\nu_c)=5.7 \times 10^{-5}$ K$^2$ at $\ell=1000$ and $\nu=130$ MHz; $\alpha=1.1$, and $\beta_c=2.07$ \cite{Bernardi2009}. As a quick comparison with the other two components, $C_c(\nu=150 \, \mathrm{MHz})=3.2 \times 10^{-5}$K$^2$. The total foreground fluctuation intensity is then 
\begin{equation}
    C_\ell^{\mathrm{FG}} = C_\ell^{\mathrm{Diff}}+C_\ell^{\mathrm{shot}}+C_\ell^{\mathrm{clus}}, 
\end{equation}
and is dominated by diffuse Galactic foreground emission. The Poisson term produces only a small contribution, and the clustering term is of comparable amplitude but smaller at low frequencies. Note that this is measured at low-galactic latitude with a fairly deep point-source masking depth of 0.1 mJy. Although the relative contribution from extragalactic point source clustering is small, it still remains stronger than the 21 cm signal.

For the 21 cm fluctuation signal at high-redshift ($z>6$), we adopt the bright-source model from \cite{Mesinger2016}, generated using the 21 cmFAST code\footnote{\url{https://github.com/andreimesinger/21 cmFAST}} \cite{Mesinger11}, and we plot the fluctuation power at $k=0.1$ Mpc$^{-1}$ as a function of redshift. At lower redshifts ($z<6$), we use the CLASS code \cite{CLASS2011} to compute the linear density power spectrum, $P_{\mathrm{lin}}(k, z)$, as a function of wavenumber $k$ and redshift. We use the Kaiser formula \citep{Kaiser87} to calculate the 21 cm fluctuation power spectrum as 
\begin{equation}
    P_{\mathrm{21 cm}}(k,z) = T_{\mathrm{21 cm}}^2 \left(b_{\mathrm{HI}}^2 + \frac{2}{3} b_{\mathrm{HI}} f_\Omega(z) + \frac{1}{5}f^2_\Omega(z) \right)P_{\mathrm{lin}}(k,z),
\end{equation}
where $T_{\mathrm{21 cm}}$ is the mean 21 cm brightness temperature described by Eq.~\ref{eq:t21b}, $b_{\mathrm{HI}}=1$ is the linear bias factor (which is taken to be constant), and $f_\Omega(z)\approx \Omega_{\mathrm{m}}(z)^{0.55}$ is the growth factor where $\Omega_{\mathrm{m}}(z)$ is the matter density at redshift $z$. We then compute the rms 21 cm fluctuation as $\sqrt{\frac{k^3}{2\pi^2} P_{\mathrm{21 cm}}(k,z)}$ in units of Kelvin at $k=0.1$ Mpc$^{-1}$.  
In the narrow bandwidth limit, we expect the 21 cm angular fluctuation power spectrum to be related to 
the 3D 21 cm power spectrum as 
$\ell (\ell +1) C^{\mathrm{21 cm}}_{\ell}/(2 \pi) \sim \# \, \Delta^2_{\mathrm{21 cm}}\left(\ell/\chi\right)$,
where $\#$ indicates an order unity factor, $\chi$ is the comoving distance to the redshift of interest, and $\Delta^2_{\mathrm{21 cm}}(k) = k^3 P_{\mathrm{21}}(k)/(2 \pi^2)$ \cite{Zaldarriaga:2003du}. As an example for reference, $k=0.1 \, \mathrm{Mpc}^{-1}$ corresponds to $\ell \sim 800$ at $z \sim 6$.

As with the mean intensity, the rms 21 cm fluctuations are many orders of magnitude smaller than the fluctuations in the foregrounds. As mentioned earlier, the largest contaminant is diffuse Galactic emission. Although the foregrounds can be separated using their spectral smoothness, Figure \ref{fig:21 cm_fluct_fg} helps to quantify the high fidelity required among 21 cm foreground removal algorithms. 

\paragraph{Far-IR} In the millimeter to far-infrared wavebands, continuum emission from the CMB and CIB are the dominant foregrounds. These are a few orders of magnitude brighter than the expected amplitude of LIM signals such as the redshifted [CII], [OIII], and [NII] emission lines (e.g. \cite{Bethermin2022}). Below, we discuss their mean and fluctuation intensities, respectively, in the customary brightness temperature units (taken in the Rayleigh-Jeans limit). Specifically, we relate the spectral radiance, $B_\nu$, the power emitted per unit emitting area, per steradian, and per unit frequency by a blackbody, to the temperature $T$ in Kelvins by $B_\nu(T)=\frac{2\nu^2 k_{\mathrm{B}} T}{c^2}$, where $k_{\mathrm{B}}$ is  Boltzmann's constant, $c$ is the speed of light, and $\nu$ is the frequency. 
We note that, strictly speaking, the Rayleigh-Jeans limit requires $h\nu \ll k_{\mathrm{B}} T$, but we employ Rayleigh-Jeans brightness temperatures throughout to facilitate direct comparisons across different frequencies.

In addition to Galactic and extragalactic foregrounds, atmospheric water contamination is an obstacle for sub-mm observations. This determines the accessible wavelength windows for ground-based observations \cite{Sayers2010}. 
We ignore atmospheric contamination here, as its impact depends on location and weather.

\begin{figure}
    \begin{center}
    \includegraphics[width=\textwidth]{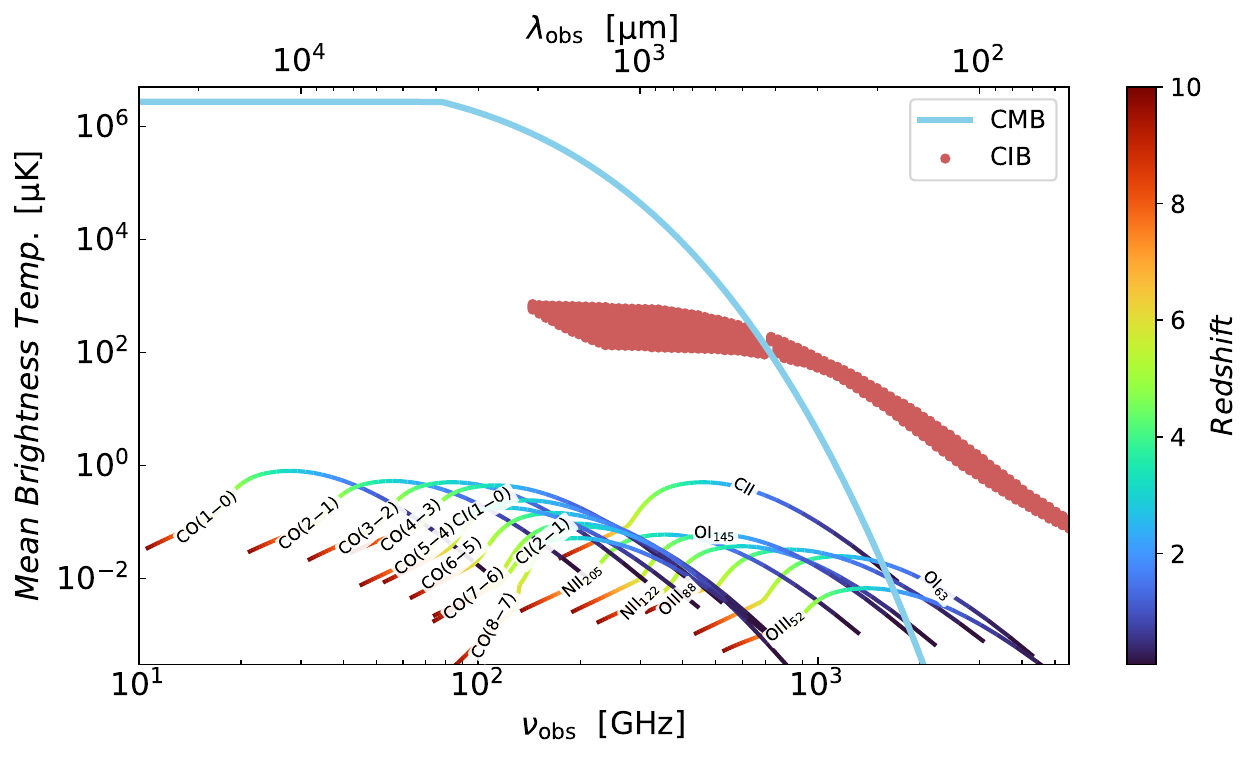}
    \caption{The mean brightness temperature of the FIR LIM signals and foregrounds as a function of frequency. We show the CMB in the Rayleigh-Jeans limit (light blue) and the CIB (red). The LIM signals are estimated based on the model of  \cite{Sun19} and the redshifts are indicated by the colorbar.} 
    \label{fig:sub-mm_fg_sig}
    \end{center}
\end{figure}

\textit{Mean Intensity} Figure \ref{fig:sub-mm_fg_sig} shows a compilation of the mean LIM signals in various lines, along with the dominant far-IR foregrounds, the CMB and CIB.  
We convert the spectral radiance of the extragalactic background light compiled by \cite{Bethermin2011} to brightness temperature in the Rayleigh-Jeans limit, as discussed above, and plot the contributions of the CMB and CIB accordingly. The LIM signals are estimated using the halo-model calculations in \cite{Sun19}\footnote{The \textsc{starter} code is authored by Guochao (Jason) Sun and is available on Github. Jason graciously incorporated a high-z SFRD model to extend the applicable redshift range and provided the plotting tool for this purpose.}. 
The LIM signal strengths are determined for multiple FIR spectral lines across redshift, including [CII], [NII]$_{122}$, [NII]$_{205}$, [OIII]$_{52}$, [OIII]$_{88}$, [OI]$_{145}$, [OI]$_{63}$, and multiple CO rotational lines CO($J+1 \rightarrow J$), where $J=[0,1,2,3,4,5,6,7]$. The style of this figure was inspired by an earlier 
comparison between LIM signal strengths in \cite{Silva2021}.

\begin{figure}
    \begin{center}
    \includegraphics[width=\textwidth]
    {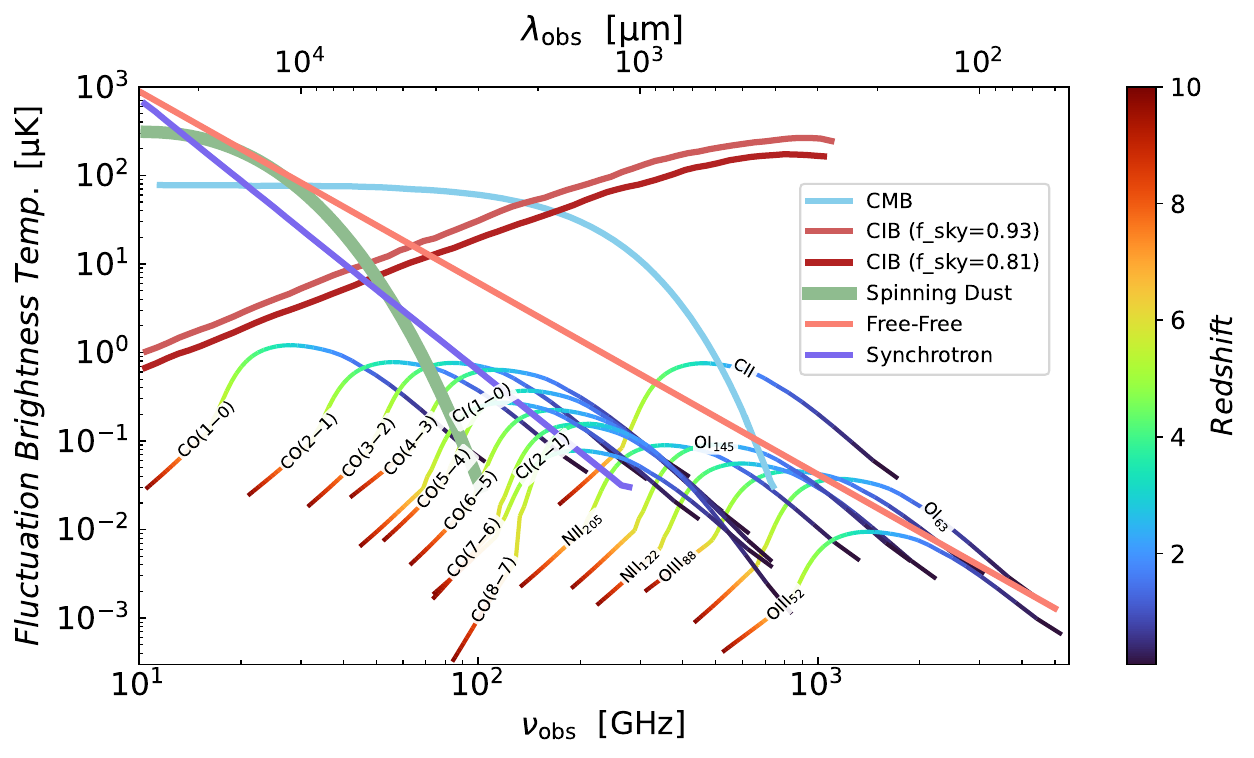}
    \caption{The fluctuation brightness temperature of the FIR LIM signals and foregrounds as a function of frequency. The foregrounds are taken from \cite{Planck2020}, smoothed at degree scale, including the CMB (light blue), CIB (red; for a sky fraction of 0.93 and 0.81, respectively), spinning dust (green), free-free emission (salmon) and synchrotron radiation (purple).  LIM signals are estimated based on the model of \cite{Sun19}, evaluated at an angular scale of $\ell=1000$ and spectral resolution of $R=100$. The redshifts are indicated by the colorbar.}
    \label{fig:sub-mm_fg_sig_clust}
    \end{center}
\end{figure}

\textit{Fluctuation Intensity} Figure \ref{fig:sub-mm_fg_sig_clust} shows the corresponding fluctuation intensity, where multiple foreground components are plotted for an angular scale of $\ell=1000$. The values are taken nominally from Figure 4 of the Planck2018 paper \cite{Planck2020}, including the CMB and CIB for two sky fractions (0.93 and 0.81), spinning dust, Galactic synchrotron, and free-free emission, measured on degree scales. There are updates to the compilations shown here in Figure 23 of reference \cite{BeyondPlanck2023}, but
the overall trends and foreground amplitudes remain the same. 

The same suite of LIM signals are similarly plotted for fluctuations on an angular scale of $\ell=1000$ and for a spectral resolution of $R=100$, as a function of redshift, using the \cite{Sun19} model. At the high frequency end, the CIB is the dominant foreground; the exact CIB amplitude depends on the sky fraction and masking strategy. Although the foreground fluctuations dominate over the LIM signals, the foreground-to-signal ratio is less severe than in the case of the 21 cm line, especially for the brightest lines near their peak frequencies.

\paragraph{Near-IR} 

In the near-infrared, the dominant foreground is the zodiacal light, arising from sunlight scattering off of interplanetary dust. Its glow is brightest along the ecliptic plane and is time-varying as the Earth moves through dust clouds in our solar neighborhood. The spatial distribution is expected to be smooth. COBE/DIRBE has provided measurements of zodiacal light in broad bands over 1.25-240 $\mu$m \cite{Hauser1998}, informing the leading Kelsall model \cite{Kelsall1998}.
SPHEREx is expected to provide an updated characterization of zodiacal light at $0.75-5 \, \mu$m with finer spectral resolution. Further, Diffuse Galactic Light (DGL), arising from starlight scattering off of interstellar dust in our own Galaxy at shorter wavelengths and from their thermal emission at longer wavelengths, known as cirrus, also makes a significant contribution. Notable thermal emission features include PAH emission features at 3.3, 6.2, 7.7, 8.6, and 11.3 $\mu$m, with widths in the range of 0.03 to 0.5 $\mu$m. These features were first observed by \cite{PAH1973} and the 3.3 $\mu$m emission is prominent in SPHEREx's Galactic map. Beyond the Milky Way, the CIB, the integrated extragalactic thermal dust emission, is a key contributor at longer wavelengths, especially in the far-IR. 

\begin{figure}
    \begin{center}
    \includegraphics[width=\textwidth]{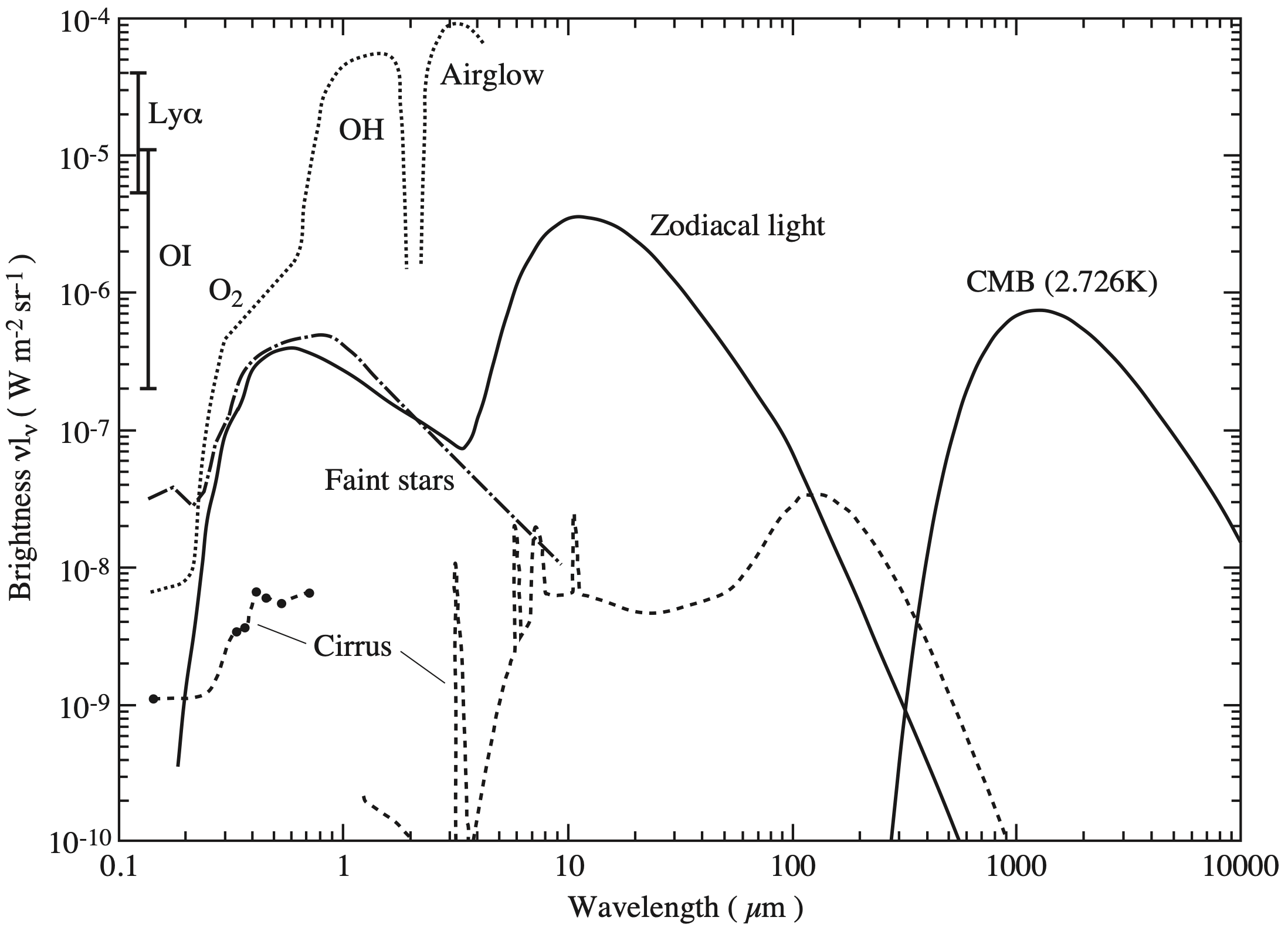}
    \caption{Overview of the sky brightness at low-Earth orbit and high ecliptic and galactic latitudes from \cite{NightSky1998}. Note that the ``faint stars'' in the figure are defined as stars fainter than a V-band magnitude V=6, so these are in fact rather bright. The zodiacal light and faint stars are measured at the South Ecliptic Pole. The Ly-$\alpha$ and OI emission are local geo-coronal intensities measured by the HST at a height of 610 km. }
    \label{fig:nightsky}
    \end{center}
\end{figure}

Figure \ref{fig:nightsky} from \cite{NightSky1998} provides a classic overview of the mean brightness of different components, showing the relative contributions from zodiacal light, cirrus, and faint stars, among others. Note that the faint star contribution depends sensitively on the masking threshold employed. We do not discuss the faint star, or the integrated star light (ISL) component in the following, assuming it can be mitigated by a combination of source masking and de-projection techniques based on star templates from the Gaia catalog\footnote{\url{https://www.cosmos.esa.int/web/gaia/dr3}} or the Trilegal simulation \cite{Trilegal2005}.

\begin{figure}
    \begin{center}
    \includegraphics[width=\textwidth]{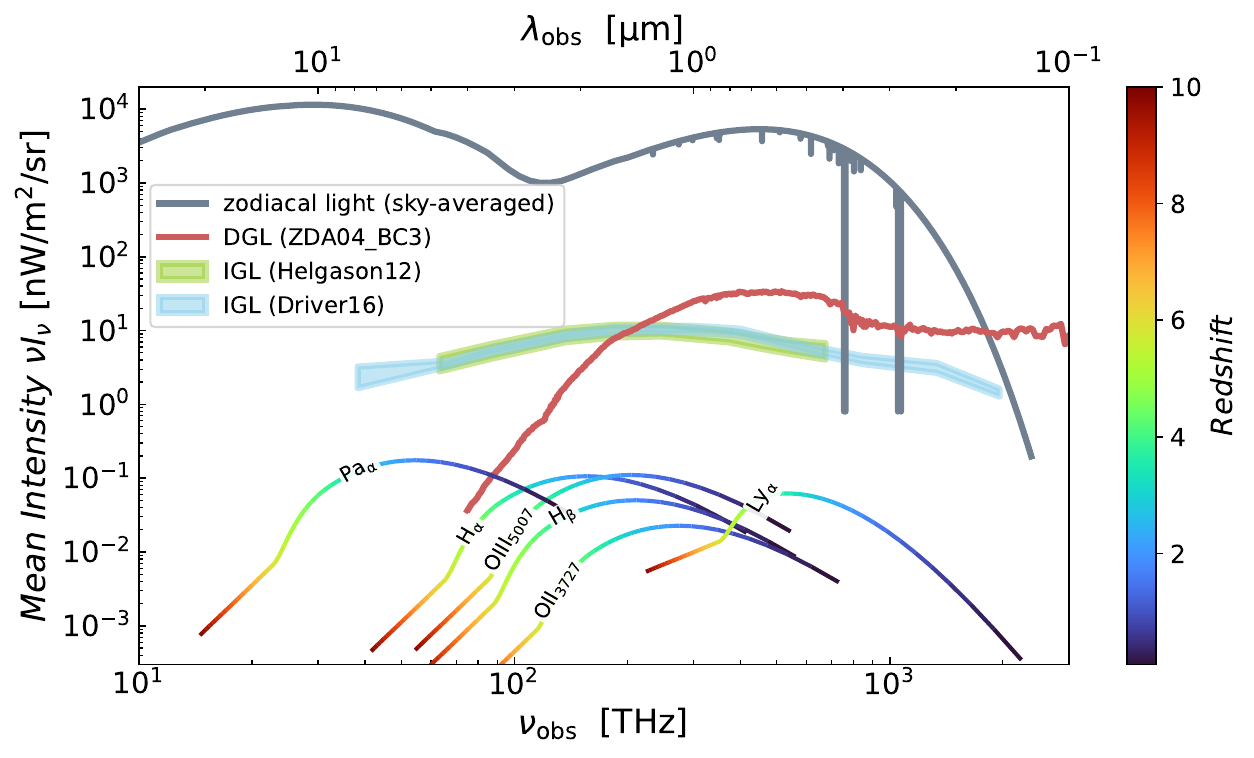}
    \caption{The mean intensity of the NIR LIM signals and foregrounds as a function of frequency. We show a snapshot in time of the zodiacal light \cite{San2024}, averaged over the entire sky (gray curve), a model of DGL \cite{Brandt2012} (red curve), and two models of the IGL \cite{Helgason2012} (green) and \cite{Driver2016} (blue). The LIM signals are estimated based on a modified model from  \cite{Sun19} combined with empirical $L-\mathrm{SFR}$ scaling relations and a simple dust attenuation treatment (see text). The redshifts are indicated by the colorbar.}
    \label{fig:nir_lim_fg_mean}
    \end{center}
\end{figure}

\textit{Mean Intensity} Figure \ref{fig:nir_lim_fg_mean} compares the near-IR LIM signals and the three key foreground components: the zodiacal light, the DGL, and the integrated galaxy light (IGL). The IGL is the combined emission from all galaxies, shifted into the near-IR, and summed over all redshifts.
The zodiacal light estimate uses the ZodiPy\footnote{\url{https://cosmoglobe.github.io/zodipy/}} code \cite{San2024}, which is based on the Kelsall model \cite{Kelsall1998}. We plot the zodiacal light averaged over the entire sky from a given (random) time snapshot. This may be an overestimate for surveys or analyses targeting high ecliptic latitude regions. We note that the mean amplitude variations in the zodiacal light over time are smaller than the changes with ecliptic latitude. 
The DGL estimate is based on the ``ZDA04-BC03'' interstellar dust scattering model presented in \cite{Brandt2012}, which is consistent with the measurement by CIBER \cite{Arai2015}. In this model, dust is composed of small grains of bare graphite, bare silicate, and PAHs, while the interstellar radiation field is modeled via stellar population synthesis. Finally, we show two models for the IGL components\footnote{We make use of the model data and uncertainties compiled by Jordan Mirocha: \url{https://github.com/mirochaj/ebl\_utils}} based on \cite{Helgason2012}
and \cite{Driver2016}; the former is an empirical model built on 233 observed UV, optical and IR luminosity functions at $z<6$. The latter derived measurements of the COB and the CIB from a compilation of galaxy number count observations ranging from the UV to FIR. 

\begin{figure}
    \begin{center}
    \includegraphics[width=\textwidth]{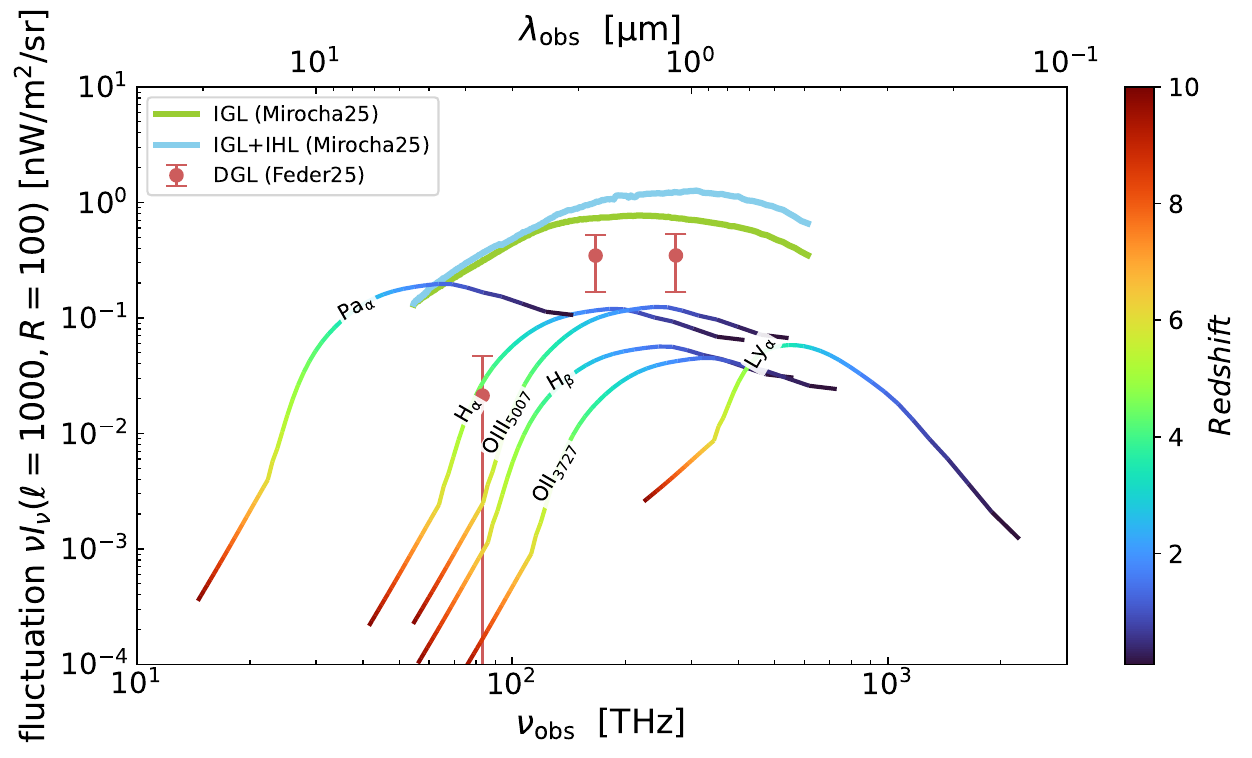}
    \caption{The NIR foreground fluctuation intensity at an angular scale of $\ell \sim 1000$. We show the DGL measurements \cite{Feder25} (red dots) and a model of the integrated galaxy continuum emission (IGL) masked at a magnitude threshold of 16.9 (green curve), as well as the intra-halo light (IHL) contribution (blue curve; J. Mirocha et al., in prep.). The 3D LIM signal fluctuation intensities at  $\ell=1000$ and $R=100$ are based on a modified model of \cite{Sun19} and the empirical scaling relations (see text), including a simple treatment of dust attenuation (and scattering for Ly-$\alpha$) effects. The LIM signal redshifts are indicated by the colorbar. }
    \label{fig:nir_lim_fg_clust}
    \end{center}
\end{figure}

 We extend the halo-model-based formalism presented in \cite{Sun2018} to model the near-IR LIM mean and fluctuation signals. Here, we adopt the scalings from \cite{Kennicutt1998}, which relate the star formation rates (SFRs) and line luminosities (L). The SFR as a function of mass and redshift is parametrized in \cite{Sun2018} based on the CIB model from \cite{Wu2017}, as constrained by Planck. Since the CIB uncertainties at $z>4$ are large, we use \cite{Sun19} to describe the halo mass functions $n(M)$ at low redshift ($z<4$) and \cite{Sun16} for the $z>6$ portion, while we interpolate between the two at $4<z<6$. The SFR and $n(M)$ are then constrained by the star formation history $\mathrm{SFRD}=\int dM~ n(M)~\mathrm{SFR}(M,z)$. These are consistent with measurements from \cite{Robertson:2015uda}.

We further adopt simple treatments to account for the attenuation effects from intervening dust. 
The L-SFR relations and dust attenuation factors assumed in these calculations are as follows. See \S \ref{S:landscape} for discussions regarding the relevant line emission physics. 
\begin{itemize}

\item H-$\alpha$: We adopt  L$_{\mathrm{H}-\alpha} [\mathrm{erg/s}] = 1.3 \times 10^{41}$ SFR [$M_{\odot} ~\mathrm{yr}^{-1}$] \cite{Kennicutt1998, Gong:2013xda}. We further include a suppression factor of $f=0.3$ to account for dust attenuation, which is found to be valid from $z=0-2.3$, with one magnitude of suppression at $z=2$ \cite{Sobral2013}. \\

\item H-$\beta$: Assuming an intrinsic ratio of H-$\beta$/H-$\alpha = 0.35$ \cite{Osterbrock06}, we adopt L$_{\mathrm{H}-\alpha} [\mathrm{erg/s}] = 4.6 \times 10^{40}$ SFR [$M_{\odot} ~\mathrm{yr}^{-1}$] \cite{Kennicutt1998, Gong:2013xda}. We similarly apply a suppression factor of $f=0.3$ to roughly account for dust attenuation \cite{Sobral2013}.\\

\item $[\mathrm{OIII}]_{5007}$: We adopt L$_{[\mathrm{OIII}]} [\mathrm{erg/s}] = 1.3 \times 10^{41}$ SFR [$M_{\odot} ~\mathrm{yr}^{-1}$] \cite{Ly2007, Gong:2013xda}, and similarly a suppression factor of $f=0.3$ to account for dust attenuation \cite{Sobral2013}.\\

\item $[\mathrm{OII}]_{3727}$: We adopt  L$_{[\mathrm{OII}]} [\mathrm{erg/s}] = 7.1 \times 10^{40}$ SFR [$M_{\odot} ~\mathrm{yr}^{-1}$] \cite{Kennicutt1998, Gong:2013xda}. To account for dust extinction, we again take the observed dust extinction at the H-$\alpha$ wavelength from \cite{Sobral2013} and apply the extinction curve from \cite{Calzetti2000} to scale to the wavelength of [OII]; the suppression factor we apply is $f=0.2$.\\

\item Pa-$\alpha$: Assuming an intrinsic ratio of Pa-$\alpha$/H-$\alpha = 0.128$ \cite{Hummer1987} and the empirical scaling with SFR, we have L$_{\mathrm{Pa}-\alpha} [\mathrm{erg/s}] = 1.7 \times 10^{40}$ SFR [$M_{\odot} ~\mathrm{yr}^{-1}$]. As Pa-$\alpha$ is less susceptible to dust extinction, we do not apply any attenuation.\\

\item Ly-$\alpha$: We adopt L$_{\mathrm{Ly}-\alpha} [\mathrm{erg/s}] = 1.1 \times 10^{42}$ SFR [$M_{\odot} ~\mathrm{yr}^{-1}$] \cite{Dijkstra2010} and a Ly-$\alpha$ photon escape fraction of $f_{\mathrm{esc}}=5.0 \times 10^{-4} \times (1 + z)^{2.8}$ \cite{Konno2016}. \\

\end{itemize}

\textit{Fluctuation Intensity} Finally, in Figure~\ref{fig:nir_lim_fg_clust}, we show the fluctuation intensity of the foregrounds and LIM signals. As the zodiacal light is expected to be spatially smooth, and the Kellsal model -- based on the COBE/DIRBE data -- has an effective angular resolution of $\sim 1^{\circ}$, we do not plot the zodiacal light fluctuation intensity inferred from ZodiPy. For the DGL, we adopt the measurements reported in \cite{Feder25}\footnote{We thank Richard Feder for kindly providing the numerical values.}, which are from cross-correlations of the CIBER 1.1 $\mu$m, 1.6 $\mu$m, and the IRAC 3.6 $\mu$m observations, respectively, with the ``CIB-cleaned SFD'' \cite{SFD1998} 100 $\mu$m dust extinction maps (``CSFD''; \cite{Chiang2023}). Estimates of the IGL fluctuation intensity are based on Mirocha et al. (in prep.)\footnote{We thank Jordan Mirocha for kindly providing the estimates of the IGL and IGL+IHL fluctuation intensities based on his model (Mirocha et al., in prep.).}, which includes an estimate of the intra-halo light (IHL). The IHL is the collective
emission from stars that orbit freely in dark matter halos after being stripped from their host galaxies during galaxy mergers or through tidal interactions.
A masking magnitude threshold is applied to mitigate the contribution from local (low-redshift) sources, including stars and galaxies, to better isolate the IGL component.
The LIM signals are shown for comparison, as a function of redshift, at $\ell=1000$ and $R=100$, using the same modeling approach described above.
Although the foregrounds are orders of magnitude higher in NIR mean intensity, the foreground spatial fluctuations are 
only one to two orders of magnitude larger in amplitude than the LIM fluctuation signals of interest (see also \cite{ChengChang2022}).
Hence, the challenges of separating LIM signals and continuum foregrounds are much less severe in the NIR than in the case of the 21 cm line at radio wavelengths.

\subsubsection{Line Interlopers}

Line interlopers, on the other hand, are a pressing issue for LIM in general; other than the fairly isolated 21 cm spectral line\footnote{
Radio recombination lines also occupy the low frequency spectrum, but those from our galaxy can be masked, while extragalactic radio recombination lines are estimated to be weak contaminants \cite{OhMack2003,Petrovic11}
}, most emission lines occupy parts of the electromagnetic spectrum that are crowded with other spectral features. 
In a LIM experiment spanning a wide frequency bandwidth to probe a large range of line-of-sight distances, emission from bright lower-redshift line interlopers can be redshifted into the same frequency range and contaminate the target signal.
We note that line interloper contamination is also relevant for traditional galaxy redshift surveys, and so this issue has been studied in that context too. For example, reference \cite{Gebhardt2019} considers joint analyses of auto- and cross-power spectra between target and interloper galaxy samples. Reference \cite{Kogut2015} explores map-space line de-confusion techniques, using multi-line information from pencil-beam spectroscopic surveys. In the LIM regime, although related approaches apply, the lower sensitivity and spectral resolution necessitate modifications to the methodologies applied in traditional redshift surveys.

\subsubsection{Summary Regarding Foreground Contaminants}

In summary, astrophysical continuum foreground contamination is a common challenge across all LIM efforts, but the severity of the contamination varies with wave band. At radio frequencies, the synchrotron radiation foreground is about four to five orders of magnitude brighter in mean intensity than the 21 cm signal. In the sub-mm, the CMB and CIB are two to three orders of magnitude larger in mean intensity than the brightest LIM line signals. Finally, in the optical and near-IR, the key foregrounds are just one to two orders of magnitude brighter than the main LIM signals.
The foreground-to-signal ratios in the rms spatial fluctuations show similar trends with frequency to the mean signals. Thus, although continuum foregrounds overwhelm the faint LIM signals across all wavelengths, the challenge is most daunting for the case of 21 cm. On the other hand, line interlopers are much less concerning for 21 cm, but pose additional obstacles for LIM measurements in many other lines.

\subsection{Mitigation Strategies: Continuum Foregrounds}

Having characterized the foregrounds, we now turn to strategies for mitigating them, beginning with continuum emission. A key property of continuum foregrounds is their smooth dependence on frequency, which helps distinguish them from spectral-line signals. By contrast, line emission along a given line of sight varies rapidly with frequency, since different frequencies probe gas at different redshifts, and the emission from widely separated redshifts is only weakly correlated. This fundamental distinction between line and continuum emission underlies many foreground mitigation techniques used in LIM.

Before discussing these methods in detail, we note several important caveats regarding continuum emission. Although foregrounds such as synchrotron and free–free emission are expected to be spectrally smooth, this smoothness must hold at the $\sim 10^{-5}$ level to robustly separate them from the much weaker 21 cm signal. In practice, small spectral features may arise from the superposition of emitting regions with different physical properties or from processes such as free–free absorption \cite{OhMack2003}. Another potential concern is that polarized foregrounds undergo frequency-dependent Faraday rotation, which can imprint spectral structure on the observed foregrounds if polarized emission leaks into total-intensity measurements.

Although the continuum foregrounds may be intrinsically extremely smooth, the mitigation challenge is complicated by the fact that LIM data is taken through an instrument which typically has its own spectral features and requires careful calibration. Thus, the combination of calibration errors and foreground residuals, convolved together, must be controlled to better than the expected signal-to-foreground ratio. This is the central analysis challenge.

Many LIM foreground mitigation strategies build on techniques developed for CMB analyses, where Galactic dust, the CIB, synchrotron radiation, and free–free emission constitute the dominant astrophysical foregrounds. However, there are two key differences between the CMB and LIM cases. First, primordial CMB anisotropies follow a nearly perfect blackbody spectrum and provide only a two-dimensional tracer of structure. In contrast, LIM probes three-dimensional fluctuations, with the frequency dimension mapping directly to redshift and encoding essential signal information. As a result, while additional frequency channels in CMB experiments primarily aid in foreground separation, in LIM they also carry new cosmological information.

Second, the relative foreground contamination is far more severe in LIM. The CMB temperature anisotropies dominate over foregrounds at frequencies near 100 GHz and at high Galactic latitudes (e.g., Figure 5 of \cite{Planck2020}), whereas in LIM the foregrounds exceed the signal across essentially all scales. In this respect, LIM foreground challenges are more analogous to large-scale CMB B-mode polarization measurements, where foregrounds overwhelm the signal by several orders of magnitude. However, an important distinction remains: CMB polarization measurements are inherently two-dimensional, whereas LIM exploits full three-dimensional information. Finally, we note that ongoing CMB instrumentation development is moving toward broader spectral coverage and higher spectral resolution, creating increasing synergies with LIM experiments.

\subsubsection{Foreground Avoidance and Delay Spectra}

One mitigation strategy is ``foreground avoidance''. In this approach, one describes the data
in a cylindrical Fourier space, with $k_\parallel$ denoting the line-of-sight wavenumber component, and $k_\perp$ describing
wavevector components that are transverse to the line-of-sight. In principle, spectrally smooth continuum foregrounds occupy only the low $k_\parallel$ modes in the $(k_\parallel,k_\perp)$ space. In contrast, the 21 cm signal resides partly at high $k_\parallel$. Therefore, one mitigation strategy is to simply remove the low $k_\parallel$ modes and use only the higher $k_\parallel$ 
regions where the signal is expected to dominate over foreground contamination. This sacrifices measuring the signal at low $k_\parallel$, yet it provides a simple and transparent approach for obtaining robust measurements at at higher $k_\parallel$. 

The foreground avoidance approach has been discussed extensively in the context of the reionization-era 21 cm signal and implemented in several interferometric experiments. 
The chromatic nature of an inteferometric observation couples spectral and spatial Fourier modes, thus expanding the foreground contamination region, and complicating the signal extraction. This gives rise to the so-called ``foreground wedge'' (e.g., \cite{Datta2010}), which consumes a significant fraction of a lower triangle in ($k_{\perp}$, $k_{\parallel}$) space, defined by the horizon. The slope of the foreground wedge (or the horizon) is only a function of redshift and the underlying cosmology \cite{Seo2016}:
\begin{equation}
    k_{\parallel} = \frac{D_{\mathrm{A,co}}(z)H(z)}{c(1+z)} k_{\perp},
    \label{eq:wedge}
\end{equation}
where $D_{\mathrm{A,co}}(z)$ is the comoving angular diameter distance, $H(z)$ the Hubble parameter and $c$ is the speed of light. The ``EoR window" then defines the region in Fourier space where the 21 cm EoR signal is expected to be free of contamination, at low-$k_{\perp}$ and high-$k_{\parallel}$.

For this approach to be effective, one needs to sample large and/or deep enough parts of the signal space to ensure that sufficient sensitivity is achieved, since data is discarded and the analysis is confined to a smaller region in Fourier space.
This can be expensive. 
Furthermore, one needs to understand the signal, foregrounds, and instrumental effects well enough to identify which regions in Fourier space are robust to contamination. 
This requires careful instrument design and calibration so that foregrounds do not leak into the signal space, as leakage can result in further loss of sensitivity and contamination. Finally, we note that the slope of this foreground wedge in Eq. \ref{eq:wedge} is shallower at low redshifts. This makes the foreground avoidance technique more favorable for post-reionization 21 cm applications than observations at higher redshifts \cite{Seo2016}.

Single-dish observations, on the other hand, do not inherently have a foreground wedge; however, one needs to be mindful of any spatial-spectral mode-coupling effects induced by the instrument or observing strategy. These effects might similarly corrupt the foreground-clean $k$-space for signal estimation. 

We also note that spectrally smooth foregrounds are often characterized in delay space. Delay space (or the delay spectrum) is obtained
by Fourier transforming the data along the frequency axis, yielding a quantity with units of time (the ``delay''), in contrast to $k_\parallel$, which is the Fourier conjugate to the line-of-sight direction in comoving coordinates. In interferometric measurements, the delay is related to the projected separation of antenna elements along the direction of wave propagation and determines the fringe pattern. Because foreground emission is spectrally smooth, it occupies a compact and finite region in delay space, allowing it to be robustly filtered (e.g., \cite{Parsons2012,EwallWice2021,CHIME2022}).

\subsubsection{Parametric vs. Non-Parametric Foreground Mitigation}

Parametric methods, such as polynomial-fitting, provide natural techniques for fitting and removing the continuum foregrounds.
Since the synchrotron, free-free, and other continuum foregrounds are spectrally smooth, their intensity is expected to 
be well-described by a power-law with small deviations, which may be characterized by a small set of parameters (e.g., \cite{Santos2005, Wang2006, McQuinn2006, Jelic2008}). For example, the foreground intensity along a line-of-sight may be described by a low-order polynomial function in $\log(\nu)$, corresponding to a Taylor expansion around a power-law dependence. Alternatively, the frequency dependence of the foregrounds can be described in terms of Legendre or Chebyshev polynomials, which are complete and orthogonal, allowing for analytic error estimates. The disadvantage of these parametric approaches is that they require priors on the number of parameters that are employed in fitting the foregrounds. Our incomplete understanding of the foregrounds, especially when observed through a chromatic instrument, makes it difficult to quantify the fidelity of the signal recovery in these methods.

In contrast, several non-parametric, blind foreground removal or projection techniques have gained popularity and 
are being widely applied.
These techniques make use of general features such as spectral smoothness, and the statistical properties of the observed data, without imposing strong priors. This helps avoid potential biases from an incomplete description of the foregrounds and the chromatic instrumental response. 
Apart from the principal component analysis approach, discussed separately below, these non-parametric techniques assume that the LIM signal, foregrounds, and other unwanted systematics have independent statistical or correlation properties, and that they are mixed together in the data in a linear manner. 
That is, the data $D$ is assumed to be a linear combination of the signal $I_{\mathrm{s}}$, foreground $I_{\mathrm{fg}}$, systematics $I_{\mathrm{sys}}$, and noise $n$:
\begin{equation}
    D = I_{\mathrm{s}} + I_{\mathrm{fg}} + I_{\mathrm{sys}} + n.
\end{equation}
The non-parametric techniques that have been applied in the 21 cm context include the Independent Component Analysis (ICA), the Generalized Morphological Component Analysis (GMCA) and the Gaussian Process Regression (GPR) methods discussed below. These algorithms prove effective on simulated data, and show some success in early applications using real data. However, there is signal loss which needs to be closely monitored and corrected in an end-to-end analysis.

\subsubsection{Foreground Mitigation Technique: PCA and SVD}
\label{S:PCA}

Principal Component Analysis (PCA) and the closely related Singular-Value Decomposition (SVD) -- which extends PCA to non-square matrices -- are widely used and demonstrably effective tools for identifying and removing foreground contamination.   
PCA and SVD are dimensionality-reduction techniques that compress large data sets by identifying the dominant modes of variance. 
In practice, these methods find a new set of mutually uncorrelated components by solving an eigenvalue (or singular value) problem for
an appropriate data matrix.

For LIM applications, the three-dimensional data set can be naturally reorganized into a 2D matrix composed of $N_x$ spatial and $N_{\nu}$ spectral pixels, which is usually a non-square matrix. The SVD is then used to decompose this data matrix (map) $\mathbf{M}$ of dimensions $N_x \times N_{\nu}$, into orthogonal spatial and spectral modes, often organized in (descending) order of eigenvalues: $\mathbf{M = U \Sigma V^T}$, where $\mathbf{\Sigma}$ is a diagonal rectangular matrix, $\mathbf{U}$ are the spectral eigenmodes and $\mathbf{V}$ describes the spatial eigenmodes. 
One can then examine the eigenvalues and identify the first few or several dominant eigenmodes as unwanted foreground modes, and project them out. 
The foreground cleaned map $\Tilde{\mathbf M}$ can be written as:
\begin{equation}
    \mathbf{\Tilde{M} = (1 - USU^T) M},    
\end{equation}
where ${\mathbf S}$ is the selection matrix with 1 along the diagonal for modes to be removed and 0 elsewhere. The obvious assumptions here are that foregrounds are the dominant, spectrally smooth components in the data, and are thus the leading eigenmodes in the frequency dimension (with large eigenvalues). In contrast, the desired LIM signal, having spectral features, appears noise-like in the frequency domain and occupies small eigenmodes. This also implicitly assumes that the data matrix is a separable function, spatially and spectrally, $\mathbf{M(x,\nu)}=\mathbf{g(x)f(\nu)}$, where the product functions $g(x)$ and $f(\nu)$ are then the singular eigenmodes.  

An alternative to applying an SVD to LIM maps is to apply it to the data covariance matrix, 
$\mathbf{C = M M^T}$, as ${\mathbf C}$ and ${\mathbf M}$ share the same frequency eigenmodes, $\mathbf{C = U \Sigma^2 U^T}$. 
Working in covariance space has the added benefit that one can construct ${\mathbf C}$ using submaps of data,
e.g., after splitting by observing time.  
As thermal noise does not correlate between the submaps, this avoids noise bias when constructing the eigenmodes (e.g. \cite{Switzer13}). 

Note that this empirical approach to foreground removal is limited by the amount of information in the maps, dictated by the number of degrees of freedom either in the frequency dimension or along the line of sight, whichever is smaller \cite{Switzer2015}. 
In other words, the number of modes occupied by foregrounds and the chromatic instrumental response must be smaller than the rank of the data matrix, otherwise there is insufficient information to capture the complexity of the foregrounds.  
This thus puts requirements on the survey size and angular resolution, as well as on the frequency bandwidth and resolution.

Since the foregrounds are not perfectly smooth, and are coupled to the chromatic instrumental response, it can be difficult to determine the optimal number of modes to remove. If a large number of modes are excised, this also removes LIM signal, and so there is a trade-off between reducing foreground contamination and minimizing signal loss. It is then important to inject simulated LIM signals into the data stream to monitor potential signal loss and make appropriate corrections (e.g. \cite{Chang10}, \cite{Masui13}, \cite{Switzer13}, \cite{Switzer2015}, \cite{Anderson2018}, \cite{Tramonte2020}, \cite{Irfan2021}). Of course, this requires a good model for the simulated LIM signal. Reference \cite{Switzer2015}
further discusses couplings -- introduced in the cleaning process -- between the foregrounds and signals, and the resulting biases in signal recovery. In another application, PCA has been used to identify correlated noise in map domain: reference \cite{ComapPCA} used PCA to suppress pointing-correlated systematics and successfully reduce large-scale variance, while reference \cite{TIME2026} used PCA to identify residual correlated noise after common-mode subtraction.

\subsubsection{Foreground Mitigation Technique: GMCA}

The Generalized Morphological Component Analysis, GMCA, is a blind source separation technique utilizing morphological diversity and sparsity to separate the signals and foregrounds. 
Reference \cite{Zibulevsky2001}
first proposed the method, which decomposes the data into a set of suitable complete basis functions in which different foreground components can be sparsely represented by a few terms. 
In the case of 21 cm, as first explored in \cite{Chapman2013} -- and other LIM applications -- one
describes the foregrounds in terms of sparse components, while the LIM signals and noise produce residuals. The LIM signals themselves are typically too weak to represent directly in the GMCA expansions. 

GMCA uses wavelet basis functions to describe each foreground source by a small number of coefficients through solving a linear mixing problem. 
Specifically, the data can be described by 
\begin{equation}
    D = A S + N,
\end{equation}
where $D$ is the data, $S$ is the signal matrix containing $n$ sources (the foregrounds) to be estimated, $A$ is the mixing matrix, and $N$ is the noise matrix. The LIM signal is assumed to reside in the residuals after subtracting the reconstructed foreground components. 
The signal matrix $S$ contains n source components, $s_j$,
where each source can be expanded using the wavelet basis functions $\Phi = [\phi_1, ... \phi_m]$ such that $s_j= \sum_{k=1}^{m} \alpha_j[k] \phi_k$. Here $s_j$ is sparse if
only a few of the $\alpha_j[k]$ are significantly non-zero. The GMCA estimates the matrix $A$ which yields the sparsest sources $S$. This can be found by solving the constrained optimization problem:
\begin{equation}
\min_{A,\alpha} \left[
\frac{1}{2} \| D - A \alpha \Phi \|_F^2
+ \lambda \sum_{j=1}^n \| \alpha_j \|_p
\right],
\end{equation}
where $\|D \|_F = \left(\mathrm{Tr}(D^T D) \right)^{1/2}$ 
is the Frobenius norm, $\| \alpha_j \|_p = (\sum_k |\alpha_j [k]|^p)^{1/p} $; $p=0$ measures the number of non-zero $\alpha$ parameters and enforces sparsity. The optimization is achieved iteratively by estimating $S$ for a fixed A and then determining $A$ for a fixed S, set to the value in the previous iteration. The desired LIM signal and noise are the residuals after foreground removal through GMCA. The GMCA technique has been studied and applied to 21 cm LIM data (e.g., \cite{Chapman2013}, \cite{Carucci2020}).

\subsubsection{Foreground Mitigation Technique: ICA}
\label{S:ICA}

Another commonly used foreground-mitigation approach is Independent Component Analysis (ICA), which has been implemented for LIM applications in algorithms such as \textsc{FASTICA} \cite{Chapman2012}. 
ICA assumes that the observed data can be modeled as a linear mixture of statistically independent components. The main idea is that a mixture of many independent contributions tends toward a Gaussian distribution by virtue of the central limit theorem, whereas the underlying source components themselves are generally non-Gaussian. By exploiting non-Gaussian statistical properties, ICA aims to separate the data into independent components that can be associated with astrophysical foregrounds, instrumental systematics, or radio frequency interference (RFI). The assumption of statistical independence is only approximate for astrophysical foregrounds, and imperfect separation can lead to residual foreground contamination or signal suppression. These effects depend sensitively on the number of components removed and the statistical properties of the signal and foregrounds.
As with other blind separation techniques, careful calibration of signal loss is required when applying ICA to LIM data sets \cite{Wolz2022}.

\subsubsection{Foreground Mitigation Technique: GPR}

Another useful technique is Gaussian Process Regression (GPR), which is a Bayesian non-parametric method that has recently been applied 
to 21 cm LIM (\cite{Rasmussen2006, Mertens2018, Soares2022, Kern2021}). 
A Gaussian process is a probability distribution over functions, an infinite-dimensional generalization of the multi-variate Gaussian distribution. It is commonly used to separate data components with distinct statistical properties, where prior assumptions about each component are encoded in a set of kernels. 
A stochastic Gaussian process can be easily sampled to determine the mean and covariance functions, and it has a well-defined likelihood that can be marginalized over.

For application to 21 cm observations, the frequency-frequency covariance function of the data K($\nu, \nu'$) is modeled as the 
sum of statistically uncorrelated contributions from the 21 cm signal, $K_{21}(\nu,\nu')$ and the foregrounds
$K_{\mathrm{fg}}(\nu,\nu')$, $K(\nu, \nu') = K_{21}(\nu,\nu') + K_{\mathrm{fg}}(\nu,\nu')$. Both components are commonly modeled using Mat\'ern kernels. The foreground kernel is assumed to have a long coherence length, reflecting the spectral smoothness of the foregrounds, while the 21 cm signal has a short coherence length owing to the spectral structure in the signal. GPR is then used to infer the posterior distribution of the foreground component given the data. The mean of the foreground posterior distribution is then subtracted to estimate the 21 cm signal.

Reference \cite{Mertens2018} demonstrates that the 21 cm power spectrum can be recovered in simulations, with a minimal and controllable impact on the signal across a range of wavenumbers.
The authors further apply it to the LOFAR-EoR data \cite{Mertens2020} with moderate success. Reference \cite{Kern2021} places the GPR foreground subtraction method into the quadratic estimator formalism. The authors show how the GPR analysis distorts window functions at low $k$, necessitating proper transfer function calculations.

\subsection{Foreground Mitigation Strategies: Line Interlopers}

We now provide an overview of techniques for cleaning line interloper foreground contamination.
First, we discuss masking analyses and targeted masking approaches (\S \ref{S:masking}), where one uses an external galaxy catalog
to mask regions in the LIM survey that are likely to be contaminated by interloper emission lines \cite{Gong:2013xda, Breysse2015, Silva:2014ira, Yue2015}. Next, we discuss the possibility of performing cross-correlations with external data sets which trace either the same cosmological volume spanned by the target emission line of interest, and/or the volume spanned by interlopers \cite{Lidz:2008ry, Visbal10, Gong2012, Gong:2013xda, Silva:2014ira}. This strategy has already 
been used to successfully extract LIM signals, even in the presence of strong foreground contamination \cite{Chang10, Masui13, Croft:2015nna, Croft:2018rwv}. Further, we explain how cross-correlations with low redshift galaxy surveys, and/or other tracers of large-scale structure, can be used to help filter out interloper contributions to LIM data cubes (\S \ref{S:nulling}). Another useful technique for separating LIM signals and interlopers is the anisotropic power spectrum method, discussed in \S \ref{S:aps}. Finally, we describe map-space reconstruction in \S \ref{S:mapspace} and 3D light-cone analyses in \S \ref{S:3dlightcone} below.

\subsubsection{Foreground Mitigation Technique: Masking}
\label{S:masking}

Source masking is arguably the most straightforward approach for directly reducing the impact of foreground interlopers.
This technique is commonly applied in CMB and CIB analyses to remove point source contamination and unwanted low-redshift objects. One complication is that pixel masking typically leads to a complex survey geometry, and coupling between Fourier modes, which must be accounted for in power spectrum measurements and other analyses, often through end-to-end simulations.

In the case of LIM, rather than masking pixels in a 2D map, one masks voxels in a 3D data cube. Several works, including \cite{Gong:2013xda, Breysse2015, Yue2015}, discuss masking strategies in which voxels above some intensity threshold are masked. The rationale here is that the bright end of the VID (see \S \ref{sec:pdf}) is often dominated by low-redshift interlopers. However, this is not always the case, and individually faint sources can sometimes constitute a significant portion of the interloper foreground. Furthermore, in some cases of interest, such as the reionization-era [CII] signal, low-z foreground interlopers -- predominantly from rotational transitions of the CO molecule in the case of the [CII] target line -- may dominate over the desired signal in fluctuation power by as much as an order of magnitude. In this case, a blind bright-voxel masking approach may be insufficient \cite{Breysse2015}.

It is often useful to combine the LIM observations with ancillary data, either a targeted survey for emitters in known 
interloper lines, and/or other traditional galaxy surveys at the redshifts of prominent interlopers. These can be used to guide the masking process, as the ancillary data helps identify which voxels to mask. The ideal survey for targeted masking may be impractical, however, since one desires a deep survey that matches the volume probed by the LIM observations. 
In cases where the ancillary data are traditional galaxy surveys, then there are also uncertainties in relating observed galaxy properties to the expected interloper line emission.  

Nevertheless, this guided masking approach will likely help in removing interloper contamination. 
For example, reference \cite{Sun2018} considers using infrared luminosity as a proxy for CO luminosity, since
both IR and CO luminosity are well-correlated with star formation rate, although these correlations are well-established
mainly for relatively luminous galaxies \cite{Carilli:2013qm}. One challenge here is obtaining deep-enough far-IR catalogs. For instance, although Spitzer MIPS data helps trace IR luminosity at $0.5 < z < 2$ \cite{Bavouzet2008}, the IR source density required to mitigate the CO foreground is around $\sim 10^5$ deg$^{-2}$, about twice that of the deepest MIPS catalog. 
Reference \cite{Sun2018} thus uses ultra-deep, near-IR selected source catalogs and cross-correlates these with far-IR/sub-millimeter maps to measure the mean infrared luminosities of galaxies and the scatter across the source catalog population.  
This approach helps guide the masking process and successfully mitigates CO interloper contamination in simulated reionization-era [CII] LIM power spectrum measurements \cite{Sun2018}.

\subsubsection{Foreground Mitigation Technique: Cross-Correlations}
\label{S:xcorr_fg}

In addition to the scientific benefits discussed previously, especially in \S \ref{S:xcorr}, cross-correlations can help avoid line interloper contamination.
As emphasized previously, the central idea is that on sufficiently large scales all line emitters -- and discrete objects such as galaxies and quasars --
are expected to trace the same underlying large-scale structure, provided they come from common redshifts. Hence, the emission in different lines and/or discrete objects should be well-correlated on large scales, regardless of the emission mechanisms and independent of the detailed source properties. In contrast, Galactic foregrounds and/or the emission from extragalactic sources at disparate redshifts, will not correlate with the large-scale structure at the redshift of interest. 
Here, this idea can be used to isolate contributions to LIM data cubes which come from the redshifts of prominent interlopers. In this case, one correlates with large-scale structure tracers at the interloper redshifts, where the tracer may be another emission line (landing at a different observed frequency) \cite{Visbal10,Gong2012,LidzTaylor2016,Gong17,Sun:2020mco}, or a catalog of discrete objects. Alternatively, one can cross-correlate with large-scale structure tracers at the redshift of the target line. In each case, other contributions to the LIM data act as noise and do not influence the cross-correlations on average, with some caveats. The uncorrelated pieces still impact the cross-correlation variance (see \cite{Switzer2019} and \cite{ChengChang2022} for relevant expressions).

The main caveat is that some tracers may share common foregrounds, and/or the foregrounds for one tracer may be correlated with the signal in the second tracer.  As an example of the former, galactic synchrotron emission will be a foreground for both 21 cm and CO LIM observations: if one observes a common sky region, the spectrally smooth synchrotron foreground will lead to correlations between the 21 cm and CO data. Similarly, cross-correlations between two CO transitions will suffer from common continuum foregrounds, while systematics will be partly shared as well, especially if both lines are measured by a single experiment. Another illustrative example is the case of [CII]-quasar cross-correlations \cite{Pullen:2017ogs,Switzer2019}. Here, the CIB is an important foreground for [CII] LIM measurements, but the CIB also correlates with the quasars. Hence, in some cases one must be careful to account for shared foreground contamination. In any case, future cross-correlation analyses will surely be an important technique for mitigating interloper contamination in LIM surveys.

\subsubsection{Foreground Mitigation Technique: Cross-Correlation ``Nulling''}
\label{S:nulling}

Recently, reference \cite{Bernal2024} proposed using cross-correlation techniques to eliminate (null) contributions from line interlopers, 
in analogy to techniques developed to remove lensing effects from CMB anisotropy measurements \cite{Huterer2005, Smith2012, Sherwin2015}.
The authors first consider a simplified scenario where a single foreground line interloper is widely separated in rest-frame frequency from the target spectral line, so that there is no overlap in the volume they trace. In this limit, the intensity fluctuations are the sum of three uncorrelated fields:
\begin{equation}
    \delta I(\mathbf{x}) = \delta I_{\mathrm{int}}(\mathbf{x}) + \delta I_{\mathrm{t}}(\mathbf{x}) + \delta \mathcal{N}(\mathbf{x}),
\end{equation}
where $\delta I_{\mathrm{int}}$ is the interloper component, 
$\delta I_{\mathrm{t}}$ is the target line of interest, and $\delta \mathcal{N}$ is the noise fluctuation field. 
Additionally, the authors suppose that a galaxy survey traces the galaxy distribution across the same volume as the interloper lines,
$\delta_\mathrm{g}^{\mathrm{int}}$. One can then use cross-correlations between the interloper galaxy sample and the measured intensity fluctuations
to identify and null the interloper contributions, $\delta I_{\mathrm{int}}$.

In Fourier space, an interloper-``cleaned'' version of the line-intensity fluctuations can be written as
\begin{equation}
    \delta \hat{I}(\mathbf{k}) = \delta I (\mathbf{k}) - \mathcal{F}(\mathbf{k}) \delta_\mathrm{g}^{\mathrm{int}}(\mathbf{k}), 
    \label{eqn:nulling}
\end{equation}
where $\mathcal{F}$ is chosen such that $\delta \hat{I}(\mathbf{k})$ is unbiased and the variance of its power spectrum is minimized:
\begin{equation}
    \mathcal{F}(k) = \frac{\tilde{P}_{Ig^{\mathrm{int}}}(k)}{\tilde{P}_{g^{\mathrm{int}}}(k)} = \frac{\tilde{P}_{I_{\mathrm{int}}g^{\mathrm{int}}}(k)}{\tilde{P}_{g^{\mathrm{int}}}(k)}. 
\end{equation}
Here $\tilde{P}$ is the measured power spectrum including noise. Using this weight $\mathcal{F}$, the resulting power spectrum of $\delta \tilde{I}$ is then given by
\begin{eqnarray}
     \tilde{P}_{\hat{I}} = P_{I_\mathrm{t}} + P_{I_{\mathrm{int}}}\left(1-\rho^2_{I_{\mathrm{int}}g^{\mathrm{int}}} \right) + \mathcal{N}_I 
     = \tilde{P_{I}} \left(1 - \tilde{\rho}^2_{Ig^{\mathrm{int}}}  \right),
\end{eqnarray}
where $\rho_{ab}$ is the cross-correlation coefficient of the two fields defined as 
\begin{equation}
    \rho_{ab} = \frac{P_{ab}}{\sqrt{(P_{aa}+\mathcal{N}_a)(P_{bb}+\mathcal{N}_b)}}. 
\end{equation}

The authors point out that an advantage of this approach is that the actual realization of the contaminant present in the data is eliminated (at least partially), along with the sample variance associated with it, and therefore the measurement noise is reduced. This can be seen in the Gaussian covariance of the power spectrum: 
\begin{equation}
    C_{\hat{I}\hat{I}} = \frac{2}{N_{\mathrm{modes}}} \left[P_{I_{\mathrm{t}}} + P_{I_{\mathrm{int}}}\left(1-\rho^2_{I_{\mathrm{int}}g^{\mathrm{int}}} \right) + \mathcal{N}_I \right]^2,
\end{equation}
as well as in the covariance of the cross-power spectrum of the target line fluctuations with the galaxies in the same volume:
\begin{equation}
    C_{\hat{I}g^{\mathrm{t}}} = \frac{ \left[P_{I_{\mathrm{t}}} + P_{I_{\mathrm{int}}}\left(1-\rho^2_{I_{\mathrm{int}}g^{\mathrm{int}}} \right) + \mathcal{N}_I \right]P_{g^{\mathrm{t}}} + P_{I^{\mathrm{t}} g^{\mathrm{t}}}^2}{N_{\mathrm{modes}}}, 
\end{equation}
where $N_{\mathrm{modes}}$ is the number of modes in a particular power spectrum $k$-bin. In both cases,
the interloper noise term, $P_{I_{\mathrm{int}}}$, is reduced by a factor of $1-\rho^2_{I_{\mathrm{int}}g^{\mathrm{int}}}$.

The authors verify the analytical formalism outlined above using lognormal simulations generated with the \textsc{simple} code \cite{Simple2023}. They also find a reduction in covariance for the auto- and cross-power spectra of the cleaned maps, obtained from Eq. \ref{eqn:nulling}, assuming a HETDEX-like observational setup with a Ly-$\alpha$ LIM target signal at $z = 2.99$ and interloper [OII] lines at $z=0.3$.

The authors quantify the effectiveness of the nulling technique by the gain in signal-to-noise ratio (SNR) with nulling compared to without. 
The improvements from nulling depend on the cross-correlation coefficient between the interloper line emission and the large-scale structure tracer ($\rho_{I_{\mathrm{int}}g^{\mathrm{int}}}$). 
The authors also examine the gain as a function of the ratio of the interloper to target line intensity. 
For auto-power spectrum measurements, the SNR improvement can be as large as a factor of six per wavenumber bin, while a factor of two per bin is
achievable for cross-power spectra. Hence, upcoming LIM experiments should take advantage of overlapping LSS surveys as these can 
help in nulling interloper contamination.

\subsubsection{Foreground Mitigation Technique: Anisotropic Power Spectrum Method}
\label{S:aps}

LIM experiments measure intensity fluctuations as a function of observed frequency and angle on the sky. Generally, one then maps into three-dimensional comoving coordinates assuming a target reference redshift and a cosmological model.  Typically, a number of interloper lines, at a range of redshifts, may occupy the same observed frequency range as the target line of interest. The interloper emission lines will be incorrectly mapped into comoving coordinates if the target reference redshift is assumed. This makes the interloper emission appear anisotropic, as the mappings from frequency to line-of-sight distance
and between angle and transverse distance are distinct functions of redshift. The degree of anisotropy induced depends on how different the target and interloper redshifts are, and is often much stronger than the anisotropies from redshift-space distortions (\S \ref{sec:master_halo_model}). 
The redshift-space distortions also have a different angular dependence than the ones from projecting interloper contamination into comoving coordinates. This anisotropy
is noted in \cite{Visbal10,Gong2012}, while \cite{LidzTaylor2016,Cheng2016} quantify its efficacy for separating the interlopers and target lines at the power spectrum level. We generally adopt the notation of \cite{Cheng2016} in what follows.

\begin{figure}
\begin{minipage}{.5\textwidth}
  \centering
  \includegraphics[width=.95\linewidth]{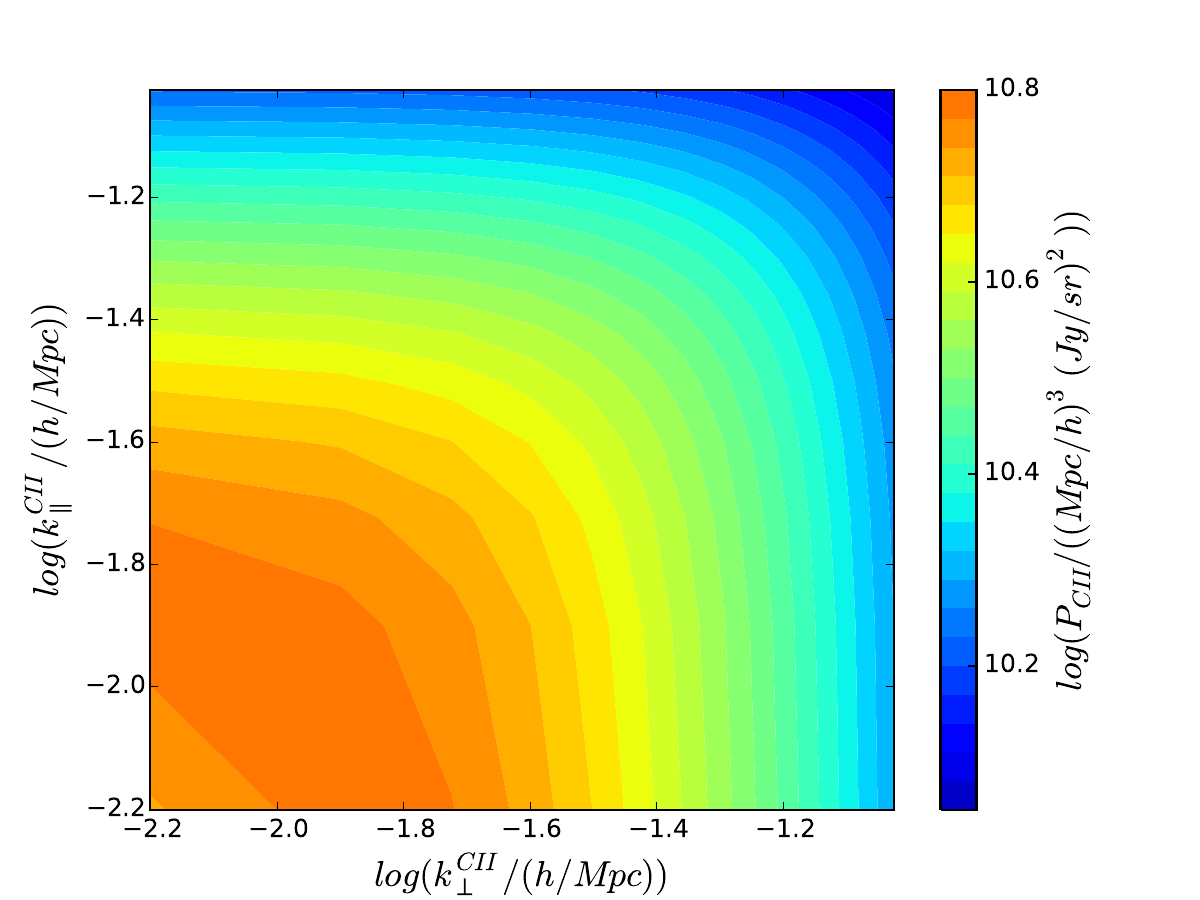}
\end{minipage}
\begin{minipage}{.5\textwidth}
  \centering
  \includegraphics[width=.95\linewidth]{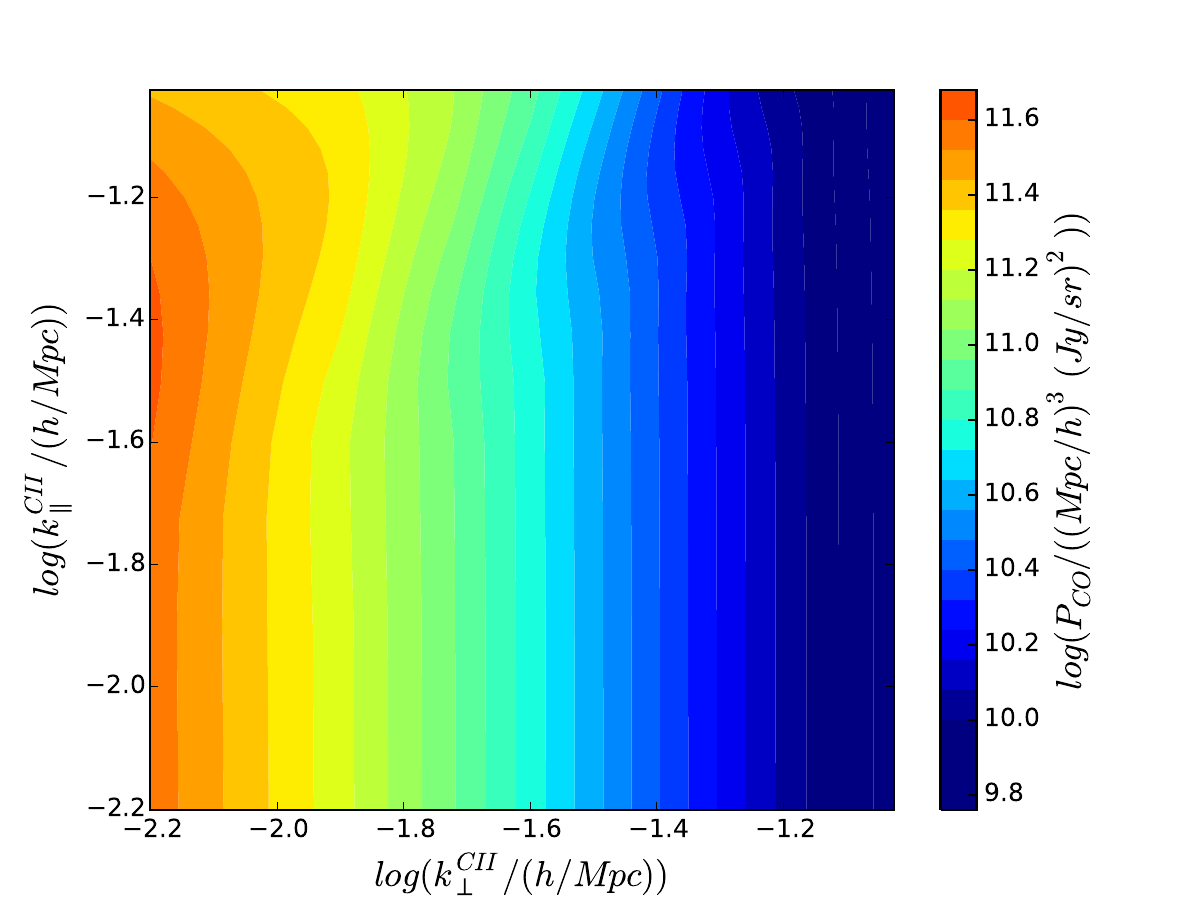}
\end{minipage}
\caption{[CII] (left) and (projected) CO (right) cylindrical power spectra from $z=6$ and $z=3$, respectively, both shown in the comoving frame of [CII] at $z=6$.  The CO power spectrum is distinctively anisotropic, and this can be used to separately fit for the target [CII] emission power spectrum and the
CO interloper contributions.
From \cite{Cheng2016}. 
\label{fig:anisotropic} }
\end{figure}

For example, consider the [CII] signal from the EoR at redshift $z_{\mathrm{CII}}$ as the target line of interest, while a CO line transition 
from redshift $z_{\mathrm{CO}}$ will land at the same observed frequency when
\begin{equation}
    \nu_{\mathrm{obs}} = \frac{\nu_{\mathrm{CII}}}{1+z_{\mathrm{CII}}} = \frac{\nu_{\mathrm{CO}}}{1+z_{\mathrm{CO}}},
\end{equation}
where $\nu_{\mathrm{obs}}$ is the observed frequency, and $\nu_{\mathrm{CII}}$ and $\nu_{\mathrm{CO}}$ are the rest-frame frequencies of the [CII] and CO lines, which are 1902 GHz and 115 $\times \, J$ GHz for the $J$ to $J - 1$ transition, respectively. Figure \ref{fig:anisotropic} illustrates the anisotropy
induced in the CO interloper power spectrum from assuming the target line redshift (here $z_{\mathrm CII}=6$) in mapping from angles and frequencies into comoving lengths. This contrasts with the more isotropic [CII] power spectrum (in the absence of redshift-space distortions the [CII] power is perfectly isotropic). The distinctive anisotropy can be used to simultaneously determine the intensity fluctuation power spectra for multiple interloper lines as well as the power spectrum of the target line.

Further details regarding this methodology are as follows. 
First, the observed frequency range $d\nu_{\mathrm{obs}}$
corresponds to a radial comoving length of $R^{\parallel}_{\mathrm{CII}}$ centered at redshift $z_{\mathrm{CII}}$ for the [C II] emission. 
The corresponding comoving length for the CO emission is
$R^{\parallel}_{\mathrm{CO}}$, centered on $z_{\mathrm{CO}}$. The relation between $R^{\parallel}_{\mathrm{CII}}$ and $R^{\parallel}_{\mathrm{CO}}$ thus follows
\begin{eqnarray}
    \frac{R^{\parallel}_{\mathrm{CO}}}{R^{\parallel}_{\mathrm{CII}}} &=& \frac{d\chi(z_{\mathrm{CO}})/d\nu_{\mathrm{obs}}}{d\chi(z_{\mathrm{CII}})/d\nu_{\mathrm{obs}}} 
    = \frac{\lambda_{\mathrm{CO}}(1 + z_{\mathrm{CO}})^2 / H(z_{\mathrm{CO}})}{\lambda_{\mathrm{CII}}(1 + z_{\mathrm{CII}})^2 / H(z_{\mathrm{CII}})} \\
    &=& \frac{(1 + z_{\mathrm{CO}}) / H(z_{\mathrm{CO}})}{(1 + z_{\mathrm{CII}}) / H(z_{\mathrm{CII}})},
\end{eqnarray}
where $\chi(z)$ is the comoving distance and $H(z)$ the Hubble parameter at redshift $z$. Similarly, in the transverse direction, the same angular scale $\theta$ will correspond to different transverse comoving scales $R^{\perp}_{\mathrm{CII}}$ and $R^{\perp}_{\mathrm{CO}}$ for [CII] and CO, respectively. The ratio is directly related to the ratio of the comoving angular diameter distances $D_{\mathrm{A,co}}(z)$:
\begin{equation}
    \frac{R^{\perp}_{\mathrm{CO}}}{R^{\perp}_{\mathrm{CII}}} = \frac{D_{\mathrm{A,co}}(z_{\mathrm{CO}})}{D_{\mathrm{A,co}}(z_{\mathrm{CII}})}
\end{equation} 
Therefore, CO fluctuations are incorrectly mapped into comoving length scales if one assumes the [CII] redshift in the mapping, while the different factors
in the radial and transverse directions lead to anisotropies. Naturally, the [CII] emission will be incorrectly mapped if one instead assumes one of the CO emission redshifts in the mapping. 
This is akin to the Alcock-Paczynski test \cite{Alcock79}, except in that case the anisotropies arise from assuming the incorrect cosmological model in the mapping to comoving units, where here the warping occurs from adopting an incorrect redshift. In the present case, note that the anisotropies can be much larger than in the traditional Alcock-Paczynski applications, especially when the target and interloper redshifts are far apart.

In Fourier space, the wavenumbers $k$ are inversely proportional to the comoving length scales, $k \propto  1/R$, and we have
\begin{eqnarray}
    k^{\parallel}_{\mathrm{CO}} &=& k^{\parallel}_{\mathrm{CII}} \frac{H(z_{\mathrm{CO}})(1+z_{\mathrm{CII}})}{H(z_{\mathrm{CII}})(1+z_{\mathrm{CO}})} \equiv k^{\parallel}_{\mathrm{CII}} r^{\parallel}(z_{\mathrm{CII}},z_{\mathrm{CO}}), \\
    k^{\perp}_{\mathrm{CO}} &=& k^{\perp}_{\mathrm{CII}} \frac{D_{\mathrm{A,co}}(z_{\mathrm{CO}})}{D_{\mathrm{A,co}}(z_{\mathrm{CII}})} 
    \equiv k^{\perp}_{\mathrm{CII}} r^{\perp}(z_{\mathrm{CII}},z_{\mathrm{CO}}), 
\end{eqnarray}
where $r^{\parallel}(z_{\mathrm{CII}},z_{\mathrm{CO}})$ and $r^{\perp}(z_{\mathrm{CII}},z_{\mathrm{CO}})$ describe stretching or compressions in the radial and transverse directions, respectively. In general,  $r^{\parallel}(z_{\mathrm{CII}},z_{\mathrm{CO}}) \,$ $\neq r^{\perp}(z_{\mathrm{CII}},z_{\mathrm{CO}})$ $ \neq 1$, and so the wavenumbers are shifted and by different factors in the radial and transverse directions.

In addition, the projection changes the comoving volume of each voxel and induces an amplitude change in the power spectrum, according to the volume factor of  $r^{\parallel}(r^{\perp})^2$. The (mistakenly) projected CO power spectrum $P^{\mathrm{proj}}_{\mathrm{CO}}$ in the comoving frame of [CII] is thus shifted in scale and amplitude from the true CO power spectrum $P_{\mathrm{CO}}$,
\begin{equation}
    P^{\mathrm{proj}}_{\mathrm{CO}}(k^{\perp}_{\mathrm{CII}}, k^{\parallel}_{\mathrm{CII}}, z_{\mathrm{CII}}, J) = r^{\parallel}(r^{\perp})^2 P_{\mathrm{CO}}(k_{\mathrm{CO}},z_{\mathrm{CO}}) 
\end{equation}
where $k_{\mathrm{CO}}=\sqrt{\left(k^{\perp}_{\mathrm{CO}}\right)^2 + \left(k^{\parallel}_{\mathrm{CO}}\right)^2}$. In the same coordinates, the [CII] power spectrum remains
\begin{equation}
    P^{\mathrm{proj}}_{\mathrm{CII}}(k^{\perp}_{\mathrm{CII}}, k^{\parallel}_{\mathrm{CII}}, z_{\mathrm{CII}}) = P_{\mathrm{CII}}(k_{\mathrm{CII}},z_{\mathrm{CII}}), 
\end{equation}
and $k_{\mathrm{CII}}=\sqrt{\left(k^{\perp}_{\mathrm{CII}}\right)^2 + \left(k^{\parallel}_{\mathrm{CII}}\right)^2}$.

In the halo model formalism (see \S \ref{sec:master_halo_model}) the power spectrum can be expressed as the sum of 1-halo and 2-halo contributions, e.g.,
\begin{equation}
    P_{\mathrm{CO}} = P_{\mathrm{CO (1h)}} + P_{\mathrm{CO (2h)}}. 
\end{equation} 
Here we incorporate the linear Kaiser effect \citep{Kaiser87} describing the coherent motion of structure growth on large scales, which enhances the power spectrum, and the suppression on small scales due to the nonlinear virial motion, which we write as an exponential damping term. The one-halo and two-halo power spectrum terms in redshift space \citep{White2001}, including the projection effects, can then be written as:
\begin{eqnarray}
&& P^{\mathrm{proj}}_{\mathrm{CO}\,(\mathrm{1h})}(k_{\mathrm{CII}},\mu_{\mathrm{CII}})
= r^{\parallel}(r^{\perp})^{2}
\left(1+\frac{f_\Omega(z_{\mathrm{CO}})}{\langle{b_{\mathrm{CO}}\rangle}}\mu_{\mathrm{CO}}^{2}\right)^{2} \int_{M_{\mathrm{min}}}^{M_{\mathrm{max}}} dM\; n(M) 
\nonumber \\[0.6em]
&& \hspace{-1.5em}\times
\Bigg[\left(
\frac{L_{\mathrm{CO}}(M,z_{\mathrm{CO}})}
     {4\pi D_{L}^{2}(z_{\mathrm{CO}})}
y_{\mathrm{CO}}(z) D_{\mathrm{A,co}}^{2}(z_{\mathrm{CO}})
\right)^{2}
\left|u(k_{\mathrm{CO}}|M)\right|^{2}
e^{-(k_{\mathrm{CO}}\sigma_{\nu}\mu_{\mathrm{CO}})^{2}/2}
\Bigg],
\end{eqnarray}
where $y$ maps between frequency intervals and comoving length, $u(k|M)$ describes the Fourier transform of the NFW profile, and $e^{-(k_{\mathrm{CO}}\sigma_{\nu}\mu_{\mathrm{CO}})^{2}/2}$ is a finger-of-god damping term. See \cite{Cheng2016} for details. 
\begin{eqnarray}
&& P^{\mathrm{proj}}_{\mathrm{CO}\,(\mathrm{2h})}(k_{\mathrm{CII}},\mu_{\mathrm{CII}})
= r^{\parallel}(r^{\perp})^{2}
P_{\mathrm{lin}}(k_{\mathrm{CO}})
\left(1+\frac{f_\Omega(z_{\mathrm{CO}})}{\langle{b_{\mathrm{CO}}\rangle}}\mu_{\mathrm{CO}}^{2}\right)^{2} \int_{M_{\mathrm{min}}}^{M_{\mathrm{max}}} dM\;
n(M)\,
\nonumber \\[0.6em]
&& \hspace{-1.5em}\times
\Bigg[
\frac{L_{\mathrm{CO}}(M,z_{\mathrm{CO}})}
     {4\pi D_{L}^{2}(z_{\mathrm{CO}})}
\, y_{\mathrm{CO}}(z) D_{\mathrm{A,co}}^{2}(z_{\mathrm{CO}})
\, |u(k_{\mathrm{CO}}|M)|
\, b(M,z_{\mathrm{CO}})
\, e^{-(k_{\mathrm{CO}}\sigma_{\nu}\mu_{\mathrm{CO}})^{2}/2}
\Bigg]^{2}. \nonumber \\
\end{eqnarray}
See also \S \ref{sec:master_halo_model} for further discussion, in a slightly different notation. 
In the linear regime, the shape of the power spectrum is independent of the luminosity–halo mass
relation. We can thus assume that on sufficiently large scales (e.g., $k<0.2 \, h \, \mathrm{Mpc}^{-1}$), light follows the distribution of matter, and that the shape of the LIM power spectrum in redshift space is known. The linear and projected power spectrum shapes can then be used as templates to extract the amplitude of the LIM signal ([CII]) and the interloper contaminants (CO), respectively. In practice, one sums over multiple contaminating CO rotational transitions. 

Reference \cite{LidzTaylor2016} uses the Fisher matrix formalism to quantify how well one can extract the LIM signal and interloper power spectra, while reference \cite{Cheng2016} uses Markov Chain Monte Carlo (MCMC) techniques to investigate the prospects here. Both works find that the signal and interloper power spectra can be recovered for suitable survey and instrumental parameters. We refer the reader to those works for more details.

\subsubsection{Foreground Mitigation Technique: Map-Space De-confusion}
\label{S:mapspace}

Map-space line reconstruction allows one to move beyond the two-point statistics that are recovered in the power spectrum anisotropy and cross-power spectrum measurement techniques. There is information in the {\em phases} of the LIM maps, and the intensity fields are generally non-Gaussian, especially those which 
trace the ionization field during the EoR. For example, the VID (\S \ref{sec:pdf}) contains information beyond that of the power spectrum and can help in constraining line luminosity function models \cite{Breysse:2016szq,Ihle2019}. Furthermore, if maps can be robustly reconstructed they can be used in multi-tracer analyses (\S \ref{sec:multi_tracer}) and for de-lensing the CMB (\S \ref{S:lim_lensing}), among other applications. Reconstructed maps also facilitate splitting the data into different spatial regions with varying foreground properties, allowing consistency tests.

Reference \cite{Moriwaki2020} uses machine learning techniques to reconstruct maps in each of two simulated emission lines which land at the same observed frequencies in their simulated LIM data sets. Specifically, they construct training data where the simulated LIM data 
contains a mixture of H-$\alpha$ emission from $z=1.3$ and [OIII] emission at $z=2$. The training data incorporate the expected clustering and line luminosity functions for each line. In a first test where foregrounds and noise are neglected, their algorithm can faithfully reconstruct maps in each of the separate lines. That is, machine learning techniques show promise for overcoming the line confusion problem in LIM data.

A follow-up study, \cite{Moriwaki2021}, further applies deep learning methods to the line de-confusion problem. This work develops conditional generative adversarial networks (cGANs) to extract signals from noisy maps. The cGANs successfully reconstruct H-$\alpha$ line emission maps, assuming noise at the level expected for the SPHEREx mission. The authors find that 60\% of the pixels in their reconstructed maps with intensities larger than 
3.5$\sigma_N$, where $\sigma_N$ is the rms observational noise, correspond to real sources. Making the same selection from {\em raw} simulated maps, only 20\% of such pixels are true sources. Therefore, cGANs are able to recover line emission information, even when the maps are noise-dominated. The reconstruction also robustly recovers the input one-point PDFs and power spectra. Furthermore, the authors show that suitably trained cGANs are robust to variations in the line emission models. However, the authors note that the overall reconstruction performance depends on the pixel size and survey volume of the training data, and that large volumes are required to reconstruct the large-scale line intensity power spectrum.

Reference \cite{Cheng2020} demonstrates a template-fitting technique, which extracts individual line-intensity maps from LIM data with
blended interloper lines. This approach exploits the fact that a galaxy at a given redshift will generally emit in multiple different spectral lines, which map onto distinct observed frequencies, leading to deterministic features that can be fit using redshift-dependent templates. 
The authors consider a set of template models, and adopt a sparse approximation, in which the number of sources along a given line-of-sight is small compared to the amount of information along the sightline. This approximation is reasonable, and works best in reconstructing bright
lower-redshift interloper contributions, as these tend to be the most sparse. The authors fit templates to LIM data iteratively using the Matching Pursuit algorithm. They show that CO line-intensity maps at $0.5 < z < 1.5$ can be successfully extracted from the ongoing reionization-era [CII] experiments, TIME and CONCERTO, since multiple CO rotational transitions are observable. Quantitatively, assuming
a plausible signal model and realistic noise levels, the reconstructed CO maps are $\sim 80\%$ spatially-correlated with the true maps.  
The VIDs of the individual CO lines are well-extracted, at high signal-to-noise ratio, down to the $L_\star$ knee in the line luminosity functions. The reconstruction performance is robust to realistic uncertainties in the line luminosity ratios and to the process of continuum foreground mitigation. The authors also show successful applications, across a range of rest-frame optical lines and redshifts, 
for a SPHEREx-like experiment.

\subsection{Holistic Foreground Removal Techniques}

Finally, we describe two holistic foreground mitigation strategies. The first, a 3D Lightcone Analysis method, simultaneously fits for all line emission signals and the continuum emission. The second approach discussed here is a Bayesian forward-modeling framework which accounts for line emission, foregrounds, and systematics. 

\subsubsection{Foreground Mitigation Technique: 3D Lightcone Analysis}
\label{S:3dlightcone}

Reference
\cite{Cheng:2022ani} proposes a novel analysis framework to fully exploit the LSS information in a 3D lightcone, or spectral-intensity maps. 
Regardless of signals or foregrounds, line or continuum emission, resolved or diffuse sources, the approach 
utilizes \emph{all photons} available in a lightcone. It extracts information about the statistics of the source spatial distributions, their redshift evolution, and spectral energy distributions simultaneously, assuming only the homogeneity and isotropy of cosmological perturbations. This holistic approach is attractive and aligned with the spirit of LIM, as it simultaneously characterizes the light from all emitting sources and the statistical properties of the large-scale matter distribution. 

The authors use angular cross-spectra $C_{l,\nu \nu'}$, at all combinations of observed frequencies $\nu$ and $\nu'$, as summary statistics for describing the spectral-intensity maps. The 3D spatial distribution of the emitting sources is assumed to trace the underlying LSS clustering on large-scales. 
Specifically, $C_{l,\nu \nu'}$ measures the product of the SEDs of all emitting sources, the bias-weighted luminosity density, and the 3D matter power spectra integrated along the line-of-sight; the 3D to 2D projection is linear and well specified. 
For a given set of observed angular power spectra $C_{l,\nu \nu'}$, the authors then formalize the likelihood function for the underlying source SED, redshift distribution, and the LSS clustering.

Specifically, the authors consider the angular power spectrum of all light emission in the linear regime, expressed as 
\begin{equation}\label{E:Cl_jljl}
\begin{split}
C_{\ell,\nu\nu'}^{\rm clus}&= \sum_{i=1}^{N_c}\int d\chi\, S^i(\nu_{\rm rf})M^i(\chi)A(\chi)\\
&\quad\cdot\sum_{{i'}=1}^{N_c}\int d\chi'\, S^{i'}(\nu'_{\rm rf})M^{i'}(\chi')A(\chi')\\
&\quad\cdot\int \frac{dk}{k}\, \frac{2}{\pi}k^3P(k)G(\chi)j_\ell(k\chi)G(\chi')j_\ell(k\chi'),
\end{split}
\end{equation}
where
\begin{align}\label{E:SMP_def}
S^i(\nu_{\rm rf})& \equiv \frac{\nu_{\rm rf}L^i_\nu(\nu_{\rm rf})}{L^i}=\frac{(1+z)\nu L^i_\nu((1+z)\nu)}{L^i},\\ 
M^i(\chi) &\equiv b^i(\chi) \int dL\,L\Phi^i(L,\chi,\hat{n}),\\
A(\chi)&\equiv\frac{D_{\mathrm{A,co}}^2(\chi)}{4\pi D_{\mathrm{L}}^2(\chi)}
\end{align}
are normalized SEDs and the bias-weighted luminosity density for component $i$ sources.
Here, $P(k)$ is the $z=0$ linear matter power spectrum, $G(\chi)$ is the linear growth rate, and $j_\ell$ is a spherical Bessel function.

Furthermore, with an observation of $N_\nu$ spectral bands and angular power spectra in $N_\ell$ bins, all auto/cross spectra at each $\ell$ bin, $\mathbf{C}^{\rm clus}_\ell$, can be arranged as an $N_\nu\times N_\nu$ matrix.  The total power spectrum $\mathbf{C}_\ell$, including a noise term $\mathbf{N}_\ell$, is then: 
\begin{equation}\label{E:Cl}
\mathbf{C}_\ell = \mathbf{C}^{\rm clus}_\ell+\mathbf{N}_\ell=\mathbf{B}_\ell\mathbf{P}\mathbf{B}_\ell^T+\mathbf{N}_\ell.
\end{equation}
Here $\mathbf{P}$ is an $N_k\times N_k$ diagonal matrix whose elements are the binned matter power spectrum $P(k)$, while $\mathbf{B}_\ell$ captures all other terms in Eq.~\ref{E:Cl_jljl}. $\mathbf{N}_\ell$ accounts for instrumental noise, foreground contamination, and Poisson noise from the emitting sources.

The authors describe the signal with a finite number of emission components and use the known mapping from the signal rest-frame into observed spectral-intensity maps to simultaneously constrain the LSS, as well as the SED and redshift evolution of each component. The components are specified by the data directly in a data-driven manner that does not require prior information regarding the SED of each component or regarding the noise in the data. In practice, the authors parametrize the normalized SED $S^i(\nu_{\rm rf})$, bias-weighted luminosity density $M^i(\chi)$, and the power spectrum $P(k)$ in Eq.~\ref{E:Cl_jljl} to express the clustering power spectra $C_{\ell,\nu\nu'}^{\rm clus}$, while $A(\chi)$ and $G(\chi)$ are assumed known given the cosmological model.

The authors point out that this technique is a generalization and an extension to the analytical formalism of reference 
\cite{dePutter2014}. It is also similar in spirit to the Spectral Matching Independent Component Analysis (SMICA; \cite{Delabrouille2003, Cardoso2008}) algorithm, which is widely used in CMB foreground-cleaning analyses to model the covariance of multi-frequency CMB maps as a sum of distinct components.

As a first demonstration, reference
\cite{Cheng:2022ani} successfully performed a lightcone reconstruction of continuum emission assuming the photometric survey setup of LSST and Euclid, making simplified assumptions regarding the source SEDs, the number of components, and the redshift evolution of the luminosity functions, while restricting the analysis to linear scales. Further, reference
\cite{Cheng2024} extended the algorithm to demonstrate the reconstruction of spectral line emission in the intensity-mapping regime, assuming the experimental setup of the SPHEREx deep fields. Finally, reference 
\cite{Cheng25} applies this technique to forecast the prospects for PAH intensity mapping measurements at $3.3 ~\mu m$ restframe for nearby galaxies ($z<0.5$), observable with SPHEREx. The PAH $3.3 ~\mu m$ is a broad emission feature and is spectrally resolved by SPHEREx, and therefore represents an interesting reconstruction challenge between the broad continuum and narrow spectral line regimes. 
The lightcone technique has been demonstrated to be effective in each of these three limits. The natural next steps, therefore, include treating all sources with more complex SEDs, including continuum and multiple emission lines together, and applying it to real data.
This work therefore lays the foundation for a more cohesive treatment of LIM datasets: it provides a path towards fully exploiting the rich information captured in a LIM experiment by using all of the measured photons to simultaneously fit for each different emission component.

\subsubsection{Foreground Mitigation Technique: End-to-end Bayesian Forward Modeling Approach}
\label{S:bayes}

\begin{figure}
    \begin{center}
    \includegraphics[width=\textwidth]{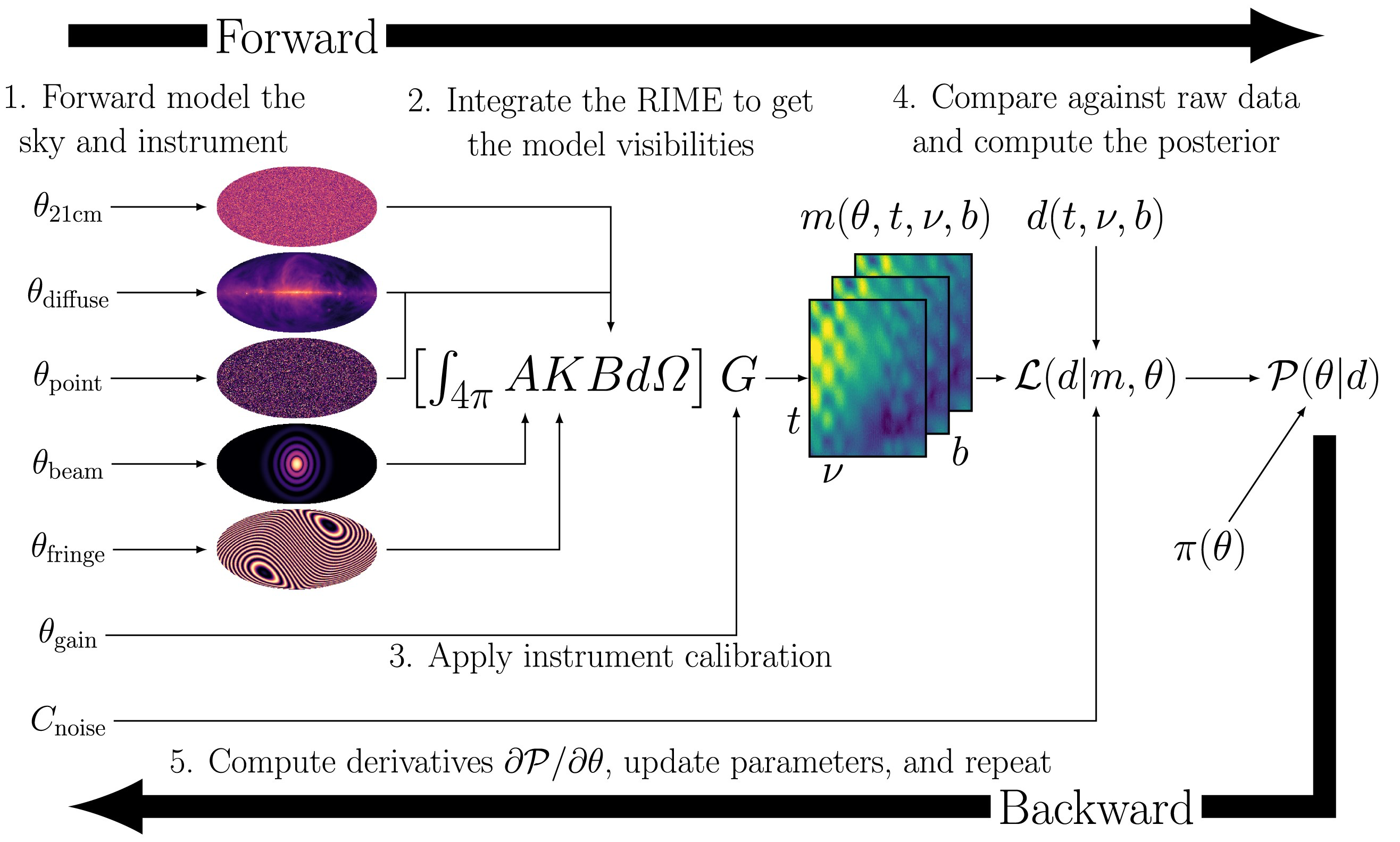}
    \caption{End-to-end Bayesian inference workflow in BayesLIM \cite{Kern2025}. Parameters describing the 21 cm signal, foregrounds, and instrumental response are propagated forward through a differentiable radio interferometric measurement equation, mapping sky emission to model visibilities $m(t, \nu, b)$ via the operators: the primary beam $A$, Fourier/fringe projection $K$, sky brightness operator $B$, and instrumental calibration operator $G$. Model and observed visibilities 
    $d(t,\nu,b)$ are processed with identical analysis operators (e.g. delay-domain high-pass filtering) and compared using a Gaussian likelihood combined with priors to form the posterior. Automatic differentiation provides gradients of the posterior through the full computational graph, enabling MAP optimization and Hessian-preconditioned Hamiltonian Monte Carlo sampling. From \cite{Kern2025}.    
    }
    \label{fig:bayselim_flow}
    \end{center}
\end{figure}

Recent years have seen substantial progress in applying Bayesian methods to 21 cm LIM, particularly to improve foreground mitigation and to quantify uncertainties associated with modeling foregrounds and systematics (e.g. \cite{Sims2019, Rapetti2020, Anstey2021, Byrne2021, Burba2023, Kennedy2023, Murphy2024, Pagano2024, Sims2025}). However, because fully end-to-end Bayesian inference that jointly models the LIM signal, foregrounds, and instrument response is computationally prohibitive, most existing studies necessarily adopt simplifying assumptions. In practice, these analyses either condition on a fixed instrument model or propagate only thermal (uncorrelated) noise, limiting their ability to robustly marginalize over instrumental and foreground uncertainties.

Reference \cite{Kern2025} introduced BayesLIM, a differentiable, end-to-end Bayesian forward-modeling framework for LIM, and demonstrated its performance in the context of 21 cm observations with the HERA interferometer. The key advance is the unification of signal, foreground, and instrument modeling within a single probabilistic framework, implemented as a fully differentiable forward model. This is made feasible by a convergence of developments: modern automatic-differentiation frameworks enable efficient computation of exact posterior gradients through complex forward models; gradient-based optimization and sampling methods scale more favorably with dimensionality than traditional Monte Carlo approaches; and contemporary GPU architectures provide sufficient memory and parallelism to evaluate and store both the forward model and its associated computational graph.

Methodologically, BayesLIM forward-simulates interferometric data using the radio interferometric measurement equation (RIME), generating model visibilities from a pixelized sky convolved with an antenna-independent primary beam and comparing them to measured visibilities via a Gaussian likelihood. To mitigate the dominant contribution from spectrally smooth foreground emission, a Gaussian-process-based high-pass delay-domain filter \cite{Kern2021}, inspired by the DAYENU operator \cite{EwallWice2021}, is applied consistently to both the data and the model prior to likelihood evaluations. This filtering suppresses low-$k_{\parallel}$ foreground modes and focuses inference on scales relevant for EoR power-spectrum measurements, typically $k \geq 0.1 $Mpc$^{-1}$ for the adopted filtering scale. Enabled by differentiable modeling and modern computing resources, this framework allows BayesLIM to perform joint inference across signal, foreground, and instrumental degrees of freedom in a regime of dimensionality that was previously inaccessible to fully end-to-end Bayesian analyses. 
Inference proceeds in two stages: posterior optimization and posterior sampling. Posterior optimization identifies a Maximum A Posteriori (MAP) solution,
\begin{equation}
    \hat{\theta}_{\mathrm{MAP}} =  \mathrm{arg \, max}_{\theta} \, P(\theta | d),
\end{equation}
which maximizes the posterior probability and is equivalent to maximum likelihood estimation with an explicit prior. This step makes use of a second-order optimization routine, the L-BFGS quasi-Newton method, which directly exploits gradients provided by automatic differentiation. The L-BGFS algorithm uses a sparse-Hessian approximation to better navigate ill-conditioned and high-dimensional parameter spaces. Given the highly degenerate and ill-conditioned nature of the parameter space, the author adopts pragmatic strategies such as removing poorly constrained modes (e.g. an $m=0$ degeneracy between sky and beam components) and staged optimization prior to a fully joint run. In the demonstrated configuration, a single gradient update requires approximately 0.3 s when distributed across four NVIDIA A100 GPUs, allowing the full optimization to converge within minutes.

Posterior sampling characterizes uncertainties. BayesLIM employs Hessian-preconditioned Hamiltonian Monte Carlo (HMC), which again relies on gradients of the posterior computed via automatic differentiation to efficiently explore a parameter space approaching $10^5$ dimensions. In the proof-of-concept analysis, effective sample sizes of order $\sim$ a few to $\mathcal O$(10) are achieved for different parameter blocks, aided by adaptive step-size control to maintain sampling efficiency in regions of high curvature.

The principal scientific result is a fully marginalized posterior on the 21 cm power spectrum that consistently propagates uncertainties from both foreground modeling and the instrumental response, rather than relying solely on thermal-noise estimates. The analysis demonstrates that neglecting these additional uncertainties can lead to severe overconfidence and potentially large biases in inferred power spectra, exceeding 10-$\sigma$ for modes between $0.1<k<0.15$ Mpc$^{-1}$ in the example presented (see Reference \cite{Kern2025} Figure 12).

Beyond the core pipeline, the author also introduces a new spherical harmonic basis that is band-limited complete and orthogonal on the spherical stripe, the spherical stripe harmonics, as well as the associated 3D generalization, the spherical stripe Fourier–Bessel basis. They are tailored to drift-scan geometries, improving parameter sparsity and computational efficiency.

Finally, the paper discusses the computational cost and practical limitations of the BayesLIM framework.
In the proof-of-concept implementation, the dominant cost is memory rather than raw compute, as storing both the forward simulation (sky, instrument, and visibilities) and the associated automatic-differentiation computational graph requires more than 200 GB of GPU memory. This challenge is mitigated through multi-GPU data parallelism and gradient accumulation. The authors estimate that extending BayesLIM to more realistic HERA Phase II data volumes is feasible with current computing resources, at an anticipated cost of order a few hundred GPU-hours.

Looking ahead, this approach opens the door to extending end-to-end Bayesian inference to larger data volumes, more realistic instrument models, and additional LIM tracers, providing a path toward fully integrated, multi-probe analyses of the universe.

\section{Current LIM Measurements}
\label{S:measurements}

The LIM field is in its early stages, but significant progress has been made via pilot measurements across a range of different emission lines and redshifts. These first analyses have primarily been made using instruments and surveys designed for other purposes. Nevertheless, these efforts are powerful for exploring LIM measurement techniques and developing analysis pipelines, understanding systematic effects, and for optimizing the instrument design and survey strategy for future, tailored LIM experiments.  They also yield early scientific results. In this section, we briefly review the current state-of-play across several different target lines, with the understanding that the best is yet to come. 

\subsection{21 cm}

As the first LIM sub-field, there are a wide range of ongoing tailored 21 cm experiments, and considerable progress has been made in recent years. The first 21 cm detections have been made in the post-reionization universe and come mainly from cross-correlations between optical galaxy surveys and 21 cm observations with single-dishes (GBT, Parkes, and MeerKAT). There are also interferometric galaxy $\times$ 21 cm measurements from CHIME and MeerKAT, and recent auto-power detections.

\begin{figure}
\begin{center}
\includegraphics[width=\textwidth]{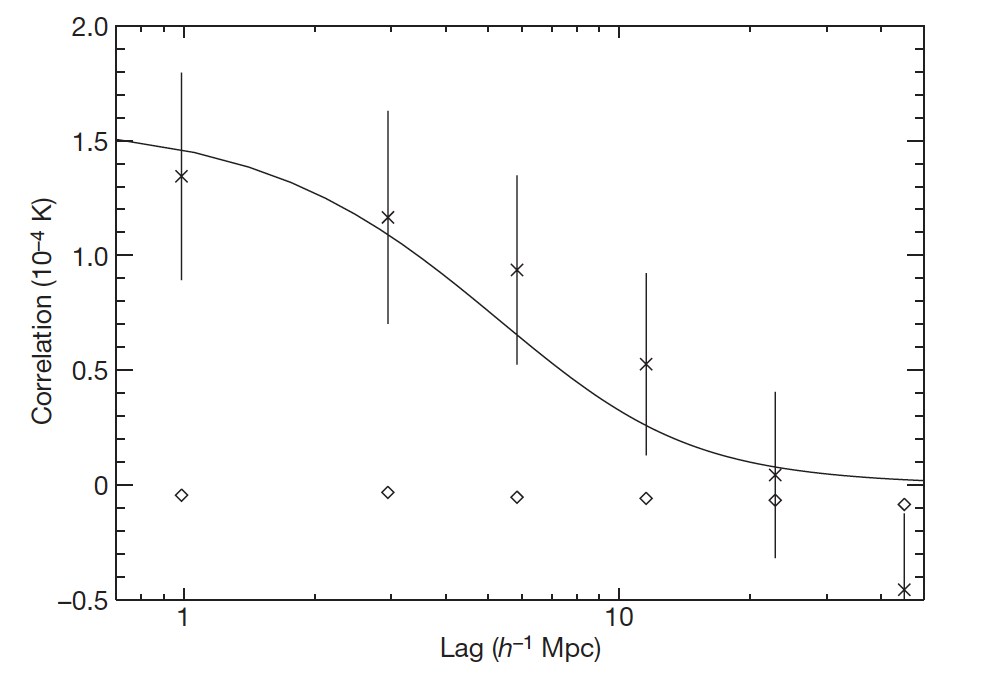}
\caption{Cross correlation between $z \sim 0.8$ redshifted 21 cm fluctuations and optically selected galaxies in the DEEP2 survey \cite{Davis03}. This measurement provided an early proof-of-concept for LIM studies of large-scale structure, 
as it verifies that the 21 cm brightness temperature fluctuations and the optical galaxy survey trace the same large-scale structure. 
From \cite{Chang10}.}
\label{fig:21cm_deep2}
\end{center}
\end{figure}

An initial 21 cm $\times$ 6dF galaxy survey cross-correlation detection using existing HIPASS data was first reported in 2009 \cite{Pen2009}.
Then in 2010, reference \cite{Chang10} presented the first cross-correlation detection using 21 cm measurements from a dedicated 21 cm LIM program. Specifically, these authors conducted a 21 cm survey that spanned the redshift range of $0.53 < z < 1.12$, corresponding to comoving distances of 1400-2600 Mpc/$h$, utilizing the 800 MHz receiver at the GBT. The GBT beam has a FWHM of 15' and so the comoving transverse spatial resolution is 9 Mpc/$h$ at $z=0.8$, while
the high spectral resolution of the radio spectrometer is binned down to 430 kHZ, or 2 Mpc/$h$. The 21 cm observations
span two survey fields, each of 120' $\times$ 30' area, which overlap with the DEEP2 optical galaxy redshift survey \cite{Davis03}, containing $10,000$ DEEP2 galaxies out to $z=1.4$.

As in all redshifted 21 cm experiments, the dominant sources of emission in the data are RFI and continuum foregrounds. 
The RFI is polarized and has sharp frequency structure, and is excised using the cross-polarization data products from
the two linearly-polarized radio signals. In this case, the astrophysical foreground contamination has a brightness
temperature of $\sim$ 125 mK, averaged over the survey fields at these frequencies, and is about $\sim 10^3$ times brighter than the expected 21 cm signals. Using the spectrally-smooth properties of the continuum foregrounds, an SVD (\S \ref{S:PCA})
mitigation approach is used to excise the dominant spectral eigenmodes. Simulated 21 cm signal are injected to account for the 21 cm signal loss. 

The authors then reported an upper limit on the residual 21 cm brightness temperature fluctuations of $464 \pm 277 \, \mu$K, for
$\sim 2 \, \mathrm{Mpc}/h$ scales at a mean redshift of $z = 0.8$. The 21 cm $\times$ DEEP2 galaxy cross-correlation measurement yields a detection out to 10 Mpc/$h$ scales, as shown in Figure \ref{fig:21cm_deep2}. The amplitude of the cross-correlation signal is $157 \pm 42 \, \mu$K at zero lag. This measurement determines the product of the HI abundance, $\Omega_{\mathrm{HI}}$, a bias parameter, $b_{\mathrm{HI}}$ and the stochasticity between the 21 cm and optical galaxy fluctuations, $r$. The measurement yields $\Omega_{\mathrm{HI}}$ = $(5.5 \pm 1.5) \times 10^{-4}$ (1/$r b_{\mathrm{HI}}$) at $z \sim 0.8$. This
measurement is less susceptible to foreground contamination and other systematic effects than inferences from 21 cm auto-correlations. This detection provided the first verification that the 21 cm brightness temperature and optical galaxies
trace the same underlying large-scale structure. This is a powerful proof-of-concept for LIM, and the measurement helped cement LIM as a viable approach for studying large-scale structure. 

Further GBT 21 cm LIM measurements expanded both the spatial coverage and sensitivity of the observations. This led to
an updated cross-power spectrum measurement at $z \sim 0.8$ \cite{Masui13} using optical galaxies from the WiggleZ survey \cite{Drinkwater10}. It also led to an upper limit on the 21 cm auto-power spectrum \cite{Switzer13}. The cross-power spectrum estimate improved the precision and range of scales compared to the earlier measurement in \cite{Chang10}. The cross-power spectrum measurement in \cite{Masui13} has a statistical significance of $4-\sigma$ and is sensitive to scales around
$k \sim 0.4 \, h \, \mathrm{Mpc}^{-1}$. Combining the cross-power spectrum measurement and the auto-power spectrum bound, 
\cite{Switzer13} found $\Omega_{\mathrm{HI}} b_{\mathrm{HI}} = 0.62^{+0.23}_{-0.15} \times 10^{-3}$.

Using the ICA technique (\S \ref{S:ICA}), as implemented in FASTICA, reference \cite{Wolz2017} re-analyzed the same
GBT data set used in \cite{Masui13}. The authors found comparable results to the earlier analysis. 

Next, reference \cite{Anderson2018} cross-correlated 21 cm data from the Parkes telescope with 2dF galaxy survey maps.
The 21 cm data spans the redshift range $0.057 < z < 0.098$, across approximately 1300 deg$^2$ on the sky. The cross-correlation detection significance is $5.7-\sigma$. Interestingly, the amplitude of the cross-power spectrum is low
relative to expectations, assuming a neutral hydrogen mass density and bias derived from the ALFALFA survey at these redshifts. The amplitude deficit is pronounced and statistically significant on small scales: at $k \sim 1.5 \, h$ Mpc$^{-1}$,
the cross-power spectrum is a factor of more than $\sim 6$ lower than expected. This either indicates smaller than expected
neutral hydrogen clustering, a small correlation coefficient between optical galaxies and neutral hydrogen, or some combination of the two. The authors found that red galaxies are more weakly correlated with HI than blue galaxies at these
scales, as expected from local observations of individual galaxies.

A further 21 cm LIM survey was conducted using a Phased-Array Feed (PAF) on the Parkes telescope \cite{Li2021}. A
cross-correlation signal was detected between the 21 cm data and the WiggleZ galaxies at $0.73 < z < 0.78$. The mean amplitude
of the detected signal is $\langle \Delta T_{\mathrm{b}} \delta_{\mathrm{opt}} \rangle = 1.32 \pm 0.42$ mK (statistical errors only).
A future Parkes cryogenic PAF is expected to yield higher significance cross-correlation detections at redshifts up to $z \sim 1$. 

A subsequent stacking analysis of Parkes 21 cm maps around the positions of $48,430$ 2dFGRS galaxies was carried out in 
reference \cite{Tramonte2020}. These results covered $\sim 1,300$ deg$^2$ across $0.06 < z < 0.10$. Ater removing
10-20 PCA modes (\S \ref{S:PCA}), the authors detected the HI signal at 12-$\sigma$ significance. 
The authors measured the HI halo content by jointly fitting for the observed halo mass $M_{\rm v}$ and the normalization $c_{0,\mathrm{HI}}$ for the H I concentration parameter. They found $\log _{10}{(M_{\mathrm{v}}/\text{M}_{\odot })}= 16.1^{+0.1}_{-0.2}$, $c_{0,\mathrm{HI}}=3.5^{+0.7}_{-1.0}$ for the 10-mode and $\log _{10}{(M_{\mathrm{v}}/\text{M}_{\odot })}= 16.5^{+0.1}_{-0.2}$ , $c_{0,\mathrm{HI}}=5.3^{+1.1}_{-1.7}$ for the 20-mode removed maps.
These detections trace the HI emission from multiple galaxies residing within massive halos. The authors found a marginally signifiant indication that satellite galaxies are more HI rich than central galaxies. This work illustrated the potential of studying the HI content within dark matter halos using 21 cm LIM.

Reference \cite{Wolz2022} reported a new cross-correlation measurement using GBT data combined with three optical galaxy samples: the Luminous Red Galaxy (LRG) and Emission Line Galaxy (ELG) samples from the eBOSS survey, and the WiggleZ galaxies. 
Foreground removal was performed with FASTICA, while mock realizations were used to correct for the resulting signal loss.
The authors reported HI-galaxy cross-power spectrum measurements at $z \simeq 0.8$ and $k_{\rm eff}=0.31 \, h\,{\rm Mpc^{-1}}$.
They found $\Omega_\mathrm{HI} b_\mathrm{HI} r_{\mathrm{HI,Wig}} = [0.58 \pm 0.09 \, {\rm (stat) \pm 0.05 \, {\rm (sys)}}] \times 10^{-3}$ for GBT-WiggleZ, $\Omega _{\mathrm{HI}} b_{\mathrm{HI}} r_{\mathrm{HI},{\rm ELG}} = [0.40 \pm 0.09 \, {\rm (stat) \pm 0.04 \, {\rm (sys)}}] \times 10^{-3}$ for GBT-ELG, and $\Omega _{\mathrm{HI}} b_{\mathrm{HI}} r_{\mathrm{HI},{\rm LRG}} = [0.35 \pm 0.08 \, {\rm (stat) \pm 0.03 \, {\rm (sys)}}] \times 10^{-3}$ for GBT-LRG. Interestingly, the cross-correlations with WiggleZ galaxies give the highest amplitudes and statistical significance. Qualitatively, this is perhaps unsurprising since the measurements trace relatively small scales, and the WiggleZ galaxies tend to be blue galaxies which typically contain more neutral hydrogen gas than red galaxies such as LRGs.
These measurements start to trace the HI content of galaxies beyond those in the local universe.

Another recent cross-correlation detection of 21 cm emission was made by the Canadian Hydrogen Intensity Mapping Experiment (CHIME) \cite{CHIME2022}. The CHIME collaboration acquired over 102 nights of data, constructed foreground-filtered 21 cm maps, and stacked these on the locations of LRGs, ELGs, and quasars (QSOs) from the eBOSS catalogs. These are the first 21 cm LIM measurement detections from an interferometer. They determined an effective amplitude of HI fluctuations, defined as $\mathcal{A}_{\rm HI}\equiv 10^{3}\,\Omega_\mathrm{HI}\left(b_\mathrm{HI}+\langle\,f_\Omega  \mu^{2}\rangle\right)$, where  $\langle\,f_\Omega \mu^{2}\rangle=0.552$ encodes the effect of redshift-space distortions at linear order. They found $\mathcal{A}_\mathrm{HI}=1.51^{+3.60}_{-0.97}$ for LRGs $(z=0.84)$, $\mathcal{A}_\mathrm{HI}=6.76^{+9.04}_{-3.79}$ for ELGs $(z=0.96)$, and $\mathcal{A}_\mathrm{HI}=1.68^{+1.10}_{-0.67}$ for QSOs $(z=1.20)$. This constraint is weaker than the previous results from cross-correlations between GBT 21 cm maps and galaxy catalogs \cite{Chang10,Masui13,Wolz2022}. The authors attribute
this to their marginalizing over uncertain non-linear modeling parameters on small scales.  With the stacking analysis in the frequency dimension, the authors also reported a bias in the spectroscopic redshifts of each tracer, and found a non-zero bias $\Delta\,v= -66 \pm 20 \mathrm{km/s}$ for the QSOs. The QSO catalog was split into three redshift bins with decisive detections in each, with the upper bin at $z=1.30$.

\begin{figure}
    \begin{center}
        \includegraphics[width=0.9 \textwidth]{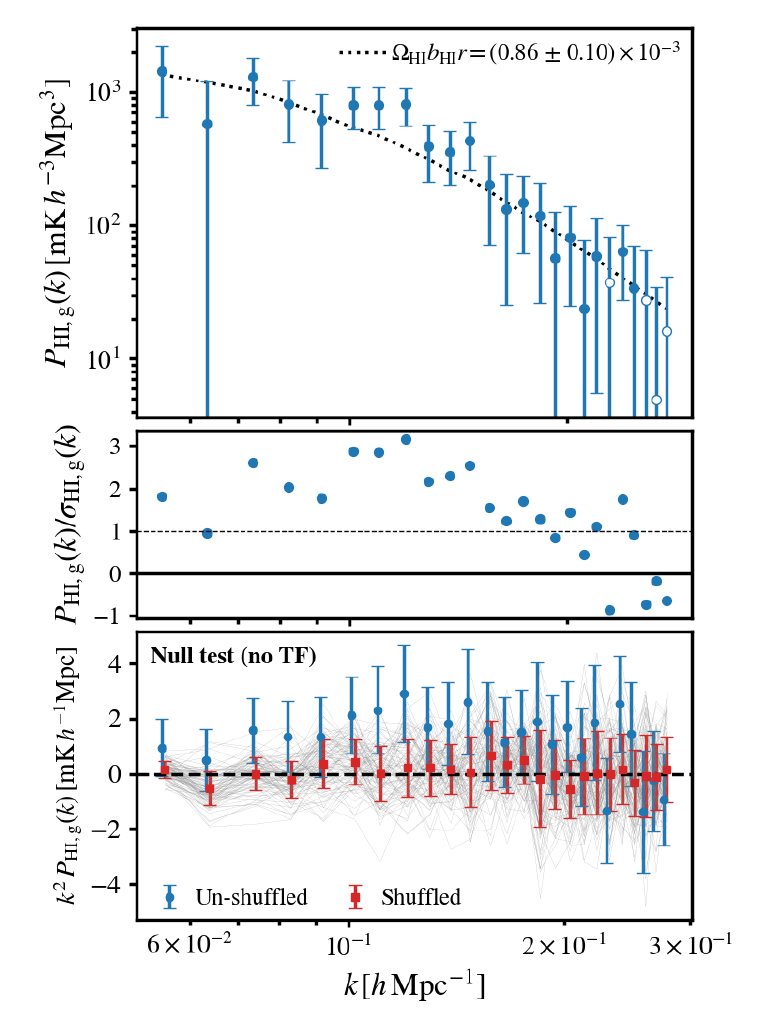}
        \caption{Cross-power spectrum between optically-selected galaxies in the Wigglez survey and HI LIM data from MeerKAT at $z = 0.4-0.459$. {\em Top:} The blue points show measurements with $1-\sigma$ error bars, while the black dotted line is a best-fit linear theory model. The product of the HI density, $\Omega_{\rm HI}$, the HI clustering bias, $b_{\rm HI}$, and the stochasticity parameter, $r$, has been adjusted to best match the measurements. The HI data has been cleaned of foregrounds and systematic effects by removing the PCA eigenmodes with the $30$ largest eigenvalues. The detected $k$-bins span from $k \sim 0.05-0.28 \, h$ Mpc$^{-1}$. {\em Middle:} The ratio of the power spectrum measurement to the estimated error bar in each $k$-bin.
        {\em Bottom:} The measurements (blue) compared to a successful null test (red). The red points show the average cross-power when the Wigglez galaxies are randomly shuffled in redshift and are consistent with zero, while the gray lines show 100 different shuffling realizations.  
        From \cite{Cunnington:2022uzo}.}
        \label{fig:meerkat_wigglez}
    \end{center}
\end{figure}

Further recent progress has been made by cross-correlating early MeerKAT data with optically-selected galaxies from the Wigglez survey \cite{Cunnington:2022uzo}. In this analysis, the authors treat each of the 64 elements from the MeerKAT array as a single dish, rather than making interferometric measurements. In a pilot survey, they combine 10.5 hours of HI LIM data at $\nu_{\rm obs} = 973-1015$ MHz, corresponding to HI redshifts from $z=0.400-0.459$, across $\sim 200$ deg$^2$ in the Wigglez survey, to detect the 21 cm-galaxy cross-power spectrum. The authors employ a PCA-based foreground subtraction technique (see \S \ref{S:PCA}) to identify and excise eigenmodes which are likely corrupted by foregrounds and other systematics. They use simulations to account for the signal loss that follows from this filtering. 

The authors carefully test the sensitivity of their cross-power spectrum measurements to the number of PCA eigenmodes that are removed from the data. After varying the number of corrupted modes removed, they find that excising $N_{\rm fg} \sim 30$ PCA
modes gives close to the maximal signal-to-noise ratio for their cross-power spectrum detection, while also yielding a reasonable goodness-of-fit with a $\chi^2$ per degree-of-freedom of around unity. 
The authors also present null tests, in which they cross-correlate the MeerKAT data after shuffling the Wigglez galaxies in redshift. Alternatively, they correlate the {\em random} catalogs from Wigglez with the redshifted 21 cm measurements: encouragingly, both tests yield
results consistent with zero. The cross-power spectrum detection, for $N_{\rm fg} = 30$, is shown in Figure \ref{fig:meerkat_wigglez}. The measurements here span $k \sim 0.05-0.28 \, h$ Mpc$^{-1}$ and the total significance is $7.7-\sigma$. 

At these spatial scales and redshifts, a linear theory analysis should suffice, at least for the current measurement precision. Assuming a standard LCDM model for the linear mattter power spectrum, and using Wigglez auto-power spectrum measurements to determine $b_{\rm gal}$, the authors find the best-fit value of $\Omega_{\rm HI} b_{\rm HI} r$ from their measurements of the angle-averaged cross-power spectrum. This gives:
$\Omega_{\rm HI} b_{\rm HI} r = (0.86 \pm 0.10) \times 10^{-3}$ (see Figure \ref{fig:meerkat_wigglez}). These results appear to be consistent with previous estimates at slightly higher redshift.
The early MeerKAT measurements suggest a hopeful outlook for future
analyses with more data both in cross-correlation with optically-selected galaxies and in auto-correlation.

In fact, using a different strategy, a first possible 
detection of the 21 cm auto-power spectrum has been achieved in an early analysis of interferometric MeerKAT data \cite{Paul:2023yrr}.
In this study, the authors measure the power spectrum from 96 hours of data across a special $\sim 2$ deg$^2$ patch of the sky with relatively weak point source foreground contamination. After choosing a ``quiet patch'' of the sky, the measurements may be less susceptible to a range of systematic challenges which generally mix spatial foreground structure into frequency variations (where they can be difficult to separate from the signal). 
The above work considers two frequency bins centered around $\nu_{\rm obs} = 1077.5$ MHz and $\nu_{\rm obs}=986$ MHz, each with a width of $\Delta \nu = 46$ MHz, capturing redshifted 21 cm radiation from $z = 0.32$ and $z=0.44$, respectively. The authors estimate the cylindrically-averaged power spectrum, $P(k_\parallel,\k_\perp)$, after taking two steps to mitigate foreground contamination and other systematic effects. First, Fourier pixels with power larger than five times that of the estimated thermal noise are masked. Second, regions in the Fourier plane within the foreground corrupted ``wedge'' -- as conservatively determined in the ``horizon limit'' (see Eq.~\ref{eq:wedge}) -- are avoided. 

After these cuts, the authors claim detections of the 21 cm auto-power spectra at $8$ and $11.5-\sigma$ in the $z=0.32$ and $z=0.44$ bins, respectively. The binned power spectrum estimates
at, e.g. $z=0.44$, span (total) wavenumbers from $k \sim 0.34-7.04$ Mpc$^{-1}$. It is important to note here, however, that the measured modes after foreground avoidance mainly lie close to the line-of-sight direction. That is, the study mainly measures high $k_\parallel$ modes, which are subject to strong finger-of-god and line-broadening effects. (Note that the line-broadening impacts the shot-noise term, while finger-of-god effects do not.)  In comparing these results with theoretical models, it is hence important to account for the precise mode-sampling in $k_\parallel$ and $\k_\perp$.  We refer the reader to the original paper for a comparison between these results and models \cite{Paul:2023yrr}, as well as an independent investigation in \cite{Padmanabhan:2023hfr}. 
These early MeerKAT auto-power and cross-power detections, along with the cross-correlation analyses from CHIME, provide an encouraging outlook for upcoming HI LIM measurements. 

\begin{figure}
    \begin{center}
        \includegraphics[width=0.9 \textwidth]{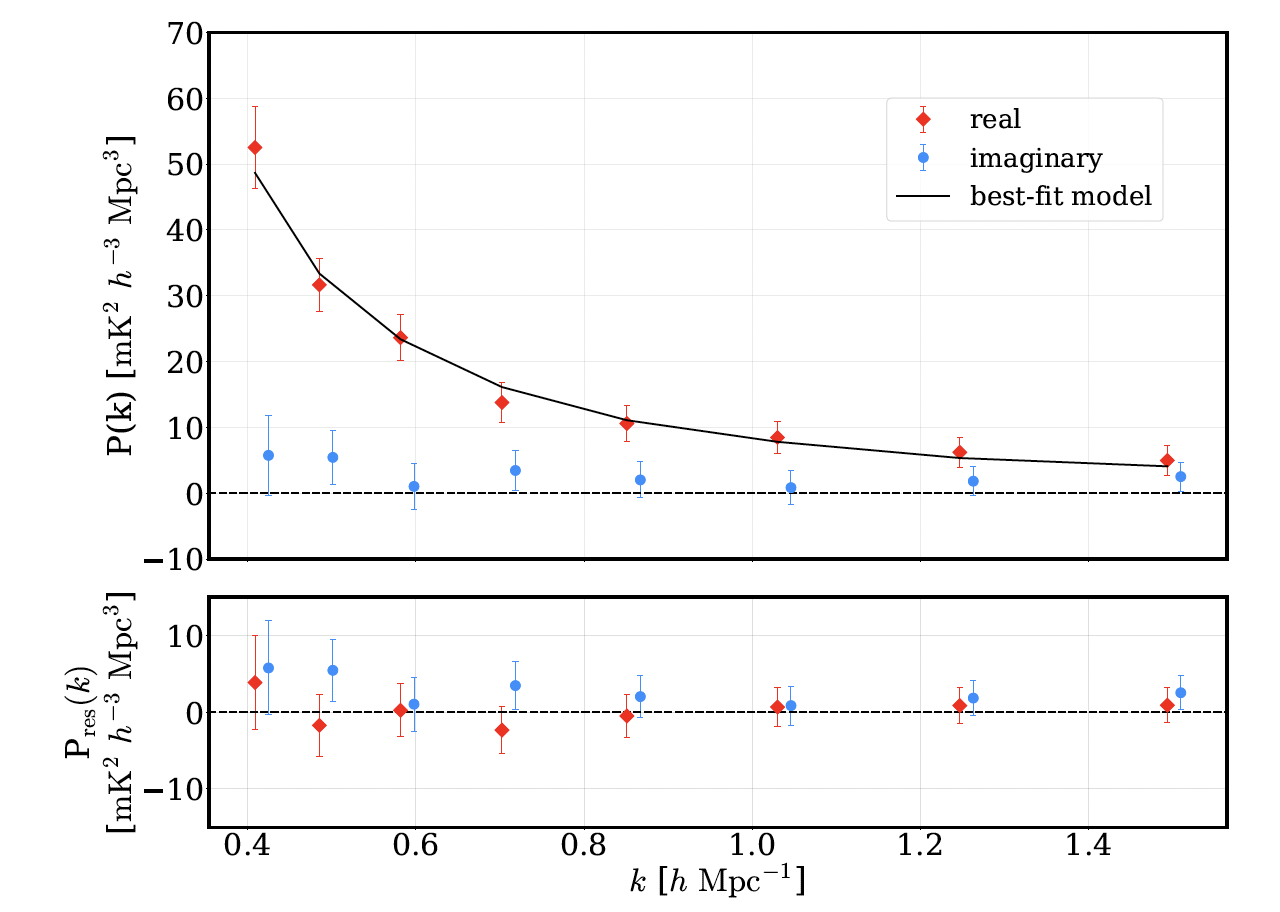}
        \caption{A detection of the 21 cm auto-power spectrum from the CHIME collaboration. {\em Top panel}: The
        red points with $1-\sigma$ error bars show CHIME measurements in a frequency band spanning redshifts from $z=1.01-1.34$, while the black curve is a best fit model. The blue points show a null test, based on the imaginary component of the cross-power spectra between different time chunks of the data. {\em Bottom panel:} The red points show the residuals around the best fit model. The blue points are again the imaginary components, offset slightly in $k$ for clarity. The model is a good fit to the measurements, and the detection significance is $12.5-\sigma$. The auto-power spectrum amplitude is consistent with the expectations from cross-correlations with galaxy surveys. From \cite{CHIME:2025cee}.}
        \label{fig:chime_auto}
    \end{center}
\end{figure}

In fact, as we were finalizing this review, the CHIME collaboration announced a $12.5-\sigma$ detection of the 21 cm auto power spectrum from their data near $z \sim 1$. Their auto-power spectrum measurements, a best-fit model, and a null test are shown in Figure~\ref{fig:chime_auto}. This work includes a number of analysis advances, including improved RFI mitigation schemes, and refined algorithms for filtering foreground contamination. 
The current detection spans moderate wavenumbers between $0.4 \, h \, \mathrm{Mpc}^{-1} < k < 1.5 \, h \, \mathrm{Mpc}^{-1}$, as larger scale measurements are still systematics-limited. Importantly, the amplitude of the 21 cm auto-power spectrum is largely controlled by $\Omega_{\rm HI} b_{\rm HI}$, as discussed previously (assuming negligible stochasticity, $r=1$), while this quantity can also be inferred from cross-correlations with galaxy surveys.
The only caveat being that the moderate wavenumbers used in the CHIME analyses require careful modeling of finger-of-god effects. {See reference \cite{Osinga:2025gpt} for a detailed comparison between \textsc{Illustris-TNG} based models for the post-reionization 21 cm signal and common analytic prescriptions, including the role of finger-of-god damping.}

Nevertheless, the auto-power spectrum in Figure~\ref{fig:chime_auto} is consistent with the expectations from earlier cross-correlation $\Omega_{\mathrm{HI}} b_{\mathrm{HI}}$ estimates. This provides a powerful validation of the robustness of the 21 cm auto-power spectrum measurements. This is an important milestone for the LIM field as it establishes that the 21 cm auto-power spectrum can be robustly measured despite the strong foreground contamination and associated systematic challenges. Indeed, as discussed in \S \ref{S:challenges}, 21 cm LIM has the largest continuum foreground-to-signal ratio among LIM target lines. Note also that the foreground-to-signal ratio for $z \sim 1$ 21 cm LIM is broadly comparable to that for reionization-era 21 cm LIM. The successful CHIME auto-power spectrum measurements should hence provide encouragement for 21 cm EoR experiments, and for the LIM field more generally. Although we will not explicitly discuss results from reionization-era 21 cm surveys here, we note that these LIM experiments have also made considerable progress recently (e.g. \cite{HERA2025,Mertens2025,Nunhokee2025}.)

\subsection{CO}

\begin{figure}
\begin{center}
    \includegraphics[width=\textwidth]{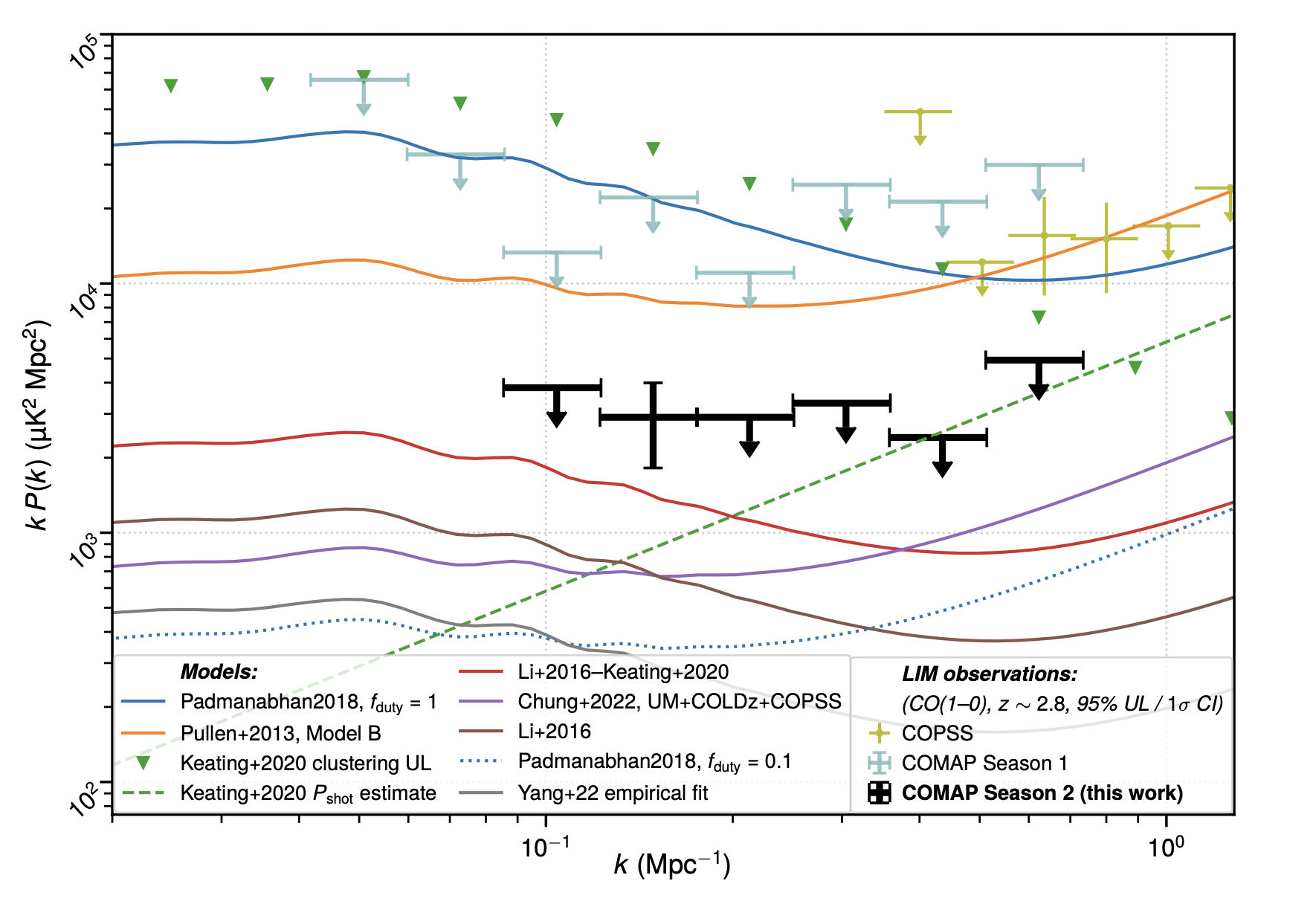}
    \caption{{CO(1-0) power spectrum results at $z \sim 3$ from COMAP Season 2 observations compared to various previous measurements and models from the literature. The solid black downward arrows show the new COMAP 2-$\sigma$ upper bounds, while the blue points are from COMAP Early Science measurements. The yellow points and bounds are the COPSS shot-noise results from \cite{Keating:2015qva}, while green triangles show COPSS bounds on the clustering term from \cite{Keating:2020wlx}. The green dashed line is a refined shot-noise estimate from \cite{Keating:2020wlx}. In the $k$-bins where COMAP 2 and COPSS results show a hint for non-zero signals, $1-\sigma$ error bars are given.
    The curves show various models for the CO(1-0) auto-power spectrum at $z \sim 3$ from \cite{Pullen13}, \cite{Padmanabhan:2017ate}, \cite{Keating:2020wlx}/\cite{Li:2015gqa}, and \cite{Chung:2022zeu}. The COMAP 2 results already disfavor two of the larger amplitude models in the literature, primarily by bounding the clustering term. From \cite{Chung:2024iob}.}
    }
    \label{fig:comap_year2}
\end{center}
\end{figure}

Several pilot studies have placed upper limits on the CO LIM signal using different data sets, redshift ranges, and various rotational transitions, while some tentative detections have been achieved. Considering first the null detections, \cite{Pullen13} combined WMAP data with SDSS photometric quasar and LRG samples to bound the bias-weighted CO brightness temperature in the CO(1-0) and CO(2-1) transitions. In the shot-noise regime, \cite{Uzgil19} placed an upper limit on the high-$k$ CO auto-power using ALMA Spectroscopic Survey Large Program (ASPECS LP) data. Reference \cite{Keenan:2021uue} bounded the CO(1-0) $\times$ optical-galaxy cross-power spectrum using data from the CO Power Spectrum Survey (COPSS) and spectroscopic galaxy catalogs in the GOODS-N field at $z \sim 3$. Finally, the COMAP collaboration placed the first upper limit on the CO(1-0) auto-power spectrum at $z \sim 3$ in the clustering regime \cite{COMAPI2022,COMAPV2022}. Turning to early detections, \cite{Keating16} made a $2-\sigma$ level measurement of the CO(1-0) auto-power spectrum in the shot-noise dominated regime near $z \sim 3$ from COPSS (see Figure \ref{fig:comap_year2}). In addition, a $\sim 4-\sigma$ detection was made in \cite{Keating:2020wlx} from the Millimeter-wave Intensity Mapping Experiment (mmIME), which combines ASPECS and Atacama Compact Array (ACA) data at $\sim$ 3 mm wavelength. Finally, \cite{Roy:2024kzc} stacked CO maps from the Planck survey around eBOSS galaxies at $z \sim 0.5$, finding a $\sim 3-\sigma$ signature of excess CO(3-2) emission. 
A still more recent CO detection from \cite{Chiang26} will be described in the next ([CII]) section. 

Here we briefly describe the current shot-noise regime detections from \cite{Keating16,Keating:2020wlx} and the COMAP upper bound. Building on the COPSS analysis, \cite{Keating:2020wlx} measures the total high-$k$ power around an observed frequency of $100$ GHz. They find that this measurement is consistent with a slightly updated version of the CO power spectrum model of reference~\cite{Li:2015gqa}. This model is then used to derive constraints on the CO shot-noise contributions from multiple rotational transitions at various redshifts, with each contributing to the observed specific intensity fluctuations at 100 GHz. Specifically, their inferred shot-noise power is $P_{\rm CO} = 120^{+80}_{-40} \, \mu {\rm K}^2$ (Mpc/$h$)$^3$ for CO(2-1) at $z=1.3$; $P_{\rm CO} = 200^{+120}_{-70} \, \mu {\rm K}^2$ (Mpc/$h$)$^3$ for CO(3-2) at $z=2.5$; and $P_{\rm CO} = 90^{+70}_{-40} \, \mu {\rm K}^2$ (Mpc/$h$)$^3$ for CO(4-3) at $z=3.6$. Using line-ratio models, the authors further convert these results into estimates of the CO(1-0) power spectrum at the above redshifts. Finally, \cite{Keating:2020wlx} derive constraints on the cosmic average mass density in molecular hydrogen assuming a constant $\alpha_{\rm CO}$ factor (see \S \ref{S:co_lum_model}) to convert between their CO models and molecular hydrogen gas density. See \cite{Keating:2020wlx} for details. 

First year CO(1-0) power spectrum results from early COMAP data already provided interesting bounds on the CO(1-0) power spectrum at $z \sim 3$.
This measurement set an upper limit on the power in a wavenumber band spanning from $k = 0.051-0.62$ Mpc$^{-1}$, where it is expected that CO fluctuations primarily trace the clustering term, as opposed to shot-noise variations. 
Reference \cite{COMAPI2022} further bounds the total CO(1-0) brightness temperature, assuming $\avg{b} \geq 2$, finding $\avg{T} \leq 3.6\, \mu {\rm K}$. Finally, adopting a constant $\alpha_{\rm CO}$ (see \S \ref{S:co_lum_model}) value of $\alpha_{\rm CO} = 3.6 \, M_\odot \left({\rm K \, km \, s^{-1} \, pc^{2}}\right)^{-1}$, the authors place an upper limit on the molecular hydrogen mass density, $\rho_{\rm H_2} \leq 2.5 \times 10^8 \, M_\odot \, {\rm Mpc}^{-3}$ (see Eq.~\ref{eq:rhoh2_from_tb}, and \S \ref{s:molecular}). 

Furthermore, the COMAP team has recently presented results from analyses of their Season 2 data which tighten these earlier constraints \cite{Lunde:2024rut,Stutzer:2024rps,Chung:2024iob}, as shown in Figure \ref{fig:comap_year2}. 
The new analyses improve the bound on the CO clustering term by a factor of two and their previous upper limit on the CO shot-noise fluctuations by a factor of five \cite{Chung:2024iob}. The COMAP Season 2 data already disfavor previously plausible models from the literature, primarily through measurements in the clustering regime. 
The new shot-noise limit disfavors (at 2.3-$\sigma$) the tentative earlier shot-noise detections from COPSS \cite{Keating:2015qva}, but the COMAP results remain consistent with the reanalysis of \cite{Keating:2020wlx}. The improved Season 2 measurements also yield a tighter bound on the molecular hydrogen gas density of $\rho_{\rm H_2} \leq 1.6 \times 10^8 \, M_\odot \, {\rm Mpc}^{-3}$ (see Figure~\ref{fig:rho_H2}). 

In addition, an improved three-dimensional stacking analysis of Season 2 data on eBOSS/DESI quasars yields a non-detection \cite{Dunne2026}. The upper limit of $<L'_{\rm CO}> = 10.0 \times 10^{10}$ K km s$^{-1}$ pc$^2$ is in mild disagreement with current CO models. The authors discuss plausible feedback and environmental quenching effects that may suppress this signal. They also point out that their stacked signal is 
likely dominated by the CO emission from correlated neighboring halos, rather than from the central quasar host galaxy itself. This complicates the interpretation of the stacked measurements (as discussed in \S \ref{sec:stacking}). In any case, the continuing progress in the auto-power and cross-correlation analyses shows that COMAP is on track to yield CO fluctuation detections within a few years \cite{Chung:2024iob}.

\subsection{[CII]}

There have been several efforts to detect [CII] LIM signals by exploiting cross-correlations with galaxy and quasar surveys \cite{Pullen:2017ogs,Yang:2019eoj,Anderson:2022svu}. First, reference~\cite{Pullen:2017ogs} cross-correlated Planck HFI maps at 353, 545, and 857 GHz with SDSS quasars and CMASS galaxies. In particular, this study searches for excess
quasar-correlated emission in the 545 GHz Planck channel -- above the expected contribution from correlated CIB emission -- owing to [CII] emitting gas at redshifts between $z=2-3.2$. That is, [CII] emission at these redshifts lands in the Planck 545 GHz band and will spatially correlate with large-scale structure, here as traced by SDSS quasars, at similar redshifts. Reference \cite{Pullen:2017ogs} measures six separate cross-power spectra (between each of three frequency bands and two spectroscopic tracer surveys), but only the 545 GHz x quasar measurement
is used to extract the $z \sim 2-3$ [CII] emission. The other channels are nevertheless important for pinning down CIB emission fluctuation models: CIB emission at 545 GHz will also correlate with the quasar distribution and so it is
important to account for this contribution accurately. 

After fitting a joint CIB and [CII] emission model to the six cross-power spectrum measurements, and further accounting for cross-correlations between the tSZ effect \cite{Sunyaev1970} and the large-scale structure catalogs, \cite{Pullen:2017ogs} finds a hint for excess emission from
[CII] at $z \sim 2-3$. In a follow-up work, reference \cite{Yang:2019eoj} refines this analysis, mainly by applying a more optimal weighting scheme to the Planck and large-scale structure maps. Assuming the excess correlated emission is entirely from [CII], these authors find 
$\avg{b}_{\rm CII} \avg{I_\nu}_{\rm CII} = 2.0^{+1.2}_{-1.1} \times 10^5$ Jy/sr at 95\% confidence level. That is, provided the correlated excess traces [CII] emission, it can be used to determine the product of the luminosity-weighted bias factor and the average specific intensity in [CII] near $z \sim 2.6$ (see the large-scale limit of Eq.~\ref{eq:px_gal}, although note that here the authors work with projected two-dimensional fields and so, in practice, use slightly different formulas). This is a statistically significant, $\sim 3.5-\sigma$, detection. The authors further show that this result agrees with some of the larger model predictions in the literature \cite{Gong11,Silva:2014ira}. The only important caveat, as emphasized in \cite{Yang:2019eoj}, is that the excess might
owe partly to quasar-correlated CIB (which might be underestimated in the model) rather than [CII]\footnote{Related to this, a possible concern is that the CIB at 545 GHz may be imperfectly correlated with that at 353 and 857 GHz. The latter frequency bands then provide imperfect templates for the 545 GHz CIB.}. 
The clear path to a more definitive detection is to improve on the spectral resolution of the Planck data: then the line emission can be more directly separated from spectrally-smooth components like the CIB. 

\begin{figure}
\begin{center}
\includegraphics[width=\textwidth]{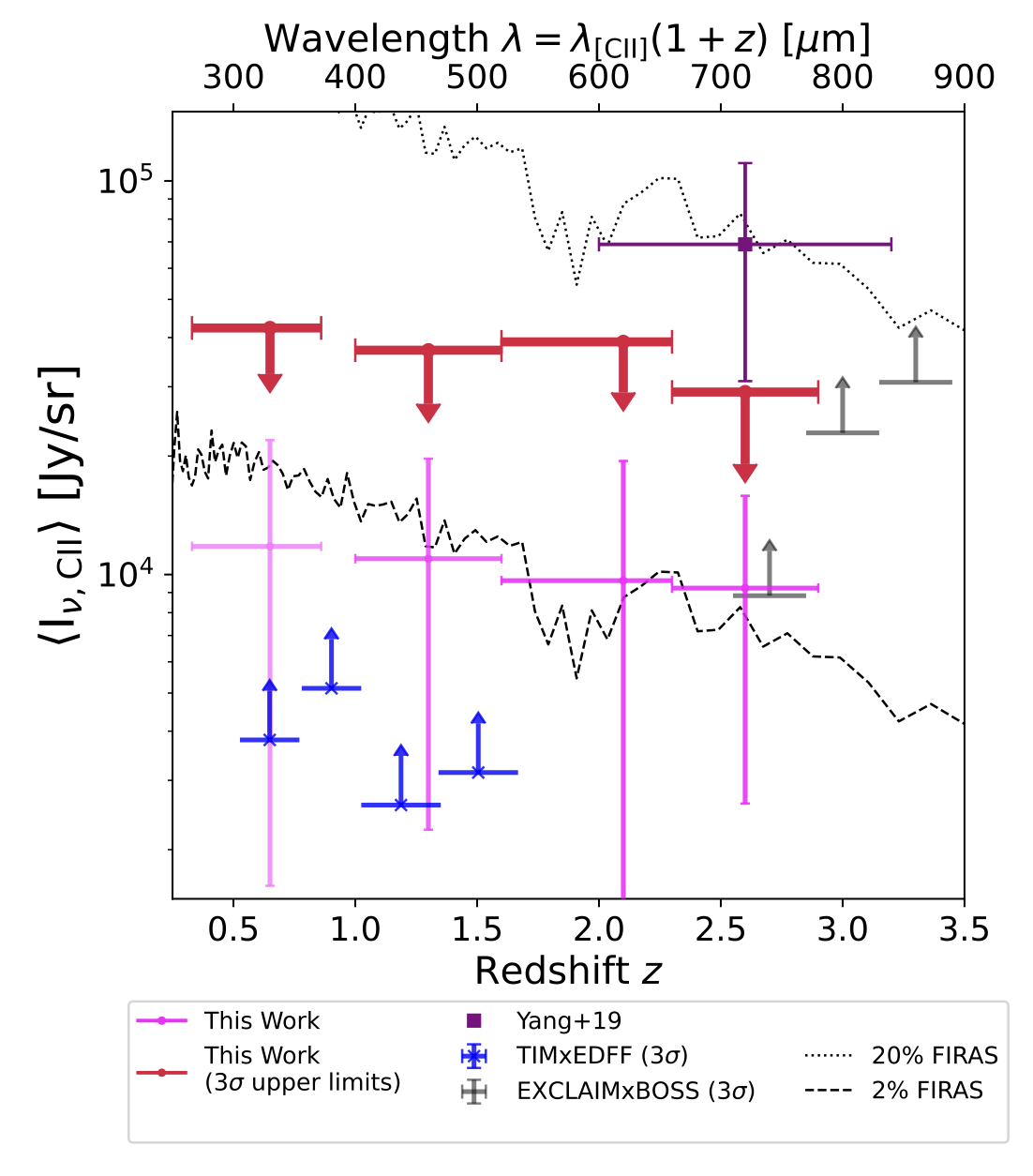}
    \caption{{The pink data points and (1-$\sigma$) error bars show best-fit estimates of the mean specific intensity of [CII] emission based on a novel stacking analysis of IR data around
    photometric galaxies in the COSMOS field. The downward pointing red arrows indicate the $3-\sigma$ upper bound on [CII] intensity implied, which disfavors a [CII] origin for the signal found in \cite{Yang:2019eoj} (purple point and error bars). For reference, the dotted and dashed lines show fractions of the COBE/FIRAS CIB monopole measurements, with the [CII] signal expected to be $\lesssim \mathcal{O}(1-2\%)$ of the CIB intensity. The blue points and upward-facing arrows show the expected $3-\sigma$ lower bounds on the [CII] signal from future TIM $\times$ Euclid galaxy cross-power spectrum measurements, while the grey points with upward arrows are for future EXCLAIM $\times$ BOSS measurements.
    From \cite{Agrawal:2025uig}.}}
    \label{fig:cii_bound_agrawal}
\end{center}
\end{figure}

Notably, reference~\cite{Agrawal:2025uig} carried out a recent stacking analysis of {\em Spitzer, Herschel} and SCUBA-2 maps around photometric galaxies from the COSMOS field.  This stacking analysis reveals a hint for [CII] emission and bounds the intensity of [CII] emission from $z \sim 0.6-2.6$, {as shown in Figure~\ref{fig:cii_bound_agrawal}.} This disfavors the larger [CII] signal tentatively inferred in the \cite{Yang:2019eoj} analysis. Quantitatively, the $3-\sigma$
{\em upper bound} from \cite{Agrawal:2025uig} at $z \sim 2.6$ exceeds the 95\% confidence {\em lower-limit} on the signal from \cite{Yang:2019eoj}. Future TIM 
and EXCLAIM measurements (see \S \ref{S:projects}) should allow decisive [CII] measurements and clarify the signal strength here. 

Reference~\cite{Anderson:2022svu} performs a cross-power spectrum analysis using FIRAS data and BOSS galaxy samples to search for [CII] emission at $0.24 < z < 0.69$. The FWHM of the FIRAS beam is around 7 degrees and so this measurement is confined to large angular scales. However, the finer spectral resolution of FIRAS, as compared to the earlier Planck measurements, helps isolate line emission contributions. Ultimately, these authors bound the bias weighted intensity to $\avg{b}_{\rm CII} \avg{I_\nu}_{\rm CII} < 3.1 \times 10^5$ Jy/sr at $z \sim 0.35$ and
$< 2.8 \times 10^5$ Jy/sr at $z \sim 0.57$ at 95\% confidence. See that work for a comparison of these bounds with current models in the literature. They also show that a related measurement with a more modern instrument, such as the proposed PIXIE experiment, would enable high-significance detections of even pessimistic [CII] models. 

\begin{figure}
\begin{center}
    \includegraphics[width=\textwidth]{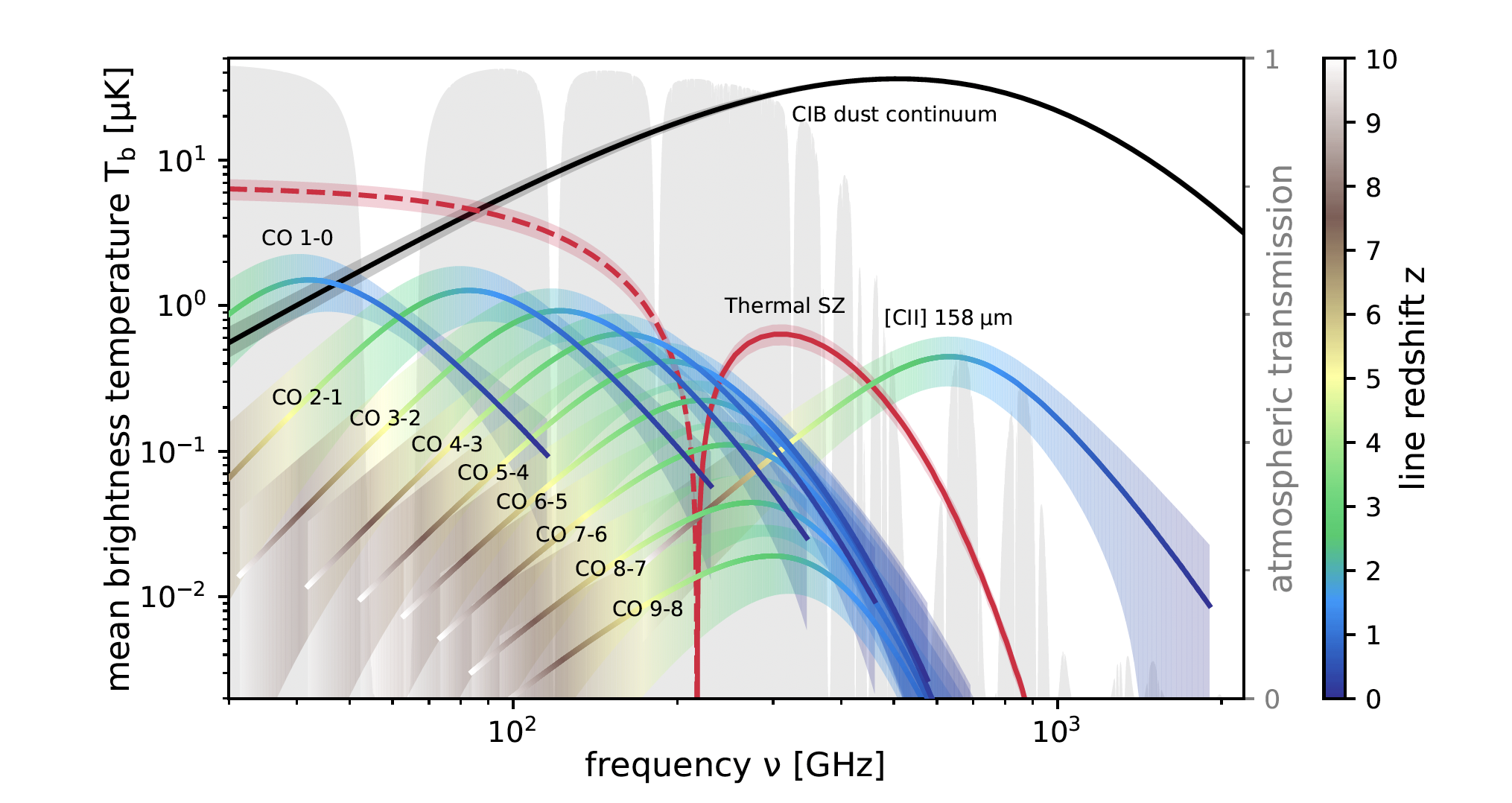}
    \caption{The mean CO and [CII] brightness temperatures inferred from cross-correlating {\em Planck, Herschel}, and {\em FIRAS} data intensity maps with SDSS galaxies and quasars in the two-halo regime. The sky-averaged CIB and tSZ signals are shown for comparison. The shaded bands indicate 68\% confidence intervals from fitting the cross-correlations, along with the {\em Planck and FIRAS} monopoles, using parameterized models for the bias-weighted CIB emissivity and emission line strengths. The emission line redshfits are indicated by the colorbar on the right.  
    For reference, the y-axis on the right and the gray regions show the atmospheric transmission as a function of frequency. These new empirical CO and [CII] constraints can help guide future LIM survey efforts.  
    From \cite{Chiang26}.}
    \label{fig:monopoles_co_cii_cib}
\end{center}
\end{figure}

Figure~\ref{fig:monopoles_co_cii_cib} shows recent reported {\em detections} of the CO and [CII] background signals, obtained by cross-correlating {\em Planck, Herschel}, and {\em FIRAS} intensity maps with galaxies and quasars from SDSS \cite{Chiang26}. 
The CO background is detected at 7-$\sigma$, while [CII] is detected at 3-$\sigma$ significance. The measurements are carried out in the two-halo regime, where the signal is set mainly by the specific intensity of the emission and a luminosity-weighted clustering bias factor (see Eq.~\ref{eq:px_gal} and \S \ref{S:px_multiline}). The author uses parameterized models to jointly fit the CIB, multiple CO rotational transitions (through a parameterized spectral line energy distribution model), and the [CII] emission. The sky-averaged (monopole) emission signals measured by {\em Planck} and {\em FIRAS} are used to help break the bias-intensity degeneracy inherent in the cross-correlation measurements. 
We refer the reader to \cite{Chiang26} for full analysis details and a discussion of the scientific implications. 
Notably, this work already yields informative empirical constraints on the CO and [CII] LIM signals, which can help guide the design of future LIM surveys. Forthcoming LIM experiments with higher spectral resolution are expected to deliver significantly higher signal-to-noise ratio measurements.

As discussed in \S \ref{S:projects}, the LIM community is advancing a broad range of special-purpose instruments to place higher significance [CII] detections. These target [CII] emission across a broad range of redshifts and should already determine the [CII] luminosity density across much of cosmic history. 

\subsection{Ly-\texorpdfstring{$\alpha$}{alpha}}
\label{sec:lya_current}

In some sense, the first Ly-$\alpha$ LIM observations were performed over a decade ago now when stacking measurements around $z \sim 2-3$ Lyman-break selected galaxies revealed extended low surface brightness ``halos'' of Ly-$\alpha$ emission \cite{Steidel2011} (see also e.g. \cite{Momose14,Wisotzki18}). This extended emission is likely the result, in part, of the scattering of Ly-$\alpha$ photons, originally sourced in recombination cascades within a galaxy's HII regions, by neutral hydrogen in the surrounding CGM (e.g. \S \ref{S:lya_rt}). In addition to these relatively small-scale Ly-$\alpha$ stacking measurements, there have been efforts to push out to large \cite{Croft:2015nna,Croft:2018rwv} and intermediate length scales \cite{Kakuma21,Kikuchihara22,Lujan22} across a range of redshifts, $z \sim 2-7$. The larger scale measurements, in the two-halo regime (see \S \ref{S:modeling}), probe the total clustered Ly-$\alpha$ line emission at the redshifts surveyed and the large-scale structure of the universe \cite{Pullen:2013dir,Silva13}. Future SPHEREx and HETDEX measurements (\S \ref{S:projects}) should improve on the current pilot investigations. 

Two pilot studies (\cite{Croft:2015nna,Croft:2018rwv}) used optical SDSS data to detect the large-scale Ly-$\alpha$ LIM
signal. In reference \cite{Croft:2015nna}, the authors presented a first measurement of the diffuse Ly-$\alpha$-quasar cross-correlation, while the second paper refined this analysis and further considered the cross-correlation between the Ly-$\alpha$ LIM signal and the Ly-$\alpha$ forest. These studies probed the Ly-$\alpha$ LIM signal at $z = 2-3.5$. More specifically, these analyses exploit SDSS fiber spectra targeted on low redshift LRGs, which
should also contain diffuse Ly-$\alpha$ emission from background objects at higher redshift. The spatially varying Ly-$\alpha$ background emission may be extracted by cross-correlating spectral pixels at the
relevant wavelengths with quasar and Ly-$\alpha$ forest samples at the high redshifts of interest, after subtracting out the LRG contribution from each spectrum to reduce measurement noise. In other words, by stacking SDSS fiber spectra around high redshift quasars, one can probe the quasar-Ly-$\alpha$ cross-correlation function, while a similar approach also yields the Ly-$\alpha$ forest-Ly-$\alpha$ emission cross-correlation.  

These pilot studies yielded a significant detection of the quasar-Ly-$\alpha$ cross-correlation function ($\xi_{q,\alpha}$) and an upper bound on the Ly-$\alpha$ forest-Ly-$\alpha$ emission ($\xi_{f,\alpha}$) cross-correlation. The initial quasar-Ly-$\alpha$ correlation detection in \cite{Croft:2015nna} was revised downward in amplitude by a factor of $\sim 2$ in \cite{Croft:2018rwv} after achieving a better understanding of systematic measurement errors. In particular, a subtle issue for these analyses is that light from quasar targets can leak into neighboring SDSS spectroscopic fibers, which are being used to measure the diffuse Ly-$\alpha$ emission, and this can lead to spurious cross-correlation signals \cite{Croft:2015nna}. Furthermore, in \cite{Croft:2018rwv} it was found that one must account for quasar clustering along with this leakage to avoid contamination in estimates of $\xi_{q,\alpha}$. Turning to the interpretation of their measurements, \cite{Croft:2018rwv} argue that their $\xi_{q,\alpha}(r)$ -- which is significant at $r \sim 1-15$ Mpc/$h$, relatively close to the quasars -- is dominated by reprocessed emission from the quasars themselves, as opposed to being sourced by surrounding star-forming galaxies. On the other hand, their $\xi_{f,\alpha}(r)$ measurement implies an interesting upper bound on the total clustered Ly-$\alpha$ emission around $z \sim 2.5$. Interestingly, their limit on the luminosity density in Ly-$\alpha$ photons is comparable to that produced by directly detected Ly-$\alpha$ emitting galaxies, and so this bounds the contribution of galaxy populations that are too faint to be observed individually. It also implies that the pilot $\xi_{f,\alpha}(r)$ measurement, although only an upper bound, is close to the detection regime. Furthermore, specialized LIM surveys may be designed to mitigate systematics, such as the fiber light leakage effect that partly limits the SDSS measurements, and so there are good prospects for improvements here.

A related study forecasts the prospects for Ly-$\alpha$ forest-Ly-$\alpha$ emission cross-correlations using DESI Ly-$\alpha$ forest data and broad-band optical images from the Dark Energy Camera Legacy Survey (DECaLS) \cite{Renard:2024efa}. DECaLS is one of the main imaging surveys that is used for DESI spectroscopic target selection. These authors find that DESI $\times$ DECaLS analyses may enable a detection of this cross-correlation signal, while future DESI $\times$ LSST measurements are extremely promising in this regard. 

There have also been two recent Ly-$\alpha$ LIM studies using the Hyper-Suprime Camera (HSC) on the Subaru telescope \cite{Kakuma21,Kikuchihara22}. These studies have used narrow-band imaging to first identify Ly-$\alpha$ emitting galaxies (LAEs) and then measure the stacked mean surface brightness around the LAEs as a function of transverse separation. That is, these studies measure the LAE-Ly-$\alpha$ two-point correlation function. Reference \cite{Kakuma21} considers LAEs in narrow bands centered around each of $z=5.7$ and $z=6.6$, and compares the stacked narrow-band images around the LAEs with a sample of non-LAE foreground galaxies to gauge background/foreground emission and systematic effects. In each redshift bin, these authors find evidence for an extended Ly-$\alpha$ surface brightness profile, with the present data probing out to about $\sim 1$ comoving Mpc (much beyond the virial radius of the LAEs). Reference \cite{Kikuchihara22} performs a similar analysis to \cite{Kakuma21} at $z=5.7$, $z=6.6$  and also adds lower redshift bins at $z=2.2$ and $z=3.3$ for comparison. In each of these redshift bins, the authors again find evidence for extended Ly-$\alpha$ surface brightness profiles out to roughly $\sim 1$ comoving Mpc. These results do not show strong redshift evolution, although the statistical uncertainties remain sizable. As discussed in \cite{Kikuchihara22}, and references therein, the extended Ly-$\alpha$ surface brightness profile may reflect some combination of Ly-$\alpha$ scattering from the central source, Ly-$\alpha$ produced in collisional excitations, Ly-$\alpha$ emission from satellite galaxies in the LAE host halo, and contributions from galaxies in clustered neighboring dark matter halos. Further modeling efforts and comparisons between the stacked Ly-$\alpha$ emission and future stacked H-$\alpha$ measurements \cite{Mas-Ribas2017} should help in disentangling the different contributions here. 

Reference \cite{Lujan22} measures stacked Ly-$\alpha$ surface brightness profiles around nearly $1,000$ LAEs at $1.9 < z < 3.5$ using early HETDEX data. These authors measure a signal out to 160 proper kpc and find fairly good agreement with radiative transfer simulation models from \cite{Byrohl2021}. This comparison suggests that the excess Ly-$\alpha$ surface brightness within $r \lesssim 100$ proper kpc is dominated by scattered Ly-$\alpha$ photons from the central galaxy, while the profile at larger scales is sourced mainly by Ly-$\alpha$ from galaxies in correlated neighboring halos. 

\begin{figure}
\begin{center}
\includegraphics[width=\textwidth]{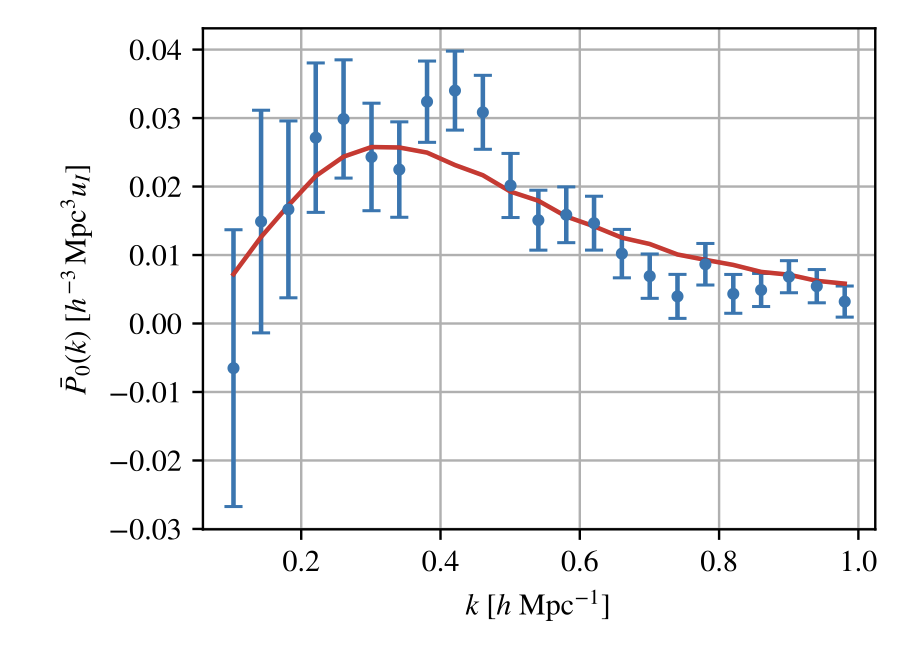}
\caption{HETDEX measurements of the cross-power spectrum between the Ly-$\alpha$ LIM signal and LAEs. The blue points and $1-\sigma$ error bars show spherically-averaged cross-power measurements, combining estimates in three fields and three redshift bins (with $\bar{z} = 2.1, 2.6,$ and $3.2$).  The red solid line is a best fit mock signal, including corrections for signal loss due to PCA foreground removal. The quantity $u_I$ in the y-axis label stands for the specific intensity units adopted of $10^{-18} \, \mathrm{erg s^{-1} cm^{-2} arcsec^{-2} \Ang^{-1}}$.
The results provide a strongly significant LAE $\times$ Ly$-\alpha$ LIM cross-power spectrum detection, yielding constraints on the total Ly-$\alpha$ intensity, and illustrating the promise of future HETDEX analyses. 
From \cite{Niemeyer:2025yvq}.}
\label{fig:hetdex_lya_lae}
\end{center}
\end{figure}

Figure \ref{fig:hetdex_lya_lae} shows the results of a recent HETDEX LAE $\times$ Ly-$\alpha$ cross-power spectrum detection \cite{Niemeyer:2025yvq}. This analysis is carried-out in Fourier space, in contrast to the earlier \cite{Lujan22} real-space analysis, and the new results extend out to much larger spatial scales. 
The authors select $\sim 137,000$ LAEs from the HETDEX LAE catalogs for use in their cross-correlation analysis. They cross-correlate the distribution of the selected LAEs with the diffuse Ly-$\alpha$ intensity field, after masking the detected LAEs in the intensity field. A PCA technique (see \S \ref{S:PCA}) is used to remove contamination from airglow, zodiacal light, foreground contamination and other systematics. In each of the three redshift bins considered in the analysis, centered on $\bar{z}=2.1, 2.6,$ and $3.2$, the authors remove $\sim 100-200$ PCA modes. They use lognormal Ly-$\alpha$ mocks to 
correct for the resulting signal loss and to help interpret the measurements.

The spherically-averaged cross-power spectrum measurement, after removing PCA modes and combining different observing fields and redshift bins, is shown in Figure \ref{fig:hetdex_lya_lae}. The statistical significance of the detection is high and the results imply interesting bounds on the Ly-$\alpha$ emissivity at $z \sim 2-3$. 
The HETDEX results are lower than the bias-weighted Ly-$\alpha$ intensity inferred from the quasar $\times$ Ly-$\alpha$ analysis of \cite{Croft:2015nna}, consistent with the possibility that the earlier measurement was influenced by Ly-$\alpha$ emission from the quasars themselves and/or residual systematics. The HETDEX detection, however, lies slightly {\em above} the bias-weighted intensity bound from the later \cite{Croft:2018rwv} Ly-$\alpha$ forest $\times$ Ly-$\alpha$ LIM analysis near $\bar{z}=2.6$, although the difference is only at the $2.6-\sigma$ level.
Reference \cite{Niemeyer:2025yvq} further discusses the implications of their results for the Ly-$\alpha$ emissivity, and presents comparisons with estimates from integrating over LAE luminosity functions and with Ly-$\alpha$ radiative transfer models. 
The HETDEX team expects to increase the size of their data sets and refine their analysis methodologies in the near future.

In summary, early Ly-$\alpha$ LIM studies have already yielded scientifically interesting results. These analyses have mainly been carried out using data intended for other purposes, and the prospects for additional Ly-$\alpha$ LIM measurements with SPHEREx, HETDEX, and future tailored surveys appear excellent. There is also further scope for Ly-$\alpha$ LIM analyses with DESI, DECaLS, LSST, and other current/forthcoming traditional galaxy survey data sets.

\section{Projects and Instruments}
\label{S:projects}

The scientific prospects of LIM have inspired a vibrant set of experimental design efforts. Impressively, within less than a decade, several such designs
have been turned into functioning instruments that have successfully seen first light. Furthermore, the community is planning a broad range of additional measurements and surveys. 
Here we provide a concise summary
of some key ongoing and forthcoming projects, spanning a range of science goals, line targets, observing wavelengths, sky coverages, and instrumental designs.

\begin{figure}
    \begin{center}
    \includegraphics[width=\textwidth]{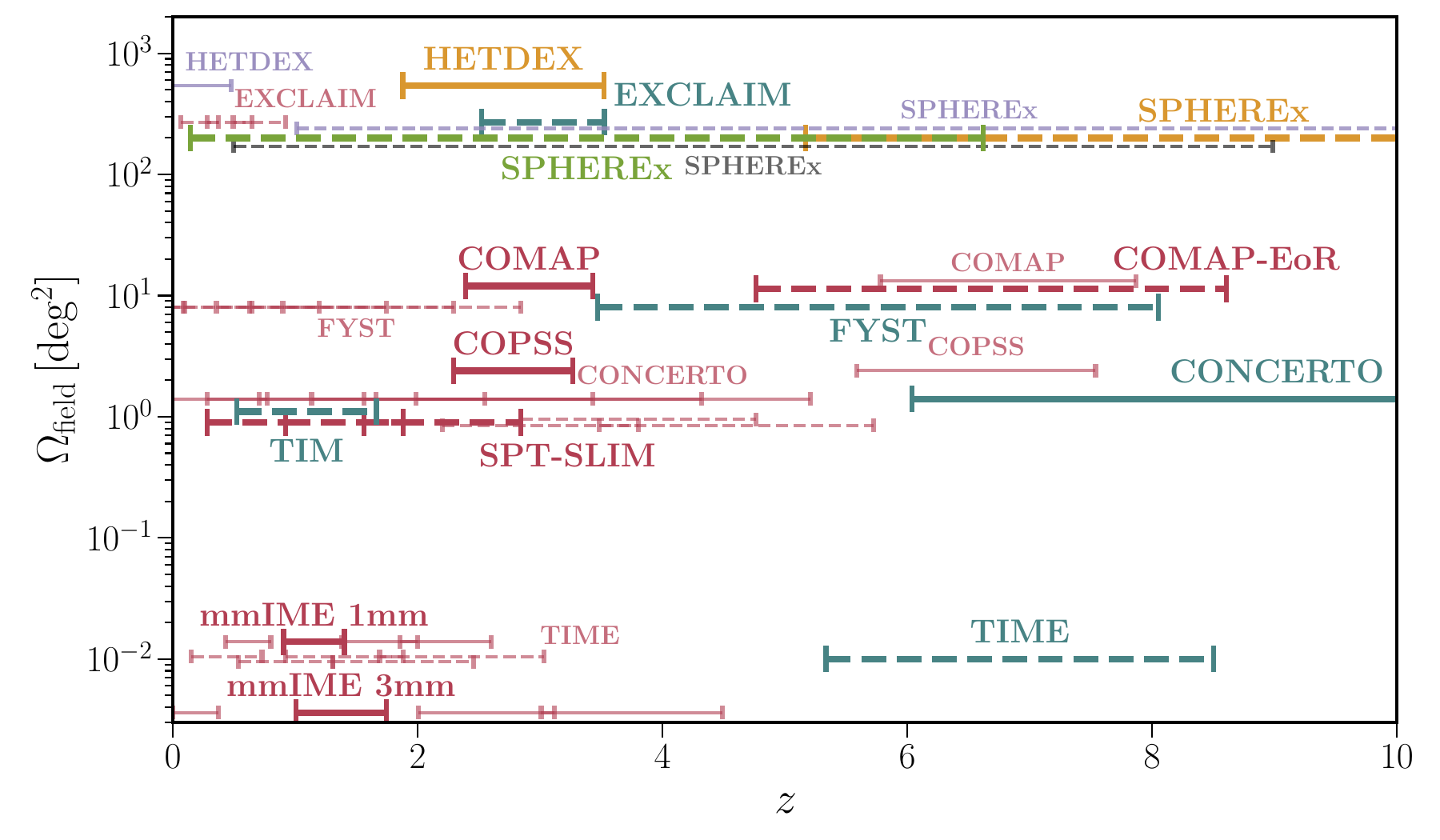}
    \includegraphics[width=\textwidth]{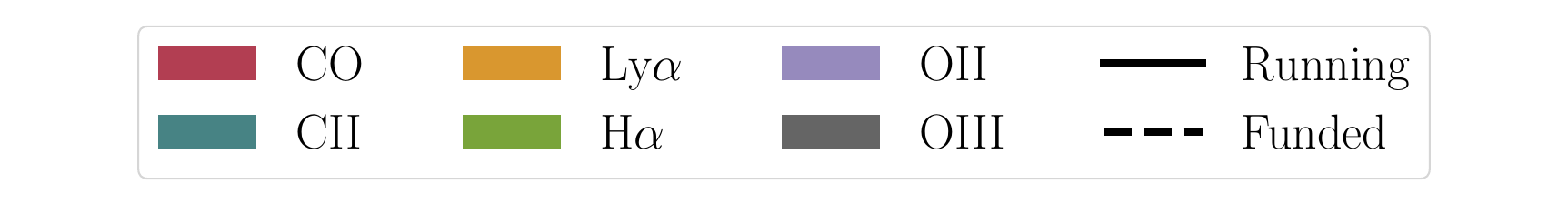}
    \caption{Spectral line, redshift, and angular scale coverage of some of the key current LIM experiments. The figure conveys the diversity of lines probed by these projects, along with the wide range in angular scales and redshifts that are covered. Many of the experiments included in this illustrative figure, and others, are discussed explicitly in this section. 
    From \cite{Bernal:2022jap}.}
    \label{fig:IMexperiments}
    \end{center}
\end{figure}

\begin{figure}
    \begin{center}
    \includegraphics[width=\textwidth]{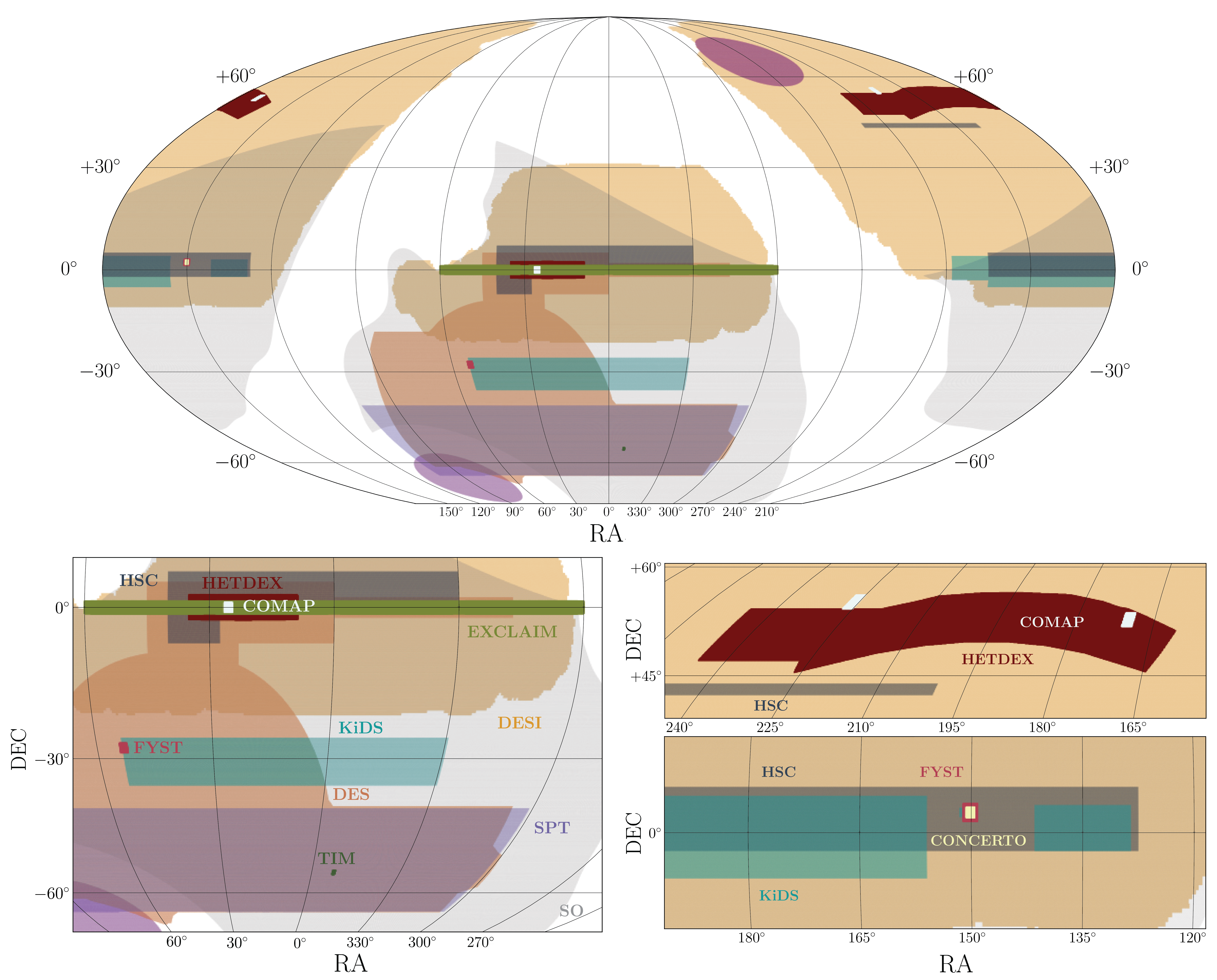}
    \caption{Survey footprints of current LIM experiments. The top panel shows a Mollweide projection of the full sky in equatorial coordinates. The bottom 
    panels zoom-in on some of the primary LIM survey regions. For comparison, galaxy and CMB survey regions are included as well (HSC, KiDS, DES, DESI, SPT, and SO). The figure helps to illustrate the near-term prospects for cross-correlations among LIM surveys and for LIM $\times$ galaxy and LIM $\times$ CMB cross-correlations. From \cite{Bernal:2022jap}.}
    \label{fig:IMexperiments_footprint}
    \end{center}
\end{figure}

Figure~\ref{fig:IMexperiments} (from \cite{Bernal:2022jap}) summarizes some of the key current LIM projects in terms of their primary emission lines probed, their angular coverage on the sky, and the redshift range spanned. The surveys included in the figure span redshifts from $z \sim 0-10$, cover up to several hundred square degrees on the sky, and probe lines from within the radio band out to rest-frame optical lines. Figure~\ref{fig:IMexperiments_footprint}, also from \cite{Bernal:2022jap}, puts some of the LIM efforts into further context by showing their planned survey footprints. Even these near-term experiments will enable a rich set of cross-correlation measurements, including cross-correlations between different pairs of lines within a given survey, cross-correlations between lines in different LIM surveys, and between LIM and
galaxy and CMB observations. 

\subsection{Radio}

The very first LIM surveys, some of which are still ongoing, targeted the redshifted 21 cm line from the EoR. As discussed previously in this review, the redshifted 21 cm line surveys now also encompass efforts to detect the post-reionization 21 cm line. Measuring the {\em pre-reionization} 21 cm signal from the Cosmic Dark Ages is a scientifically compelling yet challenging future goal. The 21 cm projects span a broad range of science targets and instrumental designs with some efforts using single-dish telescopes and others employing interferometers of varying
configurations. Hopefully, the diversity of approaches being pursued in this field will ensure further progress. Since these 21 cm surveys are well-covered in other recent reviews, we do not discuss their details here and instead refer the interested reader to \cite{Liu:2019awk} for a summary.

Along with this suite of redshifted 21 cm surveys, the radio band also features CO LIM experiments. 
Specifically, the COMAP project is a funded and operational effort to measure CO rotational transitions across a wide redshift range, with the COMAP pathfinder tracing $z \sim 2.4-3.4$ in the CO(1-0) line.

\subsubsection{COMAP}

The CO Mapping Array Project (COMAP) \cite{COMAPI2022} aims to use CO 
LIM to trace the distribution and global properties of galaxies over cosmic time, ultimately including observations into the EoR. A pathfinder instrument has been built to validate the technologies and analysis techniques for LIM and is in operation at the Owens Valley Radio Observatory located in Bishop, California. In 2022, the COMAP Collaboration published a suite of papers describing the experiment, methodology and data analysis, the science results from their first year of observations, and future plans \cite{COMAPI2022, COMAPII2022, COMAPIII2022, COMAPIV2022, COMAPV2022, COMAPVI2022, COMAPVII2022}. The team released results from a second season of observations more recently in 2024 \cite{Lunde:2024rut,Stutzer:2024rps,Chung:2024iob}.

The pathfinder instrument, with 19 receivers packed in a hexagonal focal plane on a 10.4 m antenna, has an instantaneous 26-34 GHz frequency coverage and a 2 MHz spectral resolution. It is sensitive to the CO(1–0) rotational transition from $z = 2.4–3.4$ and fainter CO(2–1) emission from $z = 6–8$. The pathfinder will survey a 12 deg$^2$ patch of sky over a 5-year observing campaign to detect the CO signal from $z \sim 3$. It is expected to yield a power spectrum detection with a signal-to-noise ratio (SNR) of $9-17$, as well as a CO–galaxy cross-correlation power spectrum measurement with an SNR of 19.

Based on data from the first 13 months of observation, the COMAP collaboration (or perhaps better referred to as a ``COllaboration'') reported a constraint on the clustering amplitude of the CO(1-0) power spectrum, finding $P_{\rm CO} (k) = -2.7 \pm 1.7 \times 10^4 \, \mu \mathrm{K}^2$ Mpc$^3$ on scales of $k = 0.051 - 0.62 \, {\rm Mpc}^{-1}$ (see \S \ref{S:measurements}). This measurement amplitude further constrains the product of the mean CO brightness temperature and the luminosity-weighted clustering bias to be $\avg{b}^2 \avg{T_b}^2 < 49  \, \mu \mathrm{K}^2$, which is nearly an order of magnitude improvement on the previous best measurement and rules out two previously viable models in the literature. As ancillary science, a 30 GHz survey of the Galactic plane is also presented. 

The next generation COMAP, COMAP-EoR, will aim to extend CO LIM measurements into the EoR. COMAP-EoR supplements the existing 30 GHz COMAP Pathfinder with two additional 30 GHz instruments and a new 16 GHz receiver. This combination of frequencies allows simultaneous observations of CO(1-0) and CO(2-1) into the EoR, at $z \sim 5-8$, in addition to providing a significant boost in signal-to-noise for the $z \sim 3$ CO(1-0) emission surveyed by the Pathfinder. 
Reference ~\cite{COMAPVII2022} compares current EoR CO models, which span more than an order of magnitude in the predicted power spectrum amplitude, and finds that COMAP-EoR can make a significant (SNR $>20$) measurement in cross-correlation with galaxy surveys in five out of six of the models considered. The highly uncertain signal strength highlights our lack of knowledge regarding molecular gas and CO during this time period: COMAP-EoR will hopefully fill-in this gap in our knowledge.

\subsection{Sub-mm and FIR}

Multiple sub-mm and far-infrared (FIR) experiments 
are underway or in the planning stages, primarily targeting the [CII] line with
some projects aiming to detect sub-mm [OIII] lines as well. The $\sim 100-300$ GHz
frequency band targeted by these surveys overlaps with the main observing windows for CMB experiments, and hence these efforts benefit from 
expertise acquired in the CMB field. In fact, [CII] LIM experiments consist
essentially of traditional CMB broad-band mapping instruments coupled with spectrometers to yield higher spectral resolution. Here we discuss two currently on-going [CII] surveys, CONCERTO and TIME, which have similar timelines, and briefly mention the planned TIFUUN, EXCLAIM, TIM, FYST/EoRSpec, and SPT-SLIM experiments.

\subsubsection{CONCERTO}

CONCERTO\footnote{\url{https://mission.lam.fr/concerto/}}, which stands for CarbON Cii line in post-rEionisation and ReionisaTiOn epoch, was installed in April 2021 on the Atacama Pathfinder EXperiment (APEX) 12-m telescope on the Chajnantor plateau in northern Chile, 5,100 m above sea level. It covers the frequency range from 130–310 GHz, has an instantaneous field-of-view of 20 arcminutes, a diffraction-limited angular resolution of 30 arcseconds, and moderate spectral resolution (R = 1–300, adjustable). A 4,300-pixel cryogenic camera, composed of lumped element kinetic inductance detectors at a base temperature of 0.06 K, is coupled to a Fourier transform spectrometer (more specifically, a Martin–Puplett interferometer) to yield varying spectral resolution: a 2.5 Hz fast-moving mirror separates spectral components from the incoming radiation in the Fourier domain for each individual pixel in the camera. CONCERTO achieved first light in April 2021 despite the COVID pandemic, and started its primary science program in July 2021. The observing campaign has now concluded.

The main CONCERTO science program, detailed in \cite{CONCERTO2020}, consists of a [CII] LIM survey and observations of galaxy clusters.
The first scientific goal is to construct three-dimensional maps of the [CII] line-intensity signal at $z > 5.3$ in the reionization and post-reionization eras. Reference \cite{VanCuyck:2023uli} presents updated CONCERTO forecasts, while \cite{Hu:2024nya} characterizes data-processing algorithms and the on-sky instrumental performance.

\subsubsection{TIME}

The Tomographic Ionized-Carbon Mapping Experiment (TIME) is designed to measure the [CII] LIM signal from the EoR \cite{TIME2014,Hunacek2016}.
TIME employs a linear array of spectrometers, each consisting of a parallel-plate diffraction grating, providing a resolving power of $R \sim 150$. The spectrometer bandwidth ranges over 185-323 GHz, covering the [CII] redshift range from $z=4.9-9.3$, as well as channels at the edges of the band for atmospheric noise mitigation. 
Feedhorns couple radiation to the waveguide spectrometer gratings, which are stacked in two blocks (for dual polarization measurements) of 16 within a single cryostat, providing a 1x16 array of beams with a 14 arcminute field-of-view. Direct absorber Transition Edge Sensors (TES) sit at the output of the grating, and the 1,840 detectors are read-out with time-domain-multiplexing microwave SQUIDS and cooled to a base temperature of 250 mK using a $^3$He adsorption refrigerator. 

TIME, originally known as TIME-Pilot, was among the first proposed [CII] LIM experiments \cite{Gong11, TIME2014}. It aims to measure [CII] emission fluctuations, and thereby constrain the integrated [CII] intensity summed over all EoR galaxies, thus shedding light on the reionization process. TIME will also produce high signal-to-noise ratio measurements of CO emission fluctuations, which trace the molecular gas in star-forming galaxies at redshifts $0.5 < z < 2$. With its unique atmospheric noise mitigation, TIME also provides significantly improved sensitivity for measuring the kSZ effect in galaxy clusters. Further details regarding TIME's science goals are given in \cite{Sun:2020mco}.

TIME has secured 1,000 hours of winter observing time at the ALMA 12 m prototype antenna (APA), located in Kitt Peak, Arizona. It has had two successful on-sky engineering runs in early 2019 and late 2021, producing source maps with partial detector arrays. The planned science survey nominally spans 1$^{\circ} \times 14'$ per field, with the observing fields to be determined. Reference \cite{TIME2026} presented Sagittarius A observations using commissioning data from 2021-2022 to demonstrate TIME's ability to recover both continuum and spectral line signals and validate its hyperspectral imaging capabilities for LIM measurements.

\subsubsection{TIFUUN}

TIFUUN (Terahertz Integral Field Unit with Universal Nanotechnology) is a funded imaging spectrograph, to be deployed on the 10-meter Atacama Submillimeter Telescope Experiment (ASTE)
telescope \cite{Kohno24}. Design studies are still underway, but a preliminary plan includes two observing bands, with the first band covering observing frequencies between
$124-176$ GHz, and the second spanning $224-301$ GHz. The second band will include [CII] emission from $z=5.3-7.4$, as well as various CO and [CI] interloper lines
at lower redshifts. In addition, the second band will contain [OIII] 88 $\mu$m line emission at $z=10.2-14.1$, which overlaps in redshift with the [CII] emission
captured in the first band. These two bands will include multiple interloper lines from common redshifts, and enable cross-correlations to help mitigate interloper contamination (see \S \ref{S:challenges}).  It will also be possible to bound or detect the [OIII] $\times$ [CII] cross-correlation signal at $z \gtrsim 10$. 
TIFUUN will extend on the Deep Spectroscopic High-redshIft MApper (DESHIMA) spectrometer technology \cite{DESHIMA2019,DESHIMA2021}. Specifically, TIFUUN will use a novel integrated superconducting spectrograph technology, employing microwave kinetic inductance detectors, with $\sim 20,000$ voxels, divided into $\sim 200$ spectral channels (across the two observing bands).  Current plans are to survey about $\sim 1.5$ deg$^2$ on the sky with a spatial resolution better
than $\sim 0.5$ arcmin in each observing band. 
The strength of the measured LIM signals will help determine the role of dust-obscured star formation in the universe and will provide a valuable handle on the chemical enrichment history of the universe into the EoR. Deep learning algorithms \cite{Moriwaki2020} will be employed to further help separate interloper contaminant lines and to retrieve the [CII] and [OIII] LIM signals (see \S \ref{S:challenges}).

\subsubsection{EXCLAIM}

The EXperiment for Cryogenic Large-Aperture Intensity Mapping (EXCLAIM) \cite{Cataldo21} is a balloon-borne far-infrared telescope that will survey the star formation history at $0 < z < 3.5$ through cross-correlations of redshifted CO and CII lines with a spectroscopic galaxy catalog. In its baseline survey, EXCLAIM will map emission over 420-540 GHz with a resolving power of $R = 512$ across a 320 deg$^2$ sky area that overlaps with BOSS. It will also cover several $\sim 100$ deg$^2$ regions in the plane of the Galaxy. EXCLAIM aims to make a definitive detection of redshifted [CII] emission in cross-correlation with BOSS quasars at redshifts $2.5 < z < 3.5$, detect two CO lines in each of the BOSS samples (MAIN, LOWZ, CMASS), and constrain CO $J=4-3$ and [CI] in the Galaxy. EXCLAIM uses a cryogenic, 90-cm telescope and probes spatial scales on the sky from the linear regime up to shot-noise-dominated scales. EXCLAIM uses six $\mu$-Spec spectrometer modules and microwave kinetic inductance detectors. EXCLAIM's cryogenic telescope and sensitive detectors allow it to reach high sensitivity in spectral windows of low emission in the upper atmosphere. Reference \cite{Pullen:2022pgp} gives detailed forecasts for EXCLAIM measurements.

\subsubsection{TIM}

The Terahertz Intensity Mapper \cite{TIM2020}, TIM, is a balloon-borne LIM experiment designed to study the cosmic star formation history at $0.5 < z < 1.7$ using the [CII] line. It will map a 3D cosmic volume over spatial scales of 1 - 50 Mpc (30" to 1$^{\circ}$) with spectroscopic information. TIM combines two long-slit grating spectrometers to cover 240-420 $\mu$m at a resolving power of R=250. It uses kinetic-inductance detector arrays with over 7,000 pixels in total and a 2-m low-emissivity carbon-fiber telescope to provide background-limited sensitivity. TIM will survey one 0.1 deg$^2$ field centered on GOODS-S and one wider field ($\sim 1$ deg$^2$) within the South Pole Telescope (SPT) Deep Field. See references~\cite{Bracks26,Agrawal:2026nzp} for recent TIM cross-correlation forecasts.

\subsubsection{FYST/EoRSpec}

The Epoch of Reionization Spectrometer (EoR-Spec) \cite{EoRSpec2020,CCAT-Prime:2021lly,Freundt:2024lsh} is an instrument module for the Prime-Cam receiver on the 6-m aperture Fred Young Submillimeter Telescope\footnote{\url{https://www.ccatobservatory.org}} (FYST), located on Cerro Chajnantor at 5,600 m in Chile. EoR-Spec will target the [CII] LIM signal at redshifts between $3.5 < z < 8$ (observing frequencies of 210-420 GHz), tracing the evolution of cosmic structure during the early phases of galaxy formation. The EoR-Spec module will have a wide 1.3 degree field-of-view and a diffraction-limited beam of 30-60 arcseconds, which is well matched to the $\sim$ Mpc clustering scale of the EoR signal. EoR-Spec will make use of a scanning, silicon substrate-based Fabry–Perot interferometer to achieve a spectral resolving power of $R=100$; it contains dichroic arrays of $\sim 6,000$ feedhorn-coupled KIDs. 
EoR-Spec will benefit from the exceptionally low water vapor site on Cerro Chajnantor, its large field-of-view, and high angular resolution.  
EoR-Spec's novel design allows simultaneous measurements of the [CII] line at two redshifts using dichroic pixels and two orders of the Fabry–Perot interferometer. FYST is currently under construction and EoRSpec first light is expected in 2027.

\subsubsection{SPT-SLIM}

The South Pole Telescope Shirokoff Line-Intensity Mapper (SPT-SLIM) is a pathfinder experiment for millimeter wavelength LIM on the SPT \cite{SPTSLIM2022}. SPT-SLIM aims to demonstrate the use of on-chip filter-bank spectrometers. Its focal plane consists of 18 spatial pixels with arcminute angular resolution, each of which connects to dual polarization filter-bank spectrometers covering 120–180 GHz with a resolving power of $R=300$, and are coupled to aluminum kinetic inductance detectors. A compact cryostat keeps the detectors cooled to temperatures of 100 mK. 

SPT-SLIM is targeting several CO rotational transition lines from galaxies at $z \sim 0.5-3$: specifically, CO(2-1) from $0.3 < z < 0.9$, CO(3-2) from $0.9 < z < 1.8$, and CO(4-3) from $1.6 < z < 2.8$. SPT-SLIM will measure the clustering amplitudes of these CO lines, informing models for the CO luminosity functions and cold molecular gas fractions over cosmic history across the full population of galaxies.

The SPT is a 10-m aperture off-axis Gregorian telescope located at the South Pole, used primarily for CMB observations. SPT-SLIM was deployed on the SPT during Nov 2024, achieved first light in Jan 2025, with science observations scheduled for 2026. SPT-SLIM will survey a $\sim 0.1^{\circ} \times 10^{\circ}$ patch of sky using constant-elevation scans.

\subsection{Optical/Near-IR}

At optical and near-infrared wavelengths, most LIM spectral lines are indirect tracers of star formation activity and their light output
across cosmic time. 

\subsubsection{HETDEX}

The Hobby-Eberly Telescope Dark Energy Experiment -- HETDEX\footnote{\url{https://hetdex.org/}} -- is in the process of conducting a spectroscopic survey dedicated to dark energy measurements using BAO signatures \cite{HETDEX2021}. It will survey over 1 million Lyman Alpha Emitters (LAEs) at $1.9<z<3.5$, and over 0.5 million [OII]-emitting galaxies at $z<0.5$ in a 540 deg$^2$ area in two separate fields. It aims to constrain the angular diameter distance, $D_{\mathrm A}(z)$, and the expansion rate, $H(z)$, to better than 1\% precision near a mean redshift of $z=2.4$.

Unlike most spectroscopic surveys aimed at detecting BAOs, HETDEX conducts a blind, unbiased survey using integral field units (IFUs) without prior target selection, directly mapping the cosmos in 3D using the Ly-$\alpha$ spectral feature. It therefore also provides a promising LIM data set, probing both Ly-$\alpha$ at $1.9 < z < 3.5$ and [OII] lines at $z < 0.5$.

HETDEX uses 74 IFUs, which populate the focal plane of the 10-meter HET, covering a 18'-diameter field-of-view. 
Each IFU contains 448 1.5" fibers, which are fed into the VIRUS spectrograph to cover wavelengths between 3500 and 5500 \AA. A three-step dithering observation scheme is used to fill in the gap between the fibers inside each IFU. The layout of the IFUs is discrete and the IFUs sparsely tile the focal plane with a fill-factor of 0.22 \cite{HETDEX2021}, while the effective fill-factor of the survey is 0.17. The survey mask thus needs to be carefully accounted for in the analysis.   

Reference~\cite{Simple2023} uses the \textsc{simple} code to generate log-normal simulations of mock galaxies and intensity maps for a HETDEX-like survey in redshift space, while including a realistic mask, selection function, and intensity noise. The authors forecast that the cross-power spectra of Ly-$\alpha$ LIM and LAEs within the survey should be detectable at the $\simeq 8 \sigma$ level in each of the two redshift bins centered near $z=2.2$ and $z=3$. See also \S \ref{sec:lya_current} for a discussion of early HETDEX measurements.

\subsubsection{SPHEREx}

SPHEREx\footnote{\url{https://spherex.caltech.edu}}, the Spectro-Photometer for the History of the Universe, Epoch of Reionization and Ices Explorer, is a NASA Medium Class Explorer (MIDEX) mission, launched on March 11, 2025 and completed its first all-sky survey (out of four, nominally) in December 2025. The data is publicly available\footnote{\url{https://irsa.ipac.caltech.edu/Missions/spherex.html}}. It is a wide-field, near-infrared, spectro-imaging telescope which is conducting four all-sky surveys during its nominal two-year prime mission. SPHEREx covers 0.75-5 $\mu$m in 102 wavelength bands with a pixel size of 6.2", and generates a rich, all-sky 3D spatial-spectral data cube, enabling a wide range of scientific applications. 

SPHEREx's mirror has an effective diameter of 20-cm and a large field-of-view of 11$^{\circ} \times 3.5^{\circ}$. A dichroic beam-splitter separates the focal plane into a short and a long wavelength detector array. It makes use of linear variable filters, sitting on top of its six 2k $\times$ 2k H2RG detectors, yielding a spatially varying wavelength response across the focal plane. By design, each of these six wavelength bands has a different spectral resolution ranging between $R=35-135$, with the two longer wavelength bands at higher resolution. There are no moving parts, and SPHEREx obtains all-sky spectra through spacecraft movements: each sky location is observed multiple times, 
each time at different positions on the detector arrays through re-pointing the spacecraft with a point-and-stare strategy. In this way, each sky pixel is measured at multiple wavelengths and spectra are built up over time.

Due to its low-Earth polar orbit and survey design, SPHEREx captures two deep fields near the north and south ecliptic poles. The deep fields are observed during every $\sim 98$-min orbit, with each spanning around 100 deg$^2$ in area, and are of particular interest for LIM. The SPHEREx point source and surface brightness sensitivities for the all-sky survey and deep fields can be found in reference \cite{Bock:2025ijl}. In the deep fields, the survey depth
is expected to reach an AB magnitude of 22. 
An extension over the nominal, two-year operating lifetime may provide further sensitivity.  

SPHEREx's wavelength coverage accesses several prominent redshifted optical/UV spectral lines, namely H-$\alpha$, [OIII], [OII], H-$\beta$, and Ly-$\alpha$, over a significant cosmological volume and up to high redshift, potentially into cosmic reionization \cite{Dore2014, Dore2018}. The SPHEREx deep fields should have sufficient sensitivity to measure fluctuation power spectra in H-$\alpha$, [OII], and [OIII] lines, over $0.8 < z < 5$ \cite{Gong17}. 
Additionally, Ly-$\alpha$ fluctuations from reionization may be measurable \cite{Feng:2018rje}, while cross-correlations with other planned galaxy and LIM surveys are also promising \cite{Gong17, Heneka:2016kss, Heneka21, ChengChang2022}, 
although we note that there are substantial model uncertainties regarding LIM signal strengths at high redshift. 

SPHEREx LIM observations can, in principle, shed light on both astrophysics and cosmology (see \S \ref{S:science}). For example, the H-$\alpha$, [OIII], and [OII] measurements at $\sim 1 < z < 5$ will trace the SFRD across a significant portion of cosmic history \cite{Gong17}, and help in understanding how feedback processes regulate galaxy formation \cite{Sun2022}. SPHEREx Ly-$\alpha$ LIM measurements
can probe the spatial structure of reionization \cite{Silva13,Pullen:2013dir,Heneka:2016kss,Visbal:2018dsi,Ambrose:2025mcg,Almualla:2025pix}, and potentially distinguish between different reionization scenarios \cite{Feng:2018rje}, as will cross-correlations with 21 cm survey data (e.g., \cite{Sun2022}). Finally, SPHEREx LIM measurements, across a range of lines, can be used to measure the geometry of the universe at high redshift \cite{Bernal:2019gfq} and constrain dark matter decay and annihilation signatures \cite{Creque-Sarbinowski:2018ebl} (see also \S \ref{S:bao_lim}-\ref{S:dm_lim}).

\section{Conclusions and Future Outlook}
\label{S:future}

LIM has generated broad interest: a large number of experiments are currently underway or in the planning stages, and targeting a wide range of scientific topics. 
The ultimate goal for LIM is to map as much of the volume of the universe in as many emission lines as possible. In some sense, the aim here is to measure the spectral energy distribution, or at least its spatially-clustered line emission components, from many individual regions across the entire universe.
This will enable a more complete census of multi-phase interstellar, circumgalactic, and intergalactic gas across cosmic time, encoding valuable information regarding structure and galaxy formation/evolution. The same data sets will also trace, in part, the power spectrum of primordial density fluctuations across unprecedented volumes and redshift ranges. This will allow for further handles on: the expansion history of the universe, the physics of inflation or alternatives, the nature of dark energy, the properties of dark matter, the abundance of light relics, and neutrino mass. The ability to trace common regions of the universe with multiple emission lines will ensure robust measurements, while studies using a range of lines with different bias factors will help cross-check that
inferences regarding the linear density power spectrum are
insensitive to uncertainties in the galaxy-halo and galaxy-line emission relationships. 

In order for LIM to live up to its full promise, a number of measurement and analysis challenges need to be overcome. The primary difficulties for LIM measurements relate to disentangling the signal from bright foreground contamination, both continuum and interloper lines, as observed through an imperfect instrument. In practice, even intrinsically smooth foreground contamination will possess unavoidable spectral structure after observations with a realistic instrument. This places stringent demands on the fidelity of instrumental calibration algorithms and necessitates exquisite control over systematic errors. As LIM experiments improve in sensitivity, one can perform jackknife tests to split the data in many different ways and cross-check against a large suite of systematic concerns. Hopefully, it will be possible to mitigate any sources of contamination that appear in such tests and avoid systematics-limited measurement ``floors''. Although there are common concerns across different LIM experiments, the details depend strongly on the continuum foreground to line emission ratios of the target lines, and the varying susceptibility to line-interloper confusion. In addition, instrumental, survey design, and analysis pipelines vary widely between different projects. The diversity of ideas and approaches being explored here should help to identify the most effective strategies, and ensure future progress.  

The LIM signals are also challenging to model: in many cases, one would ideally simulate ISM properties across entire populations of galaxies throughout large cosmological volumes.  This will require an unprecedented multi-scale simulation effort, bridging the gap between high-resolution models which capture the detailed properties of the interstellar medium in particular galaxies, out to cosmological length scales. Simulations of the interstellar medium for ensembles of individual galaxies with varying properties may be used to define sub-grid models for application on top of cosmological simulations. Further, these calculations must be supplemented with semi-analytic models to explore the vast parameter spaces of interest, and analytic calculations to isolate and elucidate the key underlying physics. 

The next several years promise to be an exciting time for the LIM field. During this period, ongoing experiments promise to move beyond increasingly stringent upper limits to achieve detections, including the first measurements of the reionization-era 21 cm signal and the $z \sim 3$ CO line-intensity fluctuations. In addition, we look forward to more precise estimates of the post-reionization 21 cm signal across a broader range in spatial scales. 
At the same time, we eagerly anticipate the first observations from tailored experiments targeting the reionization-era [CII] signal, post-reionization [CII], and SPHEREx LIM measurements of Ly-$\alpha$, H-$\alpha$, H-$\beta$, rest-frame optical [OIII]/[OII] lines, among other observations.  These pioneering surveys may already yield new scientific results, but will also undoubtedly sharpen analysis methods, and help in understanding how to optimize future instrumental and survey design. We can hence look forward to rapid progress as the field scales-up from the regime of preliminary detections to the realm of detailed measurements, and the novel insights across multiple sub-fields of astrophysics that will surely follow. 

Indeed, provided foreground contamination and other systematic concerns can be mitigated, the history of CMB anisotropy measurements provides encouragement and inspiration for LIM studies. In particular, note the rapid development in CMB angular power spectrum measurements from: the first detection of CMB anisotropies on large scales by COBE, announced in 1992, to early hints of acoustic oscillations from a number of experiments around $\sim$ 1998, confident acoustic peak detections by Boomerang in 2000, precise WMAP measurements with first-year data released in 2003, followed by early Planck results in 2013, with subsequent Planck releases mapping the primary temperature anisotropies to the cosmic-variance limit (see, e.g., the historical overview in \cite{Bucher:2015eia}). 
That is, within only about two decades the CMB field moved from an initial detection to a complete characterization of the primary temperature anisotropies. Furthermore, CMB measurements continue to thrive with ACT, SPT, and next-generation instruments mining additional insight from secondary anisotropies, polarization fluctuations, and spectral distortions. This history suggests an optimistic outlook regarding the prospects for scaling-up from early LIM efforts, 
especially if the substantial systematic worries for these measurements may be well controlled.

\clearpage

\section*{Acknowledgments}   
AL thanks JPL for support during a sabbatical year over which part of this work was carried out. We acknowledge the anonymous reviewer for helpful feedback which significantly strengthened this article.
We thank Emmanuel Schaan and Guochao Sun for helpful discussions. We thank Richard Feder, Jordan Mirocha, and Guochao Sun in particular for help with figures. We are grateful to our LIM collaborators over the years, including: James Aguirre, Christopher Anderson, Kevin Bandura, Gus Beane, Philippe Berger, Simeon Bird, Jamie Bock, Dick Bond, Justin Bracks, Matt Bradford, Patrick Breysse, Yun-Ting Cheng, Asantha Cooray, Abigail Crites, Roland de Putter, Olivier Doré, Aaron Ewall-Wice, Steve Furlanetto, Zucheng Guo, Gil Holder, Garrett Keating, Ely Kovetz, Emily Kuhn, Paul La Plante, Guilaine Lagache, Sarah Libanore, Simon Lunjun Liu, Yin-Zhe Ma, Lluis Mas-Ribas, Kiyo Masui, Jordan Mirocha, Kana Moriwaki, Peng Oh, Hamsa Padmanabhan, Ue-Li Pen, Jeff Peterson, Jonathan Pritchard, Anthony Pullen, Mahdi Quezlou, Michael Seiffert, Guochao Sun, Eric Switzer, Jessie Taylor, Bade Uzgil, Francisco Villaescusa-Navarro, Martin White, Laura Wolz, Shengqi Yang, Zi-Yan Yuwen, and Meng Zhou, as well as the SPHEREx Science Team, the TIME Collaboration, the HIRAX Collaboration, the COMAP Collaboration, the GBT-HIM team, the GMRT-EoR team, the AIM-CO team, and the CDIM Probe Study, CHIC Concept, Farside Probe Study, and Farview Science Teams for our fun work together, which has shaped the outlook presented here. Part of the research was carried out at the Jet Propulsion Laboratory (JPL), California Institute of Technology, under a contract with the National Aeronautics and Space Administration (80NM0018D0004). 
This research was supported in part by grant NSF PHY-2309135 to the Kavli Institute for Theoretical Physics (KITP).

\bibliography{lim_review_master}

@article{Inoue:2006um,
    author = "Inoue, Akio K. and Iwata, Ikuru and Deharveng, Jean-Michel",
    title = "{Escape fraction of ionizing photons from galaxies at z=0-6}",
    eprint = "astro-ph/0605526",
    archivePrefix = "arXiv",
    doi = "10.1111/j.1745-3933.2006.00195.x",
    journal = "Mon. Not. Roy. Astron. Soc.",
    volume = "371",
    pages = "L1--L5",
    year = "2006"
}

@ARTICLE{Ogata26,
       author = {{Ogata}, Erika and {Moriwaki}, Kana and {Lidz}, Adam and {Jun}, Rui Lan and {Yoshida}, Naoki},
        title = "{Signatures of Accreting Black Holes in Line Intensity Mapping}",
      journal = {arXiv e-prints},
         year = 2026,
        month = jun,
          eid = {arXiv:2606.01603},
        pages = {arXiv:2606.01603},
          doi = {10.48550/arXiv.2606.01603},
archivePrefix = {arXiv},
       eprint = {2606.01603},
 primaryClass = {astro-ph.GA},
       adsurl = {https://ui.adsabs.harvard.edu/abs/2026arXiv260601603O}
}

@article{COMAP:2025mov,
    author = "Dunne, D. A. and others",
    collaboration = "COMAP",
    title = "{COMAP Pathfinder {\textendash} Season 2 results - IV. A stack on eBOSS/DESI quasars}",
    eprint = "2510.23568",
    archivePrefix = "arXiv",
    primaryClass = "astro-ph.CO",
    doi = "10.1051/0004-6361/202557864",
    journal = "Astron. Astrophys.",
    volume = "707",
    pages = "A256",
    year = "2026"
}

@ARTICLE{ARES2012,
       author = {{Mirocha}, Jordan and {Skory}, Stephen and {Burns}, Jack O. and {Wise}, John H.},
        title = "{Optimized Multi-frequency Spectra for Applications in Radiative Feedback and Cosmological Reionization}",
      journal = {\apj},
         year = 2012,
        month = sep,
       volume = {756},
       number = {1},
          eid = {94},
        pages = {94},
          doi = {10.1088/0004-637X/756/1/94},
archivePrefix = {arXiv},
       eprint = {1204.1944},
 primaryClass = {astro-ph.CO},
       adsurl = {https://ui.adsabs.harvard.edu/abs/2012ApJ...756...94M}
}

@ARTICLE{ARES2014,
       author = {{Mirocha}, Jordan},
        title = "{Decoding the X-ray properties of pre-reionization era sources}",
      journal = {\mnras},
         year = 2014,
        month = sep,
       volume = {443},
       number = {2},
        pages = {1211-1223},
          doi = {10.1093/mnras/stu1193},
archivePrefix = {arXiv},
       eprint = {1406.4120},
 primaryClass = {astro-ph.GA},
       adsurl = {https://ui.adsabs.harvard.edu/abs/2014MNRAS.443.1211M}
}

@ARTICLE{Zeus21,
       author = {{Mu{\~n}oz}, Julian B.},
        title = "{An effective model for the cosmic-dawn 21-cm signal}",
      journal = {\mnras},
         year = 2023,
        month = aug,
       volume = {523},
       number = {2},
        pages = {2587-2607},
          doi = {10.1093/mnras/stad1512},
archivePrefix = {arXiv},
       eprint = {2302.08506},
 primaryClass = {astro-ph.CO},
       adsurl = {https://ui.adsabs.harvard.edu/abs/2023MNRAS.523.2587M}
}

@ARTICLE{oLIMpus,
       author = {{Libanore}, Sarah and {Mu{\~n}oz}, Julian B. and {Kovetz}, Ely D.},
        title = "{Effective model for line intensity mapping: Auto- and cross-power spectra in the cosmic dawn and reionization}",
      journal = {\prd},
         year = 2025,
        month = oct,
       volume = {112},
       number = {8},
          eid = {083552},
        pages = {083552},
          doi = {10.1103/xq1r-zh51},
archivePrefix = {arXiv},
       eprint = {2507.15922},
 primaryClass = {astro-ph.CO},
       adsurl = {https://ui.adsabs.harvard.edu/abs/2025PhRvD.112h3552L}
}

@ARTICLE{LIMpy,
       author = {{Roy}, Anirban and {Valent{\'\i}n-Mart{\'\i}nez}, Dariannette and {Wang}, Kailai and {Battaglia}, Nicholas and {van Engelen}, Alexander},
        title = "{LIMpy: A Semianalytic Approach to Simulating Multiline Intensity Maps at Millimeter Wavelengths}",
      journal = {\apj},
         year = 2023,
        month = nov,
       volume = {957},
       number = {2},
          eid = {87},
        pages = {87},
          doi = {10.3847/1538-4357/acf92f},
archivePrefix = {arXiv},
       eprint = {2304.06748},
 primaryClass = {astro-ph.GA},
       adsurl = {https://ui.adsabs.harvard.edu/abs/2023ApJ...957...87R}
}

@ARTICLE{SkyLine,
       author = {{Sato-Polito}, Gabriela and {Kokron}, Nickolas and {Bernal}, Jos{\'e} Luis},
        title = "{A multitracer empirically driven approach to line-intensity mapping light cones}",
      journal = {\mnras},
         year = 2023,
        month = dec,
       volume = {526},
       number = {4},
        pages = {5883-5899},
          doi = {10.1093/mnras/stad2498},
archivePrefix = {arXiv},
       eprint = {2212.08056},
 primaryClass = {astro-ph.CO},
       adsurl = {https://ui.adsabs.harvard.edu/abs/2023MNRAS.526.5883S}
}

@article{SIMPLE,
    author = "Lujan Niemeyer, Maja and Bernal, Jos{\'e} Luis and Komatsu, Eiichiro",
    title = "{SIMPLE: Simple Intensity Map Producer for Line Emission}",
    eprint = "2307.08475",
    archivePrefix = "arXiv",
    primaryClass = "astro-ph.CO",
    doi = "10.3847/1538-4357/acfef4",
    journal = "Astrophys. J.",
    volume = "958",
    number = "1",
    pages = "4",
    year = "2023"
}

@article{Richards,
  title = {Bolometers for infrared and millimeter waves},
  author = {Richards, P. L.},
  journal = {Journal of Applied Physics},
  volume = {76},
  issue = {1},
  pages = {1-24},
  numpages = {0},
  year = {1994},
  month = {July},
  publisher = {},
  doi = {10.1063/1.357128},
  url = {https://pubs.aip.org/aip/jap/article-pdf/76/1/1/18668463/1_1_online.pdf}
}

@article{SQUID,
  title = {Microwave SQUID multiplexer},
  author = {Irwin, K. D. and Lehnert, K. W.},
  journal = {Applied Physics Letters},
  volume = {85},
  issue = {11},
  pages = {2107-2109},
  numpages = {0},
  year = {2004},
  month = {September},
  publisher = {},
  doi = {10.1063/1.1791733},
  url = {https://pubs.aip.org/aip/apl/article-pdf/85/11/2107/18595393/2107_1_online.pdf}
}

@article{Dicke,
Author = {DICKE, RH},
Title = {THE MEASUREMENT OF THERMAL RADIATION AT MICROWAVE FREQUENCIES},
Journal = {REVIEW OF SCIENTIFIC INSTRUMENTS},
Year = {1946},
Volume = {17},
Number = {7},
Pages = {268-275},
DOI = {10.1063/1.1770483},
ISSN = {0034-6748},
Unique-ID = {WOS:A1946UK41300004},
}

@article{Day2003,
  title = {A broadband superconducting detector suitable for use in large arrays},
  author = {Day, Peter K. and LeDuc, Henry G. and Mazin, Benjamin A. and Vayonakis, Anastasios and Zmuidzinas, Jonas},
  journal = {Nature},
  volume = {425},
  number = {6960},
  pages = {817--821},
  year = {2003},
  publisher = {Nature Publishing Group},
  doi = {10.1038/nature02037},
  url = {https://www.nature.com/articles/nature02037}
}

@INPROCEEDINGS{CMOS,
       author = {{Loose}, Markus and {Farris}, Mark C. and {Garnett}, James D. and {Hall}, Donald N.~B. and {Kozlowski}, Lester J.},
        title = "{HAWAII-2RG: a 2k x 2k CMOS multiplexer for low and high background astronomy applications}",
    booktitle = {IR Space Telescopes and Instruments},
         year = 2003,
       editor = {{Mather}, John C.},
       series = {Society of Photo-Optical Instrumentation Engineers (SPIE) Conference Series},
       volume = {4850},
        month = mar,
        pages = {867-879},
          doi = {10.1117/12.461796},
       adsurl = {https://ui.adsabs.harvard.edu/abs/2003SPIE.4850..867L}
}

@article{CCD,
  author  = {Boyle, Willard S. and Smith, George E.},
  title   = {Charge Coupled Semiconductor Devices},
  journal = {The Bell System Technical Journal},
  volume  = {49},
  number  = {4},
  pages   = {587--593},
  year    = {1970},
  month   = {April},
  doi     = {10.1002/j.1538-7305.1970.tb01790.x}
}

@article{destripping,
  author   = {Poutanen, T. and de Gasperis, G. and Hivon, E. and Kurki-Suonio, H. and Savelainen, M. and Ashdown, M. A. J. and Borrill, J. and Cantalupo, C. and Keih{\"a}nen, E. and Natoli, P. and Rocha, G. and Stompor, R.},
  title    = {Making sky maps from Planck data},
  journal  = {Astronomy \& Astrophysics},
  year     = {2007},
  volume   = {467},
  number   = {2},
  pages    = {761--775},
  doi      = {10.1051/0004-6361:20065829},
  url      = {https://doi.org}
}

@article{MADAM2005,
  author   = {Keih{\"a}nen, E. and Kurki-Suonio, H. and Poutanen, T.},
  title    = {Madam -- a map-making method for CMB experiments},
  journal  = {Monthly Notices of the Royal Astronomical Society},
  year     = {2005},
  volume   = {360},
  number   = {1},
  pages    = {390--400},
  doi      = {10.1111/j.1365-2966.2005.09033.x},
  url      = {https://doi.org}
}

@article{MADAM2010,
  author   = {Keih{\"a}nen, E. and Keskitalo, R. and Kurki-Suonio, H. and Poutanen, T. and Sirvi{\"o}, A.-S.},
  title    = {Making cosmic microwave background temperature and polarization maps with {MADAM}},
  journal  = {Astronomy \& Astrophysics},
  year     = {2010},
  volume   = {510},
  pages    = {A57},
  doi      = {10.1051/0004-6361/200912813},
  url      = {https://doi.org}
}

@article{SANEPIC,
  author   = {Patanchon, G. and Ade, P. A. R. and Hughes, D. H. and Klein, J. and Marsden, G. and Martin, P. G. and Mauskopf, P. and Netterfield, C. B. and Olmi, L. and Pascale, E. and others},
  title    = {{SANEPIC}: A Mapmaking Method for Time Stream Data from Large Arrays},
  journal  = {The Astrophysical Journal},
  year     = {2008},
  volume   = {681},
  number   = {1},
  pages    = {708--725},
  doi      = {10.1086/588543},
  url      = {https://doi.org/10.1086/588543}
}

@misc{TOAST,
  author       = {Kisner, Theodore S. and others},
  title        = {{TOAST}: Time-Ordered Astrophysics Scalable Tools},
  howpublished = {Astrophysics Source Code Library},
  year         = {2016},
  month        = jan,
  eid          = {ascl:1601.011},
  pages        = {ascl:1601.011},
  url          = {https://ascl.net}
}

@article{Crawford2007,
    author        = "Crawford, T. M.",
    title         = "{Power spectrum sensitivity of raster-scanned CMB experiments in the presence of 1/f noise}",
    journal       = "Phys. Rev. D",
    volume        = "76",
    pages         = "063008",
    year          = "2007",
    doi           = "10.1103/PhysRevD.76.063008",
    eprint        = "0706.0125",
    archivePrefix = "arXiv",
    primaryClass  = "astro-ph"
}

@ARTICLE{Anderson2025,
       author = {{Anderson}, C.~J. and {Berger}, P. and {Chang}, T.-C. and {Dor{\'e}}, O. and {Brown}, S. and {Levin}, S. and {Seiffert}, M.},
        title = "{The Radio and Microwave Sky as Seen by Juno on its Mission to Jupiter}",
      journal = {\apj},
         year = 2025,
        month = apr,
       volume = {982},
       number = {2},
          eid = {118},
        pages = {118},
          doi = {10.3847/1538-4357/adba62},
archivePrefix = {arXiv},
       eprint = {2405.08388},
 primaryClass = {astro-ph.IM},
       adsurl = {https://ui.adsabs.harvard.edu/abs/2025ApJ...982..118A}
}

@article{Dunne2026,
  author        = {Dunne, D. A. and Cleary, K. A. and Lunde, J. G. S. and Chung, D. T. and Breysse, P. C. and Stutzer, N. O. and Bond, J. R. and Eriksen, H. K. and Gundersen, J. O. and Hoerning, G. A. and Kim, J. and Mansfield, E. M. and Mason, S. R. and Murray, N. and Rennie, T. J. and Tolgay, D. and Valentine, S. and Wehus, I. K.},
  collaboration = {COMAP Collaboration},
  title         = {{COMAP Pathfinder -- Season 2 results. IV. A stack on eBOSS/DESI quasars}},
  journal       = {Astronomy \& Astrophysics},
  volume        = {683},
  pages         = {A45},
  year          = {2026},
  month         = {mar},
  doi           = {10.1051/0004-6361/202557864},
  eprint        = {2510.23568},
  archivePrefix = {arXiv},
  primaryClass  = {astro-ph.CO},
  url           = {https://www.aanda.org/articles/aa/abs/2026/03/aa57864-25/aa57864-25.html}
}

@article{Tucker,
  title = {Quantum detection at millimeter wavelengths},
  author = {Tucker, John R. and Feldman, Marc J.},
  journal = {Rev. Mod. Phys.},
  volume = {57},
  issue = {4},
  pages = {1055--1113},
  numpages = {0},
  year = {1985},
  month = {Oct},
  publisher = {American Physical Society},
  doi = {10.1103/RevModPhys.57.1055},
  url = {https://link.aps.org/doi/10.1103/RevModPhys.57.1055}
}

@BOOK{CondonRansom,
       author = {{Condon}, James J. and {Ransom}, Scott M.},
        title = "{Essential Radio Astronomy}",
         year = 2016,
       adsurl = {https://ui.adsabs.harvard.edu/abs/2016era..book.....C}
}

@ARTICLE{HEMP,
  author={Pospieszalski, M.W.},
  journal={IEEE Transactions on Microwave Theory and Techniques}, 
  title={Modeling of noise parameters of MESFETs and MODFETs and their frequency and temperature dependence}, 
  year={1989},
  volume={37},
  number={9},
  pages={1340-1350},
  doi={10.1109/22.32217}}

@article{HausMullen,
  title = {Quantum Noise in Linear Amplifiers},
  author = {Haus, H. A. and Mullen, J. A.},
  journal = {Phys. Rev.},
  volume = {128},
  issue = {5},
  pages = {2407--2413},
  numpages = {0},
  year = {1962},
  month = {Dec},
  publisher = {American Physical Society},
  doi = {10.1103/PhysRev.128.2407},
  url = {https://link.aps.org/doi/10.1103/PhysRev.128.2407}
}

@ARTICLE{ComapPCA,
       author = {{Lunde}, J.~G.~S. and {Stutzer}, N.-O. and {Breysse}, P.~C. and {Chung}, D.~T. and {Cleary}, K.~A. and {Dunne}, D.~A. and {Eriksen}, H.~K. and {Harper}, S.~E. and {Ihle}, H.~T. and {Lamb}, J.~W. and {Pearson}, T.~J. and {Philip}, L. and {Wehus}, I.~K. and {Woody}, D.~P. and {Bond}, J.~R. and {Church}, S.~E. and {Gaier}, T. and {Gundersen}, J.~O. and {Harris}, A.~I. and {Hobbs}, R. and {Kim}, J. and {Lawrence}, C.~R. and {Murray}, N. and {Padmanabhan}, H. and {Readhead}, A.~C.~S. and {Rennie}, T.~J. and {Tolgay}, D. and {COMAP Collaboration}},
        title = "{COMAP Pathfinder -- Season 2 results: I. Improved data selection and processing}",
      journal = {\aap},
         year = 2024,
        month = nov,
       volume = {691},
          eid = {A335},
        pages = {A335},
          doi = {10.1051/0004-6361/202451121},
archivePrefix = {arXiv},
       eprint = {2406.07510},
 primaryClass = {astro-ph.CO},
       adsurl = {https://ui.adsabs.harvard.edu/abs/2024A&A...691A.335L}
}

@ARTICLE{Bromm2002,
       author = {{Bromm}, Volker and {Coppi}, Paolo S. and {Larson}, Richard B.},
        title = "{The Formation of the First Stars. I. The Primordial Star-forming Cloud}",
      journal = {\apj},
         year = 2002,
        month = jan,
       volume = {564},
       number = {1},
        pages = {23-51},
          doi = {10.1086/323947},
archivePrefix = {arXiv},
       eprint = {astro-ph/0102503},
 primaryClass = {astro-ph},
       adsurl = {https://ui.adsabs.harvard.edu/abs/2002ApJ...564...23B}
}

@ARTICLE{Abel2002,
       author = {{Abel}, Tom and {Bryan}, Greg L. and {Norman}, Michael L.},
        title = "{The Formation of the First Star in the Universe}",
      journal = {Science},
         year = 2002,
        month = jan,
       volume = {295},
       number = {5552},
        pages = {93-98},
          doi = {10.1126/science.1063991},
archivePrefix = {arXiv},
       eprint = {astro-ph/0112088},
 primaryClass = {astro-ph},
       adsurl = {https://ui.adsabs.harvard.edu/abs/2002Sci...295...93A}
}

@ARTICLE{Dunne2025,
       author = {{Dunne}, D.~A. and {Cleary}, K.~A. and {Breysse}, P.~C. and {Chung}, D.~T. and {Ihle}, H.~T. and {Lunde}, J.~G.~S. and {Padmanabhan}, H. and {Stutzer}, N.-O. and {Bond}, J.~R. and {Gundersen}, J.~O. and {Kim}, J. and {Readhead}, A.~C.~S.},
        title = "{Three-dimensional stacking as a line intensity mapping statistic}",
      journal = {\aap},
         year = 2025,
        month = oct,
       volume = {702},
          eid = {A247},
        pages = {A247},
          doi = {10.1051/0004-6361/202554780},
archivePrefix = {arXiv},
       eprint = {2503.21743},
 primaryClass = {astro-ph.CO},
       adsurl = {https://ui.adsabs.harvard.edu/abs/2025A&A...702A.247D}
}

@ARTICLE{Kramer2026,
       author = {{Kramer}, Samuel H. and {Breysse}, Patrick C. and {Pullen}, Anthony R. and {Siddique}, Faizah K. and {Switzer}, Eric R. and {Timbie}, Peter T. and {Chung}, Dongwoo},
        title = "{Comparison of cross-correlation methods for line intensity mapping}",
      journal = {\mnras},
         year = 2026,
        month = jul,
       volume = {549},
       number = {3},
          eid = {stag998},
        pages = {stag998},
          doi = {10.1093/mnras/stag998},
archivePrefix = {arXiv},
       eprint = {2603.13218},
 primaryClass = {astro-ph.CO},
       adsurl = {https://ui.adsabs.harvard.edu/abs/2026MNRAS.549ag998K}
      }

@ARTICLE{TIME2026, 
       author = {{Yang}, Selina F. and {McAtee}, Sophie M. and {Vaughan}, Benjamin J. and {Crites}, Abigail T. and {Butler}, Victoria L. and {Chung}, Dongwoo T. and {Keenan}, Ryan P. and {Pham}, Dang and {Prakash}, Shwetha and {Bock}, James J. and {Bradford}, Charles M. and {Chang}, Tzu-Ching and {Cheng}, Yun-Ting and {Dunn}, Audrey and {Emerson}, Nicholas and {Frez}, Clifford and {Hunacek}, Jonathon and {Li}, Chao-Te and {Lowe}, Ian N. and {Lau}, King and {Marrone}, Daniel P. and {Mayer}, Evan C. and {Sun}, Guochao and {Trumper}, Isaac and {Turner}, Anthony D. and {Wei}, Ta-Shun and {Zemcov}, Michael and {TIME Collaboration}},
        title = "{TIME Commissioning Observations. I. Mapping Dust and Molecular Gas in the Sgr A Molecular Cloud Complex at the Galactic Center}",
      journal = {\apj},
         year = 2026,
        month = may,
       volume = {1003},
       number = {1},
          eid = {23},
        pages = {23},
          doi = {10.3847/1538-4357/ae606c},
archivePrefix = {arXiv},
       eprint = {2511.09473},
 primaryClass = {astro-ph.IM},
       adsurl = {https://ui.adsabs.harvard.edu/abs/2026ApJ..1003...23Y}
}

@ARTICLE{Flury2022,
       author = {{Flury}, Sophia R. and {Jaskot}, Anne E. and {Ferguson}, Harry C. and {Worseck}, G{\'a}bor and {Makan}, Kirill and {Chisholm}, John and {Saldana-Lopez}, Alberto and {Schaerer}, Daniel and {McCandliss}, Stephan and {Wang}, Bingjie and {Ford}, N.~M. and {Heckman}, Timothy and {Ji}, Zhiyuan and {Giavalisco}, Mauro and {Amorin}, Ricardo and {Atek}, Hakim and {Blaizot}, Jeremy and {Borthakur}, Sanchayeeta and {Carr}, Cody and {Castellano}, Marco and {Cristiani}, Stefano and {De Barros}, Stephane and {Dickinson}, Mark and {Finkelstein}, Steven L. and {Fleming}, Brian and {Fontanot}, Fabio and {Garel}, Thibault and {Grazian}, Andrea and {Hayes}, Matthew and {Henry}, Alaina and {Mauerhofer}, Valentin and {Micheva}, Genoveva and {Oey}, M.~S. and {Ostlin}, Goran and {Papovich}, Casey and {Pentericci}, Laura and {Ravindranath}, Swara and {Rosdahl}, Joakim and {Rutkowski}, Michael and {Santini}, Paola and {Scarlata}, Claudia and {Teplitz}, Harry and {Thuan}, Trinh and {Trebitsch}, Maxime and {Vanzella}, Eros and {Verhamme}, Anne and {Xu}, Xinfeng},
        title = "{The Low-redshift Lyman Continuum Survey. I. New, Diverse Local Lyman Continuum Emitters}",
      journal = {\apjs},
         year = 2022,
        month = may,
       volume = {260},
       number = {1},
          eid = {1},
        pages = {1},
          doi = {10.3847/1538-4365/ac5331},
archivePrefix = {arXiv},
       eprint = {2201.11716},
 primaryClass = {astro-ph.GA},
       adsurl = {https://ui.adsabs.harvard.edu/abs/2022ApJS..260....1F}
}

@ARTICLE{Giovinazzo26,
       author = {{Giovinazzo}, Emma and {Oesch}, Pascal A. and {Weibel}, Andrea and {Meyer}, Romain A. and {Witten}, Callum and {Bhagwat}, Aniket and {Brammer}, Gabriel and {Chisholm}, John and {de Graaff}, Anna and {Gottumukkala}, Rashmi and {Jecmen}, Michelle and {Katz}, Harley and {Leja}, Joel and {Marques-Chaves}, Rui and {Maseda}, Michael and {Shivaei}, Irene and {Trebitsch}, Maxime and {Verhamme}, Anne},
        title = "{Breaking through the cosmic fog: JWST/NIRSpec constraints on ionizing photon escape in reionization-era galaxies}",
      journal = {\aap},
         year = 2026,
        month = mar,
       volume = {707},
          eid = {A352},
        pages = {A352},
          doi = {10.1051/0004-6361/202556204},
archivePrefix = {arXiv},
       eprint = {2507.01096},
 primaryClass = {astro-ph.GA},
       adsurl = {https://ui.adsabs.harvard.edu/abs/2026A&A...707A.352G}
}

@article{Breysse:2021ecm,
    author = {Breysse, Patrick C. and Yang, Shengqi and Somerville, Rachel S. and Pullen, Anthony R. and Popping, Gerg{\"o} and Maniyar, Abhishek S.},
    title = "{On Estimating the Cosmic Molecular Gas Density from CO Line Intensity Mapping Observations}",
    eprint = "2106.14904",
    archivePrefix = "arXiv",
    primaryClass = "astro-ph.CO",
    doi = "10.3847/1538-4357/ac5a46",
    journal = "Astrophys. J.",
    volume = "929",
    number = "1",
    pages = "30",
    year = "2022"
}

@ARTICLE{Garratt21,
       author = {{Garratt}, T.~K. and {Coppin}, K.~E.~K. and {Geach}, J.~E. and {Almaini}, O. and {Hartley}, W.~G. and {Maltby}, D.~T. and {Simpson}, C.~J. and {Wilkinson}, A. and {Conselice}, C.~J. and {Franco}, M. and {Ivison}, R.~J. and {Koprowski}, M.~P. and {Lovell}, C.~C. and {Pope}, A. and {Scott}, D. and {van der Werf}, P.},
        title = "{Cosmic Evolution of the H$_{2}$ Mass Density and the Epoch of Molecular Gas}",
      journal = {\apj},
         year = 2021,
        month = may,
       volume = {912},
       number = {1},
          eid = {62},
        pages = {62},
          doi = {10.3847/1538-4357/abec81},
archivePrefix = {arXiv},
       eprint = {2103.08613},
 primaryClass = {astro-ph.GA},
       adsurl = {https://ui.adsabs.harvard.edu/abs/2021ApJ...912...62G}
}

@article{Hu:2024nya,
    author = "Hu, W. and others",
    title = "{CONCERTO at APEX On-sky performance in continuum}",
    eprint = "2406.15572",
    archivePrefix = "arXiv",
    primaryClass = "astro-ph.IM",
    doi = "10.1051/0004-6361/202449260",
    journal = "Astron. Astrophys.",
    volume = "689",
    pages = "A20",
    year = "2024"
}

@article{VanCuyck:2023uli,
    author = "Van Cuyck, M. and others",
    title = "{CONCERTO: Extracting the power spectrum of the [C II ] emission line}",
    eprint = "2306.01568",
    archivePrefix = "arXiv",
    primaryClass = "astro-ph.CO",
    doi = "10.1051/0004-6361/202346270",
    journal = "Astron. Astrophys.",
    volume = "676",
    pages = "A62",
    year = "2023"
}

@article{Pullen:2022pgp,
    author = "Pullen, Anthony R. and others",
    title = "{Extragalactic science with the experiment for cryogenic large-aperture intensity mapping}",
    eprint = "2209.02497",
    archivePrefix = "arXiv",
    primaryClass = "astro-ph.CO",
    doi = "10.1093/mnras/stad916",
    journal = "Mon. Not. Roy. Astron. Soc.",
    volume = "521",
    number = "4",
    pages = "6124--6142",
    year = "2023"
}

@article{Yuan:2022rsc,
    author = "Yuan, Sihan and Hadzhiyska, Boryana and Bose, Sownak and Eisenstein, Daniel J.",
    title = "{Illustrating galaxy{\textendash}halo connection in the DESI era with illustrisTNG}",
    eprint = "2202.12911",
    archivePrefix = "arXiv",
    primaryClass = "astro-ph.CO",
    doi = "10.1093/mnras/stac830",
    journal = "Mon. Not. Roy. Astron. Soc.",
    volume = "512",
    number = "4",
    pages = "5793--5811",
    year = "2022"
}

@article{Wechsler:2018pic,
    author = "Wechsler, Risa H. and Tinker, Jeremy L.",
    title = "{The Connection between Galaxies and their Dark Matter Halos}",
    eprint = "1804.03097",
    archivePrefix = "arXiv",
    primaryClass = "astro-ph.GA",
    doi = "10.1146/annurev-astro-081817-051756",
    journal = "Ann. Rev. Astron. Astrophys.",
    volume = "56",
    pages = "435--487",
    year = "2018"
}

@article{Croton:2006ys,
    author = "Croton, Darren J. and Gao, Liang and White, Simon D. M.",
    title = "{Halo assembly bias and its effects on galaxy clustering}",
    eprint = "astro-ph/0605636",
    archivePrefix = "arXiv",
    doi = "10.1111/j.1365-2966.2006.11230.x",
    journal = "Mon. Not. Roy. Astron. Soc.",
    volume = "374",
    pages = "1303--1309",
    year = "2007"
}

@article{MoradinezhadDizgah:2023src,
    author = "Moradinezhad Dizgah, Azadeh and Bellini, Emilio and Keating, Garrett K.",
    title = "{Probing Dark Energy and Modifications of Gravity with Ground-based millimeter-wavelength Line Intensity Mapping}",
    eprint = "2304.08471",
    archivePrefix = "arXiv",
    primaryClass = "astro-ph.CO",
    doi = "10.3847/1538-4357/ad2078",
    journal = "Astrophys. J.",
    volume = "965",
    number = "1",
    pages = "19",
    year = "2024"
}

@ARTICLE{Bolatto13,
       author = {{Bolatto}, Alberto D. and {Wolfire}, Mark and {Leroy}, Adam K.},
        title = "{The CO-to-H$_{2}$ Conversion Factor}",
      journal = {\araa},
         year = 2013,
        month = aug,
       volume = {51},
       number = {1},
        pages = {207-268},
          doi = {10.1146/annurev-astro-082812-140944},
archivePrefix = {arXiv},
       eprint = {1301.3498},
 primaryClass = {astro-ph.GA},
       adsurl = {https://ui.adsabs.harvard.edu/abs/2013ARA&A..51..207B}
}

@article{Keating20,
    author = "Keating, Garrett K. and Marrone, Daniel P. and Bower, Geoffrey C. and Keenan, Ryan P.",
    title = "{An Intensity Mapping Detection of Aggregate CO Line Emission at 3 mm}",
    eprint = "2008.08087",
    archivePrefix = "arXiv",
    primaryClass = "astro-ph.GA",
    doi = "10.3847/1538-4357/abb08e",
    journal = "Astrophys. J.",
    volume = "901",
    number = "2",
    pages = "141",
    year = "2020"
}

@article{Chung:2017uot,
    author = "Chung, Dongwoo T. and Li, Tony Y. and Viero, Marco P. and Church, Sarah E. and Wechsler, Risa H.",
    title = "{On estimation of contamination from hydrogen cyanide in carbon monoxide line intensity mapping}",
    eprint = "1706.03005",
    archivePrefix = "arXiv",
    primaryClass = "astro-ph.GA",
    doi = "10.3847/1538-4357/aa8624",
    journal = "Astrophys. J.",
    volume = "846",
    number = "1",
    pages = "60",
    year = "2017"
}

@article{Crighton:2015pza,
    author = "Crighton, Neil H. M. and others",
    title = "{The neutral hydrogen cosmological mass density at z = 5}",
    eprint = "1506.02037",
    archivePrefix = "arXiv",
    primaryClass = "astro-ph.CO",
    doi = "10.1093/mnras/stv1182",
    journal = "Mon. Not. Roy. Astron. Soc.",
    volume = "452",
    number = "1",
    pages = "217--234",
    year = "2015"
}

@ARTICLE{Bracks26,
       author = {{Bracks}, Justin S. and {Keenan}, Ryan P. and {Agrawal}, Shubh and {Keating}, Garrett K. and {Aguirre}, James E. and {Lidz}, Adam and {Bradford}, Charles M. and {Brendal}, Brockton and {Filippini}, Jeffrey and {Fu}, Jianyang and {Garcia}, Karolina and {Groppi}, Christopher and {Hailey-Dunsheath}, Steven and {Janssen}, Reinier M.~J. and {Kang}, Wooseok and {Liu}, Lun-Jun and {Lowe}, Ian and {Manduca}, Alex and {Marrone}, Daniel P. and {Mauskopf}, Philip and {Mayer}, Evan C. and {O'Donnell}, Sydnee and {Saeid}, Talia and {Tartakovsky}, Simon and {Cuyck}, Mathilde and {Vieira}, Joaquin D. and {Zebrowski}, Jessica A.},
        title = "{Forecasting the cross correlation of Terahertz Intensity Mapper [CII] line intensity maps with Euclid galaxies}",
      journal = {arXiv e-prints},
         year = 2026,
        month = feb,
          eid = {arXiv:2602.24204},
        pages = {arXiv:2602.24204},
          doi = {10.48550/arXiv.2602.24204},
archivePrefix = {arXiv},
       eprint = {2602.24204},
 primaryClass = {astro-ph.GA},
       adsurl = {https://ui.adsabs.harvard.edu/abs/2026arXiv260224204B}
}

@article{Agrawal:2026nzp,
    author = "Agrawal, Shubh and others",
    title = "{A New Window into the Baryon Cycle at Cosmic Noon with Line Intensity Mapping: Forecasts for auto- and cross-correlations in [CII]-158$\mu$m, HI 21 cm, CO$_{J+1\rightarrow J}$, and H$\alpha$ galaxies}",
    eprint = "2602.24120",
    archivePrefix = "arXiv",
    primaryClass = "astro-ph.CO",
    reportNumber = "FERMILAB-PUB-26-0139-V",
    month = "2",
    year = "2026"
}

@article{Freundt:2024lsh,
    author = "Freundt, Rodrigo and others",
    title = "{CCAT: A status update on the EoR-Spec instrument module for Prime-Cam}",
    eprint = "2409.05979",
    archivePrefix = "arXiv",
    primaryClass = "astro-ph.IM",
    doi = "10.1117/12.3018477",
    journal = "Astronomy",
    volume = "12",
    pages = "131020",
    month = "9",
    year = "2024"
}

@article{CCAT-Prime:2021lly,
    author = "Aravena, Manuel and others",
    collaboration = "CCAT-Prime",
    title = "{CCAT-prime Collaboration: Science Goals and Forecasts with Prime-Cam on the Fred Young Submillimeter Telescope}",
    eprint = "2107.10364",
    archivePrefix = "arXiv",
    primaryClass = "astro-ph.CO",
    doi = "10.3847/1538-4365/ac9838",
    journal = "Astrophys. J. Suppl.",
    volume = "264",
    number = "1",
    pages = "7",
    year = "2023"
}

@article{Osinga:2025gpt,
    author = "Osinga, Calvin K. and Diemer, Benedikt and Villaescusa-Navarro, Francisco",
    title = "{Lost in the FoG: Pitfalls of Models for Large-scale Hydrogen Distributions}",
    eprint = "2503.01956",
    archivePrefix = "arXiv",
    primaryClass = "astro-ph.CO",
    doi = "10.3847/1538-4357/ae0e69",
    journal = "Astrophys. J.",
    volume = "994",
    number = "2",
    pages = "171",
    year = "2025"
}

@unpublished{Chiang26,
  author = {Chiang, Y.-K.},
  title  = "{Cosmic CO and [CII] Backgrounds and the Fueling of Star Formation Over 12 Gyr}",
  note   = {in preparation},
  year   = {2026}
}

@article{Padmanabhan18,
    author = "Padmanabhan, Hamsa",
    title = "{Constraining the evolution of [C II] intensity through the end stages of reionization}",
    eprint = "1811.01968",
    archivePrefix = "arXiv",
    primaryClass = "astro-ph.CO",
    doi = "10.1093/mnras/stz1878",
    journal = "Mon. Not. Roy. Astron. Soc.",
    volume = "488",
    number = "3",
    pages = "3014--3023",
    year = "2019"
}

@article{Mitchell-Wynne:2015rha,
    author = "Mitchell-Wynne, Ketron and others",
    title = "{Ultraviolet Luminosity Density of the Universe During the Epoch of Reionization}",
    eprint = "1509.02935",
    archivePrefix = "arXiv",
    primaryClass = "astro-ph.CO",
    doi = "10.1038/NCOMMS8945",
    journal = "Nature Commun.",
    volume = "6",
    pages = "7945",
    year = "2015"
}

@ARTICLE{Kovetz26,
       author = {{Kovetz}, Ely D. and {Lazare}, Hovav and {Libanore}, Sarah and {Mu{\~n}oz}, Julian B. and {Vanzan}, Eleonora},
        title = "{When galaxies burst: enhanced shot-noise for line-intensity mapping in the JWST era}",
      journal = {arXiv e-prints},
         year = 2026,
        month = may,
          eid = {arXiv:2605.13967},
        pages = {arXiv:2605.13967},
          doi = {10.48550/arXiv.2605.13967},
archivePrefix = {arXiv},
       eprint = {2605.13967},
 primaryClass = {astro-ph.GA},
       adsurl = {https://ui.adsabs.harvard.edu/abs/2026arXiv260513967K}
}

@ARTICLE{Munoz26,
       author = {{Mu{\~n}oz}, Julian B. and {Chisholm}, John and {Sun}, Guochao and {Samuel}, Jenna and {Mirocha}, Jordan and {Bregou}, Emily and {Venditti}, Alessandra and {Qezlou}, Mahdi and {Simmonds}, Charlotte and {Endsley}, Ryan},
        title = "{Relatively fast and reasonably furious: evidence for increased burstiness in smaller haloes at cosmic dawn}",
      journal = {\mnras},
         year = 2026,
        month = apr,
       volume = {547},
       number = {4},
          eid = {stag415},
        pages = {stag415},
          doi = {10.1093/mnras/stag415},
archivePrefix = {arXiv},
       eprint = {2601.07912},
 primaryClass = {astro-ph.GA},
       adsurl = {https://ui.adsabs.harvard.edu/abs/2026MNRAS.547ag415M}
}

@article{Liu:2024fti,
    author = "Liu, Lun-Jun and Sun, Guochao and Chang, Tzu-Ching and Furlanetto, Steven R. and Bradford, Charles M.",
    title = "{Effects of Bursty Star Formation on [C ii] Line Intensity Mapping of High-redshift Galaxies}",
    eprint = "2401.04204",
    archivePrefix = "arXiv",
    primaryClass = "astro-ph.GA",
    doi = "10.3847/1538-4357/ad73d5",
    journal = "Astrophys. J.",
    volume = "974",
    number = "2",
    pages = "175",
    year = "2024"
}

@ARTICLE{Yang22,
       author = {{Yang}, Shengqi and {Popping}, Gerg{\"o} and {Somerville}, Rachel S. and {Pullen}, Anthony R. and {Breysse}, Patrick C. and {Maniyar}, Abhishek S.},
        title = "{An Empirical Representation of a Physical Model for the ISM [C II], CO, and [C I] Emission at Redshift 1 {\ensuremath{\leq}} z {\ensuremath{\leq}} 9}",
      journal = {\apj},
         year = 2022,
        month = apr,
       volume = {929},
       number = {2},
          eid = {140},
        pages = {140},
          doi = {10.3847/1538-4357/ac5d57},
archivePrefix = {arXiv},
       eprint = {2108.07716},
 primaryClass = {astro-ph.GA},
       adsurl = {https://ui.adsabs.harvard.edu/abs/2022ApJ...929..140Y}
}

@article{Almualla:2025pix,
    author = "Almualla, Mouza and Smith, Aaron and Kannan, Rahul and Hernquist, Lars and Garaldi, Enrico and Lidz, Adam and Lorinc, Kevin and Chan, Jennifer Yik Ham and Vogelsberger, Mark",
    title = "{The THESAN project: Lyman-alpha intensity mapping of cosmic reionization}",
    eprint = "2512.06085",
    archivePrefix = "arXiv",
    primaryClass = "astro-ph.CO",
    month = "12",
    year = "2025"
}

@ARTICLE{Alvarez16,
       author = {{Alvarez}, Marcelo A.},
        title = "{The Kinetic Sunyaev-Zel{\textquoteright}dovich Effect from Reionization: Simulated Full-sky Maps at Arcminute Resolution}",
      journal = {\apj},
         year = 2016,
        month = jun,
       volume = {824},
       number = {2},
          eid = {118},
        pages = {118},
          doi = {10.3847/0004-637X/824/2/118},
archivePrefix = {arXiv},
       eprint = {1511.02846},
 primaryClass = {astro-ph.CO},
       adsurl = {https://ui.adsabs.harvard.edu/abs/2016ApJ...824..118A}
}

@ARTICLE{Stein20,
       author = {{Stein}, George and {Alvarez}, Marcelo A. and {Bond}, J. Richard and {van Engelen}, Alexander and {Battaglia}, Nicholas},
        title = "{The Websky extragalactic CMB simulations}",
      journal = {\jcap},
         year = 2020,
        month = oct,
       volume = {2020},
       number = {10},
          eid = {012},
        pages = {012},
          doi = {10.1088/1475-7516/2020/10/012},
archivePrefix = {arXiv},
       eprint = {2001.08787},
 primaryClass = {astro-ph.CO},
       adsurl = {https://ui.adsabs.harvard.edu/abs/2020JCAP...10..012S}
}

@ARTICLE{Lee24,
       author = {{Lee}, Jaemyoung (Jason) and {Bond}, J. Richard and {Motloch}, Pavel and {van Engelen}, Alexander and {Stein}, George},
        title = "{Exploring the non-Gaussianity of the cosmic infrared background and its weak gravitational lensing}",
      journal = {\mnras},
         year = 2024,
        month = apr,
       volume = {529},
       number = {3},
        pages = {2543-2558},
          doi = {10.1093/mnras/stae605},
archivePrefix = {arXiv},
       eprint = {2304.07283},
 primaryClass = {astro-ph.CO},
       adsurl = {https://ui.adsabs.harvard.edu/abs/2024MNRAS.529.2543L}
}

@article{Ambrose:2025mcg,
    author = "Ambrose, Abigail E. and Visbal, Eli and Kulkarni, Mihir and McQuinn, Matthew",
    title = "{Radiative transfer simulations of Ly{\ensuremath{\alpha}} intensity mapping during cosmic reionization including sources from galaxies and the intergalactic medium}",
    eprint = "2502.18654",
    archivePrefix = "arXiv",
    primaryClass = "astro-ph.GA",
    doi = "10.1088/1475-7516/2025/08/080",
    journal = "JCAP",
    volume = "08",
    pages = "080",
    year = "2025"
}

@ARTICLE{Abraham14,
       author = {{Abraham}, Roberto G. and {van Dokkum}, Pieter G.},
        title = "{Ultra-Low Surface Brightness Imaging with the Dragonfly Telephoto Array}",
      journal = {\pasp},
         year = 2014,
        month = jan,
       volume = {126},
       number = {935},
        pages = {55},
          doi = {10.1086/674875},
archivePrefix = {arXiv},
       eprint = {1401.5473},
 primaryClass = {astro-ph.IM},
       adsurl = {https://ui.adsabs.harvard.edu/abs/2014PASP..126...55A}
}

@article{ACT:2023kun,
    author = "Madhavacheril, Mathew S. and others",
    collaboration = "ACT",
    title = "{The Atacama Cosmology Telescope: DR6 Gravitational Lensing Map and Cosmological Parameters}",
    eprint = "2304.05203",
    archivePrefix = "arXiv",
    primaryClass = "astro-ph.CO",
    reportNumber = "FERMILAB-PUB-23-206-PPD",
    doi = "10.3847/1538-4357/acff5f",
    journal = "Astrophys. J.",
    volume = "962",
    number = "2",
    pages = "113",
    year = "2024"
}

@ARTICLE{Adams72,
       author = {{Adams}, Thomas F.},
        title = "{The Escape of Resonance-Line Radiation from Extremely Opaque Media}",
      journal = {\apj},
         year = 1972,
        month = jun,
       volume = {174},
        pages = {439},
          doi = {10.1086/151503},
       adsurl = {https://ui.adsabs.harvard.edu/abs/1972ApJ...174..439A}
}

@article{Agrawal:2025uig,
    author = "Agrawal, Shubh and Aguirre, James and Keenan, Ryan",
    title = "{Far-infrared lines hidden in archival deep multi-wavelength surveys: Limits on [CII]-158$\mu$m at $z \sim 0.3-2.9$}",
    eprint = "2509.24211",
    archivePrefix = "arXiv",
    primaryClass = "astro-ph.GA",
    month = "9",
    year = "2025"
}

@ARTICLE{Alcock79,
   author = {{Alcock}, C. and {Paczynski}, B.},
    title = "{An evolution free test for non-zero cosmological constant}",
  journal = {\nat},
     year = 1979,
    month = oct,
   volume = 281,
    pages = {358},
      doi = {10.1038/281358a0},
   adsurl = {http://adsabs.harvard.edu/abs/1979Natur.281..358A}
}

@ARTICLE{Alvarez12,
       author = {{Alvarez}, Marcelo A. and {Abel}, Tom},
        title = "{The Effect of Absorption Systems on Cosmic Reionization}",
      journal = {\apj},
         year = 2012,
        month = mar,
       volume = {747},
       number = {2},
          eid = {126},
        pages = {126},
          doi = {10.1088/0004-637X/747/2/126},
archivePrefix = {arXiv},
       eprint = {1003.6132},
 primaryClass = {astro-ph.CO},
       adsurl = {https://ui.adsabs.harvard.edu/abs/2012ApJ...747..126A}
}

@article{Anderson:2022svu,
    author = "Anderson, Christopher J. and Switzer, Eric R. and Breysse, Patrick C.",
    title = "{Constraining low redshift [C\,II] emission by cross-correlating FIRAS and BOSS data}",
    eprint = "2202.00203",
    archivePrefix = "arXiv",
    primaryClass = "astro-ph.CO",
    doi = "10.1093/mnras/stac1301",
    journal = "Mon. Not. Roy. Astron. Soc.",
    volume = "514",
    number = "1",
    pages = "1169--1187",
    year = "2022"
}

@article{Baldauf:2013hka,
    author = "Baldauf, Tobias and Seljak, Uro\v{s} and Smith, Robert E. and Hamaus, Nico and Desjacques, Vincent",
    title = "{Halo stochasticity from exclusion and nonlinear clustering}",
    eprint = "1305.2917",
    archivePrefix = "arXiv",
    primaryClass = "astro-ph.CO",
    doi = "10.1103/PhysRevD.88.083507",
    journal = "Phys. Rev. D",
    volume = "88",
    number = "8",
    pages = "083507",
    year = "2013"
}

@article{Barkana:2000fd,
    author = "Barkana, Rennan and Loeb, Abraham",
    title = "{In the beginning: The First sources of light and the reionization of the Universe}",
    eprint = "astro-ph/0010468",
    archivePrefix = "arXiv",
    doi = "10.1016/S0370-1573(01)00019-9",
    journal = "Phys. Rept.",
    volume = "349",
    pages = "125--238",
    year = "2001"
}

@article{Barkana:2001gr,
    author = "Barkana, Rennan and Haiman, Zoltan and Ostriker, Jeremiah P.",
    title = "{Constraints on warm dark matter from cosmological reionization}",
    eprint = "astro-ph/0102304",
    archivePrefix = "arXiv",
    reportNumber = "CITA-2001-06",
    doi = "10.1086/322393",
    journal = "Astrophys. J.",
    volume = "558",
    pages = "482",
    year = "2001"
}

@article{Bashinsky:2003tk,
    author = "Bashinsky, Sergei and Seljak, Uros",
    title = "{Neutrino perturbations in CMB anisotropy and matter clustering}",
    eprint = "astro-ph/0310198",
    archivePrefix = "arXiv",
    doi = "10.1103/PhysRevD.69.083002",
    journal = "Phys. Rev. D",
    volume = "69",
    pages = "083002",
    year = "2004"
}

@article{Baumann:2019keh,
    author = "Baumann, Daniel D and Beutler, Florian and Flauger, Raphael and Green, Daniel R and Slosar, An\v{z}e and Vargas-Maga\~na, Mariana and Wallisch, Benjamin and Y\`eche, Christophe",
    title = "{First constraint on the neutrino-induced phase shift in the spectrum of baryon acoustic oscillations}",
    eprint = "1803.10741",
    archivePrefix = "arXiv",
    primaryClass = "astro-ph.CO",
    doi = "10.1038/s41567-019-0435-6",
    journal = "Nature Phys.",
    volume = "15",
    pages = "465--469",
    year = "2019"
}

@article{Baumann:2017gkg,
    author = "Baumann, Daniel and Green, Daniel and Wallisch, Benjamin",
    title = "{Searching for light relics with large-scale structure}",
    eprint = "1712.08067",
    archivePrefix = "arXiv",
    primaryClass = "astro-ph.CO",
    doi = "10.1088/1475-7516/2018/08/029",
    journal = "JCAP",
    volume = "08",
    pages = "029",
    year = "2018"
}

@article{Bernal:2019gfq,
    author = "Bernal, Jos\'e Luis and Breysse, Patrick C. and Kovetz, Ely D.",
    title = "{Cosmic Expansion History from Line-Intensity Mapping}",
    eprint = "1907.10065",
    archivePrefix = "arXiv",
    primaryClass = "astro-ph.CO",
    doi = "10.1103/PhysRevLett.123.251301",
    journal = "Phys. Rev. Lett.",
    volume = "123",
    number = "25",
    pages = "251301",
    year = "2019"
}

@article{Bernal:2019jdo,
    author = "Bernal, Jos\'e Luis and Breysse, Patrick C. and Gil-Mar\'\i{}n, H\'ector and Kovetz, Ely D.",
    title = "{User\textquoteright{}s guide to extracting cosmological information from line-intensity maps}",
    eprint = "1907.10067",
    archivePrefix = "arXiv",
    primaryClass = "astro-ph.CO",
    doi = "10.1103/PhysRevD.100.123522",
    journal = "Phys. Rev. D",
    volume = "100",
    number = "12",
    pages = "123522",
    year = "2019"
}

@article{Beane:2018dzk,
    author = "Beane, Angus and Villaescusa-Navarro, Francisco and Lidz, Adam",
    title = "{Measuring the EoR Power Spectrum Without Measuring the EoR Power Spectrum}",
    eprint = "1811.10609",
    archivePrefix = "arXiv",
    primaryClass = "astro-ph.CO",
    doi = "10.3847/1538-4357/ab0a08",
    journal = "Astrophys. J.",
    volume = "874",
    number = "2",
    pages = "133",
    year = "2019"
}

@article{Beane:2018pmx,
    author = "Beane, Angus and Lidz, Adam",
    title = "{Extracting bias using the cross-bispectrum: An EoR and 21 cm-[CII]-[CII] case study}",
    eprint = "1806.02796",
    archivePrefix = "arXiv",
    primaryClass = "astro-ph.CO",
    doi = "10.3847/1538-4357/aae388",
    journal = "Astrophys. J.",
    volume = "867",
    number = "1",
    pages = "26",
    year = "2018"
}

@article{Bernal:2022jap,
    author = "Bernal, Jos\'e Luis and Kovetz, Ely D.",
    title = "{Line-Intensity Mapping: Theory Review}",
    eprint = "2206.15377",
    archivePrefix = "arXiv",
    primaryClass = "astro-ph.CO",
    month = "6",
    year = "2022"
}

@article{Bernal:2022wsu,
    author = "Bernal, Jos{\'e} Luis and Sato-Polito, Gabriela and Kamionkowski, Marc",
    title = "{Cosmic Optical Background Excess, Dark Matter, and Line-Intensity Mapping}",
    eprint = "2203.11236",
    archivePrefix = "arXiv",
    primaryClass = "astro-ph.CO",
    doi = "10.1103/PhysRevLett.129.231301",
    journal = "Phys. Rev. Lett.",
    volume = "129",
    number = "23",
    pages = "231301",
    year = "2022"
}

@article{Bernal:2020lkd,
    author = "Bernal, Jos\'e Luis and Caputo, Andrea and Kamionkowski, Marc",
    title = "{Strategies to Detect Dark-Matter Decays with Line-Intensity Mapping}",
    eprint = "2012.00771",
    archivePrefix = "arXiv",
    primaryClass = "astro-ph.CO",
    doi = "10.1103/PhysRevD.103.063523",
    journal = "Phys. Rev. D",
    volume = "103",
    number = "6",
    pages = "063523",
    year = "2021",
    note = "[Erratum: Phys.Rev.D 105, 089901 (2022)]"
}

@ARTICLE{Bernstein11,
       author = {{Bernstein}, Gary M. and {Cai}, Yan-Chuan},
        title = "{Cosmology without cosmic variance}",
      journal = {\mnras},
         year = 2011,
        month = oct,
       volume = {416},
       number = {4},
        pages = {3009-3016},
          doi = {10.1111/j.1365-2966.2011.19249.x},
archivePrefix = {arXiv},
       eprint = {1104.3862},
 primaryClass = {astro-ph.CO},
       adsurl = {https://ui.adsabs.harvard.edu/abs/2011MNRAS.416.3009B}
}

@book{Bethe:1957ncq,
    author = "Bethe, Hans A. and Salpeter, Edwin E.",
    title = "{Quantum Mechanics of One- and Two-Electron Atoms}",
    doi = "10.1007/978-3-662-12869-5",
    year = "1957"
}

@ARTICLE{Bethermin20,
       author = {{B{\'e}thermin}, M. and {Fudamoto}, Y. and {Ginolfi}, M. and {Loiacono}, F. and {Khusanova}, Y. and {Capak}, P.~L. and {Cassata}, P. and {Faisst}, A. and {Le F{\`e}vre}, O. and {Schaerer}, D. and {Silverman}, J.~D. and {Yan}, L. and {Amorin}, R. and {Bardelli}, S. and {Boquien}, M. and {Cimatti}, A. and {Davidzon}, I. and {Dessauges-Zavadsky}, M. and {Fujimoto}, S. and {Gruppioni}, C. and {Hathi}, N.~P. and {Ibar}, E. and {Jones}, G.~C. and {Koekemoer}, A.~M. and {Lagache}, G. and {Lemaux}, B.~C. and {Moreau}, C. and {Oesch}, P.~A. and {Pozzi}, F. and {Riechers}, D.~A. and {Talia}, M. and {Toft}, S. and {Vallini}, L. and {Vergani}, D. and {Zamorani}, G. and {Zucca}, E.},
        title = "{The ALPINE-ALMA [CII] survey: Data processing, catalogs, and statistical source properties}",
      journal = {\aap},
         year = 2020,
        month = nov,
       volume = {643},
          eid = {A2},
        pages = {A2},
          doi = {10.1051/0004-6361/202037649},
archivePrefix = {arXiv},
       eprint = {2002.00962},
 primaryClass = {astro-ph.GA},
       adsurl = {https://ui.adsabs.harvard.edu/abs/2020A&A...643A...2B}
}

@ARTICLE{Bethermin11,
       author = {{B{\'e}thermin}, M. and {Dole}, H. and {Lagache}, G. and {Le Borgne}, D. and {Penin}, A.},
        title = "{Modeling the evolution of infrared galaxies: a parametric backward evolution model}",
      journal = {\aap},
         year = 2011,
        month = may,
       volume = {529},
          eid = {A4},
        pages = {A4},
          doi = {10.1051/0004-6361/201015841},
archivePrefix = {arXiv},
       eprint = {1010.1150},
 primaryClass = {astro-ph.CO},
       adsurl = {https://ui.adsabs.harvard.edu/abs/2011A&A...529A...4B}
}

@ARTICLE{Bharadwaj05,
       author = {{Bharadwaj}, Somnath and {Pandey}, Sanjay K.},
        title = "{Probing non-Gaussian features in the HI distribution at the epoch of re-ionization}",
      journal = {\mnras},
         year = 2005,
        month = apr,
       volume = {358},
       number = {3},
        pages = {968-976},
          doi = {10.1111/j.1365-2966.2005.08836.x},
archivePrefix = {arXiv},
       eprint = {astro-ph/0410581},
 primaryClass = {astro-ph},
       adsurl = {https://ui.adsabs.harvard.edu/abs/2005MNRAS.358..968B}
}

@ARTICLE{Bond91,
       author = {{Bond}, J.~R. and {Cole}, S. and {Efstathiou}, G. and {Kaiser}, N.},
        title = "{Excursion Set Mass Functions for Hierarchical Gaussian Fluctuations}",
      journal = {\apj},
         year = 1991,
        month = oct,
       volume = {379},
        pages = {440},
          doi = {10.1086/170520},
       adsurl = {https://ui.adsabs.harvard.edu/abs/1991ApJ...379..440B}
}

@article{Bonvin:2013ogt,
    author = "Bonvin, Camille and Hui, Lam and Gaztanaga, Enrique",
    title = "{Asymmetric galaxy correlation functions}",
    eprint = "1309.1321",
    archivePrefix = "arXiv",
    primaryClass = "astro-ph.CO",
    doi = "10.1103/PhysRevD.89.083535",
    journal = "Phys. Rev. D",
    volume = "89",
    number = "8",
    pages = "083535",
    year = "2014"
}

@article{Bouwens:2022gqg,
    author = "Bouwens, Rychard and Illingworth, Garth and Oesch, Pascal and Stefanon, Mauro and Naidu, Rohan and van Leeuwen, Ivana and Magee, Dan",
    title = "{UV Luminosity Density Results at z\ensuremath{>}8 from the First JWST/NIRCam Fields: Limitations of Early Data Sets and the Need for Spectroscopy}",
    eprint = "2212.06683",
    archivePrefix = "arXiv",
    primaryClass = "astro-ph.CO",
    month = "12",
    year = "2022"
}

@ARTICLE{Bouwens22,
       author = {{Bouwens}, R.~J. and {Illingworth}, G.~D. and {Ellis}, R.~S. and {Oesch}, P.~A. and {Stefanon}, M.},
        title = "{z\raisebox{-0.5ex}\textasciitilde2-9 Galaxies Magnified by the Hubble Frontier Field Clusters II: Luminosity Functions and Constraints on a Faint-End Turnover}",
      journal = {arXiv e-prints},
         year = 2022,
        month = may,
          eid = {arXiv:2205.11526},
        pages = {arXiv:2205.11526},
archivePrefix = {arXiv},
       eprint = {2205.11526},
 primaryClass = {astro-ph.GA},
       adsurl = {https://ui.adsabs.harvard.edu/abs/2022arXiv220511526B}
}

@ARTICLE{Bouwens15,
       author = {{Bouwens}, R.~J. and {Illingworth}, G.~D. and {Oesch}, P.~A. and {Trenti}, M. and {Labb{\'e}}, I. and {Bradley}, L. and {Carollo}, M. and {van Dokkum}, P.~G. and {Gonzalez}, V. and {Holwerda}, B. and {Franx}, M. and {Spitler}, L. and {Smit}, R. and {Magee}, D.},
        title = "{UV Luminosity Functions at Redshifts z {\ensuremath{\sim}} 4 to z {\ensuremath{\sim}} 10: 10,000 Galaxies from HST Legacy Fields}",
      journal = {\apj},
         year = 2015,
        month = apr,
       volume = {803},
       number = {1},
          eid = {34},
        pages = {34},
          doi = {10.1088/0004-637X/803/1/34},
archivePrefix = {arXiv},
       eprint = {1403.4295},
 primaryClass = {astro-ph.CO},
       adsurl = {https://ui.adsabs.harvard.edu/abs/2015ApJ...803...34B}
}

@article{Breysse:2022fdi,
    author = "Breysse, Patrick C. and Chung, Dongwoo T. and Ihle, H\r{a}vard T.",
    title = "{Characteristic Functions for Cosmological Cross-Correlations}",
    eprint = "2210.14902",
    archivePrefix = "arXiv",
    primaryClass = "astro-ph.CO",
    month = "10",
    year = "2022"
}

@article{Breysse:2022alx,
    author = "Breysse, Patrick C.",
    title = "{Breaking the intensity-bias degeneracy in line intensity mapping}",
    eprint = "2209.01223",
    archivePrefix = "arXiv",
    primaryClass = "astro-ph.CO",
    month = "9",
    year = "2022"
}

@ARTICLE{Breysse19,
       author = {{Breysse}, Patrick C. and {Alexandroff}, Rachael M.},
        title = "{Observing AGN feedback with CO intensity mapping}",
      journal = {\mnras},
         year = 2019,
        month = nov,
       volume = {490},
       number = {1},
        pages = {260-273},
          doi = {10.1093/mnras/stz2534},
archivePrefix = {arXiv},
       eprint = {1904.03197},
 primaryClass = {astro-ph.GA},
       adsurl = {https://ui.adsabs.harvard.edu/abs/2019MNRAS.490..260B}
}

@ARTICLE{Cataldo21,
       author = {{Cataldo}, Giuseppe and {Ade}, Peter and {Anderson}, Christopher and {Barlis}, Alyssa and {Barrentine}, Emily and {Bellis}, Nicholas and {Bolatto}, Alberto and {Breysse}, Patrick and {Bulcha}, Berhanu and {Connors}, Jake and {Cursey}, Paul and {Ehsan}, Negar and {Essinger-Hileman}, Thomas and {Glenn}, Jason and {Golec}, Joseph and {Hays-Wehle}, James and {Hess}, Larry and {Jahromi}, Amir and {Kimball}, Mark and {Kogut}, Alan and {Lowe}, Luke and {Mauskopf}, Philip and {McMahon}, Jeffrey and {Mirzaei}, Mona and {Moseley}, Harvey and {Mugge-Durum}, Jonas and {Noroozian}, Omid and {Oxholm}, Trevor and {Pen}, Ue-Li and {Pullen}, Anthony and {Rodriguez}, Samelys and {Shirron}, Peter and {Siebert}, Gage and {Sinclair}, Adrian and {Somerville}, Rachel and {Stephenson}, Ryan and {Stevenson}, Thomas and {Switzer}, Eric and {Timbie}, Peter and {Tucker}, Carole and {Visbal}, Eli and {Volpert}, Carolyn and {Wollack}, Edward and {Yang}, Shengqi},
        title = "{Overview and status of EXCLAIM, the experiment for cryogenic large-aperture intensity mapping}",
      journal = {arXiv e-prints},
         year = 2021,
        month = jan,
          eid = {arXiv:2101.11734},
        pages = {arXiv:2101.11734},
archivePrefix = {arXiv},
       eprint = {2101.11734},
 primaryClass = {astro-ph.IM},
       adsurl = {https://ui.adsabs.harvard.edu/abs/2021arXiv210111734C}
}

@article{Breysse:2021utr,
    author = "Breysse, Patrick C. and Foreman, Simon and Keating, Laura C. and Meyers, Joel and Murray, Norman",
    title = "{Mapping the Universe in hydrogen deuteride}",
    eprint = "2104.06422",
    archivePrefix = "arXiv",
    primaryClass = "astro-ph.CO",
    doi = "10.1103/PhysRevD.105.083009",
    journal = "Phys. Rev. D",
    volume = "105",
    number = "8",
    pages = "083009",
    year = "2022"
}

@article{Breysse:2019cdw,
    author = "Breysse, Patrick C. and Anderson, Christopher J. and Berger, Philippe",
    title = "{Canceling out intensity mapping foregrounds}",
    eprint = "1907.04369",
    archivePrefix = "arXiv",
    primaryClass = "astro-ph.CO",
    doi = "10.1103/PhysRevLett.123.231105",
    journal = "Phys. Rev. Lett.",
    volume = "123",
    number = "23",
    pages = "231105",
    year = "2019"
}

@article{Breysse:2016szq,
    author = "Breysse, Patrick C. and Kovetz, Ely D. and Behroozi, Peter S. and Dai, Liang and Kamionkowski, Marc",
    title = "{Insights from probability distribution functions of intensity maps}",
    eprint = "1609.01728",
    archivePrefix = "arXiv",
    primaryClass = "astro-ph.CO",
    doi = "10.1093/mnras/stx203",
    journal = "Mon. Not. Roy. Astron. Soc.",
    volume = "467",
    number = "3",
    pages = "2996--3010",
    year = "2017"
}

@ARTICLE{Breysse16,
   author = {{Breysse}, P.~C. and {Kovetz}, E.~D. and {Kamionkowski}, M.},
    title = "{The high-redshift star formation history from carbon-monoxide intensity maps}",
  journal = {\mnras},
archivePrefix = "arXiv",
   eprint = {1507.06304},
     year = 2016,
    month = mar,
   volume = 457,
    pages = {L127-L131},
      doi = {10.1093/mnrasl/slw005},
   adsurl = {http://adsabs.harvard.edu/abs/2016MNRAS.457L.127B}
}

@ARTICLE{Bromm2009,
       author = {{Bromm}, Volker and {Yoshida}, Naoki and {Hernquist}, Lars and {McKee}, Christopher F.},
        title = "{The formation of the first stars and galaxies}",
      journal = {\nat},
         year = 2009,
        month = may,
       volume = {459},
       number = {7243},
        pages = {49-54},
          doi = {10.1038/nature07990},
archivePrefix = {arXiv},
       eprint = {0905.0929},
 primaryClass = {astro-ph.CO},
       adsurl = {https://ui.adsabs.harvard.edu/abs/2009Natur.459...49B}
}

@article{Bucher:2015eia,
    author = "Bucher, Martin",
    title = "{Physics of the cosmic microwave background anisotropy}",
    eprint = "1501.04288",
    archivePrefix = "arXiv",
    primaryClass = "astro-ph.CO",
    doi = "10.1142/S0218271815300049",
    journal = "Int. J. Mod. Phys. D",
    volume = "24",
    number = "02",
    pages = "1530004",
    year = "2015"
}

@ARTICLE{Byler17,
       author = {{Byler}, Nell and {Dalcanton}, Julianne J. and {Conroy}, Charlie and {Johnson}, Benjamin D.},
        title = "{Nebular Continuum and Line Emission in Stellar Population Synthesis Models}",
      journal = {\apj},
         year = 2017,
        month = may,
       volume = {840},
       number = {1},
          eid = {44},
        pages = {44},
          doi = {10.3847/1538-4357/aa6c66},
archivePrefix = {arXiv},
       eprint = {1611.08305},
 primaryClass = {astro-ph.GA},
       adsurl = {https://ui.adsabs.harvard.edu/abs/2017ApJ...840...44B}
}

@ARTICLE{Byler18,
       author = {{Byler}, Nell and {Dalcanton}, Julianne J. and {Conroy}, Charlie and {Johnson}, Benjamin D. and {Levesque}, Emily M. and {Berg}, Danielle A.},
        title = "{Stellar and Nebular Diagnostics in the Ultraviolet for Star-forming Galaxies}",
      journal = {\apj},
         year = 2018,
        month = aug,
       volume = {863},
       number = {1},
          eid = {14},
        pages = {14},
          doi = {10.3847/1538-4357/aacd50},
archivePrefix = {arXiv},
       eprint = {1803.04425},
 primaryClass = {astro-ph.GA},
       adsurl = {https://ui.adsabs.harvard.edu/abs/2018ApJ...863...14B}
}

@ARTICLE{Byrohl2021,
       author = {{Byrohl}, Chris and {Nelson}, Dylan and {Behrens}, Christoph and {Kostyuk}, Ivan and {Glatzle}, Martin and {Pillepich}, Annalisa and {Hernquist}, Lars and {Marinacci}, Federico and {Vogelsberger}, Mark},
        title = "{The physical origins and dominant emission mechanisms of Lyman alpha haloes: results from the TNG50 simulation in comparison to MUSE observations}",
      journal = {\mnras},
         year = 2021,
        month = oct,
       volume = {506},
       number = {4},
        pages = {5129-5152},
          doi = {10.1093/mnras/stab1958},
archivePrefix = {arXiv},
       eprint = {2009.07283},
 primaryClass = {astro-ph.GA},
       adsurl = {https://ui.adsabs.harvard.edu/abs/2021MNRAS.506.5129B}
}

@ARTICLE{Camps20,
       author = {{Camps}, P. and {Baes}, M.},
        title = "{SKIRT 9: Redesigning an advanced dust radiative transfer code to allow kinematics, line transfer and polarization by aligned dust grains}",
      journal = {Astronomy and Computing},
         year = 2020,
        month = apr,
       volume = {31},
          eid = {100381},
        pages = {100381},
          doi = {10.1016/j.ascom.2020.100381},
archivePrefix = {arXiv},
       eprint = {2003.00721},
 primaryClass = {astro-ph.GA},
       adsurl = {https://ui.adsabs.harvard.edu/abs/2020A&C....3100381C}
}

@ARTICLE{Capak15,
       author = {{Capak}, P.~L. and {Carilli}, C. and {Jones}, G. and {Casey}, C.~M. and {Riechers}, D. and {Sheth}, K. and {Carollo}, C.~M. and {Ilbert}, O. and {Karim}, A. and {Lefevre}, O. and {Lilly}, S. and {Scoville}, N. and {Smolcic}, V. and {Yan}, L.},
        title = "{Galaxies at redshifts 5 to 6 with systematically low dust content and high [C II] emission}",
      journal = {\nat},
         year = 2015,
        month = jun,
       volume = {522},
       number = {7557},
        pages = {455-458},
          doi = {10.1038/nature14500},
archivePrefix = {arXiv},
       eprint = {1503.07596},
 primaryClass = {astro-ph.GA},
       adsurl = {https://ui.adsabs.harvard.edu/abs/2015Natur.522..455C}
}

@article{Carilli:2013qm,
      author         = "Carilli, Chris and Walter, Fabian",
      title          = "{Cool Gas in High Redshift Galaxies}",
      journal        = "Ann. Rev. Astron. Astrophys.",
      volume         = "51",
      year           = "2013",
      pages          = "105-161",
      doi            = "10.1146/annurev-astro-082812-140953",
      eprint         = "1301.0371",
      archivePrefix  = "arXiv",
      primaryClass   = "astro-ph.CO",
      SLACcitation   = "%%CITATION = ARXIV:1301.0371;%%"
}

@ARTICLE{Carilli11,
   author = {{Carilli}, C.~L.},
    title = "{Intensity Mapping of Molecular Gas During Cosmic Reionization}",
  journal = {\apjl},
archivePrefix = "arXiv",
   eprint = {1102.0745},
     year = 2011,
    month = apr,
   volume = 730,
      eid = {L30},
    pages = {L30},
      doi = {10.1088/2041-8205/730/2/L30},
   adsurl = {http://adsabs.harvard.edu/abs/2011ApJ...730L..30C}
}

@ARTICLE{Carlson2025,
       author = {{Carlson}, Nathan J. and {Bond}, J. Richard and {Chung}, Dongwoo T. and {Horlaville}, Patrick and {Morrison}, Thomas},
        title = "{The $\mathtt{WebSky}$ $\mathrm{[CII]}$ Forecasts and the search for primordial intermittent non-Gaussianity}",
      journal = {arXiv e-prints},
         year = 2025,
        month = oct,
          eid = {arXiv:2510.18312},
        pages = {arXiv:2510.18312},
          doi = {10.48550/arXiv.2510.18312},
archivePrefix = {arXiv},
       eprint = {2510.18312},
 primaryClass = {astro-ph.CO},
       adsurl = {https://ui.adsabs.harvard.edu/abs/2025arXiv251018312C}
}

@article{Chang:2019xgc,
    author = "Chang, Tzu-Ching and others",
    title = "{Tomography of the Cosmic Dawn and Reionization Eras with Multiple Tracers}",
    eprint = "1903.11744",
    archivePrefix = "arXiv",
    primaryClass = "astro-ph.CO",
    month = "3",
    year = "2019"
}

@ARTICLE{Chang10,
   author = {{Chang}, T.-C. and {Pen}, U.-L. and {Bandura}, K. and {Peterson}, J.~B.
	},
    title = "{Hydrogen 21-cm Intensity Mapping at redshift 0.8}",
  journal = {ArXiv e-prints},
archivePrefix = "arXiv",
   eprint = {1007.3709},
 primaryClass = "astro-ph.CO",
     year = 2010,
    month = jul,
   adsurl = {http://adsabs.harvard.edu/abs/2010arXiv1007.3709C}
}

@article{Chang:2007xk,
      author         = "Chang, Tzu-Ching and Pen, Ue-Li and Peterson, Jeffrey B.
                        and McDonald, Patrick",
      title          = "{Baryon Acoustic Oscillation Intensity Mapping as a Test
                        of Dark Energy}",
      journal        = "Phys. Rev. Lett.",
      volume         = "100",
      year           = "2008",
      pages          = "091303",
      doi            = "10.1103/PhysRevLett.100.091303",
      eprint         = "0709.3672",
      archivePrefix  = "arXiv",
      primaryClass   = "astro-ph",
      SLACcitation   = "%%CITATION = ARXIV:0709.3672;%%"
}

@ARTICLE{Chen04,
       author = {{Chen}, Xuelei and {Miralda-Escud{\'e}}, Jordi},
        title = "{The Spin-Kinetic Temperature Coupling and the Heating Rate due to Ly{\ensuremath{\alpha}} Scattering before Reionization: Predictions for 21 Centimeter Emission and Absorption}",
      journal = {\apj},
         year = 2004,
        month = feb,
       volume = {602},
       number = {1},
        pages = {1-11},
          doi = {10.1086/380829},
archivePrefix = {arXiv},
       eprint = {astro-ph/0303395},
 primaryClass = {astro-ph},
       adsurl = {https://ui.adsabs.harvard.edu/abs/2004ApJ...602....1C}
}

@ARTICLE{Cheng25,
       author = {{Cheng}, Yun-Ting and {Hensley}, Brandon S. and {Lai}, Thomas S. -Y.},
        title = "{Feature Intensity Mapping: Polycyclic Aromatic Hydrocarbon Emission from All Galaxies Across Cosmic Time}",
      journal = {arXiv e-prints},
         year = 2025,
        month = jun,
          eid = {arXiv:2506.13863},
        pages = {arXiv:2506.13863},
          doi = {10.48550/arXiv.2506.13863},
archivePrefix = {arXiv},
       eprint = {2506.13863},
 primaryClass = {astro-ph.CO},
       adsurl = {https://ui.adsabs.harvard.edu/abs/2025arXiv250613863C}
}

@article{Cheng:2022ani,
    author = "Cheng, Yun-Ting and Wandelt, Benjamin D. and Chang, Tzu-Ching and Dore, Olivier",
    title = "{Data-driven Cosmology from Three-dimensional Light Cones}",
    eprint = "2210.10052",
    archivePrefix = "arXiv",
    primaryClass = "astro-ph.CO",
    doi = "10.3847/1538-4357/acb350",
    journal = "Astrophys. J.",
    volume = "944",
    number = "2",
    pages = "151",
    year = "2023"
}

@article{Cheng:2018hox,
    author = "Cheng, Yun-Ting and de Putter, Roland and Chang, Tzu-Ching and Dore, Olivier",
    title = "{Optimally Mapping Large-Scale Structures with Luminous Sources}",
    eprint = "1809.06384",
    archivePrefix = "arXiv",
    primaryClass = "astro-ph.CO",
    doi = "10.3847/1538-4357/ab1b2b",
    journal = "Astrophys. J.",
    volume = "877",
    number = "2",
    pages = "86",
    year = "2019"
}

@article{Chiang:2013ksa,
      author         = "Chiang, Chi-Ting and others",
      title          = "{Galaxy redshift surveys with sparse sampling}",
      journal        = "JCAP",
      volume         = "1312",
      year           = "2013",
      pages          = "030",
      doi            = "10.1088/1475-7516/2013/12/030",
      eprint         = "1306.4157",
      archivePrefix  = "arXiv",
      primaryClass   = "astro-ph.CO",
      SLACcitation   = "%%CITATION = ARXIV:1306.4157;%%"
}

@article{CHIME:2022dwe,
    author = "Amiri, Mandana and others",
    collaboration = "CHIME",
    title = "{An Overview of CHIME, the Canadian Hydrogen Intensity Mapping Experiment}",
    eprint = "2201.07869",
    archivePrefix = "arXiv",
    primaryClass = "astro-ph.IM",
    doi = "10.3847/1538-4365/ac6fd9",
    journal = "Astrophys. J. Supp.",
    volume = "261",
    number = "2",
    pages = "29",
    year = "2022"
}

@article{CHIMEFRB:2021srp,
    author = "Amiri, Mandana and others",
    collaboration = "CHIME/FRB",
    title = "{The First CHIME/FRB Fast Radio Burst Catalog}",
    eprint = "2106.04352",
    archivePrefix = "arXiv",
    primaryClass = "astro-ph.HE",
    doi = "10.3847/1538-4365/ac33ab",
    journal = "Astrophys. J. Supp.",
    volume = "257",
    number = "2",
    pages = "59",
    year = "2021"
}

@article{Chluba:2019nxa,
    author = "Chluba, J. and others",
    title = "{New horizons in cosmology with spectral distortions of the cosmic microwave background}",
    eprint = "1909.01593",
    archivePrefix = "arXiv",
    primaryClass = "astro-ph.CO",
    doi = "10.1007/s10686-021-09729-5",
    journal = "Exper. Astron.",
    volume = "51",
    number = "3",
    pages = "1515--1554",
    year = "2021"
}

@ARTICLE{Choudhury2009,
       author = {{Choudhury}, Tirthankar Roy and {Haehnelt}, Martin G. and {Regan}, John},
        title = "{Inside-out or outside-in: the topology of reionization in the photon-starved regime suggested by Ly{\ensuremath{\alpha}} forest data}",
      journal = {\mnras},
         year = 2009,
        month = apr,
       volume = {394},
       number = {2},
        pages = {960-977},
          doi = {10.1111/j.1365-2966.2008.14383.x},
archivePrefix = {arXiv},
       eprint = {0806.1524},
 primaryClass = {astro-ph},
       adsurl = {https://ui.adsabs.harvard.edu/abs/2009MNRAS.394..960C}
}

@article{Chung:2022lpr,
    author = "Chung, Dongwoo T.",
    title = "{Cross-correlations between mm-wave line-intensity mapping and weak-lensing surveys: preliminary consideration of long-term prospects}",
    eprint = "2203.12581",
    archivePrefix = "arXiv",
    primaryClass = "astro-ph.CO",
    doi = "10.1093/mnras/stac1142",
    journal = "Mon. Not. Roy. Astron. Soc.",
    volume = "513",
    number = "3",
    pages = "4090--4106",
    year = "2022"
}

@article{Chung:2022zeu,
    author = "Chung, Dongwoo T. and Bangari, Ishika and Breysse, Patrick C. and Ihle, H\r{a}vard T. and Bond, J. Richard and Dunne, Delaney A. and Padmanabhan, Hamsa and Philip, Liju and Rennie, Thomas J. and Viero, Marco P.",
    title = "{The deconvolved distribution estimator: enhancing reionisation-era CO line-intensity mapping analyses with a cross-correlation analogue for one-point statistics}",
    eprint = "2210.14890",
    archivePrefix = "arXiv",
    primaryClass = "astro-ph.CO",
    month = "10",
    year = "2022"
}

@article{Chung:2024iob,
    author = "Chung, D. T. and others",
    title = "{COMAP Pathfinder -- Season 2 results III. Implications for cosmic molecular gas content at ''Cosmic Half-past Eleven''}",
    eprint = "2406.07512",
    archivePrefix = "arXiv",
    primaryClass = "astro-ph.CO",
    month = "6",
    year = "2024"
}

@article{DES:2021gua,
    author = "Jeffrey, N. and others",
    collaboration = "DES",
    title = "{Dark Energy Survey Year 3 results: curved-sky weak lensing mass map reconstruction}",
    eprint = "2105.13539",
    archivePrefix = "arXiv",
    primaryClass = "astro-ph.CO",
    reportNumber = "FERMILAB-PUB-21-342-AE-SCD, FERMILAB-PUB-21-342-AE-SCD",
    doi = "10.1093/mnras/stab1495",
    journal = "Mon. Not. Roy. Astron. Soc.",
    volume = "505",
    pages = "4626--4645",
    year = "2021"
}

@article{Stutzer:2024rps,
    author = "Stutzer, N. -O. and others",
    title = "{COMAP Pathfinder -- Season 2 results II. Updated constraints on the CO(1-0) power spectrum}",
    eprint = "2406.07511",
    archivePrefix = "arXiv",
    primaryClass = "astro-ph.CO",
    month = "6",
    year = "2024"
}

@article{Lunde:2024rut,
    author = "Lunde, J. G. S. and others",
    title = "{COMAP Pathfinder -- Season 2 results I. Improved data selection and processing}",
    eprint = "2406.07510",
    archivePrefix = "arXiv",
    primaryClass = "astro-ph.CO",
    month = "6",
    year = "2024"
}

@article{COMAP:2021rny,
    author = "Chung, Dongwoo T. and others",
    collaboration = "COMAP",
    title = "{A Model of Spectral Line Broadening in Signal Forecasts for Line-intensity Mapping Experiments}",
    eprint = "2104.11171",
    archivePrefix = "arXiv",
    primaryClass = "astro-ph.CO",
    doi = "10.3847/1538-4357/ac2a35",
    journal = "Astrophys. J.",
    volume = "923",
    number = "2",
    pages = "188",
    year = "2021"
}

@ARTICLE{Cooray02,
   author = {{Cooray}, A. and {Sheth}, R.},
    title = "{Halo models of large scale structure}",
  journal = {\physrep},
   eprint = {astro-ph/0206508},
     year = 2002,
    month = dec,
   volume = 372,
    pages = {1-129},
      doi = {10.1016/S0370-1573(02)00276-4},
   adsurl = {http://adsabs.harvard.edu/abs/2002PhR...372....1C}
}

@ARTICLE{Cooray16,
   author = {{Cooray}, A.},
    title = "{Extragalactic background light measurements and applications}",
  journal = {Royal Society Open Science},
archivePrefix = "arXiv",
   eprint = {1602.03512},
     year = 2016,
    month = mar,
   volume = 3,
    pages = {150555},
      doi = {10.1098/rsos.150555},
   adsurl = {http://adsabs.harvard.edu/abs/2016RSOS....350555C}
}

@article{Creque-Sarbinowski:2018ebl,
    author = "Creque-Sarbinowski, Cyril and Kamionkowski, Marc",
    title = "{Searching for Decaying and Annihilating Dark Matter with Line Intensity Mapping}",
    eprint = "1806.11119",
    archivePrefix = "arXiv",
    primaryClass = "astro-ph.CO",
    doi = "10.1103/PhysRevD.98.063524",
    journal = "Phys. Rev. D",
    volume = "98",
    number = "6",
    pages = "063524",
    year = "2018"
}

@article{Croft:2018rwv,
    author = "Croft, Rupert A. C. and Miralda-Escud\'e, Jordi and Zheng, Zheng and Blomqvist, Michael and Pieri, Matthew",
    title = "{Intensity mapping with SDSS/BOSS Lyman-$\alpha$ emission, quasars, and their Lyman-\ensuremath{\alpha} forest}",
    eprint = "1806.06050",
    archivePrefix = "arXiv",
    primaryClass = "astro-ph.CO",
    doi = "10.1093/mnras/sty2302",
    journal = "Mon. Not. Roy. Astron. Soc.",
    volume = "481",
    number = "1",
    pages = "1320--1336",
    year = "2018"
}

@article{Croft:2015nna,
      author         = "Croft, Rupert A. C. and others",
      title          = "{Large-scale clustering of Lyman-alpha emission intensity
                        from SDSS/BOSS}",
      collaboration  = "BOSS",
      doi            = "10.1093/mnras/stw204",
      year           = "2015",
      eprint         = "1504.04088",
      archivePrefix  = "arXiv",
      primaryClass   = "astro-ph.CO",
      SLACcitation   = "%%CITATION = ARXIV:1504.04088;%%"
}

@article{Cunnington:2022uzo,
    author = "Cunnington, Steven and others",
    title = "{H\,i intensity mapping with MeerKAT: power spectrum detection in cross-correlation with WiggleZ galaxies}",
    eprint = "2206.01579",
    archivePrefix = "arXiv",
    primaryClass = "astro-ph.CO",
    doi = "10.1093/mnras/stac3060",
    journal = "Mon. Not. Roy. Astron. Soc.",
    volume = "518",
    number = "4",
    pages = "6262--6272",
    year = "2022"
}

@ARTICLE{Curti22,
       author = {{Curti}, M. and {D'Eugenio}, F. and {Carniani}, S. and {Maiolino}, R. and {Sandles}, L. and {Witstok}, J. and {Baker}, W.~M. and {Bennett}, J.~S. and {Piotrowska}, J.~M. and {Tacchella}, S. and {Charlot}, S. and {Nakajima}, K. and {Maheson}, G. and {Mannucci}, F. and {Arribas}, S. and {Belfiore}, F. and {Bonaventura}, N.~R. and {Bunker}, A.~J. and {Chevallard}, J. and {Cresci}, G. and {Curtis-Lake}, E. and {Hayden-Pawson}, C. and {Kumari}, N. and {Laseter}, I. and {Looser}, T.~J. and {Marconi}, A. and {Maseda}, M.~V. and {Jones}, G.~C. and {Scholtz}, J. and {Smit}, R. and {Ubler}, H. and {Wallace}, I.~E.~B.},
        title = "{The chemical enrichment in the early Universe as probed by JWST via direct metallicity measurements at z\raisebox{-0.5ex}\textasciitilde8}",
      journal = {arXiv e-prints},
         year = 2022,
        month = jul,
          eid = {arXiv:2207.12375},
        pages = {arXiv:2207.12375},
archivePrefix = {arXiv},
       eprint = {2207.12375},
 primaryClass = {astro-ph.GA},
       adsurl = {https://ui.adsabs.harvard.edu/abs/2022arXiv220712375C}
}

@article{Dalal:2007cu,
    author = "Dalal, Neal and Dore, Olivier and Huterer, Dragan and Shirokov, Alexander",
    title = "{The imprints of primordial non-gaussianities on large-scale structure: scale dependent bias and abundance of virialized objects}",
    eprint = "0710.4560",
    archivePrefix = "arXiv",
    primaryClass = "astro-ph",
    doi = "10.1103/PhysRevD.77.123514",
    journal = "Phys. Rev. D",
    volume = "77",
    pages = "123514",
    year = "2008"
}

@ARTICLE{Datta2008,
       author = {{Datta}, Kanan K. and {Majumdar}, Suman and {Bharadwaj}, Somnath and {Choudhury}, T. Roy},
        title = "{Simulating the impact of HI fluctuations on matched filter search for ionized bubbles in redshifted 21-cm maps}",
      journal = {\mnras},
         year = 2008,
        month = dec,
       volume = {391},
       number = {4},
        pages = {1900-1912},
          doi = {10.1111/j.1365-2966.2008.14008.x},
archivePrefix = {arXiv},
       eprint = {0805.1734},
 primaryClass = {astro-ph},
       adsurl = {https://ui.adsabs.harvard.edu/abs/2008MNRAS.391.1900D}
}

@INPROCEEDINGS{Davis03,
   author = {{Davis}, M. and {Faber}, S.~M. and {Newman}, J. and {Phillips}, A.~C. and 
	{Ellis}, R.~S. and {Steidel}, C.~C. and {Conselice}, C. and 
	{Coil}, A.~L. and {Finkbeiner}, D.~P. and {Koo}, D.~C. and {Guhathakurta}, P. and 
	{Weiner}, B. and {Schiavon}, R. and {Willmer}, C. and {Kaiser}, N. and 
	{Luppino}, G.~A. and {Wirth}, G. and {Connolly}, A. and {Eisenhardt}, P. and 
	{Cooper}, M. and {Gerke}, B.},
    title = "{Science Objectives and Early Results of the DEEP2 Redshift Survey}",
booktitle = {Discoveries and Research Prospects from 6- to 10-Meter-Class Telescopes II},
     year = 2003,
   series = {\procspie},
   volume = 4834,
   eprint = {astro-ph/0209419},
   editor = {{Guhathakurta}, P.},
    month = feb,
    pages = {161-172},
      doi = {10.1117/12.457897},
   adsurl = {http://adsabs.harvard.edu/abs/2003SPIE.4834..161D}
}

@article{DeLooze:2014dta,
      author         = "De Looze, Ilse and others",
      title          = "{The applicability of far-infrared fine-structure lines
                        as star formation rate tracers over wide ranges of
                        metallicities and galaxy types}",
      journal        = "Astron. Astrophys.",
      volume         = "568",
      year           = "2014",
      pages          = "A62",
      doi            = "10.1051/0004-6361/201322489",
      eprint         = "1402.4075",
      archivePrefix  = "arXiv",
      primaryClass   = "astro-ph.GA",
      SLACcitation   = "%%CITATION = ARXIV:1402.4075;%%"
}

@article{DESI:2025zgx,
    author = "Abdul Karim, M. and others",
    collaboration = "DESI",
    title = "{DESI DR2 Results II: Measurements of Baryon Acoustic Oscillations and Cosmological Constraints}",
    eprint = "2503.14738",
    archivePrefix = "arXiv",
    primaryClass = "astro-ph.CO",
    reportNumber = "FERMILAB-PUB-25-0169-PPD",
    month = "3",
    year = "2025"
}

@ARTICLE{Dijkstra17,
       author = {{Dijkstra}, Mark},
        title = "{Saas-Fee Lecture Notes: Physics of Lyman Alpha Radiative Transfer}",
      journal = {arXiv e-prints},
         year = 2017,
        month = apr,
          eid = {arXiv:1704.03416},
        pages = {arXiv:1704.03416},
archivePrefix = {arXiv},
       eprint = {1704.03416},
 primaryClass = {astro-ph.GA},
       adsurl = {https://ui.adsabs.harvard.edu/abs/2017arXiv170403416D}
}

@ARTICLE{Dijkstra2007,
       author = {{Dijkstra}, Mark and {Lidz}, Adam and {Wyithe}, J. Stuart B.},
        title = "{The impact of The IGM on high-redshift Ly{\ensuremath{\alpha}} emission lines}",
      journal = {\mnras},
         year = 2007,
        month = may,
       volume = {377},
       number = {3},
        pages = {1175-1186},
          doi = {10.1111/j.1365-2966.2007.11666.x},
archivePrefix = {arXiv},
       eprint = {astro-ph/0701667},
 primaryClass = {astro-ph},
       adsurl = {https://ui.adsabs.harvard.edu/abs/2007MNRAS.377.1175D}
}

@ARTICLE{Dijkstra2008,
       author = {{Dijkstra}, Mark and {Loeb}, Abraham},
        title = "{The polarization of scattered Ly{\ensuremath{\alpha}} radiation around high-redshift galaxies}",
      journal = {\mnras},
         year = 2008,
        month = may,
       volume = {386},
       number = {1},
        pages = {492-504},
          doi = {10.1111/j.1365-2966.2008.13066.x},
archivePrefix = {arXiv},
       eprint = {0711.2312},
 primaryClass = {astro-ph},
       adsurl = {https://ui.adsabs.harvard.edu/abs/2008MNRAS.386..492D}
}

@ARTICLE{Dijkstra04,
   author = {{Dijkstra}, M. and {Haiman}, Z. and {Loeb}, A.},
    title = "{A Limit from the X-Ray Background on the Contribution of Quasars to Reionization}",
  journal = {\apj},
   eprint = {astro-ph/0403078},
     year = 2004,
    month = oct,
   volume = 613,
    pages = {646-654},
      doi = {10.1086/422167},
   adsurl = {http://adsabs.harvard.edu/abs/2004ApJ...613..646D}
}

@article{MoradinezhadDizgah:2021upg,
    author = "Moradinezhad Dizgah, Azadeh and Keating, Garrett K. and Karkare, Kirit S. and Crites, Abigail and Choudhury, Shouvik Roy",
    title = "{Neutrino Properties with Ground-based Millimeter-wavelength Line Intensity Mapping}",
    eprint = "2110.00014",
    archivePrefix = "arXiv",
    primaryClass = "astro-ph.CO",
    reportNumber = "FERMILAB-PUB-21-527-V
",
    doi = "10.3847/1538-4357/ac3edd",
    journal = "Astrophys. J.",
    volume = "926",
    number = "2",
    pages = "137",
    year = "2022"
}

@article{MoradinezhadDizgah:2021dei,
    author = "Moradinezhad Dizgah, Azadeh and Nikakhtar, Farnik and Keating, Garrett K. and Castorina, Emanuele",
    title = "{Precision tests of CO and [CII] power spectra models against simulated intensity maps}",
    eprint = "2111.03717",
    archivePrefix = "arXiv",
    primaryClass = "astro-ph.CO",
    doi = "10.1088/1475-7516/2022/02/026",
    journal = "JCAP",
    volume = "02",
    number = "02",
    pages = "026",
    year = "2022"
}

@article{MoradinezhadDizgah:2018zrs,
    author = "Moradinezhad Dizgah, Azadeh and Keating, Garrett K. and Fialkov, Anastasia",
    title = "{Probing Cosmic Origins with CO and [CII] Emission Lines}",
    eprint = "1801.10178",
    archivePrefix = "arXiv",
    primaryClass = "astro-ph.CO",
    doi = "10.3847/2041-8213/aaf813",
    journal = "Astrophys. J. Lett.",
    volume = "870",
    number = "1",
    pages = "L4",
    year = "2019"
}

@article{MoradinezhadDizgah:2018lac,
    author = "Moradinezhad Dizgah, Azadeh and Keating, Garrett K.",
    title = "{Line intensity mapping with [CII] and CO(1-0) as probes of primordial non-Gaussianity}",
    eprint = "1810.02850",
    archivePrefix = "arXiv",
    primaryClass = "astro-ph.CO",
    doi = "10.3847/1538-4357/aafd36",
    journal = "Astrophys. J.",
    volume = "872",
    number = "2",
    pages = "126",
    year = "2019"
}

@BOOK{Dodelson17,
       author = {{Dodelson}, Scott},
        title = "{Gravitational Lensing}",
         year = 2017,
       adsurl = {https://ui.adsabs.harvard.edu/abs/2017grle.book.....D}
}

@ARTICLE{Dole06,
   author = {{Dole}, H. and {Lagache}, G. and {Puget}, J.-L. and {Caputi}, K.~I. and 
	{Fern{\'a}ndez-Conde}, N. and {Le Floc'h}, E. and {Papovich}, C. and 
	{P{\'e}rez-Gonz{\'a}lez}, P.~G. and {Rieke}, G.~H. and {Blaylock}, M.
	},
    title = "{The cosmic infrared background resolved by Spitzer. Contributions of mid-infrared galaxies to the far-infrared background}",
  journal = {\aap},
   eprint = {astro-ph/0603208},
     year = 2006,
    month = may,
   volume = 451,
    pages = {417-429},
      doi = {10.1051/0004-6361:20054446},
   adsurl = {http://adsabs.harvard.edu/abs/2006A%26A...451..417D}
}

@BOOK{Draine11,
       author = {{Draine}, Bruce T.},
        title = "{Physics of the Interstellar and Intergalactic Medium}",
         year = 2011,
       adsurl = {https://ui.adsabs.harvard.edu/abs/2011piim.book.....D}
}

@ARTICLE{Drinkwater10,
   author = {{Drinkwater}, M.~J. and {Jurek}, R.~J. and {Blake}, C. and {Woods}, D. and 
	{Pimbblet}, K.~A. and {Glazebrook}, K. and {Sharp}, R. and {Pracy}, M.~B. and 
	{Brough}, S. and {Colless}, M. and {Couch}, W.~J. and {Croom}, S.~M. and 
	{Davis}, T.~M. and {Forbes}, D. and {Forster}, K. and {Gilbank}, D.~G. and 
	{Gladders}, M. and {Jelliffe}, B. and {Jones}, N. and {Li}, I.-H. and 
	{Madore}, B. and {Martin}, D.~C. and {Poole}, G.~B. and {Small}, T. and 
	{Wisnioski}, E. and {Wyder}, T. and {Yee}, H.~K.~C.},
    title = "{The WiggleZ Dark Energy Survey: survey design and first data release}",
  journal = {\mnras},
archivePrefix = "arXiv",
   eprint = {0911.4246},
     year = 2010,
    month = jan,
   volume = 401,
    pages = {1429-1452},
      doi = {10.1111/j.1365-2966.2009.15754.x},
   adsurl = {http://adsabs.harvard.edu/abs/2010MNRAS.401.1429D}
}

@inproceedings{Dvorkin:2022jyg,
    author = "Dvorkin, Cora and others",
    title = "{The Physics of Light Relics}",
    booktitle = "{2022 Snowmass Summer Study}",
    eprint = "2203.07943",
    archivePrefix = "arXiv",
    primaryClass = "hep-ph",
    month = "3",
    year = "2022"
}

@article{Efstathiou:2025tie,
    author = "Efstathiou, George",
    title = "{Baryon Acoustic Oscillations from a Different Angle}",
    eprint = "2505.02658",
    archivePrefix = "arXiv",
    primaryClass = "astro-ph.CO",
    month = "5",
    year = "2025"
}

@article{Eisenstein:1998tu,
      author         = "Eisenstein, Daniel J. and Hu, Wayne and Tegmark, Max",
      title          = "{Cosmic complementarity: H(0) and Omega(m) from combining
                        CMB experiments and redshift surveys}",
      journal        = "Astrophys. J.",
      volume         = "504",
      year           = "1998",
      pages          = "L57-L61",
      doi            = "10.1086/311582",
      eprint         = "astro-ph/9805239",
      archivePrefix  = "arXiv",
      primaryClass   = "astro-ph",
      reportNumber   = "IASSNS-AST-98-28",
      SLACcitation   = "%%CITATION = ASTRO-PH/9805239;%%"
}

@article{Ewall-Wice:2018bzf,
    author = "Ewall-Wice, A. and Chang, T. -C. and Lazio, J. and Dore, O. and Seiffert, M. and Monsalve, R. A.",
    title = "{Modeling the Radio Background from the First Black Holes at Cosmic Dawn: Implications for the 21 cm Absorption Amplitude}",
    eprint = "1803.01815",
    archivePrefix = "arXiv",
    primaryClass = "astro-ph.CO",
    doi = "10.3847/1538-4357/aae51d",
    journal = "Astrophys. J.",
    volume = "868",
    number = "1",
    pages = "63",
    year = "2018"
}

@article{Ewall-Wice:2019may,
    author = "Ewall-Wice, Aaron and Chang, Tzu-Ching and Lazio, T. Joseph W.",
    title = "{The Radio Scream from black holes at Cosmic Dawn: a semi-analytic model for the impact of radio-loud black holes on the 21 cm global signal}",
    eprint = "1903.06788",
    archivePrefix = "arXiv",
    primaryClass = "astro-ph.GA",
    doi = "10.1093/mnras/stz3501",
    journal = "Mon. Not. Roy. Astron. Soc.",
    volume = "492",
    number = "4",
    pages = "6086--6104",
    year = "2020"
}

@ARTICLE{Fabian2000,
       author = {{Fabian}, A.~C. and {Iwasawa}, K. and {Reynolds}, C.~S. and {Young}, A.~J.},
        title = "{Broad Iron Lines in Active Galactic Nuclei}",
      journal = {\pasp},
         year = 2000,
        month = sep,
       volume = {112},
       number = {775},
        pages = {1145-1161},
          doi = {10.1086/316610},
archivePrefix = {arXiv},
       eprint = {astro-ph/0004366},
 primaryClass = {astro-ph},
       adsurl = {https://ui.adsabs.harvard.edu/abs/2000PASP..112.1145F}
}

@ARTICLE{Faisst20,
       author = {{Faisst}, A.~L. and {Schaerer}, D. and {Lemaux}, B.~C. and {Oesch}, P.~A. and {Fudamoto}, Y. and {Cassata}, P. and {B{\'e}thermin}, M. and {Capak}, P.~L. and {Le F{\`e}vre}, O. and {Silverman}, J.~D. and {Yan}, L. and {Ginolfi}, M. and {Koekemoer}, A.~M. and {Morselli}, L. and {Amor{\'\i}n}, R. and {Bardelli}, S. and {Boquien}, M. and {Brammer}, G. and {Cimatti}, A. and {Dessauges-Zavadsky}, M. and {Fujimoto}, S. and {Gruppioni}, C. and {Hathi}, N.~P. and {Hemmati}, S. and {Ibar}, E. and {Jones}, G.~C. and {Khusanova}, Y. and {Loiacono}, F. and {Pozzi}, F. and {Talia}, M. and {Tasca}, L.~A.~M. and {Riechers}, D.~A. and {Rodighiero}, G. and {Romano}, M. and {Scoville}, N. and {Toft}, S. and {Vallini}, L. and {Vergani}, D. and {Zamorani}, G. and {Zucca}, E.},
        title = "{The ALPINE-ALMA [C II] Survey: Multiwavelength Ancillary Data and Basic Physical Measurements}",
      journal = {\apjs},
         year = 2020,
        month = apr,
       volume = {247},
       number = {2},
          eid = {61},
        pages = {61},
          doi = {10.3847/1538-4365/ab7ccd},
archivePrefix = {arXiv},
       eprint = {1912.01621},
 primaryClass = {astro-ph.GA},
       adsurl = {https://ui.adsabs.harvard.edu/abs/2020ApJS..247...61F}
}

@article{Feng:2018rje,
    author = "Feng, Chang and Holder, Gilbert",
    title = "{Enhanced global signal of neutral hydrogen due to excess radiation at cosmic dawn}",
    eprint = "1802.07432",
    archivePrefix = "arXiv",
    primaryClass = "astro-ph.CO",
    doi = "10.3847/2041-8213/aac0fe",
    journal = "Astrophys. J. Lett.",
    volume = "858",
    number = "2",
    pages = "L17",
    year = "2018"
}

@article{Fermi-LAT:2012pez,
    author = "Ackermann, M. and others",
    collaboration = "Fermi-LAT",
    title = "{Anisotropies in the diffuse gamma-ray background measured by the Fermi LAT}",
    eprint = "1202.2856",
    archivePrefix = "arXiv",
    primaryClass = "astro-ph.HE",
    reportNumber = "TCC-025-11",
    doi = "10.1103/PhysRevD.85.083007",
    journal = "Phys. Rev. D",
    volume = "85",
    pages = "083007",
    year = "2012"
}

@ARTICLE{Ferrara19,
       author = {{Ferrara}, A. and {Vallini}, L. and {Pallottini}, A. and {Gallerani}, S. and {Carniani}, S. and {Kohandel}, M. and {Decataldo}, D. and {Behrens}, C.},
        title = "{A physical model for [C II] line emission from galaxies}",
      journal = {\mnras},
         year = 2019,
        month = oct,
       volume = {489},
       number = {1},
        pages = {1-12},
          doi = {10.1093/mnras/stz2031},
archivePrefix = {arXiv},
       eprint = {1908.07536},
 primaryClass = {astro-ph.GA},
       adsurl = {https://ui.adsabs.harvard.edu/abs/2019MNRAS.489....1F}
}

@ARTICLE{Fialkov13,
       author = {{Fialkov}, Anastasia and {Barkana}, Rennan and {Visbal}, Eli and {Tseliakhovich}, Dmitriy and {Hirata}, Christopher M.},
        title = "{The 21-cm signature of the first stars during the Lyman-Werner feedback era}",
      journal = {\mnras},
         year = 2013,
        month = jul,
       volume = {432},
       number = {4},
        pages = {2909-2916},
          doi = {10.1093/mnras/stt650},
archivePrefix = {arXiv},
       eprint = {1212.0513},
 primaryClass = {astro-ph.CO},
       adsurl = {https://ui.adsabs.harvard.edu/abs/2013MNRAS.432.2909F}
}

@ARTICLE{Field58,
       author = {{Field}, George B.},
        title = "{Excitation of the Hydrogen 21-CM Line}",
      journal = {Proceedings of the IRE},
         year = 1958,
        month = jan,
       volume = {46},
        pages = {240-250},
          doi = {10.1109/JRPROC.1958.286741},
       adsurl = {https://ui.adsabs.harvard.edu/abs/1958PIRE...46..240F}
}

@ARTICLE{Fixsen96,
       author = {{Fixsen}, D.~J. and {Cheng}, E.~S. and {Gales}, J.~M. and {Mather}, J.~C. and {Shafer}, R.~A. and {Wright}, E.~L.},
        title = "{The Cosmic Microwave Background Spectrum from the Full COBE FIRAS Data Set}",
      journal = {\apj},
         year = 1996,
        month = dec,
       volume = {473},
        pages = {576},
          doi = {10.1086/178173},
archivePrefix = {arXiv},
       eprint = {astro-ph/9605054},
 primaryClass = {astro-ph},
       adsurl = {https://ui.adsabs.harvard.edu/abs/1996ApJ...473..576F}
}

@article{Follin:2015hya,
    author = "Follin, Brent and Knox, Lloyd and Millea, Marius and Pan, Zhen",
    title = "{First Detection of the Acoustic Oscillation Phase Shift Expected from the Cosmic Neutrino Background}",
    eprint = "1503.07863",
    archivePrefix = "arXiv",
    primaryClass = "astro-ph.CO",
    doi = "10.1103/PhysRevLett.115.091301",
    journal = "Phys. Rev. Lett.",
    volume = "115",
    number = "9",
    pages = "091301",
    year = "2015"
}

@article{Foreman:2018gnv,
    author = "Foreman, Simon and Meerburg, P. Daniel and van Engelen, Alexander and Meyers, Joel",
    title = "{Lensing reconstruction from line intensity maps: the impact of gravitational nonlinearity}",
    eprint = "1803.04975",
    archivePrefix = "arXiv",
    primaryClass = "astro-ph.CO",
    doi = "10.1088/1475-7516/2018/07/046",
    journal = "JCAP",
    volume = "07",
    pages = "046",
    year = "2018"
}

@article{Formaggio:2021nfz,
    author = "Formaggio, Joseph A. and de Gouv\^ea, Andr\'e Luiz C. and Robertson, R. G. Hamish",
    title = "{Direct Measurements of Neutrino Mass}",
    eprint = "2102.00594",
    archivePrefix = "arXiv",
    primaryClass = "nucl-ex",
    doi = "10.1016/j.physrep.2021.02.002",
    journal = "Phys. Rept.",
    volume = "914",
    pages = "1--54",
    year = "2021"
}

@article{Fronenberg:2024olu,
    author = "Fronenberg, Hannah and Liu, Adrian",
    title = "{Forecasts and Statistical Insights for Line Intensity Mapping Cross-correlations: A Case Study with 21 cm \texttimes{} [C ii]}",
    eprint = "2407.14588",
    archivePrefix = "arXiv",
    primaryClass = "astro-ph.CO",
    doi = "10.3847/1538-4357/ad77cc",
    journal = "Astrophys. J.",
    volume = "975",
    number = "2",
    pages = "222",
    year = "2024"
}

@article{Fronenberg:2023juh,
    author = "Fronenberg, Hannah and Maniyar, Abhishek S. and Liu, Adrian and Pullen, Anthony R.",
    title = "{New Probe of the High-z Baryon Acoustic Oscillation Scale: BAO Tomography with CMB\texttimes{}LIM-Nulling Convergence}",
    eprint = "2309.07215",
    archivePrefix = "arXiv",
    primaryClass = "astro-ph.CO",
    doi = "10.1103/PhysRevLett.132.241001",
    journal = "Phys. Rev. Lett.",
    volume = "132",
    number = "24",
    pages = "241001",
    year = "2024"
}

@article{Fronenberg:2023qtw,
    author = "Fronenberg, Hannah and Maniyar, Abhishek S. and Pullen, Anthony R. and Liu, Adrian",
    title = "{Constraining cosmology with the CMB\texttimes{}line intensity mapping-nulling convergence}",
    eprint = "2309.06477",
    archivePrefix = "arXiv",
    primaryClass = "astro-ph.CO",
    doi = "10.1103/PhysRevD.109.123518",
    journal = "Phys. Rev. D",
    volume = "109",
    number = "12",
    pages = "123518",
    year = "2024"
}

@ARTICLE{Fry94,
       author = {{Fry}, J.~N.},
        title = "{Gravity, bias, and the galaxy three-point correlation function}",
      journal = {\prl},
         year = 1994,
        month = jul,
       volume = {73},
       number = {2},
        pages = {215-219},
          doi = {10.1103/PhysRevLett.73.215},
       adsurl = {https://ui.adsabs.harvard.edu/abs/1994PhRvL..73..215F}
}

@article{Furlanetto:2004nh,
    author = "Furlanetto, Steven and Zaldarriaga, Matias and Hernquist, Lars",
    title = "{The Growth of HII regions during reionization}",
    eprint = "astro-ph/0403697",
    archivePrefix = "arXiv",
    doi = "10.1086/423025",
    journal = "Astrophys. J.",
    volume = "613",
    pages = "1--15",
    year = "2004"
}

@article{Furlanetto:2006pg,
      author         = "Furlanetto, Steven and Lidz, Adam",
      title          = "{The Cross-Correlation of High-Redshift 21 cm and Galaxy
                        Surveys}",
      journal        = "Astrophys. J.",
      volume         = "660",
      year           = "2007",
      pages          = "1030",
      doi            = "10.1086/513009",
      eprint         = "astro-ph/0611274",
      archivePrefix  = "arXiv",
      primaryClass   = "astro-ph",
      SLACcitation   = "%%CITATION = ASTRO-PH/0611274;%%"
}

@article{Furlanetto:2006jb,
      author         = "Furlanetto, Steven and Oh, S. Peng and Briggs, Frank",
      title          = "{Cosmology at Low Frequencies: The 21 cm Transition and
                        the High-Redshift Universe}",
      journal        = "Phys. Rept.",
      volume         = "433",
      year           = "2006",
      pages          = "181-301",
      doi            = "10.1016/j.physrep.2006.08.002",
      eprint         = "astro-ph/0608032",
      archivePrefix  = "arXiv",
      primaryClass   = "astro-ph",
      SLACcitation   = "%%CITATION = ASTRO-PH/0608032;%%"
}

@article{Furlanetto:2006wp,
    author = "Furlanetto, Steven R. and Oh, S. Peng and Pierpaoli, Elena",
    title = "{The Effects of Dark Matter Decay and Annihilation on the High-Redshift 21 cm Background}",
    eprint = "astro-ph/0608385",
    archivePrefix = "arXiv",
    doi = "10.1103/PhysRevD.74.103502",
    journal = "Phys. Rev. D",
    volume = "74",
    pages = "103502",
    year = "2006"
}

@article{Gardner:2006ky,
    author = "Gardner, Jonathan P. and others",
    title = "{The James Webb Space Telescope}",
    eprint = "astro-ph/0606175",
    archivePrefix = "arXiv",
    doi = "10.1007/s11214-006-8315-7",
    journal = "Space Sci. Rev.",
    volume = "123",
    pages = "485",
    year = "2006"
}

@article{Giri:2018dln,
    author = "Giri, Sambit K. and D'Aloisio, Anson and Mellema, Garrelt and Komatsu, Eiichiro and Ghara, Raghunath and Majumdar, Suman",
    title = "{Position-dependent power spectra of the 21-cm signal from the epoch of reionization}",
    eprint = "1811.09633",
    archivePrefix = "arXiv",
    primaryClass = "astro-ph.CO",
    doi = "10.1088/1475-7516/2019/02/058",
    journal = "JCAP",
    volume = "02",
    pages = "058",
    year = "2019"
}

@article{Gong:2015hke,
    author = "Gong, Yan and Cooray, Asantha and Mitchell-Wynne, Ketron and Chen, Xuelei and Zemcov, Michael and Smidt, Joseph",
    title = "{Axion decay and anisotropy of near-IR extragalactic background light}",
    eprint = "1511.01577",
    archivePrefix = "arXiv",
    primaryClass = "astro-ph.CO",
    doi = "10.3847/0004-637X/825/2/104",
    journal = "Astrophys. J.",
    volume = "825",
    number = "2",
    pages = "104",
    year = "2016"
}

@article{Gong:2013xda,
      author         = "Gong, Yan and Silva, Marta and Cooray, Asantha and
                        Santos, Mario G.",
      title          = "{Foreground Contamination in Ly? Intensity Mapping
                        during the Epoch of Reionization}",
      journal        = "Astrophys. J.",
      volume         = "785",
      year           = "2014",
      pages          = "72",
      doi            = "10.1088/0004-637X/785/1/72",
      eprint         = "1312.2035",
      archivePrefix  = "arXiv",
      primaryClass   = "astro-ph.CO",
      SLACcitation   = "%%CITATION = ARXIV:1312.2035;%%"
}

@ARTICLE{Gong13,
   author = {{Gong}, Y. and {Cooray}, A. and {Santos}, M.~G.},
    title = "{Probing the Pre-reionization Epoch with Molecular Hydrogen Intensity Mapping}",
  journal = {\apj},
archivePrefix = "arXiv",
   eprint = {1212.2964},
     year = 2013,
    month = may,
   volume = 768,
      eid = {130},
    pages = {130},
      doi = {10.1088/0004-637X/768/2/130},
   adsurl = {http://adsabs.harvard.edu/abs/2013ApJ...768..130G}
}

@ARTICLE{Gong11,
   author = {{Gong}, Y. and {Cooray}, A. and {Silva}, M.~B. and {Santos}, M.~G. and 
	{Lubin}, P.},
    title = "{Probing Reionization with Intensity Mapping of Molecular and Fine-structure Lines}",
  journal = {\apjl},
archivePrefix = "arXiv",
   eprint = {1101.2892},
     year = 2011,
    month = feb,
   volume = 728,
      eid = {L46},
    pages = {L46},
      doi = {10.1088/2041-8205/728/2/L46},
   adsurl = {http://adsabs.harvard.edu/abs/2011ApJ...728L..46G}
}

@ARTICLE{Gould1996,
       author = {{Gould}, Andrew and {Weinberg}, David H.},
        title = "{Imaging the Forest of Lyman Limit Systems}",
      journal = {\apj},
         year = 1996,
        month = sep,
       volume = {468},
        pages = {462},
          doi = {10.1086/177707},
archivePrefix = {arXiv},
       eprint = {astro-ph/9512138},
 primaryClass = {astro-ph},
       adsurl = {https://ui.adsabs.harvard.edu/abs/1996ApJ...468..462G}
}

@article{Green:2022bre,
    author = "Green, Daniel",
    title = "{TASI Lectures on Cosmic Signals of Fundamental Physics}",
    eprint = "2212.08685",
    archivePrefix = "arXiv",
    primaryClass = "hep-ph",
    month = "12",
    year = "2022"
}

@article{Gruzinov:1998un,
    author = "Gruzinov, Andrei and Hu, Wayne",
    title = "{Secondary CMB anisotropies in a universe reionized in patches}",
    eprint = "astro-ph/9803188",
    archivePrefix = "arXiv",
    reportNumber = "IASSNS-AST-98-23",
    doi = "10.1086/306432",
    journal = "Astrophys. J.",
    volume = "508",
    pages = "435--439",
    year = "1998"
}

@article{Haiman:1996rc,
    author = "Haiman, Zoltan and Rees, Martin J. and Loeb, Abraham",
    title = "{Destruction of molecular hydrogen during cosmological reionization}",
    eprint = "astro-ph/9608130",
    archivePrefix = "arXiv",
    reportNumber = "CFA-4381",
    doi = "10.1086/303647",
    journal = "Astrophys. J.",
    volume = "476",
    pages = "458",
    year = "1997"
}

@ARTICLE{Haiman2000,
       author = {{Haiman}, Zolt{\'a}n and {Spaans}, Marco and {Quataert}, Eliot},
        title = "{Ly{\ensuremath{\alpha}} Cooling Radiation from High-Redshift Halos}",
      journal = {\apjl},
         year = 2000,
        month = jul,
       volume = {537},
       number = {1},
        pages = {L5-L8},
          doi = {10.1086/312754},
archivePrefix = {arXiv},
       eprint = {astro-ph/0003366},
 primaryClass = {astro-ph},
       adsurl = {https://ui.adsabs.harvard.edu/abs/2000ApJ...537L...5H}
}

@inproceedings{Hamilton:1997zq,
    author = "Hamilton, A. J. S.",
    title = "{Linear redshift distortions: A Review}",
    booktitle = "{Ringberg Workshop on Large Scale Structure}",
    eprint = "astro-ph/9708102",
    archivePrefix = "arXiv",
    doi = "10.1007/978-94-011-4960-0_17",
    month = "8",
    year = "1997"
}

@ARTICLE{Hansen2006,
       author = {{Hansen}, Matthew and {Oh}, S. Peng},
        title = "{Lyman {\ensuremath{\alpha}} radiative transfer in a multiphase medium}",
      journal = {\mnras},
         year = 2006,
        month = apr,
       volume = {367},
       number = {3},
        pages = {979-1002},
          doi = {10.1111/j.1365-2966.2005.09870.x},
archivePrefix = {arXiv},
       eprint = {astro-ph/0507586},
 primaryClass = {astro-ph},
       adsurl = {https://ui.adsabs.harvard.edu/abs/2006MNRAS.367..979H}
}

@ARTICLE{Harikane20,
       author = {{Harikane}, Yuichi and {Ouchi}, Masami and {Inoue}, Akio K. and {Matsuoka}, Yoshiki and {Tamura}, Yoichi and {Bakx}, Tom and {Fujimoto}, Seiji and {Moriwaki}, Kana and {Ono}, Yoshiaki and {Nagao}, Tohru and {Tadaki}, Ken-ichi and {Kojima}, Takashi and {Shibuya}, Takatoshi and {Egami}, Eiichi and {Ferrara}, Andrea and {Gallerani}, Simona and {Hashimoto}, Takuya and {Kohno}, Kotaro and {Matsuda}, Yuichi and {Matsuo}, Hiroshi and {Pallottini}, Andrea and {Sugahara}, Yuma and {Vallini}, Livia},
        title = "{Large Population of ALMA Galaxies at z > 6 with Very High [O III] 88 {\ensuremath{\mu}}m to [C II] 158 {\ensuremath{\mu}}m Flux Ratios: Evidence of Extremely High Ionization Parameter or PDR Deficit?}",
      journal = {\apj},
         year = 2020,
        month = jun,
       volume = {896},
       number = {2},
          eid = {93},
        pages = {93},
          doi = {10.3847/1538-4357/ab94bd},
archivePrefix = {arXiv},
       eprint = {1910.10927},
 primaryClass = {astro-ph.GA},
       adsurl = {https://ui.adsabs.harvard.edu/abs/2020ApJ...896...93H}
}

@ARTICLE{Harikane18,
       author = {{Harikane}, Yuichi and {Ouchi}, Masami and {Shibuya}, Takatoshi and {Kojima}, Takashi and {Zhang}, Haibin and {Itoh}, Ryohei and {Ono}, Yoshiaki and {Higuchi}, Ryo and {Inoue}, Akio K. and {Chevallard}, Jacopo and {Capak}, Peter L. and {Nagao}, Tohru and {Onodera}, Masato and {Faisst}, Andreas L. and {Martin}, Crystal L. and {Rauch}, Michael and {Bruzual}, Gustavo A. and {Charlot}, Stephane and {Davidzon}, Iary and {Fujimoto}, Seiji and {Hilmi}, Miftahul and {Ilbert}, Olivier and {Lee}, Chien-Hsiu and {Matsuoka}, Yoshiki and {Silverman}, John D. and {Toft}, Sune},
        title = "{SILVERRUSH. V. Census of Ly{\ensuremath{\alpha}}, [O III] {\ensuremath{\lambda}}5007, H{\ensuremath{\alpha}}, and [C II] 158 {\ensuremath{\mu}}m Line Emission with {\ensuremath{\sim}}1000 LAEs at z = 4.9-7.0 Revealed with Subaru/HSC}",
      journal = {\apj},
         year = 2018,
        month = jun,
       volume = {859},
       number = {2},
          eid = {84},
        pages = {84},
          doi = {10.3847/1538-4357/aabd80},
archivePrefix = {arXiv},
       eprint = {1711.03735},
 primaryClass = {astro-ph.GA},
       adsurl = {https://ui.adsabs.harvard.edu/abs/2018ApJ...859...84H}
}

@ARTICLE{Hauser01,
   author = {{Hauser}, M.~G. and {Dwek}, E.},
    title = "{The Cosmic Infrared Background: Measurements and Implications}",
  journal = {\araa},
   eprint = {astro-ph/0105539},
     year = 2001,
   volume = 39,
    pages = {249-307},
      doi = {10.1146/annurev.astro.39.1.249},
   adsurl = {http://adsabs.harvard.edu/abs/2001ARA%26A..39..249H}
}

@ARTICLE{Helgason16,
   author = {{Helgason}, K. and {Komatsu}, E.},
    title = "{AKARI near-infrared background fluctuations arise from normal galaxy populations}",
  journal = {ArXiv e-prints},
archivePrefix = "arXiv",
   eprint = {1611.00042},
     year = 2016,
    month = oct,
   adsurl = {http://adsabs.harvard.edu/abs/2016arXiv161100042H}
}

@ARTICLE{Heneka21,
       author = {{Heneka}, Caroline and {Cooray}, Asantha},
        title = "{Optimal survey parameters: Ly {\ensuremath{\alpha}} and H {\ensuremath{\alpha}} intensity mapping for synergy with the 21-cm signal during reionization}",
      journal = {\mnras},
         year = 2021,
        month = sep,
       volume = {506},
       number = {2},
        pages = {1573-1584},
          doi = {10.1093/mnras/stab1842},
archivePrefix = {arXiv},
       eprint = {2104.12739},
 primaryClass = {astro-ph.CO},
       adsurl = {https://ui.adsabs.harvard.edu/abs/2021MNRAS.506.1573H}
}

@article{Heneka:2016kss,
    author = "Heneka, Caroline and Cooray, Asantha and Feng, Chang",
    title = "{Probing the Intergalactic Medium with Ly$\mathrm{\alpha}$ and 21 cm Fluctuations}",
    eprint = "1611.09682",
    archivePrefix = "arXiv",
    primaryClass = "astro-ph.CO",
    doi = "10.3847/1538-4357/aa8eed",
    journal = "Astrophys. J.",
    volume = "848",
    pages = "52",
    year = "2017"
}

@article{HERA:2022wmy,
    author = "Abdurashidova, Zara and others",
    collaboration = "HERA",
    title = "{Improved Constraints on the 21 cm EoR Power Spectrum and the X-Ray Heating of the IGM with HERA Phase I Observations}",
    eprint = "2210.04912",
    archivePrefix = "arXiv",
    primaryClass = "astro-ph.CO",
    month = "10",
    year = "2022"
}

@ARTICLE{HERA2025,
       author = {{The HERA Collaboration} and {Abdurashidova}, Zuhra and {Adams}, Tyrone and {Aguirre}, James E. and {Baartman}, Rushelle and {Barkana}, Rennan and {Berkhout}, Lindsay M. and {Bernardi}, Gianni and {Billings}, Tashalee S. and {Bizarria}, Bruno B. and {Bowman}, Judd D. and {Breitman}, Daniela and {Bull}, Philip and {Burba}, Jacob and {Byrne}, Ruby and {Carey}, Steven and {Sushovan Chandra}, Rajorshi and {Chen}, Kai-Feng and {Choudhuri}, Samir and {Cox}, Tyler and {DeBoer}, David R. and {de Lera Acedo}, Eloy and {Dexter}, Matt and {Dhandha}, Jiten and {Dillon}, Joshua S. and {Dynes}, Scott and {Eksteen}, Nico and {Ely}, John and {Ewall-Wice}, Aaron and {Fagnoni}, Nicolas and {Fialkov}, Anastasia and {Furlanetto}, Steven R. and {Gale-Sides}, Kingsley and {Garsden}, Hugh and {Gorce}, Adelie and {Gorthi}, Deepthi and {Halday}, Ziyaad and {Hazelton}, Bryna J. and {Hewitt}, Jacqueline N. and {Hickish}, Jack and {Huang}, Tian and {Jacobs}, Daniel C. and {Josaitis}, Alec and {Kern}, Nicholas S. and {Kerrigan}, Joshua and {Kittiwisit}, Piyanat and {Kolopanis}, Matthew and {Lanman}, Adam and {La Plante}, Paul and {Liu}, Adrian and {Ma}, Yin-Zhe and {MacMahon}, David H.~E. and {Malan}, Lourence and {Malgas}, Cresshim and {Malgas}, Keith and {Marero}, Bradley and {Martinot}, Zachary E. and {McBride}, Lisa and {Mesinger}, Andrei and {Mirocha}, Jordan and {Mohamed-Hinds}, Nicel and {Molewa}, Mathakane and {Morales}, Miguel F. and {Mu{\~n}oz}, Julian B. and {Murray}, Steven G. and {Nikolic}, Bojan and {Nuwegeld}, Hans and {Parsons}, Aaron R. and {Pascua}, Robert and {Patra}, Nipanjana and {Pochinda}, Simon and {Qin}, Yuxiang and {Rath}, Eleanor and {Razavi-Ghods}, Nima and {Riley}, Daniel and {Rosie}, Kathryn and {Santos}, Mario G. and {Singh}, Saurabh and {Storer}, Dara and {Swarts}, Hilton and {Tan}, Jianrong and {Th{\'e}lie}, Emilie and {van Wyngaarden}, Pieter and {Wilensky}, Michael J. and {Williams}, Peter K.~G. and {Zheng}, Haoxuan},
        title = "{First Results from HERA Phase II}",
      journal = {arXiv e-prints},
         year = 2025,
        month = nov,
          eid = {arXiv:2511.21289},
        pages = {arXiv:2511.21289},
          doi = {10.48550/arXiv.2511.21289},
archivePrefix = {arXiv},
       eprint = {2511.21289},
 primaryClass = {astro-ph.CO},
       adsurl = {https://ui.adsabs.harvard.edu/abs/2025arXiv251121289T}
}

@article{Hill:2018trh,
    author = "Hill, Ryley and Masui, Kiyoshi W. and Scott, Douglas",
    title = "{The Spectrum of the Universe}",
    eprint = "1802.03694",
    archivePrefix = "arXiv",
    primaryClass = "astro-ph.CO",
    doi = "10.1177/0003702818767133",
    journal = "Appl. Spectrosc.",
    volume = "72",
    number = "5",
    pages = "663--688",
    year = "2018"
}

@article{Hill:2015tqa,
    author = "Hill, J. Colin and Battaglia, Nick and Chluba, Jens and Ferraro, Simone and Schaan, Emmanuel and Spergel, David N.",
    title = "{Taking the Universe\textquoteright{}s Temperature with Spectral Distortions of the Cosmic Microwave Background}",
    eprint = "1507.01583",
    archivePrefix = "arXiv",
    primaryClass = "astro-ph.CO",
    doi = "10.1103/PhysRevLett.115.261301",
    journal = "Phys. Rev. Lett.",
    volume = "115",
    number = "26",
    pages = "261301",
    year = "2015"
}

@article{Hopkins:2006bw,
    author = "Hopkins, Andrew M. and Beacom, John F.",
    title = "{On the normalisation of the cosmic star formation history}",
    eprint = "astro-ph/0601463",
    archivePrefix = "arXiv",
    doi = "10.1086/506610",
    journal = "Astrophys. J.",
    volume = "651",
    pages = "142--154",
    year = "2006"
}

@ARTICLE{Hopkins:2017ycn,
       author = {{Hopkins}, Philip F. and {Wetzel}, Andrew and {Kere{\v{s}}}, Du{\v{s}}an and {Faucher-Gigu{\`e}re}, Claude-Andr{\'e} and {Quataert}, Eliot and {Boylan-Kolchin}, Michael and {Murray}, Norman and {Hayward}, Christopher C. and {Garrison-Kimmel}, Shea and {Hummels}, Cameron and {Feldmann}, Robert and {Torrey}, Paul and {Ma}, Xiangcheng and {Angl{\'e}s-Alc{\'a}zar}, Daniel and {Su}, Kung-Yi and {Orr}, Matthew and {Schmitz}, Denise and {Escala}, Ivanna and {Sanderson}, Robyn and {Grudi{\'c}}, Michael Y. and {Hafen}, Zachary and {Kim}, Ji-Hoon and {Fitts}, Alex and {Bullock}, James S. and {Wheeler}, Coral and {Chan}, T.~K. and {Elbert}, Oliver D. and {Narayanan}, Desika},
        title = "{FIRE-2 simulations: physics versus numerics in galaxy formation}",
      journal = {\mnras},
         year = 2018,
        month = oct,
       volume = {480},
       number = {1},
        pages = {800-863},
          doi = {10.1093/mnras/sty1690},
archivePrefix = {arXiv},
       eprint = {1702.06148},
 primaryClass = {astro-ph.GA},
       adsurl = {https://ui.adsabs.harvard.edu/abs/2018MNRAS.480..800H}
}

@ARTICLE{Horlaville2024,
       author = {{Horlaville}, Patrick and {Chung}, Dongwoo T. and {Bond}, J. Richard and {Liang}, Lichen},
        title = "{The informativeness of [C II] line-intensity mapping as a probe of the H I content and metallicity of galaxies at the end of reionization}",
      journal = {\mnras},
         year = 2024,
        month = jul,
       volume = {531},
       number = {3},
        pages = {2958-2975},
          doi = {10.1093/mnras/stae1333},
archivePrefix = {arXiv},
       eprint = {2309.15733},
 primaryClass = {astro-ph.CO},
       adsurl = {https://ui.adsabs.harvard.edu/abs/2024MNRAS.531.2958H}
}

@article{Hu:2001kj,
    author = "Hu, Wayne and Okamoto, Takemi",
    title = "{Mass reconstruction with cmb polarization}",
    eprint = "astro-ph/0111606",
    archivePrefix = "arXiv",
    doi = "10.1086/341110",
    journal = "Astrophys. J.",
    volume = "574",
    pages = "566--574",
    year = "2002"
}

@article{Hu:2000ke,
    author = "Hu, Wayne and Barkana, Rennan and Gruzinov, Andrei",
    title = "{Cold and fuzzy dark matter}",
    eprint = "astro-ph/0003365",
    archivePrefix = "arXiv",
    doi = "10.1103/PhysRevLett.85.1158",
    journal = "Phys. Rev. Lett.",
    volume = "85",
    pages = "1158--1161",
    year = "2000"
}

@article{Hu:1997mj,
    author = "Hu, Wayne and Eisenstein, Daniel J. and Tegmark, Max",
    title = "{Weighing neutrinos with galaxy surveys}",
    eprint = "astro-ph/9712057",
    archivePrefix = "arXiv",
    reportNumber = "IASSNS-AST-97-73",
    doi = "10.1103/PhysRevLett.80.5255",
    journal = "Phys. Rev. Lett.",
    volume = "80",
    pages = "5255--5258",
    year = "1998"
}

@article{Hu:1994bz,
    author = "Hu, Wayne and Scott, Douglas and Silk, Joseph",
    title = "{Power spectrum constraints from spectral distortions in the cosmic microwave background}",
    eprint = "astro-ph/9402045",
    archivePrefix = "arXiv",
    reportNumber = "CFPA-TH-94-12",
    doi = "10.1086/187424",
    journal = "Astrophys. J. Lett.",
    volume = "430",
    pages = "L5--L8",
    year = "1994"
}

@article{Hui:2016ltb,
    author = "Hui, Lam and Ostriker, Jeremiah P. and Tremaine, Scott and Witten, Edward",
    title = "{Ultralight scalars as cosmological dark matter}",
    eprint = "1610.08297",
    archivePrefix = "arXiv",
    primaryClass = "astro-ph.CO",
    doi = "10.1103/PhysRevD.95.043541",
    journal = "Phys. Rev. D",
    volume = "95",
    number = "4",
    pages = "043541",
    year = "2017"
}

@ARTICLE{Hutsi2012,
       author = {{H{\"u}tsi}, G. and {Gilfanov}, M. and {Sunyaev}, R.},
        title = "{Angular fluctuations in the CXB: is Fe 6.4 keV line tomography of the large-scale structure feasible?}",
      journal = {\aap},
         year = 2012,
        month = nov,
       volume = {547},
          eid = {A21},
        pages = {A21},
          doi = {10.1051/0004-6361/201219796},
archivePrefix = {arXiv},
       eprint = {1206.2375},
 primaryClass = {astro-ph.CO},
       adsurl = {https://ui.adsabs.harvard.edu/abs/2012A&A...547A..21H}
}

@article{Hutter:2023rja,
    author = {Hutter, Anne and Heneka, Caroline and Dayal, Pratika and Gottl{\"o}ber, Stefan and Mesinger, Andrei and Trebitsch, Maxime and Yepes, Gustavo},
    title = "{On the general nature of 21-cm-Lyman{\,}{\ensuremath{\alpha}} emitter cross-correlations during reionization}",
    eprint = "2306.03156",
    archivePrefix = "arXiv",
    primaryClass = "astro-ph.CO",
    doi = "10.1093/mnras/stad2376",
    journal = "Mon. Not. Roy. Astron. Soc.",
    volume = "525",
    number = "2",
    pages = "1664--1676",
    year = "2023"
}

@article{Hutter:2025phq,
    author = "Hutter, Anne and Heneka, Caroline",
    title = "{The 21cm-galaxy cross-correlation: Realistic forecast for 21cm signal detection and reionisation constraints}",
    eprint = "2509.15906",
    archivePrefix = "arXiv",
    primaryClass = "astro-ph.CO",
    month = "9",
    year = "2025"
}

@article{Ivanov:2022mrd,
    author = "Ivanov, Mikhail M.",
    title = "{Effective Field Theory for Large Scale Structure}",
    eprint = "2212.08488",
    archivePrefix = "arXiv",
    primaryClass = "astro-ph.CO",
    month = "12",
    year = "2022"
}

@ARTICLE{Jones20,
       author = {{Jones}, Tucker and {Sanders}, Ryan and {Roberts-Borsani}, Guido and {Ellis}, Richard S. and {Laporte}, Nicolas and {Treu}, Tommaso and {Harikane}, Yuichi},
        title = "{The Mass-Metallicity Relation at z $\simeq$ 8: Direct-method Metallicity Constraints and Near-future Prospects}",
      journal = {\apj},
         year = 2020,
        month = nov,
       volume = {903},
       number = {2},
          eid = {150},
        pages = {150},
          doi = {10.3847/1538-4357/abb943},
archivePrefix = {arXiv},
       eprint = {2006.02447},
 primaryClass = {astro-ph.GA},
       adsurl = {https://ui.adsabs.harvard.edu/abs/2020ApJ...903..150J}
}

@article{Jones:2021mrs,
    author = "Jones, Dana and Palatnick, Skyler and Chen, Richard and Beane, Angus and Lidz, Adam",
    title = "{Fuzzy Dark Matter and the 21 cm Power Spectrum}",
    eprint = "2101.07177",
    archivePrefix = "arXiv",
    primaryClass = "astro-ph.CO",
    doi = "10.3847/1538-4357/abf0a9",
    journal = "Astrophys. J.",
    volume = "913",
    number = "1",
    pages = "7",
    year = "2021"
}

@ARTICLE{Kaiser87,
   author = {{Kaiser}, N.},
    title = "{Clustering in real space and in redshift space}",
  journal = {\mnras},
     year = 1987,
    month = jul,
   volume = 227,
    pages = {1-21},
      doi = {10.1093/mnras/227.1.1},
   adsurl = {http://adsabs.harvard.edu/abs/1987MNRAS.227....1K}
}

@ARTICLE{Kakuma21,
       author = {{Kakuma}, Ryota and {Ouchi}, Masami and {Harikane}, Yuichi and {Ono}, Yoshiaki and {Inoue}, Akio K. and {Komiyama}, Yutaka and {Kusakabe}, Haruka and {Lee}, Chien-Hsiu and {Matsuda}, Yuichi and {Matsuoka}, Yoshiki and {Mawatari}, Ken and {Momose}, Rieko and {Shibuya}, Takatoshi and {Taniguchi}, Yoshiaki},
        title = "{SILVERRUSH. IX. Ly{\ensuremath{\alpha}} Intensity Mapping with Star-forming Galaxies at z = 5.7 and 6.6: A Possible Detection of Extended Ly{\ensuremath{\alpha}} Emission at {\ensuremath{\gtrsim}}100 Comoving Kiloparsecs around and beyond the Virial-radius Scale of Galaxy Dark Matter Halos}",
      journal = {\apj},
         year = 2021,
        month = jul,
       volume = {916},
       number = {1},
          eid = {22},
        pages = {22},
          doi = {10.3847/1538-4357/ac0725},
archivePrefix = {arXiv},
       eprint = {1906.00173},
 primaryClass = {astro-ph.GA},
       adsurl = {https://ui.adsabs.harvard.edu/abs/2021ApJ...916...22K}
}

@ARTICLE{Kannan22,
       author = {{Kannan}, R. and {Garaldi}, E. and {Smith}, A. and {Pakmor}, R. and {Springel}, V. and {Vogelsberger}, M. and {Hernquist}, L.},
        title = "{Introducing the THESAN project: radiation-magnetohydrodynamic simulations of the epoch of reionization}",
      journal = {\mnras},
         year = 2022,
        month = apr,
       volume = {511},
       number = {3},
        pages = {4005-4030},
          doi = {10.1093/mnras/stab3710},
archivePrefix = {arXiv},
       eprint = {2110.00584},
 primaryClass = {astro-ph.GA},
       adsurl = {https://ui.adsabs.harvard.edu/abs/2022MNRAS.511.4005K}
}

@article{Kannan:2021ucy,
    author = {Kannan, Rahul and Smith, Aaron and Garaldi, Enrico and Shen, Xuejian and Vogelsberger, Mark and Pakmor, R\"udiger and Springel, Volker and Hernquist, Lars},
    title = "{The thesan project: predictions for multitracer line intensity mapping in the epoch of reionization}",
    eprint = "2111.02411",
    archivePrefix = "arXiv",
    primaryClass = "astro-ph.CO",
    doi = "10.1093/mnras/stac1557",
    journal = "Mon. Not. Roy. Astron. Soc.",
    volume = "514",
    number = "3",
    pages = "3857--3878",
    year = "2022"
}

@article{Kannan:2018frm,
    author = {Kannan, Rahul and Vogelsberger, Mark and Marinacci, Federico and McKinnon, Ryan and Pakmor, R\"udiger and Springel, Volker},
    title = "{AREPO-RT: Radiation hydrodynamics on a moving mesh}",
    eprint = "1804.01987",
    archivePrefix = "arXiv",
    primaryClass = "astro-ph.IM",
    doi = "10.1093/mnras/stz287",
    month = "4",
    year = "2018"
}

@article{Kamionkowski:1996ks,
    author = "Kamionkowski, Marc and Kosowsky, Arthur and Stebbins, Albert",
    title = "{Statistics of cosmic microwave background polarization}",
    eprint = "astro-ph/9611125",
    archivePrefix = "arXiv",
    reportNumber = "FERMILAB-PUB-96-426-A, CU-TP-787, CAL-617",
    doi = "10.1103/PhysRevD.55.7368",
    journal = "Phys. Rev. D",
    volume = "55",
    pages = "7368--7388",
    year = "1997"
}

@article{Kamionkowski:2022pkx,
    author = "Kamionkowski, Marc and Riess, Adam G.",
    title = "{The Hubble Tension and Early Dark Energy}",
    eprint = "2211.04492",
    archivePrefix = "arXiv",
    primaryClass = "astro-ph.CO",
    month = "11",
    year = "2022"
}

@article{Karkare:2019qla,
    author = "Karkare, Kirit S.",
    title = "{Delensing Degree-Scale $B$-Mode Polarization with High-Redshift Line Intensity Mapping}",
    eprint = "1908.08128",
    archivePrefix = "arXiv",
    primaryClass = "astro-ph.CO",
    doi = "10.1103/PhysRevD.100.043529",
    journal = "Phys. Rev. D",
    volume = "100",
    number = "4",
    pages = "043529",
    year = "2019"
}

@article{Karkare:2018sar,
    author = "Karkare, Kirit S. and Bird, Simeon",
    title = "{Constraining the Expansion History and Early Dark Energy with Line Intensity Mapping}",
    eprint = "1806.09625",
    archivePrefix = "arXiv",
    primaryClass = "astro-ph.CO",
    doi = "10.1103/PhysRevD.98.043529",
    journal = "Phys. Rev. D",
    volume = "98",
    number = "4",
    pages = "043529",
    year = "2018"
}

@ARTICLE{Kashlinsky15,
   author = {{Kashlinsky}, A. and {Mather}, J.~C. and {Helgason}, K. and 
	{Arendt}, R.~G. and {Bromm}, V. and {Moseley}, S.~H.},
    title = "{Reconstructing Emission from Pre-reionization Sources with Cosmic Infrared Background Fluctuation Measurements by the JWST}",
  journal = {\apj},
archivePrefix = "arXiv",
   eprint = {1412.5566},
     year = 2015,
    month = may,
   volume = 804,
      eid = {99},
    pages = {99},
      doi = {10.1088/0004-637X/804/2/99},
   adsurl = {http://adsabs.harvard.edu/abs/2015ApJ...804...99K}
}

@ARTICLE{Katz19,
       author = {{Katz}, Harley and {Galligan}, Thomas P. and {Kimm}, Taysun and {Rosdahl}, Joakim and {Haehnelt}, Martin G. and {Blaizot}, Jeremy and {Devriendt}, Julien and {Slyz}, Adrianne and {Laporte}, Nicolas and {Ellis}, Richard},
        title = "{Probing cosmic dawn with emission lines: predicting infrared and nebular line emission for ALMA and JWST}",
      journal = {\mnras},
         year = 2019,
        month = aug,
       volume = {487},
       number = {4},
        pages = {5902-5921},
          doi = {10.1093/mnras/stz1672},
archivePrefix = {arXiv},
       eprint = {1901.01272},
 primaryClass = {astro-ph.GA},
       adsurl = {https://ui.adsabs.harvard.edu/abs/2019MNRAS.487.5902K}
}

@ARTICLE{Keating16,
   author = {{Keating}, G.~K. and {Marrone}, D.~P. and {Bower}, G.~C. and 
	{Leitch}, E. and {Carlstrom}, J.~E. and {DeBoer}, D.~R.},
    title = "{COPSS II: The Molecular Gas Content of Ten Million Cubic Megaparsecs at $z\sim3$}",
  journal = {ArXiv e-prints},
archivePrefix = "arXiv",
   eprint = {1605.03971},
     year = 2016,
    month = may,
   adsurl = {http://adsabs.harvard.edu/abs/2016arXiv160503971K}
}

@article{Keating:2020wlx,
    author = "Keating, Garrett K. and Marrone, Daniel P. and Bower, Geoffrey C. and Keenan, Ryan P.",
    title = "{An Intensity Mapping Detection of Aggregate CO Line Emission at 3 mm}",
    eprint = "2008.08087",
    archivePrefix = "arXiv",
    primaryClass = "astro-ph.GA",
    doi = "10.3847/1538-4357/abb08e",
    journal = "Astrophys. J.",
    volume = "901",
    number = "2",
    pages = "141",
    year = "2020"
}

@article{Keating:2015qva,
      author         = "Keating, Garrett K. and others",
      title          = "{First Results from COPSS: The CO Power Spectrum Survey}",
      journal        = "Astrophys. J.",
      volume         = "814",
      year           = "2015",
      number         = "2",
      pages          = "140",
      doi            = "10.1088/0004-637X/814/2/140",
      eprint         = "1510.06744",
      archivePrefix  = "arXiv",
      primaryClass   = "astro-ph.GA",
      SLACcitation   = "%%CITATION = ARXIV:1510.06744;%%"
}

@article{Keenan:2021uue,
    author = "Keenan, Ryan P. and Keating, Garrett K. and Marrone, Daniel P.",
    title = "{An Intensity Mapping Constraint on the CO-galaxy Cross-power Spectrum at Redshift \ensuremath{\sim}3}",
    eprint = "2110.02239",
    archivePrefix = "arXiv",
    primaryClass = "astro-ph.GA",
    doi = "10.3847/1538-4357/ac4888",
    journal = "Astrophys. J.",
    volume = "927",
    number = "2",
    pages = "161",
    year = "2022"
}

@ARTICLE{Kennicutt98,
       author = {{Kennicutt}, Robert C., Jr.},
        title = "{The Global Schmidt Law in Star-forming Galaxies}",
      journal = {\apj},
         year = 1998,
        month = may,
       volume = {498},
       number = {2},
        pages = {541-552},
          doi = {10.1086/305588},
archivePrefix = {arXiv},
       eprint = {astro-ph/9712213},
 primaryClass = {astro-ph},
       adsurl = {https://ui.adsabs.harvard.edu/abs/1998ApJ...498..541K}
}

@ARTICLE{Kikuchihara22,
       author = {{Kikuchihara}, Shotaro and {Harikane}, Yuichi and {Ouchi}, Masami and {Ono}, Yoshiaki and {Shibuya}, Takatoshi and {Itoh}, Ryohei and {Kakuma}, Ryota and {Inoue}, Akio K. and {Kusakabe}, Haruka and {Shimasaku}, Kazuhiro and {Momose}, Rieko and {Sugahara}, Yuma and {Kikuta}, Satoshi and {Saito}, Shun and {Kashikawa}, Nobunari and {Zhang}, Haibin and {Lee}, Chien-Hsiu},
        title = "{SILVERRUSH. XII. Intensity Mapping for Ly{\ensuremath{\alpha}} Emission Extending over 100-1000 Comoving Kpc around z   2-7 LAEs with Subaru HSC-SSP and CHORUS Data}",
      journal = {\apj},
         year = 2022,
        month = jun,
       volume = {931},
       number = {2},
          eid = {97},
        pages = {97},
          doi = {10.3847/1538-4357/ac69de},
archivePrefix = {arXiv},
       eprint = {2108.09288},
 primaryClass = {astro-ph.GA},
       adsurl = {https://ui.adsabs.harvard.edu/abs/2022ApJ...931...97K}
}

@article{Knox95,
  title = {Determination of inflationary observables by cosmic microwave background anisotropy experiments},
  author = {Knox, Lloyd},
  journal = {Phys. Rev. D},
  volume = {52},
  issue = {8},
  pages = {4307--4318},
  numpages = {0},
  year = {1995},
  month = {Oct},
  publisher = {American Physical Society},
  doi = {10.1103/PhysRevD.52.4307},
  url = {https://link.aps.org/doi/10.1103/PhysRevD.52.4307}
}

@article{Kogut:2019vqh,
    author = "Kogut, A. and Abitbol, M. H. and Chluba, J. and Delabrouille, J. and Fixsen, D. and Hill, J. C. and Patil, S. P. and Rotti, A.",
    title = "{CMB Spectral Distortions: Status and Prospects}",
    eprint = "1907.13195",
    archivePrefix = "arXiv",
    primaryClass = "astro-ph.CO",
    month = "7",
    year = "2019"
}

@inproceedings{Kohno24,
author = {Kotaro Kohno and Akira Endo and Yoichi Tamura and Akio Taniguchi and Tatsuya Takekoshi and Shiro Ikeda and Naoki Yoshida and Kana Moriwaki and Kenichi Karatsu and Jochem J. A. Baselmans and Louis H. Marting and Arend Moerman and Bruno T. Buijtendorp and Shahab Dabironezare and Matus Rybak and Tom J. L. C. Bakx and Leon G. G. Olde Scholtenhuis and Fenno Steenvoorde and Robert Huiting and David J. Thoen and Lingyu Wang and Aurora Simionescu and Stephen J. C. Yates and Alessandro Monfardini and Martino Calvo and Paul P. van der Werf and Sten Vollebregt and Bernhard R. Brandl and Tai Oshima and Ryohei Kawabe and Kazuyuki Fujita and Shunichi Nakatsubo and Yuki Kimura and Akiyoshi Tsujita and Yuki Yoshimura and Shinji Fujita and Yuri Nishimura and Yuka Yamada and Sho Fujisawa and Kanako Narita and Tetsuhiro Minamidani and Shun Ishii and Fumiya Maeda and Adam Lidz and Denis Burgarella and Bunyo Hatsukade and Fumi Egusa and Kana Morokuma-Matsui},
title = {{Sub/millimeter-Wave Dual-Band Line Intensity Mapping Using the Terahertz Integral Field Units with Universal Nanotechnology (TIFUUN) for the Atacama Submillimeter Telescope Experiment (ASTE)}},
volume = {PC13102},
booktitle = {Millimeter, Submillimeter, and Far-Infrared Detectors and Instrumentation for Astronomy XII},
editor = {Jonas Zmuidzinas and Jian-Rong Gao},
organization = {International Society for Optics and Photonics},
publisher = {SPIE},
pages = {PC1310209},
year = {2024},
doi = {10.1117/12.3021109},
URL = {https://doi.org/10.1117/12.3021109}
}

@article{Kovetz:2019uss,
    author = "Kovetz, Ely D. and others",
    title = "{Astrophysics and Cosmology with Line-Intensity Mapping}",
    eprint = "1903.04496",
    archivePrefix = "arXiv",
    primaryClass = "astro-ph.CO",
    journal = "Bull. Am. Astron. Soc.",
    volume = "51",
    number = "3",
    pages = "101",
    year = "2020"
}

@article{Kovetz:2017agg,
    author = "Kovetz, Ely D. and others",
    title = "{Line-Intensity Mapping: 2017 Status Report}",
    eprint = "1709.09066",
    archivePrefix = "arXiv",
    primaryClass = "astro-ph.CO",
    month = "9",
    year = "2017"
}

@ARTICLE{Krumholz14,
       author = {{Krumholz}, Mark R.},
        title = "{DESPOTIC - a new software library to Derive the Energetics and SPectra of Optically Thick Interstellar Clouds}",
      journal = {\mnras},
         year = 2014,
        month = jan,
       volume = {437},
       number = {2},
        pages = {1662-1680},
          doi = {10.1093/mnras/stt2000},
archivePrefix = {arXiv},
       eprint = {1304.2404},
 primaryClass = {astro-ph.IM},
       adsurl = {https://ui.adsabs.harvard.edu/abs/2014MNRAS.437.1662K}
}

@ARTICLE{Lagache18,
       author = {{Lagache}, G. and {Cousin}, M. and {Chatzikos}, M.},
        title = "{The [CII] 158 {\ensuremath{\mu}}m line emission in high-redshift galaxies}",
      journal = {\aap},
         year = 2018,
        month = jan,
       volume = {609},
          eid = {A130},
        pages = {A130},
          doi = {10.1051/0004-6361/201732019},
archivePrefix = {arXiv},
       eprint = {1711.00798},
 primaryClass = {astro-ph.GA},
       adsurl = {https://ui.adsabs.harvard.edu/abs/2018A&A...609A.130L}
}

@article{Lauer:2022fgc,
    author = "Lauer, Tod R. and others",
    title = "{Anomalous Flux in the Cosmic Optical Background Detected with New Horizons Observations}",
    eprint = "2202.04273",
    archivePrefix = "arXiv",
    primaryClass = "astro-ph.GA",
    doi = "10.3847/2041-8213/ac573d",
    journal = "Astrophys. J. Lett.",
    volume = "927",
    number = "1",
    pages = "L8",
    year = "2022"
}

@PHDTHESIS{Laursen2010,
       author = {{Laursen}, Peter},
        title = "{Interpreting Lyman {\ensuremath{\alpha}} radiation from young, dusty galaxies}",
       school = {Niels Bohr Institute for Astronomy, Physics and Geophysics},
         year = 2010,
        month = aug,
       adsurl = {https://ui.adsabs.harvard.edu/abs/2010PhDT.......254L}
}

@ARTICLE{LeFevre20,
       author = {{Le F{\`e}vre}, O. and {B{\'e}thermin}, M. and {Faisst}, A. and {Jones}, G.~C. and {Capak}, P. and {Cassata}, P. and {Silverman}, J.~D. and {Schaerer}, D. and {Yan}, L. and {Amorin}, R. and {Bardelli}, S. and {Boquien}, M. and {Cimatti}, A. and {Dessauges-Zavadsky}, M. and {Giavalisco}, M. and {Hathi}, N.~P. and {Fudamoto}, Y. and {Fujimoto}, S. and {Ginolfi}, M. and {Gruppioni}, C. and {Hemmati}, S. and {Ibar}, E. and {Koekemoer}, A. and {Khusanova}, Y. and {Lagache}, G. and {Lemaux}, B.~C. and {Loiacono}, F. and {Maiolino}, R. and {Mancini}, C. and {Narayanan}, D. and {Morselli}, L. and {M{\'e}ndez-Hern{\`a}ndez}, Hugo and {Oesch}, P.~A. and {Pozzi}, F. and {Romano}, M. and {Riechers}, D. and {Scoville}, N. and {Talia}, M. and {Tasca}, L.~A.~M. and {Thomas}, R. and {Toft}, S. and {Vallini}, L. and {Vergani}, D. and {Walter}, F. and {Zamorani}, G. and {Zucca}, E.},
        title = "{The ALPINE-ALMA [CII] survey. Survey strategy, observations, and sample properties of 118 star-forming galaxies at 4 < z < 6}",
      journal = {\aap},
         year = 2020,
        month = nov,
       volume = {643},
          eid = {A1},
        pages = {A1},
          doi = {10.1051/0004-6361/201936965},
archivePrefix = {arXiv},
       eprint = {1910.09517},
 primaryClass = {astro-ph.CO},
       adsurl = {https://ui.adsabs.harvard.edu/abs/2020A&A...643A...1L}
}

@article{Lee:2008fm,
    author = "Lee, Samuel K. and Ando, Shin'ichiro and Kamionkowski, Marc",
    title = "{The Gamma-Ray-Flux Probability Distribution Function from Galactic Halo Substructure}",
    eprint = "0810.1284",
    archivePrefix = "arXiv",
    primaryClass = "astro-ph",
    doi = "10.1088/1475-7516/2009/07/007",
    journal = "JCAP",
    volume = "07",
    pages = "007",
    year = "2009"
}

@ARTICLE{Leung20,
       author = {{Leung}, T.~K. Daisy and {Olsen}, Karen P. and {Somerville}, Rachel S. and {Dav{\'e}}, Romeel and {Greve}, Thomas R. and {Hayward}, Christopher C. and {Narayanan}, Desika and {Popping}, Gerg{\"o}},
        title = "{Predictions of the L$_{[C II]}$-SFR and [CII] Luminosity Function at the Epoch of Reionization}",
      journal = {\apj},
         year = 2020,
        month = dec,
       volume = {905},
       number = {2},
          eid = {102},
        pages = {102},
          doi = {10.3847/1538-4357/abc25e},
archivePrefix = {arXiv},
       eprint = {2004.11912},
 primaryClass = {astro-ph.GA},
       adsurl = {https://ui.adsabs.harvard.edu/abs/2020ApJ...905..102L}
}

@article{Li:2015gqa,
      author         = "Li, Tony Y. and Wechsler, Risa H. and Devaraj, Kiruthika
                        and Church, Sarah E.",
      title          = "{Connecting CO Intensity Mapping to Molecular Gas and
                        Star Formation in the Epoch of Galaxy Assembly}",
      journal        = "Astrophys. J.",
      volume         = "817",
      year           = "2016",
      number         = "2",
      pages          = "169",
      doi            = "10.3847/0004-637X/817/2/169",
      eprint         = "1503.08833",
      archivePrefix  = "arXiv",
      primaryClass   = "astro-ph.CO",
      SLACcitation   = "%%CITATION = ARXIV:1503.08833;%%"
}

@article{Liang:2023sxx,
    author = "Liang, Lichen and others",
    title = "{$\rm [C_{II}]$ 158 $\rm \mu m$ emission as an indicator of galaxy star formation rate}",
    eprint = "2301.04149",
    archivePrefix = "arXiv",
    primaryClass = "astro-ph.GA",
    month = "1",
    year = "2023"
}

@article{Libanore:2022ntl,
    author = "Libanore, Sarah and Unal, Caner and Sarkar, Debanjan and Kovetz, Ely D.",
    title = "{Unveiling cosmological information on small scales with line intensity mapping}",
    eprint = "2208.01658",
    archivePrefix = "arXiv",
    primaryClass = "astro-ph.CO",
    month = "8",
    year = "2022"
}

@article{Lidz:2008ry,
      author         = "Lidz, Adam and Zahn, Oliver and Furlanetto, Steven and
                        McQuinn, Matthew and Hernquist, Lars and Zaldarriaga,
                        Matias",
      title          = "{Probing Reionization with the 21 cm-Galaxy Cross Power
                        Spectrum}",
      journal        = "Astrophys. J.",
      volume         = "690",
      year           = "2009",
      pages          = "252-266",
      doi            = "10.1088/0004-637X/690/1/252",
      eprint         = "0806.1055",
      archivePrefix  = "arXiv",
      primaryClass   = "astro-ph",
      SLACcitation   = "%%CITATION = ARXIV:0806.1055;%%"
}

@ARTICLE{Lidz11,
   author = {{Lidz}, A. and {Furlanetto}, S.~R. and {Oh}, S.~P. and {Aguirre}, J. and 
	{Chang}, T.-C. and {Dor{\'e}}, O. and {Pritchard}, J.~R.},
    title = "{Intensity Mapping with Carbon Monoxide Emission Lines and the Redshifted 21 cm Line}",
  journal = {\apj},
archivePrefix = "arXiv",
   eprint = {1104.4800},
     year = 2011,
    month = nov,
   volume = 741,
      eid = {70},
    pages = {70},
      doi = {10.1088/0004-637X/741/2/70},
   adsurl = {http://adsabs.harvard.edu/abs/2011ApJ...741...70L}
}

@article{Lidz:2013tra,
    author = "Lidz, Adam and Baxter, Eric J. and Adshead, Peter and Dodelson, Scott",
    title = "{Primordial Non-Gaussianity and Reionization}",
    eprint = "1304.8049",
    archivePrefix = "arXiv",
    primaryClass = "astro-ph.CO",
    reportNumber = "FERMILAB-PUB-13-207-A",
    doi = "10.1103/PhysRevD.88.023534",
    journal = "Phys. Rev. D",
    volume = "88",
    pages = "023534",
    year = "2013"
}

@article{Lidz:2018fqo,
    author = "Lidz, Adam and Hui, Lam",
    title = "{Implications of a prereionization 21-cm absorption signal for fuzzy dark matter}",
    eprint = "1805.01253",
    archivePrefix = "arXiv",
    primaryClass = "astro-ph.CO",
    doi = "10.1103/PhysRevD.98.023011",
    journal = "Phys. Rev. D",
    volume = "98",
    number = "2",
    pages = "023011",
    year = "2018"
}

@article{Linder:2005in,
      author         = "Linder, Eric V.",
      title          = "{Cosmic growth history and expansion history}",
      journal        = "Phys. Rev.",
      volume         = "D72",
      year           = "2005",
      pages          = "043529",
      doi            = "10.1103/PhysRevD.72.043529",
      eprint         = "astro-ph/0507263",
      archivePrefix  = "arXiv",
      primaryClass   = "astro-ph",
      SLACcitation   = "%%CITATION = ASTRO-PH/0507263;%%"
}

@article{Liu:2020izx,
    author = "Liu, R. Henry and Breysse, Patrick C.",
    title = "{Coupling parsec and gigaparsec scales: Primordial non-Gaussianity with multitracer intensity mapping}",
    eprint = "2002.10483",
    archivePrefix = "arXiv",
    primaryClass = "astro-ph.CO",
    doi = "10.1103/PhysRevD.103.063520",
    journal = "Phys. Rev. D",
    volume = "103",
    number = "6",
    pages = "063520",
    year = "2021"
}

@article{Liu:2015txa,
    author = "Liu, Adrian and Pritchard, Jonathan R. and Allison, Rupert and Parsons, Aaron R. and Seljak, Uro\v{s} and Sherwin, Blake D.",
    title = "{Eliminating the optical depth nuisance from the CMB with 21 cm cosmology}",
    eprint = "1509.08463",
    archivePrefix = "arXiv",
    primaryClass = "astro-ph.CO",
    doi = "10.1103/PhysRevD.93.043013",
    journal = "Phys. Rev. D",
    volume = "93",
    number = "4",
    pages = "043013",
    year = "2016"
}

@article{Liu:2019awk,
    author = "Liu, Adrian and Shaw, J. Richard",
    title = "{Data Analysis for Precision 21 cm Cosmology}",
    eprint = "1907.08211",
    archivePrefix = "arXiv",
    primaryClass = "astro-ph.IM",
    doi = "10.1088/1538-3873/ab5bfd",
    journal = "Publ. Astron. Soc. Pac.",
    volume = "132",
    number = "1012",
    pages = "062001",
    year = "2020"
}

@article{Loeb:2003ya,
      author         = "Loeb, Abraham and Zaldarriaga, Matias",
      title          = "{Measuring the small - scale power spectrum of cosmic
                        density fluctuations through 21 cm tomography prior to the
                        epoch of structure formation}",
      journal        = "Phys. Rev. Lett.",
      volume         = "92",
      year           = "2004",
      pages          = "211301",
      doi            = "10.1103/PhysRevLett.92.211301",
      eprint         = "astro-ph/0312134",
      archivePrefix  = "arXiv",
      primaryClass   = "astro-ph",
      SLACcitation   = "%%CITATION = ASTRO-PH/0312134;%%"
}

@ARTICLE{Lujan22,
       author = {{Lujan Niemeyer}, Maja and {Komatsu}, Eiichiro and {Byrohl}, Chris and {Davis}, Dustin and {Fabricius}, Maximilian and {Gebhardt}, Karl and {Hill}, Gary J. and {Wisotzki}, Lutz and {Bowman}, William P. and {Ciardullo}, Robin and {Farrow}, Daniel J. and {Finkelstein}, Steven L. and {Gawiser}, Eric and {Gronwall}, Caryl and {Jeong}, Donghui and {Landriau}, Martin and {Liu}, Chenxu and {Cooper}, Erin Mentuch and {Ouchi}, Masami and {Schneider}, Donald P. and {Zeimann}, Gregory R.},
        title = "{Surface Brightness Profile of Lyman-{\ensuremath{\alpha}} Halos out to 320 kpc in HETDEX}",
      journal = {\apj},
         year = 2022,
        month = apr,
       volume = {929},
       number = {1},
          eid = {90},
        pages = {90},
          doi = {10.3847/1538-4357/ac5cb8},
archivePrefix = {arXiv},
       eprint = {2203.04826},
 primaryClass = {astro-ph.GA},
       adsurl = {https://ui.adsabs.harvard.edu/abs/2022ApJ...929...90L}
}

@article{Ma:2015ota,
    author = "Ma, Xiangcheng and Hopkins, Philip F. and Faucher-Giguere, Claude-Andr\'e and Zolman, Nick and Muratov, Alexander L. and Keres, Dusan and Quataert, Eliot",
    title = "{The origin and evolution of the galaxy mass\textendash{}metallicity relation}",
    eprint = "1504.02097",
    archivePrefix = "arXiv",
    primaryClass = "astro-ph.GA",
    doi = "10.1093/mnras/stv2659",
    journal = "Mon. Not. Roy. Astron. Soc.",
    volume = "456",
    number = "2",
    pages = "2140--2156",
    year = "2016"
}

@article{Madau:2014bja,
    author = "Madau, Piero and Dickinson, Mark",
    title = "{Cosmic Star Formation History}",
    eprint = "1403.0007",
    archivePrefix = "arXiv",
    primaryClass = "astro-ph.CO",
    doi = "10.1146/annurev-astro-081811-125615",
    journal = "Ann. Rev. Astron. Astrophys.",
    volume = "52",
    pages = "415--486",
    year = "2014"
}

@ARTICLE{Madau99,
       author = {{Madau}, Piero and {Haardt}, Francesco and {Rees}, Martin J.},
        title = "{Radiative Transfer in a Clumpy Universe. III. The Nature of Cosmological Ionizing Sources}",
      journal = {\apj},
         year = 1999,
        month = apr,
       volume = {514},
       number = {2},
        pages = {648-659},
          doi = {10.1086/306975},
archivePrefix = {arXiv},
       eprint = {astro-ph/9809058},
 primaryClass = {astro-ph},
       adsurl = {https://ui.adsabs.harvard.edu/abs/1999ApJ...514..648M}
}

@ARTICLE{Madau97,
       author = {{Madau}, Piero and {Meiksin}, Avery and {Rees}, Martin J.},
        title = "{21 Centimeter Tomography of the Intergalactic Medium at High Redshift}",
      journal = {\apj},
         year = 1997,
        month = feb,
       volume = {475},
       number = {2},
        pages = {429-444},
          doi = {10.1086/303549},
archivePrefix = {arXiv},
       eprint = {astro-ph/9608010},
 primaryClass = {astro-ph},
       adsurl = {https://ui.adsabs.harvard.edu/abs/1997ApJ...475..429M}
}

@article{Majumdar:2017tdm,
    author = "Majumdar, Suman and Pritchard, Jonathan R. and Mondal, Rajesh and Watkinson, Catherine A. and Bharadwaj, Somnath and Mellema, Garrelt",
    title = "{Quantifying the non-Gaussianity in the EoR 21-cm signal through bispectrum}",
    eprint = "1708.08458",
    archivePrefix = "arXiv",
    primaryClass = "astro-ph.CO",
    doi = "10.1093/mnras/sty535",
    journal = "Mon. Not. Roy. Astron. Soc.",
    volume = "476",
    number = "3",
    pages = "4007--4024",
    year = "2018"
}

@ARTICLE{Malloy2013,
       author = {{Malloy}, Matthew and {Lidz}, Adam},
        title = "{Identifying Ionized Regions in Noisy Redshifted 21 cm Data Sets}",
      journal = {\apj},
         year = 2013,
        month = apr,
       volume = {767},
       number = {1},
          eid = {68},
        pages = {68},
          doi = {10.1088/0004-637X/767/1/68},
archivePrefix = {arXiv},
       eprint = {1212.2656},
 primaryClass = {astro-ph.CO},
       adsurl = {https://ui.adsabs.harvard.edu/abs/2013ApJ...767...68M}
}

@article{Maniyar:2023cuj,
    author = "Maniyar, Abhishek S. and Gkogkou, Athanasia and Coulton, William R. and Li, Zack and Lagache, Guilaine and Pullen, Anthony R.",
    title = "{Extragalactic CO emission lines in the CMB experiments: a forgotten signal and a foreground}",
    eprint = "2301.10764",
    archivePrefix = "arXiv",
    primaryClass = "astro-ph.CO",
    month = "1",
    year = "2023"
}

@article{Maniyar:2021arp,
    author = "Maniyar, Abhishek S. and Schaan, Emmanuel and Pullen, Anthony R.",
    title = "{New probe of the high-redshift Universe: Nulling CMB lensing with interloper-free line intensity mapping pair lensing}",
    eprint = "2106.09005",
    archivePrefix = "arXiv",
    primaryClass = "astro-ph.CO",
    doi = "10.1103/PhysRevD.105.083509",
    journal = "Phys. Rev. D",
    volume = "105",
    number = "8",
    pages = "083509",
    year = "2022"
}

@ARTICLE{Mao2012,
       author = {{Mao}, Yi and {Shapiro}, Paul R. and {Mellema}, Garrelt and {Iliev}, Ilian T. and {Koda}, Jun and {Ahn}, Kyungjin},
        title = "{Redshift-space distortion of the 21-cm background from the epoch of reionization - I. Methodology re-examined}",
      journal = {\mnras},
         year = 2012,
        month = may,
       volume = {422},
       number = {2},
        pages = {926-954},
          doi = {10.1111/j.1365-2966.2012.20471.x},
archivePrefix = {arXiv},
       eprint = {1104.2094},
 primaryClass = {astro-ph.CO},
       adsurl = {https://ui.adsabs.harvard.edu/abs/2012MNRAS.422..926M}
}

@article{Marsh:2015xka,
    author = "Marsh, David J. E.",
    title = "{Axion Cosmology}",
    eprint = "1510.07633",
    archivePrefix = "arXiv",
    primaryClass = "astro-ph.CO",
    reportNumber = "KCL-PH-TH-2015-50",
    doi = "10.1016/j.physrep.2016.06.005",
    journal = "Phys. Rept.",
    volume = "643",
    pages = "1--79",
    year = "2016"
}

@article{Mas-Ribas:2020wkz,
    author = "Mas-Ribas, Llu\'\i{}s and Chang, Tzu-Ching",
    title = "{Lyman-$\alpha$ Polarization Intensity Mapping}",
    eprint = "2002.04107",
    archivePrefix = "arXiv",
    primaryClass = "astro-ph.GA",
    doi = "10.1103/PhysRevD.101.083032",
    journal = "Phys. Rev. D",
    volume = "101",
    number = "8",
    pages = "083032",
    year = "2020"
}

@ARTICLE{Mas-Ribas2017,
       author = {{Mas-Ribas}, Llu{\'\i}s and {Dijkstra}, Mark and {Hennawi}, Joseph F. and {Trenti}, Michele and {Momose}, Rieko and {Ouchi}, Masami},
        title = "{Small-scale Intensity Mapping: Extended Ly{\ensuremath{\alpha}}, H{\ensuremath{\alpha}}, and Continuum Emission as a Probe of Halo Star Formation in High-redshift Galaxies}",
      journal = {\apj},
         year = 2017,
        month = may,
       volume = {841},
       number = {1},
          eid = {19},
        pages = {19},
          doi = {10.3847/1538-4357/aa704e},
archivePrefix = {arXiv},
       eprint = {1703.02593},
 primaryClass = {astro-ph.GA},
       adsurl = {https://ui.adsabs.harvard.edu/abs/2017ApJ...841...19M}
}

@article{Mashian:2016bry,
    author = "Mashian, Natalie and Loeb, Avi and Sternberg, Amiel",
    title = "{Spectral Distortion of the CMB by the Cumulative CO Emission from Galaxies throughout Cosmic History}",
    eprint = "1601.02618",
    archivePrefix = "arXiv",
    primaryClass = "astro-ph.CO",
    doi = "10.1093/mnrasl/slw027",
    journal = "Mon. Not. Roy. Astron. Soc.",
    volume = "458",
    number = "1",
    pages = "L99--L103",
    year = "2016"
}

@ARTICLE{Mashian16,
   author = {{Mashian}, N. and {Oesch}, P.~A. and {Loeb}, A.},
    title = "{An empirical model for the galaxy luminosity and star formation rate function at high redshift}",
  journal = {\mnras},
archivePrefix = "arXiv",
   eprint = {1507.00999},
     year = 2016,
    month = jan,
   volume = 455,
    pages = {2101-2109},
      doi = {10.1093/mnras/stv2469},
   adsurl = {http://adsabs.harvard.edu/abs/2016MNRAS.455.2101M}
}

@ARTICLE{Mashian15,
   author = {{Mashian}, N. and {Sternberg}, A. and {Loeb}, A.},
    title = "{Predicting the intensity mapping signal for multi-J CO lines}",
  journal = "JCAP",
archivePrefix = "arXiv",
   eprint = {1507.02686},
     year = 2015,
    month = nov,
   volume = 11,
      eid = {028},
    pages = {028},
      doi = {10.1088/1475-7516/2015/11/028},
   adsurl = {http://adsabs.harvard.edu/abs/2015JCAP...11..028M}
}

@ARTICLE{Masui13,
   author = {{Masui}, K.~W. and {Switzer}, E.~R. and {Banavar}, N. and {Bandura}, K. and 
	{Blake}, C. and {Calin}, L.-M. and {Chang}, T.-C. and {Chen}, X. and 
	{Li}, Y.-C. and {Liao}, Y.-W. and {Natarajan}, A. and {Pen}, U.-L. and 
	{Peterson}, J.~B. and {Shaw}, J.~R. and {Voytek}, T.~C.},
    title = "{Measurement of 21 cm Brightness Fluctuations at z \~{} 0.8 in Cross-correlation}",
  journal = {\apjl},
archivePrefix = "arXiv",
   eprint = {1208.0331},
 primaryClass = "astro-ph.CO",
     year = 2013,
    month = jan,
   volume = 763,
      eid = {L20},
    pages = {L20},
      doi = {10.1088/2041-8205/763/1/L20},
   adsurl = {http://adsabs.harvard.edu/abs/2013ApJ...763L..20M}
}

@article{McBride:2023exl,
    author = "McBride, Lisa and Liu, Adrian",
    title = "{A statistical framework for recovering intensity mapping autocorrelations from crosscorrelations}",
    eprint = "2308.00749",
    archivePrefix = "arXiv",
    primaryClass = "astro-ph.CO",
    month = "8",
    year = "2023"
}

@article{McDonald:2008sh,
    author = "McDonald, Patrick and Seljak, Uros",
    title = "{How to measure redshift-space distortions without sample variance}",
    eprint = "0810.0323",
    archivePrefix = "arXiv",
    primaryClass = "astro-ph",
    doi = "10.1088/1475-7516/2009/10/007",
    journal = "JCAP",
    volume = "10",
    pages = "007",
    year = "2009"
}

@article{Mellema:2006pd,
    author = "Mellema, Garrelt and Iliev, Ilian T. and Pen, Ue-Li and Shapiro, Paul R.",
    title = "{Simulating cosmic reionization at large scales. 2. the 21-cm emission features and statistical signals}",
    eprint = "astro-ph/0603518",
    archivePrefix = "arXiv",
    doi = "10.1111/j.1365-2966.2006.10919.x",
    journal = "Mon. Not. Roy. Astron. Soc.",
    volume = "372",
    pages = "679--692",
    year = "2006"
}

@ARTICLE{Mertens2025,
       author = {{Mertens}, F.~G. and {Mevius}, M. and {Koopmans}, L.~V.~E. and {Offringa}, A.~R. and {Zaroubi}, S. and {Acharya}, A. and {Brackenhoff}, S.~A. and {Ceccotti}, E. and {Chapman}, E. and {Chege}, K. and {Ciardi}, B. and {Ghara}, R. and {Ghosh}, S. and {Giri}, S.~K. and {Hothi}, I. and {H{\"o}fer}, C. and {Iliev}, I.~T. and {Jeli{\'c}}, V. and {Ma}, Q. and {Mellema}, G. and {Munshi}, S. and {Pandey}, V.~N. and {Yatawatta}, S.},
        title = "{Deeper multi-redshift upper limits on the epoch of reionisation 21 cm signal power spectrum from LOFAR between z = 8.3 and z = 10.1}",
      journal = {\aap},
         year = 2025,
        month = jun,
       volume = {698},
          eid = {A186},
        pages = {A186},
          doi = {10.1051/0004-6361/202554158},
archivePrefix = {arXiv},
       eprint = {2503.05576},
 primaryClass = {astro-ph.CO},
       adsurl = {https://ui.adsabs.harvard.edu/abs/2025A&A...698A.186M}
}

@ARTICLE{Mertens2020,
       author = {{Mertens}, F.~G. and {Mevius}, M. and {Koopmans}, L.~V.~E. and {Offringa}, A.~R. and {Mellema}, G. and {Zaroubi}, S. and {Brentjens}, M.~A. and {Gan}, H. and {Gehlot}, B.~K. and {Pandey}, V.~N. and {Sardarabadi}, A.~M. and {Vedantham}, H.~K. and {Yatawatta}, S. and {Asad}, K.~M.~B. and {Ciardi}, B. and {Chapman}, E. and {Gazagnes}, S. and {Ghara}, R. and {Ghosh}, A. and {Giri}, S.~K. and {Iliev}, I.~T. and {Jeli{\'c}}, V. and {Kooistra}, R. and {Mondal}, R. and {Schaye}, J. and {Silva}, M.~B.},
        title = "{Improved upper limits on the 21 cm signal power spectrum of neutral hydrogen at z {\ensuremath{\approx}} 9.1 from LOFAR}",
      journal = {\mnras},
         year = 2020,
        month = apr,
       volume = {493},
       number = {2},
        pages = {1662-1685},
          doi = {10.1093/mnras/staa327},
archivePrefix = {arXiv},
       eprint = {2002.07196},
 primaryClass = {astro-ph.CO},
       adsurl = {https://ui.adsabs.harvard.edu/abs/2020MNRAS.493.1662M}
}

@ARTICLE{Mesinger11,
   author = {{Mesinger}, A. and {Furlanetto}, S. and {Cen}, R.},
    title = "{21CMFAST: a fast, seminumerical simulation of the high-redshift 21-cm signal}",
  journal = {\mnras},
archivePrefix = "arXiv",
   eprint = {1003.3878},
     year = 2011,
    month = feb,
   volume = 411,
    pages = {955-972},
      doi = {10.1111/j.1365-2966.2010.17731.x},
   adsurl = {http://adsabs.harvard.edu/abs/2011MNRAS.411..955M}
}

@ARTICLE{Momose14,
       author = {{Momose}, Rieko and {Ouchi}, Masami and {Nakajima}, Kimihiko and {Ono}, Yoshiaki and {Shibuya}, Takatoshi and {Shimasaku}, Kazuhiro and {Yuma}, Suraphong and {Mori}, Masao and {Umemura}, Masayuki},
        title = "{Diffuse Ly{\ensuremath{\alpha}} haloes around galaxies at z = 2.2-6.6: implications for galaxy formation and cosmic reionization}",
      journal = {\mnras},
         year = 2014,
        month = jul,
       volume = {442},
       number = {1},
        pages = {110-120},
          doi = {10.1093/mnras/stu825},
archivePrefix = {arXiv},
       eprint = {1403.0732},
 primaryClass = {astro-ph.CO},
       adsurl = {https://ui.adsabs.harvard.edu/abs/2014MNRAS.442..110M}
}

@article{Mondal:2015oga,
    author = "Mondal, Rajesh and Bharadwaj, Somnath and Majumdar, Suman",
    title = "{Statistics of the epoch of reionization 21-cm signal \textendash{} I. Power spectrum error-covariance}",
    eprint = "1508.00896",
    archivePrefix = "arXiv",
    primaryClass = "astro-ph.CO",
    doi = "10.1093/mnras/stv2772",
    journal = "Mon. Not. Roy. Astron. Soc.",
    volume = "456",
    number = "2",
    pages = "1936--1947",
    year = "2016"
}

@article{Moodley:2023lmu,
    author = "Moodley, Kavilan and Naidoo, Warren and Prince, Heather and Penin, Aurelie",
    title = "{A cross-bispectrum estimator for CMB-HI intensity mapping correlations}",
    eprint = "2311.05904",
    archivePrefix = "arXiv",
    primaryClass = "astro-ph.CO",
    month = "11",
    year = "2023"
}

@ARTICLE{Moriwaki18,
       author = {{Moriwaki}, Kana and {Yoshida}, Naoki and {Shimizu}, Ikkoh and {Harikane}, Yuichi and {Matsuda}, Yuichi and {Matsuo}, Hiroshi and {Hashimoto}, Takuya and {Inoue}, Akio K. and {Tamura}, Yoichi and {Nagao}, Tohru},
        title = "{The distribution and physical properties of high-redshift [O III] emitters in a cosmological hydrodynamics simulation}",
      journal = {\mnras},
         year = 2018,
        month = nov,
       volume = {481},
       number = {1},
        pages = {L84-L88},
          doi = {10.1093/mnrasl/sly167},
archivePrefix = {arXiv},
       eprint = {1805.07062},
 primaryClass = {astro-ph.GA},
       adsurl = {https://ui.adsabs.harvard.edu/abs/2018MNRAS.481L..84M}
}

@article{Munoz:2021psm,
    author = "Mu\~noz, Julian B. and Qin, Yuxiang and Mesinger, Andrei and Murray, Steven G. and Greig, Bradley and Mason, Charlotte",
    title = "{The Impact of the First Galaxies on Cosmic Dawn and Reionization}",
    eprint = "2110.13919",
    archivePrefix = "arXiv",
    primaryClass = "astro-ph.CO",
    doi = "10.1093/mnras/stac185",
    month = "10",
    year = "2021"
}

@article{Munoz:2019hjh,
    author = "Mu\~noz, Julian B. and Dvorkin, Cora and Cyr-Racine, Francis-Yan",
    title = "{Probing the Small-Scale Matter Power Spectrum with Large-Scale 21-cm Data}",
    eprint = "1911.11144",
    archivePrefix = "arXiv",
    primaryClass = "astro-ph.CO",
    doi = "10.1103/PhysRevD.101.063526",
    journal = "Phys. Rev. D",
    volume = "101",
    number = "6",
    pages = "063526",
    year = "2020"
}

@article{Munoz:2019fkt,
    author = "Mu\~noz, Julian B.",
    title = "{Standard Ruler at Cosmic Dawn}",
    eprint = "1904.07868",
    archivePrefix = "arXiv",
    primaryClass = "astro-ph.CO",
    doi = "10.1103/PhysRevLett.123.131301",
    journal = "Phys. Rev. Lett.",
    volume = "123",
    number = "13",
    pages = "131301",
    year = "2019"
}

@article{Munoz:2013tv,
      author         = "Munoz, Joseph A. and Furlanetto, Steven R.",
      title          = "{Molecular Cloud Properties and CO Line Emission in z >~
                        6 Galaxies}",
      journal        = "Mon. Not. Roy. Astron. Soc.",
      volume         = "435",
      year           = "2013",
      pages          = "2676-2692",
      doi            = "10.1093/mnras/stt1480",
      eprint         = "1301.0619",
      archivePrefix  = "arXiv",
      primaryClass   = "astro-ph.CO",
      SLACcitation   = "%%CITATION = ARXIV:1301.0619;%%"
}

@article{Naidu:2021ryj,
    author = "Naidu, Rohan P. and others",
    title = "{The synchrony of production and escape: half the bright Ly\ensuremath{\alpha} emitters at z~\ensuremath{\approx} 2 have Lyman continuum escape fractions \ensuremath{\approx}50}",
    eprint = "2110.11961",
    archivePrefix = "arXiv",
    primaryClass = "astro-ph.GA",
    doi = "10.1093/mnras/stab3601",
    journal = "Mon. Not. Roy. Astron. Soc.",
    volume = "510",
    number = "3",
    pages = "4582--4607",
    year = "2022"
}

@ARTICLE{Nakazato23,
       author = {{Nakazato}, Yurina and {Yoshida}, Naoki and {Ceverino}, Daniel},
        title = "{Simulations of High-redshift [O III] Emitters: Chemical Evolution and Multiline Diagnostics}",
      journal = {\apj},
         year = 2023,
        month = aug,
       volume = {953},
       number = {2},
          eid = {140},
        pages = {140},
          doi = {10.3847/1538-4357/ace25a},
archivePrefix = {arXiv},
       eprint = {2301.02416},
 primaryClass = {astro-ph.GA},
       adsurl = {https://ui.adsabs.harvard.edu/abs/2023ApJ...953..140N}
}

@ARTICLE{Navarro97,
       author = {{Navarro}, Julio F. and {Frenk}, Carlos S. and {White}, Simon D.~M.},
        title = "{A Universal Density Profile from Hierarchical Clustering}",
      journal = {\apj},
         year = 1997,
        month = dec,
       volume = {490},
       number = {2},
        pages = {493-508},
          doi = {10.1086/304888},
archivePrefix = {arXiv},
       eprint = {astro-ph/9611107},
 primaryClass = {astro-ph},
       adsurl = {https://ui.adsabs.harvard.edu/abs/1997ApJ...490..493N}
}

@ARTICLE{Neufeld91,
       author = {{Neufeld}, David A.},
        title = "{The Escape of Lyman-Alpha Radiation from a Multiphase Interstellar Medium}",
      journal = {\apjl},
         year = 1991,
        month = apr,
       volume = {370},
        pages = {L85},
          doi = {10.1086/185983},
       adsurl = {https://ui.adsabs.harvard.edu/abs/1991ApJ...370L..85N}
}

@ARTICLE{Nunhokee2025,
       author = {{Nunhokee}, C.~D. and {Null}, D. and {Trott}, C.~M. and {Barry}, N. and {Qin}, Y. and {Wayth}, R.~B. and {Line}, J.~L.~B. and {Jordan}, C.~H. and {Pindor}, B. and {Cook}, J.~H. and {Bowman}, J. and {Chokshi}, A. and {Ducharme}, J. and {Elder}, K. and {Guo}, Q. and {Hazelton}, B. and {Hidayat}, W. and {Ito}, T. and {Jacobs}, D. and {Jong}, E. and {Kolopanis}, M. and {Kunicki}, T. and {Lilleskov}, E. and {Morales}, M.~F. and {Pober}, J.~C. and {Selvaraj}, A. and {Shi}, R. and {Takahashi}, K. and {Tingay}, S.~J. and {Webster}, R.~L. and {Yoshiura}, S. and {Zheng}, Q.},
        title = "{Limits on the 21 cm Power Spectrum at z = 6.5{\textendash}7.0 from Murchison Widefield Array Observations}",
      journal = {\apj},
         year = 2025,
        month = aug,
       volume = {989},
       number = {1},
          eid = {57},
        pages = {57},
          doi = {10.3847/1538-4357/adda45},
archivePrefix = {arXiv},
       eprint = {2505.09097},
 primaryClass = {astro-ph.CO},
       adsurl = {https://ui.adsabs.harvard.edu/abs/2025ApJ...989...57N}
}

@ARTICLE{Obreschkow09,
       author = {{Obreschkow}, D. and {Heywood}, I. and {Kl{\"o}ckner}, H. -R. and {Rawlings}, S.},
        title = "{A Heuristic Prediction of the Cosmic Evolution of the Co-luminosity Functions}",
      journal = {\apj},
         year = 2009,
        month = sep,
       volume = {702},
       number = {2},
        pages = {1321-1335},
          doi = {10.1088/0004-637X/702/2/1321},
archivePrefix = {arXiv},
       eprint = {0907.3091},
 primaryClass = {astro-ph.CO},
       adsurl = {https://ui.adsabs.harvard.edu/abs/2009ApJ...702.1321O}
}

@article{Oh:2000sg,
    author = "Oh, S. Peng and Haiman, Zoltan and Rees, Martin J.",
    title = "{HeII recombination lines from the first luminous objects}",
    eprint = "astro-ph/0007351",
    archivePrefix = "arXiv",
    doi = "10.1086/320650",
    journal = "Astrophys. J.",
    volume = "553",
    pages = "73",
    year = "2001"
}

@ARTICLE{Olsen17,
       author = {{Olsen}, Karen and {Greve}, Thomas R. and {Narayanan}, Desika and {Thompson}, Robert and {Dav{\'e}}, Romeel and {Niebla Rios}, Luis and {Stawinski}, Stephanie},
        title = "{S{\'I}GAME Simulations of the [CII], [OI], and [OIII] Line Emission from Star-forming Galaxies at z $\simeq$ 6}",
      journal = {\apj},
         year = 2017,
        month = sep,
       volume = {846},
       number = {2},
          eid = {105},
        pages = {105},
          doi = {10.3847/1538-4357/aa86b4},
archivePrefix = {arXiv},
       eprint = {1708.04936},
 primaryClass = {astro-ph.GA},
       adsurl = {https://ui.adsabs.harvard.edu/abs/2017ApJ...846..105O}
}

@ARTICLE{Olsen21,
       author = {{Olsen}, Karen Pardos and {Burkhart}, Blakesley and {Mac Low}, Mordecai-Mark and {Tre{\ss}}, Robin G. and {Greve}, Thomas R. and {Vizgan}, David and {Motka}, Jay and {Borrow}, Josh and {Popping}, Gerg{\"o} and {Dav{\'e}}, Romeel and {Smith}, Rowan J. and {Narayanan}, Desika},
        title = "{S{\'I}GAME v3: Gas Fragmentation in Postprocessing of Cosmological Simulations for More Accurate Infrared Line Emission Modeling}",
      journal = {\apj},
         year = 2021,
        month = nov,
       volume = {922},
       number = {1},
          eid = {88},
        pages = {88},
          doi = {10.3847/1538-4357/ac20d4},
archivePrefix = {arXiv},
       eprint = {2102.02868},
 primaryClass = {astro-ph.GA},
       adsurl = {https://ui.adsabs.harvard.edu/abs/2021ApJ...922...88O}
}

@ARTICLE{Osterbrock62,
       author = {{Osterbrock}, Donald E.},
        title = "{The Escape of Resonance-Line Radiation from an Optically Thick Nebula.}",
      journal = {\apj},
         year = 1962,
        month = jan,
       volume = {135},
        pages = {195},
          doi = {10.1086/147258},
       adsurl = {https://ui.adsabs.harvard.edu/abs/1962ApJ...135..195O}
}




\end{document}